\documentclass[aps, prd, a4paper, superscriptaddress, onecolumn, amsmath, preprintnumbers, nofootinbib, floatfix]{revtex4-1}

\pdfoutput=1
\usepackage{graphicx,subfigure,bm,color,psfrag}
\usepackage{amsfonts}
\usepackage{mathtools}
\usepackage[arrowdel]{physics}
\usepackage{xspace}
\usepackage[dvipsnames]{xcolor}
\usepackage[pscoord]{eso-pic}
\usepackage[normalem]{ulem}
\usepackage[colorlinks = true,
            linkcolor = blue,
            urlcolor  = blue,
           citecolor = blue,
            anchorcolor = blue]{hyperref}
\usepackage{amssymb, amsmath, amsfonts}
\usepackage{cancel} 
\usepackage{siunitx}

\makeatletter
\def\label@firstpage{\label{firstpage}}
\makeatother

\definecolor{lucacol}{rgb}{0,0.75, 1.}

\definecolor{valecol}{rgb}{0,0.5, 1.}

\definecolor{brickred}{rgb}{0.8, 0.25, 0.33}

\definecolor{cbcol}{rgb}{0.6,0.9, 0.4}

\definecolor{rcbcol}{rgb}{0.7,0.3,0}

\definecolor{darkspringgreen}{rgb}{0.09, 0.45, 0.27}

\def\nbox#1#2{\vcenter{\hrule \hbox{\vrule height#2in
\kern#1in \vrule} \hrule}}
\def\sq{\,\raise.5pt\hbox{$\nbox{.09}{.09}$}\,}

\begin{document}

\begin{center}
  \rule[-0.2in]{\hsize}{0.01in}\\
  \rule{\hsize}{0.01in}\\
  \vskip 0.1in
  Submitted to the Proceedings of the US Community Study\\ 
  on the Future of Particle Physics (Snowmass 2021)\\
  \vskip 0.1in
  \rule{\hsize}{0.01in}\\
  \rule[+0.2in]{\hsize}{0.01in}\\[-2em]
\end{center}

\title{Cosmology Intertwined:\\ A Review of the Particle Physics, Astrophysics, and Cosmology\\ Associated with the Cosmological Tensions and Anomalies}

\let\mymaketitle\maketitle
\let\myauthor\author
\let\myaffiliation\affiliation
\author{Elcio Abdalla}
\affiliation{Instituto de F\'isica, Universidade de S\~ao Paulo - C.P. 66318, CEP: 05315-970, S\~ao Paulo, Brazil}

\author{Guillermo Franco Abell\'an}
\affiliation{
 Laboratoire Univers \& Particules de Montpellier (LUPM), Universit\'e de Montpellier (UMR-5299) \\ Place Eug{\`e}ne Bataillon, F-34095 Montpellier Cedex 05, France 
}

\author{Amin Aboubrahim}
\affiliation{Institut f\"ur Theoretische Physik, Westf\"alische Wilhelms-Universit\"at M\"unster, Wilhelm-Klemm-Stra{\ss}e 9, 48149 M\"unster, Germany}

\author{Adriano Agnello}
\affiliation{DARK, Niels Bohr Institute, University of Copenhagen, Jagtvej 128, 2200 Copenhagen, Denmark}

\author{\"{O}zg\"{u}r Akarsu}
\affiliation{Department of Physics, Istanbul Technical University, Maslak 34469 Istanbul, Turkey}

\author{Yashar Akrami}
\affiliation{CERCA/ISO, Department of Physics, Case Western Reserve University, 10900 Euclid Avenue, Cleveland, OH 44106, USA}
\affiliation{Astrophysics Group \& Imperial Centre for Inference and Cosmology,
Department of Physics, Imperial College London, Blackett Laboratory,
Prince Consort Road, London SW7 2AZ, United Kingdom}
\affiliation{Laboratoire de Physique de l'\'Ecole Normale Sup\'erieure, ENS, Universit\'e PSL, CNRS, Sorbonne Universit\'e, Universit\'e de Paris, F-75005 Paris, France}
\affiliation{Observatoire de Paris, Universit\'e PSL, Sorbonne Universit\'e, LERMA, 75014 Paris, France}

\author{George Alestas}
\affiliation{Department of Physics, University of Ioannina, GR-45110, Ioannina, Greece}

\author{Daniel Aloni}
\affiliation{Physics Department, Boston University, Boston, MA 02215, USA}

\author{Luca Amendola}
\affiliation{Institute of Theoretical Physics, Heidelberg University,
Philosophenweg 16, 69120 Heidelberg, Germany}

\author{Luis A. Anchordoqui}
\affiliation{Department of Physics and Astronomy,  Lehman College, City University of New York, NY 10468, USA}
\affiliation{Department of Physics, Graduate Center, City University of New York,  NY 10016, USA}
\affiliation{Department of Astrophysics, American Museum of Natural History, NY 10024, USA}

% ...

\begin{abstract}
The standard $\Lambda$ Cold Dark Matter ($\Lambda$CDM) cosmological model provides a good description of a wide range of astrophysical and cosmological data. However, there are a few big open questions that make the standard model look like an approximation to a more realistic scenario yet to be found. In this paper, we list a few important goals that need to be addressed in the next decade, taking into account the current discordances between the different cosmological probes, such as the disagreement in the value of the Hubble constant $H_0$, the $\sigma_8$--$S_8$ tension, and other less statistically significant anomalies. 
While these discordances can still be in part the result of systematic errors, their persistence after several years of accurate analysis strongly hints at cracks in the standard cosmological scenario and the necessity for new physics or generalisations beyond the standard model. 
In this paper, we focus on the $5.0\,\sigma$ tension between the {\it Planck} CMB estimate of the Hubble constant $H_0$ and the SH0ES collaboration measurements. After showing the $H_0$ evaluations made from different teams using different methods and geometric calibrations, we list a few interesting new physics models that could alleviate this tension and discuss how the next decade's experiments will be crucial. Moreover, we focus on the tension of the {\it Planck} CMB data with weak lensing measurements and redshift surveys, about the value of the matter energy density $\Omega_m$, and the amplitude or rate of the growth of structure ($\sigma_8,f\sigma_8$). We list a few interesting models proposed for alleviating this tension, and we discuss the importance of trying to fit a full array of data with a single model and not just one parameter at a time. Additionally, we present a wide range of other less discussed anomalies at a statistical significance level lower than the $H_0$--$S_8$ tensions which may also constitute hints towards new physics, and we discuss possible generic theoretical approaches that can collectively explain the non-standard nature of these signals.
Finally, we give an overview of upgraded experiments and next-generation space missions and facilities on Earth that will be of crucial importance to address all these open questions.
\end{abstract}

\maketitle
\newpage

\tableofcontents


\newpage

\section*{Conventions}

\begin{table*}[ht]
\begin{center} 
\begin{tabular}{|c|l|}
\hline
Definition & Meaning \\
\hline
$\hbar=c=k_B=1$ & Natural units\\
$\kappa^2\equiv 8\pi G_N= M_{\rm Pl}^{-2}$ & Gravitational constant\\
$(-\,+\,+\,+)$ & Metric signature\\
  $g_{\mu\nu}$ & Metric tensor\\
   $G_{\mu \nu} \equiv R_{\mu \nu} - \frac{1}{2} g_{\mu \nu} R$ & Einstein tensor\\
$\Lambda$ & Cosmological constant\\
  ${\rm d}s^2 = -{\rm d}t^2 + a^2 (t) \left[ \frac{{\rm d}r^2}{1-k r^2} + r^2  \left(  {\rm d} \theta^2 + \sin^2 \theta {\rm d} \phi^2 \right)\right]$ & 
Friedmann-Lema\^{i}tre-Robertson-Walker (FLRW) spacetime metric\\
$a(t)$ & Scale factor \\
$a_0=1$ & Scale factor today (set to unity) \\
$t$ & Cosmic (proper) time\\
$\tau(t) = \int_{0}^{t}\frac{{\rm d}t'}{a(t')}$ & Conformal time\\
$\dot{\;} \equiv \frac{\rm d}{{\rm d}t}$ & Cosmic time derivative\\
$\;^{\prime} \equiv \frac{\rm d}{{\rm d}\tau}$ & Conformal time derivative\\
$T^{\mu \nu} = \frac{2}{\sqrt{-g}} \frac{\delta \mathcal{L}_m}{\delta
  g_{\mu \nu}}$ & Energy-momentum tensor of the Lagrangian density $\mathcal{L}$\\
$z=-1+\frac{1}{a}$ & Cosmological redshift\\
$H(z)=\frac{\dot a}{a}$ & Hubble parameter\\
$H_0$ & Hubble constant\\
$h\equiv H_0/100\, {\rm km\,s}^{-1}{\rm Mpc}^{-1}$ & Dimensionless reduced Hubble constant \\
$\rho_m$, $\rho_b$, $\rho_r$ & Energy density of total matter, baryonic matter, and radiation\\
$\rho_{\rm DM}$, $\rho_{\rm DE}$ & Energy density of dark matter and dark energy \\
$\Omega_m$ & Present-day matter density parameter \\
$\Omega_r=2.469\times10^{-5}h^{-2}(1+0.2271N_{\rm eff})$ & Present-day radiation density parameter \\
$\Omega_{\rm DM}$, $\Omega_{\rm DE}$ & Present-day density parameters of dark matter and dark energy\\
$\Omega_{\rm CDM}$ & Present-day density parameters of cold dark matter\\
$\Omega_m(z)=\frac{\kappa^2\rho_m}{3H^2}$ & Matter density parameter \\
$\Omega_r(z)=\frac{\kappa^2\rho_r}{3H^2}$ & Relativistic content density parameter \\
$\Omega_{\rm DE}(z)=\frac{\kappa^2\rho_{\rm DE}}{3H^2}$ & Dark energy density parameter \\
$w\equiv\frac{p}{\rho}$ & Equation of state (EoS) parameter \\
$c_s$ & Sound speed\\
$r_s\equiv\int_{0}^{\tau}c_s(\tau') {\rm d}\tau'$ & Sound horizon\\
$r_d \equiv r_s(\tau_d)$ & Sound horizon at drag epoch\\
$\sigma_8$ & Amplitude of mass fluctuations on scales of $8\, h^{-1}$ Mpc \\
$S_8=\sigma_8 \sqrt{\Omega_m/0.3}$ & Weighted amplitude of matter fluctuations \\
\hline
\end{tabular}
\label{tab:notation}
\caption{List of notation conventions used in this paper (unless otherwise stated).}
\end{center}
\end{table*}


\newpage 

\section*{Terminology}
\label{sec:terminology}

\begin{table*}[ht]
\begin{center}
\begin{tabular}{|c|l|||c|l|}
\hline
Acronym & Definition & Acronym & Definition \\
\hline
$\Lambda$CDM & $\Lambda$ (Cosmological Constant)-Cold Dark Matter & KSZ & Kinetic Sunyaev-Zel'dovich\\
$\Lambda$LTB & $\Lambda$-Lema\^{i}tre-Tolman-Bondi & LHC & Large Hadron Collider\\
AP & Alcock–Paczynsk & LMC & Large Magellanic Cloud\\
ATLAS & A Toroidal LHC ApparatuS & LoI & Letter Of Interest\\
BAO & Baryon Acoustic Oscillations & LSS & Large Scale Structure \\
BBN & Big Bang Nucleosynthesis& LSST & Legacy Survey of Space and Time \\
BSM & Beyond Standard Model& MGS & Main Galaxy Sample  \\
bTFR & baryonic Tully-Fisher &MSTOP & main-sequence turn off point \\
CC & Cosmic Chronometers & NGC & New General Catalog\\
CDM & Cold Dark Matter &NIR & Near infrared \\
CL & Confidence Level & PMF & Primordial Magnetic Fields  \\
CMB & Cosmic Microwave Background Radiation & PPS &  Primordial Power Spectrum \\
DDE & Dynamical Dark Energy &  QFT & Quantum Field Theory \\
DDM & Dynamical Dark Matter & QSO & Quasi-stellar object\\
DE & Dark Energy & RSD & Redshift-Space Distorsion \\
DM & Dark Matter & RVM & Running Vacuum Models  \\
DR & Data Release &  SBF & Surface Brightness Fluctuations \\
EDE & Early Dark Energy & SIDR & Strongly Interacting Dark Radiation\\
EDR & Early Data Release & SD & Spectral Distortion\\
EoS & Equation of State & SM & Standard Model \\
FLRW & Friedmann-Lema\^{i}tre-Robertson-Walker & SN & Supernovae \\
FS & Full shape & sta & Statistical \\
gDE & Graduated Dark Energy & sys & Systematical \\
GDR & Gaia Data Release & TG & Teleparallel Gravity \\
GR & General Relativity &  TRGB & Tip of the Red Giant Branch\\
GRB & Gamma Ray Burst &UVES & Ultraviolet and Visual Echelle Spectrograph \\
IDE & Interacting Dark Energy& VED & Vacuum Energy Density \\
GW & Gravitational Wave & WL & Weak Lensing \\
 && ZP & Zero-point  \\
\hline
\end{tabular}
\label{tab:notation2}
\caption{List of the acronyms/terminology used in the paper.}
\end{center}
\end{table*}


\newpage 

\section*{Acronyms and References of Experiments/Missions}
\label{sec:acronyms}

\begin{table*}[ht]
\begin{center}
\scalebox{0.6}{
\begin{tabular}{|c|l|l|l|}
\hline
Acronym & Experiment & Website & Status \\
\hline
4MOST &  4-metre Multi-Object Spectroscopic Telescope & \href{https://www.eso.org/sci/facilities/develop/instruments/4MOST.html}{https://4MOST} & expected 2023\\
ACT & Atacama Cosmology Telescope &
\href{https://act.princeton.edu}{https://act.princeton.edu} & ongoing \\
ANDES & ArmazoNes high Dispersion Echelle Spectrograph & \href{https://elt.eso.org/instrument/ANDES/}{ https://ANDES} & planned\\
ATLAS Probe & Astrophysics Telescope for Large Area Spectroscopy Probe & \href{https://atlas-probe.ipac.caltech.edu/}{https://atlas-probe} & proposed \\
BAHAMAS & BAryons and HAloes of MAssive Systems & \href{https://www.astro.ljmu.ac.uk/~igm/BAHAMAS}{https://BAHAMAS} & 2017-2018\\
BICEP & Background Imaging of Cosmic Extragalactic Polarization & \href{http://bicepkeck.org/}{http://bicepkeck.org} & ongoing
\\
BINGO & Baryon Acoustic Oscillations & \href{https://bingotelescope.org/}{https://bingotelescope.org} & planned
\\
 & from Integrated Neutral Gas Observations & &
\\
BOSS & Baryon Oscillations Spectroscopy Survey & \href{https://cosmology.lbl.gov/BOSS/}{https://BOSS} & ongoing \\
CANDELS &  Cosmic Assembly Near-infrared Deep  &\href{https://www.ipac.caltech.edu/project/candels}{https://candels}&\\
 &   Extragalactic Legacy Survey  & & \\
CCHP & Carnegie-Chicago Hubble Project & \href{https://carnegiescience.edu/projects/carnegie-hubble-program}{https://carnegiescience.edu}&\\
CE & Cosmic Explorer & \href{https://cosmicexplorer.org}{https://cosmicexplorer.org}& planned\\
CFHT & Canada-France-Hawaii Telescope & \href{https://www.cfht.hawaii.edu}{https://cfht.hawaii.edu}& ongoing\\
CHIME & Canadian Hydrogen Intensity Mapping Experiment & \href{https://chime-experiment.ca/en}{https://chime-experiment.ca} & ongoing \\
CLASS & Cosmology Large Angular Scale Surveyor & \href{https://sites.krieger.jhu.edu/class/}{https://class} & ongoing \\
CMB-HD & Cosmic Microwave Background-High Definition & \href{https://cmb-hd.org}{https://cmb-hd.org} & proposed\\
CMB-S4 & Cosmic Microwave Background-Stage IV & \href{https://cmb-s4.org}{https://cmb-s4.org} & planned 2029-2036\\
COMAP & CO Mapping Array Pathfinder & \href{https://comap.caltech.edu}{https://comap.caltech.edu} & ongoing\\
DECIGO & DECi-hertz Interferometer Gravitational wave Observatory & \href{https://decigo.jp/index_E.html}{https://decigo.jp} & planned \\
DES & Dark Energy Survey & \href{https://www.darkenergysurvey.org}{https://darkenergysurvey.org} & ongoing \\
DESI & Dark Energy Spectroscopic Instrument & \href{https://www.desi.lbl.gov}{https://desi.lbl.gov}& ongoing\\ 
dFGS & 6-degree Field Galaxy Survey & \href{http://www.6dfgs.net}{http://6dfgs.net} & 2001-2007\\
eBOSS & Extended Baryon Oscillations Spectroscopy Survey & \href{https://www.sdss.org/surveys/eboss/}{https://eboss} & 2014-2019\\
ELT & Extremely Large Telescope & \href{https://elt.eso.org}{https://elt.eso.org} & planned 2027 \\
ESPRESSO & Echelle SPectrograph for Rocky Exoplanets & \href{https://www.eso.org/sci/facilities/paranal/instruments/espresso.html}{https://espresso.html} &ongoing \\
& and Stable Spectroscopic Observations & &\\
ET & Einstein Telescope & \href{http://www.et-gw.eu}{http://www.et-gw.eu} & planned \\
{\it Euclid} & {\it Euclid} Consortium & \href{https://www.euclid-ec.org}{https://www.euclid-ec.org} & planned 2023 \\
Gaia& Gaia & 
\href{https://sci.esa.int/web/gaia/}{https://gaia} & ongoing\\
GBT& Green Bank Telescope & 
\href{https://greenbankobservatory.org/science/telescopes/gbt/}{https://greenbankobservatory.org} & ongoing\\
GRAVITY& General Relativity Analysis via VLT InTerferometrY& 
\href{https://www.mpe.mpg.de/ir/gravity}{https://gravity} & ongoing\\
GRAVITY+& upgrade version of GRAVITY& \href{https://www.mpe.mpg.de/ir/gravityplus }{https://gravityplus} & planned\\
HARPS & High Accuracy Radial-velocity Planet Searcher & \href{https://www.eso.org/sci/facilities/lasilla/instruments/harps.html}{https://harps.html} & ongoing\\
HIRAX & Hydrogen Intensity and Real-time Analysis eXperiment & \href{https://hirax.ukzn.ac.za}{https://hirax.ukzn.ac.za} & planned \\
HIRES & HIgh Resolution Echelle Spectrometer &\href{https://www2.keck.hawaii.edu/inst/hires/}{https://hires}&ongoing\\
H0LiCOW & $H_0$ Lenses in Cosmograil's Wellspring & \href{https://shsuyu.github.io/H0LiCOW/site/}{https://H0LiCOW} &\\
HSC & Hyper Suprime-Cam & \href{https://hsc.mtk.nao.ac.jp/ssp/survey}{https://hsc.mtk.nao.ac.jp} & finished\\ 
HST & Hubble Space Telescope & \href{https://www.nasa.gov/mission_pages/hubble}{https://hubble} & ongoing\\ 
KAGRA&Kamioka Gravitational wave detector&\href{https://gwcenter.icrr.u-tokyo.ac.jp/en/organization}{https://kagra}&expected 2023\\
KiDS & Kilo-Degree Survey & \href{http://kids.strw.leidenuniv.nl}{http://kids} & ongoing\\
JWST & James Webb Space Telescope & \href{https://jwst.nasa.gov/content/webbLaunch/index.html}{https://jwst.nasa.gov} & ongoing\\
LIGO & Laser Interferometer Gravitational Wave Observatory & \href{https://www.ligo.caltech.edu}{https://ligo.caltech.edu} & ongoing \\
LIGO-India &Laser Interferometer Gravitational Wave Observatory India&\href{https://www.ligo-india.in}{https://ligo-india.in}& planned\\
LiteBIRD & Lite (Light) satellite for the studies of B-mode polarization  & \href{https://www.isas.jaxa.jp/en/missions/spacecraft/future/litebird.html}{https://litebird.html} & planned \\
& and Inflation from cosmic background Radiation Detection & &  \\
LISA & Laser Interferometer Space Antenna & \href{https://lisa.nasa.gov}{https://lisa.nasa.gov} & planned\\
LGWA & Lunar Gravitational-Wave Antenna &\href{http://socrate.cs.unicam.it/index.php}{http://LGWA} & proposed\\
MCT & CLASH Multi-Cycle Treasury & \href{https://www.stsci.edu/~postman/CLASH/}{https://CLASH} & \\
MeerKAT & Karoo Array Telescope & \href{https://www.sarao.ac.za/science/meerkat/}{https://meerkat} & ongoing\\
NANOGrav & North American Nanohertz Observatory for Gravitational Waves & \href{http://nanograv.org/}{http://nanograv.org/} & ongoing\\
OWFA & Ooty Wide Field Array & \href{http://rac.ncra.tifr.res.in/ort.html}{http://ort.html} & planned\\
OWLS & OverWhelmingly Large Simulations & \href{https://virgo.dur.ac.uk/2010/02/12/OWLS}{https://OWLS} &\\
Pan-STARRS & Panoramic Survey Telescope and Rapid Response System & \href{https://panstarrs.stsci.edu}{https://panstarrs.stsci.edu} & ongoing \\
PFS & Subaru Prime Focus Spectrograph & \href{https://pfs.ipmu.jp}{https://pfs.ipmu.jp} & expected 2023
\\
{\it Planck} & {\it Planck} collaboration & \href{https://www.esa.int/Science_Exploration/Space_Science/Planck}{https://www.esa.int/Planck} & 2009-2013
\\
POLARBEAR & POLARBEAR & \href{http://bolo.berkeley.edu/polarbear/}{http://polarbear} & finished \\
PUMA &Packed Ultra-wideband Mapping Array & \href{http://puma.bnl.gov}{http://puma.bnl.gov} & planned
\\
{\it Roman}/WFIRST & Nancy Grace {\it Roman} Space Telescope & \href{http://roman.gsfc.nasa.gov}{http://roman.gsfc.nasa.gov} & planned\\
{\it Rubin}/LSST & {\it Rubin} Observatory Legacy Survey of Space and Time & \href{https://www.lsst.org}{https://lsst.org} & expected 2024-2034\\
SDSS & Sloan Digital Sky Survey & \href{https://www.sdss.org}{https://sdss.org} & ongoing\\
SH0ES & Supernovae $H_0$ for the Equation of State & \href{https://archive.stsci.edu/proposal_search.php?id=10802\&mission=hst}{https://SH0ES-Supernovae} & \\
SKAO & Square Kilometer Array Observatory & \href{https://www.skatelescope.org}{https://skatelescope.org} & planned\\
Simons Array & Simons Array & \href{http://bolo.berkeley.edu/polarbear/}{http://simonarray} & in preparation\\
SLACS & Sloan Lens ACS & \href{https://web.physics.utah.edu/~bolton/slacs/What\_is\_SLACS.html}{https://SLACS.html} & \\
SO & Simons Observatory & \href{https://simonsobservatory.org}{https://simonsobservatory.org} & expected 2024-2029\\
SPHEREx  & Spectro-Photometer for the History of the Universe, &\href{https://www.jpl.nasa.gov/missions/spherex}{https://spherex} & expected 2025\\
  &  Epoch of Reionization, and Ices Explorer & &\\
SPIDER & SPIDER & \href{https://spider.princeton.edu/}{https://spider} & planned \\
SPT & South Pole Telescope & \href{https://pole.uchicago.edu}{https://pole.uchicago.edu} & ongoing\\
STRIDES & STRong-lensing Insights into Dark Energy Survey & \href{https://strides.astro.ucla.edu}{https://strides.astro.ucla.edu} & ongoing \\
TDCOSMO & Time Delay Cosmography & \href{http://www.tdcosmo.org}{http://tdcosmo.org} & ongoing\\
uGMRT & Upgraded Giant Metre-wave Radio Telescope & \href{https://www.gmrt.ncra.tifr.res.in/}{https://gmrt.ncra.tifr.res.in} & ongoing \\
UNIONS & The Ultraviolet Near- Infrared Optical Northern Survey & \href{https://www.skysurvey.cc}{https://skysurvey.cc} &  \\
UVES & Ultra Violet Echelle Spectrograph & \href{https://www.eso.org/public/teles-instr/paranal-observatory/vlt/vlt-instr/uves/}{https://uves} & ongoing\\
VIKING & VISTA Kilo-degree Infrared Galaxy Survey & \href{http://horus.roe.ac.uk/vsa/}{http://horus.roe.ac.uk/vsa/} & ongoing\\
Virgo & Virgo& \href{https://www.virgo-gw.eu}{https://virgo-gw.eu}& ongoing\\ 
VLA & Very Large Array & \href{https://public.nrao.edu/telescopes/vla/}{https://vla} & ongoing \\
VLBA & Very Long Baseline Array & \href{https://public.nrao.edu/telescopes/vlba/}{https://vlba} & ongoing \\
VLT  &Very Large Telescope & \href{https://www.eso.org/public/teles-instr/paranal-observatory/vlt/}{https://vlt} & ongoing \\
WFC3 & Wide Field Camera 3 & \href{https://www.stsci.edu/hst/instrumentation/wfc3}{https://wfc3} & ongoing \\
WMAP & Wikilson Microwave Anisotropy Probe & \href{https://map.gsfc.nasa.gov}{https://map.gsfc.nasa.gov} & 2001-2010\\
YSE & Young Supernova Experiment & \href{https://yse.ucsc.edu}{https://yse.ucsc.edu} & ongoing \\
ZTF & Zwicky Transient Facility & \href{https://www.ztf.caltech.edu}{https://ztf.caltech.edu} & ongoing \\
\hline
\end{tabular}}
\end{center}
\caption{List of acronyms of the astronomical missions and projects mentioned in the paper. A division of the facilities for science topic is instead reported in Table~\ref{timeline}.}
\label{tab:acronyms}
\end{table*}


\newpage

\section{Executive Summary}

This White Paper has been prepared to fulfill SNOWMASS 2021 requirements and it extends the material previously summarized in the four Letters of Interest~\cite{DiValentino:2020vhf,DiValentino:2020zio,DiValentino:2020vvd,DiValentino:2020srs}. 
The Particle Physics Community Planning Exercise (a.k.a.\ \textit{SNOWMASS}) is organized by the Division of Particles and Fields of the American Physical Society, and this is an effort to bring together the community of theoretical physicists and cosmologists and identify promising opportunities to address the questions described.

This White Paper was initiated with the aim of identifying the opportunities in the cosmological field for the next decade, and strengthening the coordination of the community. It is addressed to identifying the most promising directions of investigation, and rather than attempting a long review of the current status of the whole field of research, it focuses on the upcoming theoretical opportunities and challenges described. The White Paper is a collaborative effort led by Eleonora Di Valentino and Luis Anchordoqui, and it is organized in the topics listed below, each of them coordinated by the scientists indicated (alphabetical order):
\begin{itemize}
    \item Sec.~\ref{sec:WG-comparison}. Models comparison: Alan Heavens and Vincenzo Salzano.
    
    \item Sec.~\ref{sec:WG-H0measurements}. $H_0$ measurements/systematics: Simon Birrer, Adam Riess, Arman Shafieloo, and Licia Verde.
    
    \item Sec.~\ref{sec:WG-S8measurements}. $S_8$ measurements/systematics: Marika Asgari, Hendrik Hildebrandt, and Shahab Joudaki.
    
    \item Sec.~\ref{systematics}. Early Universe measurements/systematics: Mikhail M. Ivanov and Oliver H.\,E. Philcox.
    
    \item Sec.~\ref{sec:WG-BothSolutions}. $H_0$-$S_8$ proposed solutions: Celia Escamilla-Rivera, Cristian Moreno-Pulido, Supriya Pan, M.M.\ Sheikh-Jabbari, and Luca Visinelli.
   
    \item Sec.~\ref{sec:WG-challenges}. Challenges for $\Lambda$CDM beyond $H_0$ and $S_8$: Leandros Perivolaropoulos.
    
    \item Sec.~\ref{sec:WG-perspectives}. Stepping up to the new challenges (Perspectives): Wendy Freedman, Adam Riess, and Arman Shafieloo.
\end{itemize}

Each section begins with a list of contributors who made particular, and in many cases, substantial contributions to the writing of that section.
Furthermore, this White Paper is supported by $\sim 203$ scientists, who participated in brainstorming sessions from August 2020, and provided feedback via regular Zoom seminars and meetings, and the SLACK platform.

This White Paper has been reviewed by Luca Amendola, Spyros Basilakos, Ruth Lazkoz, Eoin \'O Colg\'ain, Paolo Salucci, Emmanuel Saridakis and Anjan Sen.


\section{Introduction}
\label{sec:introduction}

The discovery of the accelerating expansion of the Universe~\cite{SupernovaSearchTeam:1998fmf,SupernovaCosmologyProject:1998vns} has significantly changed our understanding on the dynamics of the Universe and thrilled the entire scientific community. To explain this accelerating expansion, usually two distinct routes are considered. The simplest and the most traditional one is to introduce some exotic fluid with negative pressure in the framework of Einstein's theory of GR, mostly restricted to the FLRW class of solutions, which drives the late-time accelerating expansion of the Universe~\cite{Copeland:2006wr,Bamba:2012cp}. 
Alternatively, generalizations of the FLRW cosmologies by taking into account the role of regional inhomogeneities on large-scale volume dynamics within the full theory of GR provide a possible route, e.g.\ Refs.~\cite{Ellis:2005uz,Buchert:2007ik}. 
One can also modify GR leading to a modified class of gravitational theories or introduce a completely new gravitational theory~\cite{Capozziello:2002rd,Nojiri:2006ri,Capozziello:2007ec,Capozziello:2008ch, Sotiriou:2008rp,DeFelice:2010aj,Harko:2011kv,Capozziello:2011et,Clifton:2011jh,Cai:2015emx,Nojiri:2017ncd}. In the latter approaches, the gravitational sector of the Universe is solely responsible for this accelerating expansion where the extra geometrical terms (arising due to inhomogeneities or the modifications of GR) or the new geometrical terms (coming from a new gravitational theory) play the role of DE.  
Since the discovery of the accelerating phase of the Universe we are witnessing the appearance and disappearance of a variety of cosmological models motivated from these frameworks, thanks to a large amount of observational data; however, the search for the actual cosmological model having to ability to explain the evolution of the Universe correctly is still in progress. Among a number of cosmological models introduced so far in the literature, the $\Lambda$CDM cosmological model, the mathematically simplest model with just two heavy  ingredients (equivalently, two assumptions), namely, the positive cosmological constant ($\Lambda >0$) and the CDM came into the picture of modern cosmology.

There is definitely no doubt in the community that the $\Lambda$CDM cosmology has been very successful in the race of cosmological models since according to the existing records in the literature, the standard $\Lambda$CDM cosmological model provides a remarkable description of a wide range of astrophysical and cosmological probes, including the recent measurement of the CMB temperature at high redshift from H$_2$O absorption~\cite{Riechers:2022tcj}. The parameters governing the $\Lambda$CDM paradigm have been constrained with unprecedented accuracy by precise measurements of the CMB~\cite{Planck:2018nkj,Planck:2018vyg,Polarbear:2020lii,ACT:2020gnv,SPT-3G:2021eoc,BICEPKeck:2021gln}. However, despite its marvellous fit to the available observations, there is no reason to forget that $\Lambda$CDM cannot explain the key concepts in the understanding of our Universe, namely, Dark Energy~\cite{SupernovaSearchTeam:1998fmf,SupernovaCosmologyProject:1998vns},  Dark Matter~\cite{Trimble:1987ee} and Inflation~\cite{Guth:1980zm}. 

In the context of the $\Lambda$CDM paradigm, DE assumes its simplest form, that is the cosmological constant, without any strong physical basis. The nature of DM is still a mystery except for its gravitational interaction with other sectors, as suggested by the observational evidences. We know, however, that DM is essential for structure formation in the late Universe, so most of it (though not all) must be pressure-less and stable on cosmological time scales. Although attempts have been made to understand the origin of DE and DM in different contexts, such as the alternative gravitational theories beyond Einstein's GR, no satisfactory answers have been found yet that can significantly challenge Einstein's gravitational theory at all scales. Moreover, despite the significant efforts in the last decades to investigate DM and the physics beyond the SM of particle physics, in laboratory experiments and from devised astrophysical observations, no evidence pointing to the dark matter particle has been found~\cite{MarrodanUndagoitia:2015veg,Gaskins:2016cha,Buchmueller:2017qhf}.

On the other hand, even though the theory of inflation~\cite{Snowmass2021:Inflation} has solved a number of crucial puzzles related to the early evolution of the Universe, it cannot be considered to be the only theory of the early Universe. For example, we do not have a definitive answer to the initial singularity problem,\footnote{Many of the possible alternatives leading to regular cosmological models have been analyzed in~\cite{Novello:2008ra}.} and some of the well known inflationary models have been put before a question mark~\cite{Ijjas:2013vea}. Moreover, alternative theories to inflation can also not be ruled out in light of the observational evidences~\cite{Lehners:2013cka,Ijjas:2016tpn}. In addition, the possibilities of unification of the early time inflationary epoch with late time cosmic acceleration captured the geometrical grounds, avoiding the introduction of additional degrees of freedom for inflation or DE (see for instance~\cite{Benisty:2020xqm,Benisty:2020qta}). Finally, the study of the primordial non-Gaussianity contributions to the cosmological fluctuations and other cosmological observables~\cite{Planck:2019kim,Buchert:2017uup} will become an important probe for both the early and the late Universe. Here, one of the many open problems is how to precisely compute the statistical significance of a tension in the presence of significant non-Gaussianity in the posterior distributions of cosmological parameters. While the issue is conceptually straightforward, the large number of dimensions usually involved when comparing surveys of interest makes the calculation challenging. 

Undoubtedly, observational data play a very crucial role in this context and we have witnessed a rapid development
in this sector. As a consequence, with the increasing sensitivity in the experimental data from various astronomical and cosmological surveys,  the heart of modern cosmology $-$ i.e.\ the physics of the Universe $-$ is getting updated and we are approaching  more precise measurements of the cosmological parameters. The progress in observational cosmology has introduced new directions that will drive us to find a suitable cosmological theory in agreement with the observational evidence. In fact, starting from the early evolution of the Universe to the present DE dominated Universe, there are a large number of open issues that require satisfactory explanations by thorough investigations.
Therefore, in the next decade, the primary challenges would be to answer the following questions: 

\begin{itemize}

\item What is the nature of dark matter? How can multi-messenger~\cite{Engel:2022yig} (i.e.\ photons, neutrinos, cosmic rays, and gravitational waves) cosmology help with determining characteristics (mass spectrum and interactions) of the dark matter sector?

\item What is the nature of dark energy? Can dark energy be dynamical? Can the cosmic acceleration slow down in future? 

\item Did the Universe have an inflationary period? How did it happen? What are the physical observables from the inflationary era? 

\item Does gravity behave like General Relativity even at present day horizon scale $H_{0}^{-1}$? Is there a more fundamental, Modified Gravity, which includes General Relativity as a particular limit?

\end{itemize}

Even if the performance of $\Lambda$CDM is undoubtedly stunning in fitting the observations with respect to the other cosmological theories and models, due to the assumptions and uncertainties listed before, the $\Lambda$CDM standard cosmological model is likely to be an approximation to a first principles scenario yet to be found. And this underlying model, which is approximated by $\Lambda$CDM so well, is not expected to be drastically different. For this reason, the first step is to look for small deviations or extensions of the minimal $\Lambda$CDM model. Throughout we should be conscious that this may not suffice and any new cosmology could be disconnected from the $\Lambda$CDM model. 

There are a few theoretical unanswered questions, which could indicate the direction to follow to discover this first principles scenario beyond $\Lambda$CDM. For instance it will be important to answer the following questions:

\begin{itemize}

\item Why the dark energy and the dark matter densities are of the same order (coincidence problem)? Is this coincidence suggesting an interaction between the DE and the DM?

\item Are dark energy and inflation connected (as for example in Quintessential Inflation models)? Can we have dark energy with AdS vacua (presence of a negative $\Lambda$)?

\item How well have we tested the Cosmological Principle? Is the Universe at cosmic scales homogeneous and isotropic?

\item Can local inhomogeneity or anisotropy replace the need for dark energy?

\item What is the level of non-Gaussianity? 

\item Do we need quantum gravity, or a unified theory for quantum field theory and GR to complete the standard cosmological model? How does pre-inflation physics impact our observations today? How can we resolve the big bang singularity?

\item Can theoretical frameworks, like effective (quantum) field theory have further implications for the dark sector, especially DE?

\item How much can we learn from cosmological dark ages and how does its physics impact our models of cosmology?

\item How crucial is physics beyond the SM of particle physics for precision cosmology?

\item How can we explain the matter-antimatter asymmetry in the observed Universe? There has been observational evidence for a matter–antimatter asymmetry in the early Universe, which leads to the remnant matter density we observe today. The bounds on the presence of antimatter in the present-day Universe include the possibility of a large lepton asymmetry in the cosmic neutrino background.

\item What are the mutual implications for cosmology and Quantum Gravity of hypotheses like the swampland conjectures? 

\end{itemize}

Finally, there are a few hints in the data suggesting that $\Lambda$CDM should be extended to accommodate the emerging discrepancies with different statistical significance between observations at early and late cosmological time~\cite{Verde:2019ivm}. For this reason, it is timely to focus our attention on the strong and robust disagreement at more than $5\sigma$ on the Hubble constant $H_0$ present between the value inferred by the {\it Planck} CMB experiment~\cite{Planck:2018vyg} assuming a $\Lambda$CDM model and the latest value measured by the SH0ES collaboration~\cite{Riess:2021jrx} (see also the reviews~\cite{DiValentino:2020zio,Knox:2019rjx,Jedamzik:2020zmd,CANTATA:2021ktz,DiValentino:2021izs,Perivolaropoulos:2021jda,Shah:2021onj}). This is followed by the tension at $\sim 3\sigma$ on $\sigma_8 - S_8$~\cite{DiValentino:2020vvd,Perivolaropoulos:2021jda}, or by the anomalies in the {\it Planck} CMB experiment results, such as the excess of lensing~\cite{Addison:2015wyg} at $> 3\sigma$, which have put a question mark on the geometry of the Universe~\cite{Planck:2018vyg,Handley:2019tkm,DiValentino:2019qzk,DiValentino:2020hov} and its age. 
Meanwhile, several statistically independent measures of the  violation of statistical isotropy have been identified with at least 3$\sigma$ significance in the WMAP and {\it Planck} CMB temperature maps -- suppression of large angular correlations, differences between opposite hemispheres,  alignment of the quadrupole and octupole.  
These are complemented by the statistically less significant studies of the cosmic accelerated expansion rate using supernovae as standard candles, that have indicated that an anisotropic expansion rate fit by means of a dipole provides a better fit to the surveys at hand than an isotropic expansion rate at the 2$\sigma$ level. This dark energy dipole is aligned with the fine structure constant dipole~\cite{Mariano:2012wx}. Moreover, as we discuss later in section \ref{sec:beyond_FLRW} a number of intriguing dipoles appear to be aligned with the CMB dipole.

The accumulation of these tensions may suggest the addition of new physics ingredients in the vanilla $\Lambda$CDM model, based on cosmological constant and Dark Matter. Therefore, in the next decade, we aim to address these discrepancies by answering the following key questions:

\begin{itemize}

\item Can these tensions, individually or together, be systematic errors in the current measurements? Are these tensions statistical flukes or are they pointing to new physics?

\item If not due to systematics, what is the origin of the sharpened tension in the observed and inferred values of $H_0$, $f\sigma_8$, and $S_8$?

\item Are these tensions uncorrelated, or connected and different manifestations of a single tension?

\item Is it possible to explain the tensions within the standard $\Lambda$CDM cosmology?

\item Is the Universe open, flat or closed? 

\item Are cosmic tensions pointing beyond the FLRW framework?

\item Can the pure geometric generalizations of the $\Lambda$CDM model, viz., allowing spatial curvature and/or anisotropic or inhomogeneous expansion on top of the spatially flat FLRW space-time metric assumption, while preserving the physics that underpins it, address the tensions related to the $\Lambda$CDM model?

\end{itemize}

Thus, based on the above untold stories, it is difficult to believe that $\Lambda$CDM cosmology can be considered the final description of the Universe that can accommodate all the queries raised over the last couple of years. Since the observational sensitivity is increasing with time, it is  important and timely to focus on the above-mentioned limitations of this concordance cosmological paradigm, aiming to find a most suitable cosmological theory that can explain some/all the above points.


\section{Quantification of "Tensions" and Model Comparison}
\label{sec:WG-comparison}

\noindent \textbf{Coordinators: } Alan Heavens and Vincenzo Salzano.
\\

\noindent \textbf{Contributors: } M.\ Benetti, D.\ Benisty, R.~E.\ Keeley, N.\ Sch{\"o}neberg.
\bigskip

\subsection{Tensions and the Relation to Model Comparison}

The preponderance of different cosmological probes allows the cosmological model to be tested in various ways, and a situation may arise when the different probes appear to give incompatible results. There is currently a debate about the compatibility of results from probes of the Hubble constant $H_0$ and to a lesser extent from the degree of clustering of matter, often measured by the quantity $S_8 = \sigma_8(\Omega_m/0.3)^{1/2}$ (see e.g.\ Refs.~\citep{Riess:2021jrx,DES:2020hen,Lin:2017bhs}). "Tension" is the term used to describe results that appear to be discrepant, the interesting question being whether such tensions are indications of the need for a revision to the standard cosmological model, or whether they are due to statistical fluctuations, errors or approximations in analysis, or unmodelled systematic effects in the data. Finding the reasons for the apparent discrepancies is a major driver of cosmological research in the near future.

Ideally, we would like to quantify the significance of the disagreement between the outcomes of two different experiments, and this is often expressed as the difference between estimates (typically maximum marginalised posterior values) of cosmological parameters, expressed in units of the uncertainty, usually taken to be the posterior errors of the two experiments added in quadrature (if independent). The interpretation is often based on the probability to exceed the given shift, assuming a Gaussian distribution. For example a ``2 sigma tension" corresponds to a $4.6\%$ probability of a larger discrepancy, assuming the cosmological model is correct. However, the ensuing interpretation that the cosmological model only has a $4.6\%$ probability of being correct is unfortunately wrong, and a more sophisticated analysis is necessary. Here we discuss the linked issues of tensions and model comparison - to decide when the data in tension favour an alternative to the standard cosmological model.

From a Bayesian perspective, the question of ``goodness of fit" is a thorny one, and the question of whether the data are somehow compatible with a theory is not usually addressed. Without an alternative model, one often can not exclude a statistical fluke. What can be addressed is the relative posterior probabilities of two or more different data models, through computation of the Bayesian Evidence. The first Bayesian interpretations of cosmological tension were published in Refs.~\citep{Verde:2013wza, Verde:2014qea}. This approach transcends the question of what the model parameters are, and asks for the relative model probabilities, conditioned on the data. The most natural way to apply this in the case of tensions is to compute the Bayes factor (the ratio of model likelihoods) for the standard cosmological model in comparison with an alternative model, and interpret it as the posterior odds for the models. A rather hybrid alternative that does not require a new physics model is to postulate two data models: one is the standard cosmological model where the cosmological parameters are common to all experiments; the other allows the cosmological parameters to be different in each experiment. This is a well-posed comparison (albeit with an arguably unrealistic physics model) and Bayesian methods can be used to address it. Implicitly, this is related to the $R$ statistic of \citep{Handley:2019wlz}. In either of these comparisons, the outcome is a relative probability of the models, and this is often interpreted using the Jeffreys scale~\cite{Jeffreys:1939xee}, or the Kass and Raftery revision~\cite{Kass:1995loi}. The latter arguably provides better descriptive words for the relative probabilities, but in the end the probabilities speak for themselves. 

Before discussing the Bayesian approach in more detail, a number of methods of measuring the significance of parameter shifts between experiments have recently been proposed in the cosmology literature. These are frequently referred to as ``tension metrics" and seek to answer often subtly different questions about the data, and each has advantages and disadvantages (see for example~\cite{Lin:2017ikq, Raveri:2018wln,DES:2020hen}). As measurements of the tension in cosmological parameters improve it will be crucial for cosmologists to continue furthering understanding of how to interpret these measurements, as well as to understand the interplay of nuisance parameters (such as those from photometric redshift estimation) and model mis-specification (e.g., uncertain baryonic physics) with the conclusions.

In addition, cosmological tensions fundamentally involve disagreement between results of experiments and analyses. Each of these experiments and analyses takes part within a given community and with a given set of tools and models. As new data arrive it will be crucial to avoid unconscious biases towards confirmation of previous measurements, or a desire to confirm or deny a tension as real. As such, significant efforts needs to be made in ensuring analyses are performed ``blindly", see e.g.\ Refs.~\citep{DES:2019zlw, Wong:2019kwg, Sellentin:2019stv, DES:2021esc}, in ways in which the result is not known to the experimenters until after the analysis choices have been fixed.

Ensuring that it is possible to re-analyse and jointly analyse data sets is also essential, which means both that efforts towards data sharing and replication studies should be highly valued, and that it is necessary to test the data individually to check if they can be combined. In particular, joint analyses can be carried out between data that independently yield compatible results. Although the combination of complementary cosmological probes makes it possible to break degeneracies between parameter estimates, and also test systematic effects, data that independently point to different parameter spaces will give misleading and unphysical results when combined~\cite{Efstathiou:2020wem,Gonzalez:2021ojp,vagnozzi:2020dfn,Moresco:2022phi}.

When addressing tensions we are concerned with whether or not a given shift in the preferred parameter space between experiments can most probably be explained by a stochastic fluctuation~\cite{Jenkins:2011va, Joachimi:2021ffv} within a given model, or by a mis-specification of the model used to describe the data in each experiment. The proposed ``solutions" in the literature and discussed here are attempts to extend the model to describe better the data, either by including new fundamental physics or better understanding the mediating astrophysics and instruments. 

An important aspect in assessing these models is the number of model parameters, since an arbitrarily complex model with a sufficient number of parameters can fit any given data set perfectly, so ``goodness-of-fit" (however defined) is not in itself a good measure of model probability - the predictions made from such a model will generally not agree with new data. As such, it is important to penalize models with many free parameters in order to ensure that the considered models generalize robustly to new data. Many criteria include an ``Occam factor" which penalizes complex models with many free parameters, either explicitly or by considering the increase in the prior volume of the model parameter space (although for the Bayesian Evidence the penalty is not always obvious $-$ a completely unconstrained additional parameter is not penalized at all). In passing,  we notice that it is possible that the  Physics underlying Dark Matter and the Dark Energy phenomena  may be so disconnected with respect to that presently known that it  may  violate the Occam Paradigm.

At this point it is important to note that the choice of fixing certain parameters while allowing others to vary is often taken arbitrarily in a given model. As such, an important aspect to consider when comparing models is not only whether they reduce the tension metric of choice, but also whether they overall remain a good fit to the data. A simple example of why this is important is a $\Lambda$CDM model with $\Omega_{\rm CDM} h^2$ fixed to 0.11, which does give a reduction of the tension between {\it Planck} and SH0ES to $\sim 1 \sigma$, but is also not a good fit to {\it Planck} data~\cite{Schoneberg:2021qvd}. Some models proposed to ease the Hubble tensions in the past have restricted their parameter space to regions in which the tension is reduced, but so is the model's ability to fit the data. Whether or not such models should be considered equally or less valid is an important aspect of a given model comparison study. Bayesian Evidence could certainly be used here to compare data models with $\Omega_{\rm CDM} h^2$ fixed and variable, respectively.

\subsection{List of Statistical Tools}

Expansion of parameter dimensions can also have other costs in terms of interpretations of tensions. Very high dimensional probability spaces are often projected down to only two (for plotting posterior contours) or one (for quoting error bars), and such projections can mask higher dimensional tensions. As such, it is usually preferable to derive the tension between data sets using the entire parameter space of the given model. However, this should be done with caution because where additional parameters are included which are poorly constrained by a given experiment, the posteriors on other correlated parameters can become dramatically skewed by the ``prior volume effect" in which probability mass is used to fill this extra volume~\cite{Handley:2020hdp}. 

The ``explosion" of the tension problems in the latest years has made even more critical an already almost saturated aspect of cosmological research: the number of proposed models has grown steadily (and will continue to) and it has become quite hard to sort out the \textit{more successful} options from \textit{less successful} ones. 
At this stage, resorting only to the simplest rule of ``let us fit the data and see what happens" is obviously not enough and definitively it is not the ultimate tool, but just the first step. Even if we all played with an established \textit{consensus} set of data, we would still have too much degeneracy and biases to assess a hierarchy of reliable models.

This is not a novelty, of course; indeed, we do have tools to provide such a statistical hierarchy. But recent developments have quite clearly exacerbated this problem. It is thus not pointless to try to apply to the realm of statistical comparison tools the same principles which led to~\cite{Schoneberg:2021qvd}. Can we make a selection of the best/preferred statistical tools we have and we use nowadays to state if and when a given cosmological model is better than another one?

A provisional and surely non-exhaustive list would include:
\begin{itemize}
\item $\chi^2$ and reduced $\chi^2$. The first ``obvious" criterion, but which is not satisfactory for a number of  reasons: a lower $\chi^2$ does not necessarily point to a better model, since adding more flexibility with extra parameters will normally lead to a lower $\chi^2$. At best, it can be considered only as \textit{one} (among many) or the \textit{first} indicator to check, but absolutely not exhaustive and definite. Even the reduced $\chi^2$ cannot take into account the complexity of the problem and the degeneracy among the parameters, and from a Bayesian perspective, the minimum $\chi^2$ contains no information on how finely-tuned the parameters have to be - an important consideration when assessing the model probability. Moreover, still in the literature the $\chi_{\rm min}^2$ is often misused: in a Bayesian approach, to focus or to center a discussion on the specific set of values which correspond to $\chi_{\rm min}^{2}$ does not reflect the fact that these values are uncertain. Some of the statistics discussed below are prone to this criticism.
\item Information Criteria~\cite{WU19981} - AIC~\cite{Akaike:1974}, BIC~\cite{Schwarz:1978}, DIC~\cite{Spiegelhalter:2002yvw}, RIC~\cite{Leng:2007}, WAIC~\cite{2013arXiv1307.5928G}. These are widely used in literature for their intrinsic simplicity: typically they consist in adding to the $\chi_{\rm min}^{2}$ some algebraic factors which should resume all the complex information about the parameter space. From a theoretical point of view, only the BIC has some Bayesian justification, but makes simplifying assumptions that often do not hold. Thus, in general, they should not be considered as the ultimate tools but as ``some among many" and should be used in a complementary way with others methods.
\item There also exist some simple goodness of fit measures as for example defined in Ref.~\cite{Raveri:2018wln}. These include the trivial Gaussian tension (which quantifies the probability of agreement under the simple assumption of a Gaussian posterior), the difference of maximum likelihood points (a posteriori, DMAP~\cite{Raveri:2018wln}) which computes $\sqrt{\chi^2_{\mathrm{min},A+B}-\chi^2_{\mathrm{min},A}-\chi^2_{\mathrm{min},B}}$ for two experimental sets $A$ and $B$ (and naturally generalizes the Gaussian tension to non-Gaussian posteriors), and more complicated parameter difference measures such as the DM (first proposed in~\citep{Lin:2017ikq}) and UDM defined in Ref.~\cite{Raveri:2018wln}, which generalize the Gaussian tension to cases of non-Gaussian posteriors in a way that still relies on covariance matrices and means
\item Bayesian Evidence, Savage-Dickey ratio, Evidence ratio.  These are firmly based on a principled Bayesian statistical approach which, for its intrinsic properties and philosophy seems to be the most appropriate for cosmological and astrophysical purposes. The relative posterior probabilities of competing models $M$ and $M'$, given data ${\bf d}$ and model parameter sets $\theta$ and $\theta'$ is (see Ref.~\citep{Trotta:2008qt} for more details)
\[
\frac{p(M'|{\bf d})}{p(M|{\bf d})}=\frac{\pi(M')}{\pi(M)}\frac{\int
{\rm d}\theta'\,p({\bf d}|\theta' , M')\pi(\theta'|M')}{\int
{\rm d}\theta\,p({\bf d}|\theta ,M)\pi(\theta|M)},
\]
where $\pi(M)$ is a model prior, and $\pi(\theta|M)$ the parameter prior.  For equal model priors, this simplifies to the ratio of Bayesian Evidences, called the {\em Bayes Factor}, being the second ratio on the right hand side.  It is this that is often interpreted on the Jeffreys~\cite{Jeffreys:1939xee} or Kass-Raftery~\citep{Kass:1995loi} scale. Note. that the multi-dimensional integrals may be challenging to compute, but algorithms like~\citep{10.1214/06-BA127,Mukherjee:2005wg} and tools such as multiNest~\citep{Feroz:2013hea}, PolyCHORD~\citep{Handley:2015fda} and MCEvidence~\citep{Heavens:2017afc} exist, and for nested models, tricks such as the Savage-Dickey density ratio~\citep{SavageDickey} may be employed. 
            
By choosing a model that relieves the tensions, the maximum of the likelihood term for the joint experiment will increase, which will tend to increase the Evidence.  However, if the parameter space is expanded, then the additional prior volume in parameter space counteracts this increase, and the simpler model may still be favoured, even if the maximum of the likelihood of the more complex models goes up.  This encompasses to some degree Occam's razor. A drawback of this approach is that the Bayes factor is dependent on the parameter priors~\citep{Nesseris:2012cq}, even (unlike in parameter inference) in the limit of highly constraining data. Thus, there is some uncertainty unless we have a good physical reason to specify a particular prior. As a result one might want a more stringent condition than implied by the Jeffreys or Kass-Raftery scale before reaching strong conclusions in favour of one model or another, to reflect prior uncertainty. However, when $\Lambda$CDM is compared with models that involve introducing one or two extra parameters, it is a very effective approach~\citep{Martin:2013nzq, Heavens:2017hkr}, and the Bayes factor may be so large or small as to give conclusions that are robust to reasonable changes in the prior.

What comes out from studies of the Bayes factor is that tensions of \emph{at
least} 3 or 4$\sigma $ are needed before a more complex model is favoured
with reasonably high odds (say 20:1) (see Fig.~\ref{Trotta}), and it can be higher~\citep{Trotta:2008qt}. For example, Ref.~\citep{Heavens:2017hkr} showed that the $w$CDM model (see Sec.~\ref{sec:quintessence}) was favoured only by about 7:1 over $\Lambda$CDM with an apparently very significant $3.4\sigma$ Hubble tension at the time~\citep{Riess:2016jrr}. Using tail probabilities would give a far smaller and erroneous conclusion.  
            
\begin{figure}[ht]
\centering
\includegraphics[width=0.4\textwidth]{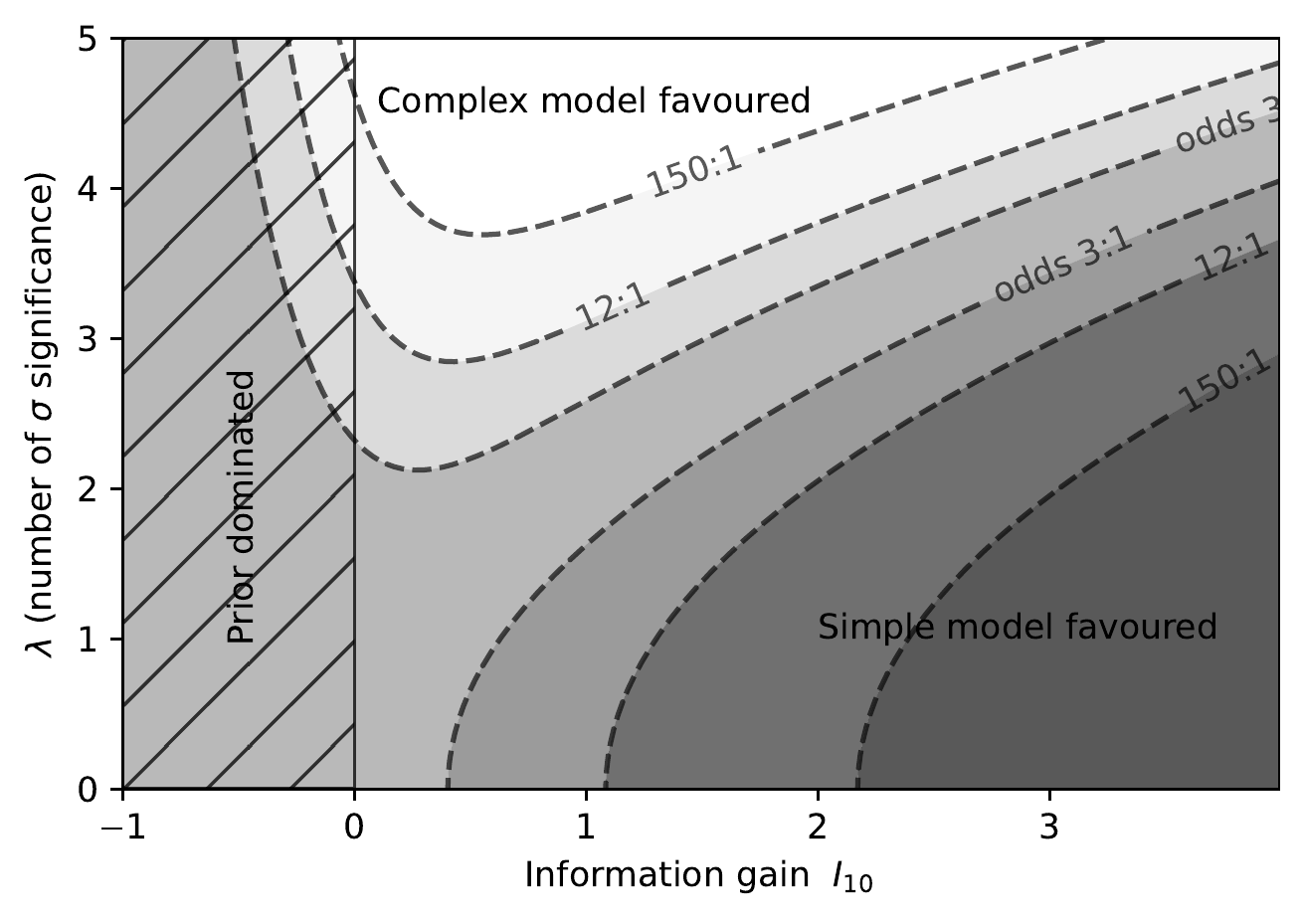}
\caption{Posterior odds ratio for nested Gaussian models vs. $\log_{10}$(prior width/likelihood width), plotted against the "significance" of the tension (number of $\sigma$).  We see that as a minimum, 3-4$\sigma$ tension is needed before the more complex model is favoured with odds of more than 20:1 (from Trotta, private comm.).} 
\label{Trotta}
\end{figure}
            
One limitation of the Bayesian Evidence approach is that if many extra parameters are added, the uncertainty in the added prior volume may become extremely large, and the Bayes factor may be highly uncertain, see e.g. Ref.~\citep{Lin:2017ikq}. This has prompted the investigation of other techniques, such as the {\em Index of Inconsistency}~\citep{Lin:2017ikq,Lin:2017bhs}, the {\em suspiciousness}~\citep{Handley:2019pqx, Lemos:2019txn} and {\em $\log\mathcal{I}$ statistic}~\citep{Joudaki:2016mvz,Joudaki:2020shz}, which are not dependent on the prior. Other measures have been considered by Refs.~\citep{Raveri:2018wln, Seehars:2014ora, Adhikari:2018wnk, Kunz:2006mc, Grandis:2015qaa, Grandis:2016fwl, Nicola:2018rcd}.
\item Likelihood distribution~\cite{Koo:2020wro,Koo:2021suo}. This data-driven, frequentist, model validation technique that uses the method of the iterative smoothing of residuals. In other words, this method seeks to test whether the data is generated from a model, independent of how well other models fit the data. The basic procedure to validate a model on a given dataset is to start with the best-fit model to that dataset. Then the iterative smoothing method is applied to that best-fit model; that best-fit model serves as the initial guess for the iterative smoothing method. The iterative smoothing method works by, at each iteration, altering the current function such that the residuals are reduced, as in smoothed, thus improving the $\chi^2$ of the current iterated function. The likelihood distribution then calculates how much over-fitting the iterative smoothing method will perform, i.e.\ if the data is generated from some fiducial parameters of some model, then how much improvement in the $\chi^2$ is expected when the best-fit of the model is used as an initial guess of the iterated smoothing method.  The distribution of the improvement from the iterative smoothing method over different realizations of the data is the likelihood distribution. If the improvement on the real data is outside of the likelihood distribution, then that model is invalidated, ruled out. If the improvement on the real data is within the likelihood distribution, then that model is validated.
\end{itemize}

In summary, there are many tools that have been proposed to quantify tensions, and these are used to indicate whether that new physics is needed.  Quantitatively, the principled Bayesian approach is to use the Bayesian Evidence, which gives posterior odds on models, and this to some degree incorporates an element of Occam's razor. Arguably, this is the best approach when comparing models with one or two extra parameters over $\Lambda$CDM, but the prior-dependency of the Bayes factor may become a problem when many extra parameters are added and their prior is not easily constrained, as the Bayes factor may then be determined largely by the prior choice.   There are many other instruments that may lack a principled basis, but do not suffer from this limitation.  The use and interpretation of such instruments raises interesting questions, some of which are similar to the discussion of p-values elsewhere in science.\footnote{\url{https://www.nature.com/articles/d41586-019-00874-8}; \url{https://www.nature.com/articles/d41586-019-00857-9}} Are our instruments fit for purpose, and are they interpreted correctly?  The alternative models considered almost always involve additional model complexity, and in circumstances where one cannot conveniently use the Bayes factor, the extent to which tools can properly penalise additional complexity (assuming we accept Occam's view) is not always clear.  Another general concern is with the complexity of the data that we collect. Our data models are inevitably simplifications and do not capture all of the astrophysics that is relevant for the real Universe. The influence of such model mis-specification on our inference is one that is rarely explored, and requires further investigation in future, whatever the statistics used.


\section{The $H_0$ Tension}
\label{sec:WG-H0measurements}

\noindent \textbf{Coordinators:} Simon Birrer, Adam Riess, Arman Shafieloo, and Licia Verde.
\\

\noindent \textbf{Contributors: } Adriano Agnello, Eleonora Di Valentino, Celia Escamilla-Rivera, Valerio Marra, Michele Moresco, Raul Jimenez, Eoin \'O Colg\'ain, Tommaso Treu, Cl{\'e}cio R.\ Bom, Mikhail M.\,Ivanov, Leandros Perivolaropoulos, Oliver H.\,E. Philcox, Foteini Skara, John Blakeslee.
\bigskip

The 2018 legacy release from the {\it Planck} satellite~\cite{Planck:2018nkj} of the Cosmic Microwave Background (CMB) anisotropies, together with the latest Atacama Cosmology Telescope (ACT-DR4)~\cite{ACT:2020gnv} and South Pole Telescope (SPT-3G)~\cite{SPT-3G:2021wgf} measurements, have provided a confirmation of the standard $\Lambda$CDM cosmological model. However, the improvement of the methods and the reduction of the uncertainties on the estimated cosmological parameters has seen the emergence of statistically significant tensions in the measurement of various quantities between the CMB data and late time cosmological model independent probes. While some proportion of these discrepancies may eventually be due to the systematic errors in the experiments, their magnitude and persistence across probes strongly hint at a possible failure in the standard cosmological scenario and the necessity for new physics. 

The most statistically significant and long-standing tension is in the estimation of the {\it Hubble constant} $H_0$ between the CMB data, that are cosmological model dependent and are obtained assuming a vanilla $\Lambda$CDM model, and the direct local distance ladder measurements. The Hubble constant $H_0$ is the present expansion rate, defined as $H_0=H(z=0)$ where $H(t) \equiv a^{-1}\mathrm{d}a/\mathrm{d}t$ and $a^{-1}=1+z$. 

\begin{figure}[ht]
\centering
\includegraphics[width=0.9\textwidth]{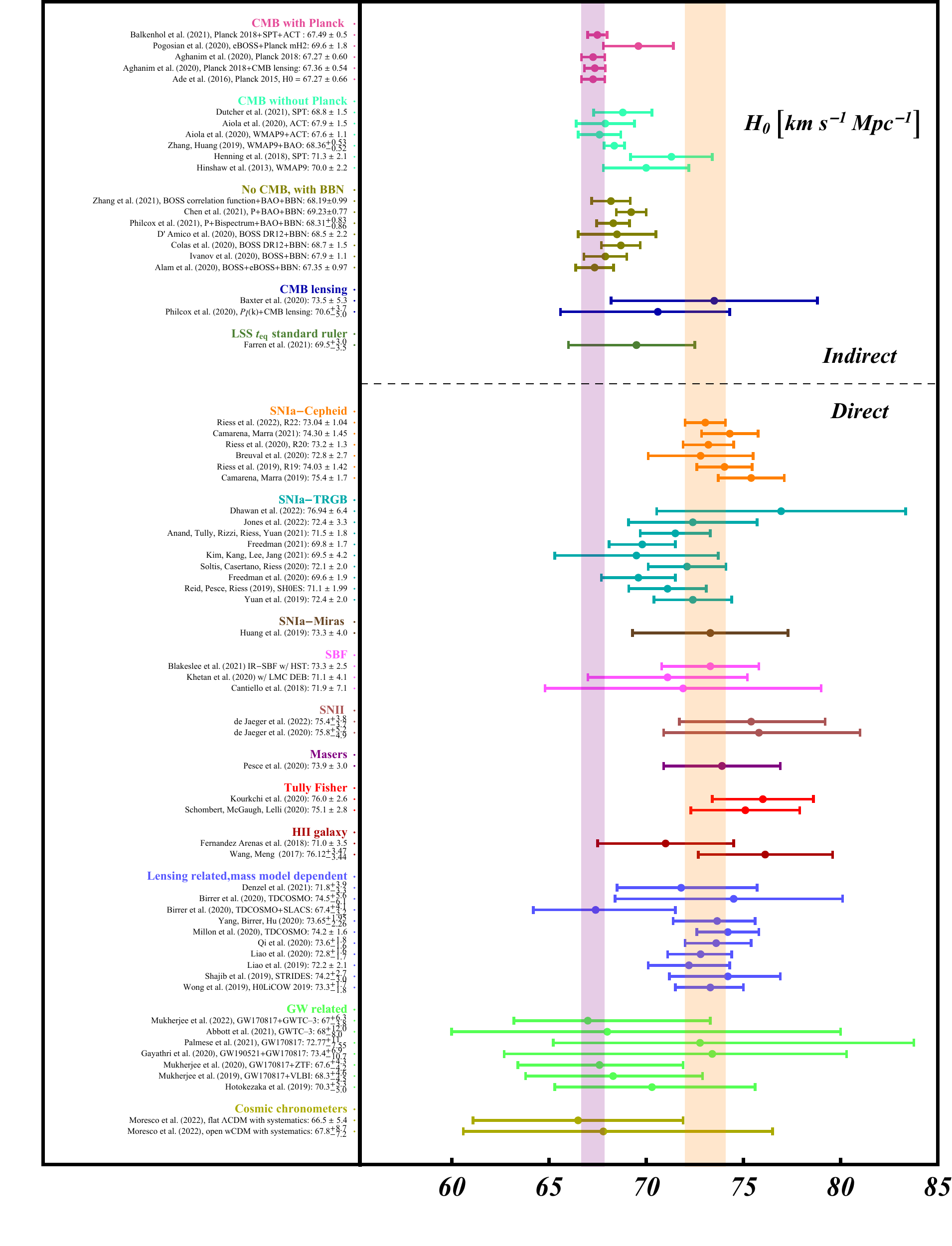}
\caption{68\% CL constraint on $H_0$ from different cosmological probes (based on Refs.~\cite{DiValentino:2021izs,Perivolaropoulos:2021jda}).}
\label{whisker_H0}
\end{figure}

In particular, we refer to the {\it Hubble tension} as the disagreement at $5.0\sigma$ between the {\it Planck} collaboration~\cite{Planck:2018vyg} value, $H_0=\left(67.27\pm0.60\right){\rm \,km\,s^{-1}\,Mpc^{-1}}$ at 68\% confidence level (CL), and the latest 2021 SH0ES collaboration (R21~\cite{Riess:2021jrx}) constraint, $H_0=(73.04 \pm 1.04){\rm \,km\,s^{-1}\,Mpc^{-1}}$ at 68\% CL, based on the Supernovae calibrated by Cepheids. However, there are not only these two values, but actually two sets of measurements, and all of the indirect model dependent estimates at early times agree between them, such as CMB and BAO experiments, and the same happens for all of the direct late time $\Lambda$CDM-independent measurements, such as distance ladders and strong lensing. We will see a collection of $H_0$ estimates in the next subsections, and a summary in the whisker plot of Fig.~\ref{whisker_H0}.

\subsection{Late $H_0$ Measurements}
\label{sec:direct_H0_measurement}

We consider as "Late $H_0$ measurements" those that are independent of the standard $\Lambda$CDM model. In general all of these measurements are in agreement with a higher value for the Hubble constant, and are in tension with the CMB estimate with a different statistical significance depending on the observable.

Here we find for example the best-established and empirical method of the "distance ladder", which allows one to measure $H_0$ locally, measuring the distance-redshift relation. In this case the geometry, e.g.\ the parallax, is used to calibrate the luminosities of specific star types which can be seen at great distances. Supernovae calibrated by Cepheids, as it has been done by the SH0ES collaboration, belong to this category. The latest SH0ES measurement gives
$H_0=(73.04 \pm 1.04){\rm \,km\,s^{-1}\,Mpc^{-1}}$ at 68\% CL~\cite{Riess:2021jrx}, a $5\sigma$ disagreement with the CMB one, while previous estimates include $H_0=(73.2 \pm 1.3){\rm \,km\,s^{-1}\,Mpc^{-1}}$ at 68\% CL~\cite{Riess:2020fzl}, improved with the latest parallax measurements provided by ESA Gaia mission Early Data Release 3 (EDR3)~\cite{gaiacollaboration2020gaia}, $H_0=(74.03 \pm 1.42){\rm \,km\,s^{-1}\,Mpc^{-1}}$ at 68\% CL~\cite{Riess:2019cxk}, $H_0=(73.5 \pm 1.4){\rm \,km\,s^{-1}\,Mpc^{-1}}$ at 68\% CL~\cite{Reid:2019tiq}, $H_0=(73.48 \pm 1.66){\rm \,km\,s^{-1}\,Mpc^{-1}}$ at 68\% CL~\cite{Riess:2018uxu}, and $H_0= 73.24 \pm 1.74{\rm\,km\,s^{-1}\,Mpc^{-1}}$ at 68\% CL~\cite{Riess:2016jrr}. Finally, we have to consider also the final result of the Hubble Space Telescope Key Project, that was $(72 \pm 8){\rm \,km\,s^{-1}\,Mpc^{-1}}$~\cite{HST:2000azd}, or the result obtained using an improved geometric distance calibration to the Large Magellanic Cloud (LMC), i.e. $(74.3 \pm 2.1){\rm \,km\,s^{-1}\,Mpc^{-1}}$~\cite{Freedman:2012ny}.

There has been many independent works trying to re-analyze the SH0ES collaboration result in the last few years, using different formalisms, statistical methods or parts of the dataset, but there is no evidence for a drastic change of the value of the Hubble constant. We have for example the re-analysis of Ref.~\cite{Cardona:2016ems} using Bayesian hyper-parameters, i.e.\ $H_0=(73.75 \pm 2.11){\rm \,km\,s^{-1}\,Mpc^{-1}}$ at 68\% CL, or of Ref.~\cite{Camarena:2021jlr} using the cosmographic expansion of the luminosity distance, i.e.\ $H_0=(74.30\pm 1.45){\rm \,km\,s^{-1}\,Mpc^{-1}}$ at 68\% CL (see Ref.~\cite{Camarena:2021jlr} for the $H_0$-$q_0$ covariance) or the previous estimate~\cite{Camarena:2019moy}. We have also the $H_0$ measurement obtained by Ref.~\cite{Dhawan:2017ywl} using the measured near-infrared (NIR) Type Ia supernovae (SNIa), i.e.\ $H_0=(72.8 \pm 1.6\,({\rm stat}) \pm 2.7\,({\rm sys})){\rm \,km\,s^{-1}\,Mpc^{-1}}$ at 68\% CL, or by Ref.~\cite{Burns:2018ggj} using a different method for standardizing SNIa light curves, i.e.\ $H_0 = (73.2 \pm 2.3){\rm \,km\,s^{-1}\,Mpc^{-1}}$ at 68\% CL. We have still the Hubble constant measured by Ref.~\cite{Follin:2017ljs} leaving the reddening laws in distant galaxies uninformed by the Milky Way, i.e.\ $H_0=(73.3 \pm 1.7){\rm \,km\,s^{-1}\,Mpc^{-1}}$ at 68\% CL, or by Ref.~\cite{Feeney:2017sgx} using a Bayesian hierarchical model of the local distance ladder, i.e.\ $H_0=(73.15\pm1.78){\rm \,km\,s^{-1}\,Mpc^{-1}}$ at 68\% CL. Then we have the measurement based on the Cepheids from the second Gaia data release (GDR2), that is $H_0=(73.0 \pm 1.9\, ({\rm stat + sys}) \pm 1.9\, ({\rm ZP})){\rm \,km\,s^{-1}\,Mpc^{-1}}$ at 68\% CL~\cite{Breuval:2020trd}, where ZP is the GDR2 parallax zero-point. Finally, there is the re-analysis of the Cepheid calibration used to infer the local value of the $H_0$ from SNIa, where one does not enforce a universal color-luminosity relation to correct the near-IR Cepheid magnitudes, which gives $H_0=(71.8\pm1.76){\rm \,km\,s^{-1}\,Mpc^{-1}}$, or $H_0=(66.9\pm2.5){\rm \,km\,s^{-1}\,Mpc^{-1}}$, at 68\% CL depending on the approach~\cite{Mortsell:2021nzg}, and $H_0=(70.8\pm2.1){\rm \,km\,s^{-1}\,Mpc^{-1}}$ at 68\% CL in Ref.~\cite{Mortsell:2021tcx}.

At the same time, the impact of known and several previously neglected or unrecognized systematic effects on the center values of the SH0ES results has been constrained to be well below the $1\%$ level~\citep{AndersonRiess:2018,Anderson:2019lbz,Riess:2020xrj,Javanmardi:2021viq,Anderson:2021fsp}. Interestingly, correcting reported $H_0$ values for such systematic effects can result in an increased value of $H_0$ (e.g., due to time dilation of observed variable star periods, cf.\ Ref.~\cite{Anderson:2019lbz}), and thus, to an increased significance of the discord between the distance ladder and the early Universe-based $H_0$ values. 

An independent determination of $H_0$ is based on calibration of the Type Ia supernovae (SNIa) using the Tip of the Red Giant Branch, where one finds $H_0=(71.17 \pm 1.66\,({\rm stat}) \pm 1.87\, ({\rm sys})){\rm \,km\,s^{-1}\,Mpc^{-1}}$ at 68\% CL~\cite{Jang:2017dxn}, $H_0=(69.8 \pm 0.8\,({\rm stat}) \pm 1.7\,({\rm sys})){\rm \,km\,s^{-1}\,Mpc^{-1}}$ at 68\% CL~\cite{Freedman:2019jwv}, $H_0=(72.4\pm2.0){\rm \,km\,s^{-1}\,Mpc^{-1}}$ at 68\% CL~\cite{Yuan:2019npk}, and the final $H_0=(69.6 \pm 0.8\,({\rm stat}) \pm 1.7\,({\rm sys})){\rm \,km\,s^{-1}\,Mpc^{-1}}$ at 68\% CL~\cite{Freedman:2020dne}. Then we have $H_0=(71.1 \pm 1.9){\rm \,km\,s^{-1}\,Mpc^{-1}}$ at 68\% CL~\cite{Reid:2019tiq}, and the $H_0$ estimate using velocities and TRGB distances to 33 galaxies located between the Local Group and the Virgo cluster, $H_0=(69.5 \pm 3.5\,({\rm stat}) \pm 2.4\,({\rm sys})){\rm \,km\,s^{-1}\,Mpc^{-1}}$ at 68\% CL~\cite{Kim:2020gai}. Moreover, it is important to mention the Freedman result $H_0=(69.8 \pm 0.6\,({\rm stat}) \pm 1.6\,({\rm sys})){\rm \,km\,s^{-1}\,Mpc^{-1}}$ at 68\% CL~\cite{Freedman:2021ahq}, or the re-analysis of Ref.~\cite{Anand:2021sum} that gives $H_0=(71.5 \pm 1.8){\rm \,km\,s^{-1}\,Mpc^{-1}}$ at 68\% CL. Finally, there are the latest measurements $H_0=(72.4 \pm 3.3){\rm \,km\,s^{-1}\,Mpc^{-1}}$ at 68\% CL of Ref.~\cite{Jones:2022mvo} and $H_0=(76.94 \pm 6.4){\rm \,km\,s^{-1}\,Mpc^{-1}}$ at 68\%~CL of Ref.~\cite{Dhawan:2022yws}.

A larger value for $H_0$ is preferred also by the analysis of the near-infrared HST WFC3 observations of the Mira variable red giant stars in NGC 1559, that gives $H_0=(73.3 \pm 4.0){\rm \,km\,s^{-1}\,Mpc^{-1}}$ at 68\% CL~\cite{Huang:2019yhh}.

A measurement of the Hubble constant can be obtained with the Surface Brightness Fluctuations (SBF) method calibrated from Cepheids, with $H_0=(71.9 \pm 7.1){\rm \,km\,s^{-1}\,Mpc^{-1}}$ at 68\% CL~\cite{Cantiello:2018ffy} from a single galaxy (the host of GW170817) at 40 Mpc and $H_0=(70.50 \pm 2.37\,({\rm stat}) \pm 3.38\,({\rm sys})){\rm \,km\,s^{-1}\,Mpc^{-1}}$ at 68\% CL~\cite{Khetan:2020hmh} from using SBF as an intermediate rung between Cepheids and SNIa. A re-analysis of the latter performed by Ref.~\cite{Blakeslee:2021rqi}, improving the LMC distance, gives instead $H_0=(71.1 \pm 2.4 ({\rm stat}) \pm 3.4 ({\rm sys})){\rm \,km\,s^{-1}\,Mpc^{-1}}$ at 68\% CL.  Moreover, if SBF is used as another long-range indicator calibrated by Cepheids and TRGB, Ref.~\cite{Blakeslee:2021rqi} finds $H_0=(73.3 \pm 0.7\,({\rm stat}) \pm 2.4\,({\rm sys})){\rm \,km\,s^{-1}\,Mpc^{-1}}$ at 68\% CL from a new sample of HST NIR data for 63 galaxies out to 100 Mpc.

Another possible observable to constrain $H_0$ is given by the Standardized Type II supernovae. Using 7 SNe II with host-galaxy distances measured from Cepheid variables or the TRGB, the Hubble constant is equal to $H_0=75.8^{+5.2}_{-4.9}{\rm \,km\,s^{-1}\,Mpc^{-1}}$ at 68\% CL~\cite{deJaeger:2020zpb}, while with 13 SNe II Ref.~\cite{deJaeger:2022lit} finds $H_0=75.4^{+3.8}_{-3.7}{\rm \,km\,s^{-1}\,Mpc^{-1}}$ at 68\%~CL.

An important independent result was obtained with the Megamaser Cosmology Project, which gives an $H_0$ estimate by using the geometric distance measurements to megamaser-hosting galaxies, finding $H_0=(66.6 \pm 6.0){\rm \,km\,s^{-1}\,Mpc^{-1}}$ at 68\% CL~\cite{Gao:2015tqd}, and $H_0=(73.9 \pm 3.0){\rm \,km\,s^{-1}\,Mpc^{-1}}$ at 68\% CL~\cite{Pesce:2020xfe}, independent of distance ladders. 

In addition, there are the estimates of the Hubble constant based on the Tully-Fisher Relation, i.e.\ the correlation between the rotation rate of spirals and their absolute luminosity used to measure the galaxy distances. Using optical and infrared calibrated Tully-Fisher Relations, the Hubble constant is equal to $H_0=(76.0 \pm 1.1\,({\rm stat})\pm2.3\,({\rm sys})){\rm \,km\,s^{-1}\,Mpc^{-1}}$ at 68\% CL~\cite{Kourkchi:2020iyz}. Considering the baryonic Tully-Fisher relation (bTFR) as a new distance indicator, instead, the Hubble constant is found to be $H_0 = (75.1 \pm 2.3\,({\rm stat}) \pm 1.5\,({\rm sys})){\rm \,km\,s^{-1}\,Mpc^{-1}}$ at 68\% CL~\cite{Schombert:2020pxm}, and $H_0 = (75.5 \pm 2.5){\rm \,km\,s^{-1}\,Mpc^{-1}}$ at 68\% CL~\cite{Kourkchi:2022ifq}.

An estimate of $H_0$ can also be obtained by modeling the extragalactic background light and its role in attenuating $\gamma$-rays, but this is challenging and the uncertainties may be underestimated. In this case there is Ref.~\cite{Dominguez:2013mfa} that finds $H_0=71.8_{-5.6}^{+4.6}({\rm stat})_{-13.8}^{+7.2}({\rm sys}){\rm \,km\,s^{-1}\,Mpc^{-1}}$ at 68\% CL, the updated value $H_0=67.4_{-6.2}^{+6.0}{\rm \,km\,s^{-1}\,Mpc^{-1}}$ at 68\% CL~\cite{Dominguez:2019jqc}, and $H_0=64.9_{-4.3}^{+4.6}{\rm \,km\,s^{-1}\,Mpc^{-1}}$ at 68\% CL~\cite{Zeng:2019mae}.

The HII galaxy measurement can also act as a good distance indicator independent of SNIa to probe the background evolution of the Universe. Using 156 HII galaxy measurements, we obtain the constraint $H_0=76.12^{+3.47}_{-3.44}{\rm \,km\,s^{-1}\,Mpc^{-1}}$ at 68\% CL~\cite{Wang:2016pag}, which is more consistent with the $H_0$ value from the SH0ES Team than that from the {\it Planck} collaboration (see also~\cite{FernandezArenas:2017dux})  

Moreover, we have the strong lensing time delays estimates, that are not $\Lambda$CDM model dependent but still astrophysical model dependent, because of the imperfect knowledge of the foreground and lens mass distributions. Here we have the H0LiCOW inferred values $H_0=71.9_{-3.0}^{+2.4}{\rm \,km\,s^{-1}\,Mpc^{-1}}$ at 68\% CL in 2016~\cite{Bonvin:2016crt},
$H_0=72.5_{-2.3}^{+2.1}{\rm \,km\,s^{-1}\,Mpc^{-1}}$ at 68\% CL in 2018~\cite{Birrer:2018vtm}, and $H_0=73.3^{+1.7}_{-1.8}{\rm \,km\,s^{-1}\,Mpc^{-1}}$ at 68\% CL in 2019~\cite{Wong:2019kwg},\footnote{Ref.~\cite{Wong:2019kwg} and the follow-up paper~\cite{Millon:2019slk} find a descending trend of $H_0$ with lens redshift. This can be interpreted as an expected signature of a late-time resolution to $H_0$ tension in line with the explanation given in Ref.~\cite{Krishnan:2020vaf} (see also Ref.~\cite{Krishnan:2020obg, Dainotti:2021pqg}). Alternatively, switching to orientation on the sky, one can interpret it as a signature of an emergent dipole in $H_0$~\cite{Krishnan:2021dyb}.} based on strong gravitational lensing effects on quasar systems under standard assumptions on the radial mass density profile of the lensing galaxies. A reanalysis of H0LiCOW's four lenses is performed in Ref.~\cite{Yang:2020eoh} finding $H_0=73.65_{-2.26}^{+1.95}{\rm \,km\,s^{-1}\,Mpc^{-1}}$ at 68\% CL. Under the same assumptions, the STRIDES collaboration measures from the strong lens system DES J0408-5354 $H_0=74.2^{+2.7}_{-3.0}{\rm \,km\,s^{-1}\,Mpc^{-1}}$ at 68\% CL~\cite{Shajib:2019toy}. Their combination (6 lenses from H0LiCOW and 1 from STRIDES, i.e. TDCOSMO) gives instead $H_0=74.2\pm1.6 {\rm \,km\,s^{-1}\,Mpc^{-1}}$ at 68\% CL~\cite{Millon:2019slk}.
Relaxing the assumptions on the mass density profile for the same sample of lenses, the TDCOSMO collaboration obtains $H_0=74.5^{+5.6}_{-6.1} {\rm \,km\,s^{-1}\,Mpc^{-1}}$ at 68\% CL~\cite{Birrer:2020tax}, and
$H_0=67.4^{+4.1}_{-3.2}{\rm \,km\,s^{-1}\,Mpc^{-1}}$ at 68\% CL~\cite{Birrer:2020tax}, by combining the time-delay lenses with non time-delay lenses from SLACS~\citep{Bolton:2008xf}, assuming they are drawn from the same parent sample.

An independent determination of $H_0$ has been obtained from strongly lensed quasar systems from the H0LiCOW program and Pantheon SNIa compilation using Gaussian process regression. This gives $H_0=72.2\pm2.1{\rm \,km\,s^{-1}\,Mpc^{-1}}$ at 68\% CL~\cite{Liao:2019qoc}, or the updated result $H_0=72.8^{+1.6}_{-1.7}{\rm \,km\,s^{-1}\,Mpc^{-1}}$ at 68\% CL using six lenses of the H0LiCOW dataset~\cite{Liao:2020zko}.
Moreover, Ref.~\cite{Qi:2020rmm} finds $H_0 = 73.6^{+1.8}_{-1.6}{\rm \,km\,s^{-1}\,Mpc^{-1}}$ at 68\% CL, taking the derived products of H0LiCOW, and another time-delay strong lensing measurement has been obtained analysing 8 strong lensing systems in~\cite{Denzel:2020zuq} giving $H_0=71.8 ^{+3.9}_{-3.3}{\rm \,km\,s^{-1}\,Mpc^{-1}}$ at 68\% CL.

Finally, a promising probe is the angular diameter distance measurements from galaxy clusters. Using the sample consisting of 25 data points, Ref.~\cite{Wang:2017fcr} finds $H_0=(70.1\pm0.8){\rm \,km\,s^{-1}\,Mpc^{-1}}$ at 68\% CL, or using another sample containing 38 data points $H_0=(74.6\pm2.1){\rm \,km\,s^{-1}\,Mpc^{-1}}$ at 68\% CL.
This measurement is very sensitive to the temperature calibration and a more conservative analysis gives $H_0=67.3^{+21.3}_{-13.3}{\rm \,km\,s^{-1}\,Mpc^{-1}}$ at 68\% CL, where errors are dominated by uncertainties of the temperature calibration~\cite{Wan:2021umh}, and $H_0=72.2^{+6.7}_ {-6.7}{\rm \,km\,s^{-1}\,Mpc^{-1}}$ at 68\% CL~\cite{Mantz:2021lcj}. 

When the late Universe estimates are averaged in different combinations, these $H_0$ values disagree between 4.5$\sigma$ and 6.3$\sigma$ with those from {\it Planck}~\cite{Riess:2020sih,Verde:2019ivm,DiValentino:2020vnx}.

\subsection{Cosmological $H_0$ Inferences from Modeled Galaxy Aging and BAO}
\label{sec:GP}

In recent years, we have witnessed an increased use of reconstructions of cosmological parameters directly from data. While Gaussian Process (GP) regression~\cite{Holsclaw:2010nb, Holsclaw:2010sk, Shafieloo:2012ht, Seikel:2012uu} is the most commonly used technique, there are a host of other approaches~\cite{Crittenden:2005wj, Crittenden:2011aa}, including most recently the use of machine learning algorithms~\cite{Arjona:2019fwb, Arjona:2020kco, Mehrabi:2021cob}. The upshot of these reconstruction approaches is evident: one can reconstruct the Hubble parameter $H(z)$ directly from cosmological data without assuming a cosmological model, at least in the traditional sense. Moreover, in the context of the Hubble constant $H_0$, one can determine $H_0$ by simply extrapolating $H(z)$ to $z= 0$. These approaches are commonly labelled "model independent" or "non-parametric", since they do not make use of a traditional "parametric" cosmological model, such as $\Lambda$CDM, $w$CDM, and so on.\footnote{There are some caveats to the "model independent" moniker. One fairly obvious caveat is that reconstructions typically assume correlations, which can lead to interesting results e.g.\ suggestive wiggles in parameters~\cite{Zhao:2017cud, Wang:2018fng}, but may not be tracking effective field theories~\cite{Colgain:2021pmf}. See Ref.~\cite{Pogosian:2021mcs, Raveri:2021dbu} for the pronounced difference a theory prior can make. A secondary concern with GP is that the results depend on the assumed covariance matrix and there may be a tendency to underestimate errors~\cite{OColgain:2021pyh} (see also Ref.~\cite{Dhawan:2021mel}).}

One can construct $H(z)$ directly using Cosmic Chronometers (CC) data~\cite{Jimenez:2001gg,Moresco:2012jh, Moresco:2015cya, Moresco:2016mzx, Moresco:2022phi}, before extrapolating to $z=0$, as explained above.\footnote{The novelty and added value of the CC with respect to other cosmological probes is that it can provide a direct estimate of the Hubble parameter without any cosmological assumption (beyond that of an FLRW metric). From this point of view, the strength of this method is its (cosmological) model independence: no assumption is made about the functional form of the expansion history or about spatial geometry; it only assumes homogeneity and isotropy, and a metric theory of gravity. Constraints obtained with this method, therefore, can be used to constrain a plethora of cosmological models~\cite{Jimenez:2019cll}.} The advantage of the cosmic chronometer method is that it infers $H(z)$ directly from the differential age evolution of the most massive and passively evolving galaxies, selected to minimize any possible contamination by star-forming objects. A current comprehensive review on this cosmological probe is provided in Ref.~\cite{Moresco:2022phi}, and a summary discussed in Sect.~\ref{sec:WG-age}. Current datasets constrain $H_0$ to $67.8^{+8.7}_{-7.2}$ and $66.5\pm5.4$ ${\rm\,km\,s^{-1}\,Mpc^{-1}}$, respectively for a generic open $w$CDM and for a flat $\Lambda$CDM cosmology~\cite{Moresco:2022phi}. Moreover, CC observations are also ideal to derive $H_0$ from GP extrapolations, since not only is GP model agnostic, at least subject to some caveats, but the CC method itself is fully cosmological model independent.

Using 30 CC data points, the Hubble constant is constrained to $H_0=(67.38\pm 4.72){\rm\,km\,s^{-1}\,Mpc^{-1}}$ at 68\% CL~\cite{Wang:2016iij}. Analysing 31 $H(z)$ data measured by the cosmic chronometers technique and 5 $H(z)$ data by BAO observations, the Hubble constant is equal to $H_0= (67 \pm 4){\rm\,km\,s^{-1}\,Mpc^{-1}}$ at 68\% CL~\cite{Yu:2017iju}. Given recent criticism~\cite{Kjerrgren:2021zuo}, some caveat has to be taken when considering the preliminary analysis obtained with CC, while the treatment of both statistical and systematic errors have been fully studied and taken into account in subsequent analyses~\cite{Moresco:2018xdr,Moresco:2020fbm, Borghi:2021rft} However, at $H_0 \sim (67 \pm 4){\rm\,km\,s^{-1}\,Mpc^{-1}}$, any discrepancy with the SH0ES determination~\cite{Riess:2020fzl} is negligible, despite (systematic) errors being possibly underestimated.  

An interesting extension of this analysis, including the SNIa data from the Pantheon compilation and the Hubble Space Telescope (HST) CANDELS and CLASH Multy-Cycle Treasury (MCT) programs, gives instead $H_0=(67.06 \pm 1.68){\rm\,km\,s^{-1}\,Mpc^{-1}}$ at 68\% CL~\cite{Gomez-Valent:2018hwc}, and $H_0=(68.90 \pm 1.96){\rm\,km\,s^{-1}\,Mpc^{-1}}$ when use is made of the Weighted Polynomial Regression technique~\cite{Gomez-Valent:2018hwc}. Observe that the inclusion of SNIa has dramatically reduced the errors and the discrepancy with the SH0ES result~\cite{Riess:2020fzl} becomes $2.9 \sigma$ and $1.8 \sigma$, respectively. A novel feature of this analysis is that since SNIa require a calibrator to determine $H_0$, CC data are being used to calibrate SNIa in an iterative procedure~\cite{Gomez-Valent:2018hwc}. That being said, a discrepancy of $2.9 \sigma$ between two \textit{a priori} "model independent" $H_0$ determinations, one local ($z \ll 1$) and one making use of GP, is an obvious problem if one neglects i) the tendency of GP to underestimate errors and ii) CC data may not have properly propagated some systematic uncertainties. 
 
Through an extension of the standard GP formalism, and utilising joint low-redshift expansion rate data from SNIa, BAO and CC data, the Hubble constant is found to be $H_0=68.52^{+0.94+2.51\,({\rm sys})}_{-0.94}{\rm\,km\,s^{-1}\,Mpc^{-1}}$ at 68\% CL~\cite{Haridasu:2018gqm}, in full agreement with~\cite{Gomez-Valent:2018hwc}. This marks another interesting result, since an effort has been made to better account for systematic uncertainties, which improves on earlier results. Observe that the inclusion of systematic uncertainties reduces another alarming "model independent" $2.9 \sigma$ discrepancy with SH0ES to a much more manageable $1.6 \sigma$. This underscores the importance of systematic uncertainties, especially in the context of CC data as currently addressed in Ref.~\cite{Moresco:2020fbm}. Making use of the same combination of data, i.e.\ SNIa+BAO+CC, while relying only on few seminal assumptions at the basis of the distance duality relation and using a combination of uncalibrated geometrical data, the authors of Ref.~\cite{Renzi:2020fnx} find $H_0=(69.5\pm1.7){\rm\,km\,s^{-1}\,Mpc^{-1}}$, showing the possibility of measuring $H_0$ at the percentage level without assuming a cosmological model. This determination, being midway between SH0ES and {\it Planck} is consistent with both. Whereas current data do not allow for a complete resolution of the Hubble tension, this method hints at a twofold reconciliation for the values of $H_0$ from SH0ES, TRGB~\cite{Freedman:2019jwv,Freedman:2020dne} and {\it Planck}~\cite{Planck:2018vyg}. An adjustment in the calibration of the SNIa, i.e.\ setting $M_B =-19.355 \pm 0.054$, $~2.5\sigma$ lower than SH0ES calibration, bringing local measurements in agreement; and a mild deviation from $\Lambda$CDM in the expansion history at intermediate redshift to bring the latter in agreement with the early time measurement of {\it Planck}.

Interestingly, the $H_0$ inferences from GP are not restricted below $H_0 = 70{\rm\,km\,s^{-1}\,Mpc^{-1}}$. In particular, one can remove SNIa and replace standard BAO with a combination of transversal BAO scale $\theta_{\rm BAO}$, with BBN and CC, to arrive at the value $H_0=72.1^{+1.2}_{-1.3}{\rm\,km\,s^{-1}\,Mpc^{-1}}$ at 68\% CL~\cite{Nunes:2020uex}, which remarkably is completely consistent with the latest SH0ES determination.

Finally, a combined analysis of SNIa, CC, BAO, and H0LiCOW lenses gives $H_0 = (73.78 \pm 0.84){\rm\,km\,s^{-1}\,Mpc^{-1}}$ at 68\% CL~\cite{Bonilla:2020wbn}. Once again, this estimation is in agreement with SH0ES and H0LiCOW collaborations within 68\% CL. This may not be so surprising given that H0LiCOW~\cite{Wong:2019kwg} prefers a value for the Hubble constant $H_0 \sim 73{\rm\,km\,s^{-1}\,Mpc^{-1}}$, but observe that all data are still cosmological in nature.

Clearly, GP regression reconstructions of $H(z)$, and by extrapolation $H_0$, can lead to different results. In particular, the central values range from $H_0 \sim 67{\rm\,km\,s^{-1}\,Mpc^{-1}}$~\cite{Gomez-Valent:2018hwc}, consistent with {\it Planck}~\cite{Planck:2018vyg}, all the way to $H_0 \sim 74{\rm \,km\,s^{-1}\,Mpc^{-1}}$, consistent with SH0ES~\cite{Riess:2020fzl}. Of course, the final outcome depends on the data, but it is worth emphasising that all the observational data are purely cosmological in nature, so the determinations are independent of the local ($z \ll 1$) determinations outlined in Sec.~\ref{sec:direct_H0_measurement}. For this reason, comparison is meaningful. Throughout CC data is playing a special role in calibrating SNIa and BAO, so it is imperative to unlock the potential in CC by fully accounting for systematic uncertainties and adding new data points. In particular, it will be fundamental to improve and validate galaxy stellar population modeling - which currently represents the main source of systematic uncertainty~\cite{Moresco:2020fbm} - and to increase the CC statistics with upcoming galaxy spectroscopic surveys. With these improvements, one distinct possibility is that data reconstructions may converge to a {\it Planck} value~\cite{Gomez-Valent:2018hwc, Haridasu:2018gqm}. If they do, given the model independent setup, then either higher $H_0 > 70{\rm\,km\,s^{-1}\,Mpc^{-1}}$ local determinations are wrong, the cosmological data are wrong, or the Universe is not FLRW, see Sec.~\ref{sec:beyond_FLRW} and Sec.~\ref{sec:cosmic_dipoles}.

\subsection{Gravitational Waves Standard Sirens}

It is possible to have Hubble constant measurements from the GW standard sirens: modelling early data on the GRB170817 jet Ref.~\cite{Guidorzi:2017ogy} gives $H_0= 74^{+11.5}_{-7.5}{\rm \,km\,s^{-1}\,Mpc^{-1}}$ at 68\% CL; late-time GRB170817 jet superluminal motion gives $H_0= 70.3^{+5.3}_{-5.0}{\rm \,km\,s^{-1}\,Mpc^{-1}}$ at 68\% CL~\cite{Hotokezaka:2018dfi}; from the dark siren GW170814 BBH merger Ref.~\cite{DES:2019ccw} finds $H_0= 75^{+40}_{-32}{\rm \,km\,s^{-1}\,Mpc^{-1}}$ at 68\% CL; from GW170817 and 4BBH from O1 and O2 one obtaines $H_0= 68^{+14}_{-7}{\rm \,km\,s^{-1}\,Mpc^{-1}}$ at 68\% CL~\cite{LIGOScientific:2018hze}; finally, from GW190814 and GW170817 one finds $H_0= 70^{+17}_{-8}{\rm \,km\,s^{-1}\,Mpc^{-1}}$ at 68\% CL~\cite{LIGOScientific:2020zkf}.
The most recent result on this observable reported by the LIGO-Virgo-KAGRA Collaboration yields $H_0= 68^{+12}_{-7}{\rm \,km\,s^{-1}\,Mpc^{-1}}$~\cite{LIGOScientific:2021aug}. This was obtained using a method that constrains the source population properties of BBHs without galaxy catalogue, but still combines with GW170817 Bright Siren (without GW170817 the result is $H_0= 50^{+37}_{-30}{\rm \,km\,s^{-1}\,Mpc^{-1}}$). Other recent dark siren results from Ref.~\cite{Palmese:2021mjm}, using only catalogues from Legacy Survey and no galaxy weighting,  presents the value $H_0= 72.77^{+11.0}_{-7.55}{\rm \,km\,s^{-1}\,Mpc^{-1}}$ when combined with GW170817, and $H_0=79.8^{+19.1}_{-12.8}{\rm \,km\,s^{-1}\,Mpc^{-1}}$ without. Finally, we have the latest $H_0= 67^{+6.3}_{-3.8}{\rm \,km\,s^{-1}\,Mpc^{-1}}$ at 68\% CL from Ref.~\cite{Mukherjee:2022afz} cross-correlating dark-sirens and galaxies.

We will discuss the expected improvement in the future using GW in Sec.~\ref{GWsubsub}.

\subsection{Early $H_0$ Measurements}
\label{sec:indirect_H0_measurement}

We consider as "Early $H_0$ measurements" those based on the accuracy of a number of assumptions, such as the model used to describe the evolution of the Universe, i.e.\ the standard $\Lambda$CDM scenario, the properties of neutrinos or the Dark Energy (a cosmological constant), the inflationary epoch and its predictions, the number of relativistic particles, the Dark Matter properties, etc. For this reason the Hubble constant tension can be the indication of a failure of the assumed vanilla $\Lambda$CDM scenario, particularly the form it takes in the pre-recombination Universe. In general all of these measurements are in agreement with a lower value for the Hubble constant.

As we can see in Fig.~\ref{whisker_H0}, there are a variety of CMB probes preferring smaller values of $H_0$. For example, we can see the satellite experiments, such as the Wilkinson Microwave Anisotropy Probe (WMAP9)~\cite{WMAP:2012nax} that assumes the $\Lambda$CDM model, gives a value for the Hubble constant $H_0=(70.0 \pm 2.2){\rm \,km\,s^{-1}\,Mpc^{-1}}$ at 68\% CL in its nine-year data release. Moreover, the {\it Planck} 2018 release~\cite{Planck:2018vyg} predicts $H_0=\left(67.27\pm0.60\right){\rm\,km\,s^{-1}\,Mpc^{-1}}$ at 68\% CL or in combination with the CMB lensing~\cite{Planck:2018vyg}, i.e.\ the four-point correlation function or trispectrum analysis, that estimates $H_0=\left(67.36 \pm 0.54\right){\rm\,km\,s^{-1}\,Mpc^{-1}}$ at 68\% CL. However, in perfect agreement there are the ground based telescopes, such as the ACT-DR4 estimate $H_0=(67.9 \pm 1.5){\rm \,km\,s^{-1}\,Mpc^{-1}}$ at 68\% CL~\cite{ACT:2020gnv}. Alternatively, SPT-3G~\cite{SPT-3G:2021eoc} returns the value of $H_0 = (68.8 \pm 1.5){\rm \,km\,s^{-1}\, Mpc^{-1}}$ at 68\% CL. Finally, they can be combined together so that ACT+WMAP finds $H_0=(67.6 \pm 1.1){\rm\,km\,s^{-1}\,Mpc^{-1}}$ at 68\% CL~\cite{ACT:2020gnv}. 

Yet, also on the low side, we have combinations of complementary probes, such as Baryon Acoustic Oscillation (BAO) data~\cite{Beutler:2011hx,Ross:2014qpa,Alam:2016hwk}, Big Bang Nucleosynthesis (BBN) measurements of the primordial deuterium~\cite{Cooke:2017cwo}, weak lensing and cosmic shear measurements from the Dark Energy Survey~\cite{DES:2021wwk}. For example, the final combination of the BAO measurements from galaxies, quasars, and Lyman-$\alpha$ forest (Ly-$\alpha$) from BOSS and eBOSS prefers $H_0=(67.35 \pm 0.97){\rm\,km\,s^{-1}\,Mpc^{-1}}$ at 68\% CL for inverse distance ladder analysis involving a BBN prior on the physical density of the baryons $\Omega_b h^2$~\cite{eBOSS:2020yzd}. 

Galaxy power spectra can be used to constrain indirectly the Hubble parameter, via the indirect effect of $h$ on the shape of the matter transfer function, in principle without requiring a calibration of the sound-horizon  from the CMB as in inverse-distance-ladder approaches (see Sec.~\ref{sec:soundhorizon}), but using effectively the matter radiation equality horizon calibrated from the CMB, and thus using a different but closely related  inverse-distance-ladder approach.
A reanalysis of the Baryon Oscillation Spectroscopic Survey (BOSS) data on the full-shape (FS) of the anisotropic galaxy power spectrum gives $H_0=(67.9 \pm 1.1){\rm\,km\,s^{-1}\,Mpc^{-1}}$ at 68\% CL~\cite{Ivanov:2019pdj} when the BBN prior on $\Omega_b h^2$ is used, whereas a similar analysis with the baryon-to-dark matter density ratio $\Omega_b/\Omega_{\rm DM}$ fixed to the {\it Planck} baseline $\Lambda$CDM best-fit value gives $H_0=(68.5 \pm 2.2){\rm\,km\,s^{-1}\,Mpc^{-1}}$ at 68\% CL~\cite{DAmico:2019fhj}.
Adding the post-reconstructed BAO signal to the FS + BBN data gives $H_0=(68.6\pm1.1){\rm\,km\,s^{-1}\,Mpc^{-1}}$ at 68\% CL~\cite{Philcox:2020vvt} (fixing the spectral slope $n_s$).
This measurement was recently updated by adding the information from the power spectrum of the eBOSS emission line galaxies~\cite{Ivanov:2021zmi} and the BOSS DR12 bispectrum~\cite{Ivanov:2021haa,Ivanov:2021kcd,Philcox:2021kcw}, yielding $H_0=(68.6\pm 1.1){\rm\,km\,s^{-1}\,Mpc^{-1}}$ and $H_0=(68.3\pm 0.85){\rm\,km\,s^{-1}\,Mpc^{-1}}$, respectively.
Similar results were obtained by other groups, in particular $H_0=(69.2\pm 0.8){\rm\,km\,s^{-1}\,Mpc^{-1}}$ from the anisotropic BOSS power spectrum combined with BAO and BBN measurements~\citep{Chen:2021wdi}, or $H_0=(68.2\pm1.0){\rm\,km\,s^{-1}\,Mpc^{-1}}$ using the galaxy correlation function, BBN priors and BAO measurements from BOSS and eBOSS~\citep{Zhang:2021yna}.
By analyzing the BOSS DR12 full-shape data combined with a prior on $\Omega_m$ from the Pantheon Supernovae, $H_0=(69.5 ^{+3.0}_{-3.5}){\rm\,km\,s^{-1}\,Mpc^{-1}}$ is found at 68\% CL~\cite{Philcox:2020xbv,Farren:2021grl}.

These measurements work within the $\Lambda$CDM model and Standard Model for the early Universe, to set the power spectrum,  with  pros and cons as  discussed in~\cite{Brieden:2021edu, Brieden:2021cfg} where it is shown that without assuming standard early time physics and a $\Lambda$CDM model the $H_0$ constraint likely gets significantly weakened.
If the common fitting formulae of Eisensten \& Hu (1998) are used, as done e.g. for the 6dFGS~\citep[section 5.2 of ][]{Beutler:2011hx}, only a limited range for the sound horizon is allowed. Therefore, the choice of $\Lambda$CDM+SM already sets the range of possible $H_0$ values to $H_0=(67.5\pm2.5){\rm \,km\,s^{-1}\,Mpc^{-1}}$ even when purely angular measurements of $H_{0}r_{\rm{d}}$ from BAO are used.
A reappraisal of the BAO measurement of $H_0$ with different constraints on $\Omega_{m}h^{2}$ yielded $H_{0}=(69\pm2){\rm\,km\,s^{-1}\,Mpc^{-1}}$ (BAO +SN +DES +CMBlensing), without constraints on $\Omega_{b}h^{2}$ from BBN, with a difference of $\approx2.5{\rm\,km\,s^{-1}\,Mpc^{-1}}$ between measurements using CMB lensing power spectra from SPTPol or {\it Planck}~\cite{Pogosian:2020ded}.
Statistically consistent, albeit more uncertain results can also be obtained from CMB lensing alone, $H_0 =(73.5\pm5.3){\rm\,km\,s^{-1}\,Mpc^{-1}}$~\citep{Baxter:2020qlr}. 
Work by the DES-y1 combined with BAO and BBN and obtained $H_0=(67.2^{+1.2}_{-1.0}){\rm\,km\,s^{-1}\,Mpc^{-1}}$ at 68\% CL~\cite{DES:2017txv}. This measurement was carried out within $\Lambda$CDM+SM ($N_{\rm eff}$ fixed to $\approx3.05$) and used the 6dFGS BAO inference, and so it indirectly anchored $H_0.$ It also adopted a compromise BBN estimate so that it would not be in immediate tension with Planck, which however would require a model with free $N_{\rm eff}\approx3.4$~\cite{Cooke:2017cwo}.

\subsection{Supernovae Absolute Magnitude} 

In many of the papers discussing the extended cosmology resolution to the Hubble tension, the distance ladder measurements of $H_0$ are incorporated into the analysis as a Gaussian likelihood. Recently several works~\cite{Camarena:2019moy,Benevento:2020fev,Camarena:2021jlr,Efstathiou:2021ocp,Greene:2021shv,Cai:2021weh,Benisty:2022psx} warned against this practice due to possible caveat on extremely late Universe transitional models. The reason is that local late Universe measurements of $H_0$ depend on observations of astrophysical objects that extend into the Hubble flow, for example the SH0ES~\cite{Riess:2016jrr} results that have been most frequently quoted uses Pantheon supernovae sample in the redshift range $0.023<z<0.15$. Hence the distance ladder $H_0$ actually calibrates the absolute magnitude of the supernovae sample that includes higher redshifts objects. If a model predicts a higher Hubble constant to be achieved by extremely late transitional effect at $z<0.02$,   like the example of red curve in Fig.~1 of Ref.~\cite{Benevento:2020fev} or the hockey-stick dark energy of Ref.~\cite{Camarena:2021jlr}, it is not actually resolving the Hubble tension. Ref.~\cite{Raveri:2019gdp,Camarena:2019moy} give the simplest method to correctly combine the distance ladder measurement of $H_0$:
\begin{equation}
    \mathcal{L} = \mathcal{L}_{\rm SN} \times \mathcal{N}(M,\bar{M}(H_0),\sigma^2_M(\sigma_{H_0})),
    \label{eq:Mcalibration}
\end{equation}
i.e.\ to include the distance ladder $H_0$ measurement as the supernovae absolute magnitude prior. It is possible to marginalize analytically Eq.~\eqref{eq:Mcalibration}, similarly to the analytical marginalization that is performed when a flat prior on $M$ is assumed, see Section~4.6 of Ref.~\cite{Camarena:2021jlr} for details.

This detail should be especially taken care of for extended cosmological models which deviate from $\Lambda$CDM phenomenology at late time (i.e. at low redshift). However, note that the consequence is not as severe as implied by Ref.~\cite{Efstathiou:2021ocp} if the transition is not restrained to $z<0.02$. The main point of the discussion above is that the $H_0$ measured at late time is not exactly the value at $z=0$, thus should be related to the supernovae absolute magnitude calibration at the same range of low redshifts. This means models that hide all the new physics at such low redshifts, which are irrelevant to the $0.02 \lesssim z \lesssim 0.2$ Hubble flow measurements, are naturally off the mark. For example, consider the red curve in Fig.~1 of Ref.~\cite{Benevento:2020fev}. If a model has no drastic jump in the Hubble constant at $z<0.02$, but a gradual change throughout several redshifts, the low redshift Hubble measurement $z\sim \mathcal{O}(0.1)$ could still be a reasonable anchor of current time $H_0$, and should be equivalent to Eq.~\eqref{eq:Mcalibration}.

Besides the previous reasons, Eq.~\eqref{eq:Mcalibration} is recommended for any future analysis using distance ladder $H_0$ combined with supernovae samples because it avoids adopting an unnecessary cosmographic expansion: when performing cosmological inference one should use the luminosity distance of the cosmological model that is being tested. Furthermore, Eq.~\eqref{eq:Mcalibration} avoids double counting low-redshift supernovae belonging to the sample that was used to perform cosmography in order to obtain the $H_0$ constraint.\footnote{The up-to-date prior on $M$ is maintained at \href{https://github.com/valerio-marra/CalPriorSNIa}{github.com/valerio-marra/CalPriorSNIa} and has been implemented in Monte Python 3.5.}
Recently, the SH0ES and Pantheon+ teams have released the SH0ES Cepheid host-distance likelihood,\footnote{\href{https://pantheonplussh0es.github.io}{pantheonplussh0es.github.io}} which allows for the optimal inclusion of the local calibration of the supernova absolute magnitude in the cosmological analysis of a given theoretical model.

\subsection{Distance Ladder Systematics}
\label{sec:H0-sys}

\begin{figure}
\centering
\includegraphics[width=0.5\textwidth]{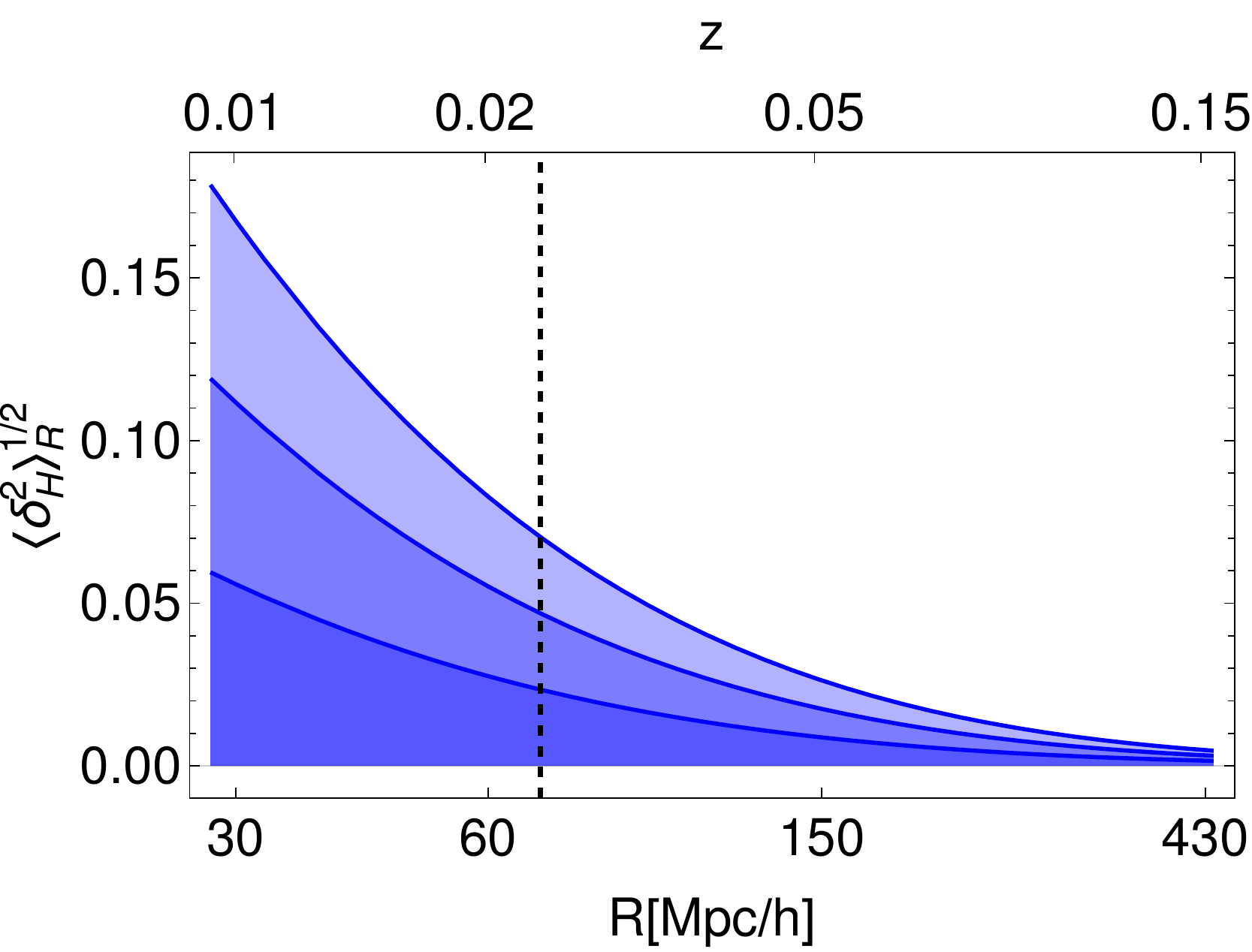}
\caption{1, 2 and 3 times the standard deviation $\left\langle \delta_{H}^2 \right\rangle_R^{1/2}$ as a function of redshift $z$ and scale $R(z)$. The dashed line marks the redshift $z=0.023$. The Hubble constant is estimated by SH0ES in the redshift range $0.023<z<0.15$. Plot from Ref.~\citep{Camarena:2018nbr}.}
\label{cosmic-var}
\end{figure}

For the sake of completeness, this subsection summarizes the outstanding sources of uncertainties in the SNIa distance ladder.
The determination of the local Hubble constant is affected by the large-scale structure around us. Considering a spherical inhomogeneity, one expects that an adiabatic perturbation in density causes a perturbation in the expansion rate given by:
\begin{equation}
\frac{\delta H_0}{H_0} = - \frac{1}{3} f(\Omega_m) \frac{\delta \rho(t_0)}{\rho(t_0)} \,,
\end{equation}
where $f\simeq 0.5$ is the present-day growth function for the concordance $\Lambda$CDM model. Fig.~\ref{cosmic-var} shows the expectation for the standard model: the effect of the local structure quickly decreases by considering larger scales. This is the motivation for placing a limiting lower redshift when fitting the supernova data in order to obtain $H_0$. The SH0ES collaboration adopts $0.023<z<0.15$~\citep{Riess:2020xrj}, for which estimates based on theoretical computations (see Ref.~\cite{Camarena:2018nbr} and references therein) and numerical simulations (see Ref.~\cite{Odderskov:2017ivg} and references therein) suggest that cosmic variance on $H_0$ amounts at 0.5--1\%. It is worth remarking that the SH0ES collaboration tries to correct for this systematic effect by correcting the velocity field via local cosmic flow maps. TRGB estimates adopt instead the redshift range $0.004 < z < 0.083$~\citep{Freedman:2021ahq}. In this case one expects a larger cosmic variance of about 2\%.

Finally, one residual contribution to systematic uncertainties, in determining both $r_s$ and $H_0$ is the uncertainty in the product $H_{0}r_{s}$ from the combination of BAO and Supernovae in the "cosmological" redshift range. Recently, the Pantheon collaboration has re-analyzed the calibration used for the photometry of the SNIa sample, including the calibration used to obtain the SH0ES results~\cite{Brout:2021mpj, Scolnic:2021amr}. The contribution to the Hubble constant from the systematic SN calibration is smaller than $\sim 0.2{\rm \,km\, s^{-1}\, Mpc^{-1}}$, so that the tension cannot be attributed to this type of systematics.


\section{The $S_8$ Tension}
\label{sec:WG-S8measurements}

\noindent \textbf{Coordinators:} Marika Asgari, Hendrik Hildebrandt, and Shahab Joudaki.
\\

\noindent \textbf{Contributors:} Erminia Calabrese, Eleonora Di Valentino, Dominique Eckert, Mustapha Ishak, Mikhail M. Ivanov, Leandros Perivolaropoulos, Oliver H.\,E. Philcox, Foteini Skara.
\bigskip

Recent observations of probes of the large-scale structure have allowed us to constrain
the strength with which matter is clustered in the Universe. These constraints on the  strength of matter clustering differ from that inferred by probes of the early Universe. 
In particular, the primary anisotropies of the cosmic microwave background (CMB) as measured by the {\it Planck} satellite exhibit a tension in the matter clustering strength at the level of $2-3\sigma$ when compared to lower redshift probes such as weak gravitational lensing and galaxy clustering (e.g.~\cite{Asgari:2019fkq,Asgari:2020wuj,Joudaki:2019pmv, DES:2021wwk,DES:2021bvc, DES:2021vln, KiDS:2021opn,  Joudaki:2016kym,Heymans:2020gsg,Hildebrandt:2018yau,DES:2020ahh,Macaulay:2013swa,Skara:2019usd,Kazantzidis:2019dvk,Joudaki:2016mvz,Bull:2015stt,Kazantzidis:2018rnb,Nesseris:2017vor,Philcox:2021kcw}).

\begin{figure*}
\centering
\includegraphics[width=0.7\textwidth]{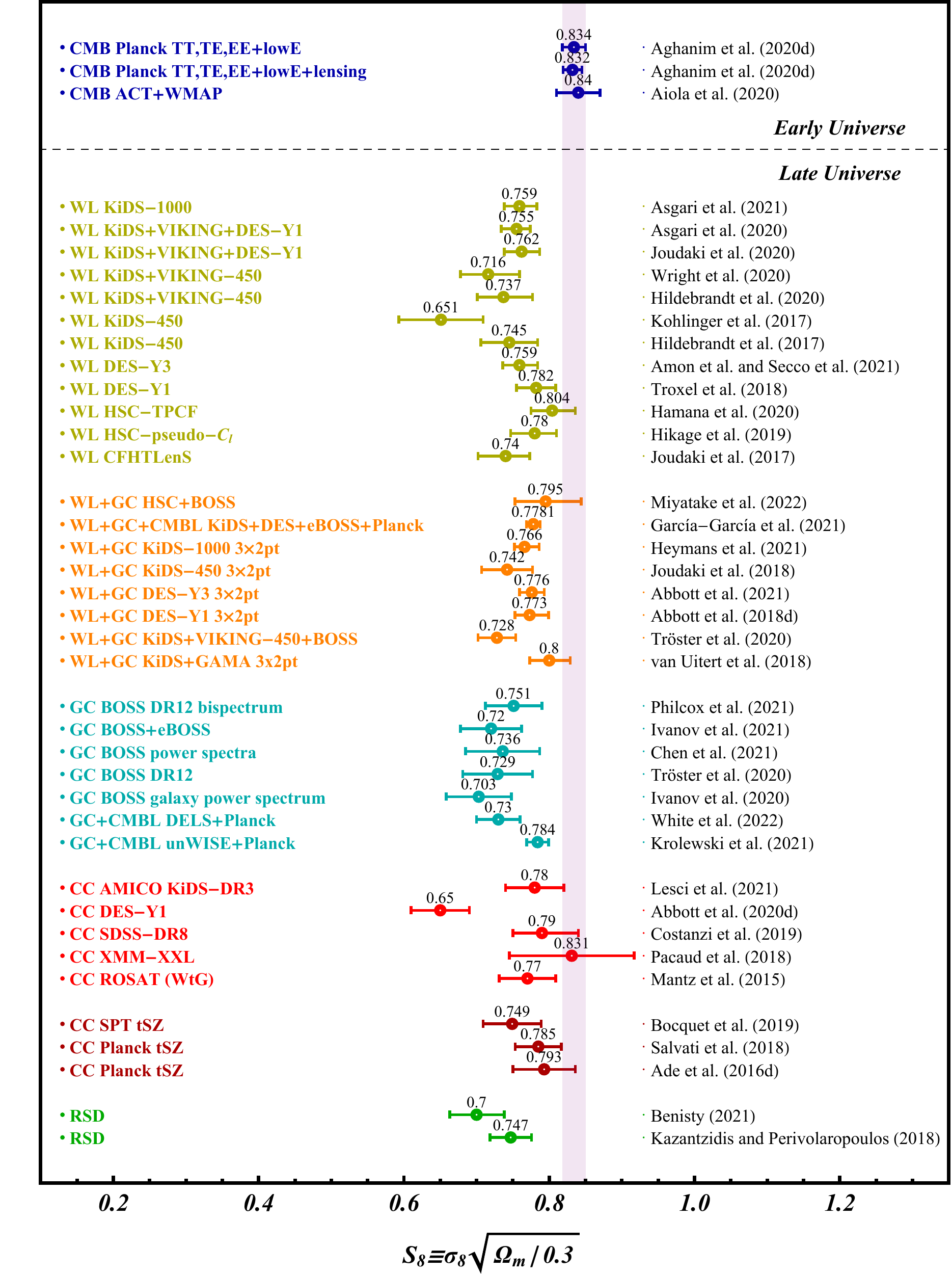}
\caption{Constraints on $S_8$ and its corresponding 68\% error (updated from Ref.~\cite{Perivolaropoulos:2021jda}). We show the nominal reported values by each study, which may differ in their definition of the constraints. 
The definition $S_8= \sigma_8 (\Omega_{\rm m}/0.3)^\alpha$ with $\alpha=1/2$ has been uniformly used for all points. In those cases where $\alpha \neq 1/2$ has been used in some references, the value of $S_8$ with $\alpha =1/2$ was recalculated (along with the uncertainties) using the constraints on $\sigma_8$ and $\Omega_{\rm m}$ shown in those references, assuming their errors are Gaussian. This concerns only 5 CC points where the published value of $\alpha$ was different from $1/2$ and the difference from the published $S_8$ (with different $\alpha$) is very small. The rest of the points are taken directly from the published values. }
\label{whisker_S8}
\end{figure*}

This tension is often quantified using the $S_8 \equiv \sigma_8 \sqrt{\Omega_{\rm m}/0.3}$ parameter, which modulates the amplitude of the weak lensing measurements.
The $S_8$ parameter is closely related to $f\sigma_8(z=0)$ measured by redshift space distortions (RSD)~\cite{Li:2016bis,Gil-Marin:2016wya}, where $f=[\Omega_{\rm m}(z)]^{0.55}$ approximates the growth rate in GR as a function of the matter density parameter, $\Omega_{\rm m}(z)$, at redshift $z$. The lower redshift probes, as seen in Fig.~\ref{whisker_S8}, generally prefer a lower value of $S_8$ compared to the high redshift CMB estimates. Measuring $S_8$ is model dependent and in all cases here the underlying model is the standard flat $\Lambda$CDM model. This model provides a good fit to the data from all probes, but predicts a lower level of structure formation compared to what is expected from the CMB observations. 

The high redshift data from {\it Planck} TT,TE,EE+lowE finds $S_8=0.834\pm0.016$. Combining this data with secondary CMB anisotropies, in the form of CMB lensing, serves to tighten the constraint to $S_8=0.832\pm0.013$. Constraints from CMB lensing alone are fully consistent with both high and other low redshift measurements as seen in Fig.~\ref{whisker_S8}. Combining the high-$\ell$ data from ACT with the low-$\ell$ measurements of WMAP yields larger errors on $S_8=0.840\pm0.030$~\cite{ACT:2020gnv}, but is consistent with the {\it Planck} results. In the following sections, we briefly describe the analyses and data used in Fig.~\ref{whisker_S8} and moreover discuss the possible systematic effects associated with each probe. 

In reporting parameter estimates and constraints, we note that there might be differences in the choices that enter each analysis. For instance, a statistical property of the $S_8$ marginal distribution might be chosen, such as its mean or mode together with the asymmetric $68\%$ region around them or the standard deviation of the sampled points. By contrast, in some analyses, the statistics relevant to the full posterior distribution have been adopted, such as the maximum a posteriori point or the best fitting values and their associated errors. These choices can impact the estimated values of the parameters, in particular when the posterior distributions are significantly non-Gaussian or when the parameter estimates are prior dominated~(see e.g.~Ref.~\cite{Joachimi:2020abi}). For simplicity, we will use the nominal values reported in each analysis, but caution the reader that the methodology used may differ from case to case (see Sec.~\ref{sec:WG-comparison} for a more detailed discussion). 

\subsection{Weak Gravitational Lensing}
\label{sec:WLmeasurements}

The images of distant galaxies are gravitationally lensed by the intervening matter. On cosmological scales we use this effect to analyse and understand the large scale structures (LSS) in the Universe. The gravitational lensing effect of the LSS only slightly distorts the shapes of these galaxies (weak lensing). The two main strands of weak lensing that we focus on in this section are cosmic shear, the study of correlations between the shapes of pairs of galaxies, and galaxy-galaxy lensing, the study of the correlation between the position of foreground galaxies and background galaxy shapes. In addition, galaxy cluster abundance measurements (cluster counts, CC) also utilise gravitational lensing to estimate cluster masses. This will be covered in Sec.~\ref{sec:CC}.

Using cosmic shear we can probe the LSS more directly compared to galaxy-galaxy lensing, where we need to make assumptions about the connection between galaxies and the underlying matter distribution. In addition, cosmic shear studies provide tighter constraints on $S_8$ and can be used as an independent probe. In Sec.~\ref{sec:3x2pt}, we discuss analyses that use galaxy-galaxy lensing in combination with other probes of the large scale structure, as they do not currently provide independent competitive constraints on cosmological parameters. 

The tension in $S_8$ between weak gravitational lensing data and {\it Planck} CMB measurements was first seen when the cosmic shear analysis of CFHTLenS~\citep[Canada France Hawaii Lensing Survey,][]{Heymans:2013fya} was compared to the first {\it Planck} data release~\citep{Planck:2013win}. Back then, the tension was still mild at the level of $\sim2\sigma$. Subsequent analysis of the CFHTLenS data with improved methodology~\cite{Joudaki:2016mvz} validated these initial observations.
This mild tension motivated the weak lensing community to obscure their data by blinding the team to the constraints on cosmological parameters and only unblinding once the analysis has been finalized. The first blinded analysis of this kind was performed by the Kilo Degree Survey on their first 450 square degrees of data~\citep[KiDS-450:][]{Kuijken:2015vca,FenechConti:2016oun,Hildebrandt:2016iqg}, which found a $2.3\sigma$ lower value of $S_8$ compared to the {\it Planck} 2015 constraints~\cite{Planck:2015bpv}. Cosmic shear analysis has seen significant improvements in the recent years and therefore we limit ourselves to results published in the last 5 years (since 2017).

All published results in the last 5 years have found $S_8$ values that are lower than the early Universe estimates. The first year data from the Dark Energy Survey~\citep[DES-Y1,][]{DES:2017qwj} and Hyper Suprime Cam~\citep[HSC][]{HSC:2018mrq,Hamana:2019etx} found slightly higher but consistent values with the KiDS-450 results. Subsequent analysis of KiDS included photometric data from VIKING which reduced the redshift errors and outlier fractions. The analyses of KiDS-450 + VIKING~\cite[KV450,][]{Hildebrandt:2018yau,Wright:2020ppw} found consistently low $S_8$ results. The combination of DES-Y1 and KV450 data were analysed in Refs.~\cite{Joudaki:2019pmv,Asgari:2019fkq}, finding tighter constraints in agreement with previous results. The latest cosmic shear analysis of KiDS~\citep[KiDS-1000,][]{Asgari:2020wuj} and DES~\citep[DES-Y3,][]{DES:2021bvc,DES:2021vln} show the same trend with better precision. Comparing their $S_8$ values with the fiducial {\it Planck} analysis, both surveys find  values that are $\sim2-3\sigma$ smaller.\footnote{Note that the tension level changes depending on the assumed method, see Sec.~\ref{sec:WG-comparison} for details.}

\subsection{Weak Lensing Systematics}
\label{sec:WLsys}

Low-redshift measurements of large-scale structure evolution require sophisticated analysis pipelines to correct for a large number of known systematic effects. Here, we discuss the most prominent systematic uncertainties affecting the weak lensing measurements.

\subsubsection{Shape Measurements} 
    
The basic observable in any weak gravitational lensing measurement is the ellipticity of a galaxy. The measured ellipticity is a combination of the intrinsic galaxy ellipticity and the shear effect of gravitational lensing. In the weak lensing regime the small shear is completely dominated by the large intrinsic ellipticity making it necessary to average over many galaxies. The idea behind this is that galaxies are intrinsically randomly oriented on the sky and the intrinsic ellipticities average to zero leaving only the shear (but see intrinsic alignments below). Under the assumption of random intrinsic orientation, a galaxy ellipticity is an unbiased but extremely noisy estimator of the shear.

In cosmic shear, pairs of galaxies are analyzed for coherent shear contributions to their ellipticities as a function of their distance. This typically involves many millions and soon billions of faint galaxy images. In practice, one has to estimate the ellipticities of these very large samples of galaxies from fuzzy, noisy images that make up a few pixels on a detector each. With weak lensing being a statistical measurement, it is not crucially important to estimate a highly accurate ellipticity measurement for every single galaxy. Rather it is important to control the average bias in the shear estimates to very low values. For current Stage-III surveys such as KiDS, DES, and HSC, biases in the shear need to be controlled to a level of $\sim 10^{-2}$, with future Stage-IV experiments requiring almost an order of magnitude better systematics control~\cite{EUCLID:2011zbd,LSSTDarkEnergyScience:2018jkl}.

Several effects make this a very delicate problem. The atmosphere (in case of ground-based observations) and the telescope and camera optics introduce distortions that are significantly larger than typical shear values. These have to be corrected to high accuracy using PSF (point-spread-function) information from stars before the galaxy-based shear estimates can be used in cosmological measurements~\cite{Kaiser:1994jb}. In the low signal-to-noise ratio regime, the measurements further suffer from noise bias that needs to be corrected~\cite{Melchior:2012un}. Depending on the actual algorithm, other biases like model bias can play a role~\cite{Voigt:2009mj,Miller:2012am}. Furthermore, any sample selection (e.g.~through source detection~\cite{Hoekstra:2020zyl} and tomographic binning via photometric redshifts) can introduce further biases. The electronics of CCDs also play a crucial role and must be understood at a deep level~\cite{Rhodes:2010yg,Gruen:2015pcc,Boone:2018iqq,coulton18,Massey:2009wv}.

In order to reach the required accuracy, it is usually necessary to simulate the whole process with a dedicated suite of image simulations, often based on or informed by high-resolution HST (Hubble Space Telescope) observations~\cite{Rowe:2014cza,FenechConti:2016oun,Kannawadi:2018moi,Hoekstra:2016kaj,DES:2020lsz,Mandelbaum:2017ctf,Bruderer:2015bsa,Li:2021mvq}. Only in this way can the crucial multiplicative bias, which describes the systematic over-/under-estimation of shear values, be calibrated. The community has engaged in several blind challenges over the past two decades to objectively determine the remaining multiplicative biases in weak lensing measurements~\cite{Heymans:2005rv,Massey:2006ha,Bridle:2009gg,Kitching:2012rt,Mandelbaum:2014fta}. State-of-the-art algorithms suppress this bias to very small levels at runtime~\cite{Huff:2017qxu,Sheldon:2017szh,Sheldon:2019uxq,Kannawadi:2020rok,Hoekstra:2021plm}. But it is really the robustness to small differences between the image simulations that are ultimately used for calibration and the real Universe that determines the quality of a shape measurement method.

A residual bias in the shear estimates scales almost linearly with $S_8$. Hence, the community has put a lot of effort over the past two decades (arguably more effort than into any other weak lensing systematic) into controlling these biases to the percent level in current surveys and is working towards an even better control for stage-IV experiments. As such, the current $\sim 5-10\%$ tension in $S_8$ between cosmic shear measurements and {\it Planck} is most probably not due to uncontrolled systematics in the shear measurement.

\subsubsection{Photometric Redshifts} 
    
The cosmic shear signal measured from wide-field imaging surveys scales strongly with redshift~\cite{VanWaerbeke:2006qt}. The higher the redshift of the sources, the stronger the signal due to the increased light path and integrated deflection and distortion of the light-bundles along their way. As the measurement is inherently statistical, averaging the signal from millions of source galaxies, it is the ensemble redshift distribution of all sources that needs to be known to high accuracy in order to obtain unbiased model predictions for the cosmic shear signal. This accuracy is traditionally summarised in the uncertainty of the mean redshift of all sources, as the integrated nature of the cosmic shear effect means that higher-order moments of the redshift distribution are less important for accurate predictions. In order to allow measurements at different cosmic epochs and increase the constraining power for various cosmological parameters, the galaxy source sample is typically split up into tomographic redshift bins via individual photometric redshift estimates.

Both tasks, determining \emph{precise} individual galaxy redshifts for the tomographic binning as well as estimating \emph{accurate} redshift distribution(s) for the ensemble(s), require photometric techniques because of the large number of faint sources that makes spectroscopic measurements of complete weak lensing source samples infeasible. Typically, a subset of sources is observed spectroscopically and can be used to validate, train, and calibrate the photometrically determined redshifts and redshift distributions.

The individual photometric redshifts used for binning the galaxies along the line-of-sight are unimportant for the accuracy of $S_8$ measurements from cosmic shear. Rather, their precision influences the signal-to-noise ratio of those measurements and the constraining power of a survey. Thus, these individual photometric redshifts cannot cause a bias in $S_8$ and will not be discussed in the following.
By contrast, the accuracy of the redshift distribution  directly impacts the accuracy of cosmological parameter estimates~\cite{Huterer:2005ez}. Any bias in the mean redshifts of the tomographic bins causes a bias in $S_8$.\footnote{This is strictly only true in the non-tomographic case or in case of coherent biases of all tomographic bins. Incoherent redshift biases of different tomographic bins can cause e.g. spurious IA signals.} If the mean redshift estimate is biased high the $S_8$ estimate will be biased low and vice versa. In practice, one marginalises over the uncertainty in the mean redshifts.

For contemporary Stage-III weak lensing surveys, the requirement on the accuracy of the mean redshifts of the tomographic bins is on the order of 0.01~\cite{Hildebrandt:2016iqg,Tanaka:2017lit,DES:2017ndt}. This is necessary for the marginalisation over the redshift uncertainty to not compromise the statistical power of the surveys. Stage-IV surveys such as {\it Euclid} and LSST will require a calibration that is more accurate by a factor of $\sim5-10$ to reach that goal~\cite{EUCLID:2011zbd,LSSTDarkEnergyScience:2018jkl}. The importance of this redshift calibration has triggered a lot of work that aims at minimising biases in the determination of the source redshift distributions by different techniques that rely on different types of (typically spectroscopic) calibration samples. Broadly, these can be categorised as either colour-based or position-based techniques.

The colour-based techniques typically employ a deep spectroscopic calibration/reference sample that is as representative of the sources as possible. Any mismatch in redshift between the source and the reference samples needs to be removed before a reliable redshift distribution can be estimated. This is usually done by some re-weighting technique that takes the relative densities of both samples in high-dimensional colour/magnitude space into account~\cite{Lima:2008kn,Gruen:2016jmj,DES:2020lea}. Early attempts used $k$-nearest-neighbour counting to estimate these densities whereas most contemporary surveys have employed self-organising maps~\cite{Masters:2015asa,DES:2019bxr,Wright:2019fwm,DES:2020ebm,Wright:2020ppw,Nishizawa:2020gac}. This latter technique has the advantage of easy visualisation and allows culling of the source sample to remove parts of colour space that are not covered by the spectroscopic reference sample. Covering the colour/magnitude space of future stage-IV weak lensing experiments with deep spectroscopic observations is a huge task. Dedicated spectroscopic surveys on the biggest available optical and near-infrared telescopes are being conducted to secure this essential calibration data~\cite{Newman:2013cac,Masters:2017fho,Masters:2019ptz,Euclid:2021ueh,Euclid:2020xbm}.

Complementary to the colour-based redshift calibration are techniques that exploit the fact that galaxies are clustered~\cite{Newman:2008mb,Hildebrandt:2016iqg,Hildebrandt:2020rno,Morrison:2016stl,DES:2020rlj,vandenBusch:2020lur}. Using a wide-area spectroscopic reference sample that covers the whole redshift range of interest, the redshift distributions of the weak lensing source sample can be determined via an angular cross-correlation approach. The amplitude of the angular cross-correlation between thin redshift slices of the reference sample and the unknown target sample (e.g.~tomographic bins) is directly related to the target sample's redshift distribution. The great advantage of this technique is that the reference samples does not have to be representative of the target sample, e.g.~one can use a bright reference sample to estimate the redshift distribution of a faint target sample since both samples cluster with each other. Correcting for the redshift evolution of galaxy bias in both samples is crucial because it is degenerate with the amplitude of the redshift distribution. This can be done via angular auto-correlation measurements for the reference sample and consistency checks for the target sample.

Both techniques of redshift calibration have been thoroughly tested against each other and on increasingly complex simulations over the past few years. So far, there is no indication that the $S_8$ tension is caused by a bias in the redshift calibration. A substantial amount of work still needs to be carried out to get those techniques ready for Stage-IV experiments, possibly combining colour and position information~\cite{Alarcon:2019jyt,Sanchez:2018wuh,LSSTDarkEnergyScience:2021ozf}, but the current $S_8$ results from weak lensing seem quite robust in the light of these tests.

\subsubsection{Intrinsic Galaxy Alignments} 
    
The intrinsic alignment (IA) of galaxies constitutes one of the most significant sources of systematic uncertainties in weak lensing (e.g.~Refs.~\cite{Hirata:2004gc,Troxel:2014dba,Heymans:2013fya,Mandelbaum:2017jpr,Bridle:2007ft,Heymans:2003pz,Joudaki:2016mvz,Vlah:2019byq,Leonard:2018aow,Blazek:2011xq,Blazek:2015lfa,Krause:2015jqa,Kirk:2010zk,Kirk:2011aw,Blazek:2017wbz,Joachimi:2013una,Chisari:2015qga,Johnston:2018nfi,Fortuna:2020vsz,Samuroff:2020gpm}). It was shown in Refs.~\cite{Bridle:2007ft,Heymans:2003pz} that if intrinsic alignments are ignored in Stage-IV weak lensing analyses (such as {\it Euclid} and LSST), the equation of state of dark energy will be biased by up to $50\%$ and the amplitude of matter fluctuations will be biased by up to $30\%$. Intrinsic alignments of galaxies refer to correlations between galaxies that arise due to the tidal gravitational field where these galaxies formed. These intrinsic alignments produce a spurious signal that contaminates the genuine lensing signal and any cosmological information inferred from it. Interestingly, there are two types of intrinsic alignments that affect lensing on large scales. The first is due to relatively close galaxies being radially aligned by the same dark matter structure, known as the "intrinsic shear -- intrinsic shear", or II, correlation. The second type is due to the fact that a dark matter structure aligns radially neighbouring galaxies and at the same time tangentially shears the background galaxy images, thus creating an anti-correlation between the two known as the "gravitational shear -- intrinsic shear" correlation, or GI, term. See reviews in Refs.~\cite{Troxel:2014dba,Kiessling:2015sma,Kirk:2015nma} and references therein.
    
After two decades of developing methods to mitigate the effects of intrinsic alignment contamination to the lensing signal in photometric galaxy surveys, two methods have emerged as being the most efficient at addressing the problem. One is the common marginalization method where an IA model is assumed and its parameters are constrained along with the cosmological parameters (see e.g.~Ref.~\cite{Joudaki:2017zdt}). The other method is the self-calibration method that is based on calculating additional correlations from the same lensing survey by taking into account the positions of the lens with respect to the sources within the same bin and then using a scaling relation that relates such a correlation to cross-correlations between bins. This method strongly complements the marginalization method as it does not need to assume an IA model and is able to separate the IA signal from the lensing signal (see e.g.\ Ref.~\cite{Troxel:2011za}). It was shown in Ref.~\cite{Yao:2017dnt} that not correcting for intrinsic alignments can shift cosmological parameters as determined from weak lensing and that correlations between the IA amplitude and other parameters are present. Similarly, it was shown earlier in Ref.~\cite{Dossett:2015nda} that the IA amplitude has some correlation with $S_8$ but that it is not significant enough to explain the $S_8$ tension. This was confirmed in Ref.~\cite{Yao:2017dnt}. It is fair to assert that while intrinsic alignments need to be corrected for in an era or precision cosmology, these cannot be responsible for a significant tension in $S_8$ between lensing and the CMB.

\subsubsection{Matter Power Spectrum Including Baryonic Feedback} 
 
The matter power spectrum can be accurately obtained on linear scales via Boltzmann codes such as CAMB~\cite{Lewis:1999bs} and CLASS~\cite{Blas:2011rf}. In order to obtain the matter power spectrum on nonlinear scales, numerical simulations need to be carried out at multiple cosmologies spanning the full parameter space of interest (see e.g.\ Refs.~\cite{Lawrence:2009uk,Schaye:2009bt,Heitmann:2015xma,McCarthy:2016mry,Euclid:2018mlb,Euclid:2020rfv}). An emulator for the simulations or a fitting function calibrated to the simulations can then be created. Examples of emulators include \textsc{Cosmic Emu}~\cite{Lawrence:2009uk,Heitmann:2013bra,Lawrence:2017ost} and \textsc{Euclid Emulator}~\cite{Euclid:2018mlb,Euclid:2020rfv}, while the two most popular fitting functions are \textsc{Halofit}~\cite{Smith:2002dz,Takahashi:2012em} and \textsc{HMCODE}~\cite{Mead:2015yca, Mead:2016zqy, Mead:2020vgs}. On nonlinear scales, the impact of baryonic processes such as radiative cooling, star formation, and feedback from supernovae and active galactic nuclei (AGN) also needs to be taken into account (see e.g.~Ref.~\cite{Chisari:2019tus} for a review). A particular benefit of \textsc{HMCODE} is that, in addition to calibrating against dark matter simulations, it accounts for these baryonic effects by calibrating against hydrodynamical simulations (originally to the OverWhelmingly Large Simulation, or \textsc{OWLS}, suite~\cite{Schaye:2009bt,vanDaalen:2011xb,Brun:2013yva} and most recently to the BAryons and HAloes of MAssive Systems, or \textsc{BAHAMAS}, suite~\cite{McCarthy:2016mry}). 
    
The uncertainties in the dark matter simulations and emulators of the matter power spectrum are at the few-percent level down to scales of $k \sim 10 \, h \, {\rm Mpc}^{-1}$~\cite{Lawrence:2017ost,Euclid:2020rfv}. 
These uncertainties grow to 5\%-10\% for the fitting functions for the same highly nonlinear scales~\cite{Takahashi:2012em,Mead:2020vgs}. Meanwhile, the uncertainties in the hydrodynamical simulations boil down to the choices that need to be made in regards to the stellar and hot gas content. The uncertainty in the simulations themselves are at the few-percent level (see e.g.\ Refs.~\cite{vanDaalen:2011xb,Chisari:2018prw}), but the impact of baryonic feedback on the matter power spectrum differs substantially between different simulations (suppressing the power between $10$--$30\%$ for scales in the range of $k\sim$~few and $20 \, h \, {\rm Mpc}^{-1}$) due in particular to their different choices for the sub-grid model, resolution, and calibration strategy~\cite{Chisari:2018prw,Huang:2018wpy,Chisari:2019tus}.
Further work is therefore needed to capture the impact of baryonic feedback on the nonlinear matter power spectrum to the precision needed for next-generation cosmological surveys. While the KiDS~\cite{Hildebrandt:2016iqg,Hildebrandt:2018yau,Asgari:2020wuj,Joudaki:2016kym,Kohlinger:2017sxk,Joudaki:2017zdt,vanUitert:2017ieu,Heymans:2020gsg} and HSC~\cite{HSC:2018mrq,Hamana:2019etx} analyses have attempted to account for the uncertainty in the matter power spectrum due to baryonic feedback by introducing one or more nuisance parameters, the fiducial approach in the DES analyses~\cite{DES:2017qwj,DES:2021vln,DES:2021bvc,DES:2017myr,DES:2021wwk} has been to avoid the uncertainty altogether by imposing more conservative scale cuts.

\subsubsection{Small-Angle Approximations}

The Limber and flat-sky approximations~\cite{Limber:1954zz,LoVerde:2008re} are commonly employed in cosmic shear analyses and are known to induce an uncertainty in the computation of the angular power spectra and correlation functions on large scales~\cite{Joudaki:2011nw, Kitching:2016zkn,Kilbinger:2017lvu,Lemos:2017arq}. However, as the impact of these approximations on the observables is smaller than the size of the sample variance, they do not contribute to the $S_8$ tension and will even remain sufficient approximations for next-generation surveys such as {\it Euclid} and {\it Rubin}/LSST~\cite{Kilbinger:2017lvu,Lemos:2017arq}. We note that in the the case of galaxy clustering and galaxy-galaxy lensing, the Limber approximation cannot be reliably used for these surveys (see e.g.\ Refs.~\cite{Giannantonio:2011ya, Fang:2019xat}).

\subsection{Redshift-Space Galaxy Clustering}

The $S_8$ parameter can also be measured from redshift-space galaxy clustering statistics, such as the galaxy power spectrum and bispectrum. These statistics are commonly used to extract the angular diameter and Hubble distances relative to the sound horizon along with the growth rate parameter, $f\sigma_8(z)$, from redshift-space distortions, all at the effective redshift of the galaxy sample. In combination with CMB surveys, which calibrate the expansion history, such measurements can be used to constrain $\sigma_8$ and $S_8$. Combined analyses of SDSS and {\it Planck} data found $\sigma_8 = 0.829\pm 0.016$ in $\nu\Lambda$CDM~\citep{Alam:2016hwk}, or $\sigma_8 = 0.8115^{+0.009}_{-0.007}$ when further including $f\sigma_8(z)$ measurements from eBOSS~\citep{eBOSS:2020yzd}. The former is equivalent to $S_8 = 0.843\pm0.016$, and both are consistent with the CMB-only analyses. We caution that these measurements are CMB-dominated, thus this is not a rigorous test of the purported $S_8$ tension.

Recently, it has been shown that $S_8$ can be measured from spectroscopic surveys without external calibration, by fitting the full shape of the observed power spectrum and bispectrum with an accurate theoretical model, analogous to that performed in CMB analyses~\citep{DAmico:2019fhj,Ivanov:2019pdj,DAmico:2020kxu,Wadekar:2020hax,Philcox:2021kcw,Ivanov:2021zmi,Chen:2021wdi,Kobayashi:2021oud,Zhang:2021yna,Colas:2019ret,Hamann:2010pw,Schoneberg:2019wmt,Troster:2019ean,Chen:2020zjt}. Natively, this measures the $\sigma_8$ parameter rather than $S_8$; however, $\Omega_{\rm m}$ can be measured from the power spectrum shape (or alternative datasets such as supernovae), allowing an $S_8$ constraint to be obtained. Furthermore, $\Omega_{\rm m}$ is generally well constrained and takes values consistent with analyses of supernovae and the CMB, such that any $S_8$ tension naturally translates into a $\sigma_8$ tension.

Recent measurements of $S_8$ from BOSS full-shape power spectra are consistent with the value found from other low-redshift probes: $S_8=0.703\pm0.045$~\cite{Ivanov:2019pdj} or $S_8 = 0.704\pm0.051$~\citep{DAmico:2019fhj}.\footnote{These measurements and a number of other results based on SDSS-III Fourier-space statistics are $\sim1\sigma$ too low due to a misnormalization of the public BOSS power spectra~\citep{Philcox:2021kcw,Beutler:2021eqq}. This occurred due to incorrect treatment of small-scale power in the survey window function, resulting in a $10\%$ suppression of the power spectrum multipoles, and affects any analyses using the BOSS DR12 Fourier-space data prior to~\citep{Beutler:2021eqq}.} This was recently refined by adding information present in the power spectrum of the eBOSS emission-line galaxies~\cite{Ivanov:2021zmi} as well as the BOSS DR12 bispectrum~\cite{Philcox:2021kcw}, yielding $S_8=0.720\pm 0.042$ and $S_8=0.751\pm 0.039$, respectively. These results are similar across various analysis choices and statistics; for example, Ref.~\citep{Chen:2021wdi} found $S_8 = 0.736\pm0.051$ using the BOSS power spectra alone, Ref.~\citep{Kobayashi:2021oud} found $S_8 = 0.740^{+0.043}_{-0.041}$ in an emulator-based analysis, and Ref.~\citep{Troster:2019ean} found $S_8 = 0.729\pm0.048$ using the BOSS correlation function. In all cases, the results are somewhat lower than {\it Planck}, though not at particularly large significance (typically less than 3$\sigma$). Systematic uncertainties affecting these measurements are described in Sec.~\ref{rsdsec}. No major source of systematic that might significantly affect the measurements
has been identified. 

Finally, the clustering density can be measured from other probes such as the one-dimensional Ly$\alpha$ power spectrum~\citep{Palanque-Delabrouille:2019iyz}
and peculiar velocity surveys~\citep{Boruah:2019icj}. These find similarly low values of $S_8$ in mild tension with {\it Planck}. 

\subsection{Combined Weak Lensing and Galaxy Clustering}
\label{sec:3x2pt}

The self-consistent combined analysis of cosmic shear, galaxy-galaxy lensing, and galaxy clustering at the level of their two-point statistics is commonly referred to as a ``$3\times{\rm 2pt}$'' analysis. The self-consistency of this analysis allows for an improved self-calibration of the systematic uncertainties that affect the observables, such as galaxy bias and intrinsic galaxy alignments. The first three $3\times{\rm 2pt}$ analyses were published approximately simultaneously, using the datasets of
KiDS-450$\times$\{2dFLenS+BOSS\}~\cite{Joudaki:2017zdt}, KiDS-450$\times$GAMA~\cite{vanUitert:2017ieu}, and
DES-Y1~\cite{DES:2017myr}. We now also have the $3\times{\rm 2pt}$ analyses of
KiDS-1000$\times$\{2dFLenS+BOSS\}~\cite{Heymans:2020gsg} and
DES-Y3~\cite{DES:2021wwk}. Here, the KiDS-1000 analysis~\cite{Heymans:2020gsg} did not only contain additional imaging data, but further included the full 2dFLenS and BOSS datasets, which had been previously restricted to the areas overlapping with KiDS in Ref.~\cite{Joudaki:2017zdt}. This had the benefit of improving the constraining power from galaxy clustering and the drawback of diminishing the importance of the galaxy-galaxy lensing~\cite{Joachimi:2020abi,Heymans:2020gsg}.

Despite the same $3\times{\rm 2pt}$ terminology, there are subtle differences between the different analyses. In particular, the KiDS Collaboration has performed combined analyses of overlapping imaging and spectroscopic surveys, while the DES Collaboration has taken the approach of performing ``internal'' $3\times{\rm 2pt}$ analyses where the photometric galaxies are used as both lenses and sources. As a result, the analyses in Refs.~\cite{Joudaki:2017zdt,Heymans:2020gsg} are the only to consider redshift-space galaxy clustering instead of angular galaxy clustering.

The $3\times{\rm 2pt}$ constraints on $S_8$ and the matter density are driven by the cosmic shear and galaxy clustering, respectively. Here, we focus on the former, where $S_8 = 0.742^{+0.035}_{-0.035}$ in KiDS-450$\times$\{2dFLenS+BOSS\}~\cite{Joudaki:2017zdt}, $S_8 = 0.800^{+0.029}_{-0.027}$ in KiDS-450$\times$GAMA~\cite{vanUitert:2017ieu}, and $S_8 = 0.794^{+0.029}_{-0.027}$ in DES-Y1~\cite{DES:2017myr}, which were followed by the updated constraints on $S_8 = 0.766^{+0.020}_{-0.014}$ in KiDS-1000$\times$\{2dFLenS+BOSS\}~\cite{Heymans:2020gsg} and $S_8 = 0.775^{+0.026}_{-0.024}$ in DES-Y3~\cite{DES:2021wwk}. The KiDS-450 and KiDS-1000 analyses using the 2dFLenS and BOSS galaxies are highly consistent with one another (to well within a standard deviation). These analyses are also consistent with Ref.~\cite{vanUitert:2017ieu}, where the clustering of the GAMA galaxies favored a larger value of $S_8$. Likewise, the DES-Y1 and DES-Y3 constraints on $S_8$ are highly consistent with one another (to well within a standard deviation). Meanwhile, in comparing the constraints on $S_8$ from KiDS$\times$\{2dFLenS+BOSS\} and DES, the posteriors seem to have approached one another in the latest analyses and are both in tension with {\it Planck} at the $2$--$3\sigma$ level. However, we note that a comparison of the published constraints on $S_8$ should be carried out with caution as the two collaborations employ a set of different assumptions, which can have a substantial impact on the parameter constraints~\cite{Joudaki:2019pmv}.

We further highlight the $2\times{\rm 2pt}$ analysis of galaxy-galaxy lensing and galaxy clustering with the HSC-Y1 and BOSS datasets in Ref.~\cite{Miyatake:2021sdd}, where $S_8 = 0.795^{+0.049}_{-0.042}$ is in agreement with both {\it Planck} and all of the $3\times{\rm 2pt}$ analyses.
Additional multi-probe analyses have resulted in low values of the $S_8$ parameter. In particular, the combined analysis of unWISE galaxy clustering and the cross-correlation with the {\it Planck} CMB lensing reconstruction resulted in $S_8 = 0.784 \pm 0.015$~\cite{Krolewski:2021yqy}. 
The corresponding analysis using the luminous red galaxies of the DESI Legacy Imaging Surveys (DELS) and {\it Planck} CMB lensing reconstruction resulted in $S_8 = 0.73 \pm 0.03$~\citep{White:2021yvw}.
Further, the self-consistent combined analysis of KiDS-1000 cosmic shear, Dark Energy Survey Year 1 cosmic shear and galaxy clustering, eBOSS quasars, {\it Planck} CMB lensing reconstruction, and photometric galaxies in DELS, including the different cross-correlations, gives $S_8 = 0.7781 \pm 0.0094$~\cite{Garcia-Garcia:2021unp}. A consistent picture therefore seems to be emerging that combined analyses of weak lensing and galaxy clustering are in tension with the {\it Planck} CMB temperature measurement of the $S_8$ parameter.

\subsection{Galaxy Cluster Counts}
\label{sec:CC}

The number density of the most massive dark matter halos in the Universe is highly sensitive to the growth of structure and hence to $S_8$ through the cosmological dependence of the halo mass function (see Ref.~\citep{Allen:2011zs} for a review). Comparing the CMB data with Sunyaev-Zel'dovich (SZ) number counts, Ref.~\cite{Planck:2013lkt} was the first to note a tension at the level of $\sim3\sigma$ between the predictions based on the fiducial {\it Planck} CMB $\Lambda$CDM model and the observed cluster counts. This tension was confirmed in the full mission data~\citep{Planck:2015lwi} and other cluster count experiments based on SZ (SPT~\cite{SPT:2018njh}; ACT~\cite{Hasselfield:2013wf}), X-ray (400d~\cite{Vikhlinin:2008ym}; RASS-WtG~\cite{Mantz:2014paa}; XMM-XXL~\cite{XXL:2018ryw}), or optical selection (SDSS~\cite{DES:2018crd}; DES~\cite{DES:2020ahh}). The fiducial CMB $\Lambda$CDM model implies there should be about twice as many very massive clusters in the local Universe compared to most studies. The measurements of $S_8$ from various cluster count experiments are summarized in Fig.~\ref{whisker_S8}. In Ref.~\cite{Pratt:2019cnf}, it is shown that the various cluster count results obtained by different groups and different detection techniques agree well with one another and with the cosmic shear results, providing an unweighted mean $S_8=0.789\pm0.012$ which is formally different from the CMB expectation (for example {\it Planck} gives $S_8=0.834\pm0.016$ assuming $\Lambda$CDM) at more than $2\sigma$ level.

The leading source of systematic uncertainty in cluster count experiments is the calibration of the relation between mass and observable (see Ref.~\cite{Pratt:2019cnf} for a review). While the halo mass function and its cosmological dependence are usually based on DM-only simulations (e.g.\ Ref.~\cite{Tinker:2008ff}), current galaxy cluster detection techniques rely on selection through the cluster's baryonic content, either through the hot gas (SZ, X-ray) or the clustering of galaxies (Friends-of-Friends, red sequence, match filter). The relation between the survey observable (e.g.\ the integrated Compton parameter $Y_{\rm SZ}$ in the case of SZ experiments) and the mass of the host halo needs to be accurately understood to infer cosmological parameters. For instance, the {\it Planck} SZ cluster counts may be reconciled with the CMB predictions in case the estimated halo masses were biased by a factor $1-b = M_{\rm obs}/M_{\rm true} = 0.58\pm0.04$~\cite{Planck:2015lwi}. While early cluster count experiments relied on external observable-mass scaling relations~\cite{Planck:2013lkt,Planck:2015lwi,Vikhlinin:2008ym}, modern experiments use internally-calibrated relations with available WL data for a subset of systems and attempt to marginalize over the uncertainties in the mass bias by performing a joint inference of the cosmological parameters and the scaling relations~\cite{Pierre:2016mfc,SPT:2018njh,Garrel:2021sgq}. Future cluster count experiments such as eROSITA~\cite{eROSITA:2012lfj} and SPT-3G~\cite{SPT-3G:2014dbx} will take advantage of upcoming wide-field WL data from e.g.\ {\it Euclid} to improve both on the statistical precision by including a much larger number of sources and on the accuracy by limiting the systematic uncertainties in the mass calibration~\cite{Grandis:2018mle}.

For the $S_8$ constraints in Fig.~\ref{whisker_S8}, the errors are propagated according to $\sigma_{S_8}^2=(\Omega_{\rm m}/0.3)^{2\alpha}\sigma_{\sigma_8}^2+\sigma_8^2\alpha^2(\Omega_{\rm m}/0.3)^{2\alpha-2}\sigma_{\Omega_{\rm m}}^2$, with the index $\alpha=1/2$. Here, $\sigma_8$ and $\Omega_{\rm m}$ are assumed to be Gaussian distributed and are considered at their published best-fit values.

\subsection{CMB $S_8$ Measurements}

The amplitude of the CMB power spectrum and even more so the lensing of the CMB set tight bounds on the matter density and on $\sigma_8$, which is usually a derived parameter in CMB analyses. The CMB estimates of these parameters and of their combination, $S_8$, strongly depend on the assumptions made in the fits and on the specific data combination~\cite{Calabrese:2013jyk,Calabrese:2017ypx,Henning:2017nuy,ACT:2020gnv}. Lately, with uniformity in analyses and with new observations, the CMB estimates have converged on values larger than the galaxy-based estimates of $S_8$, as shown in Fig.~\ref{whisker_S8}. 

As already mentioned above and as shown later in Sec.~\ref{sec:WG-BothSolutions}, this measurement is model dependent. In particular, the value of $S_8$ will fluctuate as other parameters affecting the clustering such as the amplitude and growth of the matter fluctuations are varied. Two examples of these are the sum of the neutrino masses, $\Sigma m_\nu$, and the optical depth to reionization, $\tau$, which both affect the overall power spectrum amplitude~\cite{Planck:2015mym}. We note that $\tau$ is the most uncertain of the $\Lambda$CDM parameters while $\Sigma m_\nu$ is not yet directly measured and is commonly set to the minimum expected value. In every $\Lambda$CDM fit then, these two parameters and their uncertainty will carry some weight on the derived $S_8$ constraints. These correlations are also visualized in the specific case of {\it Planck} with the excess of lensing anomaly, which Ref.~\cite{DiValentino:2018gcu} showed can mimick a larger $S_8$. However, {\it Planck} and its lensing excess is not the cause -- or not the only cause -- of this CMB-LSS tension, the CMB preference for larger amount of matter clustering is also confirmed by the latest ACT+WMAP analysis~\cite{ACT:2020gnv} which finds a large value of $S_8=0.840\pm0.030$ but differently from {\it Planck} no anomalous value for the lensing amplitude.

Given the significant importance of CMB lensing in these measurements, improvements in small-scale lensing theory, such as non-linear, foreground and baryonic effects which generate noise biases in lensing will play a major role~\cite{Beck:2018wud}.


\section{Other Measurements and Systematics}
\label{systematics}

\noindent \textbf{Coordinators:} Mikhail M. Ivanov, Oliver H.\,E. Philcox. \\
\textbf{Contributors:} Micol Benetti, Anton Chudaykin, Eleonora Di Valentino, Raul Jimenez, Michele Moresco, Levon Pogosian, Denitsa Staicova, Licia Verde, and Luca Visinelli.

\subsection{Planck CMB Data and the Excess of Lensing}
\label{sec:WG-Alens}

It is well known that CMB observations from {\it Planck} constrain the cosmological parameters with an extraordinary accuracy. However, it is quite natural to understand that, similar to any experimental measurements, {\it Planck} could be affected by systematic errors. Here, we shall briefly discuss the possible systematic errors in {\it Planck} that might be propagated in the determination of the cosmological parameters under the assumption of the $\Lambda$CDM paradigm as the background cosmological model.

The {\it Planck} collaboration~\cite{Planck:2019nip} presents the constraints on the model parameters using two different likelihood pipelines for the data at multipoles $\ell>30$, namely, {\tt Plik} and {\tt CamSpec} (see the updated version in Ref.~\cite{Efstathiou:2019mdh}). Whilst in principle, both the likelihood pipelines refer to the same measurements, however, they actually consider different sky masks and chunks of data, and additionally, the likelihoods handle foregrounds in a different way, especially for polarization. 
As a consequence, the observational constraints on the $\Lambda$CDM model parameters extracted using {\tt Plik} likelihood and the {\tt CamSpec} likelihood thus differ at most by $0.5 \sigma$ in case of the baryon density, just by $0.2 \sigma$ in case of the Hubble constant~\cite{Planck:2019nip} or by $0.3 \sigma$ in case of the $S_8$ parameter. Even though the choice between {\tt Plik} likelihood and the {\tt CamSpec} likelihood makes a very mild effect on the Hubble constant or $S_8$ tensions, however, it should be emphasized that a choice of a different likelihood may shift the constraint on some parameters coming from CMB by $0.5 \sigma$, and this should not be completely forgotten as we approach precise measurements of the cosmological parameters.

A possible indication for a systematic error in {\it Planck} is given by the excess of lensing, i.e. the so called "$A_{\rm lens}$ anomaly", see subsection~\ref{sec:Alens} for a detailed explanation. Interestingly, if this parameter is included in the analysis, then both the Hubble constant tension and the $S_8$ tension are slightly reduced. In fact, due to the inclusion of $A_{\rm lens}$ in the analysis, then the {\it Planck} and {\it Planck}+BAO constraints on $H_0$ are slightly increased leading to  $H_0=(68.3\pm0.7){\rm \,km\,s^{-1}\,Mpc^{-1}}$ (at 68\% CL for {\it Planck} data alone) and $H_0=(68.22\pm0.49){\rm \,km\,s^{-1}\,Mpc^{-1}}$ (at 68\% CL for {\it Planck}+BAO), using either {\tt Plik} likelihood or {\tt CamSpec} likelihood.

Analogously, when $A_{\rm lens}$ is free to vary, {\it Planck} gives $S_8=0.804 \pm 0.019$ at 68\% CL, from $S_8=0.834 \pm 0.016$ at 68\% CL when $A_{\rm lens}=1$. Therefore, for {\it Planck} constraints, the inclusion of $A_{\rm lens}$ in the analysis reduces the Hubble tension from $5 \sigma$ to $3.9 \sigma$ with the R21 value $H_0=(73.04 \pm 1.04){\rm \,km\,s^{-1}\,Mpc^{-1}}$ at 68\% CL~\cite{Riess:2021jrx}, and reduces the $S_8$ tension with the weak lensing experiments below $2\sigma$. If $A_{\rm lens}$ is due to systematic errors in the {\it Planck} data, then we need to understand how the systematics could affect the constraints on the Hubble constant and the $S_8$ parameter and hence its effect on the cosmological tensions. It is worthwhile to note here that in any case, the increase of the Hubble constant mean value within the $\Lambda$CDM paradigm from {\it Planck} is not enough to cancel out completely the Hubble tension, and that a lower $H_0$ value is supported by the complementary ground-based CMB experiments, such as ACT and SPT, that are not affected by the excess of lensing. The same conclusion is valid for the $S_8$ parameter.

Alternatively, the lensing information can be reconstructed from CMB power spectra in a model-independent way by modelling the principal components of the gravitational lensing potential~\cite{Motloch:2018pjy}. This setup allows for amplitude and shape variations in the lensing power spectrum beyond $\Lambda$CDM which can be parametrized by free lensing parameters $\Theta^{(i)}$. This analysis should be contrasted with the common $A_{\rm lens}$ approach which considers only changes in the amplitude of the gravitational lensing potential. Measuring the lensing principal components presents a direct and model-independent consistency test of the lensing information encoded in the CMB data. In this analysis, the amount of lensing determined from the smoothing of the acoustic peaks in the {\it Planck} power spectra is $2.8\sigma$ higher when compared with $\Lambda$CDM expectation based on the "unlensed" temperature and polarization power spectra (after marginalizing over the lensing information $\Theta^{(i)}$)~\cite{Motloch:2019gux}. Allowing for an arbitrary gravitational lensing potential still leads to anomalously high lensing power in the {\it Planck} power spectra, which supports the "$A_{\rm lens}$ anomaly". This result points to possible systematic errors in the {\it Planck} data or a statistical fluke that can be resolved by more CMB data.

\subsection{BAO and the Sound Horizon Problem}
\label{sec:soundhorizon}

It is important to discuss a possible cosmology-dependence of the BAO measurements. This cosmology 
dependence can come through two channels: (a) underlying cosmology assumptions when converting angles and 
redshifts into distances, (b) assumptions on fiducial cosmology when creating a post-reconstructed BAO template,
which is then used to fit the data. Refs.~\cite{Heinesen:2018hnh,Heinesen:2019phg} have shown that the standard underlying cosmology assumptions are accurate as long as the Universe can be locally described with the isotropic and homogeneous FLRW metric. This is certainly true for most of the cosmological models that are aimed to fit all available cosmological data. However, the caveat on the FLRW metric should be kept in mind when considering more exotic cosmological models with large metric gradients. 
The second concern is about the choice of the fiducial template. Unlike the standard RSD method, the BAO extraction is not very sensitive to the underlying fiducial template cosmology assumptions. This is because the BAO pattern is, essentially, a quasi-harmonic oscillation function. The BAO method just extracts its frequency,
and therefore, any reasonable harmonic-type template is expected to perform well in this situation
because the frequency does not depend on the shape of the template. 
This point was explicitly proven with regards to modifications of the early Universe cosmology in Ref.~\cite{Bernal:2020vbb}.
Finally, in terms of non-linear effects, the possible systematic shift in the post-reconstructed 
BAO is bound to be less than $0.2\%$~\cite{Blas:2016sfa}, which proves the common lore that the nonlinear effects are not an issue for the BAO measurements.  

With the advent of {\it Planck} precision, observations on the distributions of galaxies have become increasingly important to studying the late-time evolution of the Universe. Indeed, this type of measurement encodes not only information on the history of cosmic expansion but also on the growth of structure, a fundamental aspect for probing different mechanisms of cosmic acceleration as well as for distinguishing between competing gravitational theories. 

A tool for identifying the distribution of galaxies at large scales is the two-point spatial correlation function (2PCF) of large galaxy catalogues, which has shown a tiny excess probability of finding pairs of galaxies separated by a characteristic scale $r_s$ - the comoving acoustic radius at the drag epoch. This signature of Baryon Acoustic Oscillations arises from competing effects of radiation pressure and gravity in the primordial plasma, which is well described by the Einstein-Boltzmann equations in the linear regime, and defines a statistical standard ruler and provides independent estimates of the Hubble parameter $H(z)$ and the angular diameter distance $D_{A}(z)$. It is worth mentioning that these measures are not model-independent, since they use the radial ($dr_{\parallel} = c\delta z/H(z)$) and transverse ($dr_{\perp} = (1+z)D_A\theta_{BAO}$), respectively. Analyses of this type assume a fiducial cosmology to transform the measured angular positions and redshifts into comoving distances, such a conversion may distort the constraints on the parameters~\cite{SDSS:2005xqv, Sanchez:2012sg,BOSS:2016lsx}.

Another possibility is to use the 2-point angular correlation function (2PACF), $w(\theta)$, which involves only the angular separation $\theta$ between pairs, providing almost model-independent information on $D_{A}(z)$, as long as the comoving sounding horizon $r_s$ is known (see Refs.~\cite{BOSS:2016lsx,Carvalho:2015ica, Alcaniz:2016ryy,Carvalho:2017tuu,deCarvalho:2020ftb} for details and Refs.~\cite{Benetti:2017juy, Santos:2021ypw, Menote:2021jaq, Bengaly:2021wgc,Mukherjee:2020mha,Arjona:2021hmg,Ferreira:2017yby,Brinckmann:2018cvx, Cid:2018ugy, Dutta:2018vmq, Camarena:2019rmj, vonMarttens:2018bvz, Gonzalez:2018rop} for cosmological analysis results using this type of measurements). Although the latter would theoretically be more suitable for alternative gravity model analyses, since it does not use a fiducial model, the associated errors of these BAO angular correlation measurement are an order of magnitude larger than those of 2PCF. Furthermore, the technique used for BAO estimation of 2PACF uses a very narrow redshift shell, with $\delta z \sim 0$ making it a sufficiently complex procedure to apply.

We now turn to the problem of the sound horizon. For this we define the sound horizon at redshift $z$, here $r_s(z)$, as the distance travelled by an acoustic wave in the baryon-photon plasma in terms of the sound speed of the baryon-photon plasma $c_s(z)$, as
\begin{equation}
    \label{eq:soundspeed}
    r_s(z) = \int_{z}^{\infty} \frac{c_s(z')}{H(z')} {\rm d}z'\,.
\end{equation}
With the energy densities for the baryon $\rho_b(z)$ and the photon $\rho_\gamma(z)$, the speed of sound in the baryon-photon fluid is approximated as $c_s \approx c \left(3 + 9\rho_b /(4\rho_\gamma) \right)^{-0.5}$. Of particular importance is the drag epoch, occurring at redshift $z_d$, at which the baryons are released from the Compton drag of the photons~\cite{Zeldovich:1969ff, Sunyaev:1970eu, Peebles:1970ag}. The corresponding distance at drag epoch is $r_d \equiv r_s(z_d)$. Here, one has to take into account the difference between the sound horizon at the CMB last-scattering defined at $z_* \sim 1090.30$~\cite{Planck:2018vyg} when photons decouple and the BAO sound horizon at drag epoch $z_d \sim 1059.39$~\cite{Planck:2018vyg} when baryons stop feeling the photon drag.

The problem of the sound horizon refers to the different preferred values for the sound horizon by early and late-time measurements. It can be traced to the fact that BAO surveys measure the projected quantities $\Delta z= r_d H/c$ and $\Delta \theta=r_d /(1+z) D_A(z)$, where $\Delta z$ and $\Delta \theta$ are the redshift and the angular separation. From here, it follows that without further assumptions, one may derive from BAO only the quantity $r_d \times H$. In order to decouple these quantities, one needs either a model independent measurement of $H_0$ or an independent evaluation of $r_d$. 
The Hubble constant $H_0$ can be obtained from model independent late-time probes such as cosmic chronometers~\cite{Moresco:2020fbm}, Type Ia supernova~\cite{Scolnic:2021amr, Pan-STARRS1:2017jku}, the Cepheids in LMC~\cite{Riess:2019cxk}, or indirect probes such as the matter-radiation equality scale~\cite{Baxter:2020qlr,Philcox:2020xbv} (see Section~\ref{sec:WG-H0measurements}). Eq.~\eqref{eq:soundspeed} imposes additional assumptions on the baryon and radiation load of the Universe at decoupling. Furthermore, calculating the angular distance $D_A$ by definition depends on the implied cosmology, and thus on the implied theory of gravity. This problem is known as the $H_0$--$r_d$ tension, considered in detail in Ref.~\cite{Knox:2019rjx}, where it is found that the spread in $r_d$ depends mostly on the matter density $\Omega_m$ and that it is not possible to reconcile the tension in early and late time Universe measurements by changing $\Omega_m$ alone.

As shown in Ref.~\cite{Pogosian:2020ded}, complementing the BAO data with a prior on $\Omega_mh^2$ makes it possible to determine both $r_d$ and $H_0$ while treating them as independent parameters, {\it i.e.} without the need to calculate $r_d$ from theory. One can also decouple $r_d$ and $H_0$ by combining the BAO data with galaxy and/or CMB weak lensing, which provide a constraint on $\Omega_mh^2$ that is largely independent of the physics before and during recombination. It is interesting that the current BAO data (eBOSS DR16+), combined with the Gaussian prior of $\Omega_mh^2 =0.143 \pm 0.0011$ corresponding to the value measured by {\it Planck} within the $\Lambda$CDM model, gives $r_d=(143.8 \pm 2.6)\,$Mpc and $H_0=(69.6 \pm 1.9){\rm\,km\,s^{-1}\,Mpc^{-1}}$, thus preferring a somewhat larger $H_0$ and a smaller $r_d$ than { \it Planck}~\cite{Pogosian:2020ded}. Applying this method to future BAO data will offer a stringent consistency test against the value of $r_d$ derived in a model-dependent way from CMB. In particular, as forecasted in Ref.~\cite{Pogosian:2020ded}, the upcoming DESI BAO data combined with the {\it Planck} prior on $\Omega_mh^2 =0.143 \pm 0.0011$ will measure $r_d$ and $H_0$ with uncertainties of $\sigma_{r_d} = 0.64\,$Mpc and $\sigma_{H_0} = 0.32{\rm\,km\,s^{-1}\,Mpc^{-1}}$, respectively.

Possible alternatives that have been considered in the literature are working with $H_0 \times r_d$ as a combined parameter~\cite{LHuillier:2016mtc,Shafieloo:2018gin, Arendse:2019hev, Benisty:2021gde}, replacing the prior on $H_0$ with a prior on the peak absolute magnitude $M_B$~\cite{Camarena:2019rmj, Camarena:2021jlr, Efstathiou:2021ocp, Benevento:2020fev, Perivolaropoulos:2021bds} or using a modified gravity or early dark energy model that changes the value of $r_d$ before recombination (see the discussion in the sections below and Ref.~\cite{DiValentino:2021izs}). However, as noted in Refs.~\cite{Jedamzik:2020zmd, Benisty:2020otr, Staicova:2021ajb}, a dark energy model that merely changes the value of $r_d$ would not completely resolve the tension, since it will affect the inferred value of $\Omega_m$ and transfer the tension to it. The sound horizon problem should be considered not only in the plane $H_0$--$r_d$, but it should be extended to the parameters triplet $H_0$--$r_d$--$\Omega_m$.

\subsection{Redshift-Space Galaxy Clustering Systematics}
\label{rsdsec}

RSD measurements suffer from a number of systematic effects, which can be loosely categorized as observational, modelling, and analysis systematics. From the observational side one can mention stellar contamination, atmospheric extinction and blurring~\cite{BOSS:2012coo}, fiber collisions~\cite{Hahn:2016kiy}, integral constraints, angular and radial modes' systematics~\cite{deMattia:2019vdg}.These effects have been thoroughly studied in the past, and none of them is expected to affect the RSD results in a statistically significant way. Another observational effect that is currently under active investigation is selection bias~\cite{Hirata:2009qz}, which might directly affect the RSD clustering amplitude measurement. The claims in the literature on the presence of this effect are currently controversial: they vary from significant evidence for selection bias~\cite{Obuljen:2020ypy} to no detection thereof~\cite{Singh:2020cvu}. 

From the theory side, the largest uncertainty has been the accuracy of theoretical models for nonlinear galaxy clustering in redshift space. This uncertainty is now removed after the progress in the effective field theory of large scale structure (see~\cite{Baumann:2010tm,Carrasco:2012cv, Porto:2013qua, Senatore:2014via, Senatore:2014vja, Perko:2016puo, Lewandowski:2015ziq, Desjacques:2016bnm, Blas:2015qsi,Blas:2016sfa,Ivanov:2018gjr,Ivanov:2019pdj,Vlah:2015sea, DAmico:2019fhj,Chen:2020fxs,Chen:2020zjt,Chudaykin:2020aoj} and references therein), which provides an accurate and systematic description of the galaxy clustering statistics on large scales. This is especially important in analyses of extended cosmological scenarios, e.g.\ massive neutrinos~\cite{Chudaykin:2019ock,Ivanov:2019hqk}, dynamical dark energy~\cite{DAmico:2020kxu,Chudaykin:2020ghx}, non-zero spatial curvature~\cite{Chudaykin:2020ghx}, axion dark matter~\cite{Lague:2021frh}, light relics~\cite{Xu:2021rwg}, etc., where the use of approximate phenomenological 
models or fits to $\Lambda$CDM-based simulations (e.g.\ Halofit) becomes inadequate. 

It is also worth mentioning that there has been significant progress in analyzing the redshift space clustering with the fully simulation-based techniques ("emulators"), see e.g.\ Refs.~\cite{Heitmann:2009cu,Kobayashi:2020zsw,Hahn:2019zob}, which nominally allow one to use information from short scales that cannot be described with perturbation theory. The current results from this program show some scatter between different teams~\cite{Lange:2021zre,Chapman:2021hqe,Kobayashi:2021oud,Zhai:2022yyk}, which calls for further investigations. 

In terms of analysis systematics, another source of uncertainty in the standard RSD measurements are the so-called ``shape priors"~\cite{Ivanov:2019pdj,Ivanov:2020ril}. Most of the RSD analyses fix the shape of the linear matter power spectrum by assuming $\Lambda$CDM cosmological parameters, and only capture the cosmology dependence by means of the so-called scaling parameters. The shape priors make the fixed template analysis intrinsically cosmology-dependent, see e.g.\ Refs.~\cite{deMattia:2020fkb,Smith:2020stf}. Moreover, while the scaling parameters capture some non-standard late-time cosmological scenarios, they do not account for all possible variations of cosmological parameters.

It is especially important to take into account for the models that modify the physics at high redshifts (with respect to the observed galaxy clustering), e.g.\ the early dark energy, see Refs.~\cite{Ivanov:2019pdj,Ivanov:2020ril} and Sec.~\ref{sec:ede} for further discussion. The use of the full-shape method~\cite{Ivanov:2019pdj,DAmico:2019fhj,Chen:2021wdi} allows one to remove any uncertainty associated with the shape priors. This method currently finds a mild tension with the $Planck$ $\Lambda$CDM cosmology in the late-time clustering amplitude $\sigma_8$, and this result is consistent across different teams~\cite{Ivanov:2019pdj,DAmico:2019fhj,Chen:2021wdi}. Finally, it is worth mentioning that the current RSD results have been shown to be stable under covariance matrix assumptions~\cite{Wadekar:2020hax,Philcox:2020zyp}.

\subsection{The Age of the Universe Problem } 
\label{sec:WG-age}

The age of the Universe is not just a prediction of the $\Lambda$CDM model, that for {\it Planck}   2018 is $t_U = (13.800 \pm0.024)\,$Gyr, but can also be measured using very old objects. In particular, the age of HD 140283 equal to $t_* = (14.46 \pm 0.8)\,$Gyr~\cite{Bond:2013jka} leads to some mild tension with the {\it Planck} 2018 predictions corresponding to somewhat lower age. However, the age of HD 140283 becomes $t_* = (13.5 \pm 0.7)\,$Gyr~\cite{Jimenez:2019onw} using the new Gaia parallaxes instead of original HST parallaxes and thus becomes fully consistent with the $\Lambda$CDM predictions. In addition, using populations of stars in globular clusters Ref.~\cite{Valcin:2020vav} finds $t_U = (13.35 \pm 0.16({\rm stat})\pm0.5({\rm sys}))\,$Gyr which is also marginally consistent with $\Lambda$CDM. Thus, at present there is no confirmed tension between the different $t_U$ determinations in the context of the {\it Planck} 2018 data, in contrast to discrepancies in the past~\cite{Verde:2013wza}.

This may change in the future if significantly older objects are found in the Universe. In such a case, a possibility to increase the age of the Universe, to be larger than the age of oldest stars, would be to lower the Hubble constant value, because of the anti-correlation between these parameters~\cite{deBernardis:2007jnf}. However, such an effect would tend to increase rather than decrease the Hubble tension. For example, a positive curvature for the Universe, as suggested by {\it Planck} 2018, preferring a lower $H_0$ and worsening significantly the $H_0$ tension, predicts an older Universe $t_U = (15.31 \pm 0.47)\,$Gyr.

Considering the constraints coming from all low and high redshift data together with that of the age of Universe, leaves very little room for late resolutions to the $H_0$ tension~\cite{Krishnan:2021dyb}, unless it is an extremely recent (after $z\sim0.1$) departure from $\Lambda$CDM~\cite{Bernal:2021yli}, requiring a great deal of fine-tuning. Thus, the constraints from the age of the oldest objects in the Universe puts in disfavor a high value of $H_0$ and late time solutions of the Hubble tension that are based on late time deformation of $H(z)$ that increase $H_0$ and thus decrease the age of the Universe $t_U$. For this reason, model-independent measurements of $t_U$ are very important, since they either support or disfavor alternative proposed models that alleviate the Hubble tension~\cite{Bernal:2021yli,Vagnozzi:2021tjv}. In particular, the use of new cosmic triangle plots to simultaneously represent independent constraints on key quantities related to the Hubble parameter, i.e.\ age of the Universe, sound horizon and matter density~\cite{Bernal:2021yli} is a welcome development.

Given its direct connection with the Hubble constant and the cosmological parameters driving the expansion history, the age of the Universe is a very important quantity.
In particular, the use of extremely old local objects to measure the current age of the Universe (for example, see the reviews in Refs.~\cite{2018IAUS..334...11C, Soderblom:2010gu, VandenBerg:1996tm} and the recent determinations in Refs.~\cite{2017ApJ...838..162O, Valcin:2020vav,Valcin:2021jcg}), as well as the determination inferred from the oldest objects at higher redshifts~\cite{Dunlop:1996mp,Spinrad:1997md} have played a significant role supporting the standard cosmological model.

Since the look-back time is directly determined by the expansion rate of the Universe, age measurements can constrain the cosmological parameters determining the background evolution. While, as discussed later, differential ages can be used to measure the Hubble parameter with the cosmic chronometers method, also absolute ages can play an important role, as they have been used for at least 70 years to provide constraints on the cosmological model.

The look-back time of the Universe $t$ can be expressed as function of redshift as:
\begin{equation}
    t(z) =\frac{977.8}{H_0}\int_0^z \frac{{\rm d}z^\prime }{(1+z^\prime)E(z^\prime)}\,{\rm Gyr} \;\; ,
    \label{eq:tz}
\end{equation}
with $E(z)$ is the normalized Hubble parameter $H(z)/H_0$, and the constant provides the conversion to ${\rm km\,s^{-1}\,Mpc^{-1}}$. The age of the Universe can be expressed from Eq.~\eqref{eq:tz} as $t_{U} =t(\infty)$.

The age of the oldest star in our Galaxy, as also in many nearby galaxies, can be estimated from the color-magnitude diagram (CMD) of coeval stellar populations. Alternatively, if the metallicity and distance are known, it is also possible to determine the age of individual stars. It is important to underline that the full morphology of the CMD of resolved stellar population can constrain the absolute age without necessarily needing a distance determination. For a more detailed review on the topic, we refer the reader to Refs.~\cite{2018IAUS..334...11C, Soderblom:2010gu, VandenBerg:1996tm}. 

The most accurate method currently available to determine the ages of stars is based on the analysis of the observed CMD of Globular Clusters (GC), but there are also different alternatives. A first possibility is to use the abundances of radioactive elements (e.g., Uranium and Thorium), in an approach named nucleo-cosmochronology~\cite{2016AN....337..931C}. A second possibility is to exploit the cooling luminosity function of white dwarfs (see e.g.\ Ref.~\cite{2018IAUS..334...11C} and references therein). 

Interestingly, the measurement of old stellar ages was one of the first indications challenging the Einstein-de Sitter cosmological model~\cite{Ostriker:1995su, Jimenez:1996at,Spinrad:1997md}, since the stellar ages were larger than the age of the Universe within that model. Clearly, the ages of local objects at $z=0$ purely represent a strict lower limit to the age of the Universe, and degeneracies between cosmological parameters, such as $H_0$~and $\Omega_\mathrm{m}$, as can be seen by Eq.~\ref{eq:tz}, make it difficult to precisely disentangle different models. 
A crucial point that allowed to break this degeneracy was the discovery and measurement of the age of extremely old galaxies at $z \gg 0$ (see e.g.\ Ref.~\cite{Dunlop:1996mp} and Fig.~18 in Ref.~\cite{Spinrad:1997md}).

If we consider the absolute age determined only from the main-sequence turn off point (MSTOP) luminosity, we find that other GC properties are degenerate with it. 
The most significant source of errors in this approach is the uncertainty on the distance to the GC, which scales almost linearly in percentage with the uncertainty in the measured age. In addition, further contributions to the error budget are given by theoretical systematics involved in stellar evolution models, namely the metallicity content, the dust absorption~\cite{VandenBerg:1996tm}, and the Helium fraction.

However, as first discussed in Ref.~\cite{1996ApJ...463L..17J, Padoan:1996dc}, an important point to stress is that multiple additional information can be exploited in a GC CMD that can complement the MSTOP, since it presents features with which it is possible to jointly determine its age and distance scale. In particular, if we consider Fig.~2 from Ref.~\cite{Jimenez:1997rb} it is possible to notice how different parts of the CMD are sensible to different physical quantities, and in Fig.~1 from Ref.~\cite{Padoan:1996dc} and in Fig.~3 from Ref.~\cite{Jimenez:1997rb}, it was demonstrated that the luminosity function of the GC can provide information about various physical parameters. With this approach it was possible to determine the absolute ages of GCs M68~\cite{1996ApJ...463L..17J}, M5 and M55~\cite{Jimenez:1997rb}. In addition, it has also been found that the horizontal branch morphology can be analyzed to obtain a measurement of GCs age independent of constraints on their distance~\citep{Jimenez:1996at}.

In this context, the exploitation of Bayesian methods to fit the CMD of GCs has been only recently explored by a few works, either focusing only on some of its features~\cite{2017MNRAS.468.1038W}, or considering it entirely~\cite{Valcin:2020vav,Valcin:2021jcg}. This approach provided a combined measurement of distances, metallicities and ages for 68 GCs observed by the HST/ACS project. One of the most important benefits of this techniques is that it allows to minimize part of the systematic error, providing a larger accuracy. It is important to notice that even considering different methods and observables, the ages determined are all consistent with each other at $13.5 \pm 0.27$ Gyr. 

The determination of the absolute ages of old stellar objects has succeeded in revealing important insights in the Hubble tension debate. In Ref.~\cite{Bernal:2021yli} in particular a new approach has been proposed by highlighting the importance of three quantities that can be constrained independently even if they are interconnected: the age of the Universe $t_U$ measurable from the absolute ages (and CMB), the Hubble constant $H_0$ obtained from the cosmic distance ladder (and CMB), and the combination $H_0t_U$ from standard rules and candles (BAO and SNIa). This triad of measurements, named "new cosmic triangle", can be represented as $\ln t_U+\ln H_0\equiv \ln (H0 t_U)$. In this framework, we have some measurements depending on a cosmological model (namely CMB, SNIa and BAO), while some other being independent (the stellar ages and the distance ladder), but interestingly all the measurements were in agreement with a $\Lambda$CDM model.

To further improve the accuracy of absolute age estimates in the future it will be crucial to improve the stellar modelling by decreasing the systematic errors, exploiting better determinations of distances and metallicities for GC, and reducing the degeneracies between parameters. From this point of view, the final release of Gaia will be fundamental since it will provide distances with a percent (and even smaller) accuracy. On the other hand, better constraints on the chemical composition of individual GC stars (in particular below MSTOP) that can be obtained from advanced spectroscopic analyses (e.g.\ with JWST) will allow to appreciably reduce the priors on the metallicity of the system, leading to better constraints. In this scenario (as also reported by Ref.~\cite{Valcin:2021jcg}), the main systematic left in the analysis of the full CMD will be the nuclear reaction rates uncertainty, which in turn can be addressed from a combination of theoretical and laboratory works.

With a similar approach, it is also possible to constrain the expansion history of the Universe from the ages of the oldest objects in the Universe at larger redshifts.
This method, known as \textit{cosmic chronometers}, provides a cosmological probe able to directly estimate the Hubble parameter $H(z)$ in a direct and cosmology-independent way (see section \ref{sec:GP}). This technique, first proposed by Ref.~\cite{Jimenez:2001gg}, only relies on the direct relation between the scale factor $a(t)$ and the redshift in a universe described by a FLRW metric. Considering this only assumption, the Hubble parameters $H(z)$ can be derived without any other cosmological assumption from the differential age evolution d$t$ of the universe in a specific interval of redshift d$z$ as:
\begin{equation}
    \label{eq:CC}
    H(z)=-\frac{1}{(1+z)}\frac{{\rm d}z}{{\rm d}t}\,.
\end{equation}
In this equation, since the redshift can be directly measured in astrophysical objects, the only unknown is the differential age d$t$, and therefore the key points are to obtain an homogeneous population of cosmic chronometers, and to robustly measure their difference in age.

The strength of this approach in comparison to other cosmological probes is that it is based on minimal cosmological assumptions, considering only a FLRW metric and not relying on any explicit functional form of a cosmological model, or implicitly assuming a spatial geometry. For this reason, measurements obtained with the CC method are ideal to test a wide variety of different cosmologies. A comprehensive review on the method is provided in Ref.~\cite{Moresco:2022phi}, with a detailed discussion on the selection of the optimal CC tracers, current measurements, and  systematics involved. In the following, we will just briefly summarize the main points.

Cosmic chronometers are astrophysical objects that can accurately trace the differential age evolution of the Universe as a function of redshift. The ideal tracers, therefore, need to be an extremely homogeneous population of the oldest objects in the Universe at each redshift whose intrinsic evolution takes place on timescales much longer than the evolution due to pure passive aging. For this reason, the ideal candidates are massive and passively evolving galaxies, carefully selected in order to minimize the possible contamination by star-forming outliers. To maximize the purity of a CC sample, as highlighted in Ref.~\cite{Moresco:2013wsa,Moresco:2018xdr}, it is fundamental to exploit and combine different selection criteria, based on photometric, spectroscopic, morphological data, while also including a cut to select the most massive systems.

Once a proper sample of CC is obtained, the fundamental step is to robustly estimate the differential age d$t$ in a redshift bin d$z$. Since its first application by Ref.~\cite{Simon:2004tf}, the determination of the differential age d$t$ has been performed with different approaches. The first one relies on exploiting the information contained in the full spectrum of CC through a full-spectrum fitting. This method has the benefit of taking advantage of most of the information that can be extracted from the observation of CC, the drawback being the need for accurate models able to accurately and robustly reproduce all the spectroscopic features observed. This method has been successfully applied in Refs.~\cite{Simon:2004tf, Zhang:2012mp, Ratsimbazafy:2017vga}, obtaining 13 $H(z)$ measurements at $0<z<1.75$ from the analysis of different spectroscopic surveys, among which SDSS luminous red galaxies~\cite{SDSS:2001wju}, 2dF–SDSS luminous red galaxies~\cite{Cannon:2006qh}, GDDS~\cite{Abraham:2004ra}, and archival data~\cite{Simon:2004tf}. 

A second approach, introduced in Ref.~\cite{Moresco:2010wh} and then applied in Refs.~\cite{Moresco:2012jh, Moresco:2015cya, Moresco:2016mzx}, instead of relying on the fit of the entire spectrum, relies on the analysis of a specific spectroscopic feature, the $D4000$, which is demonstrated to be linearly dependent on the age of the stellar population (at given metallicity, for the population considered). This approach was found to be particularly valuable since it allows to separate in the analysis the contribution coming from statistical and systematic effects, making it easier to model them. With this revised approach, 15 additional $H(z)$ measurements in the redshift range $0.15<z<2$ have been obtained by considering more than 140 000 CC from new independent SDSS catalogs (Data Release 6 main galaxy sample, Data Release 7 luminous red galaxy sample, Data Release 9 BOSS sample), different spectroscopic surveys at higher redshifts (zCOSMOS, K20, UDS), and new archival data.
Finally, a last approach only recently exploited in Ref.~\cite{Borghi:2021zsr, Borghi:2021rft} relies on the estimate of d$t$ based on the analysis of several independent spectroscopic features, the Lick indices, with which it is possible to determine the stellar age and metal content of CC. Within this context, analyzing the LEGA-C survey a new $H(z)$ measurement at $z\sim0.7$ has been derived.

The total covariance matrix for the CC approach, as presented in Refs.~\cite{Moresco:2020fbm, Moresco:2022phi}, can be expressed as the sum of a statistical and a systematic part, where, in turn, the systematic part is mainly composed by four terms:
\begin{equation}
    {\rm Cov}_{ij}= {\rm Cov}_{ij}^{\rm stat}+ {\rm Cov}_{ij}^{\rm sys} \;\; ,  \;\; \;\; \;\; {\rm Cov}_{ij}^{\rm sys}= {\rm Cov}_{ij}^{\rm met}+{\rm Cov}_{ij}^{\rm SFH}+{\rm Cov}_{ij}^{\rm young}+ {\rm Cov}_{ij}^{\rm model} \;\; .
\end{equation}
The first term in the systematic error is related to the fact that in estimating the age of a stellar population, an uncertainty on the estimate of its metallicity might affect as a consequence the measurement of d$t$, and therefore the derived $H(z)$. This factor was found to scale linearly with the uncertainty on the metallicity~\cite{Moresco:2020fbm}. The second term takes into account the possible uncertainty in the estimated SFH of CC; even if those tracers are expected to form on extremely short timescales, assuming the SFH as instantaneous is an nonphysical underestimate that might bias the results. Their SFH needs to be, therefore, properly accounted for, introducing an error $\sim$2-3\%~\cite{Moresco:2012jh}. The third term includes in the analysis the possible effect that might be caused by a minor, but still non-negligible young stellar component below the otherwise old and passive population of a CC. In Ref.~\cite{Moresco:2018xdr} it was estimated that a  10\% contamination by a star-forming young component affects $H(z)$ with a systematic error of 5\%, and with a 0.5\% error if it is of the order of 1\%.
For current measurements~\citep{Moresco:2012jh, Moresco:2015cya,Moresco:2016mzx,Borghi:2021zsr}, this contribution has been found to be completely negligible due to the very stringent CC selection criteria implemented. Finally, the last term is, with the first one, the currently dominating one. The current uncertainty on galaxy spectral modelling has an impact on the estimate of the differential age $dt$, or on the calibration of observational proxies to the age, since different models have been proposed in the literature.
This term introduces also non-diagonal terms in the covariance matrix, and can be further decomposed in many different ingredients, depending on the assumed initial mass function, on the stellar library adopted, and on the actual recipe of the model considered. It has been fully estimated in Ref.~\cite{Moresco:2020fbm}, obtaining errors depending on the model assumed of the order of $\sim$4.5\%

As discussed above, an extensive discussion about how to build a full covariance for CC is provided in Ref.~\cite{Moresco:2022phi}. It is worth underlining that currently provided datasets include in the errors most of the error components but the systematic terms ${\rm Cov}_{ij}^{\rm mod}$, which need to be included following the indication provided in the reference.\footnote{Note that a tutorial on how to estimate covariance matrix for CC is provided at \url{https://gitlab.com/mmoresco/CC_covariance}.} It is worth recalling that the strict CC selection criterion and the differential approach helps to mitigate several other possible effects that might bias this kind of approach, such as, e.g., a possible dependence of the star formation rate on redshift or of the IMF on stellar mass (see~\cite{Moresco:2022phi}). Having selected CC as the most massive and passive galaxies ensures to minimize any SFH-dependent possible effect, since the estimated SF timescales for these objects is always very short by definition, and the differences $dt$ are estimated in redshift intervals very close expressly to further minimize any residual effect. Likewise, the limited extension of the range of CC stellar masses, and the analysis further performed in even smaller bins, is designed to make the analysis extremely robust w.r.t. mass dependent effects (as shown, e.g., in~\cite{Moresco:2012jh,Borghi:2021rft}).

While currently the main limitations of this method are the uncertainty on metallicity and models that at the moment dominate the error budget and the absence of dedicated surveys to properly map CC as a function of redshift (as, for example, for BAO and SNe), a path to improve these points in the future is clear. On one hand, better spectroscopic data (in terms of signal-to-noise ratio and resolution) that could be obtained by present and future instruments (X-Shooter, MOONS, etc.) could be crucial to better constrain the metal content of CC and to discriminate between the various models, significantly reducing the systematic error budget. On the other hand, future spectroscopic surveys will significantly increase the statistics of CCs in the redshift range $0.2<z<2$, even if not directly being a target of the various mission~\cite{EUCLID:2011zbd,Wang:2019jig,SDSS-IV:2019txh}.


\section{Cosmological Models Proposed to Solve the $H_0$ and the $S_8$ Tensions}
\label{sec:WG-BothSolutions}

\noindent \textbf{Coordinators:} Celia Escamilla-Rivera, Cristian Moreno-Pulido, Supriya Pan, M.M.\ Sheikh-Jabbari, and Luca Visinelli.  \\

\noindent \textbf{Contributors: } Guillermo F. Abell{\'a}n, Amin Aboubrahim, \"Ozg\"ur Akarsu, George Alestas, Daniel Aloni, Luis Anchordoqui, Ronaldo C. Batista, Micol Benetti, David Benisty, Asher Berlin, Thomas Buchert, David Camarena, Anton Chudaykin, Javier de Cruz Perez, Francis-Yan Cyr-Racine, Keith R.\ Dienes, Eleonora Di Valentino, Noemi Frusciante, Adri\`a G\'omez-Valent, Asta Heinesen,  Karsten Jedamzik, Raul Jimenez, Melissa Joseph, Lavrentios Kazantzidis, Michael Klasen, Suresh Kumar, Matteo Lucca, Valerio Marra, Laura Mersini-Houghton, Pran Nath, Florian Niedermann, Eoin \'O Colg\'ain, Leandros Perivolaropoulos, Levon Pogosian, Vivian Poulin, Joan Sol\`a Peracaula,  Emmanuel N. Saridakis, Martin Schmaltz, Nils Sch{\"o}neberg, Martin S.\ Sloth, Brooks Thomas, Shao-Jiang Wang, Scott Watson, Neal Weiner.
\bigskip

\subsection{Addressing the $H_0$ Tension}

Cosmological models addressing the $H_0$ tension are extremely difficult to concoct. Generally speaking, and maybe counter-intuitively, the reason for this relies in the extremely high precision with which the flat $\Lambda$CDM model is able to fit at the same time the multitude of data sets we dispose, ranging from BBN to BAO and LSS data. Indeed, despite the presence of the aforementioned tensions in the cosmological landscape, the minimal 6 parameter flat $\Lambda$CDM model is able to deliver a number of accurate predictions for a variety of different effects which impact both the expansion and thermal history of the Universe. Modifying the standard cosmological model without compromising its many successes explaining the current data has proven to be a rather difficult task.

In the context of the Hubble tension, combined analyses of cosmological data yields strong bounds and restrictions on a wide class of models. Formulating the $H_0$ tension in a physically more meaningful setting in the $H_0$--$r_s$ plane, where $r_s$ is the sound horizon at recombination time, which is only affected by the early physics, it has been argued that the late-time solutions intrinsically struggle to solve the problem, favoring early-time solutions as the preferred option~\cite{Bernal:2016gxb, Poulin:2018zxs, Aylor:2018drw, Knox:2019rjx,Arendse:2019hev}. A similar conclusion was reached by considering all low redshift data, including constraints from the age of Universe~\cite{Krishnan:2021dyb, Bernal:2021yli,Vagnozzi:2021tjv,Cai:2021weh,Cai:2022dkh}. On the other hand, it has been argued that early resolutions typically exacerbate cosmic shear tension and that they do not provide a complete resolution~\cite{Vagnozzi:2021gjh, Jedamzik:2020zmd,Gomez-Valent:2021cbe,Lin:2021sfs}. Moreover, it was argued that there might be a late-time variation of the Hubble parameter~\cite{Wong:2019kwg, Krishnan:2020obg}, favouring a late-time solution over the early-time alternatives. At a more fundamental level, one can define a time dependent $H_0$ ("running Hubble parameter") within a given cosmology model, e.g.\ $\Lambda$CDM, and define a $H_0$ diagnostic to precisely mark departures from the given model~\cite{Krishnan:2020vaf}. It has also been suggested that the Hubble tension may be a symptom (of the FLRW framework) and not the main malaise~\cite{Krishnan:2021dyb, Krishnan:2021jmh, Luongo:2021nqh}.

The recipe for a successful extension of or modification to the $\Lambda$CDM model appears to be far from straightforward and in the past decade many possibilities have been considered in the literature, as we briefly review below. Given that we do not have fully settled theoretical and experimental understanding, the literature does not yet unambiguously converge on a new concordance model when all data and parameters are taken into account. See the review articles~\cite{Buchert:2015wwr,DiValentino:2021izs, Schoneberg:2021qvd, Anchordoqui:2021gji, Perivolaropoulos:2021jda, Clark:2021hlo} and references therein for more detailed discussions.

The goal of this section is to systematically summarize the promising models that try to address the $H_{0}$ tension. We discuss their general features without elaborating the detailed analysis. For that, readers may refer to the original papers that we refer to in this section.

To make the comparison between proposed scenarios, we have divided them into two broad early-time and late-time proposals. The former affect the history of the Universe prior to recombination, while the latter after recombination. This allows us to generalize some of the common features, such as the ones mentioned at the beginning of this section, to a whole class of models focusing only on the model-specific features in the dedicated sub-sections. A brief summary of the models (in alphabetical order) we will consider is given below.

\subsection{Addressing the $S_8$ Tension}

The tension in the $S_{8}$ parameter (or in the related $\sigma_{8}$ parameter) has added yet another question mark over the validity of the standard $\Lambda$CDM cosmology. Both theoretical and observational attempts have been made by various researchers to solve the $S_{8}$ tension. From the theoretical side, the approach consists in modifying the matter sector or the gravity sector of the $\Lambda$CDM model resulting in a plethora of alternative cosmological scenarios (we list them in the next subsections)\footnote{Additionally to the models listed, there are many other proposals, and the interested reader can find more details in the corresponding papers~\cite{Troxel:2018qll,Anand:2017wsj,Lambiase:2018ows,DiValentino:2016ucb,Burenin:2018nuf,Lin:2017txz,Wang:2020dsc,Graef:2018fzu}}. Even though, these models are successful in relieving the $S_{8}$ tension, they, in most situations, fail to provide satisfactory solutions when all the available cosmological probes are considered~\cite{DiValentino:2020vhf, DiValentino:2020zio, DiValentino:2020vvd, DiValentino:2020srs}. In particular, due to the specific correlation between $H_{0}$ and $S_{8}$ parameters, models solving $S_{8}$ tension usually exacerbate the $H_{0}$ tension and vice versa~\cite{Planck:2015koh,Planck:2015lwi,deHaan:2016qvy}. For example, late time transitions in the dark sector preferring a higher $H_0$ value, if they match the CMB data, prefer a lower value of $\Omega_m$ as well, to preserve the well measured value of $\Omega_m h^2$; this is known as the geometric degeneracy. This in turn, produces a modified distances to sources and of the sound horizon, modifies the growth of structures and CMB anisotropies~\cite{Arendse:2019hev}, and usually results in higher $\sigma_8$ than for $\Lambda$CDM because of an extended era of matter domination.  Similarly, early-time dark energy solutions of the $H_0$ tension increase $\sigma_8$ because they need a higher primordial curvature perturbation amplitude to offset the damping effect of the unclustered component. Therefore, because of the mutual effects and correlations, it is important to perform a conjoined analysis, i.e. fitting with a single model a full array of data~\cite{Hill:2020osr,Benevento:2020fev,Knox:2019rjx,Evslin:2017qdn}, and not just one parameter alone. At the same time, if a model solves the $S_8$ tension (the $z=0$ value), it is important to confirm that it is consistent with growth history (usually studied through parameter $f\sigma_8(z)$) as observed~\cite{Linder:2016xer, DiValentino:2020kha}. Hence, any solution to the $S_8$ tension should pass other cosmological tests, i.e.\ it should simultaneously fit the expansion and growth histories probed by Baryon Acoustic Oscillations (BAO), RSD-lensing cross correlations, galaxy power spectrum shape, and void measurements~\cite{Hamaus:2020cbu}.

Moreover, the use of datasets is an additional key issue in this context since some of the non-CMB observations that are often used to estimate the values  of $S_{8}$ or $\sigma_{8}$ also assume $\Lambda$CDM as the background cosmological model, similar to what {\it Planck} does and as a consequence the results might be biased. Thus, in order to find a proper solution to the $S_8$ tension in a specific cosmological model, one needs to match the estimated values of $S_8$ (or $\sigma_8$) from CMB observations and the low redshift probes such as weak gravitational lensing and galaxy clustering.
Therefore, solving the  $S_8$ tension is indeed one of the critical challenges for the $\Lambda$CDM cosmology given the fact it is related with large span observational probes to both background and perturbed Universe, and hence needs to be studied with great care.
Among the variety of cosmological models (in alphabetical order) presented below, some of them have been proposed by many researchers to either alleviate or solve the $S_8$ tension

\subsection{Addressing Both the $H_0$ and $S_8$ Tensions}

Thus, looking for a cosmological model accommodating both the tensions is not a cup of tea. Nevertheless, curious minds never stop at any point and as a result it has been argued that in some extended cosmological models beyond $\Lambda$CDM, one can simultaneously tackle this problem~\cite{Berezhiani:2015yta,DiValentino:2019ffd,DiValentino:2019jae,Kumar:2019wfs,Kumar:2021eev,SolaPeracaula:2021gxi,Akarsu:2021fol,Yan:2019gbw}. However, even though we have a series of alternative possibilities beyond the standard $\Lambda$CDM model, we are still left with some vital issues that are usually overlooked while discussing the simultaneous solutions to both the tensions. When all the cosmological probes are taken into account, the models do not offer satisfactory solutions to both the tensions. In addition, the choice of the observational data is also a very important issue, since some of the non-CMB cosmological probes that are often used to estimate the value of the $S_8$ parameter also assume $\Lambda$CDM as the background cosmological model in a manner similar to Planck. This, as a consequence, may influence the estimations in $S_8$ and the conclusion towards the solution or alleviation of the $S_8$ tension has a high chance to be biased.
As a result, finding a viable cosmological platform for the simultaneous solutions to both $H_0$ and $S_8$ tensions keeping all the constraints open, becomes very costly and challenging while this is incredibly exciting on the other hand. Below we shall present a number of cosmological scenarios (in alphabetical order) that have been proposed aiming to solve or alleviate the $H_0$ and $S_8$ tensions simultaneously.

\subsection{Early-Time Alternative Proposed Models}
\label{early}

\subsubsection{Axion Monodromy} 

The axion monodromy model can be an interesting candidate to alleviate the $S_8$ tension. This model is a realization of inflation~\cite{Snowmass2021:Inflation} in which pseudo-Nambu–Goldstone boson axions arise from symmetry breaking in the shift symmetry, also called the natural inflation~\cite{Freese:1990rb}. The potential of the axion monodromy is~\cite{Meerburg:2014bpa}
\begin{eqnarray}
    V (\phi) = V_0 (\phi) + \Lambda^4 \cos(\phi/f)\,,
\end{eqnarray}
where $\phi$ is a canonically normalized scalar field and $V_0(\phi)$ denotes  the slow-roll potential in the absence of modulations; $\Lambda$ and $f$ have the dimensions of mass. In Ref.~\cite{Meerburg:2014bpa}, the model is confronted to a combination of data coming from {\it Planck} 2013~\cite{Planck:2013win}, ACT~\cite{Das:2013zf}, SPT~\cite{Keisler:2011aw, Reichardt:2011yv}, BICEP2~\cite{BICEP2:2014owc} and low-$\ell$ WMAP polarization data from {\it Planck} Likelihood. As discussed in Ref.~\cite{Meerburg:2014bpa}, within this context, the suppression of the matter power spectrum can lead to a lower value of $S_8$ and thus, this model can alleviate the $S_8$ tension as a result.

\subsubsection{Early Dark Energy}
\label{sec:ede}
 
Early dark energy (EDE) behaves like a cosmological constant for $z \geq 3000$ and decays away as radiation or faster at later times~\cite{Karwal:2016vyq, Poulin:2018cxd, Agrawal:2019lmo, Lin:2019qug}. 
Related models include: {\it (i)}~coupling of the EDE scalar to neutrinos~\cite{Sakstein:2019fmf} and to dark matter~\cite{Karwal:2021vpk,McDonough:2021pdg}; {\it (ii)}~a first-order phase transition in a dark sector before recombination, which leads to a short phase of (N)EDE (see later subsection)~\cite{Niedermann:2019olb}; 
{\it (iii)}~an EDE model with an Anti-de Sitter phase around recombination~\cite{Akarsu:2019hmw,Ye:2020btb,Jiang:2021bab,Ye:2021iwa}; 
{\it (iv)}~an evolving scalar field asymptotically oscillating or with a non-canonical kinetic term~\cite{Agrawal:2019lmo,Lin:2019qug}, 
{\it (v)}~an axion-like particle sourcing dark radiation~\cite{Berghaus:2019cls}, 
{\it (vi)}~a scalar field with a potential inspired by ultra-light axions~\cite{Smith:2019ihp,Lucca:2020fgp}, 
{\it (vii)}~an axion field tunneling through a chain of energy metastable minima~\cite{Freese:2021rjq}, {\it (viii)}~$\alpha$-attractors~\cite{Braglia:2020bym}. 
EDEs with scaling solutions in the matter- and radiation-dominated eras (see Ref.~\cite{Wetterich:1987fm,Copeland:1997et,Sabla:2021nfy}) are not capable of relieving the $H_0$ tension in a significant way~\cite{Pettorino:2013ia,Gomez-Valent:2021cbe}. For a more model-independent reconstruction of the EDE fraction see Ref.~\cite{Gomez-Valent:2021cbe}.

Let us briefly discuss the main results of the original EDE proposal resolving the Hubble tension~\cite{Karwal:2016vyq, Poulin:2018cxd,Smith:2019ihp}, its limitations~\cite{Hill:2020osr, Ivanov:2020ril, DAmico:2020kxu, Murgia:2020ryi, Smith:2020rxx}, and the recent hint for EDE in ACT data~\cite{Hill:2021yec,Poulin:2021bjr,Jiang:2022uyg} (which is also consistent with SPT data~\cite{LaPosta:2021pgm, Smith:2022hwi,Jiang:2022uyg}). The model proposes a new scalar field $\phi$ with a potential of the form
\begin{equation}
    \label{eq:potential}
    V(\theta) = m^2 f^2\,\left(1-\cos \theta\right)^n,
\end{equation} where $m$ represents the axion mass, $f$ the axion decay constant, and $\theta \equiv \phi/f$ is a re-normalized field variable defined such that $-\pi \leq \theta \leq \pi$.

At early times, Hubble friction ensures that the field is held fixed at its initial value until a critical redshift $z_c$ (typically fixed when the approximate condition $9 H^2(z_c) \lesssim \abs{\partial^2V(\theta)/\partial\theta^2} $ is met~\cite{Marsh:2010wq,Smith:2019ihp}). Subsequently, the field starts rolling down its potential and oscillates about the minimum. As a result, the field dilutes at a rate controlled by the exponent $n$, with an approximate equation of state (EoS) parameter $w(n)=(n-1)/(n+1)$. This oscillatory behavior can be captured through a cycle-averaged evolution of the background and perturbative field dynamics~\cite{Poulin:2018dzj,Poulin:2018cxd} or followed exactly provided the oscillation period is not much faster than the Hubble rate~\cite{Agrawal:2019lmo,Smith:2019ihp,Hill:2020osr}.

To make the exploration of parameter space easier, most studies traded the "theory parameters" $m$ and $f$ for the "phenomenological parameters", namely the critical redshift $z_c$ at which the field becomes dynamical and the fractional energy density $f_{\rm EDE} \equiv f_{\rm EDE}(z_c)$ contributed by the field at this redshift. Therefore, in full generality, the model has four free parameters: $z_c$, $f_{\rm EDE}(z_c)$, $w(n)$ (or the exponent $n$), and the initial field value $\theta_i$ which controls the effective sound speed $c_s^2$ and thus most of the dynamics of perturbations. Note, that most studies assume that the field always starts in a slow-roll regime, as enforced by the very high value of the Hubble rate at early times. It was also shown that data have little constraining power on $n$ provided $ n > 1$ (per construction- the fluid must dilute faster than matter) and $\lesssim 5$ (at 95\% CL from {\it Planck} satellite data~\cite{Agrawal:2019lmo,Smith:2019ihp}), such that past works have often simply considered $n=3$ to restrict the parameter space.\footnote{Strictly speaking, there is some constraining power on the region $1<n<2$ which disfavors values too close to $n=1$~\cite{Agrawal:2019dlm}.}

The latest fit of this model to {\it Planck}, BAO, Pantheon, and SH0ES data~\cite{Murgia:2020ryi,Schoneberg:2021qvd} indicates $f_{\rm EDE}(z_c) \simeq 0.1\pm0.03 $ with $z_c \simeq 4070^{+400}_{-840} $. Moreover, it was shown that {\it Planck} polarization data are very sensitive to the dynamics of perturbations in the EDE fluid, leading to a tight constraint on $\theta_i\simeq 2.6\pm0.3$, and therefore on the shape of the potential close to the initial field value. A pure power-law potential is indeed disfavored by {\it Planck} satellite data~\cite{Agrawal:2019lmo,Smith:2019ihp}. In this model, {\it Planck}+BAO+Pantheon and SH0ES are in agreement at $\sim1.6\sigma$.

Nevertheless, the combination of {\it Planck}+BAO+Pantheon data alone does not show signs of non-zero EDE contribution, resulting in an 95\% CL upper limit on $f_{\rm EDE}<0.08$. This apparent contradiction is partially, but not entirely, driven by the highly non-Gaussian nature of the posteriors, with tails in the $f_{\rm EDE}$- and $H_0$-posteriors extending to high values, in the EDE model with three free parameters. Indeed, in models where $f_{\rm EDE}$ is too small to be detected, any parameter combination $\{z_c,\theta_i\}$ leads to a cosmology identical to $\Lambda$CDM, which greatly enhances the $\Lambda$CDM-like volume of parameters. On the other hand, only a narrow region of $\{z_c,\theta_i\}$ values allow for the $H_0$-degeneracy within CMB data to clearly appear. In fact, when the degrees of freedom are restricted to $f_{\rm EDE}$ only by fixing $z_c \simeq z_{\rm eq}$ (where $z_{\rm eq}$ is the redshift at matter-radiation equality) and $\theta_i\simeq 2.8$, {\it Planck} data alone lead to a non-zero detection of EDE with $f_{\rm EDE} \simeq 0.08\pm 0.04$, the Hubble constant $H_0\simeq(70\pm1.5){\rm\,km\,s^{-1}\,Mpc^{-1}}$, and $\Delta\chi^2\simeq 6$ in favor of the EDE model~\cite{Murgia:2020ryi,Smith:2020rxx}. Although this somewhat na\"{i}ve exercise illustrates that some features within {\it Planck} data are better fit by the EDE cosmology, the improvement in $\chi^2$ is too mild to drive $f_{\rm EDE}$ away from zero when three EDE parameters are considered. See Ref.~\cite{Herold:2021ksg} for a frequentist analysis along this line using a profile likelihood with three EDE parameters, which found $f_{\rm EDE} \simeq 0.072\pm 0.036$. 

Recently, the EDE model has been confronted with ACT data leading to a (perhaps surprising) {\em preference for EDE at $\sim 2\sigma$ from ACT alone}. Once combined with WMAP (or {\it Planck} TT at $\ell < 650$), BAO, and Pantheon data, the presence of EDE is favored at $3\sigma$, with $f_{\rm EDE}(z_c)\simeq 0.16^{+0.05}_{-0.09}$ (note that posteriors are non-Gaussian) and $H_0 \simeq (73.4\pm 3){\rm\,km\,s^{-1}\,Mpc^{-1}}$, without including any information from SH0ES~\cite{Hill:2021yec,Poulin:2021bjr,Jiang:2022uyg}. The significance of the preference and the parameter estimation remain consistent with this picture even after the inclusion of SPT-3G~\cite{LaPosta:2021pgm,Smith:2022hwi,Jiang:2022uyg} and {\it Planck} polarization data~\cite{Poulin:2021bjr,Smith:2022hwi}. However, the combination of the full {\it Planck} and ACT data leads instead to an upper limit $f_{\rm EDE}(z_c)<0.11$, which is weaker than what obtained from using {\it Planck} data alone. These somewhat contradicting results were shown to be due to the fact that the ACT best-fit EDE cosmology is disfavored by high-$\ell$ {\it Planck} TT data because of an amplitude offset in the ACT TE power spectrum compared to that predicted by the best-fit model to {\it Planck} TT data, and a (mild) discrepancy between ACT and {\it Planck} high-$\ell$ TT data. For discussions about ACT and {\it Planck} statistical consistency, see Refs.~\cite{ACT:2020gnv, Handley:2020hdp}.

The mismatch between ACT and {\it Planck} TT data on small angular scales gives rise to the different predictions of CMB lensing effect, see Sec.~\ref{sec:WG-Alens} for additional detail. For the full {\it Planck} data it leads to the so-called $A_{\rm lens}$ anomaly introduced in Sec.~\ref{sec:WG-Alens}. It pulls the late-time amplitude to a higher value~\cite{Planck:2016tof} being in conflict with the LSS measurements. Since there is no physical justification behind this anomaly, it is warranted to consider alternative CMB measurements specifically on small scales~\cite{Chudaykin:2020acu}. Combining the large-scale {\it Planck} temperature, SPTPol polarization and lensing measurements along with the full-shape BOSS data and information from photometric LSS surveys one finds $H_0\sim(69.8\pm1){\rm \,km\,s^{-1}\,Mpc^{-1}}$~\cite{Chudaykin:2020igl}. The inclusion of SH0ES result leads to $H_0\sim(71.8\pm1.2){\rm \,km\,s^{-1}\,Mpc^{-1}}$ and does not substantially worsen the fit to the galaxy clustering and weak lensing measurements. An alternative strategy to deal with the {\it Planck} lensing anomaly is to marginalize over the lensing information in the {\it Planck} CMB power spectra. The lensing-marginalized {\it Planck} data favor non-zero $f_{\rm EDE}$ at the $2\sigma$ level~\cite{Murgia:2020ryi}. Compared to the EDE cosmology reconstructed from the full {\it Planck} data, the tensions with $H_0$ and $S_8$ have decreased by $\sim1\sigma$ due to a shift in the mean of the reconstructed posterior in the unlensed cosmology.

There are a number of legitimate issues with the EDE model, that certainly indicate that this model cannot be the "end of the story". First, the degeneracy between $f_{\rm EDE}$ and $H_0$ only appears for somewhat tuned choices of parameters $\{z_c,\theta_i\}$, with {\it Planck} and SH0ES data favoring in particular $z_c\simeq z_{\rm eq}$ (although ACT suggests $z_c$ closer to the recombination epoch~\cite{Hill:2021yec}). This is reminiscent of the coincidence problem for "late" dark energy, and could either be the indication of a fine-tuning issue~\cite{Pettorino:2013ia}, or that a more complicated dynamics between the dark matter and dark energy sectors is at play. Second, the choice of potential itself, that assumes that lower-order terms in the axion-like potential do not play a role, requires some level of fine-tuning. Finally, in this model (as is true for most early Universe solutions), the DM density must be increased in order to compensate for the early integrated Sachs-Wolfe effect introduced by the gravitational potential decay for modes that enter the horizon when the field is dynamically important (i.e.\ around $z_c$)~\cite{Hill:2020osr,Vagnozzi:2021gjh}. This degeneracy between $f_{\rm EDE}$ and $\Omega_{\rm DM}$ allows to essentially keep the peak heights as measured by {\it Planck}~\cite{Poulin:2018cxd}, through the early integrated Sachs-Wolfe effect~\cite{Vagnozzi:2021gjh}. As a consequence, $\sigma_8$ increases ($\Omega_m$ is kept fixed in this model) worsening slightly the $S_8$ tension, typically by $\sim0.5\sigma$~\cite{Poulin:2018cxd,Hill:2020osr,Murgia:2020ryi}. Moreover, it was shown that when BOSS data are analyzed using the effective field theory of large-scale structure, they can further constrain EDE because they break the aforementioned $f_{\rm EDE}-\Omega_{\rm DM}$ degeneracy~\cite{Ivanov:2020mfr,DAmico:2020ods,Ivanov:2020ril}.

Including both the BOSS full-shape and weak lensing $S_8$ data, Refs.~\cite{Ivanov:2020ril,DAmico:2020ods} have shown that the parameters of the early dark energy model addressing the Hubble tension are disfavored at more than 99.95\% confidence level in the standard Bayesian analysis. This result is obtained without imposing the SHOES prior on $H_0$. However, if the $S_8$ data from at least DESy1 and HSC are excluded, and the EDE parameters $\{z_c,\theta_i\}$ are fixed instead of marginalizing, the early dark energy model is not disfavored anymore~\cite{Smith:2020rxx,Murgia:2020ryi}. Besides, it has been shown that the BOSS results depend, in part, on a mismatch in the late-time clustering amplitude $\sigma_8$ between BOSS and Planck. Although the two agree at the $95\%$ CL, this non-statistically significant mismatch still contributes to the $\sigma_8$ tension. These results suggests that further systematic studies of the EDE model in the context of the large-scale structure data are required.

Nevertheless, it is clear that LSS data have a potentially strong constraining power on the EDE cosmology (including future measurements of the halo mass function at high-redshift by the James Webb Space Telescope (JWST)~\cite{Klypin:2020tud}), and that the EDE cosmology cannot resolve the $S_8-$ and $H_0-$tensions simultaneously.

\subsubsection{Extra Relativistic Degrees of Freedom} 
 
Extra relativistic degrees of freedom at recombination, parametrized by the number of equivalent light neutrino species $N_{\rm eff}$~\cite{Steigman:1977kc}, provide one of the simplest extensions of $\Lambda$CDM. For three active massless neutrino families, $N_{\rm eff}^{\rm SM} \simeq 3.044$~\cite{Mangano:2005cc,deSalas:2016ztq, Akita:2020szl,Froustey:2020mcq,Bennett:2020zkv}. For the well-known degeneracy, we can increase $H_0$ at the price of additional radiation at recombination. 
Models of dark radiation can generically be classified into two distinct categories:
(1) Free streaming radiation and (2) Self interacting Dark Radiation (SIDR). The canonical example of (1) features the addition of 3 superweakly-interacting right-handed (Dirac) neutrinos, which decouple around the QCD phase transition~\cite{Anchordoqui:2011nh}. This model can be framed within the context of intersecting D-brane models of vibrating strings, and the extra-gauge boson could be within reach of the LHC Run-3~\cite{Anchordoqui:2012qu,Anchordoqui:2020znj}. On the other hand, SIDR models are in local thermal equilibrium and behave as perfect fluids~\cite{Baumann:2015rya,Brust:2017nmv,Blinov:2020hmc,Ghosh:2021axu}. 
Other generic examples that would enhance the value of $N_{\rm eff}$ include Majorana sterile neutrinos, Goldstone bosons, axions, vector fields, and neutrino asymmetry~\cite{Jacques:2013xr, Weinberg:2013kea, Allahverdi:2014ppa,DiValentino:2015sam, Barenboim:2016lxv, Carneiro:2018xwq, Paul:2018njm, Green:2019glg, Ferreira:2018vjj, DEramo:2018vss,Anchordoqui:2019yzc, Gelmini:2019deq, DiValentino:2015wba, Poulin:2018dzj,Baumann:2016wac, Zeng:2018pcv,Giare:2020vzo,Seto:2021tad, Gu:2021lni, Fernandez-Martinez:2021ypo, Franchino-Vinas:2021nsf,Feng:2021ipq,Cuesta:2021kca}. Future surveys will detect deviations from $N_{\rm eff}^{\rm SM}$ within $\Delta N_{\rm eff} \lesssim 0.06$ at 95\% CL, allowing us to probe a vast range of light relic models, where $\Delta N_{\rm eff} = N_{\rm eff} - N_{\rm eff}^{\rm SM}$~\cite{CMB-S4:2016ple, Abazajian:2019eic}. However, high-$\ell$ polarization measurements of the CMB disfavor models involving enough additional radiation to substantially increase $H_0$. More concretely, the correlation between $H_0$ and $\Delta N_{\rm eff}$ has been estimated numerically using data from {\it Planck} 2018 TT,TE,EE + lowE + BAO + Pantheon, 
\begin{equation}
\Delta H_0 = H_0 - \left. H_0 \right|_{\Lambda{\rm CDM}} \simeq 5.9 \, \Delta N_{\rm eff} \, ,
\end{equation}
where $\left. H_0 \right|_{\Lambda{\rm CDM}} = (67.27 \pm 0.60){\rm\,km\,s^{-1}\,Mpc^{-1}}$~\cite{Vagnozzi:2019ezj,Anchordoqui:2020djl}. Using the 95\% CL bound $\Delta N_{\rm eff} < 0.214$ reported by the {\it Planck} Collaboration~\cite{Planck:2018vyg}, it is straightforward to see that the data disfavor a solution of the $H_0$ tension in terms of $N_{\rm eff}$. Even when considering multi-parameter models, the number of extra relativistic degrees of freedom at recombination remains close to $N_{\rm eff}^{\rm SM}$~\cite{Anchordoqui:2021gji}.

A special class of SIDR models are {\it stepped fluids} which were studied in~\cite{Aloni:2021eaq}. In such models the SIDR fluid consists of a mix of massless and at least one massive particles. When the temperature of the dark fluid drops below the mass of a massive particles species the dark radiation undergoes a "step" in which its relative energy density increases as the massive particles deposit their entropy into the lighter species. If this transition occurs while CMB-observable modes are inside the horizon, high- and low-$\ell$ peaks are impacted differently, corresponding to modes that enter the horizon before or after the step. Such dynamics already occurs in the simplest supersymmetric theory, the Wess-Zumino model with a trilinear interactions and soft supersymmetry breaking. In this model the scalar mass can naturally be near the eV-scale which would lead to CMB-observable effects. Ref.~\cite{Aloni:2021eaq} investigates the cosmological signatures of such "Wess-Zumino Dark Radiation" (WZDR) and finds that it provides an improved fit to the CMB and BAO data alone, while favoring values of $H_0$ between 68-72${\rm\,km\,s^{-1}\,Mpc^{-1}}$ at 90\% confidence. If supernovae measurements from the SH0ES collaboration are also included in the analysis, the preferred values of $H_0$ become larger yet, but the preference for dark radiation and the location of the transition is left nearly unchanged. It was shown in Ref.~\cite{Aloni:2021eaq} that WZDR is among the most successful models at resolving the $H_0$ tension and the best of those with a Lagrangian formulation.

Another model where extra relativistic degrees of freedom appear to alleviate the Hubble tension is the one proposed in Ref.~\cite{Aboubrahim:2022gjb}. This model involves a long-lived massive scalar which  decays to  a pseudo-scalar at late times. These  along with a dark fermion and a dark photon make up the dark sector of a Stueckelberg extension of the SM. The dark sector which is feebly coupled to the SM via kinetic mixing is populated by energy injection from the visible sector. The feeble couplings forbid the dark particles from developing an equilibrium number density. This is a key feature which allows evading the constraint on $\Delta N_{\rm eff}^{\rm BBN}$. In this model the dark fermion is a DM candidate whose relic density is determined by solving a set of coupled Boltzmann equations for the comoving number density and taking into consideration the evolution of the visible and dark sectors in different heat baths~\cite{Aboubrahim:2020lnr}. After chemical and kinetic decoupling of the dark sector species, the massive scalar decays out of equilibrium into the massless pseudo-scalars increasing their energy density and contributing to $\Delta N_{\rm eff}^{\rm CMB}$ as needed to alleviate the tension between high $z$ and low $z$ measurements of the Hubble parameter.  The analysis of~\cite{Aboubrahim:2022gjb} provides a  complete model of particle cosmology from very high temperatures down to the  temperatures at CMB  consistent with all astrophysical and  laboratory constraints on the dark sector.

\subsubsection{Modified Recombination History} 

Shifting the sound horizon of baryonic acoustic oscillations has been shown to be a rather successful way of easing the Hubble tension. This can either be accomplished through a variation of the early time expansion history (see above), or through modifying the redshift of recombination. Solutions that modify the recombination history usually attempt to have an earlier recombination in order to infer a smaller sound horizon, which is compatible with a larger Hubble parameter. In particular, a route to explain the Hubble tension can be found by modifying the recombination history through heating processes~\cite{Chiang:2018xpn}, variation of fundamental constants~\citep{Hart:2017ndk, Hart:2019dxi, Sekiguchi:2020teg, Hart:2021kad}, the effects of primordial magnetic fields~\cite{Jedamzik:2020krr}, or a non-standard CMB temperature-redshift relation~\cite{Bose:2020cjb}.

The different proposed models have different mechanisms of accomplishing such a shift in recombination. 
\begin{itemize}
\item The inclusion of a variation of the effective rest mass of an electron, for example, shifts the atomic energy levels of hydrogen in the early Universe, allowing for a larger energy gap between the various orbitals, which also increases the required photo-ionization temperature. As such, recombination will already be possible at higher temperatures and correspondingly higher redshift (earlier times), while also altering the reionization history. Similarly, adopting a non-standard CMB temperature-redshift relation allows for the usual photo-ionization temperature to correspond to higher redshifts and thus earlier recombination. The model of baryon clumping (usually assumed to originate from a primordial magnetic field) instead allows recombination to be an inhomogeneous process. Given the non-linear dependence on the baryon abundance of the recombination problem, this allows for earlier recombination when $\langle n_b^2 \rangle \geq \langle n_b \rangle^2$.

Assessing the details of recombination against the {\it Planck} 2018 + BAO data, while allowing for a varying effective electron rest mass, increases the Hubble constant to $H_0 = (69.1 \pm 1.2){\rm \,km\, s^{-1}\, Mpc^{-1}}$~\cite{Hart:2019dxi}. Including additional curvature (as initially proposed in Ref.~\cite{Sekiguchi:2020teg}) allows the model to reach $H_0 = (69.3 \pm 2.1){\rm \,km\, s^{-1}\, Mpc^{-1}}$~\cite{Schoneberg:2021qvd}. Instead, When SNIa data are included, a similar analysis in the $\Lambda$CDM framework gives $H_0 = (68.7\pm 1.2){\rm \,km\, s^{-1}\, Mpc^{-1}}$~\cite{Sekiguchi:2020teg} in a flat Universe, and $H_0 = (72.3\pm2.7){\rm \,km\, s^{-1}\, Mpc^{-1}}$ in a curved Universe.

\item Primordial magnetic fields (PMFs) have been studied for many decades as a possible ingredient needed for explaining the observed galactic, cluster and extragalactic fields (see Refs.~\cite{Durrer:2013pga, Subramanian:2015lua, Vachaspati:2020blt} for reviews). A PMF present in the plasma before and during recombination would leave a variety of imprints in the CMB~\cite{Planck:2015zrl}. In particular, as first pointed out in Ref.~\cite{Jedamzik:2013gua} and later confirmed by detailed magnetohydrodynamics (MHD) simulations~\cite{Jedamzik:2018itu}, the PMF induces baryon inhomogeneities (clumping) on scales below the photon mean free path, enhancing the process of recombination and making it complete at an earlier time. This lowers the sound horizon at last scattering, which helps to alleviate the Hubble tension~\cite{Jedamzik:2020krr,Thiele:2021okz,Rashkovetskyi:2021rwg} and, as a byproduct, also slightly relieves the $\sigma_8$ tension. The required PMF comoving strength is $\sim 0.05\,$nG, which happens to be of the right order to naturally explain all the observed galactic, cluster and extragalactic fields.

Like all other solutions that reduce the sound horizon, baryon clumping is limited by how much it can increase the value of $H_0$ while remaining consistent with both CMB and BAO data~\cite{Jedamzik:2020zmd}. Current constraints on baryon clumping are based on a simple three-zone model first introduced in Ref.~\cite{Jedamzik:2013gua}, which contains one free parameter $b \equiv (\langle\rho_b^2\rangle-\langle\rho_b\rangle^2)/\langle\rho_b\rangle^2$. The most up to date constraints on the $b+\Lambda$CDM model are provided in Ref.~\cite{Galli:2021mxk}, where the CMB data from {\it Planck}, ACT DR4~\cite{ACT:2020frw} and SPT-3G Year 1 data~\cite{SPT-3G:2021eoc} was combined with the latest BAO, SN and distance ladder measurements. Constraints from ACT DR4 and {\it Planck} data were also presented earlier by Ref.~\cite{Thiele:2021okz}.  Interestingly, ACT and SPT were found to be in mild tension when it comes to $b$, with ACT tightening the {\it Planck} bound, i.e., tending to disfavor the PMF model, and SPT weakly preferring non-zero clumping. Combining the CMB data together with BAO and SNIa provides an upper limit of $b<0.4$ at 95\% CL ($b<0.5$ without ACT). Adding a SH0ES-based prior on the Hubble constant gives $b = 0.31^{+0.11}_{-0.15}$ and $H_0=69.28 \pm 0.56 {\rm\,km\,s^{-1}\,Mpc^{-1}}$ ($b = 0.41^{+0.14}_{-018}$ and $H_0=69.70 \pm 0.63{\rm\,km\,s^{-1}\,Mpc^{-1}}$ without ACT).  Further work is needed to clarify the potential (in)consistency of constraints on this model from various CMB data sets.

Future high resolution CMB temperature measurements, such as the full-survey Advanced ACT data~\cite{Henderson:2015nzj}, full-survey SPT-3G data~\cite{SPT-3G:2014dbx}, Simons Observatory~\cite{SimonsObservatory:2018koc}, and CMB-S4~\cite{Abazajian:2019eic}, along with better modelling of magnetically assisted recombination informed by comprehensive MHD simulations, should be able to conclusively confirm or rule out this proposal. Even if baryon clumping did not fully resolve the Hubble tension, a conclusive detection of evidence for a PMF would be a major discovery in its own right, opening a new observational window into the processes that happened in the very early Universe. 
\end {itemize}

\subsubsection{New Early Dark Energy}

While Early Dark Energy (EDE)~\cite{Poulin:2018dzj,Poulin:2018cxd,Smith:2019ihp} shows potential to resolve the Hubble tension, there are challenges at the theoretical level. The EDE potential can be motivated by higher order terms in the axion potential, but such a potential does not follow naturally from the effective field theory expansion and requires some fine-tuning for the lowest order terms in the potential not to dominate. In addition, it is disputed whether EDE leads to a full resolution of the Hubble tension at the phenomenological level~\cite{Hill:2020osr,Ivanov:2020ril, DAmico:2020kxu, Gomez-Valent:2021cbe,Benisty:2019vej} as a tension remains with LSS, although not at a significantly higher level than in the $\Lambda$CDM model~\cite{Murgia:2020ryi, Smith:2020rxx}.

A promising way forward to resolve these theoretical and phenomenological issues with EDE is the "New Early Dark Energy" (NEDE)~\cite{Niedermann:2019olb,Niedermann:2020dwg}. Instead of a slow roll-over phase transition, leading to the aforementioned problems with EDE, it is natural to think that the dark sector instead has undergone an almost instantaneous \textit{first order phase transition}.\footnote{An effectively instantaneous second order transition is also possible, which has been labeled a "hybrid NEDE model"~\cite{Niedermann:2020dwg}.} While this idea is different both at the theoretical and phenomenological level, it maintains the gist of the EDE proposal, and also holds in it the possibility of addressing its weaknesses.

At the background level, it is assumed that a NEDE phase transition took place in the dark sector {at zero temperature} shortly before recombination, triggered by a sub-dominant scalar field. Such a phase transition lowers an initially high value of the cosmological constant in the early Universe down to its present value as inferred from the measurement of $H_0$, potentially resolving the Hubble tension in analogy with the motivation behind the EDE model.

However, similarities between EDE and NEDE end at the perturbation level. The initial condition for the NEDE fluid {perturbations} after the instantaneous transition are fixed by the covariant junction conditions~\cite{Israel:1966rt, Deruelle:1995kd} across the space-like transition surface, relating them to adiabatic perturbations of the sub-dominant trigger field. Subsequently, NEDE perturbations evolve in a semi-stiff fluid describing the collision phase of newly nucleated bubbles on large scales. At the perturbative level, the NEDE and the EDE models are therefore very different, which is also exhibited in fits to CMB and LSS data. This will enable future CMB and LSS experiments to further discriminate between the models. The two models also lead to different signatures of gravitational waves (GWs), which will be yet another testing avenue using pulsar timing arrays (compare Ref.~\cite{Niedermann:2020dwg} and~\cite{Weiner:2020sxn}).

{Microscopically,} the cold NEDE phase transition can be described by a scalar field whose potential at some critical point develops two non-degenerate minima at zero temperature. This evolution is caused by an additional {subdominant} trigger field, which at the right moment makes the tunneling rate very high when the Hubble drag is released. 
When space becomes filled with bubbles of true vacuum during the phase transition, they will expand and almost instantaneously start to collide. As a part of the collision process the $\psi$ condensate will start to decay. At the microscopic level, the released free energy is converted into small scale anisotropic stress, which sources GWs and is expected to decay as $1/a^6$ similar to a stiff fluid component. After the phase transition, the fluid is therefore effectively, on large scales, a mixture of a stiff fluid as well as scalar and gravitational radiation.

The fast decay of the anisotropic stress is key to the phenomenological success of NEDE. Setting the EoS parameter of the decaying NEDE fluid fixed at $2/3$ and fitting the basic cold NEDE model with the fraction of NEDE energy density $f_{\rm NEDE}$ and the trigger mass $m$ as two additional parameters to {\it Planck} 2018 temperature, lensing and polarization data, large and small-$z$ BAO and LSS data (using the effective field theory of large-scale structure applied to BOSS/SDSS data~\cite{DAmico:2019fhj,Ivanov:2019pdj,Colas:2019ret}), and the Pantheon data set, a preference of NEDE over $\Lambda$CDM was reported~\cite{Niedermann:2020qbw}. The phase transition takes place at $z_c=4900_{-800}^{+660}$ with the Hubble constant $H_0 = (71.2 \pm 1.0){\rm\,km\,s^{-1}\,Mpc^{-1}}$ and a non-vanishing NEDE fraction $f_\text{NEDE} = 11.7^{+3.3}_{-3.0}\,\%$, thus excluding $f_\text{NEDE}=0$ at a $\simeq 4\sigma$ significance.

Notably, in a subsequent analysis, NEDE has been ranked high among other early-time competitors, bringing the DMAP tension\footnote{This tension measure refers to "difference of the maximum a posteriori" as for example discussed in~\cite{Raveri:2018wln}. It is applicable in cases with non-Gaussian posterior.} below $2 \sigma$~\cite{Schoneberg:2021qvd}. Beyond that, a direct comparison between EDE and NEDE models using ACT and {\it Planck} data is exemplifying the phenomenological differences between both models at a perturbation level~\cite{Poulin:2021bjr}, although it is not clear how much they are driven by internal tensions between both data sets~\cite{Hill:2021yec,Poulin:2021bjr,Moss:2021obd}.

While future CMB, LSS and pulsar timing array experiments can test this model more precisely, more theoretical developments will also be needed to fully resolve the $S_8$ tension if it turns out to become larger in future measurements~\cite{Niedermann:2020qbw}. This may require to integrate the NEDE sector with other parts of the dark sector and/or the neutrino sector. As outlined in the conclusion of Ref.~\cite{Niedermann:2020qbw}, a minimal resolution might already be present in the model's simplest implementation in terms of NEDE scalar field oscillations around the true vacuum, which constitute a form of interacting dark matter and hold the potential to resolve the $S_8$ tension~\cite{Archidiacono:2019wdp,Allali:2021azp}.

As an alternative pathway, it was recently demonstrated that the NEDE phase transition can also be triggered by subsiding temperature corrections (rather than a scalar field) as the Universe expands~\cite{Niedermann:2021ijp, Niedermann:2021vgd}. This model, dubbed \textit{hot NEDE}, leads to a sizeable fraction of false vacuum energy -- doubling as an early dark energy component -- in the strongly supercooled region where the dark radiation fluid is subdominant. This phase transition can also give mass to the active neutrinos through the inverse seesaw mechanism~\cite{Abada:2014vea}, when "higgsing" a set of (super)-eV sterile neutrinos. The coupling between the neutrino sector, the NEDE field and the Goldstone boson associated with a spontaneously broken lepton symmetry (the majoron) is expected to change the Boltzmann evolution of neutrinos at late times and provide novel decay channels for the bubble wall condensate. This offers an alternative way of addressing the $H_0$ and $S_8$ tension simultaneously. Ultimately, these recent developments of NEDE suggests the possibility that the Hubble tension could be a signature of how neutrinos acquired their mass. For other models building up on NEDE see also Refs.~\cite{Freese:2021rjq,Allali:2021azp}.

We would like to point out, as done in~\cite{DAmico:2021zdd}, an interesting consequence of several models aimed at addressing the Hubble tension, in particular early dark energy models and NEDE (but also models with extra relativistic degrees of freeedom).
Fitting CMB data in terms of these models, a common feature is that the spectral index is bluer with respect to the $\Lambda$CDM value 
(see~\cite{DiValentino:2018zjj,Ye:2020btb,Ye:2021nej,Tanin:2020qjw,Takahashi:2021bti}): such a shift of $n_s$ will have very important implications for inflationary model constraints.

\begin{figure}[thb]
    \centering
    \includegraphics[scale=0.85]{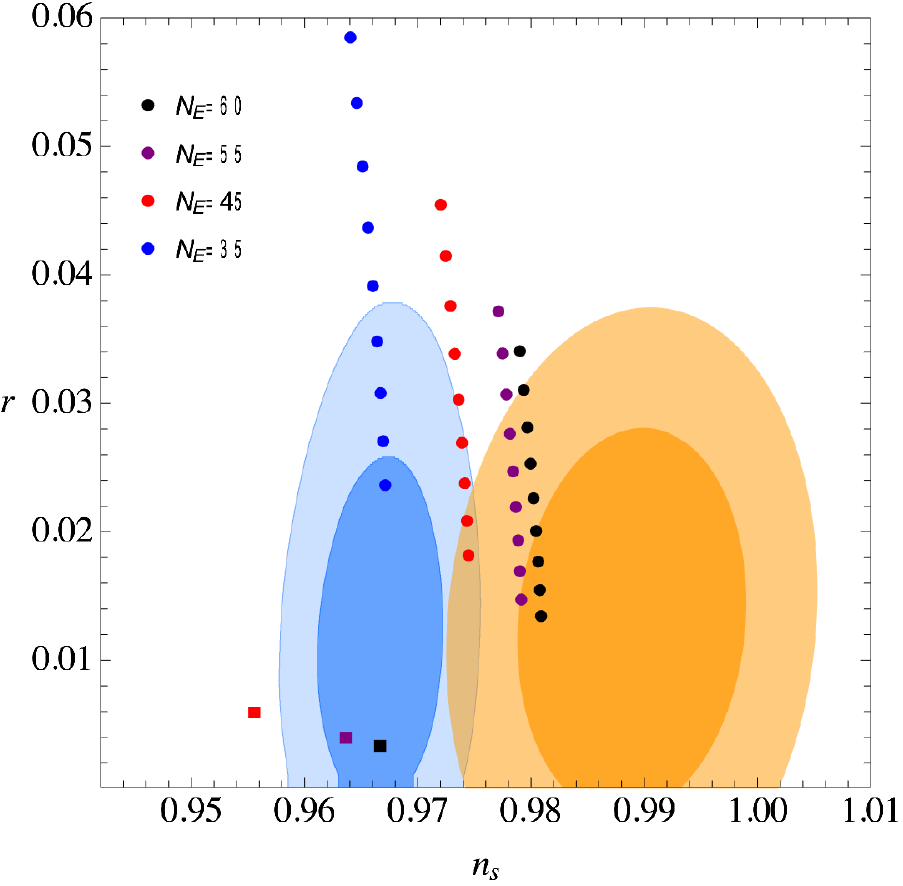}
    \caption{Tensor-to-scalar ratio $r$ versus spectral index $n_s$, for the $\Lambda$CDM model (blue) 
    and for the NEDE model (gold). To get the NEDE model constraint, we approximate the $\Lambda$CDM 
    contour as a bivariate Gaussian, and substitute the mean and error on $n_s$ by the ones got in the NEDE model. 
    This approximate procedure reproduces well a full analysis~\cite{Ye:2021nej}. The round dots are predictions of
    potentials $\sim \phi^p$ with $0.1 \leq p \leq 0.5$. The square dots are the predictions of the Starobinsky $R^2$ inflation~\cite{Starobinsky:1980te}. This figure is from~\cite{DAmico:2021zdd}.}
\label{fig:ede_ns}
\end{figure}

We illustrate this in Fig.~\ref{fig:ede_ns}, where we plot the contours obtained from the $\Lambda$CDM model, and the one we infer from the constraints on $n_s$ in the NEDE model, from the analysis of~\cite{Niedermann:2019olb,Niedermann:2020dwg}. 
This plot clearly shows how many models of inflation such as flattened monodromy models with a long stage of inflation and power-law behavior of, for example, $\sim \phi^{1/2}$ at large $\phi$, which were supposedly excluded by having too large an $n_s$, may end up being better candidates than, for example, the Starobinsky $R^2$.
Of course, it is too soon to claim this. But until the $H_0$ tension is resolved, it may also be too soon to claim the opposite.
Therefore, the solution to the (late-Universe) $H_0$ and $S_8$ tensions may have important consequences on our understanding of inflation.

\subsection{Late-Time Alternative Proposed Models}

\subsubsection{Bulk Viscous Models} 

The cosmological scenarios in which a fluid is endowed with a bulk viscosity is an effective description to alleviate the Hubble constant tension~\cite{Wang:2017cel, Yang:2019qza, daSilva:2020mvk}. A bulk viscous fluid is characterized by its energy density $\rho$ and pressure $p$ where this pressure term has two distinct components, namely, $p = p_{u}+ p_{\rm vis}$. The first component of the pressure term $p_{u}$ is the usual one $p_{u} = w \rho$, where $w$ is the EoS parameter of the fluid, and the second component $p_{\rm vis}$ characterizes the viscosity of the fluid defined by $p_{\rm vis} = -\xi (t) {u^{\mu}}_{; \mu}$ in which $\xi(t)>0$ is the coefficient of the viscosity or the bulk viscous coefficient and $u^\mu$ is the four-velocity of the fluid~\cite{Brevik:2005bj}.

For any bulk viscous fluid, one can write down its evolution in the FLRW Universe as 
\begin{equation}
    \frac{\mathrm{d}\rho}{\mathrm{d}t} + 3 H(1+w)\rho = 9H^2\xi (t)\,,
\end{equation}
where $H$ is the Hubble rate of the FLRW Universe. Thus, one can see that for different choices of the bulk viscous coefficient, $\xi(t)$, one can extract the constraints out of various cosmological scenarios. The cosmic fluid endowed by the bulk viscosity can either represent a DE fluid~\cite{daSilva:2020mvk} or a unified fluid which combines both DM and DE behaviours~\cite{Yang:2019qza}. The cosmological scenarios in which DE enjoys a viscous nature have been investigated in the literature aiming to address the Hubble tension~\cite{Wang:2017klo,Mostaghel:2016lcd, Mostaghel:2018pia,Velten:2012uv}. The bulk viscosity coefficient  characterized by  $\xi(t) \propto H$~\cite{Wang:2017klo} leads to $H_0=69.3\pm1.7{\rm\,km\,s^{-1}\,Mpc^{-1}}$ at 68\% CL for the dataset {\it Planck} 2018 CMB distance priors + Pantheon + BAO + BBN + CC~\cite{daSilva:2020mvk}. 
For the bulk viscosity coefficient $\xi (t) \propto \sqrt{\rho_{\rm DE}}/H$~\cite{Mostaghel:2016lcd, Mostaghel:2018pia}, the Hubble constant takes the value $H_0=69.3\pm1.7{\rm\,km\,s^{-1}\,Mpc^{-1}}$ at 68\% CL for the combined dataset {\it Planck} 2018 CMB distance priors + Pantheon + BAO + BBN + CC~\cite{daSilva:2020mvk}.  A different choice of the form 
$\xi (t) \propto \rho_{\rm DE}^\nu$~\cite{Velten:2012uv} can lead to $H_0=69.2\pm1.7{\rm\,km\,s^{-1}\,Mpc^{-1}}$ at 68\% CL~\cite{daSilva:2020mvk} for {\it Planck} 2018 CMB distance priors + Pantheon + BAO + BBN + CC. 
As the choice of the bulk viscous coefficient is not unique, one can therefore introduce a different parametrization to investigate the effectiveness of the models in the light of the Hubble constant tension.

Alternatively,  a unified fluid endowed with bulk viscosity can increase the Hubble constant value. In this context, the bulk viscous coefficient $\xi (t) \propto \rho^m$ ($m$ is a real number)~\cite{Velten:2012uv} leads to $H_0 = 70.2^{+1.6}_{-1.9}{\rm\,km\,s^{-1}\,Mpc^{-1}}$ at 68\% CL~\cite{Yang:2019qza} (assuming a free $w_0$ with $m =0$) for the dataset {\it Planck} 2015 + Pantheon~\cite{Yang:2019qza} and $H_0 = 68.0^{+2.7}_{-2.4}{\rm\,km\,s^{-1}\,Mpc^{-1}}$ at 68\% CL (assuming both $w_0$ and $m$ are free parameters) for the dataset {\it Planck} 2015 + Pantheon~\cite{Yang:2019qza}.

On the other hand, an inhomogeneous cosmic fluid endowed with the bulk viscosity can also be considered as a possible scenario to alleviate the Hubble constant tension~\cite{Elizalde:2020mfs}. Using the Bayesian machine learning approach, the authors of Ref.~\cite{Elizalde:2020mfs} investigated two equation of state parameters representing an inhomogeneous cosmic fluid, namely, Model 1: $p= (w_0 + w_1/(1+z)) \rho - A H^n$ ($w_0, w_1, n, A$ are free parameters) and Model 2: $p = (w_0 + w_1/(1+z)) \rho - A q H^n $ ($w_0, w_1, n, A$ are free parameters and $q \equiv -1 - \dot{H}/H^2$ is the deceleration parameter of the Universe). For Model 1, the Bayesian approach leads to $H_0 = 73.4 \pm 0.1 {\rm\,km\,s^{-1}\,Mpc^{-1}}$ at 68\% CL (for the redshift interval $z \in [0, 2.5]$) and $H_0 = 73.4 \pm 0.1 {\rm\,km\,s^{-1}\,Mpc^{-1}}$ at 68\% CL (for the redshift interval $z \in [0, 5]$). For Model 2,  the Bayesian approach leads to $H_0 = 73.5 \pm 0.2 {\rm\,km\,s^{-1}\,Mpc^{-1}}$ at 68\% CL (for the redshift interval $z \in [0, 2.5]$) and $H_0 = 73.6 \pm 0.1 {\rm\,km\,s^{-1}\,Mpc^{-1}}$ at 68\% CL (for the redshift interval $z \in [0, 5]$).

\subsubsection{Chameleon Dark Energy} 

The chameleon field~\cite{Khoury:2003aq,Khoury:2003rn,Wang:2012kj,Upadhye:2012vh,Khoury:2013yya,Vagnozzi:2021quy,Benisty:2021cmq} can be considered to be a potential candidate for resolving the Hubble constant tension as recently explored in Ref.~\cite{Cai:2021wgv}. In addition to the screening effect from changing the effective mass of the chameleon field according to the ambient matter density, the chameleon field also admits a higher potential energy at its effective potential minimum (namely the vacuum energy) in the high-density regions than in the low-density regions. Hence, the over-densities would expand locally faster than the under-densities since there is more chameleon dark energy stored in the overdense regions. Therefore, the chameleon dark energy model~\cite{Cai:2021wgv} provides an environmentally dependent determination for the local Hubble constant. For a toy model with a top-hat density profile connecting an inner FLRW region to a outer FLRW region, a 10\% matter over-density could raise the local $H_0$ value up to $70{\rm \,km\,s^{-1}\,Mpc^{-1}}$, while a 20\% matter over-density could reach a local $H_0$ value as high as $74{\rm \,km\,s^{-1}\,Mpc^{-1}}$, which could be fitted precisely by the following approximation (here $n=1,2,3,\cdots$ is the power of the Peebles-Ratra potential),
\begin{equation}
    \frac{\delta\rho_m}{\rho_m} = \left(0.65\,e^{-0.57n}+2.13\right)\left(\frac{H_0}{67.27{\rm\,km\,s^{-1}\,Mpc^{-1}}}-1\right)+\left(0.65\,e^{-0.27n}+1.07\right)\left(\frac{H_0}{67.27{\rm\,km\,s^{-1}\,Mpc^{-1}}}-1\right)^2\,.
\end{equation}
At large scales beyond $\gtrsim 1\,$Gpc, the matter homogeneity is well tested by the CMB data, therefore, the Hubble constant inferred from this scale simply recovers the CMB constraint on $H_0$. At the intermediate scales of a few $\sim\mathcal{O}(100)$ Mpc, the matter homogeneity is relatively poorly guaranteed due to the presence of some large-scale structures like voids~\cite{Keenan:2013mfa} and walls~\cite{deLapparent:1985umo,Gott:2003pf,Lietzen:2016thc}. Therefore, Ref.~\cite{Cai:2021wgv} also constructs a realistic model with a LTB metric to describe such intermediate inhomogeneities by explicitly counting the galaxy number excess from BOSS DR12 data~\cite{Reid:2015gra,Kitaura:2015uqa} in each concentric spherical shells, however, the outcome is highly dependent on the adopted galaxy bias model~\cite{Lavaux:2019fjr}. At even smaller scales below $\sim100$ Mpc, the matter homogeneity is obviously broken by computing the fractal dimension of galaxy number growth within a sphere, see e.g.\ Ref.~\cite{Scrimgeour:2012wt}. Using the distance indicators in the small-scale high-density regions (for example, SNIa in galaxies) would generally return back a higher $H_0$ value than the background. Similarly, calibrating the SNIa with Cepheids~\cite{Riess:2016jrr,Riess:2018byc,Riess:2018uxu,Riess:2019cxk,Riess:2020fzl,Riess:2021jrx} generally renders a larger $H_0$ value than that with TRGB as the calibrator~\cite{Freedman:2019jwv, Yuan:2019npk, Freedman:2020dne,Soltis:2020gpl,Freedman:2021ahq,Anand:2021sum}. This is because that most of TRGBs are specifically chosen to reside away from high matter density (for example, galaxy disk) to reduce large dust extinctions. Furthermore, the most recent improved measurements of strong lensing time delay from TDCOSMO+SLACS sample~\cite{Birrer:2020tax} favors a $H_0$ value closer to the background CMB value, which could also be explained within the picture of chameleon dark energy model by lensing objects that are specifically chosen in the under-density regions in order to diminish the effect from galaxy clusters. Note in end that since the chameleon dark energy drives the over-density region to expand faster so that the matter fluctuations smear at a larger scale until reaching a new virial equilibrium with lower matter density. This could shed the light on resolving the $S_8$ tension at the same time, which would require for a full time evolution formulation at perturbation level.

\subsubsection{Clustering Dark Energy}

In clustering DE models, the sound speed of DE perturbations is much smaller than unity, allowing for non-negligible dark energy perturbations on scales well inside the horizon, see Ref.~\cite{Batista:2021uhb} for a review. Consequently, the growth of matter perturbations can be impacted by DE perturbations. Such models have been explored as a possible way to suppress $\sigma_8$~\cite{Kunz:2015oqa}. 

In the limit $c_s \to 0$, on small scales and during the matter-dominated era, DE perturbations follow matter perturbations according to the solution~\cite{Abramo:2008ip,Sapone:2009mb,Creminelli:2008wc}
\begin{equation}
    \label{clust-de-solution}
    \delta_{\rm DE}=\frac{1+w_{DE}}{1-3w_{DE}}\delta_m\,.
\end{equation}
Thus, the impact of clustering DE on $\sigma_{8}$ depends on the $1+w_{DE}$. For non-phantom (phantom) models, DE perturbations are correlated (anti-correlated) to matter perturbations, enhancing (diminishing) $\sigma_{8}$. If at intermediate redshifts the $w_{DE}$ is not very close to $-1$, DE perturbations can be of the same order of magnitude as matter perturbations. In this case, the impact of DE on matter growth is maximal.

\begin{figure}
    \centering
    \includegraphics[width=0.5\textwidth]{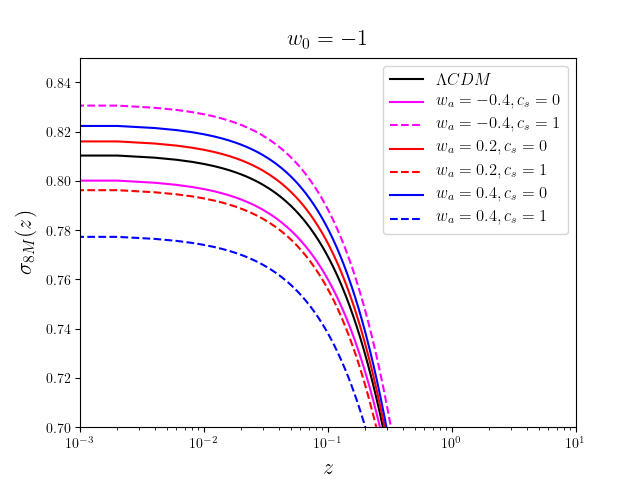}
    \caption{Evolution of $\sigma_{8M} (z)$ for clustering and homogeneous
DE models. For all models, $w_0=-1$,  $\Omega_m=0.315$, and
$\sigma_{8\Lambda}=0.811$ have been used to normalize $\sigma_{8M}
(z)$ at $z_{\rm rec}$.}
    \label{sig8-clust-de}
\end{figure}

Since DE perturbations can also contribute significantly to the gravitational potential, the associated perturbations should include its weighted contribution
\begin{equation}
    \label{delta-tot-clust-de}
    \delta_{\rm tot}=\delta_{m}+\frac{\Omega_{\rm DE}}{\Omega_m}\delta_{\rm DE}\,.
\end{equation}
The redshift evolution of $\sigma_{8}$ for a given model M, $\sigma_{8M}\left(z\right)=\sigma_{8M}D_{M}\left(z\right)$, where $D_{M}$ is the growth function computed with $\delta_{\rm tot}$ can be largely affected by clustering DE. Normalizing $\sigma_{8M}\left(z\right)$ to the $\Lambda$CDM value at $z_{\rm rec}$, Fig.~\ref{sig8-clust-de} shows the evolution of $\sigma_{8M}\left(z\right)$ for clustering and homogeneous DE models. Phantom clustering DE can potentially alleviate the $S_{8}$ tension, whereas the impact of non-clustering DE worsens it. But it is also possible that, for a given evolution of $w_{DE}$ which provides a too low $\sigma_{8}$, clustering non-phantom DE can be used to increase it.

\subsubsection{Diffusion Models}

A diffusive interaction between dark energy and dark matter was introduced in Refs.~\cite{Haba:2016swv,Koutsoumbas:2017fxp}. In Ref.~\cite{Calogero:2013zba}, the authors formulate diffusion of energy density between dark energy into dark matter by using a non-conserved stress energy tensor $T^{\mu\nu}$ with a source current $j^\mu$,
\begin{equation} 
    \nabla_\mu T^{\mu\nu}_{(\textbf{M})}=\gamma^{2} j^\nu\,,
\end{equation}
where $\gamma^{2}$ is the coupling diffusion coefficient of the fluid. The current $j^\mu$ is a time-like covariant conserved vector field $j^{\mu}_{;\mu}=0$ which describes the conservation of the number of particles in the system.

In a homogeneous expansion, the modified Friedman equations read:
\begin{equation}
 \dot{\rho}_m + 3 H \rho_m = \frac{\gamma}{a^3} , \quad  \dot{\rho}_\Lambda  = -\frac{\gamma}{a^3} 
\end{equation}
The contribution of the current goes as $\sim a^{-3}$ since the current is covariantly conserved $j^{\mu}_{;\mu}=0$. In this way, there is a compensation between the dark energy and the dark matter.  Ref.~\cite{Perez:2020cwa} claims that the diffusive interaction models can alleviate the Hubble constant tension with different types of matter fields.

Refs.~\cite{Benisty:2017eqh,Benisty:2017rbw,Benisty:2017lmt,Benisty:2018oyy} formulate an action principle for these interactions starting from actions. The theory reads:
\begin{equation} \label{nhd1}
S_{(\chi,A)}=\int d^{4}x\sqrt{-g}\chi_{\mu;\nu}T_{\left(\chi\right)}^{\mu\nu} +\frac{\sigma}{2}\int d^4x \sqrt{-g}(\chi_{\mu}+\partial_{\mu}A)^2 
\end{equation} 
where $A$ is a scalar field. From a variation with respect to the vector field $\chi_{\mu}$ we obtain:
\begin{equation} \label{nhd2}
\nabla_{\nu}T_{\left(\chi\right)}^{\mu\nu}=\sigma(\chi^{\mu}+\partial^{\mu}A)= f^\mu,
\end{equation} 
 {where $f^\mu=\sigma (\chi^{\mu}+\partial^{\mu}A)$ is a current source for the stress energy momentum tensor} $T_{\left(\chi\right)}^{\mu\nu}$. From the variation with respect to the new scalar $A$, a covariant conservation of the current emerges: 
\begin{equation}\label{nhd3}
\nabla_{\mu}f^\mu=\nabla_{\mu}(\chi^{\mu}+\partial^{\mu}A)=0
\end{equation} 
Different expressions for $T^{\mu\nu}_{(\chi)}$ which depends on different variables give the conditions between the Dynamical Spacetime vector field $\chi_\mu$ and the other variables. Cosmological implications are described in Ref.~\cite{Benisty:2016ybt,Benisty:2018qed,Benisty:2018gzx,Anagnostopoulos:2019myt,Benisty:2020nql}.

\subsubsection{Dynamical Dark Energy}

Dropping the field interpretation of DE, it is possible to explore models in which DE sector deviates from the cosmological constant to a DE fluid with an extended EoS parameter $w_{\rm DE} \neq -1$, which can generally vary with redshift. These dynamical dark energy (DDE) models have been extensively studied in the literature~\cite{Planck:2018vyg, Zhao:2017cud, Yang:2018qmz, Yang:2018prh, DiValentino:2019dzu, Yang:2018qmz, Vagnozzi:2019ezj, Keeley:2019esp, Joudaki:2016kym,DiValentino:2017gzb, Yang:2021flj, Benisty:2021gde, Roy:2022fif,Sharma:2022ifr}, based on considerations over the reconstruction of the effective $w_{\rm DE}$ from data~\cite{Zhao:2017cud}.\footnote{However, low redshift data do not necessitate DDE~\cite{Colgain:2021pmf}.} DDE models typically solve the $H_0$ tension within two standard deviations at the price of a phantom-like DE, i.e.\ $w_{\rm DE}<-1$, because of the geometrical degeneracy present with the DE EoS parameter $w_{\rm DE}$. However, these models appear to worsen the growth tension of $\Lambda$CDM~\cite{Alestas:2021xes}, while providing a worse fit to the BAO and SNIa data than $\Lambda$CDM. In addition, they favor a value of SNIa absolute magnitude that is marginally consistent with Cepheid measurements~\cite{Perivolaropoulos:2021jda,Alestas:2021xes}.

A particular Dynamical DE with a phantom crossing behaviour, as introduced in Ref.~\cite{DiValentino:2020naf}, seems to be able to solve the Hubble constant problem also when BAO and Pantheon data are considered, and at the same time alleviate the $S_8$ tension, see also Ref.~\cite{Chudaykin:2022rnl}.

\subsubsection{Emergent Dark Energy} 

A class of dark energy models in which dark energy has no effective presence in the past as it emerges at later times, usually dubbed as the \textit{emergent dark energy} models~\cite{Li:2019yem,Pan:2019hac,Rezaei:2020mrj,Liu:2020vgn,Li:2020ybr,Yang:2020tax,Benaoum:2020qsi,Yang:2021eud}, has received considerable attention in the cosmological community. In emergent dark energy models, the parametrization of the dark energy density is inserted into the Hubble equation as
\begin{equation}
    H^2(z) = H_0^2\, \Bigl[\Omega_r (1+z)^{4} + \Omega_m (1+z)^{3}+ \widetilde{\Omega}_{\rm DE}(z)\Bigr]\,,
\end{equation}
where $\widetilde{\Omega}_{\rm DE}(a) = \rho_{\rm DE} (z)/\rho_{{\rm crit},0}$ is the energy density of the DE fluid with respect to the critical energy density at present, namely $\rho_{\rm {crit},0}(z)=3M_{\rm Pl}^2H_0^2$. In Ref.~\cite{Li:2019yem}, the authors introduced a very simple emergent dark energy scenario, known as phenomenologically emergent dark energy (PEDE) scenario, where the dark energy sector evolves as
\begin{equation}
    \label{PEDE}
    \widetilde{\Omega}_{\rm DE}(z) = \Omega_{\rm DE} \left[1- \tanh (\log_{10} (1+z)) \right]\,,
\end{equation}
where $\Omega_{\rm DE} = 1-\Omega_m-\Omega_r$. It is important to mention that within the PEDE scenario, even if the dark energy sector is dynamical, this scenario has exactly six free parameters as in the $\Lambda$CDM model. For {\it Planck} 2018 alone, this model takes a very high value of the Hubble constant $H_0 = 72.35_{-0.79}^{+0.78}{\rm \,km\,s^{-1}\,Mpc^{-1}}$ at 68\% CL~\cite{Yang:2020tax} which is compatible with Ref.~\cite{Riess:2020fzl} within $1\sigma$. Recently, in Ref.~\cite{DiValentino:2021rjj}, the PEDE model has been further confronted in the  presence of the sterile neutrinos and perfectly agrees with the earlier observations~\cite{Li:2019yem,Pan:2019hac,Yang:2020tax}. 

A generalization of the PEDE scenario known as generalized emergent dark energy (GEDE) introduces a dark energy evolution in terms of a free parameter $\Delta$ as~\cite{Li:2020ybr}
\begin{equation}
    \label{GEDE}
    \widetilde{\Omega}_{\rm DE}(z)\,=\, \Omega_{\rm DE}\frac{ 1 - \tanh\left(\Delta \times \log_{10}(\frac{1+z}{1+z_t}) \right) }{{1+ \tanh\left(\Delta \times \log_{10}({1+z_t}) \right)}}\,.
\end{equation}
While the symbol $z_t$ is a derived parameter representing the epoch where the matter energy density and the DE density are equal, that means equating, $\rho_m (z_t) = \rho_{\rm DE} (z_t) \implies \Omega_m (1+z_t)^3 = \widetilde{\Omega}_{\rm DE}(z_t)$, one can derive $z_t$. Although this scenario does not offer a very large value of the Hubble constant compared with its value estimated in the PEDE scenario, however, as explored in Ref.~\cite{Yang:2021eud}, within this GEDE scenario, the Hubble constant tension is reduced to $1.8 \sigma$ with~\cite{Riess:2020fzl} (see also~\cite{Hernandez-Almada:2020uyr}).

\subsubsection{Graduated Dark Energy - AdS to dS Transition in the Late Universe}
\label{subsec:gDE}

The positive cosmological constant assumption of the $\Lambda$CDM model was investigated via the graduated dark energy (gDE)---borrowed from the proposal of graduated inflation~\cite{Barrow:1990vx}---characterised by a minimal dynamical deviation from the null inertial mass density $\varrho=0$ (where $\varrho\equiv\rho+p$) of the cosmological constant (or, the usual vacuum energy of QFT)~\cite{Akarsu:2019hmw}. This deviation is in the form of
\begin{equation}
\varrho\propto \rho^{\lambda},    
\end{equation}
for which, provided that the conditions $\rho_0>0$ (present-day energy density), $\varrho_0<0$ (present-day inertial mass density), and $\lambda=1-\frac{n}{m}<1$ with $n$ and $m$ being odd numbers are satisfied, the energy density $\rho$ dynamically takes negative values in the past~\cite{Akarsu:2019hmw} (see also~\cite{Visinelli:2019qqu}). These three conditions lead to the gDE; the first two imply that gDE is in the phantom region in the late/present-day universe, and the last one causes the energy density to take negative values for redshifts larger than a certain redshift (see Ref.~\cite{Akarsu:2019hmw} for details). During the transition from negative to positive energy density as the Universe expands, there comes a time $t_\dagger$ (or a redshift $z_\dagger$) for which the energy density passes through zero and the EoS parameter exhibits a pole. The gDE in fact exhibits a wide variety of behaviors depending on $\lambda$, but it is of particular interest that for large negative values of $\lambda$, it establishes a phenomenological model characterized by a smooth function that approximately describes a cosmological constant which switches sign in the late Universe to become positive today.\footnote{The $\lambda=0$ case is called simple-gDE and investigated in~\cite{Acquaviva:2021jov}.} The energy density of the gDE model, in terms of redshift, reads
\begin{equation}
    \begin{aligned}
    \label{eqn:greatDE}
    \rho_{\rm gDE}= \rho_{\rm gDE0}\, {\rm sgn}[1+\Psi \ln (1+z)] \,\big| 1+\Psi \ln (1+z) \big|^{\frac{1}{1-\lambda}}\,,
    \end{aligned}
\end{equation}
where $\Psi<0$ is a negative constant. Here "${\rm sgn}$" is the signum function that reads ${\rm sgn}[x]=-1,0,1$ for $x<0$, $x=0$ and $x>0$, respectively. This expression indicates if there exists a sign change in the energy density of the gDE (accompanied by a pole in its EoS parameter~\cite{Ozulker:2022slu}), it will happen at the redshift 
\begin{equation}\label{eq:redshift}
    z_{\dagger}={\rm e}^{-\Psi^{-1}}-1,
\end{equation}
around which, $H(z)$ can exhibit a non-monotonic behavior. It was shown via the gDE that the joint observational data, including but not limited to the {\it Planck} CMB and Ly-$\alpha$ BAO (BOSS DR11) data, suggest that the cosmological constant changed its sign at $z\approx 2.32$ and triggered the late-time acceleration, the behaviour of which alleviates the $H_0$ tension by predicting $H_0 \approx 69.7\pm0.9{\rm \,km\, s^{-1}\, Mpc^{-1}}$ and provides excellent fit to the Ly-$\alpha$ BAO (BOSS DR11) data~\cite{Blomqvist:2019rah} at the effective redshift $z\sim2.34$, which is at $\sim2.5\sigma$ tension with the {\it Planck} 2015 best-fit $\Lambda$CDM. Note that this tension is reduced to $\sim1.5\sigma$ when the final eBOSS (SDSS DR16) measurement, which combines all the data from eBOSS and BOSS~\cite{Blomqvist:2019rah,eBOSS:2020yzd,duMasdesBourboux:2020pck}, is considered, see Sec.~\ref{sec:lyman_alpha}.

Inspired by these observational findings, and the theoretically compelling fact that the gDE submits to the weak energy condition and the bounds on the speed of sound only in the limit $\lambda\rightarrow-\infty$ which corresponds to a cosmological constant that rapidly changes sign at redshift $z_\dagger$, this limit was dubbed $\Lambda_{\rm s}$CDM and further investigated in Ref.~\cite{Akarsu:2021fol}. The $\Lambda_{\rm s}$CDM model can be constructed phenomenologically by simply replacing the usual cosmological constant ($\Lambda$) of the standard $\Lambda$CDM model with a cosmological constant ($\Lambda_{\rm s}$) that switches its sign from negative to positive, and thus attains its present-day value ($\Lambda_{\rm s0}>0$), when the Universe reaches a certain energy scale (redshift $z_\dagger$) during its expansion,
\begin{equation}
    \Lambda\quad\rightarrow\quad\Lambda_{\rm s}\equiv\Lambda_{\rm s0}\,{\rm sgn}[z_\dagger-z].
\end{equation}
It was shown in Ref.~\cite{Akarsu:2021fol} that, when the consistency of the $\Lambda_{\rm s}$CDM model with the CMB data is guaranteed, \textbf{(i)} $H_0$ and $M_B$ (SNIa absolute magnitude) values are inversely correlated with $z_\dagger$ and reach $H_0\approx74.5~{\rm km\, s^{-1}\, Mpc^{-1}}$ and $M_B\approx-19.2\,{\rm mag}$ for $z_\dagger=1.5$, in agreement with the measurements from SH0ES~\cite{Riess:2016jrr, Riess:2019cxk}, and \textbf{(ii)} $H(z)$ presents an excellent fit to the Ly-$\alpha$ data provided that $z_\dagger\lesssim 2.34$. The assessment of the model against {\it Planck} 2018 yields $H_0 = 70.22\pm 1.78{\rm \,km\, s^{-1}\, Mpc^{-1}}$ and against {\it Planck} 2018 + SDSS DR16 yields $H_0 = 68.82\pm 0.55{\rm \,km\, s^{-1}\, Mpc^{-1}}$ with $z_\dag = 2.44\pm 0.29$~\cite{Akarsu:2021fol}. It was found that the lower and upper limits of $z_\dagger$ are controlled by the Galaxy and Ly-$\alpha$ BAO data, correspondingly, and the larger $z_{\dagger}$ values imposed by the Galaxy BAO data prevent the model from achieving the largest estimations of $H_0$ from the direct local distance ladder measurements. It is intriguing that, as long as $z_\dagger\lesssim2.34$, the model remains in excellent agreement with the Ly-$\alpha$ data even for $z_\dagger\sim 1.1$, which barely satisfies the condition that we live in an ever-expanding Universe; a good agreement with the Ly-$\alpha$ data is an intrinsic feature of the $\Lambda_{\rm s}$CDM model as long as $z_\dagger\lesssim2.34$.

Similar to the situation with the Ly-$\alpha$ data, alleviating the $S_8$ discrepancy, prevailing within the $\Lambda$CDM model and its minimal extensions, usually results in exacerbating the $H_0$ tension, see Sec.~\ref{sec:WG-S8measurements} and Ref.~\cite{DiValentino:2020vvd}. In addition to this, the constraints on $S_8$ based on the Ly-$\alpha$ data are in agreement with the weak lensing surveys that probe similar late-time redshift scales as the Ly-$\alpha$ measurements~\cite{Palanque-Delabrouille:2019iyz}. Accordingly,
it is conceivable that the $\Lambda_{\rm s}$CDM model provides a remedy for the $S_8$ discrepancy while retaining the better fit to the local measurements of $H_0$, like in the case of the Ly-$\alpha$ discrepancy. Indeed, in the CMB-only analysis, it is found that $S_8=0.8071\pm0.0210$ for the $\Lambda_{\rm s}$CDM model, whereas $S_8=0.8332\pm 0.0163$ for the $\Lambda$CDM model. Although $\sigma_8$ is smaller for the $\Lambda$CDM model, its $\Omega_m$ value greater than $0.3$ results in an increased $S_8$ value compared to its $\sigma_8$ value. In contrast, the $\Lambda_{\rm s}$CDM model has an $\Omega_m$ value lower than $0.3$ which results in a decreased $S_8$ value compared to its $\sigma_8$ value. This results in the lower value of $S_8$ for $\Lambda_{\rm s}$CDM compared to $\Lambda$CDM. The $\Lambda_{\rm s}$CDM and $\Lambda$CDM models have similar $S_8$ values when the BAO data are also included in the analysis; this is due to the preference for larger $z_\dagger$ values by the Galaxy BAO data, since $\Lambda_{\rm s}$CDM approaches $\Lambda$CDM for larger $z_\dagger$ values and the $\Omega_m$ value of $\Lambda_{\rm s}$CDM becomes greater than 0.3. Thus, the $\Lambda_{\rm s}$CDM model partially reconciles the CMB data with the low redshift cosmological probes regarding $S_8$, and can potentially resolve the discrepancy in the absence of the Galaxy BAO data; however, for a robust conclusion, the constraints on $S_8$ from low redshift probes should also be explored within the $\Lambda_{\rm s}$CDM model.

Ultimately, it turns out via the $\Lambda_{\rm s}$CDM model that sign switch in the cosmological constant, viz., transition from an Anti-de Sitter background (provided by $\Lambda<0$) to a de Sitter one (provided by $\Lambda>0$), at $z\sim2$ \textbf{(i)} relaxes the SH0ES $H_0$ tension while being fully consistent with the TRGB measurement, \textbf{(ii)} relaxes the $M_B$ tension, \textbf{(iii)} removes the discrepancy with the Ly-$\alpha$ measurements, \textbf{(iv)} relaxes the $S_8$ tension, and \textbf{(v)} finds a better agreement with the BBN constraints on the physical baryon density~\cite{Akarsu:2021fol}. These results seem to encourage looking for a phase transition from AdS vacua to dS vacua in the late-Universe. 

It is reasonable to look for a potential origin of this phenomenon, viz. a very rapid single transition or its limiting case a single instantaneous (discontinuous) transition in the value of the cosmological constant, in a theory of fundamental physics by considering it as a first-order phase transition. The phase transition approach has been used to address the $H_0$ tension; see e.g.\ Refs.~\cite{Banihashemi:2018oxo,Banihashemi:2018has,Banihashemi:2020wtb}, which consider that the DE density resembles the magnetization of the Ising model and present a realization of this behavior within the Ginzburg-Landau framework. Additionally,  Ref.~\cite{Farhang:2020sij} considers a gravitational phase transition that is justified from the effective field theory point of view (see also Ref.~\cite{Khosravi:2021csn}). The model studied in Ref.~\cite{Banihashemi:2018oxo} partially corresponds to a one-parameter phenomenological extension of $\Lambda_{\rm s}$CDM; it considers an arbitrary shift in the value of the cosmological constant, but does not allow negative values of the cosmological constant in contrast to $\Lambda_{\rm s}$CDM. It addresses the $H_0$ tension with a shift in the value of the cosmological constant, however, at very low redshifts, viz. $z_{\rm t}=0.092^{+0.009}_{-0.062}$, signaling that it could suffer from the $M_B$ tension~\cite{Camarena:2021jlr,Efstathiou:2021ocp} (see also Refs.~\cite{Alestas:2020zol,Marra:2021fvf,Alestas:2021luu} ans Sec.~\ref{sec:LMT}). Given the promising advantages of having a negative cosmological constant for $z\gtrsim2$ in light of cosmological tensions, and that negative cosmological constant is a theoretical sweet spot---AdS space/vacuum is welcome due to the AdS/CFT correspondence~\cite{Maldacena:1997re} and is preferred by string theory and string theory motivated supergravities~\cite{Bousso:2000xa}---it would be most natural to associate this phenomena with a possible phase transition from AdS to dS that is derived in string theory, string theory motivated supergravities, and theories that find motivation from them.

Besides, there is a rapidly growing literature on the cosmological models that suggest dark energy (as an actual source or an effective source from a modified theory of gravity) transitions in the late Universe from a negative (or vanishing) cosmological constant-like behavior to a positive cosmological constant-like behavior, accompanied by a pole in its EoS parameter. These models can relax the $H_0$ and $S_8$ tensions and/or the Ly-$\alpha$ discrepancy, and the cosmological data either prefer or are fully consistent with the presence of a negative cosmological constant at high redshifts and the presence of a negative cosmological constant accompanying quintessence/phantom dark energy, see e.g.\ Refs.~\cite{Sahni:2014ooa,Poulin:2018zxs,Wang:2018fng,Mortsell:2018mfj,Dutta:2018vmq,Akarsu:2019ygx,Bonilla:2020wbn,Calderon:2020hoc,LinaresCedeno:2021aqk,Bag:2021cqm,Escamilla:2021uoj,Sen:2021wld}.

\subsubsection{Holographic Dark Energy} 

DE motivated from the Holographic principle~\cite{Li:2004rb,Huang:2004mx,Zhang:2014ija} is widely known as Holographic DE (HDE). The model has been intensively investigated for its ability to explain the late-time cosmic acceleration (see the review~\cite{Wang:2016och}). In the simplest (and original) HDE model, the DE density is given by
\begin{equation}
    \label{HDE-rho}
    \rho_{\rm DE} = 3c^2 M_{\rm Pl}^2 L^{-2},\qquad L= a(t) \int_{t}^\infty \frac{{\rm d}t'}{a(t')}\,,
\end{equation}
where $L$ is the radius of the future event horizon of the Universe and $c$ is the free parameter of the model. The holographic nature comes in the fact that $L^{-2}$ appears in the dark energy density. In the simplest case therefore, such models have three parameters. There are many models based on variants of this theme which typically have more parameters (see Ref.~\cite{Wang:2016och} for a review). At late times when the DE is expected to dominate, the above HDE behaves like a cosmic fluid of EoS, 
\begin{equation}
    w_{\rm HDE} =-\frac{1}{3}-\frac{2}{3c}.
\end{equation}
So, for $c\gtrless 1$, $w_{\rm HDE} \gtrless -1$. It has been argued that HDE can also alleviate the Hubble constant tension, see e.g.\ Refs.~\cite{Guo:2018ans,Dai:2020rfo} (see Ref.~\cite{vanPutten:2017bqf} for an earlier realisation), but as discussed in Ref.~\cite{Colgain:2021beg} this comes at the expense of violating the null energy condition, or having a turning point in $H(z)$. There are many extensions of the previous holographic model following either the Tsallis non-extensive entropy~\cite{Tsallis:2012js,Saridakis:2018unr}, the quantum-gravitational modified Barrow entropy~\cite{Barrow:2020tzx,Saridakis:2020zol}, or the Kaniadakis relativistic entropy~\cite{Kaniadakis:2002zz,Drepanou:2021jiv,Hernandez-Almada:2021rjs}. It has been discussed that the Hubble constant tension can be alleviated with Tsallis holographic dark energy~\cite{daSilva:2020bdc} or with Kaniadakis holographic dark energy~\cite{Hernandez-Almada:2021aiw}.

\subsubsection{Interacting Dark Energy} 

A cosmological scenario in which dark matter (DM) and DE share interactions other than gravitational leads to various interesting features~\cite{Amendola:1999er,Chimento:2003iea,Cai:2004dk,Barrow:2006hia,Chen:2008ft,Pettorino:2013oxa,Pan:2012ki,Kumar:2016zpg,Nunes:2016dlj,Pan:2016ngu,Sharov:2017iue,DiValentino:2017iww,Kumar:2017dnp,vandeBruck:2017idm,Yang:2017zjs,Yang:2017ccc,Pan:2017ent,Yang:2018pej,Yang:2018euj,Barros:2018efl,Yang:2018xlt,Yang:2018uae,Yang:2018qec,Yang:2019uzo,Martinelli:2019dau,Pan:2020zza,DiValentino:2019ffd,Teixeira:2019hil,DiValentino:2019jae,Teixeira:2019tfi,Pan:2019jqh,Paliathanasis:2019hbi,Benevento:2020fev,Gomez-Valent:2020mqn,vandeBruck:2020fjo,Lucca:2020zjb,Yang:2020uga,Yang:2019uog,Pan:2020mst,Yang:2018ubt, Kumar:2021eev, Lucca:2021eqy}. In Ref.~\cite{DiValentino:2019ffd,DiValentino:2019jae,Pan:2019gop,Yang:2019uog,Pan:2020bur,Kumar:2021eev,Nunes:2021zzi,Gariazzo:2021qtg,Guo:2021rrz}, the authors have found that the IDE model can solve the tension with SH0ES within one standard deviation, leading to a preference for a non-zero DE-DM coupling at more than $5$ standard deviations, while fixing the DE EoS to a cosmological constant. However, this category can be further extended into two classes~\cite{DiValentino:2019jae}: {\it (i)}~models with $w_{\rm DE} < - 1$ in which energy flows from DE to DM, {\it (ii)}~models with $w_{\rm DE}> -1$ in which energy flows from DM to DE. Related models can be realized in string theory~\cite{Agrawal:2019dlm,Anchordoqui:2019amx,Anchordoqui:2020sqo}.

A phenomenological description allows for the interaction between DE and another component, i.e.\ DM, which leads to an {\it effective} phantom-like EoS for DE. In more detail, we assume that the densities of DM and DE do not evolve independently but through a "dark coupling" as~\cite{Gavela:2009cy}
\begin{eqnarray}
	\nabla_\mu\,T^{\mu\nu}_{\rm DE} &=& -Q\,u^\nu_{\rm DM}/a\,,\\ 
	\nabla_\mu\,T^{\mu\nu}_{\rm DM} &=& Q\,u^\nu_{\rm DM}/a\,,
\end{eqnarray}
where $T^{\mu\nu}_{\rm DE}$ and $T^{\mu\nu}_{\rm DM}$ are the energy momentum tensors for DE and DM, respectively, while $u^\nu_{\rm DM}$ is the DM four-velocity in the DM frame. The quantity $Q$ describes the interaction rate between the two dark sectors, and we specialise it here as $Q = \xi\,H\,\rho_{\rm DE}$, where $\xi$ is a (negative) dimensionless parameter that regulates the flow of DM into DE and $H = \dot a/a$. The background evolves as
\begin{eqnarray}
    \dot \rho_{\rm DE} + 3H(1+w_{DE})\rho_{\rm DE} &=& -\xi H\rho_{\rm DE}\,,\label{eq:de_evolves}\\
    \dot \rho_{\rm DM} + 3H\rho_{\rm DM} &=& \xi H\rho_{\rm DE}\,,\\
    H^2 &=& \frac{8\pi G_N}{3}\,\rho\,,
\end{eqnarray}
where $w_{DE}$ parametrizes the EoS for the DE component and $\rho$ is the total energy content of the Universe at a given time. The solution to Eq.~\eqref{eq:de_evolves} reads $\rho_{\rm DE} \propto a^{-3(1 + w_{\rm eff})}$, where the effective EoS $w_{\rm eff} = w_{DE} + \xi/3$ leads to a phantom-like EoS for $\xi < -3 (1 + w_{DE})$. The DM-DE coupling term $Q$ impacts on the evolution of perturbations~\cite{Valiviita:2008iv, Gavela:2009cy, Gavela:2010tm} and affects the observables in the CMB. In this view, the $\Lambda$CDM model augmented with the DM-DE coupling, or $\xi-\Lambda$CDM model, can be constrained from the observations of the CMB modes. The $\xi-\Lambda$CDM model is characterized by the usual six cosmological parameters of $\Lambda$CDM plus $\xi$, namely $(\Omega_bh^2, \Omega_{\rm DM}h^2, \theta_s, A_s, n_s, \tau, \xi)$.

The observations used for constraining the model consist of various independent measurements:
\begin{itemize}
\item Measurements of the CMB temperature and polarization anisotropies, as well as their cross-correlations, from the {\it Planck} 2018 legacy data release ({\it Planck} TT,TE,EE+lowE)~\cite{Planck:2018vyg, Planck:2018lbu};
\item Measurements of the CMB lensing power spectrum reconstructed from the CMB temperature four-point function~\cite{Planck:2018lbu};
\item Measurements of the Baryon Acoustic Oscillation (BAO) from the 6dFGS~\cite{Beutler:2011hx}, SDSS-MGS~\cite{Ross:2014qpa}, and BOSS DR12~\cite{eBOSS:2020yzd,Alam:2016hwk} surveys;
\item Measurements of the Type Ia Supernovae (SNIa) distance moduli from the Pantheon sample~\cite{Pan-STARRS1:2017jku}.
\end{itemize}

A Monte Carlo analysis of the $\xi-\Lambda$CDM model that accounts for the {\it Planck} data alone leads to the value of the Hubble constant $H_0 = 72.8^{+3.0}_{-1.5}{\rm \,km\,s^{-1}\,Mpc^{-1}}$ at 68\% CL, which is a significant shift in the central value of the Hubble constant with respect to what obtained within the $\Lambda$CDM model alone~\cite{DiValentino:2019ffd}. The value obtained is within $1\sigma$ from the value $H_0 = (73.04 \pm 1.04){\rm \,km\,s^{-1}\,Mpc^{-1}}$ obtained from the observation of a sample of long-period Cepheids in the LMC by the Hubble Space Telescope~\cite{Riess:2021jrx}, and can thus be referred to as a possible solution to the $H_0$ tension. Incidentally, the method predicts a non-zero DM-DE interaction, with the parameter $\xi = -0.54^{+0.12}_{-0.28}$ differing from zero by more than $4\sigma$. When a joint analysis of the {\it Planck}+BAO or {\it Planck}+Pantheon datasets is considered, these results weaken and the tension with the HST results lies within the $3\sigma$ level. This mechanism is then not powerful enough to lead to a strong resolution to the $H_0$ tension.

Stepping away from phenomenological models, explicit examples of a DM-DE coupling can be built within models of supergravity. Of particular interest are models that admit a UV completion to a consistent theory of quantum gravity~\cite{Addazi:2021xuf}, based on satisfying a few key requirements that avoids such theories to lie in the swampland of inconsistent theories instead of in the string theory landscape of vacua~\cite{Vafa:2005ui}. Cosmological observations can come to hand in constraining swampland conjectures~\cite{Akrami:2018ylq,OColgain:2018czj, Colgain:2019joh,Kinney:2018nny, Danielsson:2018qpa,Anchordoqui:2021eox}. Swampland conjectures generally make it difficult for fundamental theories based on compactification from extra dimensions to accommodate a period of accelerated cosmological expansion~\cite{Montefalcone:2020vlu}. Such a restriction can be avoided in models whose internal space is not conformally Ricci flat, as it is the case for the Salam-Sezgin model~\cite{Anchordoqui:2020sqo}. Within this supergravity model, dark matter could acquire a mass term which depends on the value of the quintessence field, thus realizing an effective DM-DE coupling which addresses the $H_0$ tension~\cite{Anchordoqui:2019amx}, in the so-called fading dark matter model~\cite{Agrawal:2019dlm}.

Apart from providing a valid solution to the $H_0$ tension, this class of cosmological models can also play an effective role in alleviating the $S_8$ tension.

For instance, one viable possibility to largely reduce the significance of the tension $S_8$ is to allow for DE-DM interactions in a scenario where the DE energy density can be transferred to the DM fluid via a coupling function proportional to the DE energy density itself (unlike the model able to address the $H_0$ tension, where the energy transfer is from the DM to the DE)~\cite{DiValentino:2019jae, Lucca:2021dxo} (see also~\cite{DiValentino:2019ffd,Kumar:2019wfs,Kumar:2021eev}. Indeed, in this scenario, because of the additional energy flow from the DE to the DM over the cosmic history, a lower DM energy density with respect to the $\Lambda$CDM model is allowed at recombination without a sizable effect on the shape of the CMB anisotropy power spectra. As a consequence, the redshift of matter-radiation equality decreases and since the peak of the matter power spectrum and its amplitude depend on this redshift, it follows that the whole matter power spectrum shits horizontally to larger scales and vertically to lower values (see Ref.~\cite{Lucca:2021dxo} and Fig.~1 therein for a more detailed discussion about this mechanism). As a result, after the inclusion of data from {\it Planck} 2018, BAO and Pantheon the model reduces the significance of the tension below the 2.5$\sigma$ level even for very discordant estimates. The addition of late-time data from weak-lensing surveys such as DES and KV450 further improves the concordance between early and late-time probes, without exacerbating the $H_0$ tension nor the overall fit to the data (feature also present when excluding late-time measurements).
	
Other possible DE-DM interaction scenarios involve for instance elastic interactions between the two fluids that only produce a momentum exchange~\cite{Pourtsidou:2016ico, Amendola:2020ldb,Jimenez:2021ybe}. In this scenarios, as is common in models where the DM has such interaction channels with other particles, the additional drag effect prevents the DM component from clustering, which results in a suppression of the matter power spectrum at relatively small scales. By comparing this model to data from {\it Planck} 2018, BAO and Pantheon one obtains that the mean value of the $\sigma_8$ parameter does not change with respect to the $\Lambda$CDM equivalent but the error bars are enlarged to the point where the tension largely disappears. The inclusion of late-time measurements such as from the {\it Planck} SZ cluster counts can then pick up the values of the distributions that bring the datasets in agreement, resulting in a clear detection of the interaction parameter, see Fig.~6 of Ref.~\cite{Jimenez:2021ybe}.

Finally, another possibility is an IDE model with a sign-changing interaction, as in Ref.~\cite{Pan:2019jqh}.

Note that in this Section the DM component enters only in relation to its interaction with DE. Models in which the interaction of DM with either the visible or other dark sectors affects the cosmological evolution are discussed below in Secs.~\ref{sec:cannibalDM} (cannibal DM),~\ref{sec:CDMdarkdecay} and~\ref{sec:dDM} (decaying DM models),~\ref{sec:dynDM} (dynamical DM), and~\ref{sec:IDM} (interacting DM).

\subsubsection{Quintessence Models and their Various Extensions} 
\label{sec:quintessence}

The present accelerating expansion rate of the Universe can be brought forth by the presence of a light scalar field known as quintessence. The simplest of such models consists of minimally coupled scalar fields with a flat enough potential which, contrary to a cosmological constant, leads to a dynamical EoS parameter, $w_{\rm DE} \neq -1$~\cite{Ratra:1987rm, Caldwell:1997ii}, see also Refs.~\cite{Zlatev:1998tr, Peebles:2002gy, Copeland:2006wr, Tsujikawa:2013fta} for early reviews. The model building of quintessence fields as DE models with a field theory backend has recently featured extensively through string theory swampland conjectures~\cite{Agrawal:2018own, Palti:2019pca}. Various extensions of the simple quintessence models such as coupled quintessence, in which DE sector couples to (dark) matter sector~\cite{Amendola:1999er, Agrawal:2019dlm, Gomez-Valent:2020mqn} or non-minimally coupled quintessence, in which the scalar field is coupled to gravity sector through non-minimal~\cite{Capozziello:1999xt} or non-minimal derivative~\cite{Saridakis:2010mf} couplings, and many other extensions~\cite{Frieman:1995pm, Choi:1999xn, Nomura:2000yk, Visinelli:2018utg, Choi:2021aze,Akrami:2017cir,Akrami:2020zxw,Akrami:2020zfz,Eskilt:2022zky} have been analysed. As quintessence models generally lead to $w_{\rm DE} > -1$ and exacerbate the Hubble constant tension~\cite{Vagnozzi:2019ezj, Banerjee:2020xcn}, these are generally not viable DE models. More generally, DE models with $w_{DE} > -1$ can be shown to be at odds with $H_0$ and $S_8$ tension~\cite{Heisenberg:2022lob, Heisenberg:2022gqk, Lee:2022cyh}.

\subsubsection{Running Vacuum Models} 
\label{sec:RRVM}

The persisting tensions between the $\Lambda$CDM and some cosmological data (cf.\ Secs.~\ref{sec:WG-H0measurements} and~\ref{sec:WG-S8measurements}) suggest that the rigid cosmological constant, $\Lambda$, might be performing insufficiently at the observational level. It is tempting to check if we could alleviate some of these problems on assuming  that the dark (or just vacuum) energy density might be (slowly) dynamical. A related option is to assume that the gravitational coupling $G_N$ can be mildly time-evolving. These possibilities may occur at a time or individually, but each has different implications on the local conservation laws. Rather than introducing some \textit{ad hoc} phenomenological dependence of $\Lambda$ and $G_N$ as a function, say, of the cosmic time, we would like to have some more fundamental motivation for such a dependence. The Running Vacuum Model (RVM), see Refs.~\cite{Sola:2013gha,Sola:2015rra} and references therein, may provide such a fundamental dynamics, emanating either from quantum field theory (QFT) or string theory. The RVM can implement any of the mentioned options and is characterized by a vacuum energy density, $\rho_{\rm vac}$, which at the background level carries a mild dependence on the expansion rate and its time derivative. The relevant structure of the RVM for the current Universe reads as follows:
\begin{equation}
    \label{eq:RVMvacuumdadensity}
    \rho_{\rm vac} (H,\dot{H}) = \frac{3}{8\pi G_N}(c_0+\nu H^2+\tilde{\nu}\dot{H})+\mathcal{O}(H^4)\,,
\end{equation}
with $\nu,\tilde{\nu}\lesssim\mathcal{O}(10^{-3})$, $G_N$ the local value of Newton's constant, and the additive constant $c_0$ is fixed by the boundary condition $\rho_{\rm vac}(H_0) = \rho^{0}_{\rm vac}$ (current value). The RVM at low energies has been tested repeatedly (and successfully) in the past, see e.g.\ Refs.~\cite{Sola:2015wwa,Sola:2016jky,Sola:2017znb,SolaPeracaula:2016qlq,Sola:2017jbl, Sola:2017jbl, Sola:2018sjf,Gomez-Valent:2017idt,Gomez-Valent:2018nib,Banerjee:2019kgu}. The state-of-the-art phenomenological performance of the RVM has been presented in a recent study~\cite{SolaPeracaula:2021gxi}. The $\mathcal{O}(H^4)$ terms in Eq.~\eqref{eq:RVMvacuumdadensity} can only be important for the very early Universe as they can trigger inflation, but we shall not address the implications for the early Universe here (see Refs.~\cite{Sola:2013gha,Lima:2013dmf,Sola:2015rra,Sola:2015csa,SolaPeracaula:2019kfm} and references therein).

Theoretically, the structure of the RVM can be motivated from different perspectives, to wit: from the renormalization group in curved space-time~\cite{Shapiro:2000dz,Sola:2007sv,Shapiro:2009dh}, from scalar-tensor theories such as the Brans-Dicke theory of gravity with a non-vanishing $\Lambda$~\cite{SolaPeracaula:2018dsw, deCruzPerez:2018cjx, Sola:2019jek, Sola:2020lba, Singh:2021jrp}, see Sec.~\ref{sec:BDLCDM}, and also from QFT in curved space-time using adiabatic regularization and renormalization techniques~\cite{Moreno-Pulido:2020anb,Moreno-Pulido:2022phq}. In addition, beyond the QFT approach, there recently exists an intriguing "stringy" version of the RVM, which is based on the effective action of the bosonic part of the gravitational supermultiplet in string theory, see Ref.~\cite{Mavromatos:2020kzj} and references therein.  At low energies, the stringy version makes the important prediction that DM should be axion-like but otherwise it behaves as the standard RVM. This means that the DE associated to the RVM is in both cases dynamical through a $\sim \nu H^2$ component (with $|\nu|\ll 1$). In contrast, at high energies the stringy version nicely connects inflation  with the physics of the primordial GWs, see Ref.~\cite{Mavromatos:2020kzj} for a comprehensive exposition and Ref.~\cite{Mavromatos:2021urx} for new developments in this direction.

The phenomenological performance of the RVM has last been discussed using the most up-to-date dataset available in Ref.~\cite{SolaPeracaula:2021gxi}, under the assumption $\tilde{\nu}=\nu/2$. Neglecting the higher order terms, which are only relevant during inflation, Eq.~\eqref{eq:RVMvacuumdadensity} takes the form
\begin{equation}
    \rho_{\rm vac}(R) =\frac{3}{8\pi{G_N}}\left(c_0 + \frac{\nu}{12} {R}\right)\,,    
\end{equation}
where $R$ is the Ricci scalar. For this reason, we may call this form of the vacuum energy density (VED) the Ricci RVM, or just "RRVM".

\begin{figure}[t!]
\centering
\begin{subfigure}
  \centering
  \includegraphics[width=0.5  \linewidth]{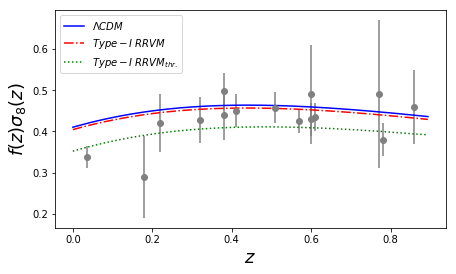}
  \label{fig:fs8_TI}
\end{subfigure}
\begin{subfigure}
    \centering
    \includegraphics[width=0.308\linewidth]{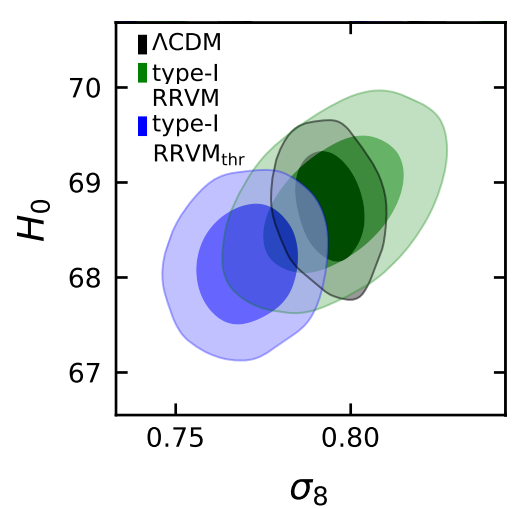}
    \label{fig:RRVMI}
\end{subfigure}
\caption{{\it Left panel}\/: Theoretical curves of $f(z)\sigma_8(z)$ for the $\Lambda$CDM and type-I RRVM with and without threshold redshift $z_{*}$, together with the data points employed in the analysis of Ref.~\cite{SolaPeracaula:2021gxi}; {\it Right plot:} $1\sigma$ and $2\sigma$ CL regions in the ($\sigma_8$--$H_0$)-plane. $H_0$ is expressed in units of ${\rm km\,s^{-1}\,Mpc^{-1}}$. See also Fig.~\ref{RRVM2_BD} in Sec.~\ref{sec:BDLCDM}, and the numerical tables in Ref.~\cite{SolaPeracaula:2021gxi}.}
\label{RRVMI}
\end{figure}

This implementation has the double advantage of using one single parameter and provides a safe path to the early epochs of the cosmological evolution. In fact, when we approach the radiation dominated era we have $R/H^2\ll 1$ and, hence, the impact of the running is highly suppressed in that epoch. We may consider two types of RRVM scenarios~\cite{SolaPeracaula:2021gxi}: type-I, in which the vacuum interacts with matter; and  type-II,  where matter is conserved at the expense of an exchange between the vacuum and a mildly evolving gravitational coupling $G (H)$. For type-I models we assume that the vacuum ("vac") exchanges energy with CDM only: $\dot{\rho}_{\rm CDM} + 3H\rho_{\rm CDM} = -\dot{\rho}_{\rm vac}$. Solving for the CDM and vacuum energy densities in terms of the scale factor $a$ yields
\begin{equation}
    \label{eq:LocalConsLaw}
    \rho_m(a) = \rho^0_m{a^{-3\xi}}\,,\quad \quad \quad \rho_{\rm vac}(a) = \rho^{0}_{\rm vac} + \left(\frac{1}{\xi} -1\right)\rho^0_m\left(a^{-3\xi} -1\right)\,,
\end{equation}
where $\xi \equiv \frac{1 -\nu}{1 - \frac{3}{4}\nu}$ and $\rho_m = \rho_{\rm CDM} + \rho_b$ (i.e.\ includes the densities of CDM and baryons). As expected, for $\xi=1$ ($\nu=0$) we recover the $\Lambda$CDM. For type-I models we  admit also the  possibility that the dynamics of vacuum is activated only very recently, namely at a threshold redshift $z_{*}\simeq 1$, so that $\rho_{\rm vac}(z) = {\rm const.}$ for $z>z_{*}$. This possibility has been deemed plausible in the literature, see e.g.\ Ref.~\cite{Martinelli:2019dau}. We compare this option with the situation when there is no such threshold, see Fig.~\ref{RRVMI}. As for type-II models, matter is conserved, but the vacuum can still evolve as long as the effective gravitational coupling also evolves with the expansion $G_{\rm eff}=G_{\rm eff}(H)$, starting from an initial value (which enters our fit). The approximate behavior of the VED for this model in the late-time Universe is:
\begin{equation}
    \label{eq:VDEm}
    \rho_{\rm vac}(a)=\frac{3c_0}{8\pi G_N}(1+4\nu)+\nu\rho_m^{0}a^{-3}+\mathcal{O}(\nu^2)\,.
\end{equation}
The effective gravitational coupling evolves very mildly as $G_{\rm eff}(a)\propto (1+\epsilon\ln\,a)$ in the current epoch (with $0<\epsilon\ll 1$ of order $\nu$). The vacuum dynamics has also an important impact on the large scale structure (LSS) through the $f\sigma_8$ data. For a list of the main equations at the linear perturbation level, see Ref.~\cite{SolaPeracaula:2021gxi}.

To compare the RRVM's with the $\Lambda$CDM, we have defined a joint likelihood function ${\cal L}$, where the total $\chi^2$ to be minimized in our case is given by $\chi^2_{\rm tot}=\chi^2_{\rm SNIa}+\chi^2_{\rm BAO}+\chi^2_{ H}+\chi^2_{\rm f\sigma_8}+\chi^2_{\rm CMB}$. In particular, the $\chi^2_{H}$ part may contain or not the local $H_0$ value measured by the SH0ES team, but here we only present the results obtained by including it. Fig.~\ref{RRVMI} shows that the effect of the redshift threshold $z_{*}$ in type-I RRVM can be very important and suggests that a mild dynamics of the vacuum is welcome, especially if it gets activated at around the  epoch when the vacuum dominance appears, namely at $z\simeq 1$.  Unfortunately, type-I models without such a threshold  do not help to solve the tensions since the values of $H_0$ and $\sigma_8$ remain very close to the CMB values~\cite{SolaPeracaula:2021gxi}. Nevertheless, this might not be a problem if the current expansion is slower than measured by SH0ES, as suggested e.g.\ in Refs.~\cite{Freedman:2021ahq,Mortsell:2021nzg,Gomez-Valent:2018hwc,Haridasu:2018gqm}. In stark contrast, Fig.~\ref{RRVM2_BD} shows that type-II models can alleviate the two tensions at a time. This variant of the RRVM provides values of $H_0$ markedly higher as compared to type-I models (specifically $H_0=70.93^{+0.93}_{-0.87}{\rm \,km\,s^{-1}\,Mpc^{-1}}$~\cite{SolaPeracaula:2021gxi}) along with $\sigma_8$ and $S_8$ values in the needed moderate range ($\sigma_8=0.794^{+0.013}_{-0.012}$ and $S_8=0.761^{+0.018}_{-0.017}$)~\cite{SolaPeracaula:2021gxi}. The values of $S_8$ in all RRVM's are perfectly compatible with recent weak lensing and galaxy clustering measurements (cf. Ref.~\cite{SolaPeracaula:2021gxi} and Sec.~\ref{sec:WLmeasurements}). The net outcome of this analysis is that of the two RRVM's considered, only the type-II ones are capable of alleviating the two tensions ($H_0$ and $\sigma_8)$, whereas the type-I models with threshold redshift can (fully) solve the $\sigma_8$ tension but have no bearing on the $H_0$ one. It is remarkable that the purely cosmographic analyses (which are more model-independent) also favor the RVM option as compared to other models of the dark energy~\cite{Rezaei:2021qwd}.

\subsubsection{Time-Varying Gravitational Constant}
\label{sec:GLambdaz}

Seminal work of Weinberg suggests that an effective quantum theory of gravity would be asymptotically safe, with the renormalization group equations possessing a fixed point~\cite{Weinberg1978, Weinberg:2009wa}. A consequence of this framework is an extension of the $\Lambda$CDM model in which Newton's gravitational constant $G_N$ could depend on time~\cite{Xue:2014kna, Begue:2017lcw}. In more detail, in this model the gravitational coupling is a function that varies with redshift $z$ as $G_N(z) = G_{N,0}(1+z)^{-\delta_G}$, such that $G_N(z=0)\equiv G_{N,0}$ is the value today, the suffix "0" refers to the present time values, and $\delta_G$ parametrizes the change in the gravitational constant. As a consequence, the cosmological constant would also be redshift-dependent and follows a scaling law, $\Lambda(z)= \Lambda_0(1+z)^{\delta_\Lambda}$. The matter "$m$" and radiation "$r$" contents would satisfy the scaling relations
\begin{eqnarray}
    (G_N/G_{N,0})\,\rho_m &=& \rho_{m,0}\,(1+z)^{3(1+w_m)-\delta_G}\,,\\
    (G_N/G_{N,0})\,\rho_r &=& \rho_{r,0}\,(1+z)^{3(1+w_r)-\delta_G}\,,\\
    (G_N/G_{N,0})\,\rho_\Lambda &=& \rho_{\Lambda,0}\,(1+z)^{\delta_\Lambda}\,.
\end{eqnarray}
The requirement for a flat Universe leads to the relation
\begin{equation}
    \label{eq:EFT_closure}
    \delta_\Lambda = \delta_G\,\frac{\rho_{r,0}+\rho_{m,0}}{\rho_{\Lambda,0}}\,.
\end{equation}
The assessment of the model against cosmological data is performed in Ref.~\cite{Gao:2021xnk}. In particular, two realizations of this model are considered, namely the $\tilde\Lambda$CDM model in which Eq.~\eqref{eq:EFT_closure} is enforced, and the e$\tilde\Lambda$CDM model in which $\delta_G$ and $\delta_\Lambda$ are kept as independent parameters. Both models cannot ease the tension in the Hubble constant when using an analysis that uses the combination of {\it Planck} 18 TT,TE,EE+lowE+lensing data~\cite{Planck:2018vyg} plus BAO plus data from SN1e Pantheon compilation~\cite{Pan-STARRS1:2017jku}. The tension is reduced when the measurement of $H_0$ by the SH0ES team is used as a Gaussian prior. Recently, it has been speculated that the Hubble tension could be resolved within the stochastic quantization formalism of gravity~\cite{Lulli:2021bme}, although more work is required along this direction.

Other approaches considering varying gravitational constant are analysed in~\cite{Sakr:2021nja}, that considering $G_N$ as a free parameter, 
and in~\cite{Maeda:2022ozc} in the context of a cuscuta-galileon gravity theory. Finally, Ref~\cite{Benevento:2022cql} implements a sharp transition in the value of the effective gravitational constant prior recombination, effectively lowering the sound horizon at CMB last scattering and addressing both the $H_0$ and the $S_8$ tensions. Assessing the model against {\it Planck} 2018 TT,TE,EE + LowE and BAO data, the value inferred for the Hubble constant is $H_0 = 69.22_{-0.86}^{+0.67}{\rm\,km\,s^{-1}\,Mpc^{-1}}$~\cite{Benevento:2022cql}.

\subsubsection{Vacuum Metamorphosis} 

The Vacuum Metamorphosis (VM) model is motivated by the fact that quantum gravitational effects~\cite{Parker:2003as, Parker:2000pr, Caldwell:2005xb} can significantly increase the Hubble constant value leading to a solution for the Hubble tension~\cite{DiValentino:2017rcr,DiValentino:2020kha,DiValentino:2021zxy,DiValentino:2021rjj}. This model is a first principles theory based on a late-time gravitational phase transition that takes place when the Ricci scalar curvature $R$ is of the order of the squared mass of the scalar field, $m^2$, and is related to the matter density today, $\Omega_m$. For this reason, the VM model has the same number of degrees of freedom as the $\Lambda$CDM scenario.

The expansion rate before and after the phase transition result in~\cite{DiValentino:2020kha}:
\begin{equation}
    \frac{H^2(z)}{H_0^2} =
    \begin{cases}
        \Omega_m (1+z)^3 \!+\! \Omega_r (1+z)^4  \!+\! \Omega_k (1+z)^2\!-\! M\left\{1-\left[3\left(\frac{4}{3\Omega_m}\right)^4 M(1-M-\Omega_k-\Omega_r)^3\right]^{-1}\right\}, &\, \hbox{for $z > z_t^{\rm ph}$}\,,\\
        (1-M-\Omega_k)(1+z)^4+\Omega_k(1+z)^2+M\, &\, \hbox{for $z\leq z_t^{\rm ph}$}\,,
    \end{cases}
\end{equation}
with $M = m^2/(12H_0^2)$, $\Omega_k=-k/H_0^2$, and the redshift of the phase transition
\begin{equation}
    z_t^{\rm ph} = -1+\frac{3\Omega_m}{4(1-M-\Omega_k-\Omega_r)} \,.
\end{equation}
In other words, after the phase transition the Universe effectively has a radiation component that rapidly redshifts away leaving the Universe in a de Sitter phase, and therefore it is not nested with the $\Lambda$CDM scenario.

In the original Vacuum Metamorphosis scenario described above, there is not a cosmological constant at high redshifts and therefore
\begin{equation}
    \Omega_m=\frac{4}{3}\left[3M(1-M-\Omega_k-\Omega_r)^3\right]^{1/4}\,.
\end{equation}
It is also possible to extend this VM model assuming a possible cosmological constant at high redshifts, as the vacuum expectation value of the massive scalar field. In this case, we have to impose the additional conditions $z_t^{\rm ph}\ge0$, i.e.\ the transition happening in the past, and $\Omega_{\rm DE}(z>z_t^{\rm ph})\ge0$. With both the VM model and its extension, the Hubble constant $H_0 \sim 73-74{\rm\,km\,s^{-1}\,Mpc^{-1}}$ is completely in agreement with the SH0ES value, without assuming a Gaussian prior on it. Unfortunately this model fails to fit in a good way the low redshift data, such as BAO and Pantheon: $H_0$ remains in the SH0ES range but with a much worse $\chi^2$ than $\Lambda$CDM.

\subsection{Modified Gravity Models}

 Modified gravity (MG) models~\cite{Clifton:2011jh,Planck:2015bue} in which some feature of gravity (such as the gravitational coupling) changes with redshift, can lead to an $H_0$ estimate from CMB larger
than that obtained from late-time probes
~\cite{Umilta:2015cta,Ballardini:2016cvy,Raveri:2019mxg,Yan:2019gbw,Frusciante:2019puu,Frusciante:2020gkx,Sola:2019jek,Sola:2020lba,Ballesteros:2020sik,Braglia:2020iik,Ballardini:2020iws,Rossi:2019lgt,Joudaki:2020shz,Braglia:2020auw,Akarsu:2019pvi,Benetti:2018zhv}. To solve (or at least relax) the $H_0$ tension in the context of MG models, instead of modifying the matter content, the gravitational sector is modified in a manner that  current cosmic dynamics is derived. In particular, it is easy to argue that Newton's constant needs to decrease as the Universe evolves in order to stand a chance of resolving cosmological tensions~\cite{Heisenberg:2022lob, Heisenberg:2022gqk,Lee:2022cyh}. There has been several proposals in GR that follows this line of thought~\cite{Clifton:2011jh,Ishak:2018his}. One interesting approach in this direction has been the $f(R)$ theories~\cite{Sotiriou:2008rp, Faraoni:2008mf,DeFelice:2010aj}, in which an arbitrary function of the Ricci scalar replaces the standard Einstein-Hilbert Lagrangian density. In the metric version, the Ricci scalar and the metric tensor evolve in time in an independent fashion. Thus, there is a scalar degree of freedom (absent in GR), which helps to explain the late cosmic acceleration and structure formation without the need of the dark matter sector.\footnote{In the Palatini version, the same result is achieved due to the new terms that arise from the affine connection when written in of the Christoffel symbols and of the quantities that describe the matter~\cite{Sotiriou:2008rp}.} Along with various cosmological models in Einstein gravity, the modified gravity models can  also reconcile the $S_8$ tension~\cite{Planck:2015bue,DiValentino:2015bja,Joudaki:2016kym,Joudaki:2017zdt,Sola:2019jek,Sola:2020lba,Wen:2021bsc}.

\subsubsection{Effective Field Theory Approach to Dark Energy and Modified Gravity}

Above we have discussed several alternatives to a purely cosmological constant in late-time cosmology, e.g., modified gravity. Much of these alternatives are often very model specific and must confront precise observations. Given the current wealth of data another theoretical approach – motivated by techniques utilized in particle and condensed matter physics – is to utilize methods of effective field theory as applied to cosmology~\cite{Park:2010cw,Gubitosi:2012hu,Bloomfield:2012ff}. This can also lead to an effective method to confront a large, and much general set of models with observations~\cite{Wen:2021bsc,Frusciante:2013zop,Raveri:2014cka}.

This approach is to establish how gravity may, or may not, be modified to explain the current acceleration of the expansion of the Universe and at what scales this is relevant (e.g. it is well known that gravity cannot be modified at solar system scales). In the effective field theory approach, one can establish groups of models to collect model dependent approaches and establish these into groups of models independent (universality classes) based on their symmetries (or lack thereof). This allows for a method to connect a large model specific set of predictions to be confronted with observations. For example, Horndeski models (mentioned above) are of broad interest in the community, but in this formalism, they form a universality class that can be viable or ruled out by current and near-term observations.

\subsubsection{$f(T)$ Gravity}

Recently, teleparallel gravity (TG) and its extensions have started to take advantage of their mathematical description to solve the $H_0$ tension  and describe the cosmic late-time dynamics with current data. In these kind of theories, gravity is still described in geometrical terms, but using torsion instead of curvature, namely using the Weitzenb{\"o}ck connection \cite{Weitzenbock1923} instead of the Levi-Civita one. Hence, the actions of GR and TG differ only by a boundary term,  and thus they lead to the same equations, that is why the corresponding theory is called Teleparallel equivalent of General Relativity (TEGR). There are several ways to modify the TEGR proposal: (1) through the so-called torsion scalar $T$~\cite{Nunes:2018xbm}. In this form, we can generalise the Lagrangian to arbitrary functions of the torsion scalar to produce $f(T)$ gravity~\cite{Ferraro:2006jd,Linder:2010py,Chen:2010va,Benisty:2021sul,Bahamonde:2022ohm}, analogous to $f(R)$ theories (for reviews see~\cite{Cai:2015emx,Bahamonde:2021gfp}). (2) Furthermore, the importance of the boundary term, $f(T,B)$ gravity models has also been well-studied~\cite{Escamilla-Rivera:2019ulu,Bahamonde:2015zma,Farrugia:2018gyz} as a possible extension to TEGR to infer $H_0$~\cite{Escamilla-Rivera:2019ulu}.
Performing consistency tests with current astrophysical data allows to identify a gravity theory and deal at the same time with systematic effects in the observational data. In some cases, data samples are sensitive to the geometry and dynamics of the Universe and other samples are sensitive to the growth of LSS. In such cases, these two sets of observations must be consistent with one another in order to solve the cosmological tension inside a specific model. At late-times, any deviations between models can be measured through an effective EoS that mimics a dark energy component near to the $\Lambda$CDM EoS value ($w=-1$). In Ref.~\cite{Escamilla-Rivera:2020ges} it was provided an insightful test via examples in $f(R)$ and $f(T,B)$ gravity theories that can solve the $H_0$ tension using current late-time surveys (see also Refs.~\cite{Lambiase:2018ows, Benetti:2019smr, Cai:2019bdh, DAgostino:2020dhv, Wang:2020zfv, Parashari:2021qjg}). Theories that deviate from GR can be delineated into categories according to what principle or requirement they violate. 
Theoretical efforts to find a dynamic model describing the data have been placed side by side to kinematic models, as the cosmography, where the current expansion is a function of the cosmic time~\cite{Sahni:2014ooa, Capozziello:2019cav, Benetti:2019gmo,ElHanafy:2020pek}.

In its most simple formulation, it is possible to consider $f(T)$ extensions of teleparallel  gravity intended as corrections to TEGR where only the torsion scalar $T$ is considered. The action takes the form of
\begin{equation}\label{actionfT}
  S = \frac{M_{\rm Pl}^2}{2}\int{{\rm d}^4x \,e \left[T+f(T)\right]} +S_m\,,
\end{equation}
where $M_{\rm Pl}$ is the reduced Planck mass, $f(T)$ is a generic function of the torsion scalar $T$, $S_m$ is the action of matter fields, and the tetrad $e={\rm det}(e^i_\mu)=\sqrt{-g}$ is the metric determinant.

For a flat FLRW background, it is possible to derive the relation between the torsion $T$ and
the Hubble parameter, $T=6H^2$. 
The sign of such a torsion element depends on the signature of the metric, which is identified by  
the vierbien fields $e_{\mu}^a={\rm diag}(-1,a,a,a)$. 
Assuming that matter sector is described by a perfect fluid with energy density $\rho$ and pressure $p$, the field equations give 
\begin{eqnarray}
\label{fdme}
H^{2} = \frac{8 \pi G_N}{3} \rho_{m}-\frac{f(T)}{6}+\frac{T f_{T}}{3} \label{friedmann}\\
\dot{H} = -\frac{4 \pi G_N\left(\rho_{m}+p_{m}\right)}{1+f_{T}+2 T f_{T T}}. \label{acceleration}
\end{eqnarray}
Moreover, the set of equations is closed with the equation of continuity for the matter sector $\dot{\rho}+3H(\rho+p)=0$. Eqs.~\eqref{friedmann}-\eqref{acceleration} can be rewritten in terms of the effective energy density $\rho_T$ and pressure $p_T$ arising from $f(T)$,

\begin{eqnarray}
\rho_{f(T)}=\frac{M_{\rm Pl}^{2}}{2}\left[2 T f_{T}-f(T)\right] \label{rhofT1} \\
p_{f(T)}=\frac{M_{\rm Pl}^{2}}{2}\left[\frac{f(T)-Tf_{T}+2 T^2 f_{T T}}{1+f_{T}+2 T 
f_{T T}}\right],
\end{eqnarray}
and define the effective torsion  equation-of-state
\begin{equation}
 w \equiv \frac{p_{f(T)}}{\rho_{f(T)}}=\frac{f(T)-Tf_{T}+2 T^2 f_{T T}}{\left[1+f_{T}+2 T f_{T T}\right]\left[2T f_{T}-f(T)\right]}. 
\end{equation}
These effective models are hence responsible for the accelerated phases of the early or/and late Universe~\cite{Cai:2015emx}.

In order to study the background evolution, we can rewrite the first FLRW equation, making explicit the form of the torsional energy density~\cite{Nesseris:2013jea,Nunes:2018xbm,Nunes:2016qyp}
\begin{equation}
    \label{eq:Ha}
    \frac{H(a)^2}{H_0} \equiv E(a)^2 = \left[\Omega_m a^{-3}+\Omega_r a^{-4}+ \frac{1}{T_0} [-f+2Tf'] \right],
\end{equation}
where we considered the relation $T=6H^2$ and $T_0$ is the present value of $T$. 

At this point, specific forms of $f(T)$ functions must be chosen and several in the literature have passed basic observational tests~\cite{Nesseris:2013jea,Nunes:2016qyp,Capozziello:2017bxm,Nunes:2018xbm,Briffa:2021nxg}. Between them, the simplest form of power-law proved to be the one that predicts an $H_0$ compatible with Cepheids observations~\cite{Riess:2019cxk}. This model is described by~\cite{Nesseris:2013jea}
\begin{equation}
    \label{fT_powerlaw}
    f (T) = \beta \left(-T \right)^{b}.
\end{equation}
Inserting this $f(T)$ form into (\ref{fdme})   and for $b=0$ and $\beta=-2\Lambda $, the scenario reduces to $\Lambda$CDM cosmology, namely $T+f(T)=T-2\Lambda$.

Analysing this model considering both the background and the linear perturbation evolution, in the light of both large and small scale data, a value of $H_0 = (73.85 \pm 1.05){\rm\,km\,s^{-1}\,Mpc^{-1}}$ has been constrained using CMB+lensing+BAO+R19+Pth+DES data~\cite{Benetti:2020hxp} (see Ref.~\cite{Briffa:2021nxg} for for analysis with several other combinations of large-scale data). Also, a deviation from the GR was detected at more than $3\sigma$. Noteworthy, this allows to significantly relax the Hubble constant tension, but worsens $\sigma_8$ tension~\cite{MacCrann:2014wfa,Battye:2014qga,Benetti:2017juy} since the correlation of the two parameters does not seem to be removed from this treatment of $f(T)$ models.

Let us now try to suitably extract an $f(T)$ form that could alleviate bothe tensions simultaneously.
In order to avoid the $H_0$ tension  one needs a positive correction to the first Friedmann equation at late times that could yield an increase in $H_0$ compared to the $\Lambda$CDM scenario. As for the $\sigma_8$ tension, since at sub-Hubble scales and through the matter epoch, the equation that governs the evolution of matter perturbations in the linear regime is
\begin{equation}
    \ddot{\delta}+2 H \dot{\delta} = 4 \pi G_{\mathrm{eff}} \rho_m \delta\,,
\end{equation}
where $G_{\mathrm{eff}}$ is the effective gravitational coupling. Thus, alleviation of $\sigma_8$ tension may be obtained if $G_{\mathrm{eff}}$ becomes smaller than $G_{N}$ during the growth of matter perturbations.
In the case of   $f(T)$ gravity the effective gravitational coupling is given by~\cite{Nesseris:2013jea}
\begin{equation}
    G_{\mathrm{eff}}=G_{N}\,\left(1+\frac{\partial f}{\partial T}\right)^{-1}\,.
\end{equation}

According to the Effective Field Theory approach to $f(T)$ gravity~\cite{Li:2018ixg,Yan:2019gbw,Ren:2022aeo} one can choose $f(T)$ in order to obtain a Hubble function evolution of the form 
\begin{equation}
    H(z) = -\frac{d(z)}{4} +\sqrt{\frac{d^2(z)}{16} +H_{\Lambda \text{CDM}}^2(z)}\,,    
\end{equation}
where $H_{\Lambda \text{CDM}}(z) \equiv H_0 \sqrt{\Omega_m(1+z)^3+\Omega_\Lambda}$ is the Hubble rate in $\Lambda$CDM after recombination, $\Omega_m$ is the present matter density parameter, and primes denote derivatives with respect to $z$. The function $d(z)$ can be selected to be $d<0$ in order to have $H(z\rightarrow z_{\text{CMB}}) \approx H_{\Lambda\text{CDM}}(z\rightarrow z_{\text{CMB}})$, therefore the $H_0$ tension is solved. One should choose $|d(z)| < H(z)$, and thus, since $H(z)$ decreases for smaller $z$, the deviation from $\Lambda$CDM will be significant only at low redshift]. Additionally, since $G_{\mathrm{eff}}$ becomes smaller than $G_{N}$ during the growth of matter perturbations, the $\sigma_8$ tension is also solved.
 
We follow Ref.~\cite{Yan:2019gbw} and we consider the ansatz $f(T)=-[6H_0^2(1-\Omega_m)+F(T)]$, where $F(T)$ describes the deviation from GR (mind the difference in the conventions). Under these assumptions, the first Friedmann equation becomes
\begin{align}
    \label{eq:bg3}
    T(z)+2\frac{F'(z)}{T'(z)} T(z)-F(z)= 6H^2_{\Lambda{\rm CDM}}(z)\,.
\end{align}
In order to solve the $H_0$ tension, we need $T(0) = 6 H_0^2$, with $H_0 = 74.03{\rm\,km\,s^{-1}\,Mpc^{-1}}$ following results with local measurements~\cite{Riess:2019cxk},
while in the early era of $z\gtrsim 1100$ we require the Universe expansion to evolve as in $\Lambda$CDM, namely $H(z\gtrsim 1100) \simeq H_{\Lambda\text{CDM}}(z\gtrsim 1100)$. This implies $F(z)|_{z\gtrsim 1100}\simeq c T^{1/2}(z)$ (the value $c=0$ 
corresponds to standard GR, while for $c\neq0$ we obtain $\Lambda$CDM too). Therefore, the perturbation equation at linear order becomes
\begin{equation}
    \label{eq:delta-z}
    \delta" + \left[ \frac{T'(z)}{2T(z)} -\frac{1}{1+z} \right] \delta' = \frac{9 
H_0^2 \Omega_m (1+z) }{[1+F'(z)/T'(z)] T(z)} \delta\,.
\end{equation}
Since around the last scattering moment $z\gtrsim 1100$ the Universe should be matter-dominated, we impose $\delta'(z)|_{z\gtrsim 1100} \simeq -\frac{1}{1+z}\delta(z)$, while at late times we look for $\delta(z)$ that leads to an $f\sigma_8$ in agreement with redshift survey observations.

By solving Eqs.~\eqref{eq:bg3} and~\eqref{eq:delta-z} with initial and boundary conditions at $z \sim 0$ and $z \sim 1100$, we can find the $f(T)$ form that can alleviate both $H_0$ and $\sigma_8$ tensions. In particular, we find that we can  fit the numerical solutions very efficiently through  $F(T) \approx 375.47\,[T/(6 H_0^2)]^{-1.65}$, dubbed Model-1, and   
 $F(T) \approx 375.47\,[T/(6 H_0^2)]^{-1.65} + 25 T^{1/2}$, dubbed  Model-2. We examine $G_{\mathrm{eff}}$, and as expected we find that at high redshifts in both models, $G_{\mathrm{eff}}$ becomes $G_N$, recovering the $\Lambda$CDM paradigm.  We have checked that both models can easily pass the BBN constraints, which demand $|G_{\mathrm{eff}}/G_N-1|\leq0.2$~\cite{Copi:2003xd}, as well as the ones from the Solar System, which demand $|G_{\mathrm{eff}}'(z=0)/G_N|\leq10^{-3}h^{-1}$ and $| G_{\mathrm{eff}}"(z = 0) / G_N | \leq 10^{5} h^{-2}$~\cite{Nesseris:2006hp}.

\begin{figure}[ht]
\centering
\includegraphics[width=3.in]{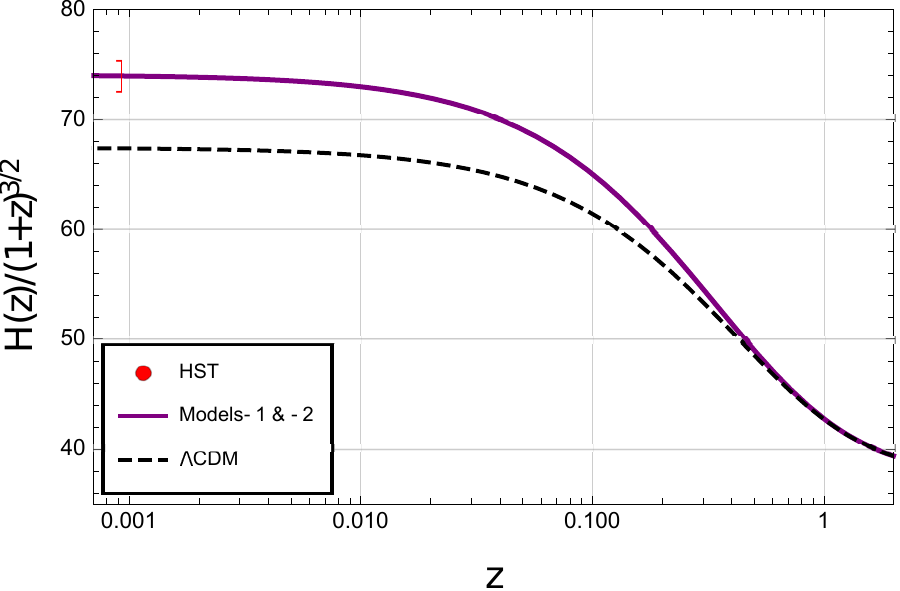}
\caption{Evolution of the Hubble parameter $H(z)$ in the two $f(T)$ models (purple solid line) and in $\Lambda$CDM cosmology (black dashed line). The red point (with error bars) represents the latest data from extragalactic Cepheid-based local measurement of $H_0$~\cite{Riess:2019cxk}. The figure is from~\cite{Yan:2019gbw}.}
\label{fig:H0&fs81111}
\end{figure}

\begin{figure}[ht]
\centering
\includegraphics[width=3.1in]{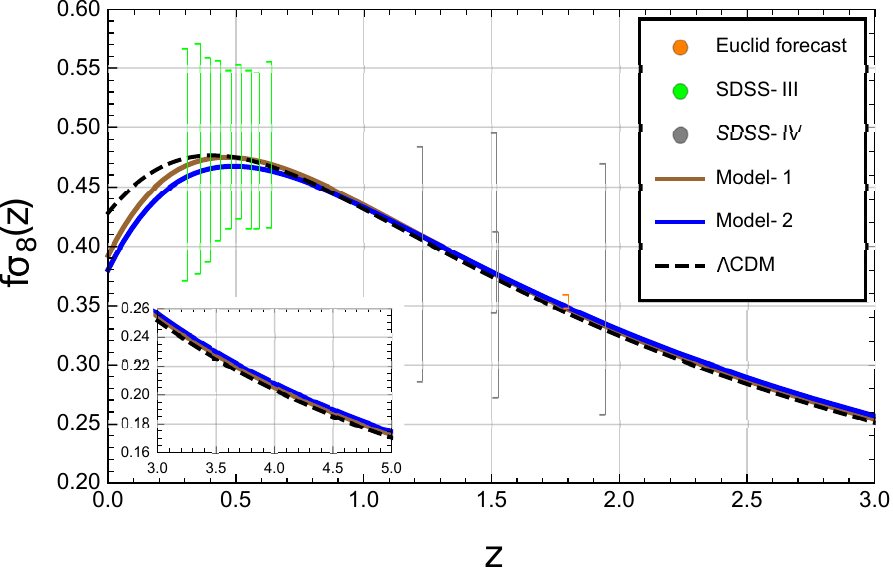}
\caption{The evolution of $f\sigma_8$ in Model-1 (brown solid line) and Model-2 (blue solid line) of $f(T)$ gravity and in $\Lambda$CDM cosmology (black dashed line). The green data points (with error bars) are from BAO observations in SDSS-III DR12~\cite{Wang:2017wia}, the gray data points (with error bars) at higher redshift are from SDSS-IV DR14~\cite{Gil-Marin:2018cgo}, while the red point (with error bars) around $\sim 1.8$ is the forecast from Euclid~\cite{Taddei:2016iku}. The subgraph in the left bottom displays $f \sigma_8$ at high redshift $z = 3 \sim 5$, which shows that the curve of Model-2 is above the one of Model-1 and $\Lambda$CDM scenario and hence approaches $\Lambda$CDM slower than Model-1. The figure is from~\cite{Yan:2019gbw}.}
\label{fig:H0&fs8}
\end{figure}

In Fig.~\ref{fig:H0&fs81111} we present the evolution of $H(z)$, while in Fig.~\ref{fig:H0&fs8} the evolution of $f\sigma_8$, for two $f(T)$ models, and we compare them with $\Lambda$CDM. We stress that the $H_0$ tension can be alleviated as $H(z)$ remains statistically consistent for all CMB and CC measurements at all redshifts. We remind the reader that the two $f(T)$ models differing merely by a term $\propto T^{1/2}$, which does not affect the background as explained before, are degenerate at the background level. However, at the perturbation level, the two models behave differently as their gravitational coupling $G_{\mathrm{eff}}$ differs. 

We further stress that both models can alleviate the $\sigma_8$ tension, and fit efficiently to BAO and LSS measurements. Note that at high redshifts ($z\geq2$), Model-2 approaches $\Lambda$CDM slower than Model-1, but in a way that is statistically indistinguishable for present-day data. Nevertheless, future high-redshift surveys such as eBOSS for quasars and Euclid~\cite{EUCLID:2011zbd} for galaxies have the potential to discriminate among the predictions of $f(T)$ gravity and the $\Lambda$CDM scenario. Moreover, the clusters and CMB measurements on $\sigma_8$ are in good agreement in our models, as the CMB preferred values in $\Lambda$CDM get further lowered than local ones from rescaling $\sigma_8$ by the ratio of the growth factors in $f(T)$ gravity and $\Lambda$CDM. In more detail,
\begin{equation}
    \sigma_8^{f(T)}(z=0) =\frac{D^{f(T)}(z=0)}{D^{\Lambda}(z=0)} \frac{D^{\Lambda}(z_{\rm eff})}{D^{f(T)}(z_{\rm eff})} \sigma_8^{\Lambda}(z=0)\,,
\end{equation}
where $D(z)$ is the growth factor, $f(T)$ and $\Lambda$ denote the models for modified gravity and $\Lambda$CDM, respectively, and $z_{\rm eff}$ is the effective redshift of the measurements, with $z_{\rm eff} \sim 0.1$ for clusters experiments and $z_{\rm eff} \sim 1100$ for CMB temperature fluctuations observations. At high redshifts $z_{\rm eff} \sim 1100$, the growth factor is the same in both the $f(T)$ model and in $\Lambda$CDM, while at low redshifts $z_{\rm eff} \sim 0.1$, the growth factor is approximately 1.03 times bigger in the latter one. For this  reason, cluster $\sigma_8$-measurements increase by about the same factor, reducing the gap with CMB preferred value in this modified gravity scenario.
 
In summary, we conclude that the class of $f(T)$ gravity described by
\begin{equation}
    f(T) = -2\Lambda/M_{\rm Pl}^2 +\alpha T^\beta\,,
    \label{fTgeneralmodel}
\end{equation}
where only two out of the three parameters $\Lambda$, $\alpha$, and $\beta$ are independent (the third one is eliminated using $\Omega_m$), can alleviate both $H_0$ and $\sigma_8$ tensions with suitable parameter choices. Moreover, these type of $f(T)$ gravity models could also be examined through galaxy-galaxy lensing effects~\cite{Chen:2019ftv}, strong lensing effects around black holes~\cite{Yan:2019hxx}, and GW experiments~\cite{Cai:2018rzd}.

\subsubsection{Horndeski Theory} 

\paragraph{Shift-Symmetric Horndeski Theories.} After the detection of the gravitational-wave event GW170817~\cite{LIGOScientific:2017vwq} and its electromagnetic counterpart, the speed of GWs $c_t$ is constrained to be very close to that of light $c$. Without the shift symmetry condition, namely that the theory is invariant under scalar shifts, $\phi \rightarrow \phi + c$ with constant $c$, Horndeski would reduce to three functions, $G_2(\phi, X)$, $G_{3}(\phi, X)$ and $G_{4}(\phi)$ (for example, see Ref.~\cite{Kase:2018aps}). Further imposing the shift symmetry condition, $G_{2}$ and $G_{3}$ are reduced to functions of $X$ only, while $G_{4}$ becomes a constant. The Lagrangian of the Horndeski theories~\cite{Horndeski:1974wa,Kobayashi:2011nu, DeFelice:2011bh,Kobayashi:2019hrl} realizing this condition is strongly constrained~\cite{Ezquiaga:2017ekz}, in particular the cubic-order shift-symmetric Horndeski theories falls back in this class of theories. The cubic-order shift-symmetric Horndeski action is given by~\cite{Kobayashi:2019hrl}
\begin{equation}
    \label{actionHorn0}
     S=\int {\rm d}^4 x \sqrt{-g} \left[\frac{M_{\rm Pl}^2}{2}R+G_2(X)+G_3(X) \square \phi \right]+ S_i (\chi_i, g_{\mu \nu})\,,
\end{equation}
where $g$ is the determinant of the metric tensor $g_{\mu \nu}$, $M_{\rm Pl}$ is the reduced Planck mass, $R$ is the Ricci scalar, and $G_2$, $G_3$ are functions of $X=\nabla^{\mu} \phi \nabla_{\mu} \phi$\,, with $\phi$ being the scalar field. Finally, $ S_i$ corresponds to the action of standard $i$th fluid fields $\chi_i$. The continuity equations of fluids reads on the FLRW background,
\begin{equation}
    \dot{\rho}_i+3H(1+w_i) \rho_i=0\,,
\end{equation}
where $\rho_i$ is the fluid density, $w_i\equiv p_i/\rho_i$ is the constant barotropic coefficient relating the fluid density and the pressure,  $H=\dot{a}/a$ is the Hubble expansion rate, and a dot stands for the derivative with respect to $t$. We recall that the modified Friedmann and scalar-field equations at at the background level are given by
\begin{eqnarray}
&& 3M_{\rm Pl}^2 H^2 = \rho_{\rm \phi}+\rho_m +\rho_r\,,\label{Frid1}\\
&& 2 M_{\rm Pl}^2 \dot{H} = -\rho_{\rm \phi}-P_{\rm \phi}-\rho_m-\frac43 \rho_r\,,\label{Frid2}\\
&& (G_{2,X}-2 \dot{\phi}^2 G_{2,XX}-6H \dot{\phi} G_{3,X}+6H \dot{\phi^3}G_{3,XX}) \ddot{\phi}+3(H G_{2,X}- \dot{H} \dot{\phi} G_{3,X} -3H^2 \dot{\phi} G_{3,X})\dot{\phi}=0\,.\label{field}
\end{eqnarray}
The density and pressure of the scalar field are obtained as 
\begin{eqnarray}
    \rho_{\rm \phi} &=& -G_2-2\dot{\phi}^2 G_{2,X} +6 H \dot{\phi}^3 G_{3,X}\,,\\
    P_{\rm \phi} &=& G_2-2\ddot{\phi} \dot{\phi}^2 G_{3,X}\,,
\end{eqnarray}
with $G_X\equiv {\rm d}G/{\rm d}X$. From this, we can define a scalar field EoS as
\begin{equation}
    \label{wphi}
    w_{\rm \phi} \equiv \frac{P_{\rm \phi}}{\rho_{\rm \phi}}=-\frac{G_2-2\ddot{\phi} \dot{\phi}^2 G_{3,X}}{G_2+2\dot{\phi}^2 G_{2,X}-6 H \dot{\phi}^3 G_{3,X}}\,.
\end{equation}
From Eq.~\eqref{Frid1}, there is the constraint $\Omega_\phi+\Omega_m +\Omega_r=1$, where $\Omega_i \equiv \rho_i/(3M_{\rm Pl}^2 H^2)$. Finally, by solving Eqs.~\eqref{Frid2}-\eqref{field} for $\dot{H}$ and $\ddot{\phi}$  one can know the dynamics of the Universe. For linear perturbations we consider  the perturbed FLRW line element: ${\rm d}s^2=-\left( 1+2\Psi \right) {\rm d}t^2+a^2(t) \left(1-2\Phi \right) \delta_{ij} {\rm d}x^i {\rm d}x^j$, where $\Psi$ and $\Phi$ are the gravitational potentials and for the perturbations in the matter sector follow the standard perturbation theory~\cite{Ma:1995ey}. In the case of modified gravity, in order to derive an explicit relation between the gravitational potentials and the matter perturbation $\delta_m$, one has to use the quasi-static approximation (QSA)~\cite{Boisseau:2000pr,Tsujikawa:2007gd,DeFelice:2011hq,Sawicki:2015zya} which, within the Horndeski class of models, it is a valid assumption for the wavenumber in the range $k > 10^{-3}\,h{\rm \,Mpc^{-1}}$~\cite{Peirone:2017ywi, Frusciante:2018jzw}. In this approximations one has:
\begin{eqnarray}
    \label{mudef}
    -k^2\Psi &=& 4\pi G_N a^2\mu(a,k)\rho_m\delta_m\,, \\
    -k^2(\Psi+\Phi) &=& 8\pi G_N a^2\Sigma(a,k)\rho_m\delta_m\,,
\end{eqnarray}
where $G_N$ is Newton's gravitational constant. The dimensionless quantities $\mu$ and $\Sigma$ characterize 
the effective gravitational couplings felt by matter and light, respectively. For the cubic-order shift-symmetric Horndeski theories one has $\Phi=\Psi$ and~\cite{DeFelice:2011hq,Kase:2018aps}
\begin{equation}
    \label{muSigma}
    \mu=\Sigma=1+\frac{4\dot{\phi}^4 G_{3,X}^2}{Q_s c_s^2}\,,
\end{equation}
where the kinetic term $Q_s$ and the speed of propagation of the scalar field $c_s$ are given by
\begin{eqnarray}
    Q_s &=& 4M_{\rm Pl}^2 \left(-G_{2,X}+2\dot{\phi}^2 G_{2,XX}+6 H \dot{\phi} G_{3,X}-6 H \dot{\phi}^3 G_{3,XX} \right)+12 \dot{\phi}^4 G_{3,X}^2\,,\label{qsg}\\
    c_s^2 &=& \frac{-4M_{\rm Pl}^2 (G_{2,X}-2\ddot{\phi} G_{3,X}-4H \dot{\phi}G_{3,X}+2\ddot{\phi}\dot{\phi}^2 G_{3,XX})-4\dot{\phi}^4 G_{3,X}^2}{Q_s}\,.\label{csg}
\end{eqnarray}
Both the above quantities need to be positive defined to avoid Ghost and Gradient instabilities. Finally the matter perturbation equations can be written as 
\begin{equation}
    \label{delmse2}
    \ddot{\delta}_m+2H \dot{\delta}_m-4\pi G_N \mu \rho_m \delta _m\simeq 0\,.
\end{equation}
In the following we specialize to two models which are of interest for alleviating the $H_0$ tension.

\begin{itemize}
\item Galileon Ghost Condensate (GGC):  We consider the Galileon Ghost Condensate (GGC)  model which is specified  by the functions~\cite{Deffayet:2010qz,Kase:2018iwp}
\begin{equation}
    \label{actionGGC}
    G_2(X)=a_1 X+a_2 X^2\,,\qquad G_3(X)=3a_3X\,,
\end{equation}
where $a_1$, $a_2$, $a_3$ are constants. This model departs from the Cubic Galileon model~\cite{Deffayet:2009wt} by the inclusion of the term $a_2X^2$, which affects the cosmic expansion history and growth of structures at both linear~\cite{Kase:2018iwp,Peirone:2019aua} and non-linear scales~\cite{Frusciante:2020zfs} compared to $G_3$. The presence of nonvanishing $a_2X^2$ allows to prevent the solutions from approaching the tracker at background level. In this case it is possible to realize the phantom behaviour for the scalar field  ($w_{\rm \phi}<-1$) without having ghosts~\cite{Kase:2018iwp,Peirone:2019aua}. At linear perturbation $\mu$ and $\Sigma$ in Eq.~\eqref{muSigma} are larger than 1, so both $\Psi$ and $\Psi+\Phi$ are enhanced compared to those in GR, showing a stronger gravitational interaction. In particular, it has been found that the model suppresses the large-scale CMB temperature anisotropy in comparison to the $\Lambda$CDM model. These features lead to the statistical preference of GGC over $\Lambda$CDM when data from CMB, BAO, SNIa, and RSDs are used~\cite{Peirone:2019aua}. Furthermore the CMB temperature and polarization data from {\it Planck} 2015 release  lead to an estimation for today’s Hubble parameter $H_0$ with is higher than the $\Lambda$CDM and consistent with its direct measurements at 2$\sigma$,  $H_0=\left(69.3^{+3.6}_{-3.0}\right){\rm\,km\,s^{-1}\,Mpc^{-1}}$ at 95\% CL with best fit $H_0^{bf}=70{\rm\,km\,s^{-1}\,Mpc^{-1}}$~\cite{Peirone:2019aua}.

\item Generalized Cubic Covariant Galileon (GCCG): 
Let us consider the Generalized Cubic Covariant Galileon (GCCG) model~\cite{DeFelice:2011bh}, defined by the functions
\begin{equation}
G_2(X)=-c_2 M_2^{4(1-p)} (-X/2)^{p}\,,
\qquad
G_3(X)=-c_3 M_3^{1-4p_3} (-X/2)^{p_3}\,,
\end{equation}
where $c_2$, $c_3$, $p$, $p_3$ are dimensionless constants, and $M_2$, $M_3$ are constants having dimension of mass. For simplicity another parameters are used defined by $q=p_3-p+\frac{1}{2}$ and $s=p/q$. The GCCG model is an extension of the Covariant Galileon ($s=2$, $q=1/2$), and the same as the latter, it also allows for  the existence of a tracker solution following the relation $H \dot{\phi}^{2q}={\rm constant}$ with $q>0$. For this model it is possible to verify that $\mu \geq 1$ for any viable value of $q$ and $s$ and hence the gravitational interaction is  always stronger than in GR.

Observational constraints on $H_0$, in the case of {\it Planck} 2015 alone are: $H_0= 72^{+8}_{-5}{\rm\,km\,s^{-1}\,Mpc^{-1}}$ at 95\% CL, finding a consistency with its determination from Cepheids at 1$\sigma$~\cite{Frusciante:2019puu}. Such result on $H_0$ is also compatible with the estimation obtained with the tip of the red giant branch in LMC, $H_0=(72.4\pm 2){\rm\,km\,s^{-1}\,Mpc^{-1}}$~\cite{Yuan:2019npk}. The main feature behind the eased $H_0$ tension in the GCCG model is due to a modified low redshift expansion history  which is enhanced with respect to $\Lambda$CDM. In this model is also interesting to note that if the tension between CMB data and low redshift measurements of $H_0$ disappears for the GCCG model, another tension between the latter and BAO data appears~\cite{Frusciante:2019puu} (see also~\cite{Renk:2017rzu}).
\end{itemize}


\paragraph{Horndeski Theories with Non-Minimal Derivative Coupling.} Let us now examine one interesting subclass of Horndeski gravity, or equivalently  generalized Galileon theory, that contains the $G_5$ term of the corresponding action, which is called "non-minimal derivative coupling term". Thus, instead of Eq.~\eqref{actionHorn0}, we consider the action~\cite{Saridakis:2010mf,Petronikolou:2021shp}
\begin{eqnarray}
    \label{actionHorn11}
     S&=&\int {\rm d}^4 x \sqrt{-g} \bigg\{\frac{M_{\rm Pl}^2}{2}R+K(\phi,X)  +G_5(\phi,X)\,G_{\mu\nu}\,(\nabla^{\mu}\nabla^{\nu}\phi)\nonumber\\
    && -\frac{1}{6}\, G_{5,X}\,[(\Box\phi)^{3}-3(\Box\phi)\,(\nabla_{\mu}\nabla_{\nu}\phi)\,(\nabla^{\mu}\nabla^{\nu}\phi)+2(\nabla^{\mu}\nabla_{\alpha}\phi)\,(\nabla^{\alpha}\nabla_{\beta}\phi)\,(\nabla^{\beta}\nabla_{\mu}\phi)]\bigg\}
    + S_i (\chi_i, g_{\mu \nu})\,,
\end{eqnarray}
where $G_{\mu\nu}$ is the Einstein tensor. The Friedmann equations in FLRW geometry become~\cite{DeFelice:2011bh}
\begin{eqnarray}
    2XK_{,X}-K-3H^2 M_{\rm Pl}^2+2H^3X \dot{\phi} (5G_{5,X}+2X G_{5,XX}-6H^2X (3G_{5,\phi}+2X G_{5,\phi X})&=&-\rho_{m}\,,\label{Fr1gen}\\
    K+ M_{\rm Pl}^2 (3H^2+2\dot{H}) -4H^2X^2\ddot{\phi} G_{5,XX}-2X (2H^3\dot{\phi}+2H\dot{H} \dot{\phi}+3H^2\ddot{\phi}) G_{5,X}&&\nonumber\\
    +4H X(\dot{X}-HX)G_{5,\phi X}+4H X\dot{\phi} G_{5,\phi\phi}+2 [ 2(\dot{H}X+H\dot{X})+3H^2X ] G_{5,\phi}&=&-p_m\,,
\label{Fr2gen}
\end{eqnarray}
while the scalar-field equation is
$    \frac{1}{a^3}\frac{\rm d}{{\rm d}t}(a^3J)=P_{\phi}$,
    with 
  $J\equiv \dot{\phi} K_{,X} +2H^3X (3G_{5,X}+2X G_{5,XX})+6H^2\dot{\phi} (G_{5,\phi}+X G_{5,\phi X})$
and $P_{\phi}\equiv K_{,\phi} -6H^2X G_{5,\phi\phi}+2H^3X \dot{\phi} G_{5,\phi X}$. Finally, the conditions in Eqs.~\eqref{qsg}-\eqref{csg} for absence of Laplacian and ghost instabilities become:
\begin{eqnarray}
    Q_{s} &\equiv& \frac{w_{1}(4w_{1}w_{3}+9w_{2}^{2})}{3w_{2}^{2}}>0\,,\label{Qscon2}\\
    c_{s}^{2} &\equiv& \frac{3(2w_{1}^{2}w_{2}H-w_{2}^{2}w_{4}+4w_{1}w_{2}\dot{w}_{1}-2w_{1}^{2}\dot{w}_{2})}{w_{1}(4w_{1}w_{3}+9w_{2}^{2})} \geq 0\,,\label{cscon2}
\end{eqnarray}
with 
\begin{eqnarray}
    w_{1} &\equiv& M_{\rm Pl}^2-2X(G_{{5,X}}{\dot{\phi}}H-G_{{5,\phi}})\,,\label{w1def}\\
    w_{2} &\equiv& 2M_{\rm Pl}^2H+8 {X}^{2}HG_{{5,\phi X}}+2H  X (6G_{{5,\phi}}-5 G_{{5,X}}\dot{\phi}{H}) -4G_{{5,{XX}}}{\dot{\phi}}X^{2}{H}^{2}\,,\\
    w_{3} &\equiv& 3X(K_{,{X}}+2  XK_{,{  XX}}) -9 M_{\rm Pl}^2 H^2\,\nonumber \\
    && +6{H}^{2}X\!\left(2  H\dot{\phi}G_{{5,{XXX}}}{X}^{2}\!-\!6 {X}^{2}G_{{5,\phi{  XX}}}\!-\!18G_{{5,\phi}}+13XH\dot{\phi}G_{{5,{XX}}}\!-\!27G_{{5,\phi X}}X\!+\!15  H\dot{\phi}G_{{5,X}}\right),\\
    w_{4} &\equiv& M_{\rm Pl}^2-2XG_{5,\phi}\!-\!2XG_{5,X}\ddot{\phi}\,.
\end{eqnarray}
In order to alleviate the $H_0$ tension using the above construction we need to obtain a weakening of gravity at low redshifts by the terms depending on the scalar field's kinetic energy. We impose a simple scalar field potential and standard kinetic term, thus  $K= -V(\phi) + X$, and we will consider the $G_5$ term to be shift-symmetric, i.e.\ $G_5(\phi,X)=G_5(X)$. We consider two models.
\begin{itemize}
\item {Model I: $G_5(X)=\xi X^{2}$}
In this case, the Friedmann Eqs.~\eqref{Fr1gen}-\eqref{Fr2gen} become
\begin{eqnarray}
    \label{FR1111}
    3 M_{\rm Pl}^2 H^{2}&=&\frac{\dot{\phi}^2}{2}+V_0\phi+7\xi H^3\dot{\phi}^5+\rho_{m}\,, \\
    \label{FR2111}
    -2 M_{\rm Pl}^2\dot{H}&=&\dot{\phi}^2 +7\xi H^3\dot{\phi}^5  -\xi\dot{\phi}^4 \left(2H^3\dot{\phi}+2H\dot{H}\dot{\phi}+5H^2\ddot{\phi}\right)+\rho_{m}+p_{m}\,.
\end{eqnarray}
We choose the model parameter $V_0$ and the initial conditions for the scalar field in order to obtain $H(z_{\rm CMB})=H_{\Lambda\text{CDM}}( z_{\rm CMB})$ and $\Omega_{m}=0.31$, in agreement with the results from {\it Planck} 2018~\cite{Planck:2018vyg}, and we leave $\xi$ as the parameter that determines the late-time deviation from $\Lambda$CDM cosmology necessary to alleviate the $H_0$ tension. Choosing $1.3 \leq\xi \leq1.5$ in Planck units leads to $H_0 \approx 74{\rm\,km\,s^{-1}\,Mpc^{-1}}$ and thus the $H_0$ tension is indeed alleviated, while $c_{s}^{2}$ remains almost equal to 1 and $Q_{s}$ remains positive, and therefore the 
scenario is free from pathologies~\cite{Petronikolou:2021shp}.

\item {Model II: $G_5(X)=\lambda X^{4}$}
In this case the Friedmann Eqs.~\eqref{Fr1gen}-\eqref{Fr2gen} become
\begin{eqnarray}
    \label{FR1111b}
    3M_{\rm Pl}^2 H^{2}&=&\frac{\dot{\phi}^2}{2}+V_0\phi+\frac{11}{2}\lambda H^3 \dot{\phi}^9 +\rho_{m}\,,\\
    \label{FR2111b}
    -2 M_{\rm Pl}^2\dot{H}&=&\dot{\phi}^2+\frac{11}{2}\lambda H^3\dot{\phi}^9 -\frac{\lambda \dot{\phi}^8}{2}\left(2H^3\dot{\phi}+2H\dot{H}\dot{\phi}+9H^2 \ddot{\phi}\right) +\rho_{m}+p_{m}\,.
\end{eqnarray}
Choosing $0.9 \leq\lambda \leq1.1$ in Planck units leads to $H_0 \approx 74{\rm\,km\,s^{-1}\,Mpc^{-1}}$, while $c_{s}^{2}$ remains almost equal to 1 and $Q_{s}$ remains positive~\cite{Petronikolou:2021shp}. Hence the Hubble tension is alleviated. 
\end{itemize}
In summary, we showed that the above particular sub-class of Horndeski/generalized Galileon gravity can alleviate the $H_0$ tension due to the effect of the kinetic-energy-dependent $G_5$ term. In particular, at early times the field's kinetic term is negligible and hence the $G_5(X)$ terms do not introduce any deviation from $\Lambda$CDM scenario, nevertheless as time passes they increase in a controlled and suitable way in order to make the 
Hubble function, and thus $H_0$ too, to increase. 
However, see Refs.~\cite{Banerjee:2020xcn,Lee:2022cyh}, where it is argued that DE models in a general class of scalar-tensor theories, including quintessence and a good class of Horndeski theories, perform worse than $\Lambda$CDM model as far as the $H_0$ increasing is concerned. Moreover, a similar result is discussed in Refs.~\cite{Heisenberg:2022gqk,Heisenberg:2022lob} where it is argued that the family of scalar-tensor DE models generically exacerbate $H_0$ and $S_8$ anomalies. A further problem of the Horndeski models with non-minimal derivative coupling, i. e. $G_{5} \neq 0$, is that to evade constraints from gravitational waves~\cite{LIGOScientific:2017vwq}, some fine tuning is required. \\

\paragraph{Minimal Scalar-Tensor Theories of Gravity.} The action for the simplest scalar-tensor theories of gravity within the most general Horndeski model~\cite{Horndeski:1974wa} with a scalar field $\sigma$ in units of the reduced Planck mass $M_{\rm Pl}$ and non-minimally coupled (NMC) to the Ricci curvature $R$ is~\cite{Capozziello:1994du,Boisseau:2000pr,Esposito-Farese:2000pbo,Nesseris:2006jc,Cai:2009zp}
\begin{equation}
    \label{eqn:ST_action}
      S = \int {\rm d}^{4}x \sqrt{-g} \left[ \frac{F(\sigma)}{2}R - \frac{g^{\mu\nu}}{2} \partial_\mu \sigma \partial_\nu \sigma - V(\sigma) + {\cal L}_m \right] \,,
\end{equation}
where $F(\sigma)$ is a generic function of $\sigma$. Examples include induced gravity (IG), i.e.\ $F(\sigma) = \xi\sigma^2$ where $\xi>0$ is the coupling to the Ricci scalar, and conformally coupled scalar field (CCSF), i.e.\ $F(\sigma) = N_{\rm Pl}^2 -\sigma^2/6$ where $N_{\rm Pl}$ is a dimensionless parameter. A scalar-tensor theory allows to easily attain values for $H_0$ that are larger than in $\Lambda$CDM. This effect is mainly due to a degeneracy between the coupling to the Ricci curvature and $H_0$ and is connected to the rolling of the scalar field, which regulates the gravitational strength, at the onset of the matter dominated era driven by pressureless matter. Therefore, larger values for $H_0$ can appear already in archetypal models of scalar-tensor theories of gravity, namely Jordan-Brans-Dicke (JBD) models~\cite{BransDicke1961,Dicke1962}, see Sec.~\ref{sec:BDLCDM} for additional detail. 
This effect is {\it a)} largely independent on the part of the scalar field potential which acts as an effective cosmological constant, {\it b)} robust with respect to uncertainties in the neutrino sector, and {\it c)} possibly enhanced by allowing a difference between the gravitational constant relevant for cosmology and Newton's constant as measured in a Cavendish-like experiment.

The possibility to connect a larger value of $H_0$ with an early-time modification of GR in the context of scalar-tensor gravity has been highlighted since the first {\it Planck} 2013 data release, where one of the simplest scalar-tensor gravity model, such as induced gravity (equivalent to JBD) with a quartic potential $V(\sigma) = \lambda \sigma^4/4$ has been studied~\cite{Umilta:2015cta}, showing the degeneracy between the coupling to the Ricci scalar and $H_0$. Subsequent studies updated and generalized this result for different simple potentials and couplings~\cite{Ballardini:2016cvy,Rossi:2019lgt,Ballesteros:2020sik,Ballardini:2020iws,Ballardini:2021eox}.

To sum up, for an effectively massless ($V \propto F^2$) scalar field $\sigma$ at rest in the radiation era and with adiabatic initial condition for the scalar field perturbations~\cite{Paoletti:2018xet}, the current constraints on the Hubble parameter from {\it Planck} 2018 data (temperature, E-mode polarization, and CMB lensing) at 68\% CL are $H_0 = \left(69.6^{+0.8}_{-1.7}\right){\rm \,km\,s^{-1}\,Mpc^{-1}}$ for IG and $H_0 = \left(69.0^{+0.7}_{-1.2}\right){\rm \,km\,s^{-1}\,Mpc^{-1}}$ for CCSF, respectively~\cite{Ballardini:2020iws}. When BAO data from BOSS DR12 are added, the constraint at 68\% CL becomes $H_0 = \left(68.78^{+0.53}_{-0.78}\right){\rm \,km\,s^{-1}\,Mpc^{-1}}$ for IG and $H_0 = \left(68.62^{+0.47}_{-0.66}\right){\rm \,km\,s^{-1}\,Mpc^{-1}}$ for CCSF. Once a Gaussian likelihood based on the determination of the Hubble constant from the SH0ES team is also included, i.e.\ $H_0 = (74.03 \pm 1.42){\rm \,km\,s^{-1}\,Mpc^{-1}}$~\cite{Riess:2019cxk}, we obtain $H_0 = (70.1 \pm 0.8){\rm \,km\,s^{-1}\,Mpc^{-1}}$ at 68\% CL for IG and $H_0 = \left(69.64^{+0.65}_{-0.73}\right){\rm \,km\,s^{-1}\,Mpc^{-1}}$ at 68\% CL for CCSF. These constraints have been obtained by fixing the value of the scalar field today $\sigma_0\equiv \sigma(z=0)$ to
\begin{equation}
	\frac{1}{8 \pi F_0}\frac{2F_0+4F_{0,\sigma}^2}{2F_0+3F_{0,\sigma}^2} = G_N\,,
	\label{eq:EffGNST}
\end{equation}
where $F_0 \equiv F(\sigma_0)$ and $F_{0,\sigma} = \partial F/\partial \sigma|_{\sigma=\sigma_0}$, in order to guarantee that the effective gravitational constant today corresponds to the value of the bare gravitational constant~\citep{Boisseau:2000pr}.\footnote{See Refs.~\cite{Avilez:2013dxa,Akarsu:2019pvi,Joudaki:2020shz, Ballardini:2021evv} for studies of these models without imposing this condition.}

NMC models with $F(\sigma) = N_{\rm Pl}^2 + \xi\sigma^2$ and a generic negative value of the coupling $\xi < 0$ have been explored in~\cite{Rossi:2019lgt,Ballesteros:2020sik,Braglia:2020iik,Ballardini:2020iws,Abadi:2020hbr}, where the scalar field decreases from an initial value $\sigma_I$ to a final value close to zero, so that the Newton's constant changes after BBN from a value $G_N^{\rm BBN}$ to the present value $G_N^0$. Such models were shown to lead to $H_0=(69.08_{-0.71}^{+0.60}){\rm \,km\,s^{-1}\,Mpc^{-1}}$, with $1-G_N^{\rm BBN}/G^0_N=-0.05403^{+0.044}_{-0.019}$, $\sigma_I= 0.2789^{ +0.097}_{-0.054}$ and a marginal improvement of $\Delta \chi^2=-3.2$ compared to $\Lambda$CDM, at the cost of 2 extra parameters (with {\it Planck} 2018, BAO, SH0ES 2019 and Pantheon, plus Solar system constraints)~\cite{Ballesteros:2020sik}, or $H_0=(69.65_{-0.78}^{+0.8}){\rm \,km\,s^{-1}\,Mpc^{-1}}$ with $1-G_N^{\rm BBN}/G^0_N=-0.02239^{+0.0082}_{-0.0087}$, $\sigma_I= 0.6214^{+0.33}_{-0.11}$ and a more significant improvement $\Delta \chi^2=-5.4$ with 2 extra parameters (with {\it Planck} 2018, BAO, SH0ES 2019 and Pantheon), when ignoring Solar system constraints~\cite{Ballesteros:2020sik}.

Finally, NMC models have been studied in presence of an effective mass in Ref.~\cite{Braglia:2020auw}. In this Early Modified Gravity (EMG), the scalar field, which is frozen during radiation era, grows around the time of recombination driven by the coupling to pressureless matter and is subsequently driven into damped oscillations around its minimum at $\sigma = 0$ by the small effective mass induced by a quartic potential. In addition to providing a better fit to cosmological datasets compared to the $\Lambda$CDM model and EDE models, in the EMG model the positive branch for $\xi >0$ also satisfies automatically the tight constraints on the gravitational constant from laboratory experiments and Solar System measurements on post-Newtonian parameters thanks to the fast rolling of the scalar field towards the bottom of the potential.
For this model with $F(\sigma) = M_{\rm Pl}^2 + \xi\sigma^2$ and 
$V(\sigma) = \Lambda + \lambda \sigma^4/4$, we obtain 
$H_0=(71.00_{-0.79}^{+0.81}){\rm \,km\,s^{-1}\,Mpc^{-1}}$ and $\xi < 0.42$
combining {\it Planck} 2018, BAO and the full shape information from BOSS DR12, SH0ES 2019, 
and Pantheon SNe.

Minimally and non-minimally coupled scalar field models can also potentially decrease the $S_8$ tension.
Minimally coupled models can be classified via their barotropic EoS parameter, $w=P_{\phi}/\rho_{\phi}$. The models with $-1 < w <-1/3$ are referred to as quintessence models, while the models with $w < -1$ are referred to as phantom models. When a coupling between the scalar field $\sigma$ and the Ricci scalar is allowed one has the so-called NMC quintessence models or scalar-tensor theories. In the literature, scalar-tensor theories have been studied as possible solutions to the $S_8$ tension, since under certain circumstances they can produce a reduced growth rate compared to the standard one in the context of GR that seems to be required by different dynamical data~\cite{Macaulay:2013swa,Hildebrandt:2016iqg,Nesseris:2017vor,Kohlinger:2017sxk,Joudaki:2017zdt,DES:2017myr,Kazantzidis:2018rnb,Perivolaropoulos:2019vkb,Kazantzidis:2019dvk,Skara:2019usd,Asgari:2019fkq,Joudaki:2020shz,Heymans:2020gsg}. This behaviour can be attributed either to the reduction of the $\Omega_m$ parameter or to the existence of an evolving Newton's constant giving $G_{\rm eff}\leq G_N$ at low $z$.

Ref.~\cite{Davari:2019tni} showed that the constraints on the $\Omega_m$--$S_8$ plane are affected when considering these classes of models. In particular, while minimally coupled models simply feature a different background evolution, NMC models possess a qualitatively different growth of perturbations as the perturbation equation is modified by replacing Newton constant $G_{\rm N}$ that appears in the Einstein action of GR with  the effective gravitational constant that would be measured in Cavendish-like in the context of ST theories expressed as~\cite{Boisseau:2000pr,Esposito-Farese:2000pbo,Nesseris:2006jc},
\begin{equation}
    \label{eq:Geffsctns}
    G_{\rm eff}=\frac{G_N}{8 \pi F}\left(\frac{2F+4F_{,\sigma}^2}{2F+3F_{,\sigma}^2}\right)\,,
\end{equation}
which is a generalization of Eq. \eqref{eq:EffGNST}.
In Ref.~\cite{Davari:2019tni} it is found that minimally and NMC models give similar constraints on the $\Omega_m$--$S_8$ parameter plane and lower values of $\Omega_m$ as compared to the standard $\Lambda$CDM model. However, this is done considering specific forms of the scalar field potential and non-minimal coupling function, a more general ansatz could relieve the $S_8$ tension.

In the context of an evolving Newton's constant at low $z$, it has been shown that various mechanisms can produce weaker gravity ($G_{\rm eff}< G_N$), such as a scalar-tensor theory in a $w$CDM background with $w>-1$~\cite{Gannouji:2018ncm,Perivolaropoulos:2019vkb,Kazantzidis:2019dvk} as well as more general modified gravity theories involving scalar fields such as Horndeski theories~\cite{Linder:2018jil,Kennedy:2018gtx,Gannouji:2020ylf} and beyond Horndeski theories~\cite{DAmico:2016ntq}. In Refs.~\cite{Nesseris:2017vor,Kazantzidis:2018rnb,Perivolaropoulos:2019vkb,Kazantzidis:2019dvk} the following purely phenomenological parametrization that takes into account the solar system~\cite{Pitjeva:2021hnc,Will:2005va,Uzan:2002vq} and Big Bang Nucleosynthesis constraints~\cite{Alvey:2019ctk,Uzan:2002vq}, has been proposed as a possible solution to the growth tension
\begin{equation}
    \label{eq:geffansanz}
    G_{\rm{eff}}(z,g_a,n)= G_N \left[1+g_a\left(\frac{z}{1+z}\right)^n - g_a\left(\frac{z}{1+z}\right)^{2n} \right]\,,
\end{equation}
where $g_a$ is an extra parameter and $n$ is an integer constrained at $n\geqslant2$ due to solar system tests. It is straightforward to see that for $g_a<0$ one can reproduce the required behavior of the evolving Newton's constant. In fact, it has been shown that in the context of the parametrization in Eq.~\eqref{eq:geffansanz}, negative values of this parameter are supported by RSD~\cite{Nesseris:2017vor,Kazantzidis:2018rnb,Perivolaropoulos:2019vkb, Kazantzidis:2019dvk} and $E_G$ data~\cite{Skara:2019usd}. Another quite interesting alternative parametrization that has the potential to address both the Hubble and growth tensions simultaneously is briefly discussed in Sec.~\ref{sec:LMT}. It includes a late-time abrupt transition of the absolute magnitude $M$ at an ultra low redshift $z_t$ of the following form~\cite{Alestas:2020zol,Marra:2021fvf,Alestas:2021luu}
\begin{equation}
    \label{eq:Mtrans}
    M_B(z) =\left\lbrace \!\!
    \begin{array}{lll}
        M_B^{\rm R20}\,,  & &\text{ if } z \le z_t\,,  \\
        M_B^{\rm R20} + \Delta M_B\,, & &\text{ if }  z> z_t\,,
    \end{array}
    \right.
\end{equation}
possibly coupled with an equally abrupt transition in the dark energy EoS. The transition Eq.~\eqref{eq:Mtrans}, corresponds to a shift in $M_B$ from the Cepheid measurement $M_B^{\rm R20}$ to the $M_B^{\rm P18}$ at $z=z_t$~\cite{Alestas:2020zol,Marra:2021fvf,Alestas:2021luu} and according to Eqs.~\eqref{muztrans} and~\eqref{demu} is induced by a reduction of the ratio $\frac{G_{\rm eff}}{G_N}$. This condition leads immediately to $G_{\rm eff}<G_N$ in the low redshift regime, alleviating as a result the $S_8$ tension.

\paragraph{Generalized Brans-Dicke Theories.}
\label{sec:BDLCDM} 
Brans-Dicke (BD) theory~\cite{BransDicke1961,Dicke1962} represents the first historical attempt of incorporating a dynamical gravitational coupling to GR. It is the simplest scalar-tensor theory which can be embedded within the broader class of Horndeski models~\cite{Horndeski:1974wa}. Apart from the metric tensor, the BD theory involves a new (scalar) degree of freedom $\varphi$ that controls the strength of the gravitational interaction: $G=G_N/\varphi$, with $G_N$ the (locally measured) Newton constant. The scalar field is non-minimally coupled to gravity via the Ricci scalar and its dynamics is controlled by a new dimensionless parameter $\omega_{\rm BD}$ (or equivalently $\epsilon_{\rm BD}\equiv 1/\omega_{\rm BD}$). The action can be written, in the Jordan frame, as
\begin{equation}
\label{BD:action}
S=\int {\rm d}x^4 \sqrt{-g}\left[\frac{1}{16\pi}\left(\varphi R-\frac{\omega_{BD}}{\varphi}g^{\mu \nu}\partial_\mu \varphi \partial_\nu \varphi\right)-\rho_\Lambda\right]+\int {\rm d}x^4 \sqrt{-g}\mathcal{L}_m (\chi_i, g_{\mu \nu}),
\end{equation}
where $\rho_\Lambda$ is the energy density (constant) of the usual vacuum energy.\footnote{It is possible to consider generalizing the original BD theory by introducing a potential $V(\varphi)$ for the scalar (Jordan) field $\varphi$~\cite{Faraoni:2009km}. Two exceptions are particularly important, as they provide us with the simplest BD extensions of the $\Lambda$CDM model. The case with a constant potential, $V(\varphi)=\rm const$, which, within the original BD theory, corresponds to including the usual vacuum energy (described by  $p=-\rho$) as a source, is the one considered here, see the action given in Eq.~\eqref{BD:action}. The other option, the case with a potential proportional to the scalar field, $V(\varphi)\propto\varphi$, corresponds to the inclusion of a bare cosmological constant along with the Ricci scalar, viz., $\varphi R\rightarrow \varphi(R-2\Lambda)$, in the original BD action~\cite{Uehara:1981nq,Boisseau:2010pd}. This type of BD extension of $\Lambda$CDM has been recently investigated in light of observational data~\cite{Akarsu:2019pvi}. It was found that it exhibits no significant deviations from $\Lambda$CDM all the way to the BBN epoch, and does not help with the $H_0$ tension.} The last term stands
for the matter action $S_m$, constructed from the Lagrangian density of the matter fields, denoted by $\chi_i$. The wave equation reads $\Box{\varphi} = 8\pi{G_N}T/(3 + 2\omega_{\rm BD})$, with $T$ the trace of the total energy-momentum tensor. GR is recovered by considering the simultaneous limits $\varphi\rightarrow 1 $ {\it and} $\epsilon_{\rm BD}\rightarrow 0$, i.e.\ by suppressing the dynamics of $\varphi$ and matching its constant value with $G_N$. It follows from the
action that only the first condition is needed to actually get
GR$+\Lambda$. The second condition by itself guarantees that BD goes to
GR in the weak field regime, but not for higher orders in the parametrization of post-Newtonian 
expansion, see for instance~\cite{Faraoni:2019sxw}.

\begin{figure}[t!]
\centering
\begin{subfigure}
  \centering
  \includegraphics[width=0.44  \linewidth]{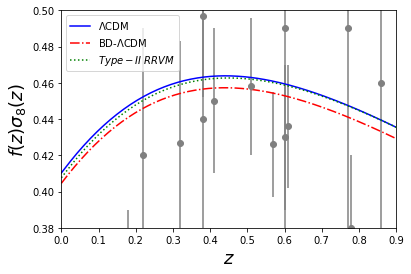}
  \label{fig:fs8_BDTII}
\end{subfigure}
\begin{subfigure}
  \centering
 \includegraphics[width=0.31\linewidth]{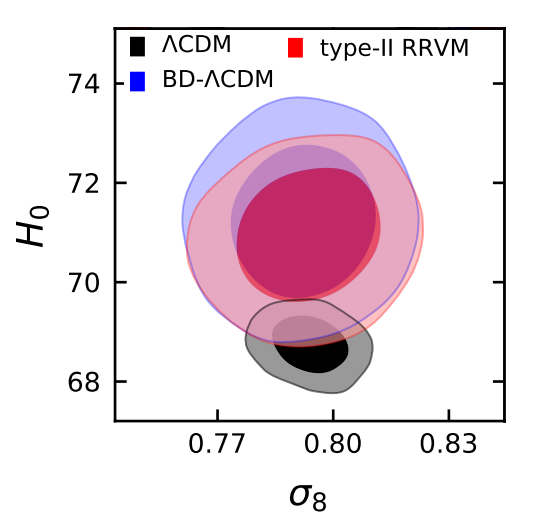}
  \label{fig:RRVMII_BD}
\end{subfigure}
\caption{{\it Left panel}\/: Theoretical curves of $f(z)\sigma_8(z)$ for the GR-$\Lambda$CDM, type-II RRVM and BD-$\Lambda$CDM, together with the data points used in the fitting analysis of Ref.~\cite{SolaPeracaula:2021gxi}; {\it Right panel}\/: $1$ and $2\sigma$ CL regions in the ($\sigma_8$--$H_0$)-plane for the same models. The type-II RRVM and BD-$\Lambda$CDM are able to alleviate the $H_0$ tension without worsening the $\sigma_8$ one. See also Sec.~\ref{sec:RRVM} and Refs.~\cite{SolaPeracaula:2021gxi,Sola:2019jek,Sola:2020lba}. $H_0$ is expressed in units of ${\rm \,km\,s^{-1}\,Mpc^{-1}}$.}
\label{RRVM2_BD}
\end{figure}

We denote by BD-$\Lambda$CDM the Brans-Dicke counterpart of the standard $\Lambda$CDM, i.e.\ the BD model with the usual vacuum energy. The latter was not present in the original BD theory and is needed to trigger the late-time acceleration of the Universe. This theory has been recently tested in the light of a large string of cosmological data~\cite{Sola:2020lba}: SNIa+BAO+$H(z)$+CMB+LSS, where $H(z)$ may include or not the SH0ES prior on $H_0$. Here we discuss only the results obtained including the latter, see Fig.~\ref{RRVM2_BD}. We find that the BD-$\Lambda$CDM model is capable of reducing the $H_0$ and $\sigma_8$ tensions to an inconspicuous level of $\sim 1.5 \sigma$ (when the high-$l$ polarization and lensing data from {\it Planck} 2018 are not considered)~\cite{Sola:2020lba}. This is pretty similar to the outcome of type-II RRVM, which the BD-$\Lambda$CDM model mimics,  see Sec.~\ref{sec:RRVM} and Ref.~\cite{SolaPeracaula:2021gxi}. The BD-$\Lambda$CDM fits the cosmological data significantly better than the standard GR-$\Lambda$CDM model (based on GR), and this is confirmed by different statistical criteria (see Ref.~\cite{Sola:2019jek, Sola:2020lba,SolaPeracaula:2021gxi} for details). Values of the effective cosmological gravitational strength $\sim 7-9\%$ larger than $G_N$ are preferred at $\sim 3\sigma$ CL. This is possible thanks to the fact that $\varphi$ remains below 1 throughout the entire cosmic evolution while keeping the matter and radiation energy densities very close to the typical GR-$\Lambda$CDM values. This leads to higher $H(z)$ values during all the stages of the cosmic history without changing dramatically the abundances of the matter species in the pre- and post-recombination epochs. The lowering of the sound horizon at the baryon-drag epoch, $r_d$, is accompanied by an increase of the Hubble parameter. This helps to alleviate the $H_0$ tension (cf. Fig.~\ref{RRVM2_BD}, right plot) in perfect accordance with other relevant background data sets, such as e.g.\ BAO and SNIa. In order not to spoil the correct fit of the CMB temperature data the model needs to accommodate values of the spectral index $n_s$ considerably larger than the GR-$\Lambda$CDM. This causes no problem since the dynamics of $\varphi$ can compensate the changes in the matter power spectrum introduced by this fact (cf. Sec. 3 of Ref.~\cite{Sola:2020lba}). Remarkably, the model is able at the same time to cut back the $\sigma_8$ tension (or $S_8$, if preferred), as can also be appraised in the right plot of Fig.~\ref{RRVM2_BD}. The loosening of this tension can be attained through a negative value of $\epsilon_{\rm BD}$, or a positive one if sufficiently massive neutrinos are included~\cite{Sola:2020lba}. In both cases values  $|\epsilon_{\rm BD}|\lesssim\mathcal{O}(10^{-3})$ are found. This parameter has a direct impact on the LSS through the linear perturbation equations. For instance,  for $\epsilon_{\rm BD}<0$ the friction term in the equation of the matter density contrast grows, and the source term decreases. Both effects contribute to the lowering of $\sigma_8$. The compatibility between the larger cosmological value obtained for the gravitational coupling  and the value measured locally, $G_N$, should be possible provided one can find an appropriate screening mechanism. Note, the model appears to be consistent with the findings of Refs.~\cite{Heisenberg:2022lob, Heisenberg:2022gqk}, once a screening mechanism is invoked. Different studies show that these mechanisms are possible, although other works find that their implementation may not be so straightforward~\cite{Gomez-Valent:2021joz}.\footnote{The extensions of the BD theory may lead to modifications in the limits on $\epsilon_{\rm BD}$ from solar system observations, see~\cite{Perivolaropoulos:2009ak}.}

As pointed out above, the BD-$\Lambda$CDM model mimics the type-II RVM (cf. Sec.~\ref{sec:RRVM}), and hence it is perceived from the GR perspective as a running vacuum model with a (very) mild dynamical gravitational coupling~\cite{SolaPeracaula:2018dsw,deCruzPerez:2018cjx,Sola:2020lba}. The effective RVM-behaviour of the BD-$\Lambda$CDM can be made apparent if one writes the Friedmann and  pressure equations in GR-like fashion, from which one can identify the characteristic RVM form of the vacuum energy density: $\rho_{\rm DE}\sim A+B\Delta\varphi H^2$, with  $A$ and $B$ constants and $\Delta\varphi$ tracking the departure of the effective gravitational coupling from the local $G_N$ value~\cite{Sola:2020lba}. In Refs.~\cite{Sola:2019jek, Sola:2020lba} it is shown that the effective EoS of the combined fluid of $\varphi$ and usual vacuum energy mimics quintessence at more than $3\,\sigma$ CL. This feature could be the smoking gun signaling the departure from GR.

\paragraph{A Word of Caution.} Using Horndeski models to address the $H_0$ and/or $S_8$ tensions comes with a significant caveat, however. The general form of the Lagrangian may be strongly constrained by requiring that the ensuing equation of motion admits a well-posed initial value formulation. In the strongly coupled regime, this issue was studied in~\cite{Papallo:2017qvl}, while the weakly coupled regime was considered in Ref.~\cite{Kovacs:2020ywu}. In the latter, some examples of loss of uniqueness through shock formation and evolution with change in character of the equation of motion (from hyperbolic to elliptic through a parabolic stage) were presented in~\cite{Bernard:2019fjb}.

\subsubsection{Quantum Conformal Anomaly Effective Theory and Dynamical Vacuum Energy}
\label{sec:Anom} 

Classical General Relativity receives corrections from the quantum fluctuations of massless conformal fields, through the conformal trace anomaly of the energy-momentum tensor in curved space~\cite{Duff:1977ay,Birrell:1982ix}
\begin{equation}
    \big\langle \hat T^a_{\ \ a} \big\rangle\  =  \ b\, C^2 + b'\, \left( E - \tfrac{2}{3} \sq R\right) \equiv \ {\cal A}\,.
\end{equation}
This leads to additional gravitational interactions at macroscopic distance scales~\cite{Mottola:2006ew,Giannotti:2008cv,Mottola:2010gp,Mottola:2011ud}, as becomes clear from the local form of the effective action of the conformal anomaly,
\begin{equation}
    \label{Sanom}
    S_{\cal A}[g; \varphi]=  \frac{b'\!}{2}\int\!\!{\rm d}^4\!x\sqrt{-g}\,\bigg\{\!-\left(\sq \varphi\right)^2  + 2\,\Big(R^{ab}\! - \!\tfrac{1}{3} Rg^{ab}\Big)\,(\nabla_{\! a} \varphi)\,(\nabla_{\! b}\varphi)\bigg\}  + \frac{1}{2} \!\int\!\!{\rm d}^4\!x\,\sqrt{-g}\,{\cal A}\,\varphi\,,
\end{equation}
in terms of an additional weakly coupled long range scalar field $\varphi$~\cite{Antoniadis:1991fa,Mazur:2001aa,Mottola:2010gp,Mottola:2016mpl,Coriano:2017mux}. This "conformalon" scalar $\varphi$ allows vacuum energy to change, leading to a dynamical cosmological dark energy, with potentially observable effects imprinted in the CMB, GWs, and the growth of Large Scale Structure (LSS)~\cite{Antoniadis:1996dj,Antoniadis:2006wq,Mottola:2011ud,Antoniadis:2011ib}.

Specifically, by recognizing that the linear coupling in the anomaly action involves the topological Euler density $E$, and $\sq R$, both of which are total derivatives, $\left(E - \tfrac{2}{3}\sq R\right) \propto \varepsilon^{abcd}\nabla_d A_{abc}$ naturally defines a $3$-form Abelian potential $A_{bcd}$. Hence its linear coupling to the scalar $\varphi$ in (\ref{Sanom}) can be written
\begin{equation}
\frac{\,b'\!}{2}\! \int \!{\rm d}^4x \sqrt{-g}\, \left(E - \tfrac{2}{3}\sq R\right)\,\varphi = -\frac{b'\!}{2}\, \frac{1}{3!\!} \int\!{\rm d}^4x \sqrt{-g} \,\varepsilon^{abcd}\,\nabla_{\! d} \varphi \,A_{abc}\,  \equiv -\frac{b'}{2\times 3!}\int\!{\rm d}^4x \sqrt{-g} \, J^{abc} A_{abc}\,,
\label{linterm}
\end{equation}
by an integration by parts. The totally anti-symmetric tensor $F_{abcd} = \nabla_{[a} A_{bcd]}$ is the $4$-form field strength corresponding to $A_{abc}$. If the "Maxwell" term 
\begin{equation}
S_{F} = -\frac{1}{2\times 4!\ \kappa^4\!} \int {\rm d}^4x \sqrt{-g} \ F_{abcd}F^{abcd}\,,
\label{Maxw}
\end{equation}
for this field strength is added to the effective action, where $\kappa$ is a coupling constant with dimensions of mass, then the "Maxwell" equation $\nabla_d F^{abcd} =0$ in the absence of any sources implies that $F_{abcd}  = \varepsilon_{abcd} \widetilde F$ is a constant. Since the stress-tensor for this constant $4$-form Maxwell field matched to the number of $4$ spacetime dimensions is $-\widetilde F^2 g_{ab}/2\kappa^2$, (\ref{Maxw}) is entirely equivalent to the addition of a positive cosmological constant $\Lambda$ term to the classical Einstein-Hilbert action~\cite{Aurilia:1978qs,Brown:1988kg,Aurilia:2004cb}--in the absence of sources.

With the $J\cdot A$ coupling (\ref{linterm}) from the anomaly, the Maxwell eq.~acquires a source, and becomes
\begin{equation}
\nabla_{\! d} F^{abcd} =  -\frac{b'\kappa^4\!}{2}\,J^{abc} = -\frac{b'\kappa^4\!}{2}\,\varepsilon^{abcd}\, \nabla_{\! d}\varphi,
\label{Maxeq}
\end{equation}
with the derivative of the scalar conformalon $\varphi$ acting as current source. Hence $\widetilde F = -b'\kappa^4 \varphi/2 + $const, and the value of the cosmological "constant" vacuum energy, determined by the value of $\widetilde F^2$ can change as the scalar field $\varphi$ does.

This effective field theory of the conformal anomaly (\ref{Sanom}) coupled to the $3$-form potential and $4$-form Maxwell field strength term is different from either scalar-tensor theories, or other modified gravity theories of the Hordenski type, but is well-founded on fundamental quantum field theory principles. The fluctuations of $\varphi$ can lead to the scale invariance of the CMB being promoted to full conformal invariance, with the implication of a well-defined non-Gaussian bispectrum differing from slow roll single field inflation models~\cite{Antoniadis:1996dj,Antoniadis:2011ib}. 

The scalar $\varphi$ also couples to the conformal part of the metric and leads to  scalar "breather" mode polarization GWs, which should be produced in the early Universe~\cite{Mottola:2016mpl}. Since the vacuum energy can vary in general in both space and time, this effective theory also allows the possibility of spatially inhomogenous non-FLRW cosmologies. A detailed study of dynamical dark energy based on this theory is now just beginning, in order to derive and test its predictions against the trove of LSS and other observational data expected in this decade.

\subsubsection{Ultra-Late Time Gravitational Transitions}
\label{sec:LMT}

The extremely tight constraints on the form of $H(z)/H_0$ in the range $z\in [0.01,1100]$~\cite{Lemos:2018smw,Efstathiou:2020wxn} combined with the local measurement of $H_0$, namely $H^{\rm R20}_0 = (73.2 \pm 1.3){\rm\,km\,s^{-1}\,Mpc^{-1}}$~\cite{Riess:2009pu}, which is inconsistent with the CMB constraint $H^{\rm P18}_0 = (67.36 \pm 0.54){\rm\,km\,s^{-1}\,Mpc^{-1}}$~\cite{Planck:2018vyg} 
implies that a possible transition event may have taken place either just before $z=1100$, see Sec.~\ref{early}, or just after $z=0.01$ (gravitational transition~\cite{Marra:2021fvf}). Such a transition event would leave intact the standard model physics in the range $z\in [0.01,1100]$ while introducing new physics before or after this redshift (time) range. In this subsection we briefly describe the latter possibility.

The absolute luminosity of SNIa may be calibrated using two distinct independent calibrators:
\begin{enumerate}
    \item The sound horizon at recombination ($z\sim 1100$) used as a standard ruler to set the scale $H^{\rm P18}_0$ for the $\Lambda$CDM $H(z)$ and thus the SNIa absolute magnitude $M_B=M_B^{\rm P18}$\cite{Camarena:2019rmj,Camarena:2021jlr} and luminosity (inverse distance ladder approach).
    \item The locally calibrated  standard candles like Cepheid stars which measure the distance to SNIa host galaxies at low redshifts ($z<0.01$, $D<40$Mpc) and thus calibrate the absolute luminosity $L^{R20}$ and magnitude ($M_B=M_B^{\rm R20}$) of SNIa leading to the best fit value $H^{\rm R20}_0$~\cite{Riess:2009pu} (distance ladder approach).
\end{enumerate}
The two approaches lead to values of the SNIa absolute magnitude $M_B$ that are inconsistent at a level more than $4\sigma$ with $\Delta M_B=M_B^{\rm R20}-M_B^{\rm P18}\simeq-0.2$~\cite{Camarena:2021jlr}. This inconsistency is the basis of the Hubble tension because the difference in the values of $M_B$ is easily translated to a difference in the values of the Hubble constant, since $H_0$ and $M_B$ are degenerate~\cite{Perivolaropoulos:2021bds,Theodoropoulos:2021hkk} (observable for $z>0.01$ using SNIa):
\begin{equation}
\mathcal{M}=M_B+5\log_{10}\left[\frac{c/H_0}{\rm Mpc}\right]+25 \,,
\label{combM}
\end{equation} 
which implies that a shift of $M_B$ of
\begin{align}
    \label{rel}
    \Delta M_B \equiv M_B^{\rm P18} - M_B^{\rm R20} \approx 5 \log_{10} \frac{H^{\rm P18}_0}{H^{\rm R20}_0} \approx - 0.2 \,,
\end{align}
is needed for a decrease of the locally (SNIa) measured $H_0$ by about $10\%$ and thus a resolution of the Hubble tension.

The two measurements of $M_B$ however (CMB sound horizon and local calibrators/SNIa) are performed at different redshift ranges: the Cepheid measurement $M_B^{\rm R20} = -19.244 \pm 0.037$ is performed at $z<0.01$ where Cepheids and other local distance calibrators are detectable while the measurement $M_B^{\rm P18} = - 19.401 \pm 0.027$ using the sound horizon is performed at $z>0.01$ where the Hubble flow and the $\Lambda$CDM $H(z)$ are applicable. Thus, the two measurements can be made consistent if a fundamental physics transition changes the SNIa absolute luminosity (or equivalently $M_B$) at a redshift $z_t \lesssim 0.01$~\cite{Marra:2021fvf}. 

An abrupt step-like shift of Newton's constant $G_{\rm eff}$ in the context of a gravitational spatial or temporal transition could play the role of the required fundamental physics transition~\cite{Alestas:2020zol,Marra:2021fvf,Alestas:2021luu}. In this context we have 
\begin{equation}
\mu_G(z)\equiv \frac{G_{\rm eff}}{G_N} =(1+\Delta \mu_G \; \Theta (z-z_t))\,,
\label{muztrans}
\end{equation}
where $G_N$ is the locally measured Newton's constant. A very recent false vacuum decay~\cite{Coleman:1977py,Callan:1977pt,Patwardhan:2014iha} in the context of a scalar tensor theory~\cite{Lee:2006vka} with vacua energy scale $M_\Lambda \sim 0.002\,$eV and with the present horizon scale would naturally produce true vacuum bubbles on scales of~\cite{Perivolaropoulos:2021bds}
\begin{equation}
    R_b \simeq \frac{\ln(M_{\rm Pl}/ M_{\Lambda})}{4 H_0}\sim (15\textrm{--}20){\rm  \,Mpc}\,.    
\end{equation}
Thus, if we were inside such a true vacuum bubble we would observe a transition of $G_{\rm eff}$ at a distance of about $20\,$Mpc. Surprisingly, such a transition is consistent with current $G_{\rm eff}$ constraints~\cite{Uzan:2010pm,PhysRev.73.801} (which constrain mainly the time derivative of $G_{\rm eff}$ and much less a $G_{\rm eff}$ abrupt transition). In fact, hints for such a transition have recently been pointed out in Cepheid~\cite{Perivolaropoulos:2021bds,Mortsell:2021nzg} and Tully-Fisher data~\cite{Alestas:2021nmi}. It has also been shown that such a transition is consistent with the observed number of galaxies per redshift bin in low redshift galaxy surveys~\cite{Alestas:2022xxm}.

Assuming a SNIa absolute luminosity $L\sim G_{\rm eff}^{-\alpha}$~\cite{Amendola:1999vu,Garcia-Berro:1999cwy}, where $\alpha$ is a parameter determined by the detailed physics of SNIa explosions, it is found that the required transition of $M_B$ is connected with a gravitational transition of $G_{\rm eff}$ as
\begin{align}
    \Delta M_B = - \frac{5}{2} \log_{10}\left( \frac{L^{\rm P18}}{L^{\rm R20}}\right) = \alpha \frac{5}{2} \log_{10} \mu_G^> \,,
\end{align}
where $L^{\rm P18}$ and $L^{\rm R20}$ are the CMB-calibrated and Cepheid-calibrated SNIa absolute luminosities, respectively. Since the Chandrasekhar mass $M_C$, which plays a crucial role in the SNIa explosion, is $M_C\sim G_{\rm eff}^{-3/2}$, it is clear that the simplest assumption is that $L\sim M_C\sim G_{\rm eff}^{-3/2}$ which leads to $\alpha=3/2$. In the context of this simplest generic assumption we have 
\begin{align} \label{demu}
\Delta \mu_G = 10^{\frac{4}{15} \Delta M_B} -1\simeq -0.1\,, 
\end{align}
which is marginally consistent with current constraints on a possible transition of the Newton constant and implies the presence of weaker gravity by about $10\%$ at times earlier than the transition time corresponding to $z_t$. Such weaker gravity would lead to weaker growth rate of cosmological matter fluctuations $\delta(z) \equiv \frac{\delta \rho}{\rho}(z)$ according to the dynamical linear growth equation~\cite{Skara:2019usd,Kazantzidis:2018rnb,Kazantzidis:2019dvk}:
\begin{equation}
    \delta" \!+\! \left[ \frac{(H^2)'}{2~H^2} \!-\! \frac{1}{1+z}\right]\delta' \approx  \frac{3 H_0^2}{2 H^2} (1+z) \, \mu_G(z) \, \Omega_m  \, \delta \,,
\label{eq:odedeltaz}
\end{equation}
where a prime denotes a derivative with respect to redshift $z$. In fact, a $10\%$ reduction of $\mu_G$ would fit the same data with a 10\% larger value for $\Omega_m$. In the context of a gravitational transition with a $\Lambda$CDM background $H(z)$, RSD and WL data are well fit by~\cite{DES:2017rfo,Skara:2019usd,Alestas:2020mvb}
\begin{equation}
    \Omega_m^{\rm growth}=0.256^{+0.023}_{-0.031} = (1+\Delta \mu_G)\, \Omega_m^{\rm P18}\,,    
\end{equation}
which implies $\Delta \mu_G=-0.19\pm 0.09$ when setting $\Omega_m^{\rm P18}=0.3153\pm 0.0073$~\cite{Planck:2018vyg} and leads to a mechanism for the resolution of the growth $S_8$ tension.

An ultra-late gravitational transition could also have implications on geo-chronology and solar-system chronology addressing corresponding long standing puzzles. For example observational evidence from terrestrial and lunar cratering rates indicates that the impact flux of kilometer sized objects increased by at least a factor of 2 over that last $100\,$Myrs compared to the long term average~\cite{Alestas:2021luu}. It has been demonstrated that such an increase could be explained in the context of the ultra-late gravitational transition required for the resolution of the Hubble and $S_8$ tensions.

\subsection{Specific Solutions Assuming FLRW}

In this subsection, we list some specific solutions of interest to the community that cannot be classified in the previous sections, whether because they are broad enough to enclose very different scenarios, or they simply have an impact along the whole cosmic history. We refer the reader to the more complete review~\cite{DiValentino:2021izs} or the original papers for a more detailed discussion of proposed solutions of the Hubble tension in: inflationary models~\cite{Tram:2016rcw,DiValentino:2016ziq,Rodrigues:2020dod,Rodrigues:2021txa}, self-interacting dark matter~\cite{Binder:2017lkj}, \"{U}ber-gravity~\cite{Khosravi:2017hfi}, Nonlocal gravity~\cite{Belgacem:2017cqo}, Modified Gravity~\cite{Lin:2018nxe,Benetti:2018zhv}, unified fluid~\cite{Yang:2019jwn}, quintessential models~\cite{DiValentino:2019exe}, screened fifth forces~\cite{Desmond:2019ygn}, Generalized Chaplygin gas models~\cite{Yang:2019nhz,Benetti:2021div,Benetti:2019lxu,Salahedin:2020emw}, Viscous Generalized Chaplygin gas~\cite{Hernandez-Almada:2021osl}, addition of extra SNIa systematic uncertainties with a varying DE~\cite{Martinelli:2019krf}, metastable DE~\cite{Shafieloo:2016bpk,Li:2019san,Yang:2020zuk}, Super-$\Lambda$CDM~\cite{Adhikari:2019fvb,Adhikari:2022moo}, DM-photon coupling~\cite{Yadav:2019jio}, Local Inhomogeneity~\cite{Kasai:2019yqn}, Interaction in the anisotropic Universe~\cite{Amirhashchi:2020qep}, Heisenberg uncertainty~\cite{Capozziello:2020nyq, Spallicci:2021kye}, Enhanced Early gravity model~\cite{Zumalacarregui:2020cjh, Joudaki:2020shz}, CMB monopole temperature~\cite{Ivanov:2020mfr}, Decaying Ultralight Scalar~\cite{Gonzalez:2020fdy}, Decaying coupled fermions~\cite{Benisty:2019pxb}, Early mass varying neutrino DE~\cite{Gogoi:2020qif}, Neutrino-majoron interactions~\cite{Escudero:2019gvw,EscuderoAbenza:2020egd,Escudero:2021rfi}, Light gravitino scenarios~\cite{Choi:2019jck}, Degenerate decaying fermion DM~\cite{Choi:2020tqp}, Lifshitz cosmology~\cite{Berechya:2020vcy}, a frame dependent, scale invariant DE~\cite{Adler:2019fnp}, light gravitino dark matter with a small non-thermal fraction~\cite{Gu:2020ozv}, cosmology with a Bianchi type-I metric~\cite{Akarsu:2019pwn}, Finslerian models~\cite{Wang:2018est}, Dark Energy model from Generalized Proca Theory~\cite{Geng:2021jso}, Axi-Higgs Universe~\cite{Fung:2021fcj,Luu:2021yhl}, transitional dark energy~\cite{Zhou:2021xov}, interaction between dark matter and the baryonic component in a modified gravity scenario~\cite{Adil:2021zxp}, Minimal Theory of Massive Gravity~\cite{deAraujo:2021cnd}, Bound Dark Energy~\cite{delaMacorra:2021hoh}, Chameleon Early Dark Energy~\cite{Karwal:2021vpk}, interacting dark energy axions~\cite{Mawas:2021tdx}, the mirror twin Higgs model~\cite{Bansal:2021dfh}, Milgromian gravity~\cite{Banik:2021woo}, a quantum origin for the DE~\cite{Belgacem:2021ieb}, Emergent DE from unparticles~\cite{Artymowski:2020zwy,Artymowski:2021fkw}, General Relativistic Entropic Acceleration theory~\cite{Arjona:2021uxs}.

\subsubsection{Active and Sterile Neutrinos} 

The existence of massive neutrinos (either active or sterile) has been confirmed by various observational surveys~\cite{KamLAND:2004mhv,Super-Kamiokande:2005mbp,KamLAND:2008dgz} and their presence can affect the cosmological parameters (see Refs.~\cite{Lesgourgues:2006nd,Conrad:2013mka} for more discussions on the massive neutrinos). According to the observational surveys, the sum of the active neutrino masses, $\sum m_{\nu}$, must be at least $0.06\,$eV~\cite{Gonzalez-Garcia:2012hef}. Additionally, some observational experiments indicate that there could be a sterile neutrino which does not interact with the standard model~\cite{Conrad:2013mka}, but it still contributes a mass $m_{\nu, {\rm sterile}}^{\rm eff}$ and in a model dependent way, an increase in the total number of effective relativistic degrees of freedom, $N_{\rm eff}= 3.046 + \Delta N_{\rm eff}$. It has been found that the active and sterile neutrinos in the Universe sector can be effective to alleviate the $S_8$ tension~\cite{Battye:2013xqa,Boehringer:2016bzy}.
A Pseudo-Dirac sterile neutrino has been instead proposed as a model to alleviate the Hubble tension in Ref~\cite{Chao:2021grp}, as well as an extended parameter space with ground-based CMB data and a total active neutrino mass different from zero~\cite{DiValentino:2021imh}.

\subsubsection{Cannibal Dark Matter} 
\label{sec:cannibalDM}

Cannibalistic dark matter models were initially proposed by Refs.~\cite{Dolgov:1980uu,Carlson:1992fn,Dolgov:2017ujf} in the context of general self-interacting dark matter~\cite{Dolgov:1995rm,deLaix:1995vi,Machacek:1994abc}. Such models have received a resurgence of interest within the last decade~\cite{Hochberg:2014kqa,Kuflik:2016abc,Pappadopulo:2016pkp,Farina:2016llk,Erickcek:2020wzd} given their promising ability to impact both the small-scale crisis~\cite{Spergel:1999mh,deBlok:2009sp,Boylan-Kolchin:2011qkt,Salucci:2018hqu} and the $S_8$ tension~\cite{Buen-Abad:2018mas,Heimersheim:2020aoc}.

If the Lagrangian of a massive dark matter candidate permits $2\leftrightarrow3$ interactions, such as in Refs.~\cite{Bernal:2015ova,Chu:2017msm,Bernal:2018hjm,Heeba:2018wtf}, it is possible that it exhibits a cannibalistic behavior. When the species' thermal energy is high compared to its mass, both the $3 \to 2$ and $2 \to 3$ interactions will be balanced, and the dark matter will behave as an ordinary relativistic species. However, as the species cools and the thermal energy becomes insufficient to overcome the mass gap inherent in the $2 \to 3$ interaction, a cannibalistic phase begins, in which the mass energy of particles is converted through an efficient $3 \to 2$ interaction into thermal energy. As soon as the $3 \to 2$ interaction freezes out (when the number density is sufficiently diluted due to the expansion of the Universe), the species continues to behave like ordinary cold dark matter. In its cannibalistic regime the species is close to (but not quite) relativistic for an extended time, allowing the species to keep a large Jeans scale (suppressing growth on small scales) while avoiding coupling to other relativistic species through the Einstein equations (and thus impacting CMB observables). As such, the cosmological evolution is very similar to that of warm dark matter, except for the intermediate cannibalistic phase, which allows the model to have a reduced impact on the CMB while still strongly suppressing the power spectrum.

While the initial investigations~\cite{Buen-Abad:2018mas,Heimersheim:2020aoc} in terms of cosmological evolution have shown a promising parameter region, more work is required both at the level of theoretical modeling and complimentary experimental constraints. In particular, joint investigations of constraints from CMB, weak lensing, Ly-$\alpha$ data, and/or small scale probes of clustering would be required to further investigate the validity of this model.

\subsubsection{Decaying Dark Matter}
\label{sec:dDM}

The $\Lambda$CDM model postulates the additional form of Dark Matter (DM) which clusters gravitationally but is (almost) immune to other interactions. The nature of DM remains elusive so far. From a theoretical point of view, DM can be easily multi-component itself.\footnote{Idea of multi-component DM is inspired by the apparent complexity of the "visible" sector. For instance, the current cosmological data resolve at least three different components: photons, neutrinos, and baryons. It seems to be natural that DM which dominates the matter content of the Universe also consists of several different species.}

One interesting approach that provides a composite content of DM is decaying DM (dDM).\footnote{Not to be confused with the "Dynamical Dark Matter" (DDM) model discussed in Sec.~\ref{sec:dynDM}.}  The motivation of this scenario arises in purely theoretical considerations~\cite{Ibarra:2013cra}, it also can explain some interesting experimental results~\cite{Ibarra:2013zia,Cirelli:2012ut,Anchordoqui:2015lqa,Lovell:2014lea,Iakubovskyi:2014yxa} and address the small-scale problems of the cold DM paradigm~\cite{Wang:2014ina,Cheng:2015dga}. The decaying DM model was developed in the 1980s~\cite{Flores:1986jn,Doroshkevich:1989bf} and has recently gained renewed interest because of the appearance of persistent tensions between low-redshift measurements and predictions of the $\Lambda$CDM cosmology. 

There exist several varieties of the decaying DM proposal. DM can be a single unstable component decaying into invisible radiation with a lifetime exceeding the age of the Universe~\cite{Audren:2014bca}. This scenario was suggested as a resolution of the $S_8$ tension, however, the current cosmological data is not constraining enough to unambiguously favor a non-zero DM decay rate~\cite{Enqvist:2015ara}. Alternatively, the unstable DM may decay to the massive particles and invisible radiation at late times~\cite{Aoyama:2014tga,DelNobile:2015bqo} or even undergo a many-body decay~\cite{Blackadder:2014wpa,Blackadder:2015uta}. However, this decaying DM scenario is unlikely to alleviate the $H_0$-tension~\cite{Haridasu:2020xaa,DES:2020mpv}.\footnote{The somewhat related model in which self-interacting dark matter may be converted into a non-interacting form of radiation with rate proportional to the cosmic expansion~\cite{Bjaelde:2012wi,Pandey:2019plg} is also disfavored by data~\cite{Liu:2021mkv}.} More progress towards reconciling discrepancies between different cosmological measurements can be achieved if one assumes an earlier decay of the DM constituents. 

In Ref.~\cite{Berezhiani:2015yta} it was argued that the subdominant DM component decaying after recombination may alleviate the cosmological tensions. This decaying DM setup has two extra parameters with respect to the $\Lambda$CDM scenario -- a fraction of decaying component in the total DM abundance, $f_{\rm dDM}=\omega_{{\rm dDM},i}/(\omega_{{\rm dDM},i}+\omega_{{\rm CDM},i})$ where $\omega_i\equiv\Omega_ih^2$ denotes {\it initial} densities of decaying and cold DM components, and its inverse lifetime, or width, $\Gamma$. The background dynamics for unstable DM ($\rho_{\rm dDM}$) and dark radiation ($\rho_{\rm DR}$) abundances is driven by
\begin{eqnarray}
    \label{ddm}
    \dot{\rho}_{\rm dDM} + 3H\rho_{\rm dDM} &=& -\Gamma\rho_{\rm dDM}\,,\\
    \dot{\rho}_{\rm DR} + 4H\rho_{\rm DR} &=& \Gamma\rho_{\rm dDM}\,.
\end{eqnarray}
A slightly reduced matter content in the late Universe helps accommodate a low $S_8$ in full accordance with local probes of Large Scale Structure (LSS). Remarkably, the downward shift in the total matter density must be accompanied by the upward shift of $H_0$ to keep the angular acoustic scale of the CMB intact. However, the lack of DM at low redshifts reduces the power of the CMB lensing effect which is in odds with the {\it Planck} observations. Thus, the CMB data alone put the stringent limits on this proposal, $f_{\rm dDM} < 4\,\%\,(2\sigma)$~\cite{Chudaykin:2016yfk,Poulin:2016nat} from the {\it Planck} 2015 data and $f_{\rm dDM}<2\,\%\,(2\sigma)$~\cite{Nygaard:2020sow} using the newest {\it Planck} 2018 likelihood. Including the BAO measurements from BOSS DR12 induces even tighter constraints on the decaying DM fraction by a factor of $\sim1.3-1.5$~\cite{Chudaykin:2017ptd,Nygaard:2020sow}. 
\begin{figure*}[!t]
    \centerline{
    \includegraphics[width=0.43\textwidth]{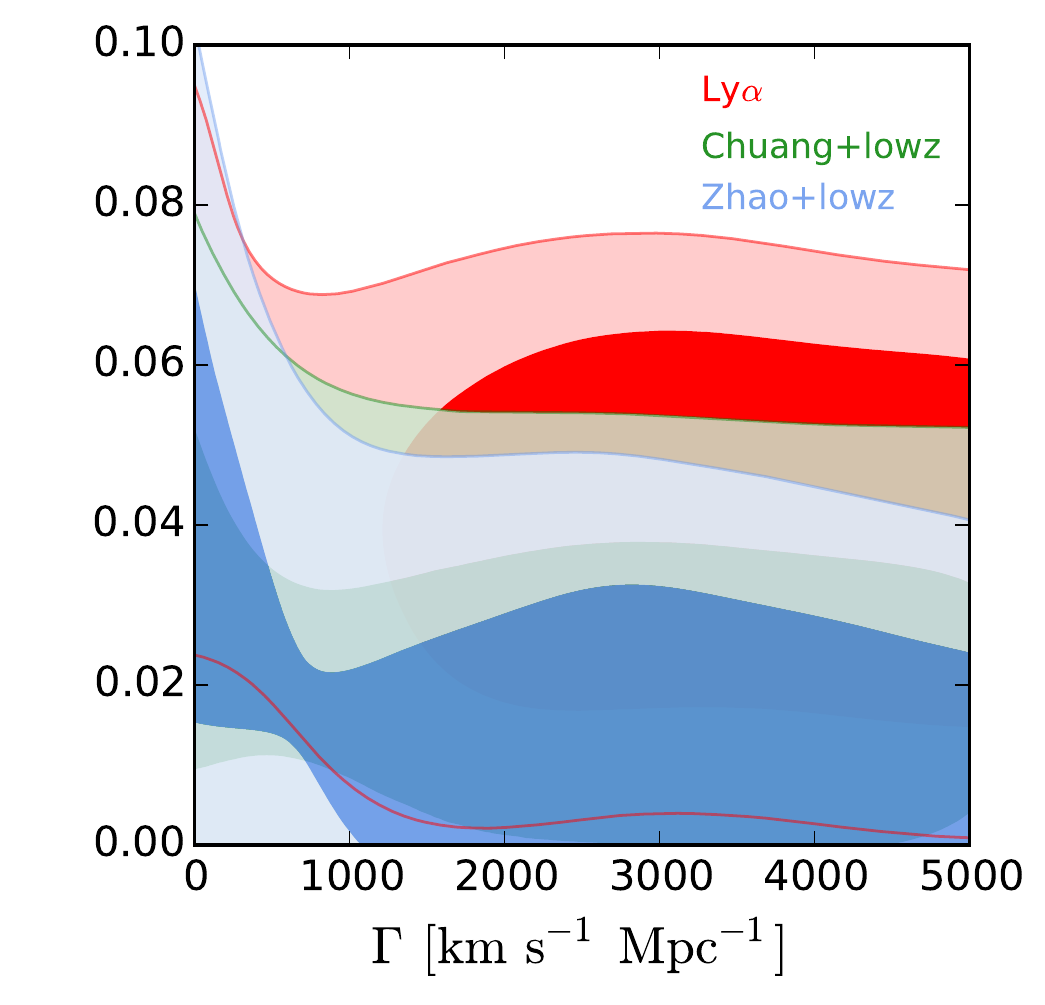}
    \includegraphics[width=0.41\textwidth]{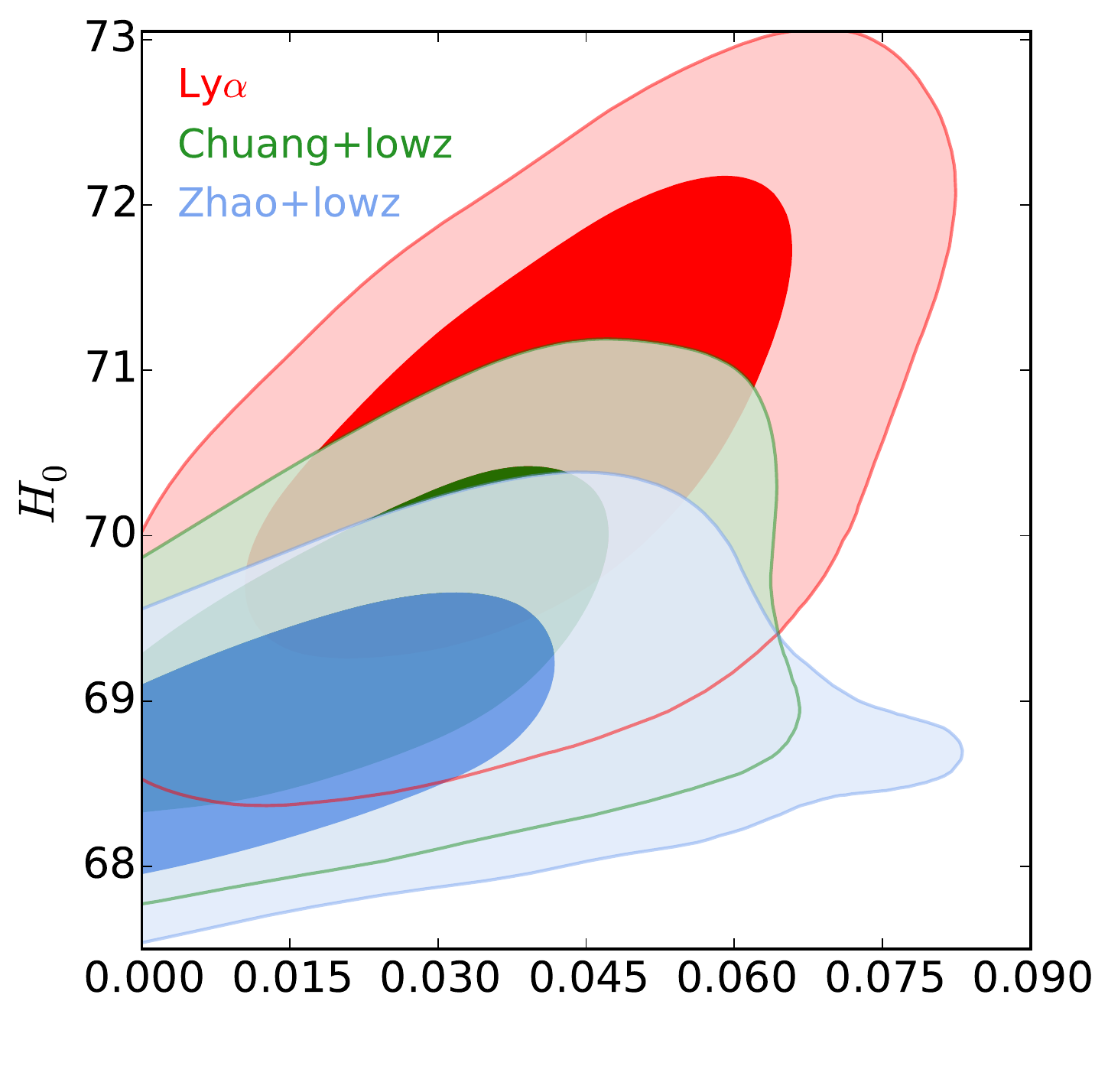}
    \put(-105,5){\large$f_{\rm dDM}$}
    \put(-430,105){\rotatebox{90}{\large$f_{\rm dDM}$}}}
    \caption{Posterior distributions ($1\sigma$ and $2\sigma$ contours) in the $\Gamma-f_{\rm dDM}$ ({\it left panel}) and $f_{\rm dDM}$-$H_0$ ({\it right panel}) planes. This analysis includes the full {\it Planck} likelihood for TT,TE,EE and lensing power spectra, the galaxy cluster counts from {\it Planck} catalogues, the local measurement of $H_0$ from SH0ES combined with galaxy BAO ("Zhao"), galaxy RSD ("Chuang") or Ly-$\alpha$ BAO ("Ly$\alpha$") measurements from the BOSS DR12 samples. The panels are taken from Ref.~\cite{Chudaykin:2017ptd}.}
    \label{fig_ddm}
\end{figure*}
Corresponding constraints in the $\Gamma-f_{\rm dDM}$ and $f_{\rm dDM}-H_0$ subspaces for a different choice of large-scale structure data are shown in Fig.~\ref{fig_ddm}. Since a full resolution of the Hubble tension requires $f_{\rm dDM}\sim10\,\%$~\cite{Berezhiani:2015yta} the CMB and BAO analysis makes the decaying DM solution of $H_0$-tension increasingly unlikely. The main culprit of these strong constraints is a so-called "lensing-like anomaly" in the {\it Planck} data that over-predicts the CMB lensing signal~\cite{Planck:2016tof,Planck:2018vyg} and thus strongly disfavors the decaying DM solution which, in turn, favors a weaker lensing effect. In this regard, it is curious to see what fraction of decaying DM is available upon marginalizing over the lensing information in the {\it Planck} power spectra. By allowing the CMB lensing power amplitude $A_{\rm lens}$ to be a free fitting parameter, the CMB and BAO data along with the {\it Planck} cluster counts and SH0ES measurement yield $f_{\rm dDM}=0.06\pm0.02\,\%$~\cite{Chudaykin:2017ptd}. Overall, this analysis indicates a mild $1.6\sigma$ preference in favor of the decaying DM scenario over $\Lambda$CDM~\cite{Chudaykin:2017ptd}. Therefore, before making strong conclusions it has to be understood first why the $\Lambda$CDM prediction is in tension with the {\it Planck} CMB lensing effect.

One way to mitigate the tight CMB constraints is to assume a short-lived decaying DM component which decays {\it before} recombination. Recent investigations~\cite{Poulin:2016nat,Nygaard:2020sow} reveal interesting implementations of the subdominant unstable component of DM decaying between matter-radiation equality and recombination epochs. The tight {\it Planck} constraints on $f_{\rm dDM}$ in this case can be significantly alleviated~\cite{Poulin:2016nat,Nygaard:2020sow}. Moreover, a decay near matter-radiation equality reduces the sound horizon at the drag epoch, $r_d$, alleviating the tensions with the late-time model-independent probes of $r_d$, see Ref.~\cite{Bernal:2016gxb}. However, the Hubble tension in this scenario is only marginally alleviated~\cite{Poulin:2016nat,Nygaard:2020sow}. Like an EDE proposal, in the short-lived decaying DM model, the growth of both metric and density perturbations is amplified that exacerbates (although weakly) the $S_8$ tension. 

It was recently proposed that a fraction of DM particles which decay into a much lighter stable relic and photon can solve the long-standing Lithium problem~\cite{DiBari:2013dna}. The electromagnetic injection in the specific energy range would destroy enough Lithium without affecting the abundance of other elements. Since the light decay products act as dark radiation in the early Universe, this scenario was also suggested to alleviate the $H_0$ tension~\cite{Alcaniz:2019kah}. However, this proposal is tightly constrained by the {\it Planck} data~\cite{Anchordoqui:2020djl}. Model-independent constraints from BBN on unstable DM which decay into photons and/or electron-positron pairs are provided in Ref.~\cite{Depta:2020zbh}.

Overall, the totality of the cosmological data put tight constraints on the decaying DM scenario as a resolution of $S_8$ and $H_0$ tensions although a few percent decaying DM fraction is still allowable~\cite{Chudaykin:2017ptd,Nygaard:2020sow}. For models with a subdominant component of DM decaying between recombination and the present epoch, the understanding of the "lensing-like anomaly" in the {\it Planck} data is highly warranted because this feature is responsible for stringent CMB limits on this proposal. To conclude, the current $S_8$ and $H_0$ tensions require either a different (and possibly major) alteration of the $\Lambda$CDM model, or derives from some yet unknown systematic effect.

The cosmological scenarios where the heavier particles decay to lighter and invisible radiation are argued to weaken the $S_8$ tension~\cite{Aoyama:2014tga,Enqvist:2015ara,Abellan:2020pmw}. These scenarios also relax the tension of CDM predictions with observation of structures at small scales~\cite{Wang:2014ina,Cheng:2015dga} and can elucidate the origin of high-energy neutrino IceCube events~\cite{Anchordoqui:2015lqa,Anchordoqui:2021dls}. Ultrahigh-energy cosmic ray experiments could be powerful probes of dDM models in which DM couples to the SM sector~\cite{Anchordoqui:2018qom}.

\subsubsection{Dynamical Dark Matter}
\label{sec:dynDM}

Although many models of decaying DM transcend the canonical WIMP or axion frameworks and involve new regions of the DM parameter space, perhaps none do so as dramatically as those that arise within the Dynamical Dark Matter (DDM) framework. The basic idea behind DDM is relatively simple~\cite{Dienes:2011ja}. Rather than focus on one or more stable DM particles, the DDM framework is built on the proposition that the DM in the Universe comprises a vast {\it ensemble}\/ of interacting fields with a variety of different masses, mixings, and cosmological abundances. Moreover, rather than imposing the stability for each field individually, the DDM framework recognizes that the decay of a DM component in the early Universe is not ruled out if the cosmological abundance of that component is sufficiently small at the time of its decay. The phenomenological viability of the DDM framework therefore requires that those states within the ensemble with larger masses and SM decay widths have correspondingly smaller relic abundances, and vice versa.

{\it In other words, DM stability is not an absolute requirement in the DDM framework, but is replaced by a balancing of lifetimes against cosmological abundances across the entire ensemble.}\/ As a result, individual constituents of the DDM ensemble are decaying {\it throughout}\/ the evolution of the Universe, from early times until late times and even today. This leads to a highly dynamical scenario in which cosmological quantities such as the total DM abundance $\Omega_{\rm DM}$ experience a non-trivial time dependence beyond those normally associated with cosmological expansion. 

Many extensions to the SM give rise to large ensembles of dark states in which such a balancing naturally occurs. Moreover, because the DM "candidate" in this framework consists of a carefully balanced DDM ensemble which cannot be characterized in terms of a single well-defined mass, decay width, or interaction cross section, this framework gives rise to many striking experimental and observational signatures which transcend those usually associated with DM and which ultimately reflect the collective behavior of the entire DDM ensemble.

The DDM framework was originally introduced in Ref.~\cite{Dienes:2011ja}, while in Refs.~\cite{Dienes:2011sa, Dienes:2012jb} explicit DDM models were constructed which satisfy all known collider, astrophysical, and cosmological constraints. Since then, a considerable body of work has focused on various aspects of the DDM framework. One major direction of research consists of analyzing the various signatures by which this framework might be experimentally tested and constrained. These include unique DDM signatures at direct-detection experiments~\cite{Dienes:2012cf}, at indirect-detection experiments~\cite{Dienes:2013lxa,Boddy:2016fds,Boddy:2016hbp}, and at colliders~\cite{Dienes:2012yz,Dienes:2014bka,Curtin:2018ees,Dienes:2019krh,Dienes:2021cxr}.

Indeed, such DDM scenarios also give rise to enhanced complementarity relations~\cite{Dienes:2014via,Dienes:2017ylr}
between different types of experimental probes. Moreover, the non-minimal dark sectors which are the cornerstone of the DDM framework can also lead to observable imprints across the cosmological timeline, stretching from structure formation~\cite{Dienes:2020bmn,Dienes:2021itb} all the way to late-time supernova recession data~\cite{Desai:2019pvs} and unexpected implications for evaluating Ly-$\alpha$ constraints~\cite{Dienes:2021cxp}. Such dark sectors also give rise to new theoretical possibilities for stable mixed-component cosmological eras and a potential re-examination of the age of the Universe~\cite{Dienes:2021woi}.

The second direction of DDM research over the past decade has focused on the numerous ways in which suitably-balanced DDM ensembles $-$ i.e., ensembles of dark states in which the widths for decays into SM states are naturally inversely balanced against cosmological abundances $-$ emerge naturally within various models of BSM physics. These include theories involving large extra dimensions, both flat~\cite{Dienes:2011ja,Dienes:2011sa} and warped~\cite{Buyukdag:2019lhh}; theories involving strongly-coupled hidden sectors~\cite{Dienes:2016vei, Buyukdag:2019lhh}; theories involving large spontaneously-broken symmetry groups~\cite{Dienes:2016kgc}; and even string theories~\cite{Dienes:2016vei}.

Indeed, the dark states within these different realizations can accrue suitable cosmological abundances in a variety of ways, including not only through non-thermal generation mechanisms such as misalignment production~\cite{Dienes:2011ja, Dienes:2011sa} but also through thermal mechanisms such as freeze-out~\cite{Dienes:2017zjq}. Mass-generating phase transitions in the early Universe can also endow collections of such states with non-trivial cosmological abundances~\cite{Dienes:2015bka,Dienes:2016zfr,Dienes:2016mte}.

In general, the mass spectra and corresponding lifetimes and abundances of the individual states within the DDM ensemble turn out to be tied together through scaling relations involving only a few scaling exponents. This is reviewed, for example, in Sect.~III of Ref.~\cite{Curtin:2018ees}. The particular physical scenario that one has in mind within the DDM framework then determines these scaling exponents. As a result, the DDM ensemble is described by only a few free parameters, rendering the DDM framework every bit as constrained and predictive as more traditional dark-matter scenarios.

DDM scenarios in which the constituents decay entirely within the dark sector  --- i.e., to final states comprising other, lighter ensemble constituents and/or dark radiation --- are particularly challenging to probe and constrain.  Nevertheless, there exist observational handles that can be used to probe and constrain DDM ensembles which decay primarily via "dark-to-dark" decay processes of this sort, and thus potentially permit us to distinguish them from traditional DM candidates. 

\begin{figure}[t]
    \begin{center}
    \includegraphics[width=0.31\textwidth, keepaspectratio]{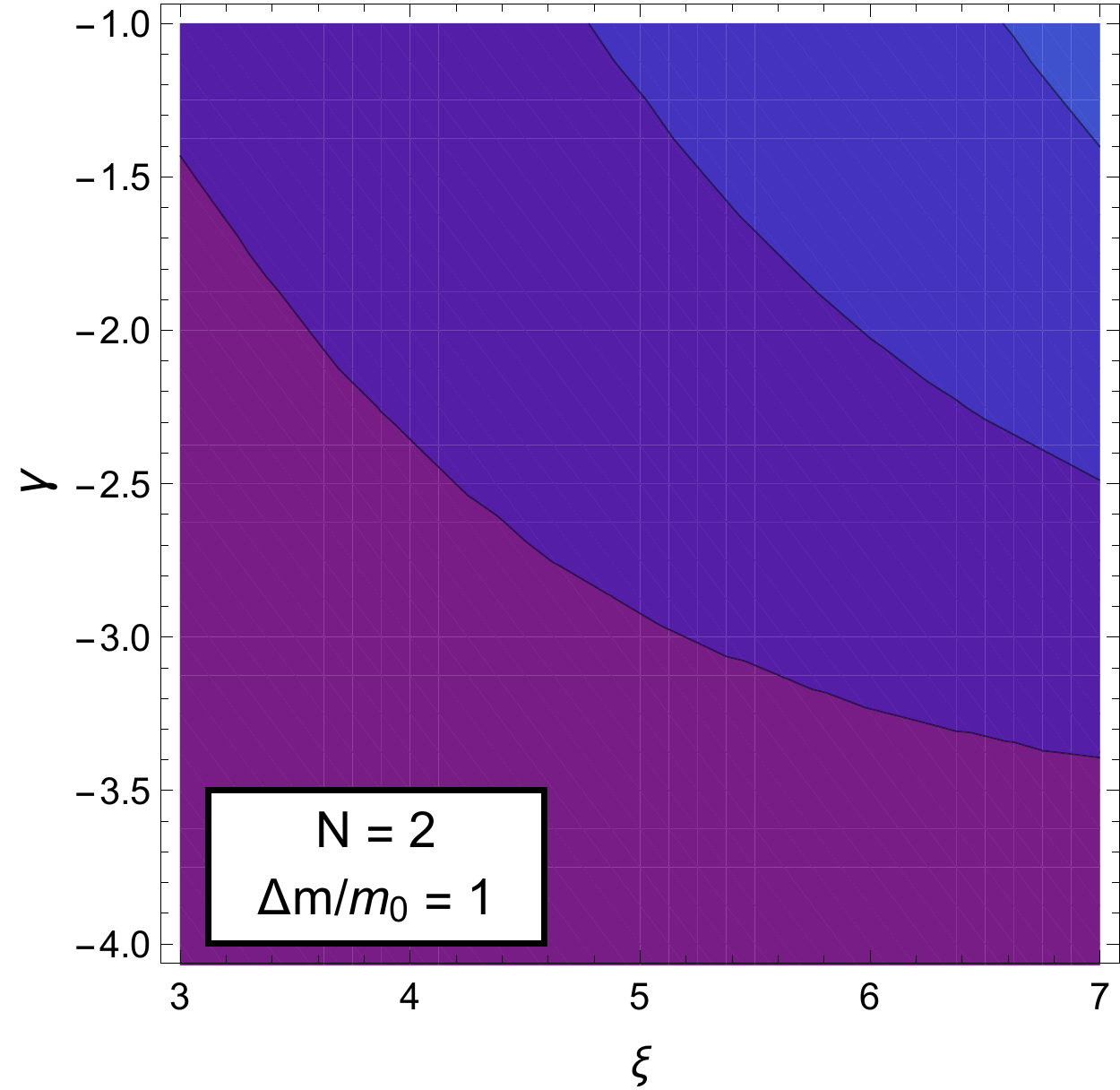}~
    \includegraphics[width=0.31\textwidth, keepaspectratio]{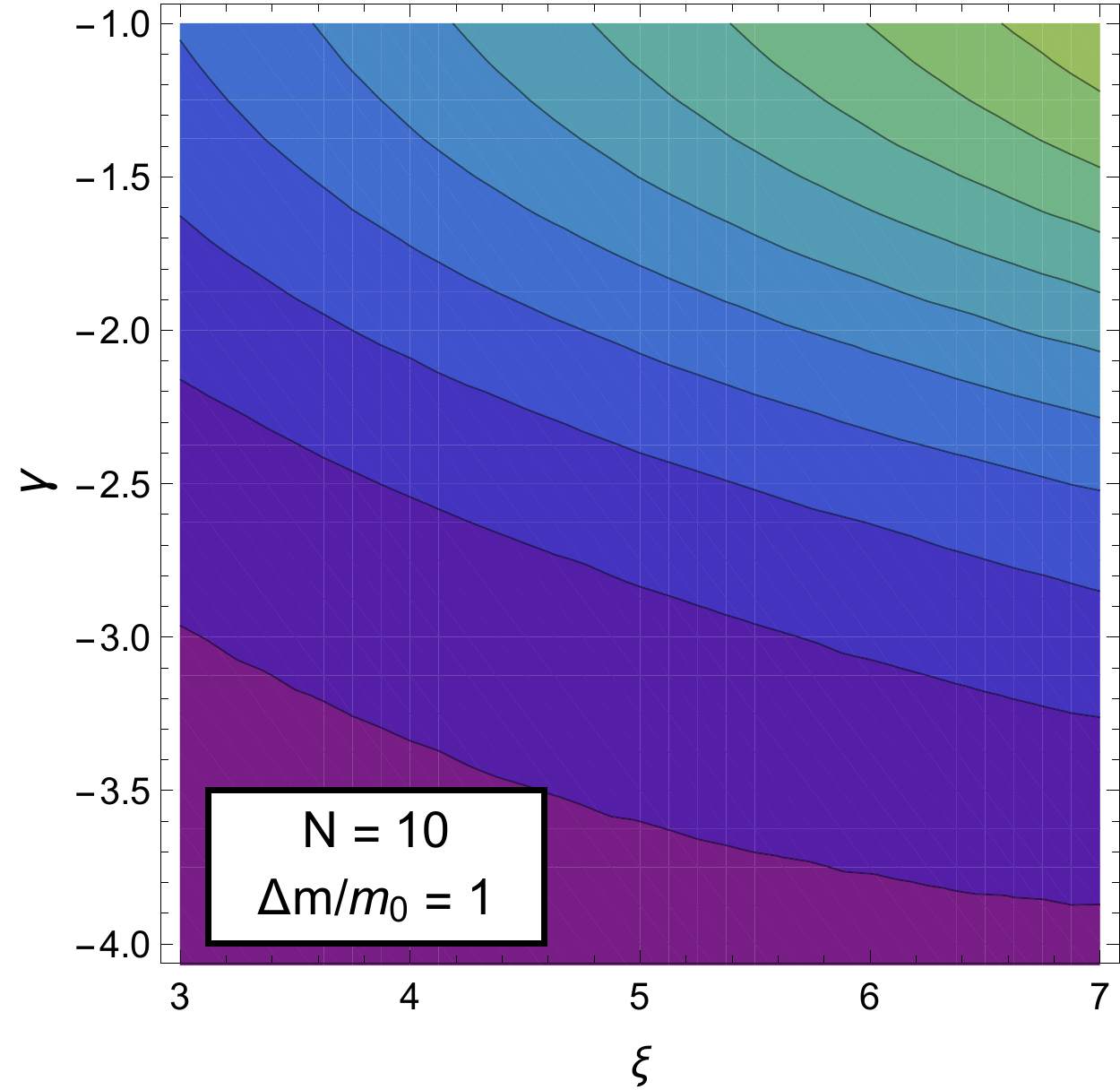}~
    \includegraphics[width=0.31\textwidth, keepaspectratio]{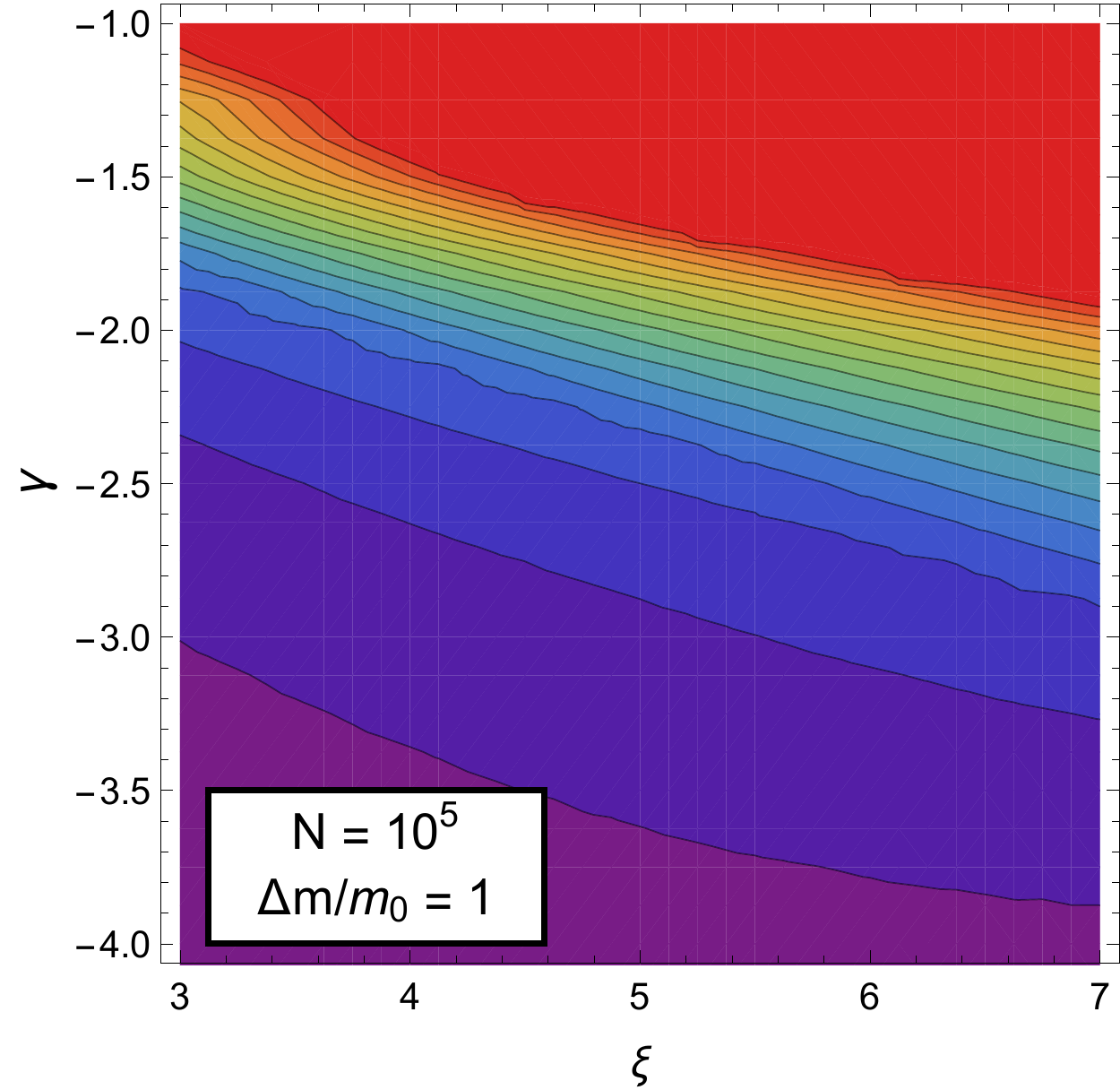}
    \includegraphics[width=0.67\textwidth, keepaspectratio]{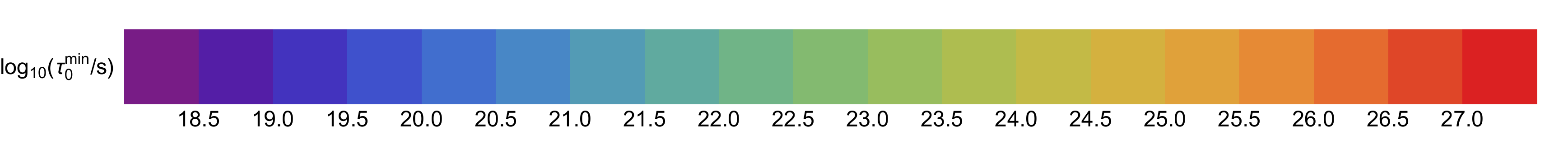}
        \caption{
            Constraints from Type-Ia supernova data on DDM ensembles whose 
            constituents $\chi_\ell$ decay directly to dark radiation.  The 
            parameters $\gamma$ and $\xi$ respectively parametrize the manner
            in which the abundances and decay widths of the $\chi_\ell$ scale 
            across the ensemble as a function of $m_\ell$.
            In each panel, the contour plot shows the $3\sigma$ lower 
            bound on the lifetime of the lightest ensemble constituent.  
            The results displayed in the left, center, and right panels 
            respectively correspond to ensembles consisting of $N=2$, $N=10$,
            and $N=10^5$ fields. In all cases, the mass splitting 
            $m_{\ell + 1} - m_\ell$ between successive states in the ensemble 
            is taken to be equal to the mass of the lightest ensemble constituent.  
            All the panels are taken from Ref.~\protect\cite{Desai:2019pvs}.
            }
            \label{fig:SNTypeIaDDM}
    \end{center}
\end{figure}

For example, dark-to-dark decays of this sort modify the way in which the expansion rate of the Universe, as described by the Hubble parameter $H(z)$, evolves with redshift.  These modifications, in turn, affect the functional relationship between the redshifts $z$ and luminosity distances $D_L(z)$ of Type-Ia supernovae. In Ref.~\cite{Desai:2019pvs}, constraints based on the observed relationship between the redshifts and luminosity distances of Type-Ia supernovae within the combined Pantheon sample~\cite{Pan-STARRS1:2017jku} were derived for DDM ensembles whose constituents decay via a two-body process of the form $\chi_\ell \rightarrow \bar{\psi}\psi$, where $\psi$ is a massless dark-radiation field. A partial summary of these constraints is provided in Fig.~\ref{fig:SNTypeIaDDM}.

Since the dark-to-dark decays of a DDM ensemble alter the dependence of $H(z)$ on $z$, the DDM framework can potentially also provide a way of addressing the $H_0$ tension.  
In this regard, the advantage of a DDM ensemble relative to a single decaying dark-matter species is that the timescale across which the decays have a significant impact on the expansion rate can be far broader. Nevertheless, the depletion of the overall DM abundance at late times due to $\chi_\ell$ decays leads to a suppression of the CMB lensing effect $-$ and therefore to tension with {\it Planck} data $-$ just as it does in other late-time decaying DM scenarios~\cite{Chudaykin:2016yfk}. DDM scenarios in which the $\chi_\ell$ decay directly into dark radiation 
are significantly constrained~\cite{Anchordoqui:2022gmw}.

By contrast, scenarios in which the $\chi_\ell$ decays primarily via intra-ensemble processes -- e.g., of the form 
$\chi_\ell \rightarrow \chi_m \bar{\psi} \psi$, where $\psi$ once again denotes a dark-radiation field -- are more  
promising~\cite{Anchordoqui:2020djl,Anchordoqui:2022gmw}. Such decays endow the final-state $\chi_m$ with non-negligible velocities, thereby modifying the equation of state $w_m(z)$ for each ensemble constituent and modifying the DM velocity distribution of the ensemble as a whole. Moreover, complementary scattering processes of the form $\chi_\ell \psi \rightarrow \chi_m \psi$ through which the different ensemble constituents interact with the dark radiation could potentially also help to ameliorate the $\sigma_8$ tension in the same way that they do in partially acoustic DM scenarios~\cite{Chacko:2016kgg}.  

\begin{figure}[t]
    \begin{center}
    \includegraphics[width=0.49\textwidth, keepaspectratio]{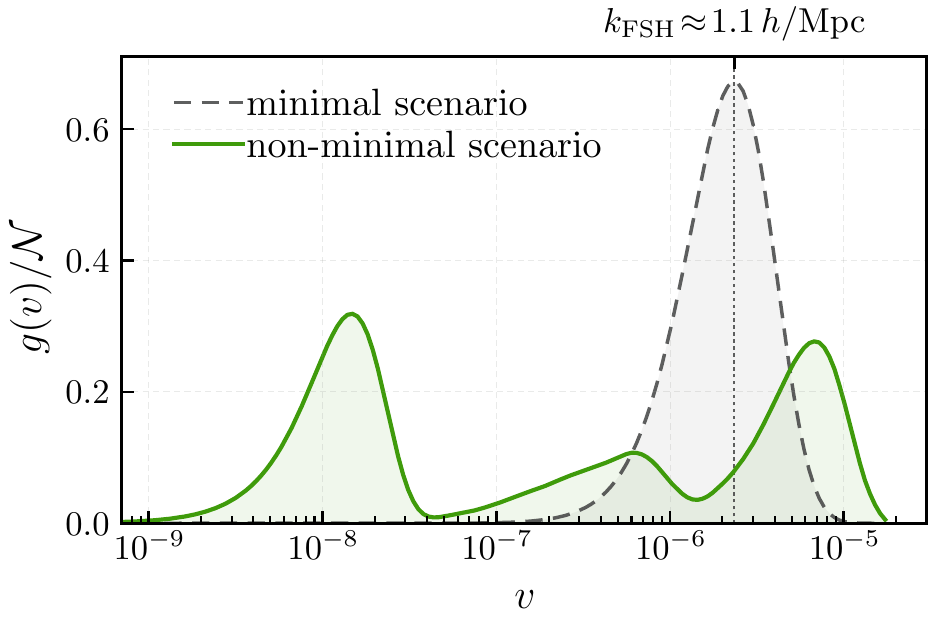}~
    \includegraphics[width=0.47\textwidth, keepaspectratio]{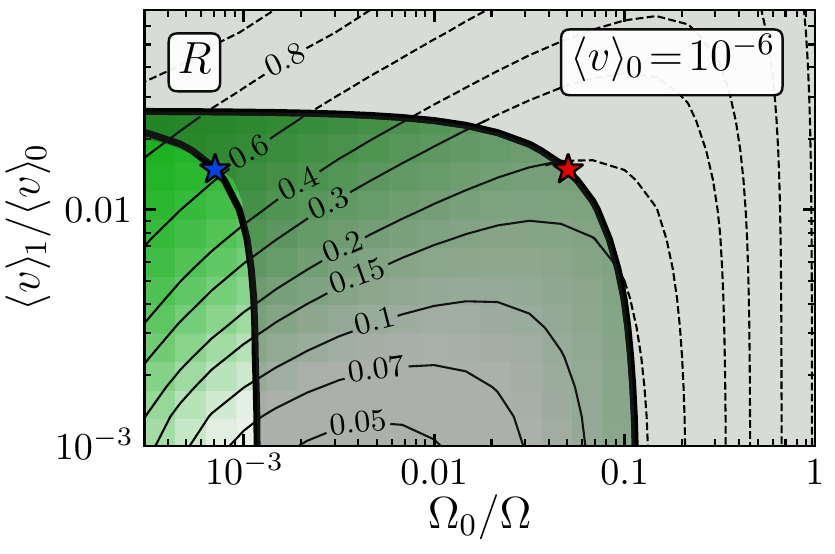}
        \caption{
            {\it Left panel}\/: 
            Two examples of DM velocity distributions 
            $g(v)\propto v^3 f(v,0)$ which have the same present-day 
            average velocity and would na\"{i}vely be characterized by the same 
            free-streaming horizon $k_{\rm FSH}\approx 1.1\, h/{\rm Mpc}$. 
            The dashed gray curve is the distribution for a warm-dark-matter model,  
            while the solid blue curve represents a distribution generated 
            by decay cascades within a non-minimal 
            dark sector~\protect\cite{Dienes:2020bmn}.
            {\it Right panel}\/:
            Contours of $R$, plotted within a slice of the 
            parameter space of a model for which $g(v)$ consists of two
            log-normal peaks, where $R$ is a ratio quantifying the difference 
            between the results of two recasting 
            methods~\protect\cite{Konig:2016dzg,Murgia:2017lwo} commonly employed 
            in estimating Ly-$\alpha$ constraints on non-cold DM models.
            The degree to which $R$ differs from unity provides a  
            measure of the degree to which the results of the recasts differ.
            The two thick black curves indicate the Ly-$\alpha$ 
            constraints obtained from these two recasts.
            The region above and to the right of each of these thick
            black curves is excluded by the corresponding recast.
            Both panels are similar to figures which appear in Ref.~\protect\cite{Dienes:2021cxp}.
            }
            \label{fig:NCDMDistsComp}
    \end{center}
\end{figure}

In this connection, it is also worth noting that it is not merely the value of $\sigma_8$ alone, but rather the detailed dependence of the amplitude of matter-density perturbations across a broad range of wavenumbers, which provides an observational handle on DM scenarios with non-cold DM velocity distributions $f(v,z)$. In such scenarios, free-streaming effects lead to a suppression of cosmic structure on small scales. For DM models such which yield relatively simple, unimodal forms for $f(v,z)$ at all $z$, assessing the impact of these effects on $P(k)$ is comparatively straightforward.

By contrast, in the context of non-minimal dark-sector scenarios wherein a significant fraction of the dark-matter abundance is produced non-thermally (e.g., from particle decays) and wherein $f(v,z)$ can be highly non-trivial and even multi-modal, the impact of free-streaming on small-scale structure is more complicated~\cite{Konig:2016dzg,Heeck:2017xbu,Dienes:2020bmn,Du:2021jcj,Decant:2021mhj}. For example, standard "recasting" procedures~\cite{Konig:2016dzg,Murgia:2017lwo} employed in deriving constraints on $f(v,z)$ from Ly-$\alpha$-forest data can become unreliable for velocity distributions of this sort. A study of the performance of these recasts for general dark-matter velocity distributions was performed in Ref.~\cite{Dienes:2021cxp}.  Results from a study of this sort are shown in Fig.~\ref{fig:NCDMDistsComp}.

On the other hand, there exist methods~\cite{Dienes:2020bmn} which permit one to reconstruct the detailed shape of $f(v,z)$ solely from information contained within the linear matter power spectrum $P(k)$ --- methods which are applicable even when that shape is highly non-trivial. In cases in which $P(k)$ differs from the matter power spectrum $P_{\rm CDM}(k)$ for CDM primarily due to free-streaming effects, these methods are quite robust in their applicability and are capable of reconstructing the DM velocity distribution with impressive fidelity. By contrast, the situation becomes more subtle in cases wherein the growth of density perturbations at late times is affected by additional complications --- such as decays of the $\chi_\ell$ after last scattering and/or dark acoustic oscillations~\cite{Cyr-Racine:2013fsa} of the sort which would arise from non-negligible interactions between $\psi$ and the $\chi_\ell$. Nevertheless, the relationship between small-scale structure and the dark-matter velocity distribution can provide a valuable cross-check on cosmological models of this sort.

\subsubsection{Extended Parameter Spaces Involving $A_{\rm lens}$}

Extended cosmological scenarios beyond the 6-parameter $\Lambda$CDM model including the $A_{\rm lens}$ parameter free to vary have been widely investigated in the literature, showing that when $A_{\rm lens}>1$ as preferred by the {\it Planck} data, see the discussion in Sec.~\ref{sec:WG-Alens}, the $S_8$ tension disappears~\cite{DiValentino:2015ola, DiValentino:2016hlg, DiValentino:2017zyq,DiValentino:2018gcu,DiValentino:2019dzu, Benisty:2020kdt, DiValentino:2020hov}. This could suggest a possible systematic error present in the {\it Planck} data not accounted for, that is producing the disagreement with the weak lensing and cluster counts measurements. Once the data analysis of the CMB data marginalizes over this $A_{\rm lens}$ phenomenological parameter, the agreement between the $S_8$ parameter obtained by the {\it Planck} data and the low redshift measurements is restored.

\subsubsection{Cosmological Scenario with Features in the Primordial Power Spectrum}
\label{sec:Features}

In~\cite{Hazra:2018opk} without considering any extension to the standard model at the background level, authors project the effect of the differences in the values of the key cosmological parameters  to the form of the primordial power spectrum (PPS). In order to realize this task, they did uncover the form of the primordial spectrum by implementing the Modified Richardson-Lucy algorithm (MRL)~\cite{Shafieloo:2003gf,Shafieloo:2006hs,Nicholson:2009pi,Hazra:2013xva,Hazra:2014jwa} that would fit the {\it Planck} temperature data as good as the case of the concordance $\Lambda$CDM model, but with a Hubble constant consistent with local measurements, as well as improving the consistency between the derived $S_8$ and $\sigma_8$ parameters with estimations of the weak lensing surveys. Recently there have been more progress in this line to address various tensions by a featured form of the primordial spectrum~\cite{Hazra:2022rdl}. In this context,~\cite{Antony:2022ert} introduces examples of single field inflationary trajectories beyond the slow-roll regime that provides improvements in the {\it Planck} data compared to the standard model -- partially mimics the effect of unphysical excess lensing and closed Universe. This model simultaneously prefers a higher $H_0$ and lower $S_8$ -- a trend that is enhanced with the addition of $H_0$ data from the SH0ES measurements.

Furthermore and in Ref.~\cite{Keeley:2020rmo} authors did explore a class of primordial spectra that can fit the observed cosmic microwave background data well and that could result in a larger value for the Hubble parameter consistent with the local measurements. This class of spectra consists of a continuous deformation between the power law primordial spectrum and the reconstructions from the MRL deconvolution algorithm. More works are needed to assign significance to such features and seek for possible theoretical implementation from inflation~\cite{Snowmass2021:Inflation}. Testing such a form of the primordial spectrum as initial conditions against other cosmological observations would be another important task in this direction.

\subsubsection{Interacting Dark Matter}
\label{sec:IDM}

Several models of interacting dark matter feature a suppression of the matter power spectrum. Some of these models have a cosmologically relevant suppression impacting the $S_8$ tension even after taking into account constraints from the CMB. In particular, dark matter interactions with photons or neutrinos~\cite{Ali-Haimoud:2015pwa,Weiner:2012cb,Wilkinson:2013kia,Diacoumis:2018ezi} have shown to impact structure formation~\cite{Boehm:2004th,Bringmann:2013vra,Cherry:2014xra,Boehm:2014vja,Wilkinson:2014ksa,DiValentino:2017oaw,Olivares-DelCampo:2017feq,Escudero:2018thh,Kumar:2018yhh,Stadler:2018jin,Stadler:2019dii,Becker:2020hzj}. This is because the small residual interactions introduce a collisional damping on small scales (for which the Fourier modes entered the Hubble horizon early enough to experience significant interactions) that suppresses the power spectrum on these scales. Similarly, interactions with a dark radiation component have also been invoked in the past~\cite{Cyr-Racine:2013fsa,Chu:2014lja,Rossi:2014nea,Buen-Abad:2015ova,Schewtschenko:2015rno,Lesgourgues:2015wza,Cyr-Racine:2015ihg,Krall:2017xcw,Archidiacono:2017slj,Buen-Abad:2017gxg,Archidiacono:2019wdp}, though these have been found to be more promising in terms of the Hubble tension~\cite{Lesgourgues:2015wza,Becker:2020hzj,Velten:2021cqj}. Interacting models and their consequences were considered in several papers reviewed in~\cite{Wang:2016lxa}.

\subsubsection{Quantum Landscape Multiverse} 

The authors of Refs.~\cite{Holman:2005ei,Holman:2005qk,Mersini-Houghton:2014jha,Mersini-Houghton:2008eyr} showed in 2005 that the mystery of the unlikely origin of our Universe can be answered and derived within the framework of a a larger phase space of initial conditions, by means of quantum cosmology, when the out of equilibrium dynamics of gravitational and matter degrees of freedom is taken into account. It has been initially proposed~\cite{Mersini-Houghton:2005axz,Kobakhidze:2004gm,Mersini-Houghton:2006phg} to allow the wavefunction of the Universe to propagate through the stochastic structure of the string theory landscape vacua, by considering the later to be a physical realization of the phase space of initial conditions from which Universes can spring into existence, and thereby {\it derive instead of postulate} the probability of the selection of our initial conditions by means of quantum cosmology.

In the second stage, the decoherence among the branches of the wavefunctions was included, thus extending the mini-superspace of three-geometries and landscape moduli fields to include an infinite number of long wavelength fluctuations of space time and of the moduli, which provided the environment and it is sufficiently weakly coupled to the system in order to not interfere with it in process of measurement. Mathematically, this complex system of the wavefunction of the Universe being a functional of an infinite dimensional superspace is similar to known condensed matter systems such as spin glass and quantum dots, and can be solved by means of random matrix theory. Solutions found for the branches of the wavefunction of the Universe~\cite{Holman:2005ei, Holman:2005qk, Mersini-Houghton:2008eyr, Mersini-Houghton:2014jha} revealed that: firstly, a whole family of wavepackets of branches of the wavefunction that settled on high energy vacua on the landscape undergo a similar expansion history like our Universe to transition from a quantum to a classical universe, they survive the backreaction of matter fluctuations (which tries to slow their growth), and therefore have the highest chance of existence, while the ones that settled on low energy vacua cannot survive the backreaction matter and remain quantum particles forever; secondly, since there is a whole family of solutions that survive and grow to classical Universes, then the answer to the question of our high energy cosmic inflation comes at the price of extending our standard model of cosmology to a quantum multiverse framework in which our Universe is embedded and is just a humble member in a vastness.

The attraction of this theory is that for the first time it provided an answer to the problem of our unlikely initial state, and it did so by deriving it from first principles, namely to: why our Universes originate from a high energy low and entropy state?  Having a formalism of well trusted quantum equations which allows us to follow consistently a coherent story of evolution of our Universe from before it inflated, when it was just a branch of the wavefunction on some landscape energy vacua, follow it through cosmic inflation, and after it transitions into a large classical Universe, opens a window from which we can glean into the larger structure on which our Universe was embedded, it allows us a way to collect evidence on the multiverse. Until this work, the conventional wisdom was that limited by the speed of light, we cannot test the existence of multiverse and therefore any theory of the multiverse cannot be scientific. Since, the multiverse research has become a mainstream field in cosmology.

In Refs.~\cite{Holman:2006an,Holman:2006ny,Mersini-Houghton:2016jno} the authors realized that quantum entanglement and decoherence are two sides of the same coin, and that the backreaction of the long wavelength fluctuations, very weakly coupled to the system being ‘watched’ (our Universe), which triggers decoherence and destroys entanglement among the different branches which becomes decohered Universe, leaves its traces on our CMB sky. The authors proposed to calculate them and thus use quantum entanglement as a way to test this theory and collect evidence for the quantum multiverse in which our Universe is embedded. In two papers named, "Avatars of the Landscape" in 2005~\cite{Holman:2006an,Holman:2006ny,Mersini-Houghton:2016jno} they derived a series of predictions leftover in our sky from the early quantum origins, including the Cold Spot, a suppressed $\sigma_{8}$, a supersymmetry breaking scale which is much higher than the expected tEV scale, a power asymmetry between the two hemispheres, suppressed multipoles at the lowest $l’s$ and so on. Later observations with WMAP and {\it Planck} confirmed the existence of these anomalies at exactly the scales and sized that were predicted in Refs.~\cite{Holman:2006an,Holman:2006ny,Mersini-Houghton:2016jno}, in the case of the Cold Spot at a significance higher than 4.

The reason for the giant void that shows as a Cold Spot in temperature CMB maps, and the suppressed value of $\sigma_8$ is that the entanglement of our Universe at its infancy with all the other branches of the wavefunction and the backreaction from the collective contribution of long wavelength fluctuations which was calculated in Refs.~\cite{Holman:2006an,Holman:2006ny,Mersini-Houghton:2016jno} provided a second source of fluctuations that contributed to CMB in addition to the typical inflationary fluctuations. Since the only scales in this problem are the energy of cosmic inflation and that of SUSY breaking scale, then the model does not allow much tweaking room for fitting data to the model. Therefore in Refs.~\cite{DiValentino:2018wum,DiValentino:2016nni,DiValentino:2016ziq} the status of this theory has been checked with the most recent and detailed {\it Planck} satellite data and found it is reassuring that the fit of anomalies predicted in the theory continues to be robust.

\subsubsection{Quantum Fisher Cosmology}

The aim of Quantum Fisher Cosmology~\cite{Gomez:2020xdb,Gomez:2020gqa,Gomez:2021yhd,Gomez:2021hbb,Gomez:2021jcl} is to use the quantum Fisher information about pure de Sitter states to derive model independent observational consequences of the existence of a primordial phase of the Universe of de Sitter accelerated expansion. These quantum features are encoded in a scale dependent quantum cosmological tilt that defines what we can call the de Sitter universality class. The experimental predictions are: i) A phase transition from red into blue tilt at a scale order $k= 1$ Mpc$^{-1}$ that naturally solves the cosmological {\it trans-Planckian problem}, ii) A spectral index for curvature fluctuations at CMB scales $k= 0.05$ Mpc$^{-1}$ equal to $0.0328$, iii) A tilt running at scale $k=0.002$ Mpc$^{-1}$ equal to $-0.0019$, iv) An enhancement of the amplitude of CMB peaks for extremely high multipoles ($l > 10^5$) that can provide a natural mechanism for primordial black hole formation as a source of dark matter, v) A lack of power at scales of $8$ Mpc with respect to the CMB scale that can explain the $\sigma_8$ tension. 

The key ingredient is the quantum Fisher information associated with the family of de Sitter invariant vacua describing scalar spectators in a pure de Sitter background. These pure states are sometimes denoted, in the literature, as $\alpha$-vacua. What the quantum Fisher information naturally defines is a metric on this set of states.  In essence, it measures the quantum distinguishability of different $\alpha$ vacua. There exists an extensive literature on both the quantum consistency of $\alpha-$vacua as well as on the physical viability of using the Bunch Davis vacuum to define the quantum fluctuations describing the CMB spectrum of fluctuations. Some of these problems are related with the trans-Planckian problem and the computation of one loop effects on these vacua. We will surpass some of these well known difficulties focusing on a well defined quantity associated with the family of $\alpha$ vacua, namely the quantum Fisher information associated with this one parameter family of pure states. As stressed before, this quantum information naturally leads to a finite quantum variance for the parameters labeling these vacua. The main message of our work is that this quantum variance can account for the anomalous scale dependence of the cosmological power spectrum normally derived after adding a quasi de Sitter deformation. Moreover, this quantum information encodes the quantum variance of the $\alpha$ parameter. The parameter $\alpha$ can be associated with a natural energy scale defined as $k\eta H$. More precisely $\alpha= \ln \tanh(r(\Lambda)) -2i\phi(\Lambda)$ with $r(\Lambda) =-\sinh^{-1}(\frac{H}{2\Lambda})$ the standard squeezing parameter and $\phi(\Lambda) =-\frac{\pi}{4} - \frac{1}{2} \tan^{-1}(\frac{H}{2\Lambda})$ with $\Lambda = Hk|\eta|$.

\begin{figure}
    \centering
    \includegraphics[width=0.6\columnwidth]{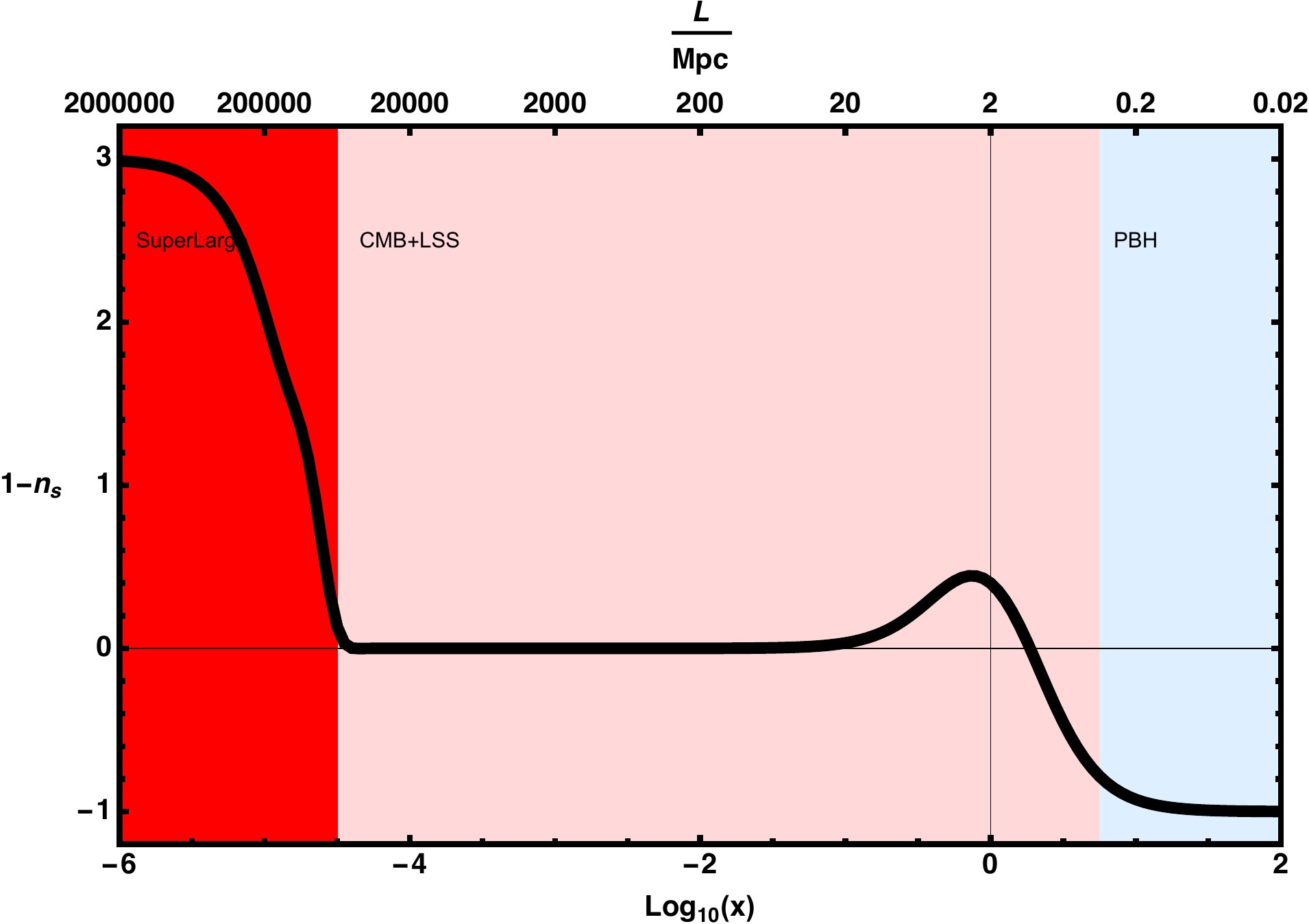}
    \caption{The quantum cosmological tilt as a function of energy scale as determined by the quantum Fisher. For values of $x=|k \eta| < 0$ the end of inflation is determined by the IR cutoff on the number of particles $\mathcal{N}$. In this plot, for illustrative purposes, we set $\mathcal{N}=10^{9}$. The different shaded regions correspond to different scales. The blue region is where the dark matter forms and the spectral index is blue. Figure taken from~\cite{Gomez:2021hbb}.
    }
    \label{fig:qtilt}
\end{figure}

The starting point of the quantum Fisher approach to Cosmology is to identify the scale transformations of this quantum Fisher information. In other words, we are interested in identifying how the information controlling the quantum variance of $\alpha$ depends on the energy scale at which we are working.  The main finding of~\cite{Gomez:2020xdb,Gomez:2021yhd,Gomez:2021jcl} is that this scale transformation of the quantum Fisher is anomalous with a scale dependent tilt defined as $\alpha_F$. This is the tilt depicted in Fig.~\ref{fig:qtilt}. As discussed in~\cite{Gomez:2020zev} this figure represents the numerical result obtained after evaluating the quantum Fisher information with an IR cutoff on the number of contributing entangled pairs. The sensitivity of the result on this IR cutoff was discussed in~\cite{Gomez:2020zev}. 

For comoving Lagrangian scales $8 > \frac{1/k}{\rm Mpc} > 1$ we predict that the value of $1-n_s$ should be larger than at scales of $20$ Mpc, which is where Planck18 measured the tilt. Therefore, there should be a lack of power with respect to the CMB inferred one given by the difference in a power-law power spectrum when adopting the CMB value $1-n_s=0.0328$ or the value from Fig.~\ref{fig:qtilt} corresponding at these scales ($\log_{10} (x) \sim -0.5$). In more detail:
\begin{equation}
    \Delta P(k)=\frac{P(k)_{\rm CMB}}{P(k)_{\rm \sim 8 Mpc}} = \frac{(k/k_0)^{(n_s-1)_{k \eta = 0.1}}}{(k/k_0)^{(n_s-1)_{k \eta \sim 0.3}}}\,,
\end{equation}
where we have used that the difference in scales between Planck18 CMB and the weak lensing surveys at $\sim 8$ Mpc is about a factor 3. Now, from Fig.~\ref{fig:qtilt} at the corresponding scale of $k \eta \sim 0.3$ we see that $1-n_s \sim 0.15$, so 
\begin{equation}
    \Delta P(k) = \Delta(k)^{(0.0328 - 0.15)}\,,
\end{equation}
where $\Delta(k) = 3$ as measured from the CMB scale of $\log_{10} (|k \eta|) \sim -1.0$ (to probe scales of 8 Mpc which is where KiDS and SPT report their measurements). This results in $\Delta P(k) = 0.85$. On the other hand the observed ratio $\Delta P(k) [S_8 ({\rm KiDS})/S_8 ({\rm Planck18})] =  0.9 \pm 0.03 $. This is in good agreement with our prediction that at small scales the Universe would be less clumpy than the one at CMB scales.

\subsubsection{Quartessence} 

An interesting class of cosmological models, known as quartessence or unified dark matter, in which $\Lambda$ and dark matter are considered to be the two faces of the same quantity, can explain the current accelerated expansion of the Universe and also can explain the growth of the large scale structure of the Universe~\cite{Camera:2017tws}.  
A family of quartessence models can be described by the following Lagrangian~\cite{Bertacca:2008uf}:
\begin{equation}
    \mathcal{L}_\mathrm{Q}=f(\varphi)g(X)-V(\varphi)\label{L_phi}\,,
\end{equation}
where $g(X)$ ($X = - (1/2) \nabla_{\mu} \varphi \nabla^{\mu} \varphi$) is a Born-Infeld type kinetic term~\citep{Born:1934gh} and
\begin{eqnarray}
f(\varphi) &=&\frac{\Lambda c_{\infty}}{1-c_{\infty}^2} \cosh(\xi\varphi) \left[\sinh(\xi\varphi)\left[1+\left(1-c_{\infty}^2\right)\sinh^2(\xi\varphi)\right]\right]^{-1}\,,\\
V(\varphi) &=&\frac{\Lambda}{1- c_{\infty}^2}\left[1+\left(1-c_{\infty}^2\right)\sinh^2\left(\xi\varphi\right)\right]^{-1}\left[\left(1-c_{\infty}^2\right)^2\sinh^2\left(\xi\varphi\right)+2(1-c_{\infty}^2)-1\right]\,,
\end{eqnarray}
in which $c_{\infty}$ is a free parameter and $\xi = \sqrt{3/[4(1-c_{\infty}^2)]}$. At the background level, these models are indistinguishable from the $\Lambda$CDM model. The only free parameter $c_{\infty}$ is dependent on the effective sound speed of quartessence, $c_{\rm s}^2$, and for $t \to \infty$, we have $c_{\rm s}^2 \to c_{\infty}^2$. In fact, the sound speed evolves with redshift as~\citep{Bertacca:2008uf}
\begin{equation}
    \label{cs2-quartessence}
    c_{\rm s}^2(z)=\frac{\Omega_{\Lambda} c_{\infty}^{2}}{\Omega_{\Lambda}+(1-c_{\infty}^{2}) \Omega_{\rm DM}(1+z)^3}\,,
\end{equation}
where $\Omega_{\Lambda}$ and $\Omega_{\rm DM}$ are respectively the present day values of the density parameter for the effective cosmological constant and dark matter. Note, that for $c_{\infty} = 0$, one recovers the $\Lambda$CDM model. For the above model labeled as $\Lambda$CDM$+ c_{\infty}$~\cite{Camera:2017tws}, it has been found that {\it Planck} 2015 alone can estimate a very lower value of $S_8 = 0.719^{+0.15}_{-0.046}$ at 68\% CL, which is close to the result $S_8=0.745\pm0.039$ from KiDS-450~\cite{Kuijken:2015vca,Hildebrandt:2016iqg, FenechConti:2016oun, Joudaki:2016kym}, thus resolving the $S_8$ tension completely within $0.2 \sigma$.

\subsubsection{Scaling Symmetry and a Mirror Sector}

Since the Hubble rate (or rather, its inverse $H^{-1}$) sets the total volume of the observable Universe at any given epoch, changing $H$ means rescaling the universe's overall size. It turns out that if such a transformation is accompanied by a rescaling of all other important cosmological distances by the exact same amount, then important observables seen projected on the celestial sphere such as the CMB and LSS (including baryon acoustic oscillation (BAO)) are left entirely unchanged from our perspective. Fundamentally, this symmetry exists because (i) most cosmological data are analyzed in terms of $n$-point angular correlation functions, which are unaffected by the above scaling since angles are invariant under such transformation; (ii) the equations of motion describing the linear evolution of fractional density fluctuations (for photons, baryons, dark matter, etc.) in the Universe only depend on \emph{ratios} of distances, which are invariant under a rescaling of all length scales in the problem; and (iii) the initial conditions for structure formation in our Universe do not have an intrinsic scale built-in (i.e.\ their spectrum is a power law), allowing us to rescale the physical size of density fluctuations without changing their amplitude (up to a small harmless correction~\cite{Zahn:2002rr}). 

In the pre-recombination Universe ($z\gtrsim1100$), there are two important length scales in cosmology: the Hubble length $H^{-1}$ and the photon mean free path  $\dot{\kappa}^{-1} = (a n_{\rm e} \sigma_{\rm T})^{-1}$, where $a$ is the scale factor, $n_{\rm e}$ is the free electron density, and $\sigma_{\rm T}$ is the Thomson cross section. Rescaling both quantities by a constant factor $f$ (at all times) leaves cosmological observables invariant. Of course, scaling up the Hubble rate $H$ implies via the Friedmann equation ($H^2 = (8\pi G/3)\sum_i\rho_i$) that we must increase the energy density of the Universe at all times. To leave the evolution of the gravitational potentials unchanged, the Einstein equations then tell us that this must be done by equally scaling the individual energy densities $\rho_i$ of \emph{all} the constituent of the Universe, $\sqrt{G\rho_i} \to f \sqrt{G\rho_i}$. In all, Ref.~\cite{Cyr-Racine:2021alc} finds that a broad array of cosmological observables are invariant under the following transformation
\begin{equation}
    \label{eq:scaling_symmetry}
    \left\{ \sqrt{G\rho_i} \to f \sqrt{G\rho_i}, \,\,\sigma_{\rm T} n_{\rm e}(a)  \rightarrow  f \sigma_{\rm T}n_{\rm e}(a),\,\,  A_{\rm s}  \rightarrow  A_{\rm s}/f^{(n_{\rm s}-1)} \right\},
\end{equation}
for some constant $f$. The third entry in the transformation is necessary to ensure the initial amplitude of fluctuations is unchanged under the transformation. Here, $A_{\rm s}$ is the amplitude of scalar fluctuations and $n_{\rm s}$ is the scalar spectral index. The existence of this symmetry opens the doors for the CMB and LSS data to accommodate a larger value of the Hubble constant without necessarily degrading the quality of the fit between model and data, and also leave $S_8$ unchanged.

We emphasize that the above is \emph{not} another model that could alleviate the Hubble tension, but rather a general paradigm to understand which models can naturally accommodate large Hubble constant values while being automatically consistent with cosmological data. As in much of theoretical physics, the existence of a symmetry broken by different physical effects can lead to key insights about the fundamental origins of the Universe. In our particular case, the scaling symmetry given in Eq.~\eqref{eq:scaling_symmetry} is broken by the COBE/FIRAS measurement of the CMB black body spectrum~\cite{Fixsen1996,fixsen09}, which fixes the energy density of photons today (and thus prohibits us from performing the transformation $\sqrt{G\rho_\gamma} \to f \sqrt{G\rho_\gamma}$). Indirectly, COBE also prohibits us from adding more baryons to the Universe, since the baryon-to-photon ratio is precisely measured by  CMB anisotropies.
Nevertheless, nothing prevents us from adding \emph{dark photons} that couples to \emph{dark baryons} in a similar way to how regular photons couple to standard baryons\cite{Ackerman:2008gi,Feng:2009mn,Agrawal:2016quu,Foot:2002iy,Foot:2003jt,Foot:2004pa,Foot:2004wz}. Such a mirror dark sector has been studied extensively in the literature since it was long recognized that it could help explain why gravity is so much weaker than other forces in the Standard Model (SM). Could the Hubble tension be revealing the existence of such a mirror sector?  

While necessary to evade the COBE/FIRAS bound, the mirror sector does not implement by itself to the second important ingredient of the scaling transformation: the necessary increase of the photon scattering rate (the second entry in Eq.~\eqref{eq:scaling_symmetry}). This is certainly the most challenging aspect of the scaling transformation since it involves SM physics at low energies, which is tightly constrained by a multitude of observations. Ref.~\cite{Cyr-Racine:2021alc} used a phenomenological approach based on varying the primordial helium abundance, which runs into serious constraints from light-element abundance measurements.\footnote{A realistic model of Dark Atoms based on it has been explored in~\cite{Blinov:2021mdk}.} A more promising approach likely requires variation of fundamental constants, especially the fine-structure constant and the electron mass. Varying these parameters has shown some promises in alleviating the Hubble tension (see e.g.\ Refs.~\cite{Sekiguchi:2020teg, Hart:2019dxi, Hart:2021kad}), and the scaling symmetry given above can help us understand why such ideas work well. Another challenge is how to reconcile BBN predictions in the presence of a rescaled Hubble rate with direct abundance measurements. 

In all, the scaling symmetry highlights the important role that the photon scattering rates and BBN light-element abundances play in the $H_0$ and $S_8$ tension. It provides a clear target for model builders that, if reached, would guarantee compatibility between CMB, LSS, and local inference of $H_0$.

\subsubsection{Self-Interacting Neutrinos}\label{sec:SInu}

The physics of neutrinos is one of the fascinating topics in modern cosmology even though we have been able to extract only a very minimal information about the neutrinos. Recently, it has been  observed that  
the introduction of new physics in the neutrino sector in terms of the interactions between themselves~\cite{Kreisch:2019yzn,Lancaster:2017ksf,Blinov:2019gcj,Barenboim:2019tux}, which is theoretically possible, might be able to offer an appealing route to alleviate the cosmological tensions~\cite{Kreisch:2019yzn,Venzor:2022hql}. After decoupling from other standard model particles, neutrinos are considered to be freely
streaming throughout the Universe, but they can interact with each other or with other cosmic species via their gravitational interactions. This gravitationally influenced interaction could affect the cosmic observables and this may result in interesting consequences, including opening up the inflation parameter space~\cite{Barenboim:2019tux}. In Ref.~\cite{Kreisch:2019yzn} the authors considered a Lagrangian describing the interaction between different neutrino mass eigenstates of the form 
\begin{eqnarray}\label{lagrangian-self-interacting-neutrinos}
\mathcal{L}_{\rm int} = \theta_{ij} \bar{\nu}_j \nu_i \phi,
\end{eqnarray}
where  $\nu_i$ is a left-handed neutrino Majorana spinor, $\theta_{ij}$ is a (generally complex) coupling matrix in which the indices $i$, $j$ labeled the neutrino mass eigenstates. This is a Yukawa-type interaction with a massive
scalar $\phi$ and this may arise in the scenarios where neutrinos interact with a Majoron~\cite{Gelmini:1980re,Simpson:2016gph,Berlin:2018ztp}. The presence of the interaction in the neutrino sector quantified through the above Lagrangian may delay the onset of neutrino free  streaming until close to the matter-radiation equality leading to a higher value of the Hubble constant and a lower value of the matter fluctuations~\cite{Kreisch:2019yzn}.

\subsubsection{Self-Interacting Sterile Neutrinos}\label{sec:SISnu}

The short baseline (SBL) neutrino oscillation experiments~\cite{Giunti:2019aiy} suggest the existence of a fourth sterile neutrino with a $\sim$ eV mass. However, the existence of sterile neutrinos is not compatible with cosmology since they suppress the structure formation of the Universe. This problem can be bypassed if by some unknown mechanism they can be either prevented from thermalizing in the early Universe or removed by subsequent annihilation. Considering  such limitation,  the authors of Ref.~\cite{Archidiacono:2014nda} suggested a possible revision in the sterile neutrino sector in which the sterile neutrino interacts with a new light pseudoscalar degree of freedom. This proposal got attention in the cosmological community because with such coupling in the sterile neutrino sector, the effective scenario can increase the Hubble constant value and consequently alleviate the Hubble constant tension~\cite{Archidiacono:2015oma,Archidiacono:2016kkh,Archidiacono:2020yey,Corona:2021qxl}.

\subsubsection{Soft Cosmology}

In soft cosmology~\cite{Saridakis:2021qxb,Saridakis:2021xqy} one allows for small deviations from the usual cosmological framework due to the effective appearance of soft-matter properties in the Universe sectors~\cite{jones2002soft,sagis2011dynamic}. Hence, as a first approach on the subject, one considers that, intrinsically or effectively, dark energy and/or dark matter may have a different EoS  at large scales, i.e. at scales entering the Friedmann equations, and a different one  at intermediate scales,  i.e. at scales entering the perturbation equations. This possible deviation of the large-scale (ls) and intermediate-scale (is) EoS can be quantified by introducing the softness function $s$. Hence, in the simplest scenario one has 
\begin{eqnarray}
    \label{wdesoft}
    w_{\rm DE,is} &=& s_{\rm DE}\cdot w_{\rm DE,ls}\,,\\
    w_{\rm DM,is}+1 &=& s_{\rm DM}\cdot (w_{\rm DM,ls}+1)\,, \label{wdmsoft}
\end{eqnarray}
where $s_{\rm DE}$ and $s_{\rm DM}$ are the softness parameters for the dark energy and dark matter sectors, respectively. Standard cosmology is recovered for $s_{\rm DE}= s_{\rm DM}=1$. Thus, although the Friedmann equations are $H^{2}=\frac{\kappa^2}{3}(\rho_b+\rho_r+\rho_{\rm DM}+\rho_{\rm DE})$ and $ 2\dot{H} + 3 H^2  = -\kappa^2 (p_b+p_r+p_{\rm DM}+p_{\rm DE})$, where $w_{i,{\rm ls}}\equiv p_i/\rho_i$ is the equation-of-state parameter of the $i$th sector at large scales, the (linear, scalar, isentropic) perturbation equations in the Newtonian gauge read
\begin{eqnarray}
    \dot{\delta}_i+(1+w_{i,{\rm is}})\left(\frac{\theta_i}{a}-3\dot{\Psi}\right)+3H\left((c_{\rm eff}^{(i)})^2 -w_{i,{\rm is}} \right) \delta_i &=& 0\,,\label{eq:line2}\\
    \dot{\theta}_i+H\left(1-3 w_{i,{\rm is}} +\frac{\dot{w}_{i,{\rm is}} }{H(1+w_{i,{\rm is}})}\right)\theta_i -\frac{k^2(c_{\rm eff}^{(i)})^2\delta_i}{(1+w_{i,{\rm is}})a}-\frac{k^2\Psi}{a} &=& 0\,, \label{eq:line4}
\end{eqnarray}
where $\delta_i\equiv \delta\rho_i/\rho_i$ is the density perturbation, $\theta_i$ the divergence of the fluid velocity, $k$ the wavenumber of the Fourier modes, and $(c_{\rm eff}^{(i)})^2\equiv \frac{\delta  p_i}{\delta\rho_i}$ the effective sound speed square for the $i$th sector which determines the clustering properties, being zero for maximal clustering and one for no clustering.

\begin{figure}[!]
    \includegraphics[width=7.5cm,height=5cm]{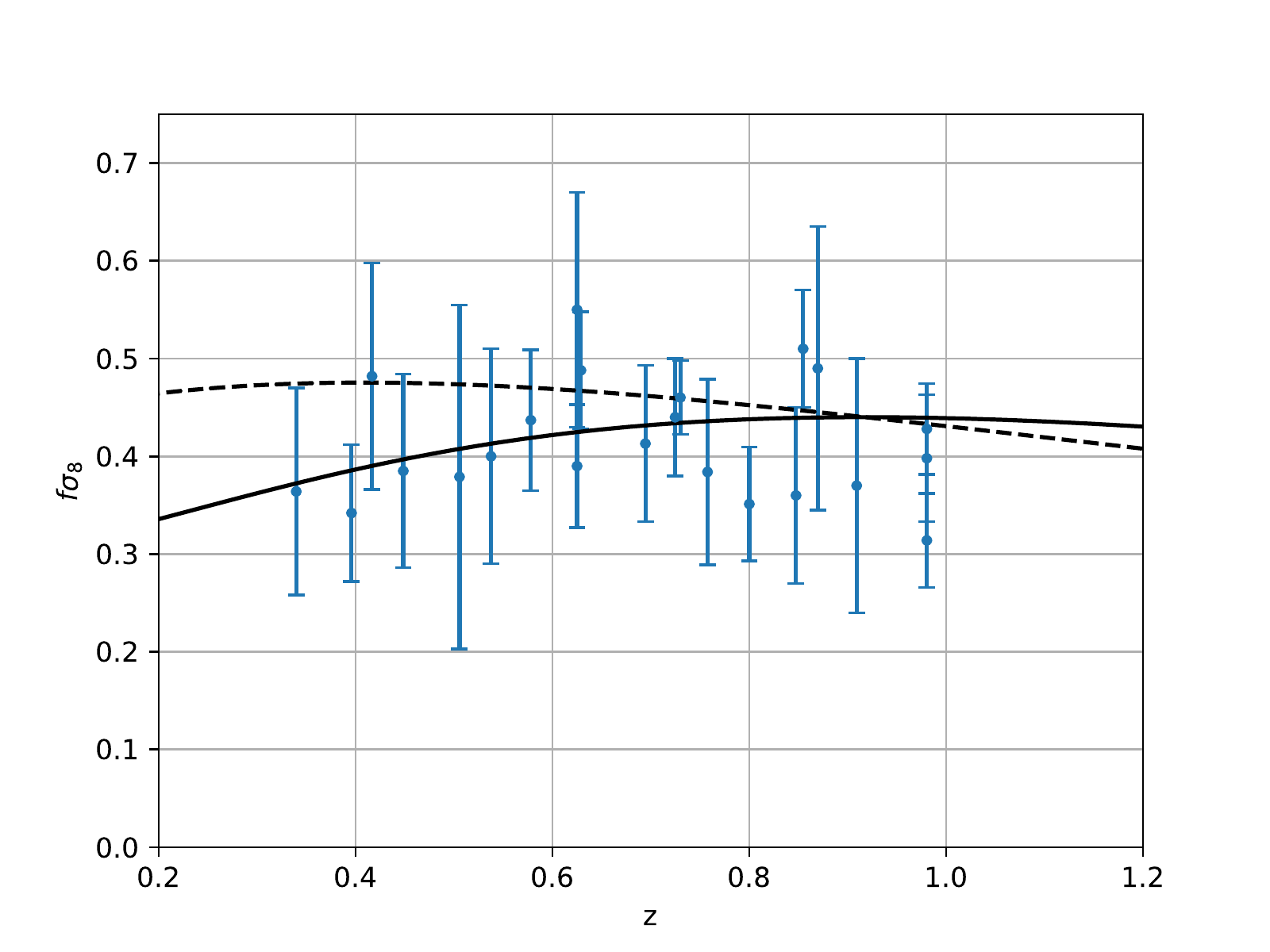}\includegraphics[width=7.5cm,height=5cm]{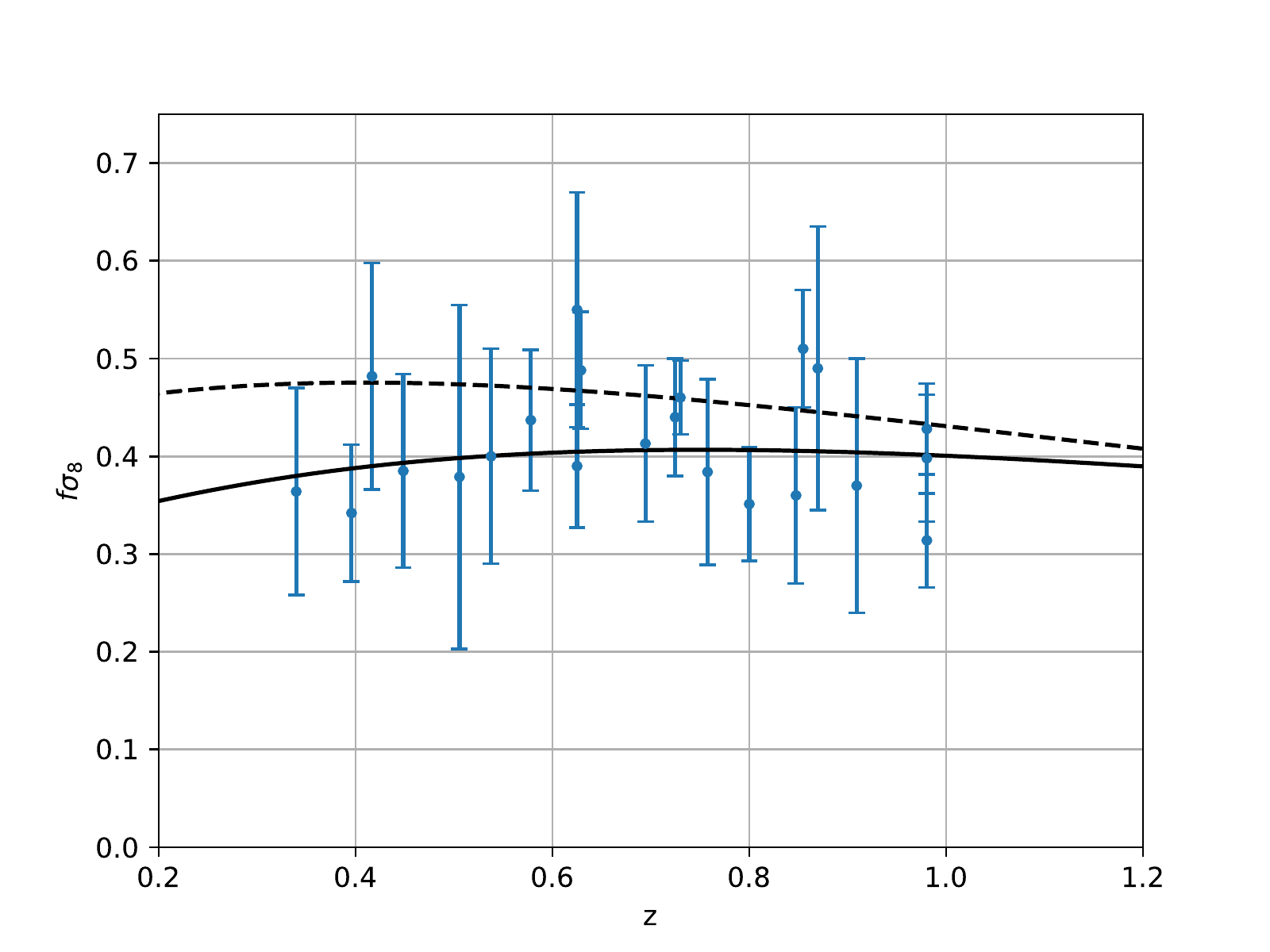}
	\caption{ Left panel: The $f\sigma_8$ as a function of $z$ in the case of soft dark energy. The dashed curve is for $\Lambda$CDM. The solid curve is for soft dark energy with $s_{\rm DE}=1.1$, i.e. with  $w_{\rm DE,ls}=-1$ and $w_{\rm DE,is}=-1.1$, and $c_{\rm eff, DE} = 0.1$, while dark matter is standard (i.e.\ not soft) with $w_{\rm DM}=0$. Right panel: The $f\sigma_8$ as a function of $z$ in the case of soft dark matter. The dashed curve is for $\Lambda$CDM. The solid curve is for soft dark matter with softness parameter $s_{\rm DM}=1.05$, i.e. for dark matter with $w_{\rm DM,ls}=0$ and $w_{\rm DM,is}=0.05$ (note that dark energy is not soft). The figures are from~\cite{Saridakis:2021qxb}. }
    \label{softfig1}
\end{figure}

{\it Soft Dark Energy} --  As the first soft extension of $\Lambda$CDM scenario we consider a model where dark matter is the usual, non-soft, dust one at all scales, while dark energy is the soft component with large-scale behavior that of a cosmological constant. Hence, we impose fixed $s_{\rm DM}=1$, namely $w_{\rm DM,ls}=w_{\rm DM,is}=0$ as in the standard dust dark matter case, while we set $w_{\rm DE,ls}=-1$ and $w_{\rm DE,is}= s_{\rm DE} w_{\rm DE,ls}=-s_{\rm DE}$, so that $s_{\rm DE}$ is the only extra free parameter comparing to $\Lambda$CDM cosmology. Standard cosmology is recovered for the value $s_{\rm DE}=1$. In the left graph of Fig.~\ref{softfig1} we depict the $f\sigma_8$ as a function of $z$, where we can see alleviation of the $S_8$ tension.

{\it Soft Dark Matter} --  As another soft extension of $\Lambda$CDM scenario we consider a model where dark energy is the usual cosmological constant, with $w_{\rm DE,ls}=w_{\rm DE,is}=-1$, however dark matter is soft with $w_{\rm DM,ls}=0$ and $w_{\rm DM,is}= s_{\rm DM}-1$, and therefore $s_{\rm DM}$ is the only extra free parameter comparing to $\Lambda$CDM cosmology (the latter is recovered for the value $s_{\rm DM}=1$). In the right graph of Fig.~\ref{softfig1} we depict the $f\sigma_8$ as a function of $z$, where we can see the $S_8$ tension alleviation.

In summary, soft cosmology  can alleviate the $S_8$ tension due to the slightly deviated perturbation-level properties it introduces. Finally, we mention that the whole consideration is independent of the underlying gravitational theory, and it can be applied both in the framework of general relativity, as well as in modified theories of gravity in which dark energy sector is of gravitational origin (nevertheless, deviating from general relativity would provide additional possibilities to induce effective soft properties to the dark sectors).

\subsubsection{Two-Body Decaying Cold Dark Matter into Dark Radiation and Warm Dark Matter}
\label{sec:CDMdarkdecay}

One of the solutions proposed to solve the $S_8$ tension is given by a two-body Decaying Cold Dark Matter (dCDM) model where all of the dark matter is assumed to decay into a massless (dark radiation) and a massive warm DM (WDM) species. See Sec.~\ref{sec:dDM} for alternative models of decaying DM. The phenomenology of such a model has recently been reviewed in great details in Ref.~\cite{Abellan:2021bpx}. The model is characterized by two parameters, the dCDM lifetime, $\Gamma^{-1}$, and the fraction of dCDM rest mass energy transferred to the dark radiation, given by~\cite{Blackadder:2014wpa}
\begin{equation}
    \label{eq:epsilon}
    \varepsilon = \frac{1}{2} \left(1-\frac{m^2_{\rm WDM}}{m^2_{\rm dCDM}}\right)\,,
\end{equation}
where $m_{\rm dCDM}$ and $m_{\rm WDM}$ denote the mass of the parent particle and massive decay product respectively. Thus, $0 \leq \varepsilon \leq 1/2$, with the lower limit corresponding to the standard CDM case, so that $\Omega_{\rm CDM} = \Omega_{\rm dCDM}+\Omega_{\rm WDM}$, and $\varepsilon = 1/2$ corresponding to DM decaying solely into dark radiation. In general, small $\varepsilon$ values (i.e.\ heavy massive decay products) and small $\Gamma$ values (i.e.\ lifetimes much longer than the age of the Universe) induce little departures from $\Lambda$CDM. In the intermediate regime, the velocity-kick received by the warm decay product imprints a characteristic suppression to the matter power spectrum, in a similar fashion as massive neutrinos or warm dark matter. One key difference with respect to these scenarios is the fact that the power suppression is less significant at high redshift, simply because the abundance of the warm decay product is smaller in the past. This allows the model to reduce the $\sigma_8$ value as compared to that inferred from the standard $\Lambda$CDM model, while preserving a good fit to CMB, BAO, growth factor and uncalibrated SNIa data. The authors of Ref.~\cite{Vattis:2019efj} suggested that such a model can provide a resolution of the Hubble tension (see also~\cite{Clark:2020miy}).

However, this conclusion was recently challenged in Refs.~\cite{Haridasu:2020xaa}, where it was shown that including CMB and SNIa data spoils the success of the model to resolve the Hubble tension. Still, in Ref.~\cite{Abellan:2020pmw}, it was shown thanks to a new fluid approximation to describe the perturbations of the massive decay products, that such a model can in fact resolve the $S_8$ tension if $\Gamma^{-1} \simeq 55\,(\varepsilon/0.007)^{1.4}\,$Gyr. As discussed in Ref.~\cite{Abellan:2020pmw}, this model could also have interesting implications for model building, the small-scale crisis of $\Lambda$CDM and the recent XENON1T excess.

\subsection{Beyond the FLRW Framework}
\label{sec:beyond_FLRW}

The Hubble constant $H_0$ has a special standing within the class of  Friedmann-Lema{\^i}tre-Robertson-Walker (FLRW) cosmological models, since it arises as an integration constant when one solves the Friedmann equations. As a result, it is generic to \textit{all} FLRW cosmologies. Now, it is no secret that our local Universe is \textit{not} an FLRW Universe and that there are well documented flows towards the Shapley supercluster at a distance of approximately $200\,$Mpc, e.g.\ Ref.~\cite{Hoffman:2017ako}. In the aftermath of the HST Key project~\cite{HST:2000azd}, it was also recognised that $H_0$ varies across the sky in the local Universe~\cite{McClure:2007vv}. This in itself is unsurprising, since the underlying Hubble flow and the rate of expansion need to be inferred from these averages. Historically, this has been an insanely difficult endeavour and even now disagreements have emerged in the distance ladder between Cepheid~\cite{Riess:2020fzl} and TRGB calibration~\cite{Freedman:2021ahq}. 

Recall that the cosmological principle 
paradigm simply states that the Universe is well modelled as being isotropic and homogeneous at \textit{suitably large scales}. 
If one further assumes the decoupling of scales in the underlying metric theory, such that the dynamical theory defined on such large scales is insensitive to the particularities of the regional matter configurations and dynamics, this implies that cosmic structures and their evolution can be ignored altogether at such scales and that the large-scale Universe can be modelled by an FLRW metric.
If one stipulates that the large-scale Universe is well described by a homogeneous-isotropic  model, one may ask what the scale of transition towards homogeneity and isotropy is? While this statement is vague, one can try to quantify what is meant by "suitably large scales", and within the flat $\Lambda$CDM model, estimates of $260 h^{-1}$ Mpc exist~\cite{Yadav:2010cc} as a scale beyond which the fractal dimension of the distribution becomes statistically indistinguishable from that of a perfectly homogeneous and isotropic distribution.
The homogeneity scale estimates depend on the statistical measure used. 
Taking into account all correlation properties in the distribution by employing the Minkowski Functionals~\cite{Mecke:1994ax} the homogeneity scale seems to be even larger~\cite{Wiegand:2013xfa}.  
There are claims of large structures in the Universe that appear challenging for the hypothesised transition to homogeneity, for example the Sloan Great Wall~\cite{Gott:2003pf}, the Huge Large Quasar Group~\cite{Clowes:2012pn} or the  Hercules–Corona Borealis Great Wall~\cite{Horvath:2015axa}. These claims of violations of the cosmological principle have quickly been  countered~\cite{Park:2012dn, Nadathur:2013mva, Ukwatta:2015rxa, Christian:2020den}, usually on statistics alone. In short, despite a host of claims, there is little evidence to support any breakdown of the cosmological principle, at least with inhomogeneities.

The strongest observational evidence for an isotropic and homogeneous Universe comes from the CMB. Even to the naked eye, {\it Planck}'s maps of micro-Kelvin temperature anisotropies provide a convincing snapshot of a statistically isotropic Universe. 
The statistically isotropic CMB temperature field is conventionally used as a heuristic argument for modelling the large-scale Universe by an FLRW cosmological model. However, the FLRW model assumption is stronger than that of statistical homogeneity and isotropy; cf. the above discussion. Care must therefore be taken in making rigorous any statements about the underlying metric model based on observations such as the CMB.  
The CMB photon energy field as viewed by the observer is a quantity which is integrated over scales comparable to the age of the Universe. It is possible that large regional departures from FLRW curvature can be present without breaking of the almost-isotropy of the CMB. Extended versions of the Ehlers-Geren-Sachs theorem~\cite{1995ApJ...443....1S,Rasanen:2009mg} that seek to constrain the departures from an FLRW space-time metric based on the level of isotropy of the CMB temperature field within general relativity, must make assumptions on individual components of the covariant \emph{derivatives} of the energy function for classes of observers to arrive at the conclusion that the Universe is "almost-FLRW".\footnote{The original Ehlers-Geren-Sachs theorem~\cite{Ehlers:1966ad} states that the space-time is FLRW or static if the Universe energy-momentum content is a \emph{perfectly} isotropic radiation fluid around every point. Almost-versions of the theorem are concerned with the realistic scenario of almost-isotropic CMBs as viewed by observers in space-times with realistic energy-momentum content.} These components are not directly observable, and are in fact expected to be large in a Universe with large density contrasts~\cite{Rasanen:2009mg}. Thus, even though appealing as an argument for FLRW modeling of the Universe as a whole, almost-isotropy of the CMB does not constitute a proof of the almost-FLRW Universe. 

The most notable anisotropy in the CMB is a dipole $(\ell =1$ multipole), the magnitude of which is in the milli-Kelvin range. This dipole is \textit{assumed} to be a Doppler effect due to relative motion. Obviously, this assumption is a loose end, and now that we find ourselves in a time of crisis, potentially confronted by a $\sim 10 \%$ discrepancy in $H_0$, it is prudent to test the assumption that the dipole is purely kinematic in origin.  
Interestingly, those tests have been pursued over the last two decades, and caveats aside, a consensus has emerged whereby the \textit{magnitude} of the cosmic dipole inferred from both radio galaxies~\cite{Blake:2002gx, Singal:2011dy, Gibelyou:2012ri, Rubart:2013tx, Tiwari:2015tba, Colin:2017juj, Bengaly:2017slg, Siewert:2020krp} and quasars~\cite{Secrest:2020has} exceeds the CMB dipole. Similar conclusions have been recently reached by studying the Hubble diagram of Type Ia SN~\cite{Singal:2021crs} and QSOs~\cite{Singal:2021kuu}. This na\"{i}vely implies that distant sources are not in the "CMB frame" and opens up the possibility that the CMB dipole may have been misinterpreted. On the flip side, a recent analysis of the high CMB multipoles is consistent with a kinematic CMB dipole~\cite{Saha:2021bay} (see also Ref.~\cite{Ferreira:2020aqa}), but the errors are large. Curiously, various CMB anomalies exist (see Ref.~\cite{Schwarz:2015cma} for a review), such as the planar alignment of the quadrupole-octopole~\cite{deOliveira-Costa:2003utu, Schwarz:2004gk}  and the anomalous parity asymmetry~\cite{Land:2005jq}, which may have some origin in the CMB dipole~\cite{Kim:2010gf, Naselsky:2011jp}. Objectively, if there is no coherence in CMB anomalies, then few conclusions can be drawn. However, if the anomalies have directional dependence, then the plot thickens.  
Assuming there is a mismatch in the cosmic dipole that can be independently verified by SKAO~\cite{Bengaly:2018ykb}, the simplest and most consequential interpretation is that the Universe is anisotropic. While current data suggests that flat $\Lambda$CDM is a good approximation, one can ask how pronounced is any anisotropy, if real? Tellingly, the CMB dipole direction also appears in a documented preferred axis for QSO polarisations~\cite{Hutsemekers:2000fv, Hutsemekers:2005iz} and scaling relations in galaxy clusters~\cite{Kaiser:1986ske} also point to an anisotropy consistent with the CMB dipole direction~\cite{Migkas:2020fza, Migkas:2021zdo}. Furthermore, observations of strong lensing time delay~\cite{Wong:2019kwg, Millon:2019slk}, Pantheon Type Ia SN~\cite{Pan-STARRS1:2017jku}, QSOs standardised through UV-X-ray relation~\cite{Risaliti:2015zla, Lusso:2020pdb} and GRBs standardised through the Amati relation~\cite{Amati:2008hq}, consistently return higher values of $H_0$ in hemispheres aligned with the CMB dipole direction, at least within the flat $\Lambda$CDM model~\cite{Krishnan:2021dyb, Krishnan:2021jmh, Luongo:2021nqh}. Obviously, variations of cosmological $H_0$ across the sky with the flat $\Lambda$CDM model make any discrepancy between {\it Planck}'s determination of $H_0$ with the \textit{same} model and local $H_0$ determinations a moot point.  
Given the rich variety of the above observations, and the differences in underlying physics, it is hard to imagine that the CMB dipole direction is not a special direction in the Universe. The status quo of simply assuming that it is kinematic in origin, especially since it is based on little or no observational evidence, may be untenable. If this claim is substantiated, not only do the existing cosmological tensions in $H_0$ and $S_8$ need revision, but so too does virtually all of cosmology. In short, great progress has been made through the cosmological principle, but it is possible that data has reached a requisite precision that the cosmological principle has already become obsolete.

\subsubsection{Cosmological Fitting and Averaging Problems}
\label{sec:fittingproblem}

The cosmological principle stipulates that there is a length scale beyond which the Universe appears as being approximately spatially homogeneous and isotropic. If one further assumes the decoupling of dynamics of volume sections beyond such a scale of homogeneity and isotropy from the local space-time configurations of matter and curvature, this implies the modelling of the large-scale Universe by an FLRW metric model. The $\Lambda$CDM framework of cosmology builds on this assumption of decoupling. However, going beyond the $\Lambda$CDM framework, one could imagine a Universe which \emph{does} obey the cosmological principle, but where the regional distributional properties of matter and curvature has importance for the large scale cosmological dynamics. 

The problem of taking into account such distributional properties in the large scale modelling of our Universe is sometimes referred to as the \emph{cosmological fitting problem}~\cite{Ellis:1984bqf,Ellis:1987zz}. Various approaches to formalising the fitting problem that build on averaging operations of Einstein's equations have been developed in e.g.\ Refs.~\cite{Zalaletdinov:1992cg, Zalaletdinov:1992cf, Zalaletdinov:1996aj, Buchert:1999er, Buchert:2001sa, Buchert:2007ik, Wiltshire:2007jk, Rasanen:2008be, Rasanen:2009uw, Korzynski:2009db, Gasperini:2009wp, Gasperini:2009mu, Buchert:2011sx, Wiltshire:2011vy, Gasperini:2011us, Ben-Dayan:2012uam, Uzun:2018yes, Buchert:2019mvq}. Within such formalisms, the departure of the large-scale dynamics from that of an FLRW Universe without structure is sometimes denoted \emph{cosmological backreaction}; see Refs.~\cite{Green:2014aga,Buchert:2015iva} for a debate on the level of importance of backreaction effects in our Universe. 

The Buchert averaging scheme~\cite{Buchert:1999er, Buchert:2001sa, Buchert:2019mvq} is the most widespread framework for quantifying backreaction effects; see also~\cite{Buchert:2022zaa} for an extension of the Buchert averaging scheme to the past lightcone of an observer. In this framework, it is made explicit how the large-scale volume evolution is affected by backreaction terms. For instance, in an irrotational dust Universe backreaction in the matter frame is given by variance in expansion rate over volume and regional shearing~\cite{Buchert:1999er}. The backreaction effects can in this case be collected in a single term, which enters in the average dynamical equations for the large-scale volume in such a way that it mimics a dark energy-like effect for large-scale isotropic averages (dominance of expansion variance between voids and matter-dominated regions), but also a dark matter-like effect on smaller scales where the backreaction term can change its sign in cases where anisotropic structures dominate the distribution~\cite{Buchert:1999er, Buchert:2007ik}. The backreaction term in turn couples to the average spatial Ricci-curvature of the space-time, thus breaking the Friedmann conservation law ${}^{(3)}\! R \, a^2(t) = $const, where ${}^{(3)}\! R$ is the spatial Ricci-curvature and $a(t)$ is the scale factor of the FLRW Universe model; see Refs.~\cite{Buchert:1999er, Heinesen:2020sre} for details. 

Backreaction as defined by Buchert~\cite{Buchert:1999er} reduces to a boundary term in Newtonian cosmology~\cite{Buchert:1995fz}; however, see Ref.~\cite{Vigneron:2021tpi} describing the possibility of volume-integrated backreaction in locally Newtonian theory defined on non-Euclidean topologies, and see also~\cite{Brunswic:2020bjx} for an assessment of topology and GR integral constraints. Non-zero backreaction effects for domains without boundaries defined on non-Euclidean topological structures must thus arise from relativistic effects; see~\cite{Rasanen:2010wz} and references therein. 
However, such relativistic effects need not be associated with astrophysical structures with large relative speeds or with structures in the "strong-field" regime,\footnote{See~\cite{Korzynski:2014nna} for an example of an exact GR solution where all of the individual structures are in the "weak-field" regime, but where the global matter distribution nevertheless exhibits manifestly relativistic effects.} in the sense of the Schwarzschild potential $2GM/R$, where $M$ is the mass and $R$ is the radius of the structure. We remark that backreaction effects could arise in Newtonian \emph{approximations} of general relativity: The small velocity and weak-gravitational potential limit of general relativity is not strictly Newtonian but subject to additional degrees of freedom and constraints~\cite{Ellis:1994md,Rainsford:2000ew}; see for instance Ref.~\cite{Senovilla:1997bw} and references therein for examples of Newtonian theories that cannot be recovered in any limit of general relativity.   
We might further add to the above mentioned subtleties that, even for space-times with negligible backreaction as measured by the backreaction functionals defined in~\cite{Buchert:1999er, Buchert:2001sa, Buchert:2019mvq}, observations might exhibit non-negligible departures from any simple FLRW model prediction, depending on the  scale of observation and the type of the cosmological measurement. The reverse is  also true: it is in principle possible to have space-times where certain backreaction terms are formally large, but where observations are close to an FLRW prediction. 

The dynamical differences of the large-scale universal dynamics from that prescribed by a structure-less FLRW model space-time would be a natural candidate for explaining the differences in $\Lambda$CDM parameters as inferred from early and late epoch Universe probes. The potential for explaining the Hubble tension with dynamical backreaction effects has been examined within crude analytical approaches~\cite{Heinesen:2020sre} and numerical relativity~\cite{Macpherson:2018akp}, but is far from being an exhausted topic of investigation.
The cosmological fitting problem and its significance for Universe observations -- including the impact of cosmic structure on the determination of "background" cosmological parameters such as the Hubble constant -- is a rich topic of investigation with many subtleties as indicated in the above review. 
Regional inhomogeneity and anisotropy and any resulting backreaction effects on the large scale Universe description are appealing possible solutions to the dark energy problem, coincidence problem, and parameter tensions in modern cosmology within the theory of general relativity and without the introduction of additional energy-momentum degrees of freedom. 
Though any physical backreaction effects must likely arise from subtle departures from strict Newtonian Universe modelling, and is therefore an involved topic of research, it might thus at the same time be argued to be one of the simplest possible solutions to the modern challenges faced in cosmology.

\subsubsection{Data Analysis in an Universe with Structure: Accounting for Regional Inhomogeneity and Anisotropy}
\label{sec:cosmography}

In a Universe that contains structures, a spatial invariance with respect to translation and rotation transformations is necessarily broken at some level. As suggested in the above introduction, such symmetry breaking might be more pronounced and important for cosmological inference than is usually accounted for in $\Lambda$CDM analysis of data. One way to take into account regional inhomogeneities and anisotropies is to employ exact solutions or perturbative models, a few examples of which are discussed in Sec.~\ref{sec:LVS}. Another option is to remain agnostic about the particularities of the regional properties of the underlying space-time metric.

For this purpose, we might use cosmographic approaches\footnote{The cosmographic frameworks listed here do \emph{not} assume the FLRW class of metrics (or any other particular form of the metric), and their applications are thus more general than those in Ref.~\cite{Visser:2003vq}.} for analysis of data that do not make assumptions about the form of the metric or the determining field equations~\cite{Kristian:1965sz, 1985PhR...124..315E, Ellis:1968vb,Clarkson:2010uz,Clarkson:2011br,Umeh:2013UCT,Heinesen:2020bej,Heinesen:2021qnl,Bargiacchi:2021fow,Benetti:2019gmo}. Such approaches rely on geometrical series expansions around the observer, and are thus most appropriate for analysing low-redshift data.\footnote{Traditionally, cosmography is an expansion in $z$, but one should expect the approximation to break down when $z$ becomes a large parameter, $z \gtrsim 1$. See Refs.~\cite{Cattoen:2007sk, OColgain:2021pyh} for discussion. } See also Refs.~\cite{Korzynski:2021aqk,Korzynski:2017nas,Heinesen:2021nrc} for non-perturbative frameworks for model independent determination of Universe properties. 

Cosmographic frameworks for model-independent analysis of different observables, such as distance--redshift data, number count statistics, and cosmic drift effects, can be formulated. 
As an illustration, we shall in the present section focus on luminosity distance cosmography relevant for analysing low-redshift standardisable candles and sirens. 
Following Ref.~\cite{Clarkson:2011br,Umeh:2013UCT,Heinesen:2020bej}, we will thus formulate luminosity distance as a Taylor series expansion in redshift in a general continuous and expanding space-time description of emitters and observers between which photons travel on null geodesics. 
Assuming that Etherington's reciprocity theorem for relating luminosity distance to angular diameter distance~\cite{Etherington1933,Ellis:1998ct} holds, the luminosity distance as a function of redshift, $z$, and position on the sky, $\bm e$, of the astrophysical source is given by its Taylor series~\cite{Heinesen:2020bej}
\begin{equation}
    \label{eq:series}
    d_L(z,\boldsymbol{e}) =  d_L^{(1)}(\boldsymbol{e}) z + d_L^{(2)} (\boldsymbol{e})z^2 +  d_L^{(3)}(\boldsymbol{e}) z^3 + \mathcal{O}( z^4),
\end{equation} 
within the radius of convergence of the series, where the anisotropic coefficients up till third order:   
\begin{equation}
\begin{aligned}
\label{eq:dLexpand2}
d_L^{(1)}(\boldsymbol{e}) &= \frac{1}{\mathfrak{H}_o (\boldsymbol{e}) } \, ; \qquad d_L^{(2)}(\boldsymbol{e}) =   \frac{1 - \mathfrak{Q}_o(\boldsymbol{e}) }{2 \mathfrak{H}_o(\boldsymbol{e})}  \, ; \qquad 
d_L^{(3)}(\boldsymbol{e}) =  \frac{- 1 +  3 \mathfrak{Q}_o^2(\boldsymbol{e}) + \mathfrak{Q}_o(\boldsymbol{e})    -  \mathfrak{J}_o(\boldsymbol{e})   + \mathfrak{R}_o(\boldsymbol{e}) }{ 6  \mathfrak{H}_o(\boldsymbol{e})}     \, ,
\end{aligned}
\end{equation} 
can be expressed in terms of the effective cosmological parameters
\begin{align} \label{eq:paramseff}
    \mathfrak{H}(\boldsymbol{e}) &\equiv - \frac{1}{E^2}     \frac{ {\rm d} E }{{\rm d} \lambda}   \, ; \quad \; \,
    \mathfrak{Q}(\boldsymbol{e})  \equiv - 1 - \frac{1}{E} \frac{     \frac{ {\rm d} \mathfrak{H}}{{\rm d} \lambda}    }{\mathfrak{H}^2}   \, ; \quad \; \,
    \mathfrak{R}(\boldsymbol{e}) \equiv  1 +  \mathfrak{Q}  - \frac{1}{2 E^2} \frac{k^{\mu}k^\nu R_{\mu \nu} }{\mathfrak{H}^2}   \, ;  \quad \; \,
    \mathfrak{J}(\boldsymbol{e})  \equiv   \frac{1}{E^2} \frac{      \frac{  {\rm d^2} \mathfrak{H}}{{\rm d} \lambda^2}    }{\mathfrak{H}^3}  - 4  \mathfrak{Q}  - 3 \, ,
\end{align} 
where the subscript $o$ denotes evaluation at the space-time event of observation.
Here, $\lambda$ is the affine parameter defined on the photon path, ${\rm d}/{\rm d}\lambda \equiv k^\mu \nabla_\mu$ is the derivative evaluated along the incoming null ray, where ${\bm k}$ is the 4--momentum of the photon, with energy $E = -u^\mu k_\mu$ as measured in the congruence frame of emitters and observers as generated by the 4--velocity field $\bm u$, and $z \equiv E/E_o - 1$ is the associated redshift function relative to the observer of interest at $o$. The Ricci curvature of the space-time, $R_{\mu \nu}$, is left unspecified in the general framework and might be given by any theory of gravity. 

We notice that $\mathfrak{H}$ replaces the Hubble parameter of the FLRW cosmography. Thus it is the relation $d_L = z/\mathfrak{H}$ that consistently determines distances in the $\mathcal{O}(z)$ vicinity of the observer for a general space-time. Similarly, the parameters $\mathfrak{Q}$, $\mathfrak{J}$, and $\mathfrak{R}$ generalise the FLRW deceleration, jerk and curvature parameters in the FLRW cosmography results~\cite{Visser:2003vq}. For this reason, we refer to $\{\mathfrak{H},\mathfrak{Q},\mathfrak{J},\mathfrak{R}\}$ as effective observational Hubble, deceleration, jerk and curvature parameters of the luminosity distance--redshift relation.

The effective cosmological parameters carry information about the space-time kinematics and curvature in the cosmic neighbourhood of the observer. We can get an idea of how such information enters in the effective Hubble parameter, by performing a multipole expansion of $\mathfrak{H}$ in the direction vector, ${\bm e}$, of the source~\cite{Clarkson:2011br,Umeh:2013UCT,Heinesen:2020bej}: 
\begin{equation}
    \label{def:generalH}
    \mathfrak{H}(\boldsymbol{e}) = \frac{1}{3}\theta  - e^\mu a_\mu + e^\mu e^\nu \sigma_{\mu \nu}   \,  , 
\end{equation} 
where $\theta$ is the expansion rate of space in the observer frame, and where $\sigma_{\mu \nu}$ is the shear tensor describing the anisotropic deformation of space. The 4--acceleration of the observer $a^\mu$ will usually be set to zero when non-gravitational forces are neglected. The expression in Eq.~\eqref{def:generalH} is exact, and the truncation of the expansion in $\bm e$ at quadrupolar order is true for any congruence description of emitters and observers. 

The Hubble law determining cosmological distances in the $\mathcal{O}(z)$ vicinity of the observer will vary between observers through spatial variations of $\theta$ and $\sigma_{\mu \nu}$ due to breaking of the translation invariance. The Hubble law will also in general vary over the individual  observer's sky on account of the anisotropy of expansion incorporated in the shear tensor, $\sigma_{\mu \nu}$. 
If, for instance, the direction, $\bm e$, of a particular source as viewed by the observer is pointing towards a direction where $ e^\mu e^\nu \sigma_{\mu \nu} > 0$, there is a positive contribution to the effective Hubble parameter prescribing the $\mathcal{O}(z)$ distance to that source. 

\begin{figure*}[ht]
    \centering
    \includegraphics[width=0.8\textwidth]{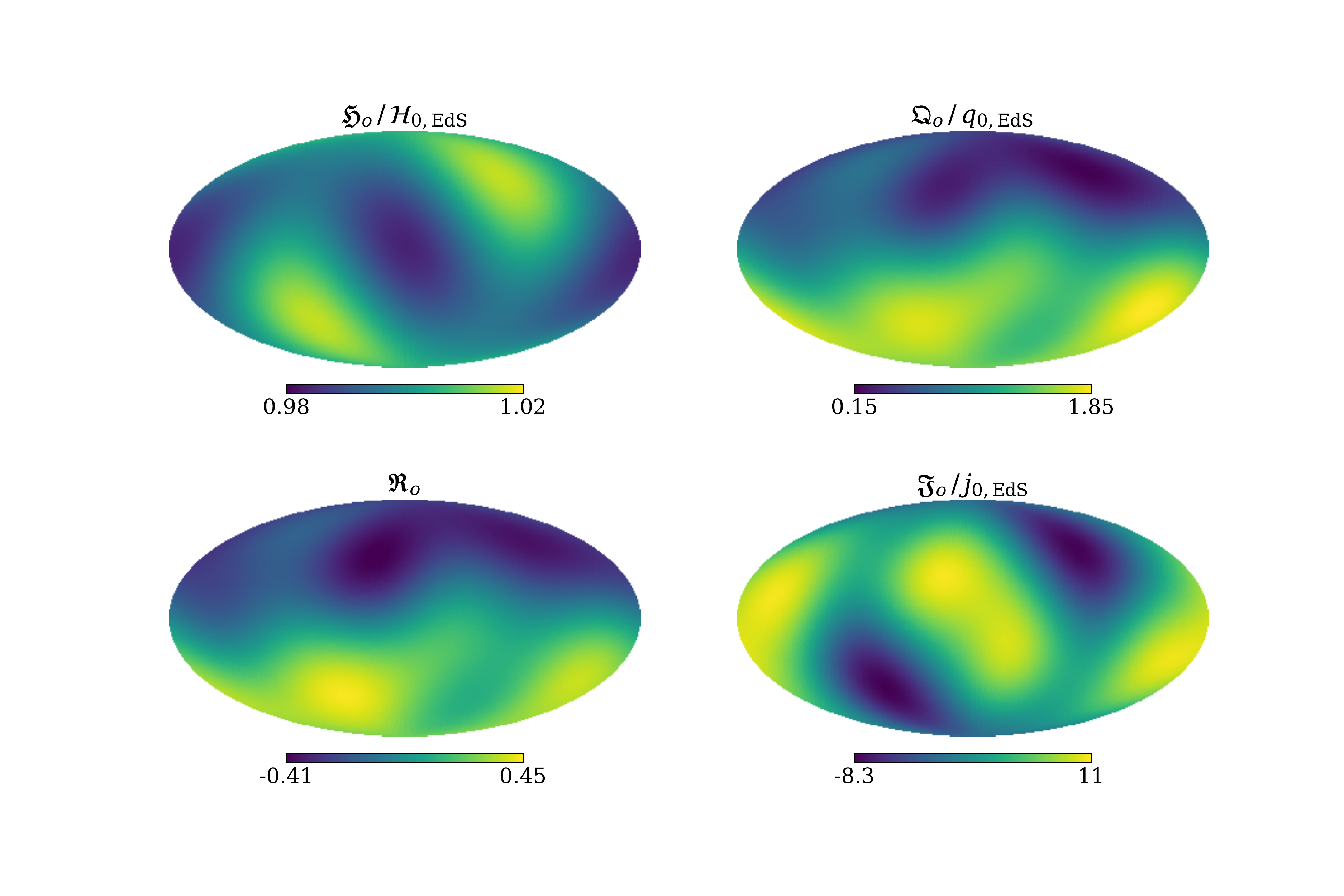}
    \caption{Sky-maps of the effective Hubble ({\it Top-left panel}), deceleration ({\it Top-right panel}), curvature ({\it Bottom-left panel}), and jerk parameters ({\it Bottom-right panel}), for an observer placed at the "present epoch" hypersurface of the $200\,$Mpc/h smoothing scale simulation. The parameters (with the exception of the curvature parameter) are normalised by their Einstein-de Sitter reference values as computed by an average over the simulation domain. This figure corresponds to Fig.~2 in Ref.~\cite{Macpherson:2021gbh}. 
    }
    \label{fig:200_params_skymap}
\end{figure*}

Similarly, the effective deceleration parameter and the effective curvature and jerk parameters can be written in terms of truncated multipole series in the direction vector, $\bm e$, of the source as in Ref.~\cite{Heinesen:2020bej} -- See also Ref.~\cite{Umeh:2013UCT} for the first decomposition of the effective deceleration parameter into multipoles. The dominant anisotropies as predicted by this formalism might be examined with data or within numerical simulations. In Refs.~\cite{Macpherson:2021gbh}, the analytical expression of the luminosity distance in Eqs.~\eqref{eq:series}--\eqref{eq:paramseff} was analysed for an ensemble of observers in fully relativistic large scale numerical simulations. A sky map of a typical observer in this simulation, as smoothed over scales of $200\,$Mpc/h, is shown in Fig.~\ref{fig:200_params_skymap}.

The signatures of the anisotropy of the luminosity distance cosmography as seen for the ensemble of observers, and the variations of the cosmography across observers, are discussed in Ref.~\cite{Macpherson:2021gbh}, in context of the Hubble tension. The dominant dipole as seen in $\mathfrak{Q}$ and $\mathfrak{R}$ for the observer in Fig.~\ref{fig:200_params_skymap}, and as found for the majority of observers in the simulation, is interesting in light of the cosmic dipoles found in data as discussed in Sec.~\ref{sec:cosmic_dipoles}.\footnote{The dipolar signatures as reported for the Hubble parameter (see Sec.~\ref{sec:H0_dipole}), might be signatures of higher order parameters in the $d_L$ cosmography, which are present in the ratio $z/d_L$.} 

The general formalism presented in this section provides a consistent way of taking into account anisotropies in standard candle and standard siren data, without the need of imposing exact symmetries at the level of the metric tensor. If inhomogeneous and/or anisotropic features in the luminosity distance--redshift signal are sufficiently large in our Universe, these could significantly impact local measurements and our inference expansion and deceleration of space. 
The formalism can be extended to include model independent analysis of other observables, such as cosmological drift effects~\cite{Heinesen:2021qnl}.

\subsubsection{Local Void Scenario}
\label{sec:LVS}

A simple explanation of the $H_0$ tension could be that we live inside a bubble with a higher expansion rate---the so-called local void scenario.\footnote{Here, we are not considering void models as alternative to dark energy, a scenario that has been strongly disfavored by kSZ constraints~\citep{Zhang:2010fa}.} For such a Hubble bubble model one expects that an adiabatic perturbation in density causes a perturbation in the expansion rate given by:
\begin{align}
    \frac{\delta H_0}{H_0} = - \frac{1}{3} f(\Omega_m) \frac{\delta \rho(t_0)}{\rho(t_0)} \,,
\end{align}
where $f\simeq 0.5$ is the present-day growth function for the concordance $\Lambda$CDM model. It follows that, to have a $\approx 9$\% change in $H_0$, one needs a Hubble bubble of contrast $\approx -0.5$. As shown by Fig.~\ref{cosmic-var}, within the standard model, such contrasts are reached only at small scales. Therefore, one does not expect that a local structure could explain away the Hubble tension as $H_0$ is measured at larger scales (see again Fig.~\ref{cosmic-var}). As discussed in Sec.~\ref{sec:H0-sys}, estimates based on theoretical computations and numerical simulations suggest that cosmic variance on $H_0$ amounts at 0.5$-$1\%.

\begin{figure}
\centering
\includegraphics[width=0.5\textwidth]{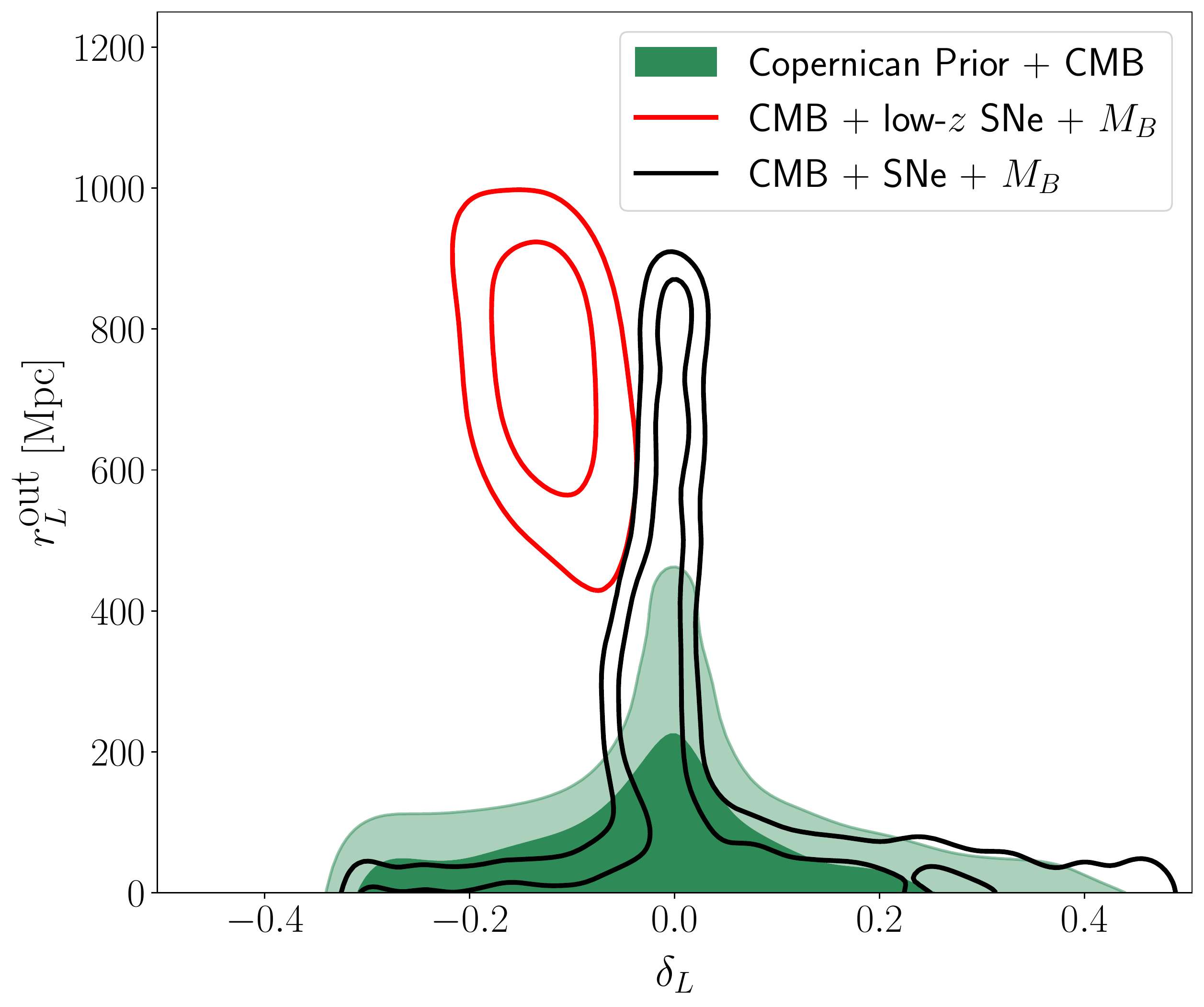}
    \caption{Marginalized constraints on the effective contrast $\delta_L$ and size $r_L^{\rm out}$ of the $\Lambda$LTB inhomogeneity at 68\% and 95\% confidence level. The empty contours show the constraints from the corresponding combination of observables. The green area shows the region of the parameter space that is allowed by the standard model, here represented via the Copernican prior convolved with the CMB likelihood. The figure has been taken from Ref.~\citep{Camarena:2021mjr}.}
    \label{LLTB}
\end{figure}

However, the previous reasoning assumes the standard model, in particular that the FLRW background can describe observations on cosmological scales. If one drops the FLRW assumption, there could be inhomogeneity around us of arbitrary depth and size.\footnote{There could also be large-scale anisotropies, but these are much more constrained from observations (see, however, recent studies on large-scale dipole anisotropies~\citep{Siewert:2020krp, Secrest:2020has}).} Perturbations around us are expected to follow the power spectrum derived from the CMB (the one that was used for Fig.~\ref{cosmic-var}) only within the standard FLRW paradigm, which implies that the observer's local Universe is statistically similar to the one at the last scattering surface. One can then study a model that features an arbitrary inhomogeneity around the observer and see how much observations can constrain such inhomogeneous model. Equivalently, one can reconstruct the metric from observations, checking if FLRW is recovered. To this end, the LTB metric with the cosmological constant (the $\Lambda$LTB model) provides a convenient formalism to constrain deviations from FLRW (see Ref.~\cite{Camarena:2021mjr} and reference therein). Contrary to the Hubble bubble model, the $\Lambda$LTB model features a smooth profile without discontinuities, allowing for a better comparison with observational data.

Ref.~\cite{Camarena:2021mjr} constrained the $\Lambda$LTB model using the latest available data from CMB, BAO, type Ia supernovae, local $H_0$, cosmic chronometers, Compton $y$-distortion and kinetic Sunyaev–Zeldovich effect.
Regarding the $H_0$ tension, it was found that, if one considers CMB, low-redshift supernovae from Pantheon ($0.023<z<0.15$) and the local calibration on the supernova magnitude $M_B$ from SH0ES~\cite{Camarena:2021jlr}, then a local underdensity of size $(700$--$800)\,$Mpc and contrast $-0.15$ is indeed preferred by the data, see the red contours in Fig.~\ref{LLTB}. This structure is clearly non-Copernican and at strong odds with the standard model, which predicts the structures depicted by the green contours. However, if we include all the supernovae, one finds that the void scenario is ruled out and the FLRW limit is basically recovered (black contours in Fig.~\ref{LLTB}).

Specifically, when including only low-$z$ supernovae one finds $H_0=(72.3 \pm 1.1){\rm \,km\,s^{-1}\,Mpc^{-1}}$. If instead, all supernovae are included, one finds $H_0 = (69.06 \pm 0.56){\rm \,km\,s^{-1}\,Mpc^{-1}}$: as the local structure is very much constrained by data, the $H_0$ value is close to the one inferred from CMB observations (see Ref.~\cite{Camarena:2021-LLTB-H0} for more details). Concluding, assuming that CMB and supernova systematics are under control, an underdensity around the observer as modelled within the $\Lambda$LTB model cannot solve the $H_0$ tension~\cite{Wojtak:2013gda,Odderskov:2014hqa,Wu:2017fpr,Kenworthy:2019qwq,Cai:2020tpy,Martin:2021wvb,Castello:2021uad}. See Sec.~\ref{sec:cosmography} for model-independent frameworks for analysing low-redshift cosmological data without imposing a particular form of the metric tensor.


\section{Challenges for $\Lambda$CDM Beyond $H_0$ and $S_8$}
\label{sec:WG-challenges}

\noindent \textbf{Coordinator:} Leandros Perivolaropoulos. \\

\noindent \textbf{Contributors: } \"Ozg\"ur Akarsu, Yashar Akrami, Luis Anchordoqui, David Benisty, Anton Chudaykin, Craig Copi, Eleonora Di Valentino, Celia Escamilla-Rivera, Pablo Fosalba, Enrique Gaztanaga, Dragan Huterer, Andrew Jaffe, Chung-Chi Lee, Benjamin L'Huillier, Matteo Lucca, Roy Maartens, C.J.A.P. Martins, Suvodip Mukherjee, Pavel Naselsky, Eoin \'O Colg\'ain, Paolo Salucci, Arman Shafieloo, M.M. Sheikh-Jabbari, Foteini Skara, Glenn Starkman, Richard Watkins, John Webb.
\bigskip

In this section we discuss in a unified manner many existing signals in cosmological and astrophysical data that appear to be in some tension ($2\sigma$ or larger) with the standard $\Lambda$CDM model as defined by the {\it Planck} 2018 parameter values. In addition to the well known tensions discussed in previous sections ($H_0$ tension and $S_8$ tension), there is a wide range of other less-discussed less-standard signals, at a statistical significance level lower than the $H_0$ tension, that may also constitute hints of new physics. The goal of this section is to collectively present the current status of these signals and their levels of significance, refer to recent resources where more details can be found for each signal and discuss possible generic theoretical approaches that can collectively explain the non-standard nature of these signals. It is worth stressing that some of the tensions are painting a coherent picture, whereas others are isolated and potentially even contradict tensions in $H_0$ and $S_8$.

According to these ideas, we adopt the following strategic questions:

\begin{itemize}
    \item What are the current cosmological and astrophysical datasets that include such non-standard signals?
    \item What is the statistical significance of each signal?
    \item Is there a common theoretical framework that may explain these non-standard signals if they are of physical origin?
\end{itemize}

Along the same lines, a recent extensive broad review of non-standard signals may be found in Ref.~\cite{Perivolaropoulos:2021jda}, which has updated earlier works on the subject~\cite{Perivolaropoulos:2008ud,Perivolaropoulos:2011hp} collecting and discussing signals in data that are at some statistical level in tension with the standard $\Lambda$CDM model.
In addition, interesting reviews focusing mainly on the Hubble tension may be found in Refs.~\cite{DiValentino:2020zio,Knox:2019rjx,Jedamzik:2020zmd,CANTATA:2021ktz,DiValentino:2021izs,Shah:2021onj}, while other reviews focusing mainly on CMB anomalies may be found in Refs.~\cite{Schwarz:2015cma,Copi:2010na}, and in Refs.~\cite{Planck:2019evm,Planck:2015igc,Planck:2013lks}.

In what follows, we provide a short example list of the non-standard cosmological signals in cosmological data. In many cases the signals are controversial and there is currently debate in the literature on the possible systematics origin of some of these signals. However, for completeness we refer to all signals we could identify in the literature referring also to references that dispute the physical origin of these signals.

\subsection{The $A_{\rm lens}$ Anomaly in the CMB Angular Power Spectrum}
\label{sec:Alens}

The $A_{\rm lens}$ parameter was first introduced in Ref.~\cite{Calabrese:2008rt}. $A_{\rm lens}$ is an "unphysical" parameter that actually rescales by hand the effects of the gravitational lensing on the CMB angular power spectra, and can be computed by the smoothing of the peaks in the damping tail. For example, $A_{\rm lens}=0$ indicates no lensing effect, whilst for $A_{\rm lens}=1$ one recovers the value expected in GR. It is quite interesting to note that for {\it Planck} CMB power spectra one finds a preference for $A_{\rm lens}>1$ at more than 95\% CL for both {\tt Plik} and {\tt CamSpec} likelihoods, with a large improvement of the $\chi^2$. Moreover, the inclusion of BAO data shows an evidence for $A_{\rm lens}>1$ at more than $99 \%$ CL for {\tt Plik} likelihood  and about 99\% for the {\tt CamSpec} likelihood. 

The evidence for $A_{\rm lens}>1$ is not easily describable in the existing theoretical frameworks and this requires some new challenges in theory, for example, we need either a closed Universe model and surely challenge several observational datasets and the simplest inflationary models, see Ref.~\cite{DiValentino:2019qzk,Handley:2019tkm,DiValentino:2020hov} and subsection~\ref{sec:WG-Curvature}, or we need more exotic cosmological theories that would lead to modifications of the GR~\cite{Planck:2015bue,DiValentino:2015bja,Planck:2018vyg}, a running of the running of the spectral index~\cite{Cabass:2016ldu}, or compensated baryon isocurvature perturbations~\cite{Munoz:2015fdv,Valiviita:2017fbx}, or a Ginzburg-Landau Theory of DE~\cite{Banihashemi:2022vfv}. Moreover, this lensing anomaly is not observed in the {\it Planck} trispectrum data (i.e.\ CMB lensing) which offer a complementary and independent measurement of the parameters. If the $A_{\rm lens}$ anomaly does not call for a new physics, then this anomaly may connect to a mild but undetected systematic error in all the releases of the {\it Planck} data~\cite{Planck:2013pxb,DiValentino:2013mt,Planck:2015fie,Planck:2018vyg} that one cannot rule out from the current picture. If $A_{\rm lens}$ is related to systematics, then we need to understand how this systematic could affect the constraints on the Hubble constant or S8 parameter, and hence on the different tensions explored in this work.

A recently developed method to directly constrain the CMB gravitational lensing potential from the CMB data is based on a principal component decomposition of the lensing power spectrum~\cite{Motloch:2018pjy}. This approach extends the usual $A_{\rm lens}$ analysis by introducing arbitrary shape variations around a fixed fiducial lensing power spectrum. Marginalizing over effective parameters $\Theta^{(i)}$ leads to interesting implications in various extensions of the $\Lambda$CDM model. In the $\Lambda$CDM+$N_{\rm eff}$ scenario marginalizing lensing substantially broadens the $H_0$ constraints, $H_0=(68.2\pm1.6){\rm \,km\,s^{-1}\,Mpc^{-1}}$ at 68\% CL~\cite{Motloch:2019gux}. The lensing-like anomaly also strengthens the constraints on neutrino masses. In fact, ignoring {\it Planck} lensing reconstruction ie allowing for an arbitrary gravitational lensing potential one finds $\sum m_\nu<0.87$ eV at $2\sigma$~\cite{Motloch:2019gux} which is three times weaker than the {\it Planck} constraint. On the other hand when including BAO, supernova constraints, and {\it Planck} lensing reconstruction, the neutrino mass constraint degrades only by $20\%$ over the {\it Planck} constraint after marginalizing over lensing information $\Theta^{(i)}$. 

The latest ACT measurements find no deviation from the standard lensing effect predicted in the $\Lambda$CDM model, $A_{\rm lens}=1.01\pm0.11$~\cite{ACT:2020gnv}. Recent measurements of TE and EE power spectra collected by SPT-3G are also consistent with the standard model prediction, $A_{\rm lens}=0.98 \pm 0.12$~\cite{SPT-3G:2021eoc}. Finally, combining {\it Planck} TT ($\ell<1000$) and SPTPol polarization and gravitational lensing measurements arrives at $A_{\rm lens}=0.99 \pm 0.03$~\cite{Chudaykin:2020acu}. These results indicate that the lensing anomaly persists only in the {\it Planck} data at small scales and is absent for other CMB experiments. Forecasts for future CMB data have been performed in Ref.~\cite{Renzi:2017cbg}.

\subsection{Hints for a Closed Universe from {\it Planck} Data} 
\label{sec:WG-Curvature}

\begin{figure*}
\centering
    \includegraphics[width=0.32\textwidth]{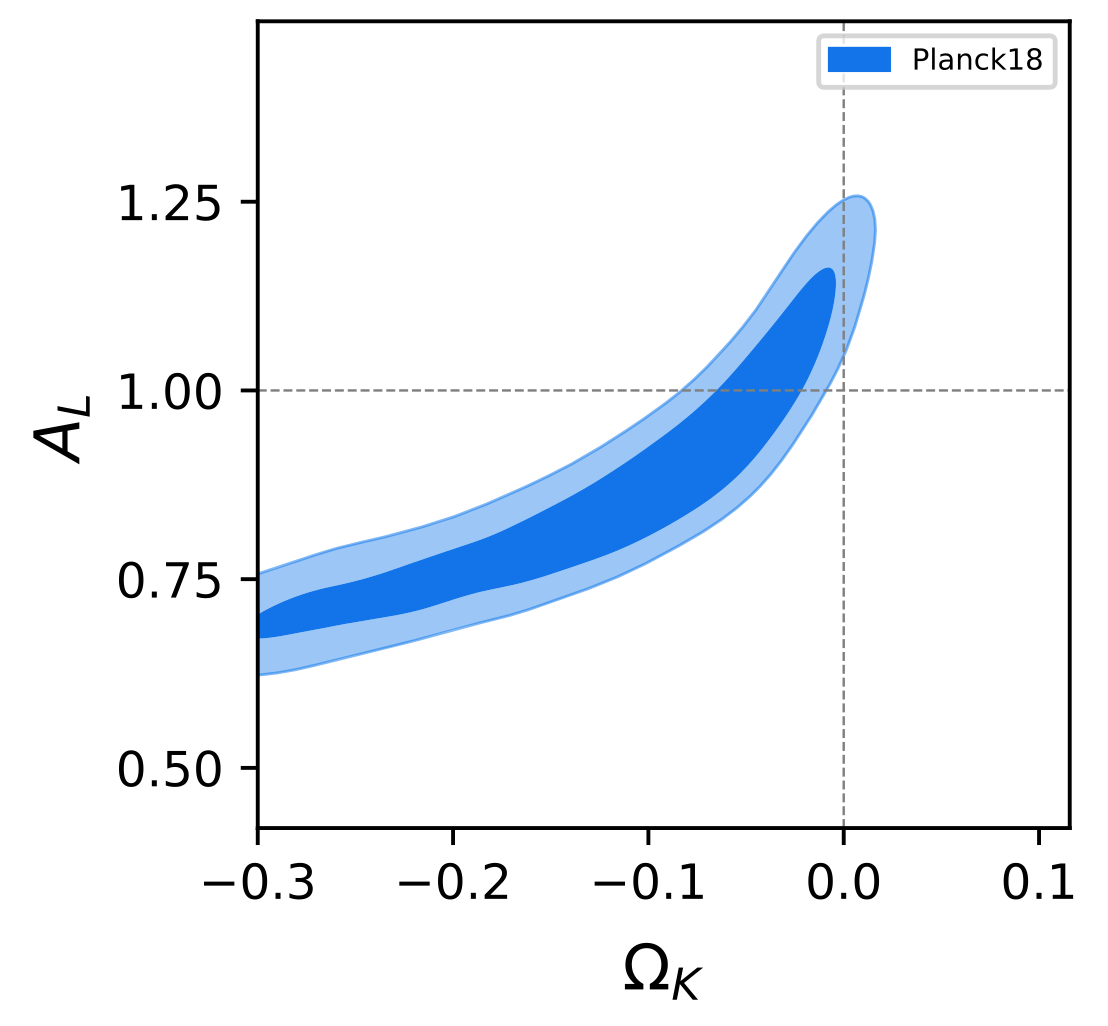}
    \includegraphics[width=0.4\textwidth]{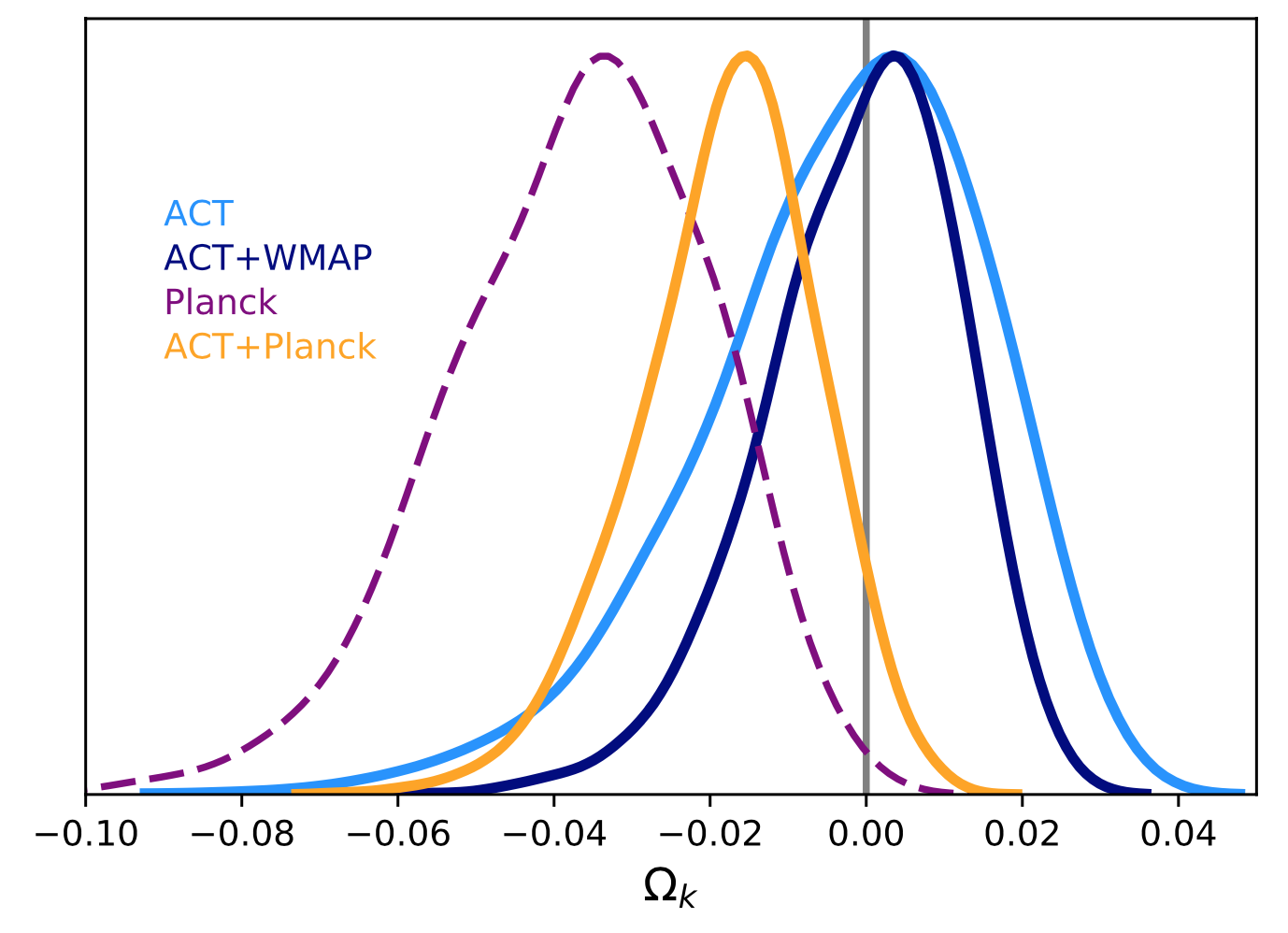}
    \caption{\textit{Left panel}\/: 68\% CL and 95\% CL contour plots for $\Omega_k$ and $A_{\rm lens}$ from Ref.~\cite{DiValentino:2019qzk}. \textit{Right panel}\/: 1D posterior distributions on $\Omega_k$ from Ref.~\cite{ACT:2020gnv}.}
    \label{2D_OkAl}
\end{figure*}

The enhanced lensing amplitude ($A_{\rm lens}$ anomaly~\cite{Calabrese:2008rt,Planck:2018vyg}) in the CMB power spectrum indicated by the different {\it Planck} data releases is a significant challenge for the standard $\Lambda$CDM model. A physical explanation for this enhancement could be provided in the context of a closed Universe which due to this anomaly, is preferred over a flat Universe at a level of $3.4\sigma$~\cite{DiValentino:2019qzk,Planck:2018vyg,Handley:2019tkm}.\footnote{There is a related preference of the {\it Planck} dataset at more than $2\sigma$ for Modified Gravity~\cite{Planck:2018vyg,DiValentino:2015bja,Planck:2015bue,Capozziello:2002rd}.} This disagreement with the predictions for a flat Universe of the standard model is due to the strong degeneracy between the $A_{\rm lens}$ parameter~\cite{Calabrese:2008rt,Planck:2018vyg}, and the spatial curvature parameter $\Omega_k$ as indicated in Fig.~\ref{2D_OkAl}. A closed Universe could also reduce the well-know tension  between the low and high $l$ multipoles  of the angular CMB power spectrum~\cite{Addison:2015wyg,Planck:2016tof,DiValentino:2019qzk}. The oscillating features of the CMB power spectrum indicating spatial curvature could be due to unresolved systematics in the {\it Planck} 2018 data, or can be simply due to statistical fluctuations.

In fact, while {\it Planck} 2018~\cite{Planck:2018vyg} finds $\Omega_k=-0.044^{+0.018}_{-0.015}$,\footnote{All the bounds are reported at 68\% confidence level in the text.} i.e.\ $\Omega_k<0$ at about $3.4\sigma$ ($\Delta \chi^2 \sim -11$) using the baseline Plik likelihood~\cite{Planck:2019nip}, the evidence is reduced when considering the alternative CamSpec~\cite{Efstathiou:2019mdh} likelihood (see discussion in~\cite{Efstathiou:2020wem}), albeit with the marginalized constraint still above the $99\%$ CL ($\Omega_k=-0.035^{+0.018}_{-0.013}$).

If $A_{\rm lens}$ anomaly demands for more lensing, and this is connected to an increase in the cold dark matter density, then it could be explained by a closed Universe. However, in this case, $H_0$ from {\it Planck} could be very small, something like $H_0 \sim 55{\rm \,km\,s^{-1}\,Mpc^{-1}}$~\cite{Planck:2018vyg, DiValentino:2019qzk, Handley:2019tkm}, exacerbating the Hubble tension as well as the $S_8$ tension.

Other combinations of data considered in the literature appear to favor (or not favor) a closed Universe at varying levels of statistical significance. In particular:

\begin{itemize}
\item  The recent results from the ground-based experiment ACT, in combination with data from the WMAP experiment, are fully compatible with a flat Universe with $\Omega_k=-0.001^{+0.014}_{-0.010}$. 
\item 
The same ACT data  when combined with a portion of the {\it Planck} dataset lead to slight preference for a closed Universe since in this case $\Omega_k=-0.018^{+0.013}_{-0.010}$~\cite{ACT:2020gnv}, see Fig.~\ref{2D_OkAl}.
\item
A closed Universe is  preferred by a combination of non-CMB data obtained by Baryon Acoustic Oscillation (BAO) measurements~\cite{Beutler:2011hx,Ross:2014qpa,Alam:2016hwk,Benisty:2020otr}, SNIa distances from the recent Pantheon catalog~\cite{Pan-STARRS1:2017jku}, and a prior on the baryon density derived from measurements of primordial deuterium~\cite{Cooke:2017cwo} assuming Big Bang Nucleosynthesis (BBN).  In the context of this combination, a much larger $H_0$~\cite{DiValentino:2019qzk} is obtained in agreement with the SH0ES value~\cite{Riess:2021jrx,Riess:2019cxk} of $H_0$.
\item A flat Universe is preferred by the combination {\it Planck} + BAO, or + CMB lensing~\cite{Planck:2018lbu} or + Pantheon data. However these dataset combinations are in disagreement at more than $3\sigma$ when the curvature is allowed to vary~\cite{DiValentino:2019qzk,Handley:2019tkm,Vagnozzi:2020rcz}. 
\item A phantom closed Universe is preferred by {\it Planck} combined with the luminosity distance probes, such as Pantheon or SH0ES, when the dark energy equation of state is let free to vary, along the curvature, at more than 99\% CL~\cite{DiValentino:2020hov}.
\item A flat Universe is also in agreement with the analysis made by~\cite{Liu:2020pfa} using the $H(z)$ sample from the cosmic chronometers (CC) and the luminosity distance $D_L(z)$ from the 1598 quasars ($\Omega_k=0.08\pm0.31$) or the Pantheon sample ($\Omega_k=-0.02\pm0.14$), in agreement with the previous analysis of~\cite{Cai:2015pia} (in these analyses however, the error bars are too large to discriminate among the models).
\item In~\cite{Nunes:2020uex} a combination of BAO+BBN+H0LiCOW provides $\Omega_k=-0.07^{+0.14}_{-0.26}$ with $H_0$ in agreement with R19, while BAO+BBN+CC gives a positive $\Omega_k=0.28^{+0.17}_{-0.28}$.
\item
The curvature can also be measured model-independently via  the $\mathcal O_k$ parameter~\cite{Clarkson:2007pz,LHuillier:2016mtc,Marra:2017pst,Shafieloo:2018gin}
\begin{align}
    \mathcal{O}_k(z) & = \frac{(h(z) \mathcal{D}'(z))^2-1}{\mathcal{D}^2(z)},
\end{align}
which is constant and equal to the curvature parameter $\Omega_k$ in an FLRW Universe. Departures from constancy could point towards departure from FLRW or the presence of systematics in the data. Current data (Supernovae from Pantheon and BAO from BOSS) are consistent with a flat FLRW~\cite{LHuillier:2016mtc,Shafieloo:2018gin}. 
\item
Another approach is to use three distances between a source, lens, and the observer to estimate the curvature independently from cosmological assumptions~\cite{Rasanen:2014mca,Denissenya:2018zcv}. This does not involve any derivative and is consistent with a flat Universe in the context of large uncertainties.
\end{itemize}

Overall, the spatial curvature remains insensitive to local $H_0$ measurements from the Cepheid distance ladder~\cite{Zuckerman:2021kgm}. However, in general, letting the curvature free to vary implies an increase of both the $H_0$ and the $S_8$ tensions~\cite{DiValentino:2019qzk}.  Therefore, the constant-curvature degree of freedom of the standard model is not sufficient to explain the tensions and anomalies seen in the data. However, within full GR, in the context of the scalar averaging scheme, the average scalar curvature can provide considerably stronger evolutionary impact~\cite{Heinesen:2020sre}.

Detecting a curvature $\Omega_k$ different from zero could be due to a local inhomogeneity biasing our bounds~\cite{Bull:2013fga}, and in this case CMB spectral distortions such as the KSZ effect and Compton-y distortions, present a viable method to constrain the curvature at a level potentially detectable by a next-generation experiment. If a curvature $\Omega_k$ different from zero is the evidence
for a truly superhorizon departure from flatness, this will have profound implication for a broad class of inflationary scenarios. While open Universes are easier to obtain in inflationary models~\cite{Gott82, Linde:1998gs, Bucher:1994gb, Kamionkowski:1994sv, Kamionkowski:1993cv}, with a fine-tuning at the level of about one percent one can obtain also a semi-realistic model of a closed inflationary Universe~\cite{Linde:2003hc, Lasenby:2003ur}. In Ref.~\cite{Leonard:2016evk} it has been shown that forthcoming surveys, even combined together, are likely to place constraints on the spatial curvature of $\sim 10^{-3}$ at 95\% CL at best, but enough for solving the current anomaly in the {\it Planck} data. Experiments like {\it Euclid} and SKAO, instead, may further produce tighter measurements of $\Omega_k$ by helping to break parameter degeneracies~\cite{DiDio:2016ykq,Vardanyan:2009ft}.

\subsection{Large-Angular-Scale Anomalies in the CMB Temperature and Polarization}
\label{sec:Largeangleanomalies}

Several features unexpected in $\Lambda$CDM cosmology have been noted in CMB temperature fluctuations at the largest angular scales.
Most of these anomalies share several common features:
\begin{itemize}
    \item they refer to the largest observable angular scales $\gtrsim 60^\circ$ -- whether directly as large-angle two-point functions (lack of large-angle correlations, quadrupole-octopole anomalies), or as large-angle modulations of smaller-angular-scale fluctuations (hemispherical asymmetries, point parity anomaly);
    \item they are not captured in a standard experimental $\Lambda$CDM likelihood function because they are not deviations of individual $C_\ell$ from the values predicted by $\Lambda$CDM with appropriate parameters; they are therefore characterized by statistics that were proposed {\it post-facto} to capture a pattern noticed in the data;
    \item WMAP and {\it Planck} are in remarkable agreement about them being real features on the sky.
\end{itemize}
An additional anomaly, that is on somewhat smaller scales ($\simeq10^\circ$), but shares the other features, is the so-called "cold spot." 

We now describe the most compelling of these anomalies:
\begin{enumerate}
    \item the lack of large-angle CMB temperature correlations;
    \item hemispherical power asymmetry;
    \item octopole planarity and alignment with the quadrupole;
    \item the point-parity anomaly;
    \item variation in cosmological parameters over the sky;
    \item the cold spot.
\end{enumerate}
In contrast to the Hubble constant tension 
($\sim5\sigma$), these anomalies are individually statistically less significant ($2\sigma$ to $4 \sigma$). However, within $\Lambda$CDM, they appear to be statistically independent, although this has been rigorously demonstrated only for the first and last of the above list.  Jointly, they are highly significant.
An understanding of their nature can point to new physics in the primary perturbations formation process or shed light on unaccounted systematic effects in both astronomical and CMB data~\cite{Planck:2015igc,Planck:2019evm}. 

We make clear below why these anomalies are difficult to explain in the context of $\Lambda$CDM. We explore the $\Lambda$CDM predictions for analogous statistics (if any) for polarization, and contrast those predictions with what might be expected from explanations of the anomalies based on new physics.
Finally, we discuss how future experiments could be valuable and why they should make the testing of these predictions a key element of their scientific programs.

\subsubsection{The Lack of Large-Angle CMB Temperature Correlations}
\label{sec:LargeCMBanomalies}

Suppressed correlations at the largest observable angular scales had been noticed in the COBE data~\cite{Bennett4766}. 
Specifically, the two-point angular correlation function 
\begin{equation}
  \mathcal{C}(\theta) \equiv \overline{T(\hat e_1) T(\hat e_2)},
\end{equation}
(where the average is over all pairs of pixels with $\hat e_1\cdot\hat e_2=\cos\theta$)
was found to be very close to zero on scales above $\sim60^\circ$~\cite{Hinshaw:1996ut}. 
The near-vanishing of the large-angle correlation function was clear in the early WMAP data~\cite{WMAP:2003elm}, and was later confirmed in the 3-yr~\cite{Copi:2006tu} and the 5-yr WMAP data in~\cite{Copi:2008hw}, as well as in the {\it Planck} data~\cite{Copi:2013cya}. 

\begin{figure}[t]
  \includegraphics[scale=0.45]{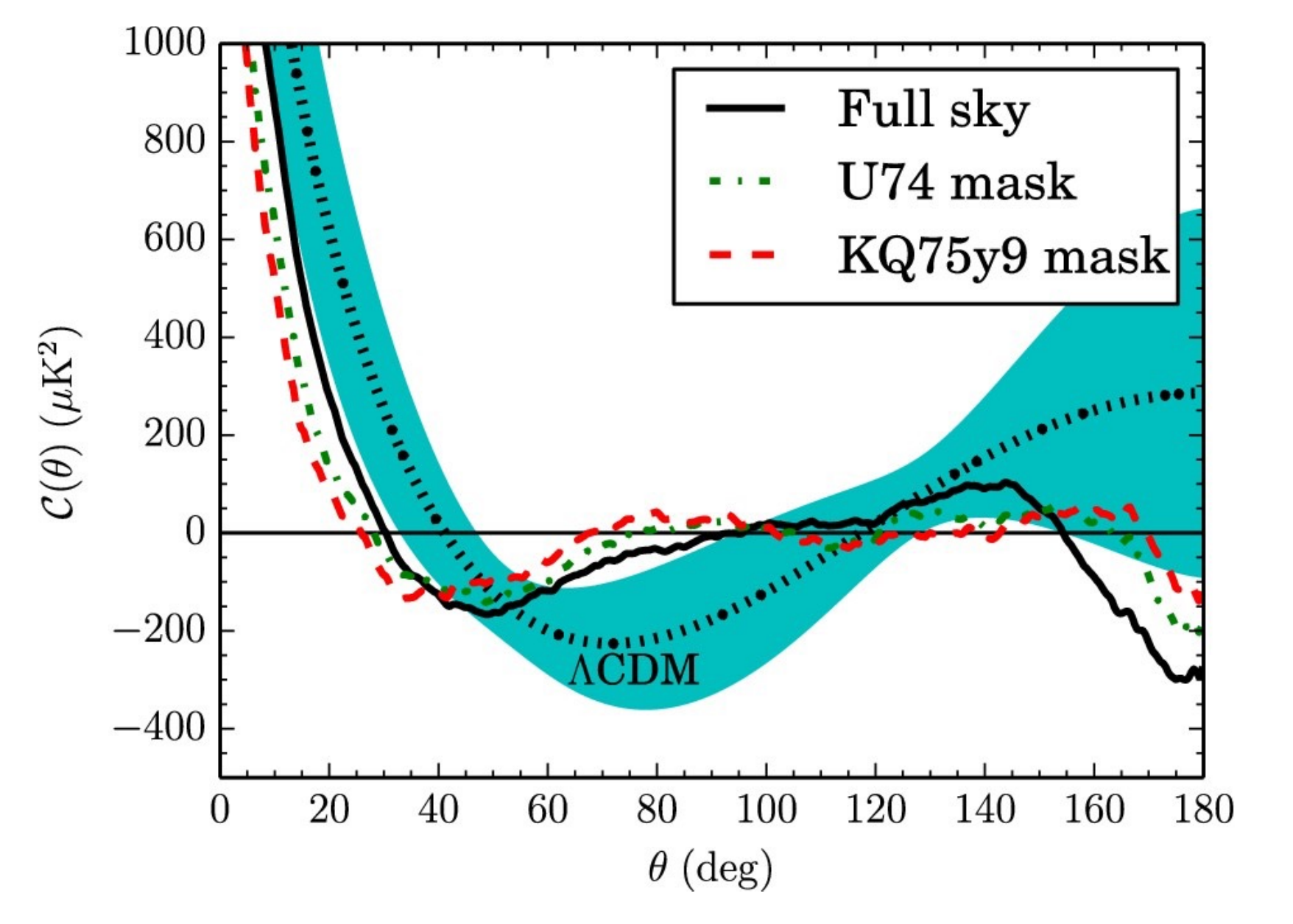}
  \includegraphics[scale=0.45]{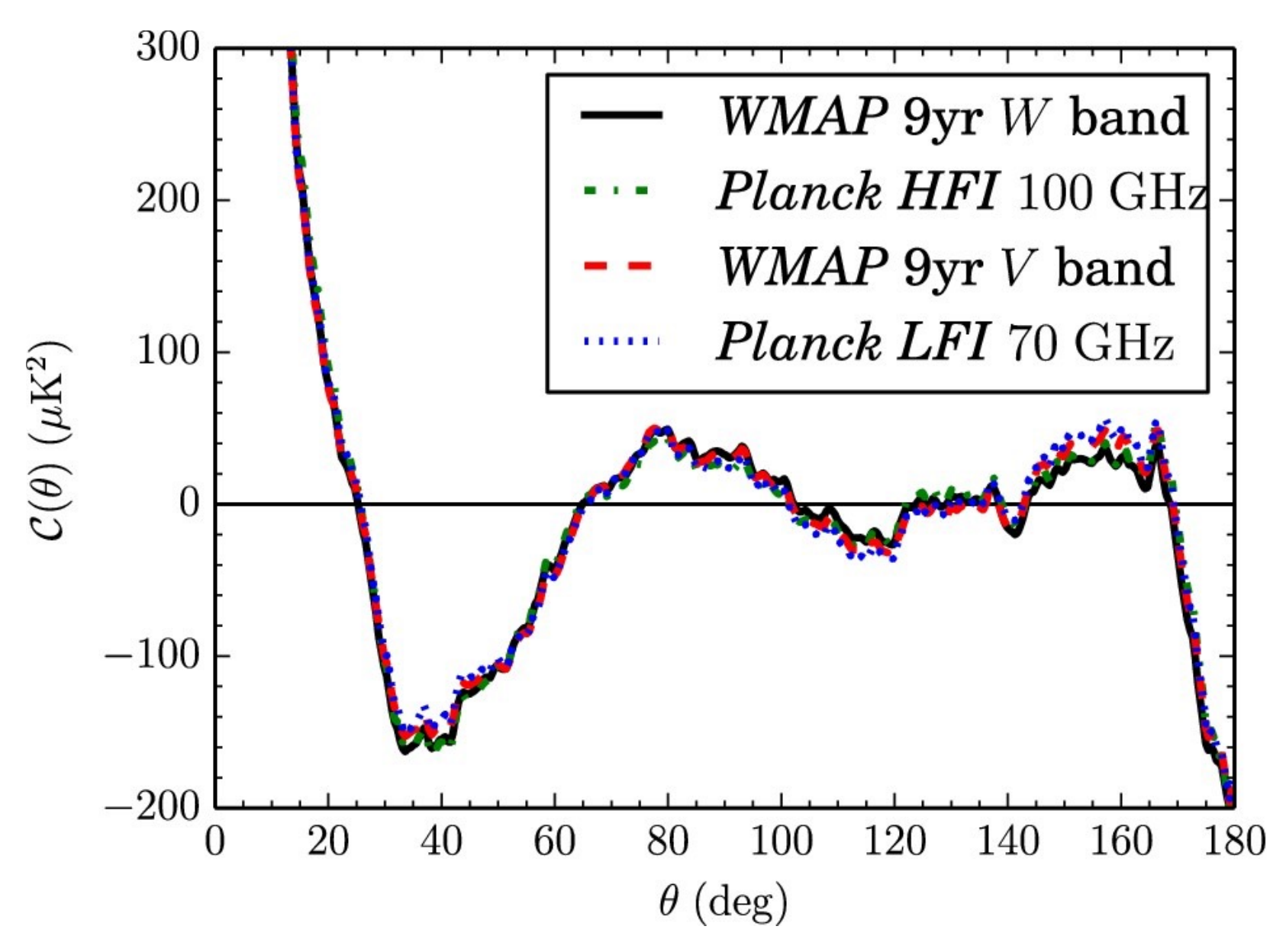}
  \caption{{\it Left panel}\/: Two-point angular correlation function, ${\mathcal C}(\theta)$, computed in pixel space, for three different bands masked with the KQ75 mask (from WMAP 5 year data). Also shown is the correlation function for the ILC map with and without the mask, and the value expected for a statistically isotropic sky with best-fit $\Lambda$CDM cosmology together with 68\% cosmic variance error bars. {\it Right panel}\/: Same, but for alternative sky cuts applied to WMAP and {\it Planck} data. Notice the remarkable mutual agreement between the inferred large-angle correlation functions. Both panels adopted from Ref.~\cite{Copi:2013cya}.}
\label{fig:ctheta}
\end{figure}

Fig.~\ref{fig:ctheta}, adopted from Ref.~\cite{Copi:2013cya}, illustrates the degree to which the correlations are suppressed on the cut sky (part of the sky not contaminated by Galactic emission) for both WMAP and {\it Planck}. The most striking feature of the cut-sky $\mathcal{C}(\theta)$ is that it is very nearly zero above about $60^\circ$, except for some anti-correlation near $180^\circ$. This is true for all reasonable Galaxy masks, for all WMAP and {\it Planck} CMB-dominated wavebands, in all data releases. It is also true for $\mathcal{C}(\theta)$ calculated from the pseudo-$C_\ell$ using $\mathcal{C}(\theta)=\sum_\ell C_\ell P_\ell(\cos\theta)$. 

In order to be more quantitative about these observations, the WMAP team~\cite{WMAP:2003elm} introduced a statistic that quantifies the deviation of $C(\theta)$ from zero,
\begin{equation}
  S_{1/2} \equiv \int_{-1}^{1/2} \left[ \mathcal{C}(\theta)\right]^2 \mathrm{d}
  (\cos\theta).
  \label{eqn:Shalf}
\end{equation}
Spergel et al.~\cite{WMAP:2003elm} found that only $0.15\%$ of the realizations in their Markov chain of $\Lambda$CDM model CMB skies had lower values of $S_{1/2}$ than the observed one-year WMAP cut sky. Extending this to later releases of WMAP data, as well as {\it Planck}, the p-value becomes ballpark $0.1\%$,  depending on the map and mask used~\cite{Copi:2013cya}; see again Fig.~\ref{fig:ctheta}. There have been some suggestions that certain three-point and four-point functions are also anomalous~\cite{Planck:2013lks,Planck:2015igc,Eriksen:2004iu}; we discuss that below in the context of hemispherical asymmetry, where this anomaly is more evident. 

While $\mathcal{C}(\theta)$ computed in pixel space over the masked sky agrees with the harmonic space calculation that uses the pseudo-$C_\ell$ estimator, it somewhat disagrees with the $C(\theta)$ obtained from a variety of  foreground-corrected reconstructed \textit{full-sky} maps. The latter however agree with the $C(\theta)$ calculated using the maximum-likelihood estimator (MLE) $C_\ell$ that are typically reported. These $C(\theta)$ have $S_{1/2}$ that are low only at the $p\simeq 0.05$ level~\cite{Copi:2011pe}.
However, problematically, the full-sky map $C(\theta>60^\circ)$ is dominated by pairs of points at least one of which is in the masked region, suggesting that the full-sky $C(\theta)$ is potentially contaminated by artefacts coming from the galactic plane. Nevertheless, it is the angular correlation function observed on unmasked, "clean" portions of the sky that is statistically low at the level of $p\simeq 0.0003$--$0.003$ (depending on the choice of the map and the sky cut). 

In Ref.~\cite{Efstathiou:2009di} it is argued that $C(\theta)$ cannot be used to exclude the $\Lambda$CDM model. The relevant low-$\ell$ $C_\ell$ have large cosmic variance as a fraction of their expected values, so inferred $C_\ell$ can differ markedly from the expected values.
This results in very large sampling covariance error predictions for $C(\theta)$ in $\Lambda$CDM.  
Moreover $C(\theta)$ could be small because the $\Lambda$CDM predictions have more power on the low $C_{\ell}$ than the actual CMB measurements. The significance of the discrepancy then depends on the model assumptions~\cite{Gaztanaga:2003ub,Camacho-Quevedo:2021bvt}. Nevertheless, using model-independent errors, based only on direct measured sampling variations, Ref.~\cite{Camacho-Quevedo:2021bvt} finds strong evidence for a homogeneity scale and lack of correlations on scales larger than  $\theta_\mathrm{CMB} \simeq 66 \pm 9$ degrees on the CMB sky. 

It is widely underappreciated, however, that the smallness of $S_{1/2}$ is {\em {not}} primarily due to the smallness of any individual $C_\ell$ nor set of $C_\ell$s, and so is {\em{not}} accurately described as a "lack of large-angle power". Rather, it is due to cancellations among the contributions of the low-$\ell$ $C_\ell$ to $S_{1/2}$ up to at least $\ell=5$ --- $C_2$ is almost optimal~\cite{Copi:2008hw} to cancel the contributions of $C_{\ell>2}$. In other words, $\langle C_{\ell}C_{\ell'}\rangle~\cancel{\propto}~\delta_{\ell\ell'}$. The fractionally large variance in the low-$\ell$ $C_\ell$ actually makes such correlations less likely to emerge accidentally. Such correlations between $C_\ell$ do not feature (at least at leading order) in vanilla $\Lambda$CDM likelihoods, because $\Lambda$CDM says that there are no such correlations. Thus parameter fitting within $\Lambda$CDM, comparisons of $\Lambda$CDM with other statistically isotropic models, and even goodness-of-fit tests of $\Lambda$CDM will not be sensitive to these correlations---they are either statistical flukes, or evidence of a violation of a fundamental assumption of vanilla $\Lambda$CDM.

Thus the $S_{1/2}$ anomaly as realized on the observed sky is evidence for the violation of statistical isotropy, which we might expect to result also in more general correlations between $a_{\ell m}$ as we discuss below. Explanations of this anomaly are relevant only if they accomplish it through such cancellation and not by reducing low-$\ell$ $C_\ell$. This cannot be accomplished by, for example, changing the primordial power spectrum $P(k)$. Non-trivial topology has been shown to be able to produce low $S_{1/2}$ with appropriate $C_\ell$~\cite{Bernui:2018wef}, albeit with a fundamental domain that is ruled out by the absence of pairs of matched circles in CMB $T$ maps~\cite{Vaudrevange:2012da}.

In Ref.~\cite{Aurich:2021ofm} it is argued that the lack of power on large scales may be simply due to the fact that our Universe is of a finite topology. Samples of 100,000 CMB maps with cubic 3-torus topology are shown to be consistent with the large-angle CMB temperature correlations, when the Dirichlet domain side length of our Universe is around three times the Hubble length. Meanwhile, the $\simeq60^\circ$ angular scale corresponds to the causal horizon of a $\Omega_\Lambda \simeq 0.7$  Universe~\cite{Camacho-Quevedo:2021bvt}, which is in agreement with the super-horizon anomalies found in the fit to the small scale CMB acoustic peaks, see below in Sec.~\ref{sec:parameter-variation}. 

\paragraph{Predictions.} Given the concern that this anomaly (and others) are based on {\it a posteriori} statistics, with unknown look-elsewhere effects, to be more fully convinced that they are due to new physics, rather than a statistical fluke, requires testable predictions. The first type of predictions come from $\Lambda$CDM conditioned on one or more anomaly (or more generally on the measured temperature maps); these can be used to test the "fluke hypothesis" that the anomalies are not signs of new physics. The second type of predictions would best come from a model of new-physics. However, even absent that one can in some cases argue generically that a physical explanation would likely have certain consequences.

\noindent\textit{Prediction for $C_1^T$:}
An important point that is widely overlooked is that $\mathcal{C}(\theta)$  as discussed above is actually the monopole-and-dipole subtracted angular correlation function, which we write $\mathcal{C}_{\ell\geq2}(\theta)$. A physical explanation of $S_{1/2}$ seems likely to predict that $\mathcal{C}(\theta)$ is small, not just $\mathcal{C}_{\ell\geq2}(\theta)$. This requires that the cosmological (i.e., non-Doppler) contribution to $C_1^T\lesssim200\,\mu{\rm K}^2\ll\langle C_1^{\Lambda{\rm CDM}}\rangle\simeq 3000\,\mu{\rm K}^2$~\cite{Copi:2013cya}. This happens by chance less than $0.5\%$ of the time in $\Lambda$CDM with standard cosmological parameters, and so if a low $C_1$ were observed it would be compelling evidence of a physical origin for low $S_{1/2}$. This suggests that measuring the CMB Doppler dipole well enough to separate out the intrinsic $C_1$ is worthwhile. Similar statements could perhaps be made about the cosmological monopole $C_0$, the fluctuation in $T_{\mathrm{CMB}}$ in our current Hubble patch about the mean calculated over an infinite, or at least much larger volume, but that is presumably not measurable.

\noindent\textit{Predictions for Polarization:}
$\Lambda$CDM realizations of temperature and $E$-mode and $B$-mode polarization are usually generated together, with $a_{\ell m}^T$, and $a_{\ell m}^E$ correlated Gaussian random variables characterized by $C_{\ell}^{\rm TT}$, $C_{\ell}^{\rm TE}$,  and $C_{\ell}^{\rm TE}$, while $a_{\ell m}^B$ is a Gaussian random variable characterized by $C_{\ell}^{BB}$. However, given measurements of $a_{\ell m}^T$, one can produce constrained realizations of $a_{\ell m}^E$~\cite{Yoho:2013tta}; realizations of $a_{\ell m}^B$ are unchanged since $B$ is uncorrelated with $E$. One can then calculate $a_{\ell m}^Q$ and $a_{\ell m}^U$, for the Q and U mode polarizations, and assemble predictions for $\mathcal{C}^{\rm QQ}(\theta)$, $\mathcal{C}^{\rm QU}(\theta)$, $\mathcal{C}^{\rm UU}(\theta)$, $\mathcal{C}^{\rm TQ}(\theta)$, and  $\mathcal{C}^{\rm TU}(\theta)$, as well as associated $S_{1/2}$ statistics,
Because $TE$ correlations are weak in $\Lambda$CDM ($(C_\ell^{\rm TE})^2\ll C_\ell^{\rm TT}C_\ell^{\rm EE}$), the suppression of $\mathcal{C}^{TT}(\theta>60^\circ)$ does not lead to the suppression of  QQ and UU angular correlation functions~\cite{Yoho:2015bla}, with, for example, $\langle S_{1/2}^{QQ}\rangle\simeq0.012\,\mu{\rm K}^4$ and $\langle S_{1/2}^{UU}\rangle\simeq0.013{\rm \,\mu K^4}$ for $r=0.1$, essentially unchanged from unconstrained $\Lambda$CDM.
A low measured value for either would be a statistically independent anomaly for $\Lambda$CDM. Meanwhile, it would be expected that a physical model would predict suppression of these correlation functions.

There is some subtlety regarding angular correlation functions for polarization~\cite{Yoho:2015bla}. Whereas, once certain conventions have been established, $Q(\hat{n})$ and $U(\hat{n})$ are fields measured on the sky, the separation into $E$ and $B$ is performed in harmonic space, and the reassembly into $E(\hat{n})$ and $B(\hat{n})$ is ambiguous. The local-E and local-B fields~\cite{Yoho:2015bla}, $\hat{E}$ and $\hat{B}$, are heavily weighted toward high-$\ell$ contributions, and in particular the usual $\theta$ to $1/\ell$ correspondence fails for $\mathcal{C}(\theta)$ for $\hat{E}\hat{E}$ or $\hat{B}\hat{B}$. This makes $S_{1/2}^{\hat{E}\hat{E}}$ and $S_{1/2}^{\hat{B}\hat{B}}$ independently interesting~\cite{Yoho:2015bla}, both as null tests for $\Lambda$CDM and as tests of physical explanations for the low $S_{1/2}^{\rm TT}$. It is also possible to put other two-point angular correlation functions to good use, such as between temperature and lensing potential, or temperature and 21-cm fluctuations. In such cases, the optimum range of $\cos\theta$ over which to integrate should be determined \textit{a priori}~\cite{Yoho:2013tta}. Future full-sky polarization observation programs could readily measure all interesting polarization $S_{1/2}$ statistics and test the $\Lambda$CDM predictions~\cite{Yoho:2015bla}.

\subsubsection{Hemispherical Power Asymmetry}

Here, we focus on a number of large-scale anomalies, which in spite of having been detected with different methods and statistics, all show some level of asymmetry in power (or related properties) between hemispheres on the CMB sky; this is why we collectively call them "hemispherical power asymmetry." We do not intend to provide an historical overview of these anomalies or an exhaustive review of the literature on detection, analysis and interpretation of the anomalies. Our aim in this subsection is rather to highlight the most statistically significant hints of the power asymmetry and their common features, with the objective of emphasising their potential importance for both theoretical cosmology and future CMB and large-scale-structure observations.

The bulk of this section discusses tests that involve the so-called dipolar power asymmetry or "dipole modulation," either directly or via measures of directionality, even though there are reasons to believe that such a dipolar description of the asymmetry oversimplifies the actual anomaly. One main reason is that most of the analyses imply that only the hemispheres which include all or most of the northern ecliptic hemisphere are anomalous and the southern ones seem to be consistent with an isotropic $\Lambda$CDM CMB sky. Nevertheless, various forms of dipole modulation have been noted since the early WMAP releases~\cite{Eriksen:2003db} and later confirmed by {\it Planck}~\cite{Planck:2013lks,Planck:2015igc,Planck:2019evm}. All the dipole-modulation tests we discuss here can be divided into two categories, amplitude-based and direction-based, and share in common the fitting of a dipole, which is done either by fitting for a dipole explicitly in a map of power on the sky, by employing Bayesian techniques in pixel space for a specific model, or by measuring the coupling of $\ell$ to $\ell\pm1$ modes in the CMB covariance matrix. Given the differences in the approaches, it is important to keep in mind that the results cannot usually be directly compared, even though all probe some aspect of dipolar asymmetry.

\paragraph{Low Northern Variance.} A low value for the variance of the CMB temperature fluctuations was originally observed in the WMAP data by Refs.~\cite{Monteserin:2007fv,Cruz:2010ud,Gruppuso:2013xba}. This low variance was then confirmed by the {\it Planck} collaboration~\cite{Planck:2013lks,Planck:2015igc,Planck:2019evm}, with a $p$-value of $0.5\%$-$1.0\%$ for the low-resolution temperature maps ($N_\mathrm{side}=16$), depending on whether the impact of a possible look-elsewhere effect is considered. {\it Planck} 2013~\cite{Planck:2013lks} showed, using both full-resolution ($N_\mathrm{side}=2048$) and low-resolution ($N_\mathrm{side}=16$) maps that the low variance was localised in the northern ecliptic hemisphere, with a $p$-value of $\sim 0.1\%$, while the $p$-value for the southern hemisphere was $\sim45\%$. Similar levels of significance were also found for the Galactic northern and southern hemispheres when the  full-resolution maps were used. No updates on these significance levels have been provided by {\it Planck} 2015~\cite{Planck:2015igc} and 2018~\cite{Planck:2019evm}. It is important to note that the map-based variance is dominated by contributions from large angular scales, whilst the cosmological parameter fits are relatively insensitive to these low-order $\ell$-modes and are instead largely dominated by scales corresponding to $\ell>50$. Therefore, the variance of the temperature CMB sky appears to be anomalous as there is a dearth of large-angular-scale power compared to the predictions of the $\Lambda$CDM model with parameters estimated using the same CMB sky. Finally, it is important to note that although the anomaly discussed here is about the variance of the CMB fluctuations, it can be seen as an indirect indication of power asymmetry, as variance is nothing but a weighted sum of the angular power spectrum of the fluctuations over all multipoles. 

\paragraph{Hemispherical Asymmetry.} Refs.~\cite{Eriksen:2003db,Hansen:2004vq} discovered that the angular power spectrum of the first year WMAP data calculated for a hemisphere centred at the Galactic coordinates $(l,b)=(273^\circ, -20^\circ)$ was larger than when calculated for the opposite hemisphere over the multipole range $\ell=2-40$. Ref.~\cite{Park:2003qd} also presented evidence for the existence of such hemispherical asymmetry by applying a Minkowski functional to the WMAP data.  
The preferred direction of Ref.~\cite{Eriksen:2003db} lies close to the ecliptic plane, and they also demonstrated that the large-angular real-space $N$-point correlation functions (for $N=2,3,4$) were different when computed on ecliptic hemispheres. Many studies then focused on hemispheres in the ecliptic coordinate system, with Ref.~\cite{Schwarz:2004gk} particularly emphasizing the connection. Finally, Refs.~\cite{Eriksen:2004df,Eriksen:2004iu,Rath:2007ti} also detected the hemispherical asymmetry with other measures of non-Gaussianity. 

In 2013, the {\it Planck} collaboration~\cite{Planck:2013lks} studied the asymmetry by computing four $N$-point correlation functions (2-point function, pseudo-collapsed 3-point function, equilateral 3-point function and rhombic 4-point function) on the northern and southern ecliptic hemispheres for the {\it Planck} 2013 $N_\mathrm{side}=64$ temperature maps. The results were in agreement with the findings of Ref.~\cite{Eriksen:2003db}. It was shown, particularly, that the northern ecliptic hemisphere correlation functions were relatively featureless, and in particular, both the 3- and 4-point functions were very close to zero. The northern hemisphere $p$-values for the $\chi^2$ statistic were found to be $>93\%$ for all the four correlation functions and $>99\%$ for the 3- and 4-point functions, with the significance level exceeding $99.9\%$ for the pseudo-collapsed 3-point function. This was all consistent with the anomalous lack of power towards the northern ecliptic pole seen by the simpler one-point statistic, i.e.\ the low northern variance anomaly discussed earlier. Similar analyses were done in reference frames set by Galactic coordinates and the CMB Doppler boost direction (see e.g.~\cite{Quartin:2014yaa}), as well as the angular-clustering and dipole-modulation directions that we will discuss below. Although the largest asymmetry was seen for the ecliptic hemispheres, a substantial asymmetry was present also for Galactic coordinate hemispheres.

The {\it Planck} 2015 analysis~\cite{Planck:2015igc} confirmed all of these high levels of significance, and additionally found that a substantial asymmetry was present also for the dipole-modulation direction. Surprisingly though, the latest {\it Planck} 2018 analysis~\cite{Planck:2019evm} no longer observed the high significance level for the pseudo-collapsed 3-point function of the temperature map in the northern ecliptic hemisphere. The {\it Planck} collaboration states~\cite{Planck:2019evm} that this discrepancy between the first two and the final releases of the data may be due to the different masks used in the analyses or a consequence of "the improved treatment of poorly determined modes in the estimated correlation matrix used for the computation of the $\chi^2$ statistic." Even if these are the reasons behind the discrepancy, the fact that the anomaly has been observed persistently and consistently in both the WMAP data and the first two releases of the {\it Planck} data (with different systematics) strongly suggests that the changes in the mask or the improvement in the analysis might have removed important information in the data that would contribute significantly to this anomaly, and the low significance seen in the {\it Planck} 2018 data should therefore be considered with caution. {\it Planck} 2018 found, however, that in the dipole-modulation reference frame the 3-point temperature correlation functions in the negative (northern) hemisphere were somewhat significant, reaching a level of $98\%-99\%$ for the pseudo-collapsed case. 

\paragraph{Angular Clustering of the Power Distribution.} In a search for dipolar power asymmetry, Ref.~\cite{Hansen:2008ym} analysed the 5-year WMAP data by computing the CMB power spectrum on a number of discs on the sky and binning them into independent blocks of 100 multipoles from $\ell=2$ to $\ell=600$. Each block was then used to look for a dipolar asymmetry in the power distribution. The six $\ell$ ranges considered showed evidence of a consistent dipole direction, and not a single realization in a set of 10,000 simulations showed a similarly strong asymmetry.

A similar analysis was then performed by the {\it Planck} collaboration in 2013~\cite{Planck:2013lks}, where a simpler approach developed by Ref.~\cite{Axelsson:2013mva} and applied to the 9-year WMAP data was used. {\it Planck} 2013 estimated the power spectrum amplitude on 12 non-overlapping patches of the sky in $\ell$-bins of 16 multipoles each, and fitted a dipole to the spatial distribution of the amplitudes for each $\ell$-bin. The alignment of the dipole directions between the different multipole blocks was then used to construct a measure of the power spectrum asymmetry. They compared the clustering of the dipole directions evaluated for the different scales to that observed in simulated maps. None of the 500 available simulations showed degrees of clustering higher than the one observed in the data for two choices of $\ell_\mathrm{max}=600$ and $\ell_\mathrm{max}=1500$. However, after deboosting the data (i.e.\ correcting for the Doppler modulation effect), the significance went down for $\ell_\mathrm{max}\gtrsim600$, and therefore, a significant power asymmetry was claimed only up to $\ell\sim600$. The $\ell_\mathrm{max}=1500$ deboosted case, however, resulted in the mean dipole direction of $(l,b) = (218^\circ,-21^\circ)$, which was intriguingly close to the direction found for the dipole modulation, as we will discuss below. Even though a power asymmetry was detected using the angular clustering of the preferred directions for different bins, the ratio of the power spectra in the two opposite hemispheres defined by the asymmetry axis for $\ell = 2-600$ was not statistically anomalous. The interpretation for this was that the power asymmetry detected through the clustering technique might be different from the one found by other methods with statistics based on the amplitude of the asymmetry, usually quantified by a dipole (as done, e.g., in the cases of dipole modulation and local-variance asymmetry discussed below).

In an updated and extended analysis, {\it Planck} 2015~\cite{Planck:2015igc} considered bin sizes between $\Delta\ell = 8$ and $\Delta\ell = 32$ for the multipole range of $\ell=2-1500$. In contrast to the {\it Planck} 2013 results where the $p$-values started to increase systematically for $\ell_\mathrm{max} \gtrsim 600$, here they remained low for $\ell_\mathrm{max} > 750$. Depending on the bin size and component separation method, $p$-values were as low as $<0.04\%$. These all suggest that, beyond a dipole modulation of power on large angular scales, some form of directional asymmetry continues to small scales. There are also indications that the directions of dipolar asymmetry are correlated between large and small angular scales.

\paragraph{Dipole Modulation (Pixel-Based Likelihood).} Even though no compelling theoretical explanation exists for the nature of the hemispherical power asymmetry, a phenomenological multiplicative dipole modulation model, suggested by Ref.~\cite{Gordon:2005ai}, has been widely used to study the asymmetry through a model-dependent, Bayesian approach. Here, the asymmetry is modeled in terms of a dipole modulation of the form
\begin{equation}
    \label{eq:dipole}
    \frac{\Delta T}{T}|_{\textrm{mod}}({\bf \hat{n}})=(1+\alpha \,{\bf \hat{n}}\cdot{\bf \hat{p}})\frac{\Delta T}{T}|_{\rm iso}({\bf \hat{n}})\,,
\end{equation}
where $\frac{\Delta T}{T}|_{iso}$ and $\frac{\Delta T}{T}|_{\textrm{mod}}$ are the isotropic and modulated CMB temperature fluctuations along a direction ${\bf \hat{n}}$ on the sky, respectively, and $\alpha$ is the amplitude of the dipole modulation and ${\bf \hat{p}}$ is the preferred direction.

It is important to note that since various other analyses of the CMB sky seemed to imply that the southern hemisphere was relatively consistent with an isotropic $\Lambda$CDM sky and it was the northern hemisphere that showed an anomalous deficit of power, the dipole modulation may not be a well justified way of modeling the asymmetry; it has been used mainly because of its simplicity. Refs.~\cite{Eriksen:2007pc,Hoftuft:2009rq} studied the dipole modulation model using the 3- and 5-year WMAP data, respectively, and the {\it Planck} collaboration then performed a similar direct pixel-space likelihood analysis of the model in their 2013~\cite{Planck:2013lks} and 2015~\cite{Planck:2015igc} investigations of the hemispherical power asymmetry by only considering large angular scales. {\it Planck} used $N_\mathrm{side}=32$ temperature maps, smoothed to angular resolutions ranging from $5^\circ$ to $10^\circ$ FWHM. {\it Planck} 2015 particularly focused on a smoothing scale of $5^\circ$ FWHM as a representative example and the highest angular resolution accessible for an $N_\mathrm{side} = 32$ map. The results of the analysis were all consistent with those derived from the 5-year WMAP ILC map in Ref.~\cite{Eriksen:2007pc}. The significance levels for the amplitude of the dipole compared to the ones from the $\Lambda$CDM simulations varied with smoothing scale and the $5^\circ$ scale showed the highest significance ($\sim 3.5\sigma$). It is important to note that the model of Ref.~\cite{Gordon:2005ai} assumes that the modulation amplitude $\alpha$ is equally strong on all scales, but it has been shown~\cite{Quartin:2014yaa,Notari:2013iva} that the data are not consistent with a simple constant-amplitude dipole modulation of the power. The preferred direction derived from the low-$\ell$ dipole modulation analysis, e.g.\ $(l, b)= (227^\circ, -15^\circ)\pm 19^\circ$ for the \texttt{Commander} component separation technique, is, however, remarkably consistent with the high-$\ell$ direction derived from the clustering of directions for $\ell\lesssim600$. Both {\it Planck} 2013 and 2015 found almost exactly the same preferred directions.

\paragraph{Dipole Modulation (QML Analysis).} In a harmonic-space analysis, {\it Planck} 2015~\cite{Planck:2015igc} used the quadratic maximum likelihood (QML) estimator introduced by Ref.~\cite{Moss:2010qa} to assess the level of dipole modulation for the full-resolution ($N_\mathrm{side} = 2048$) CMB temperature maps. This technique exploits the fact that dipole modulation is equivalent to coupling of $\ell$ to $\ell\pm1$ modes in the CMB covariance matrix to leading order. {\it Planck} computed $p$-values of the fitted modulation amplitude as a function of $\ell_\mathrm{max}$, and the $p$-values showed several peaks, at $\ell_\mathrm{max} \approx 40$, $\ell_\mathrm{max} \approx 67$ and $\ell_\mathrm{max} \approx 240$. The latter peak, while not previously emphasized, was also present in the WMAP results~\cite{Bennett:2010jb}. The dip at $\ell_\mathrm{max} \approx 67$, with a $p$-value of $0.9\%-1.0\%$ ($\sim 3\sigma$), corresponds to the low-$\ell$ dipole modulation, which has been the focus of most attention in the literature.

{\it Planck} 2018~\cite{Planck:2019evm} repeated the analysis and studied two ranges of $\ell=2-64$ and $\ell=2-220$. The preferred direction for the former range was found to be $(l,b)=(221^\circ, -22^\circ) \pm 31^\circ$, while it was $(l,b)=(213^\circ, -26^\circ) \pm 28^\circ$ for the {\it Planck} 2015 analysis~\cite{Planck:2015igc}, both of which are consistent with the directions found through the pixel-based likelihood analysis of dipole modulation. One should note that this QML analysis was a purely phenomenological one performed in multipole space, with no attempt to connect to any real-space modulation. Even though the preferred directions are close to that of a pixel-space likelihood analysis of the dipolar modulation model, the two methods do not necessarily probe the same thing.

Meanwhile, WMAP also observed a dip in the first Doppler peak ($\ell_\mathrm{max} \approx 220$), an excess at $\ell_\mathrm{max} \approx 44$, and a dip at $\ell_\mathrm{max} \approx 22$ that were present in the ecliptic polar data but not the ecliptic planar data, see Fig.~7 of the first arXiv version of their Year 1 paper on the angular power spectrum~\cite{WMAP:2003zzr}. That first peak dip was shown in Ref.~\cite{Yoho:2010pb} to be localized to the region of the north ecliptic pole, and to persist beyond WMAP1 into later WMAP data releases. This is consistent with the dip's presence in ARCHEOPS data, taken in the northern hemisphere, but in no other pre-WMAP ground-based data.

\paragraph{Dipole Modulation (Rayleigh Statistic).} The results of the analysis of the angular clustering of the power distribution performed by the {\it Planck} collaboration and discussed above suggest that, beyond a dipole modulation of power on large angular scales, some form of directional asymmetry continues to small scales. There are also indications from the same analysis that the directions of dipolar asymmetry are correlated between large and small angular scales. {\it Planck} 2015~\cite{Planck:2015igc} then performed a Rayleigh-statistic-based analysis as a generic test for directionality with minimal assumptions about the nature of the asymmetry. The method, however, takes into account other information pertaining to modulation, including its amplitude, in addition to the direction. An asymmetry was found on scales larger than $\ell\approx 240$, and the minimum $p$-value was $0.1\%-0.2\%$, to be compared to the $p$-value of $0.9\%-1.0\%$ obtained for the dipole modulation amplitude at $\ell_\mathrm{max} = 67$. The preferred direction for $\ell_\mathrm{max} \approx 240$ was found to be $(l,b) = (208^\circ,-29^\circ)$, which is approximately $20^\circ$ away from the dipole modulation direction determined by the QML analysis for the multipole range of $\ell=2-64$.

\paragraph{Generalised Dipole Modulation (Bipolar Spherical Harmonics).} The {\it Planck} collaboration studied, in both their 2013~\cite{Planck:2013lks} and 2015~\cite{Planck:2015igc} analyses, a generalisation of the dipole modulation using the Bipolar Spherical Harmonics (BipoSH) formalism of Refs.~\cite{Hajian:2003qq,Hajian:2006ud}.  Here, the CMB angular power spectrum corresponds to the $L = 0$ BipoSH coefficients and a dipole modulation corresponds to the bipolar multipole $L = 1$. The BipoSH formalism provides a unified way to systematically detect and study properties of different types of statistical anisotropy, which may go beyond the dipole modulation. A BipoSH-based simulation tool is available to model Gaussian realizations of CMB temperature and polarization maps with induced isotropy violation~\cite{Mukherjee:2013kga}. {\it Planck} 2013 applied the BipoSH formalism to $N_\mathrm{side}=32$ temperature maps and detected a dipole modulation signal with $2.9\sigma$ to $3.7\sigma$ significance, depending on the component separation method. The amplitude and direction of the dipole modulation matched those obtained via the likelihood analysis of dipole modulation. Both {\it Planck} 2013 and 2015 also used higher-resolution maps, but the dipole modulation was recovered at over $3\sigma$ significance only for the lowest multipole bin they considered, i.e.\ $\ell = 2-64$. For the full-resolution $N_\mathrm{side}=2048$ maps, {\it Planck} 2015 found a preferred direction of $(l,b)\approx(230^\circ,-18^\circ)\pm 31^\circ$, in very good agreement with the direction inferred from the likelihood analysis of dipole modulation. No significant power was detected at the higher BipoSH multipoles of the modulation field, $1<L\leq 32$. Assuming a scale-dependent dipole modulation, Ref.~\cite{Shaikh:2019dvb} has measured this scale dependence using the BipoSH formalism.   

\paragraph{Variance Asymmetry.} In an attempt to study the observed CMB power asymmetry in a model-independent way, Ref.~\cite{Akrami:2014eta} invented a pixel-space local-variance estimator of the CMB fluctuations using a set of discs of various sizes uniformly distributed on the sky. The technique was first applied to the WMAP 9-year and {\it Planck} 2013 temperature data in Ref.~\cite{Akrami:2014eta} and later used in the analyses of {\it Planck} 2015~\cite{Planck:2015igc} and 2018~\cite{Planck:2019evm} for both temperature and polarisation. Since the technique looks for deviations from statistical isotropy in terms of the variance of CMB fluctuations by fitting a dipole to a given local-variance map, it can be considered as an alternative way of detecting the hemispherical power asymmetry which had already been detected through the dipole modulation analysis and the angular clustering of dipole directions. This is because variance is effectively a weighted sum of the power spectrum over all scales (or multipoles).

\begin{figure}
    \centering
    \includegraphics[width=0.49\textwidth]{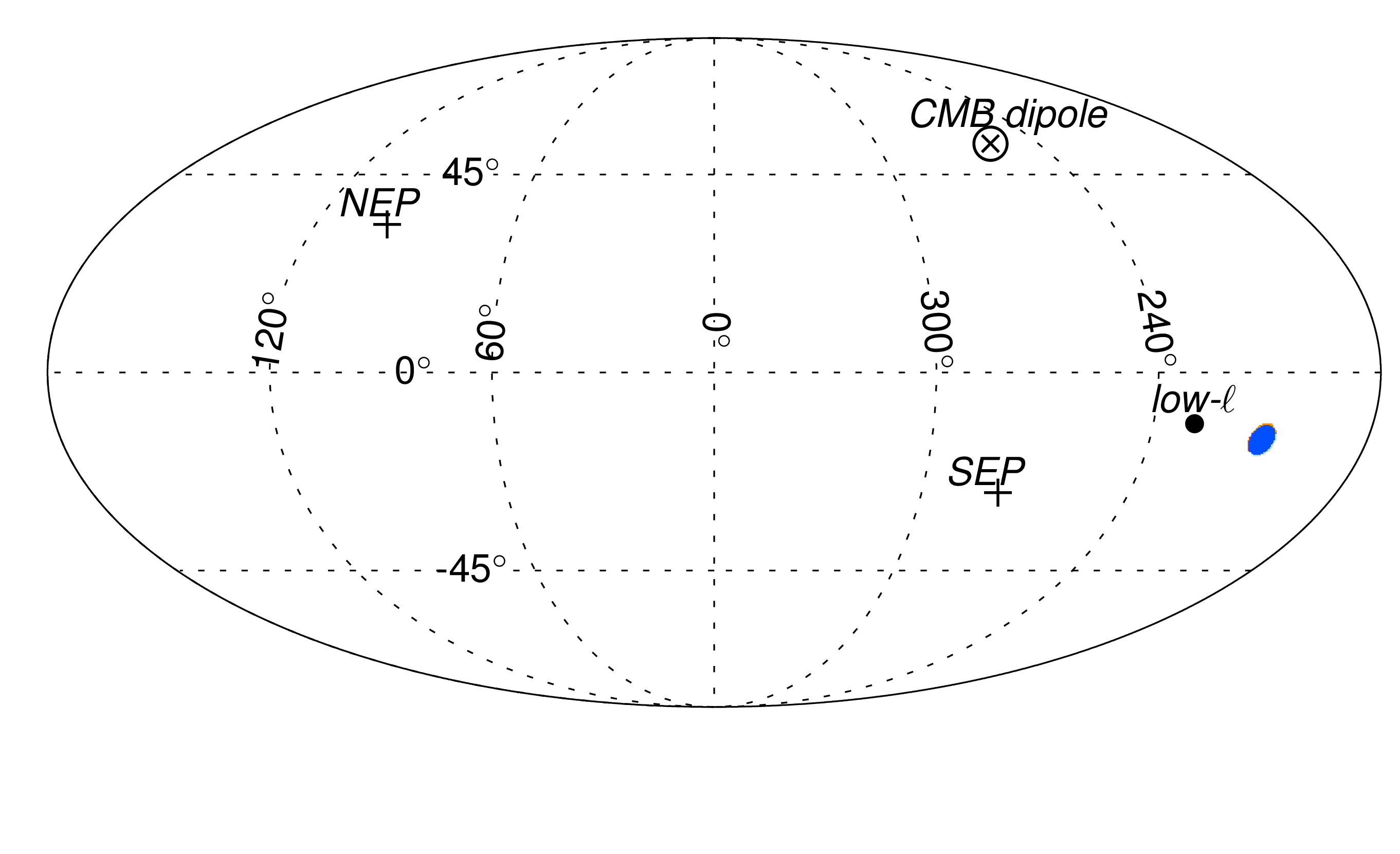}
    \includegraphics[width=0.49\textwidth]{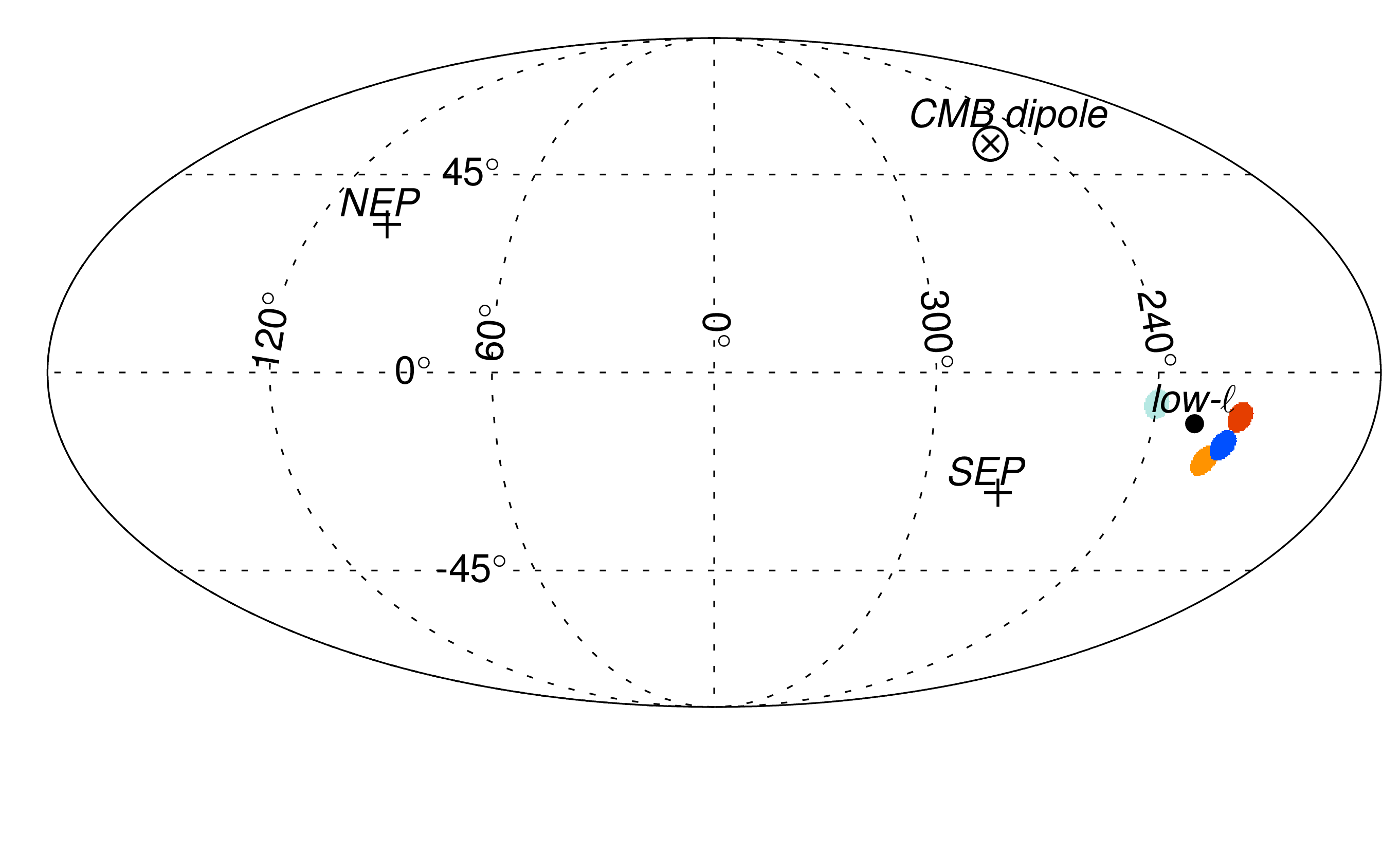}
    \caption{Variance asymmetry directions for the four low-resolution ($N_\mathrm{side}=64$) component-separated maps \texttt{Commander} (red), \texttt{NILC} (orange), \texttt{SEVEM} (green) and \texttt{SMICA} (blue), and for $4^\circ$ discs with the lowest $p$-values for {\it Planck} 2018 temperature (left) and $E$-mode polarization (right) data in Galactic coordinates. Note that the four directions for the temperature maps match almost perfectly, so that the dots with different colors essentially overlap. For reference, four additional directions are also shown in each panel: the north ecliptic pole (NEP), the south ecliptic pole (SEP), the CMB dipole direction and the preferred dipole modulation direction (labelled as "low-$\ell$") derived from the temperature data through the QML analysis. Both panels adopted from Ref.~\cite{Planck:2019evm}.}
    \label{fig:varianceasymmetry}
\end{figure}

The original analysis of Ref.~\cite{Akrami:2014eta} found that none of the $1000$ available CMB temperature simulations had a larger variance asymmetry than that estimated from the data. This suggested presence of a power asymmetry at a statistical significance of at least $3.3\sigma$, with a preferred direction of $(l, b) \approx (212^\circ, -13^\circ)$, in good agreement with other studies of the hemispherical power asymmetry. The local-variance estimator was then applied to the full-resolution $N_\mathrm{side}=2048$ {\it Planck} 2015 temperature maps~\cite{Planck:2015igc}, and significantly low $p$-values were found, although the significance levels and preferred directions showed slight differences for different disc radii. The highest statistical significance was found for $8^\circ$ discs with $p$-values less than $0.1\%$. In an analysis with enhanced sky coverage, where a mask smaller than the {\it Planck}'s common mask was used, a number of anomalies were restudied through an $N_\mathrm{side}=256$ \texttt{Commander} temperature map which included about $93\%$ of the sky. While in general the significance levels and preferred directions for the hemispherical power asymmetry inferred from different techniques did not change much with this reduced mask, the local-variance asymmetry method showed higher significance for larger disc radii. The variance asymmetry technique was finally applied to the {\it Planck} 2018 temperature data~\cite{Planck:2019evm} and $p$-values smaller than $0.1\%$ were found for disc radii up to $8^\circ$ when the full-resolution $N_\mathrm{side}=2048$ maps were analysed. The preferred direction was found to be $(l,b)=(205^\circ,-20^\circ)$ for this resolution and for $4^\circ$ discs, and changed slightly for lower-resolution maps; the direction was found to be $(l,b)=(209^\circ,-15^\circ)$ for $N_\mathrm{side}=64$; see Fig.~\ref{fig:varianceasymmetry} (left panel).

{\it Planck} 2018~\cite{Planck:2019evm} also studied the hemispherical power asymmetry, as well as a few other large-scale anomalies, for the CMB polarization data. By applying the local-variance estimator of Ref.~\cite{Akrami:2014eta} to $N_\mathrm{nside}=64$ $E$-mode polarization maps, {\it Planck} 2018 found preferred directions for polarization with the smallest $p$-values of $\sim0.7\%$ and $\sim0.4\%$ for \texttt{Commander} and \texttt{SEVEM} component-separated maps, respectively, and $\sim5.8\%$ and $\sim5.5\%$ for \texttt{NILC} and \texttt{SMICA}, when $4^\circ$ discs were used. This difference in the significance levels might reflect differences in the component separation approaches, particularly given that the former methods operate in the pixel domain, while the latter in the harmonic domain. Another interesting finding of the {\it Planck} 2018 variance asymmetry analysis is that the preferred directions for polarization are intriguingly close to those determined for the temperature data, with the smallest polarization-temperature separation angle of $\arccos{(0.99)}$ for \texttt{Commander} and \texttt{SMICA} (with $p$-values of $\sim 0.9\%$). The preferred directions are  $(l,b)=(217^\circ, -10^\circ)$ and $(l,b)=(219^\circ, -16^\circ)$, respectively, which are close to the temperature variance asymmetry direction of $(l,b)=(209^\circ,-15^\circ)$ for the same $N_\mathrm{side}$ and the dipole modulation direction of $(l,b)=(221^\circ, -22^\circ) \pm 31^\circ$ based on the QML analysis of the 2018 data; see Fig.~\ref{fig:varianceasymmetry} (right panel). While the differences between the $p$-values and preferred directions associated to the different component-separated polarization maps argue against a detection of cosmological power asymmetry in the polarization data and, therefore, the close alignment of the temperature and polarization directions could simply be a coincidence, future higher-quality polarization data may offer additional insight.

\paragraph{Peak Distribution Asymmetry.} The {\it Planck} collaboration employed, in both their 2015~\cite{Planck:2015igc} and 2018~\cite{Planck:2019evm} analyses, local extrema (or peaks) in the CMB maps to search for localized anomalies by examining how their statistical properties would vary in patches of the sky as a function of their location. This method was particularly used as a further test for large-scale asymmetries by examining the differences in the peak distribution when divided according to orientation with respect to a previously specified asymmetry direction. In the {\it Planck} 2015 analysis, a disc of radius $70^\circ$ centred on $(l, b) = (225^\circ, -18^\circ)$, the dipole modulation direction found in the \texttt{SMICA} temperature map, and its corresponding antipodal disc were used to select the peaks. For maps filtered with a $40^\prime$ FWHM GAUSS filter, as well as a number of other filtering scales, the distribution of the peaks for the positive-direction disc was in general agreement with the full sky result, while that for the negative direction (close to the northern ecliptic point) was marginally different---this again showed that the northern hemisphere is anomalous. {\it Planck} 2018 repeated the analysis using the direction determined by the \texttt{SMICA} QML estimator, $(l,b)= (213^\circ, -26^\circ) \pm 28^\circ$, and confirmed the 2015 results.

\paragraph{Predictions.} Limited work has been done to study the predictions of $\Lambda$CDM for the wide range of hemispherical power anomalies. However, Ref.~\cite{ODwyer:2016xov} showed that the $\Lambda$CDM predictions for the variance in Q and U  polarization conditioned on the low northern temperature variance are nearly unchanged from the unconditioned predictions. Any observed  suppression of variance in Q and U in the north would therefore be an independent anomaly in $\Lambda$CDM, but might be expected in a physical explanation of the temperature anomaly.  We expect that similar statements hold for other characterizations of the hemispherical power asymmetry. However, extension of the dipolar power asymmetry signal beyond $\ell>64$~\cite{Aiola:2015rqa,Shaikh:2019dvb} can lead to a power asymmetry in the lensing B-modes of polarization, as shown in Ref.~\cite{Mukherjee:2015wra}. This could be measured from the upcoming ground-based CMB experiments such as SO~\cite{SimonsObservatory:2018koc} and CMB-S4~\cite{Abazajian:2019tiv}. Other probes such as the kinematic and polarized SZ~\cite{Cayuso:2019hen} can also explore the power asymmetry signal from future surveys. A violation of statistical isotropy can also lead to direction dependence in the inferred cosmological parameters~\cite{Mukherjee:2015mma}.  While Ref.~\cite{Mukherjee:2017yxd} concludes that  {\it Planck} temperature maps do not show any significant departure from statistical isotropy in the inferred cosmological parameters, others~\cite{Fosalba:2020gls, Yeung:2022smn} claim strong evidence of such variation (see subsection~\ref{sec:parameter-variation}). In the future, this can be tested using both temperature and polarization data.

\subsubsection{Quadrupole and Octopole Anomalies}
\label{sec:quadrupoleoctopole}

\begin{figure}
    \centering
    \includegraphics[width=0.9\textwidth]{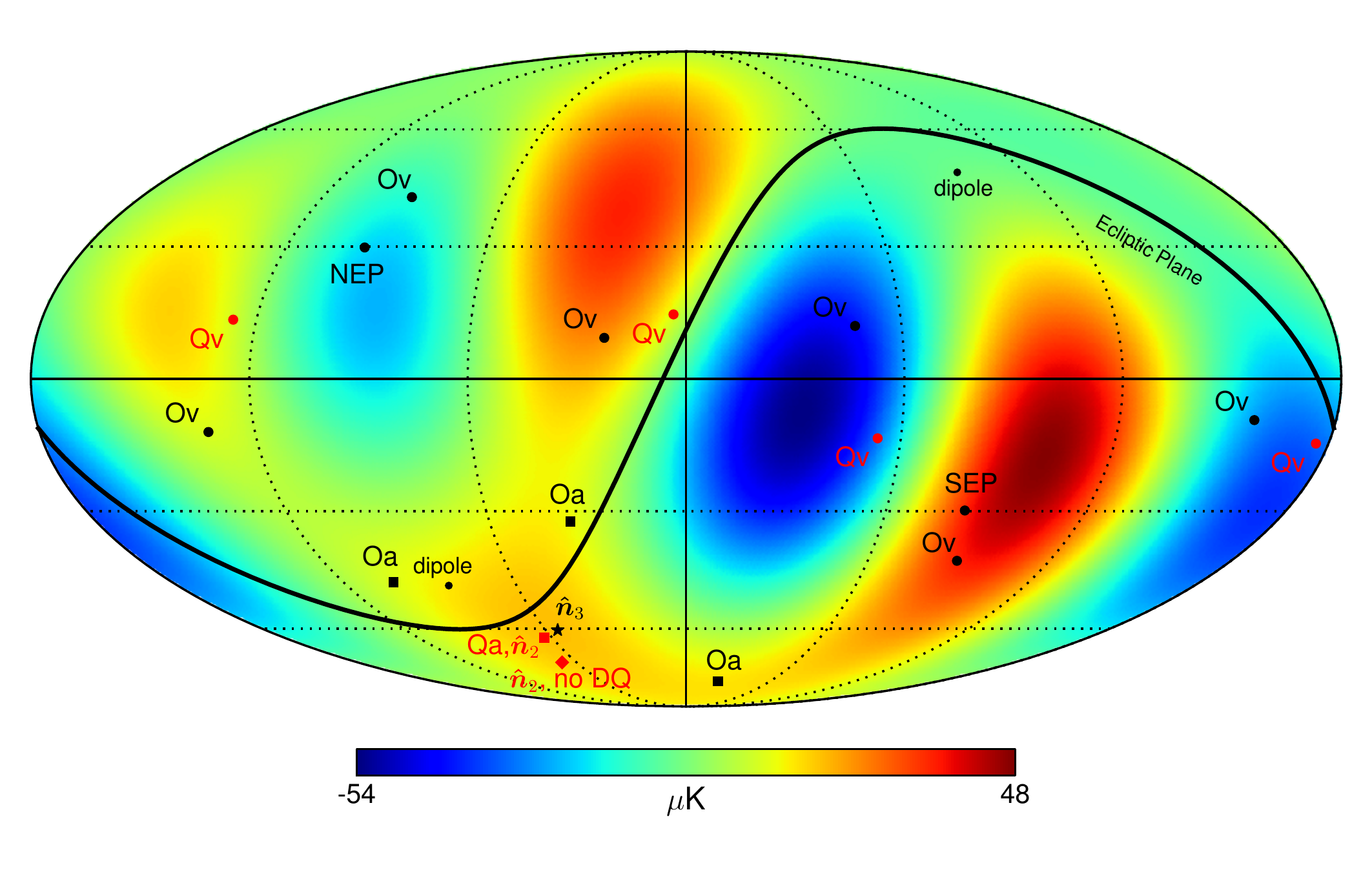}
    \caption{Fig.~5 from Ref.~\cite{Copi:2013jna}. Quadrupole and octopole multipole vectors for the Doppler-quadrupole (DQ) corrected \texttt{SMICA} map in Galactic coordinates. The background shows the quadrupole+octopole pattern. The multipole vectors are shown as circles, in red and labelled "Qv" for the quadrupole and in black and labelled "Ov" for the octopole. The directions of the area vectors defined in the text, $\hat{w}^{(\ell;i,j)}$, are shown as squares, with $\ell=2$ in red and labelled "Qa" and $\ell=3$ in black and labelled "Oa." Since the multipole vectors are only determined up to a sign each vector appears twice in the figure. The area vectors have only been plotted in the Southern hemisphere to avoid cluttering the plot. The maximum angular momentum dispersion direction for $\ell=3$, $\hat{n}^{(3)}$ is shown as the black star, Since $\hat{n}^{(2)}=\hat{w}^{(2;1,2)}$. The direction of $\hat{n}^{(2)}$ without the DQ correction is the red diamond. For reference also shown are the ecliptic plane (black line), the north (NEP) and south (SEP) Ecliptic poles, and the  CMB (dipole).}
    \label{fig:quadrupoleoctopoleanomaly}
\end{figure}

$\Lambda$CDM asserts that the $a_{\ell m}$ are statistically independent random variables. Yet, $a_{2m}$ and $a_{3m}$ show anomalous correlations~\cite{deOliveira-Costa:2003utu}, suggesting $\langle a^\star_{\ell m} a_{\ell'm'}\rangle~\cancel{\propto}~\delta_{\ell\ell'}\delta_{mm'}$.

There are two principal statistical approaches to describing these anomalous correlations. The first is in terms of the multipole vectors of the quadrupole and octopole (Fig.~\ref{fig:quadrupoleoctopoleanomaly}): $\hat{v}^{\ell,i}$, $i=1,\ldots,\ell$, $\ell=2,3$. These $\ell$ unit vectors plus a scalar $A_\ell$, are an alternative to $a_{\ell m}$ for representing real scalar functions on the sphere. In $\Lambda$CDM, the $\hat{v}^{\ell,i}$ of different $\ell$ are uncorrelated; the correlations between $\hat{v}^{\ell,i}$ of fixed $\ell$ are calculable.  

It was observed in Ref.~\cite{Schwarz:2004gk} that the oriented areas (normal) of the  plane defined by the two quadrupole multipole vectors $\vec{w}^{2;1,2}\equiv\hat{v}^{2,1}\times \hat{v}^{2,2}$, and those of the 3 planes defined by the three octopole multipole vectors ($i\neq j=1,2,3$), $\vec{w}^{3;i,j}\equiv\hat{v}^{3,i}\times \hat{v}^{3,j}$, are unexpectedly close to one another: the magnitude of the dot products of  $\hat{w}^{2;1,2}$, with the three  $\hat{w}^{3;i,j}$ are approximately $0.95$, $0.87$, and $0.84$.  
Using instead the $\vec{w}^{\ell;i,j}$ themselves and not unit vectors parallel to them, the three dot products $\vert\vec{w}^{2;2,1}\cdot\vec{w}^{3;i,j}\vert$ are  $0.85$, $0.78$, and $0.78$; in $\Lambda$CDM, the p-value of these  being at least this large is $0.02\%$! In other words, the three octopole multipole vectors  nearly share a a common plane, this is called octopole planarity, and that plane is nearly that of the quadrupole.  This is quadrupole-octopole alignment.

The quality of this alignment improved noticeably on removing (as one ought) the Doppler contribution to the quadrupole~\cite{Copi:2013jna,Notari:2015kla}, reducing the p-value from $0.1\%$ to the $0.02\%$ just quoted. This would be a separate "statistical fluke" in $\Lambda$CDM.

The original approach to seeing these alignments was through the angular momentum dispersion in each multipole~\cite{deOliveira-Costa:2003utu}, $\sum_m m^2 \vert a_{\ell m}\vert^2_{\hat{n}_\ell}$. The subscript $\hat{n}_\ell$ indicates that this is a function of the choice of $z$ axis. For each $\ell$,  $\hat{n}_\ell$ is the z-axis that maximizes the angular momentum dispersion in that $\ell$. In the WMAP 1st year ILC map, $\hat{n}_2\cdot\hat{n}_3\simeq0.98$, which has a p-value of $\simeq1.5\%$; this again is quadrupole-octopole alignment. It was noted that in this frame the magnitudes of $a_{3\pm 3}$ are much larger than those of other $a_{3m}$, at a level that has a p-value of 7\% in $\Lambda$CDM; this is octopole planarity. If the three octopole $\hat{w}^{3,i,j}$ had been exactly parallel, then they would have been parallel to $\hat{n}_3$. This common axis has been called "the axis of evil"~\cite{Land:2005jq}. Finally, there are some suggestions of correlations also between $a_{\ell m}$ of pairs of $\ell$ other than $2$ and $3$~\cite{Copi:2003kt,Land:2005jq}.

There are other unexpected correlations of the quadrupole and octopole, in particular with the ecliptic.
The great circle defined by the common plane of the quadrupole and octopole is nearly perpendicular to the ecliptic.  Near that great circle the quadrupole-plus-octopole combine to produce 6 extrema. The ecliptic great circle seems to trace out surprisingly well the zero separating three strong extrema in the ecliptic south from three weaker extrema in the ecliptic north.
This connection to the ecliptic suggests either an experimental systematic or a solar-system foreground. The WMAP and {\it Planck} satellites orbited in the plane of the ecliptic and their sampling strategies and noise maps are  "ecliptic-aware," however their systematics are very different and both see nearly identical quadrupoles and octopoles.  As far as solar-system foregrounds, the structure of the quadrupole and octopole as planar and perpendicular to the ecliptic is entirely unlike anything that would be expected from the solar system itself --- no reasonable proposal has been made that would generate the observed quadrupole and octopole. 

It has been found that if the Universe has non-trivial topology it can induce correlations between the quadrupole and octopole~\cite{Bielewicz:2008ga}, though the statistical evidence for that is weak, and the manifolds specifically consider  are ruled out by the search for matched circle pairs~\cite{Vaudrevange:2012da}. In such an explanation, the correlations with the ecliptic would be a fluke.

\paragraph{Predictions.} Because $(\mathcal{C}_{\ell}^{TE})^2\ll \mathcal{C}_\ell^{TT}\mathcal{C}_\ell^{EE}$ in $\Lambda$CDM, the peculiar correlations of the $a^T_{2m}$ and $a^T_{3m}$ has little effect on the multipoles of polarization.
This may not be the case in specific model explanations of the temperature anomaly.  For example, one would expect that in a topologically non-trivial manifold, the boundary conditions would imprint not just on the $a^T_{\ell m}$, but equally on the $a^E_{\ell m}$ and $a^B_{\ell m}$.  However, the precise prediction is expected to depend strongly on the specific manifold.

\subsubsection{Point-Parity Anomaly}

\begin{figure}[!tbh]
    \centerline{
    \includegraphics[width=0.4\textwidth]{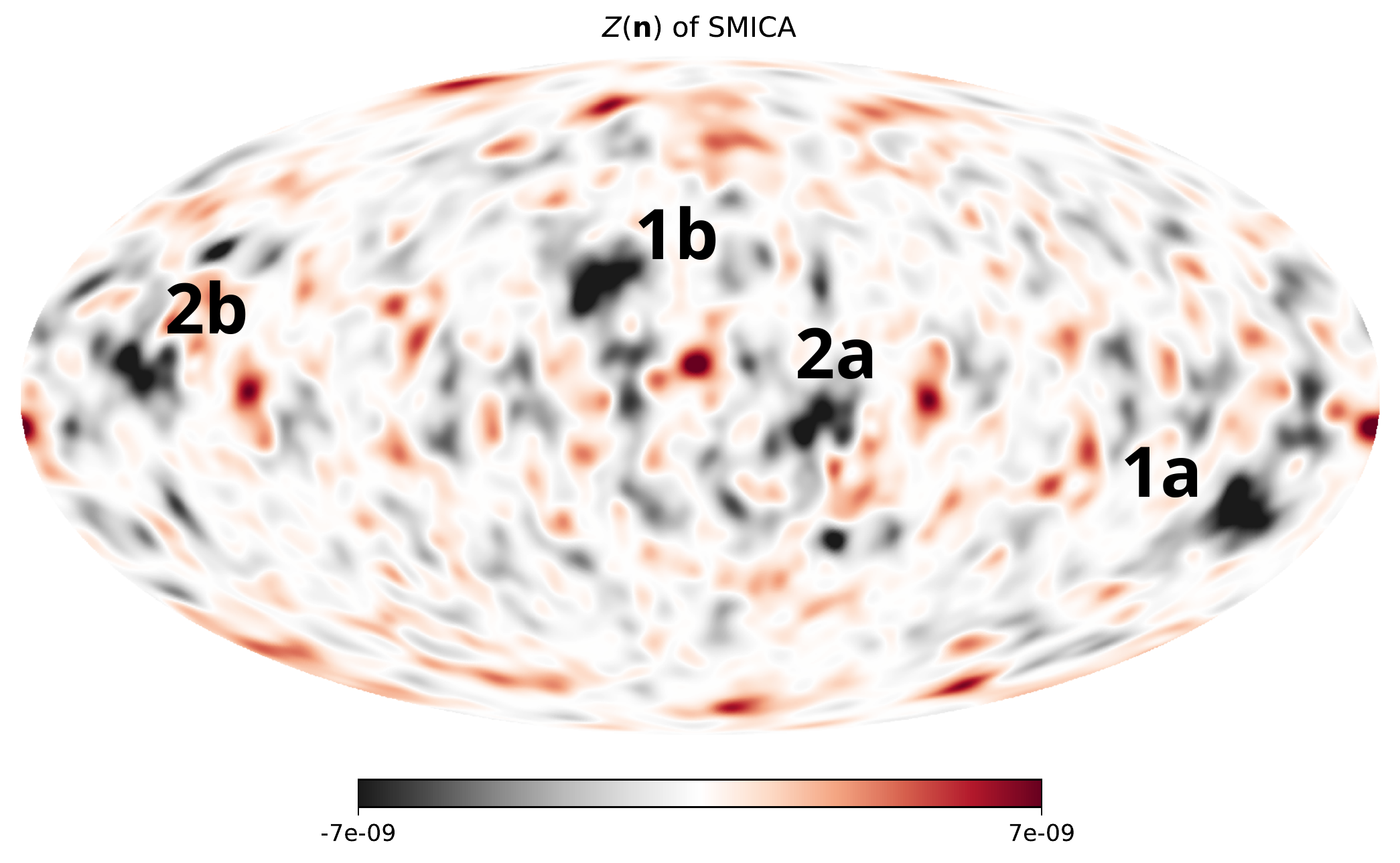}}
 \caption{Map of $Z({\bf \hat{n}})\equiv T({\bf \hat{n}})T(-{\bf \hat{n}})$ for \texttt{SMICA} with $\Theta=5^\circ$ smoothing~\cite{Creswell:2021eqi}. Note, the two pairs of very strong high negative peaks of $Z({\bf \hat{n}})$, labelled 1a/1b and 2a/2b, and about twenty negative peaks with smaller amplitudes, mainly localized within the area $|b|\le 30^\circ$ in Galactic coordinates.}
    \label{fig1naselsky}
\end{figure}
\begin{figure}[!tbh]
    \centering
    \includegraphics[width=0.4\textwidth]{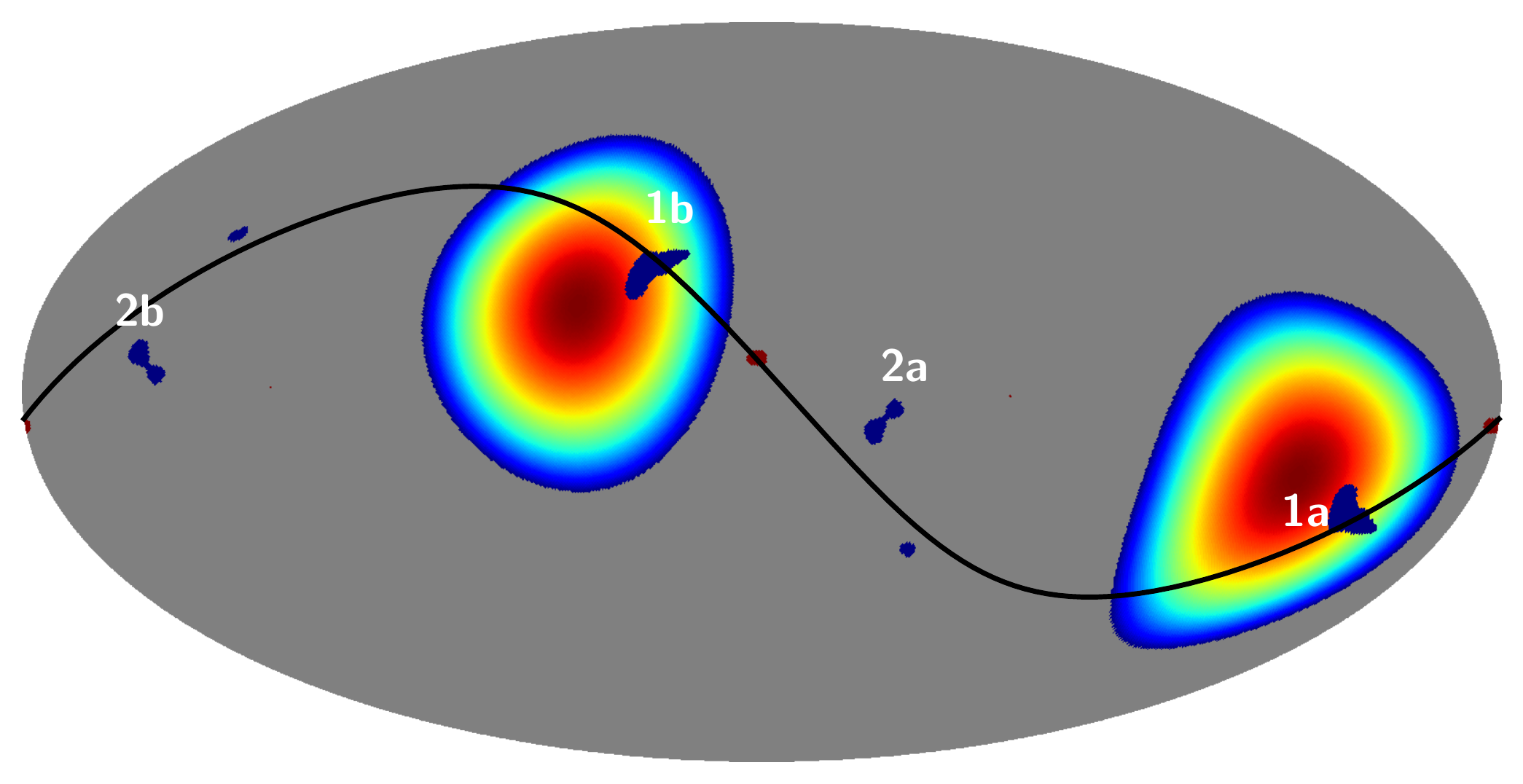}
    \includegraphics[width=0.4\textwidth]{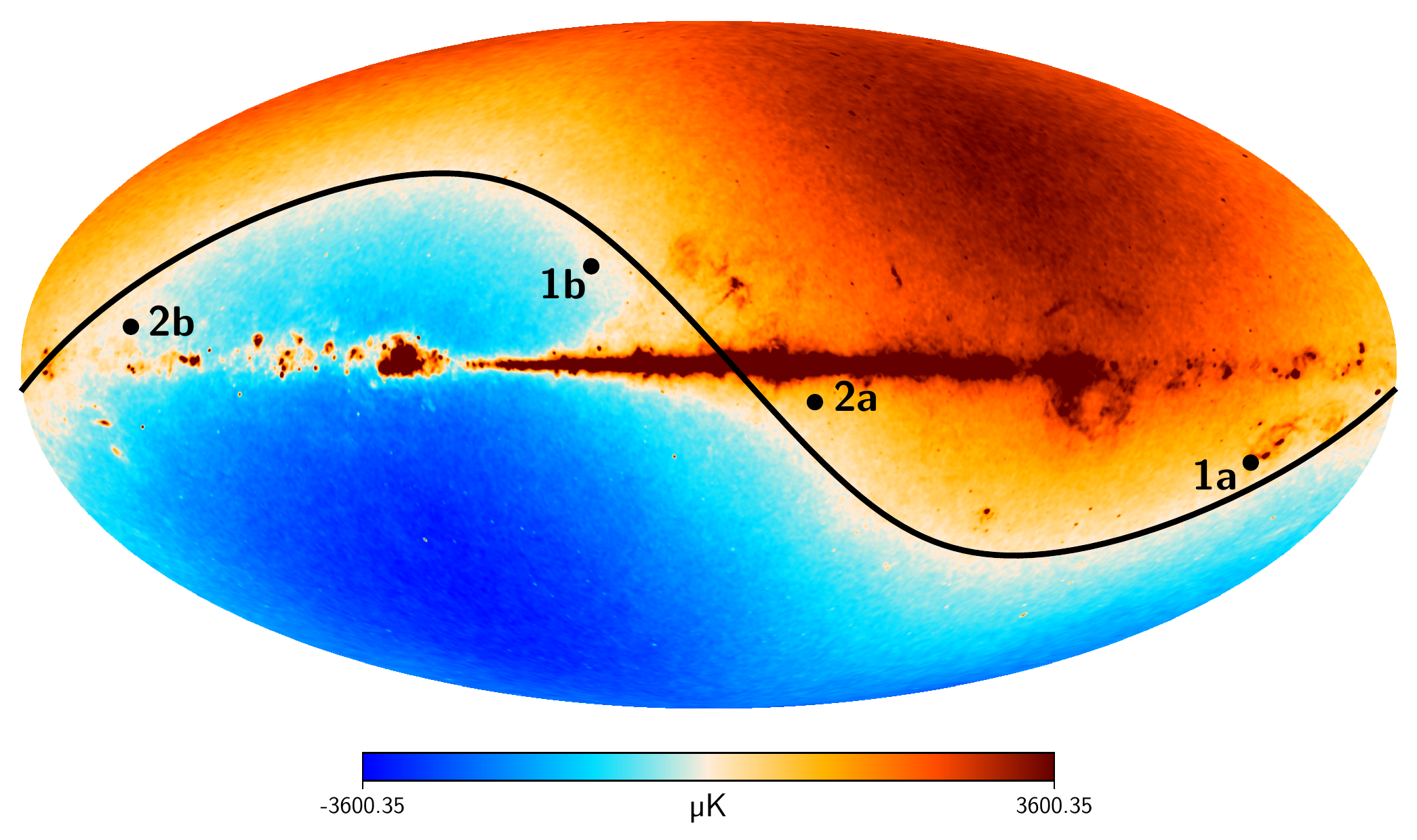}   
    \caption{Left panel: The coloured contours show the posterior distribution of the direction of the dipole modulation in combination with the ring of attraction and 1a/1b and 2a/2b peaks responsible for maximum of the parity asymmetry. The right contour corresponds to the maximum of CMB power, the left contour indicates the minimum. Right panel: The $30\,$GHz {\it Planck} map with kinematic dipole. The white ring is the zone where the total signal is vanished.}
    \label{fig2naselsky}
\end{figure}

An anomalous power excess of odd $\ell$ multipoles compared to even $\ell$ multipoles has been shown to exist in the CMB angular power spectrum on the largest angular scales ($2<\ell<30$)~\cite{Land:2005jq,Kim:2010gd,Kim:2010gf,Kim:2010st,Gruppuso:2010nd,Gruppuso:2017nap,Planck:2013lks,Planck:2015igc,Planck:2019evm}. As odd-$\ell$ (even-$\ell$) spherical harmonics possess odd (even) point-parity, this effect is known as the point-parity anomaly.\footnote{Note that this is different from the so-called "mirror-parity anomaly", which is about the properties of the CMB maps under reflection with respect to a plane. Both {\it Planck} 2013~\cite{Planck:2013lks} and {\it Planck} 2015~\cite{Planck:2015igc} reported evidence for an anti-symmetry plane in low-resolution $N_\mathrm{side}=16$ temperature maps with the direction $(l,b)=(264^\circ,-17^\circ)$ and a $p$-value ranging from $1.6\%$ to $2.9\%$ depending on the component separation method. The analysis of the higher-resolution $N_\mathrm{side}=32$ maps showed almost the same direction, $(l,b)=(264^\circ,-16^\circ)$, with lower $p$-values ($1.6\%$ to $2.9\%$). This mirror-parity anomaly was first detected in the WMAP 7-year data~\cite{Finelli:2011zs}, with an anti-symmetry direction almost identical to the one found by {\it Planck}, which is within $40^\circ$ of the dipole modulation direction.}

In order to compare even and odd multipoles, Ref.~\cite{Kim:2010gf} considered the parity asymmetry statistic  $P\equiv P^+/P^-$, where $P^+$ and $P^-$ are the mean power in even and odd multipoles respectively for the range $2\leq \ell \leq \ell_{\rm max}$,
\begin{equation}
    P^{\pm}=\sum_2^{\ell_{\rm max}}\frac{\left[1 \pm (-1)^\ell\right]\ell(\ell+1) C_\ell}{4\pi}\,.
\end{equation}
The $\Lambda$CDM model predicts $P^+=P^-$, i.e.\ $P=1$. 
Examining WMAP3 and WMAP5, Ref.~\cite{Kim:2010gf} found that $P<1$ for $\ell_{\rm max}$ up to $\sim25$, with a $p$-value in $\Lambda$CDM of $\lesssim1\%$.

The point-parity anomaly was re-examined in Refs.~\cite{Aluri:2011wv, Hansen:2011wk, Kim:2012hf, Gruppuso:2010nd, Kim:2010gd} using the 5-year and 7-year releases of WMAP data. 
Ref.~\cite{Aluri:2011wv} found that the anomaly extended out at least to $\ell\simeq100$. 
Focusing on the power spectrum in the $2 \leq \ell \leq 30$ domain, an excess of oddness with significance $p = 0.004$ was found, which was related to the correlation function at large angular scales~\cite{Kim:2012hf}. The point-parity anomaly persists in the {\it Planck} temperature data~\cite{Planck:2013lks,Planck:2015igc,Planck:2019evm, Gruppuso:2017nap,Shaikh:2019dvb} with the lowest $p$-values of $\sim 0.2\%$ found at $\ell\sim28$.

Recently, properties of the point-parity asymmetry were investigated directly for \texttt{SMICA} CMB map pixels to identify the preferred directions on the sky responsible for this type of anomaly~\cite{Creswell:2021eqi,Creswell:2021xzf}.
The method proposed in Refs.~\citep{Creswell:2021eqi, Creswell:2021xzf} is based on the signal symmetry properties at each pixel ${\bf \hat{n}}$ and its anti-pode $-{\bf \hat{n}}$ with respect to the center of the map: $T({\bf \hat{n}})=S({\bf \hat{n}})+A({\bf \hat{n}})$, where the symmetric ($S({\bf \hat{n}})$) and asymmetric ($A({\bf \hat{n}})$) parts of the temperature anisotropy map are
\begin{eqnarray}
    S({\bf \hat{n}}) &=& 0.5\left[T({\bf \hat{n}}) + T(-{\bf \hat{n}})\right]\,,\\
    A({\bf \hat{n}}) &=& 0.5\left[T({\bf \hat{n}}) - T(-{\bf \hat{n}})\right]\,.
\label{eq:eq1}
\end{eqnarray}
Here ${\bf \hat{n}}$ is a unit vector pointed to each pixel of the map. Following Refs.~\cite{Creswell:2021eqi, Creswell:2021xzf}, one can define the  estimator of symmetry or asymmetry of the temperature map for each pixel,
\begin{equation}
    \label{eq:eq2}
    Z({\bf \hat{n}}) = T({\bf \hat{n}}) T(-{\bf \hat{n}})=S^2({\bf \hat{n}})-A^2({\bf \hat{n}})\,.
\end{equation}
\begin{figure*}[!tbh]
    \centerline{
    \includegraphics[width=0.4\textwidth]{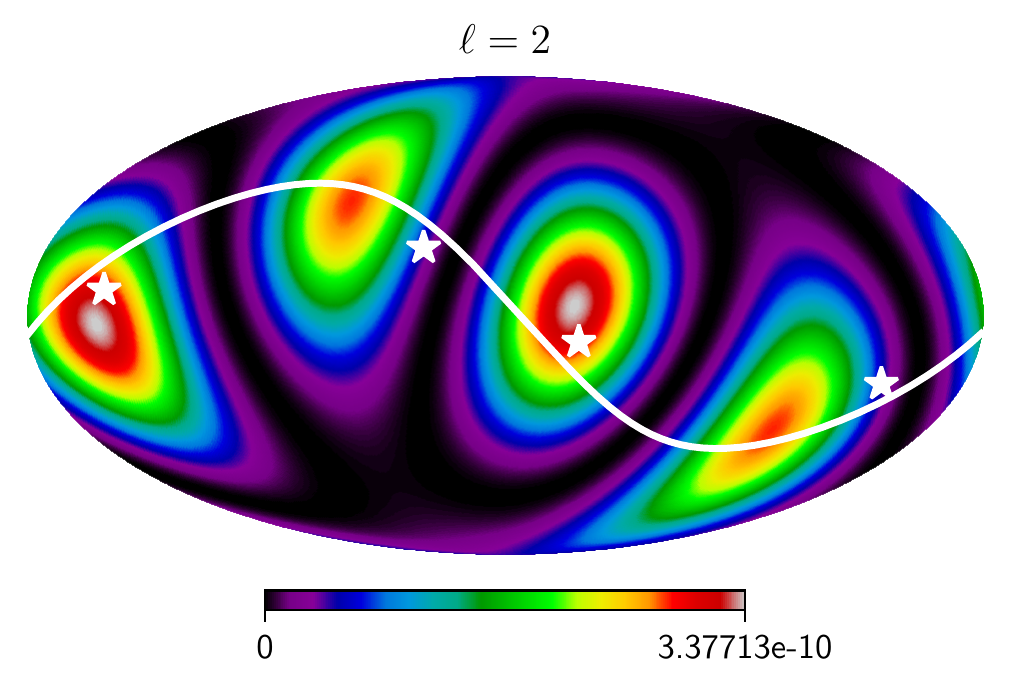}
    \includegraphics[width=0.4\textwidth]{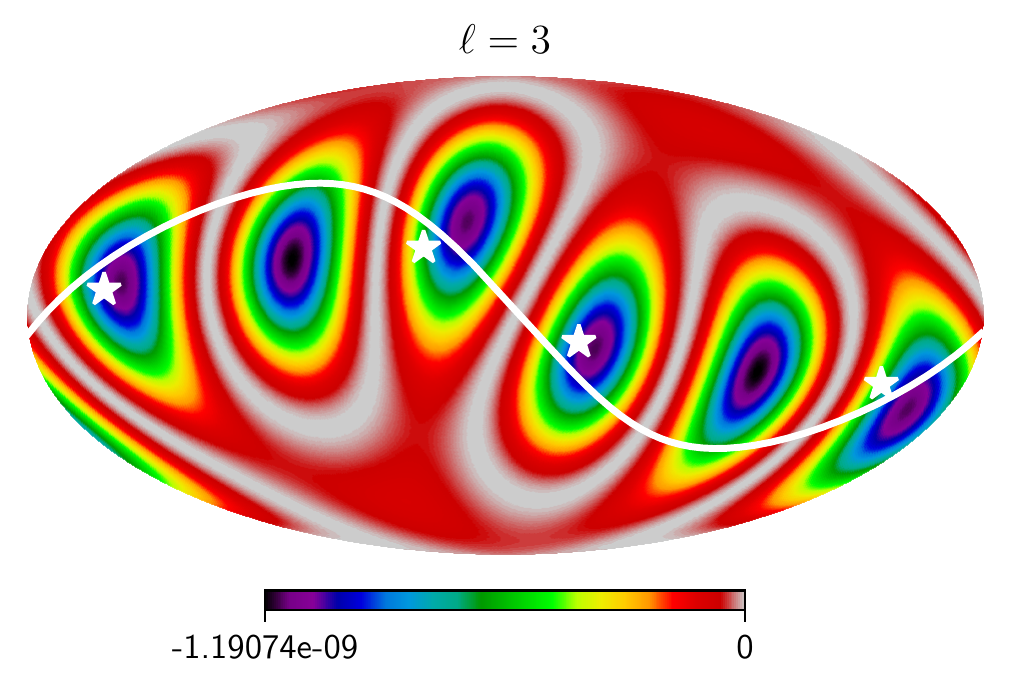}}
    \caption{The $Z_{\ell}({\bf \hat{n}})$-maps, defined individually for the quadrupole ($\ell=2$) (left) and octopole ($l=3$) (right) modes, marks the position of the peaks 1a/1b and 2a/2b along the ring of attraction (white line).}
    \label{fig3naselsky}
\end{figure*}

Thus, if $Z({\bf \hat{n}})>0$, then the symmetric component dominates the asymmetric component, and vice versa. In Fig.~\ref{fig1naselsky} we plot $Z({\bf \hat{n}})$ maps derived from the {\it Planck} 2018 \texttt{SMICA} map with Gaussian smoothing  $\mathrm{FWHM}=\Theta=5^o$~\cite{Creswell:2021eqi, Creswell:2021xzf}. An important feature visible in Fig.~\ref{fig1naselsky} is the two pairs of very strong high negative peaks of $Z({\bf \hat{n}})$, labelled by 1a/1b and 2a/2b  (a and b of each pair are antipodal from one another), and about ten negative peak-pairs with smaller amplitudes, mainly localized within the area $|b|\le 30^\circ$ in Galactic coordinates. These highest peaks (1a/1b and 2a/2b) with $Z({\bf \hat{n}})<0$ belong to the asymmetric component of the signal and have the following Galactic coordinates $(l, b)$:
\begin{align}
    &\mathrm{1a}: (212^\circ, -21^\circ), \qquad &\mathrm{1b}: (32^\circ, 21^\circ)\,,\\
    &\mathrm{2a}: (332^\circ, -8^\circ), \qquad &\mathrm{2b}: (152^\circ, 8^\circ)\,.
\end{align}
Peaks 1a/1b lie outside the area of the Union mask smoothed by a Gaussian filter with $\mathrm{FWHM}=5^\circ$, while peaks 2a/2b are inside this masked area.

The most powerful negative peaks 1a/1b and 2a/2b are located along the kinematic dipole percolation line, where $T_d(\theta,\phi)=0$ and $\theta,\phi$ are polar and azimuthal coordinates~\cite{Creswell:2021eqi, Creswell:2021xzf}. The kinematic dipole is a well-known non-cosmological component of the CMB signal induced  due to the motion of the solar system with respect to the rest frame of the CMB radiation. Observations from COBE, {\it Planck} and WMAP give a direction of the kinematic dipole $T_d({\bf \hat{n}})$ of $(l, b) = (264.00^\circ\pm 0.03^\circ, 48.24^\circ\pm 0.02^\circ)$ ({\it Planck} 2015 nominal) and $(l, b) = (263.99^\circ\pm 0.14^\circ, 48.26^\circ\pm 0.03^\circ)$ (WMAP) in the Galactic coordinates $(l,b)$~\citep{Planck:2015mrs, WMAP:2010sfg}. Denoting this direction with the unit vector ${\bf \hat{q}}$, we introduce the concept of the {\it ring of attraction} (RA) including all directions ${\bf \hat{g}}$ such that ${\bf \hat{q}} \cdot {\bf \hat{g}} = 0,$
i.e.\ ${\bf \hat{g}}$ are all unit vectors orthogonal to the direction of the kinematic dipole ${\bf \hat{q}}$. 
Fig.~\ref{fig2naselsky} shows that RA includes the directions of not only the 1a/1b
and 2a/2b parity peaks but also the CMB dipole modulation used in Refs.~\cite{Planck:2013lks,Planck:2015igc,Planck:2019evm} to explain the power asymmetry.
Analysis of the random simulations for the multidirectional dipoles shows that the coincidence of the point-parity peaks and the power asymmetry direction shown in Fig.~\ref{fig2naselsky} is represented in one of the 2000 realizations.

The presence of RA can also be seen in the Z-maps for the quadrupole and CMB octopole shown in Fig.~\ref{fig3naselsky}. For the quadrupole, the two highest positive peaks coincide with the position of peaks 2a/2b from Fig.~\ref{fig1naselsky}, while for the octopole Z-map all peaks 1a/1b, 2a/2b coincide with the position of the deepest negative peaks. Overall, the above results might suggest a causal relationship between kinematic dipole, parity anomaly, CMB spectrum power asymmetry and quadrupole and octopole features, motivating more detailed studies of the systematic and calibration effects of CMB data. It is, however, important to note that while the connection to the CMB dipole direction (or the ecliptic pole) would suggest a possible systematic, there is no known common systematic in WMAP and {\it Planck}.
Similarly, there is no known foreground that could be responsible for a connection to the CMB dipole. Additionally, if the signal is dominated by a systematic or foreground, then this suggests that the underlying cosmological signal is even smaller, exacerbating the lack of large-angle correlations.

\subsubsection{Variation in Cosmological Parameters Over the Sky}
\label{sec:parameter-variation}

Several attempts have been made to assess whether $\Lambda$CDM parameters are homogeneous and isotropic, as FLRW cosmology assumes. (See, for example,\cite{Axelsson:2013mva,Mukherjee:2015mma,Mukherjee:2017yxd,Yeung:2022smn,Camacho-Quevedo:2021bvt}.) Given the statistically significant violations of statistical isotropy described above, one might wonder whether the differences in  parameters estimated from different parts of the Universe are consistent with the expected variation in perturbed-FLRW.

A recent analysis~\cite{Fosalba:2020gls} of the {\it Planck} legacy temperature maps, considering $\ell > 30$, finds evidence for a violation of statistical isotropy. 
Three patches of the sky $40^\circ-60^\circ$ in angular diameter, over each of which the the $\Lambda$CDM parameters are approximately homogeneous, exhibit sizable differences between one another. 
For example, $H_0$ ranges from $61.3\pm2.6$ to $76.6\pm5.4$; $\Omega_ch^2$ ranges from $0.102\pm0.008$ to $0.134\pm0.008$. 
This level of variation exceeds that in any of the 300 realizations of $\Lambda$CDM to which they compare, though the authors calculate that the discrepancy has a  $10^{-9}$  probability of being a statistical fluctuation in $\Lambda$CDM.
The authors note that these patches subtend the same angular scale ($60^\circ$) above which $C^{TT}(\theta)$ is unexpectedly close to zero (see Sec.~\ref{sec:LargeCMBanomalies} above).
They are also comparable in size to the patterns of the quadrupole and octopole, which are anomalous (see Sec.~\ref{sec:quadrupoleoctopole}).

The authors connect the size of these patches to the size of the anti-trapped surface (causal horizon) due to $\Lambda$, $r_\Lambda=\sqrt{\Lambda/3}$, placed at the last scattering surface.   They suggest that the underlying physical mechanism sourcing the observed temperature anisotropies therefore encompasses scales beyond our causal Universe. 
The authors argue that these well-defined regions may reflect casually disjoint horizons across the observable Universe. 
They show that, within each patch, the observed relations between its size and the mean value of the dark-energy density or $H_0$ are in good agreement with expectations from a recently proposed model of the Universe~\cite{Gaztanaga:2020ksy,Gaztanaga:2021bgb,gaztanaga:hal-03344159,Gaztanaga:2022}.

The possible existence of similar horizons in the local Universe could provide a simple explanation for the claimed cosmological parameter tensions between the low and high redshift Universe.  Future Surveys such as {\it Rubin}/LSST, Euclid, and {\it Roman} will map similar scales at $z>1$.

More recently, similar conclusions have been arrived at independently in Ref.~\cite{Yeung:2022smn} by using hemispherical masks and fitting the {\it Planck} temperature angular power spectrum for 48 directions on the sky. It is found that the directional dependences of the cosmological parameters follow a dipole to good approximation and that the Bayes factor strongly disfavours an isotropic Universe. The reported dipole is roughly perpendicular to the fine structure dipole and within $45^{\circ}$ of the directions corresponding to the CMB kinematic dipole, the CMB parity asymmetry dipole, and the axis in QSO polarizations. The Hubble constant varies between $64.4 \pm 1.3 {\rm \,km\,s^{-1}\,Mpc^{-1}}$ and $70.1 \pm 1.4 {\rm \,km\,s^{-1}\,Mpc^{-1}}$ across the sky.

\subsubsection{The Cold Spot}

The cold (blue) spot was first found in WMAP 1-year temperature data by Ref.~\cite{Vielva:2003et} and was confirmed in {\it Planck} data~\cite{Planck:2013lks,Planck:2015igc,Planck:2019evm} in the southern hemisphere at the galactic longitude and latitude $(l,b)=(209^{\circ},-57^{\circ})$. It is a statistical anomaly of the large-angle fluctuations in the CMB indicating non-Gaussian features. The cold spot is an unusually large region of low temperature with the mean temperature decrease $\Delta T\approx -100{\rm\,\mu K}$  and is in disagreement with the prediction of Gaussianity of the standard $\Lambda$CDM model~\cite{Cruz:2004ce,Cruz:2006sv,Cruz:2006fy}.  Its significance is $2$-$4\sigma$ depending on the statistical method applied~\cite{Zhang:2009qg,Nadathur:2014tfa,Kovacs:2017hxj}. The inconsistency with Gaussian simulations has a $p$-value of $\sim1\%$~\cite{Cruz:2009nd}. 
The anomalous nature of the cold spot corresponds to a rather cold region with an angular radius in the sky of about $5^{\circ}-10^{\circ}$ from the centre surrounded by a hot ring~\cite{Zhang:2009qg,Nadathur:2014tfa,Kovacs:2017hxj}.

The origin of the cold spot anomaly could be due to the following effects: non-Gaussian feature due to a large statistical fluctuation~\cite{Vielva:2003et}, an artifact of inflation~\cite{Cruz:2004ce}, the foreground contamination~\cite{Cruz:2006sv,Hansen:2012xq}, the multiple voids~\cite{Naidoo:2015gab}, the imprint of a supervoid (about $140-200\,$Mpc radius completely empty void at $z\leq 1$) through the ISW effect~\cite{Inoue:2006rd,Inoue:2006fn,Rudnick:2007kw,Granett:2008ju, DES:2021cge,Masina:2008zv,Masina:2009wt,Masina:2010dc}, the axis of rotation of the Universe~\cite{Jaffe:2005pw}, the cosmic texture~\cite{Cruz:2004ce,Zhao:2012hz} and the adiabatic perturbation on the last scattering surface~\cite{Valkenburg:2011ty} (see Refs.~\cite{Cruz:2009nd,Vielva:2010ng} for reviews).

\subsubsection{Explaining the Large-Angle Anomalies}

The apparent connection of certain anomalies to the ecliptic plane (quadrupole-octopole, hemispherical asymmetry) -- which certainly has no role in a theory of cosmology -- suggests a local origin.  However,
the large angle anomalies are extremely difficult to ascribe to the usual suspects: experimental systematics and foregrounds.  

The systematic effects of both WMAP and {\it Planck} T measurements are thought to be well in hand at low-$\ell$, especially outside the Galactic plane. Moreover the experiments have very different observational strategies and it would be very surprising for them to independently generate the same set of anomalies, especially with such remarkable agreement. 

A foreground typically adds a signal with its own correlations to the underlying cosmological signal.  It would be challenging for a foreground to suppress angular correlations on large angles, or to reduce the variance in the fluctuations over half the sky. If the peculiar quadrupole and octopole are due to a foreground, then that foreground signal is on top of an even more highly suppressed cosmological quadrupole, and the magnitude of the underlying cosmological angular correlation function is even smaller.  Any "cancellation" would itself be a fluke.  
The spatial structure of the quadrupole and octopole -- essentially a ring of extrema perpendicular to the ecliptic plane -- make them particularly difficult to explain through foreground. It is not easily mapped into any feature of the solar system, nor of the galaxy. Explanations of quadrupole-octopole alignment that produce aligned $Y_{2 0}$ and $Y_{3 0}$ -- say from an unusual patch of foreground -- are irrelevant. Explanations that are perturbative in some small quantity, $\epsilon$  must explain why $C_2 < C_3$, rather than $C_3 \sim \epsilon C_2$.

Microphysical explanations are also challenging.  Changes in the inflationary potential will change the power spectrum of fluctuations, but will not induce correlations between $a_{\ell m}$.  Similarly, changes in the growth of fluctuations. Changes to the propagation of photons will not correlate $a_{\ell m}$s, unless it is the underlying propagation physics that itself is modulated, and even then will superpose some new signal on top of the "original" cosmological signal, not a \textit{tabula rasa}.

An attempt to ascribe the absence of large angle correlations (low $S_{1/2}$) to an absence of long distance correlations -- by replacing Fourier modes with wavelets~\cite{Copi:2018wsv}, could suppress $\vert\mathcal{C}(\theta>60^\circ)\vert$, but not $\vert\mathcal{C}_{\ell\geq2}(\theta>60^\circ)\vert$, and so not $S_{1/2}$. 

The most promising idea appears to be that the anomalies are induced by horizon-scale or super-horizon-scale properties of cosmic geometry, for example by non-trivial spatial topology. Spatial topology changes the boundary conditions of wave operators.  Their eigenspectra are altered, as are their eigenmodes. Specifically, radial functions times individual spherical harmonics are no longer eigenmodes, and so even though the background metric may remain homogeneous and isotropic, statistical isotropy is broken and $\langle a^\star_{\ell m}a_{\ell' m'}\rangle\cancel{\propto}\delta_{\ell\ell'}\delta_{mm'}$, as required for any explanation for the anomalies. (Non-FLRW metrics would change the differential operators themselves, and would have similar consequences\cite{McEwen:2013cka}.)

A number of preliminary attempts~\cite{Aurich:2007yx,Bielewicz:2008ga,Aurich:2012jc,Aurich:2014sea,Bernui:2018wef} have been made to explore the potential for topology to explain the anomalies  -- with some success in reproducing the quadrupole-octopole structure~\cite{Bielewicz:2008ga} and low $S_{1/2}$~\cite{Aurich:2007yx}, without suppressed $C_{\ell}$~\cite{Bernui:2018wef}. However, the specific topologies considered run afoul of existing limits on topology from the absence of pairs of circles on the CMB sky with matched patterns of fluctuations~\cite{Cornish:1997ab,Cornish:2003db,ShapiroKey:2006hm,Vaudrevange:2012da,Planck:2013okc,Planck:2015gmu} (widely accepted despite the claims in Refs.~\cite{Aurich:2007yx,Aurich:2014sea}), and from comparisons to the predicted two-point $T$ correlations of specific topologies (with fixed values of certain free parameters)~\cite{Aslanyan:2013lsa,Planck:2013okc,Planck:2015gmu}. More work to search for topology and explore its potential connection to CMB anomalies is ongoing.

\subsubsection{Predictions and Future Testability}
\label{sec:PredictionsfromCMBanomalies}

The key to establishing whether the observed large-angle anomalies are statistical flukes or exciting evidence for new physics is to identify testable predictions 
for properties of the CMB polarization, large scale structure, or other cosmological observables.
As discussed above, predictions  come in three types:  predictions of $\Lambda$CDM, conditioned on the existing temperature maps (or on summary statistics of those maps); generic predictions expected of physical explanations, but not tied to a specific model; predictions of specific models to explain specific anomalies.  To date, there are no specific models for which such predictions have been made. 

We summarize what was written above regarding each anomaly.

\paragraph{$\Lambda$CDM Conditional Predictions.} Largely because the correlation between temperature and polarization is weak in $\Lambda$CDM ($(C_\ell^{\rm TE})^2\ll C_\ell^{\rm TT}C_\ell^{\rm EE}$, $C_\ell^{\rm TB}=0=C_\ell^{\rm EB}$), $\Lambda$CDM conditional predictions for polarization are nearly indistinguishable from unconditioned predictions. Similar statements likely hold for predictions for other observables such as large-scale structure, lensing, other secondary CMB anisotropies, 21cm and other intensity fluctuations, and so on. This is actually a useful state of affairs because it allows us to test the fluke hypothesis (that anomalies are statistical accidents in $\Lambda$CDM) by searching for analogous anomalies in these other observables. Which anomalies are most amenable to this approach is perhaps best guided by expectations from the generic predictions expected of physical explanations, but not tied to a specific model (see below).

What can change significantly is predictions for correlations between temperature and other observables, e.g.\ Ref.~\cite{Yoho:2013tta} for the temperature-lensing-potential two-point correlation function $C^{T\phi}(\theta)$.  Here one finds that the peak of the distribution of $S_{1/2}^{T\phi}$ in constrained realizations is reduced by a factor of approximately $3$ compared to unconstrained realizations. Nevertheless, a value of $S_{1/2}^{T\phi}$ far into the low tail of the PDF for unconstrained realizations will still be in the low tail of the PDF for constrained realizations.  Similar analyses should be performed for cross correlations of  CMB $T$ with other signals that are sensitive to high redshifts (though still with $z\ll z_{\mathrm{recombination}}$).

\paragraph{Generic Predictions Expected of Physical Explanations.}
Even in the absence of a specific model, one can put forward certain reasonable expectations of a physical explanation for the anomalies:
\begin{itemize}
        \item The low value of $S^{TT}_{1/2}$ will result in a  low $S^{QQ}_{1/2}$ and $S^{UU}_{1/2}$. There is a weaker expectation for a suppressed value of  $S^{\hat{E}\hat{E}}_{1/2}$ of local-E  polarization~\cite{Yoho:2015bla}, because it is dominated by higher-$\ell$ modes; there is no clear expectation value of properties of non-local versions of the pseudo-field E.
        \item The low value of  $S^{TT}_{1/2}$ implies~\cite{Copi:2013cya} a low Doppler-subtracted (i.e.\ intrinsic) dipole $C^{TT}_1$.
        \item The hemispherical asymmetry in $T$ implies a similar asymmetry for $Q$, $U$, $E$ (local), and perhaps $B$. This should perhaps best be viewed as a low variance in those quantities in the northern (approximately ecliptic) hemisphere.
\end{itemize}

\paragraph{Generic Predictions Expected of Specific Models.}
Specific choices of topology have been argued to potentially explain low $S_{1/2}$ or quadrupole-octopole correlations. However, these specific topologies appear to violate known limits. Some physical models that seek to explain the apparent violation of statistical isotropy observed in T, may also predict independent statistical isotropy violation in polarization of the CMB (i.e. that is not a direct consequence of the observed temperature sky), which could be detected in future polarization data~\cite{Mukherjee:2014lea,Mukherjee:2015wra,Cayuso:2019hen}.

\paragraph{Future Tests.}
We have have identified three specific tests of the fluke hypothesis that are also motivated by expectations for explanations through new physics:
\begin{itemize}
    \item Measurement of $S_{1/2}^{\rm QQ}$ and $S_{1/2}^{\rm UU}$, as well as $S_{1/2}^{\hat{E}\hat{E}}$ and $S_{1/2}^{\hat{B}\hat{B}}$~\cite{Yoho:2015bla}. This would require improvements on {\it Planck} for large-angle correlations in polarization through an all-sky experiment.
    \item Measurement of the cosmological contribution to $C_1^{\rm TT}$, by subtraction of the Doppler contribution to $C_1^{TT}$. This requires sub-percent-accuracy determination of our velocity relative to the CMB rest frame from mixing between high-$\ell$ multipoles, enabling the Doppler contribution to $C_1$ to be subtracted from a similarly accurate direct measurement of $C_1$. This would require improvements on {\it Planck} for high $\ell$.
    \item Measurement of the variance in $Q$, $U$, and $\hat{E}$ in the northern hemisphere~\cite{ODwyer:2016xov,CMB-S4:2016ple}. This would require improvements on {\it Planck} for large-angle correlations in polarization over a large fraction of the north-ecliptic hemisphere, as well as an improved independent determination of the optical depth to reionization $\tau$~\cite{ODwyer:2019rfg}.
\end{itemize}

\subsubsection{Summary}

There are multiple anomalies on large angular scales in the CMB temperature maps. Each is of  $<5\sigma$ significance, each is  characterized with {\textit a posteriori} statistics, but several sets of anomalies are statistically independent (or nearly so) in standard $\Lambda$CDM. Collectively they are very troubling. They are common to WMAP and {\it Planck} (and COBE), so difficult to explain by unaccounted-for observational systematics. They are unexplained by known foregrounds, and a foreground explanation would itself imply an (even more) highly suppressed large-angle cosmological signal.

For several anomalies, one anticipates that a physical explanation would imply an analogous anomaly in the scalar (E-mode) contributions to Q and U polarization, and perhaps in the tensor (B-mode) contribution as well. Similar statements may be applied to other probes of the large-scale Universe -- e.g., intensity maps, surveys of large scale structure -- but the expectation is weaker. Meanwhile, $\Lambda$CDM predictions for polarization (and late-time observables) are nearly unchanged by conditioning on the observed  temperature maps, therefore only cross-correlations with temperature are significantly altered.

Large-angle properties of the CMB should be a matter of continued attention for the experimental and theoretical cosmology communities. Conditional predictions in $\Lambda$CDM, provide accessible null tests of $\Lambda$CDM through improved measurements of polarization on large angular scales, of improved determination of our velocity relative to the CMB, and possibly through large angle/distance auto and cross correlations between observables including those probing moderate redshift. The existence of reasonable expectations for physical models, suggest opportunities to provide compelling evidence for fundamentally important new physics. Non-trivial cosmic topology or anisotropic geometry are the only current promising frameworks for a comprehensive explanation; these should be more thoroughly explored.

\subsection{Abnormal Oscillations of Best Fit Parameter Values}

Signature of an oscillation, or unusual behavior of the data around the best fit $\Lambda$CDM model have been reported in Refs.~\cite{Riess:2017lxs,Colgain:2019pck,Camarena:2019moy,Kazantzidis:2020tko,Sapone:2020wwz,Dainotti:2021pqg}. These signatures were mainly found using the binned Pantheon data. However, assigning significance to such oscillations in the data in the residual space around a best fit model requires a large number of simulations. One should note that different realizations of a data based on the same covariance matrix given a specific model would differ from each other and in some particular realizations we might see some specific features. Hence to assign a reliable statistical significance to any specific feature beyond expectations of a given model or hypothesis, we need to generate a large number of Monte Carlo simulations and see how often we might observe such or similar features. We should also consider a "look elsewhere effect" that features due to random fluctuations might occur at different parts of the data and not necessarily at a specific range of the data. Following this approach, Ref.~\cite{Kazantzidis:2020xta} generated 1000 realizations of the Pantheon supernovae data and evaluated how likely it is to observe the oscillations we see in the Pantheon real data around the best fit $\Lambda$CDM model. Considering the full covariance matrix of the data taking into account both systematic and statistical uncertainties, they did show that at the redshifts below $z \simeq 0.5$ such oscillations can only occur in less than $5\%$ of the simulations. While statistical fluctuations can be responsible for this oscillation, this could also be a hint for some feature beyond the expectations of the standard model. Alternatively this could be a hint for some specific systematics in the data. More observations would make it soon clear if the observed effect is due to statistical fluctuations or there is some real physics or systematics behind such apparent behavior.

\subsection{Anomalously Strong ISW Effect}

The decay of cosmological large-scale gravitational potential $\Psi$ causes the integrated Sachs-Wolfe (ISW) effect~\cite{Sachs:1967er} which imprints tiny secondary anisotropies to the primary fluctuations of the CMB and is a complementary probe of dark energy, e.g.\ Ref.~\cite{Fosalba:2003iy}. Using a stacking technique~\cite{Marcos-Caballero:2015lxp,Planck:2015igc} in the CMB data, an anomalously strong integrated Sachs–Wolfe (ISW) signal has been identified for supervoids and superclusters on scales larger than $100h^{-1}Mpc$ at a $\sim 3\sigma$ level~\cite{Granett:2008ju,Granett:2008xb}. This stronger than expected within standard $\Lambda$CDM signal of the ISW effect first emphasized in Ref.~\cite{Hunt:2008wp}  has been studied by Refs.~\cite{Nadathur:2011iu,Flender:2012wu,Ilic:2013cn,Cai:2016rdk,Kovacs:2017hxj,Kovacs:2018irz,Dong:2020fqt}. 

In particular, the analysis by Ref.~\cite{Kovacs:2018irz} for DES data alone found an excess ISW imprinted profile with $A_{\rm ISW}\equiv \Delta T^{\rm data}/\Delta T^{\rm theory}\approx 4.1\pm2.0$ amplitude (where the value $A_{\rm ISW}=1$ corresponds to the $\Lambda$CDM prediction). A combination with independent BOSS data leads to  $A_{\rm ISW}=5.2\pm 1.6$, in tension at $2.6\sigma$ with the predictions from $\Lambda$CDM cosmology. The average expansion rate approximation (AvERA) inhomogeneous cosmological simulation~\cite{Racz:2016rss} indicates under the inhomogeneity assumption, about $\sim 2-5$ times higher ISW effect than $\Lambda$CDM depending on the $l$ index of the spherical power spectrum~\cite{Beck:2018owr}. Thus the large scale spatial inhomogeneities may be the plausible sources of this ISW excess signal. Using angular cross-correlation techniques Ref.~\cite{Giannantonio:2012aa} combines several tracer catalogues to find $A_{\rm ISW}\approx 1.38\pm 0.32$. In addition the early Integrated Sachs-Wolfe (eISW) effect~\cite{Bowen:2001in,Galli:2010it} ($30<z<1100$) has been studied by Refs.~\cite{Hou:2011ec,Cabass:2015xfa,Kable:2020hcw,Vagnozzi:2021gjh} and constraints were imposed on the corresponding parameter $A_{\rm eISW}$. In general, the reported $A_{\rm ISW}$ amplitude varies in the literature depending on the dataset and the assumptions of the analysis. Further investigation of this issue is needed.

\subsection{Cosmic Dipoles}
\label{sec:cosmic_dipoles}

There have been claims of signals indicating the violation of the cosmological principle. Some of these signals coherently point to the CMB dipole direction, which is \textit{assumed} to be of purely kinematic origin. Some of the dipoles have different directional dependence, and for these observations, there is no obvious pattern. A physical mechanism for producing such violation on Hubble scales is studied in Ref.~\cite{BuenoSanchez:2011wr}. See also Sec.~\ref{sec:cosmography} for an introduction to the mapping of anisotropies in cosmological observables in general space-times without exact symmetries.
The dipolar signals found include the following:

\subsubsection{The $\alpha$ Dipole }

\paragraph{Measuring $\alpha$ from Quasar Spectra.} High resolution spectra of distant quasars reveal numerous narrow absorption lines due to gaseous components of galaxies intersecting the Earth--quasar line of sight. The large number of atomic species and transitions detected allow precise measurements of the fine structure constant $\alpha_{\rm SI} = e^2/(4\pi\epsilon_0\hbar c)$ over cosmological distances, where $e$ is the electron charge, $\epsilon_0$ the vacuum permittivity, $\hbar$ the reduced {\it Planck} constant, and $c$ the speed of light. The dimensionless $\alpha$ is the ratio of the speed of an electron in the lowest energy orbit of the Bohr-Sommerfeld atom to the speed of light, and hence connects quantum mechanics (through $\hbar$) with electromagnetism (through the remaining constants in the ratio).
 
The 1999 invention of the {\it Many Multiplet method} (MM)~\cite{Dzuba:1999zz} provided an instant order of magnitude precision gain over previous methods in searches for space-time variations of $\alpha$. This is because the difference between ground-state and/or excited state relativistic corrections can be large; $s$--$p$ and $s$--$d$ transitions for example may even be of the opposite sign. The MM method can thus produce sensitive results when applied simultaneously to multiplets of the same atomic species, or to species having widely differing atomic masses. Any real change in $\alpha$ therefore generates a unique pattern of observed wavelength shifts that is not degenerate with a simple redshift.

In parallel with the development of the MM method, its first application to high-resolution quasar spectra from the HIRES spectrograph on the Keck telescope was reported in Ref.~\cite{Webb:1998cq} and suggested a possible variation. The acquisition of several further quasar samples, initially from the Keck telescope~\cite{Webb:2000mn, Webb:2002vd, MurphyPhD2002, Murphy:2003hw}, also suggested variations and led ultimately to the extensive study carried out using the UVES spectrograph on the VLT, reported in Refs.~\cite{Webb:2010hc, KingPhD2011, King:2012id}. The combined Keck and VLT samples comprised 293 independent measurements of the fractional change $\Delta\alpha/\alpha = (\alpha_z - \alpha_0)/\alpha_0$, where $\alpha_z$ is the value of the fine structure observed in a gas cloud at redshift $z$ and $\alpha_0$ is the terrestrial value. These 293 measurements cover both hemispheres and span a redshift range of $0.2<z_{\rm abs}<3.6$, permitting a search for temporal and spatial variations of $\alpha$.

The main outcome of the detailed 2011 study by King, Webb, and collaborators is the intriguing hint of a spatial variation of the electromagnetic force, consistent with an angular dipole model pointing in the direction RA $= 17.3 \pm 1.0 h$, dec $= -61^{\circ} \pm 10^{\circ}$, with amplitude $0.97^{+0.22}_{-0.20} \times 10^{-5}$. A bootstrap analysis showed a dipole model was preferred over a simple monopole model at the $4.1 \sigma$ level. The apparent dipole was captivating for several reasons: first, measurements made at lower redshift use a different set of atomic transitions to those at higher redshift, yet both measurements independently support a dipole. Second, the independent Keck and VLT samples both suggest a dipole effect, again along the same direction and with consistent dipole amplitudes. Despite these apparent coincidences, it should be noted that measurement uncertainties for subsets of the whole sample are large and not individually significant. A reasonable statistical significance is only seen when the whole sample is modelled simultaneously.

Ref.~\cite{Berengut:2010yu} made use of the 2011 King {\it et al} data to further examine the properties of the dipole model and to explore the possibility of detecting dipoles in measurements of variation in electron-to-proton mass ratio $\mu=m_e/m_p$, combinations of fundamental constants such as $x = \alpha^2 \mu g_p$ (where $g_p$ is the proton g-factor), and big bang nucleosynthesis probes of the baryonic density parameter $\Omega_b$ using measurements of the primordial deuterium abundance, D/H. Since the number of $\mu$, $x$, and D/H measurements has increased only slightly, little has been done on those dipole studies since the analysis in Ref.~\cite{Berengut:2010yu}.

Detailed studies have also been done to check whether $\alpha$ may change in strong gravitational fields~\cite{Berengut:2013dta, Bainbridge2017a, Bainbridge:2017lsj, Hu:2020zeq}. For some white dwarfs, where physical conditions permit, many narrow photospheric absorption lines are detected. The large number of Fe{\sc\,v} and Ni{\sc\,v} transitions are particularly useful since they have a broad range of sensitivities to $\alpha$, so that (as with quasar measurements) any change in $\alpha$ does not emulate redshift. At present, the most detailed analysis to date gives a best result of $\Delta\alpha/\alpha = (6.36 \pm 0.35\,({\rm stat}) \pm 1.84\,({\rm sys})) \times 10^{-5}$, so there is no clear evidence for a  change in $\alpha$. Although the quality of the astronomical data is high (indicated by $\sigma_{\rm stat}$), results are currently severely constrained by the precision of laboratory wavelengths used to compare with the detected lines, the dominant part of $\sigma_{\rm sys}$~\cite{Hu:2020zeq}. Significant improvements in the laboratory wavelength measurements are needed to make further progress.

\paragraph{Overcoming the Systematics.}
An early list of potential systematic effects associated with $\Delta\alpha/\alpha$ measurements is given in Ref.~\cite{Murphy:2000pz, Murphy:2002ve} although the subject has moved on since that time. We now have a considerably better understanding of systematics and new mathematical/statistical tools have been developed to identify and quantify them. An exhaustive discussion of systematics is not appropriate here so instead a precis of four important effects is given. The first two are essentially solved problems (but exist in previously published surveys) whilst the second two are not yet solved.

\noindent{\it Non-uniqueness in absorption system modelling:} It has long been recognised that a complex quasar absorption system can be modelled in essentially an infinite number of ways. However, interactive model fitting is a laborious process such that it is impractical to construct more than a single model for an absorption system in most cases. To constrain model options to some extent, a common approach is to gradually increase the number of model parameters until the model matches the data with a normalised $\chi^2$ of around unity. Although the $\chi^2$ method is reasonable and has been widely used, it is not the best approach, so Ref.~\cite{King:2012id} employed an {\it information criterion} (IC) to optimise the number of model components. Whilst using an IC strikes a balance between fitting too few and too many parameters, different humans modelling the same data will usually derive slightly different models, each with a slightly different value of $\Delta\alpha/\alpha$. From the point of view of understanding systematics, the important question then becomes, how different? This "non-uniqueness" problem has very recently been solved. A new "Artificial Intelligence" approach was introduced that completely removes human decision making (and time commitment) from the modelling process~\cite{Lee:2020lof}. This enables a Monte Carlo approach to modelling, emulating multiple interactive modellers. Computations are intensive but manageable using high performance computers. In conjunction with those advances, a new spectroscopic IC has been developed specific to spectroscopy, that appears to improve on previous statistics~\cite{2021MNRAS.501.2268W}. The three developments just described enable us to quantify, for the first time, the effect of non-uniqueness on $\Delta\alpha/\alpha$ measurement uncertainty~\cite{Lee:2021kjr}. To summarise, the issue of model non-uniqueness is now a solved problem, {\it provided that} AI Monte Carlo (or something equivalent) is used to derive a sample of models for each absorption system being measured.

\noindent{\it Wavelength calibration problems:} Ref.~\cite{Molaro:2007nm} first searched for wavelength distortions in high-resolution echelle spectra by correlating the reflected solar spectrum from VLT/UVES asteroid observations with absolute solar calibrations but found no evidence for long-range wavelength distortion. Subsequently, Ref.~\cite{Rahmani:2013dia} applied the same technique and indeed found that long-range wavelength distortions do occur in UVES spectra. The form of the distortion appears to be reasonably well approximated by a simple linear function of velocity shift versus observed wavelength. Ref.~\cite{Whitmore:2014ina} examined the problem in further detail confirming problem. They then attempted to estimate the impact of this distortion effect on possible $\alpha$ dipole but did not take into account the considerably more complex distortion that would apply to a quasar spectrum formed from the co-addition of multiple exposures at the telescope, some or all having different instrumental settings. That problem was resolved in Refs.\cite{Dumont:2017exu, 2021MNRAS.508.3620W}, such that any possible long-range distortion VLT/UVES and Keck/HIRES spectra can be allowed for, as discussed in detail in Ref.~\cite{Lee:2021kjr}; future $\alpha$ measurements made using archival UVES and HIRES spectra should in principle be unbiased and the total error budget realistic.

Considerable effort has gone into wavelength calibration of the new ESPRESSO spectrograph ({\it Echelle SPectrograph for Rocky Exoplanets and Stable Spectroscopic Observations}) on the VLT~\cite{2020PhDT........13M, Milakovic:2020ehq, 2020NatAs...4..603P}. In particular, laser frequency combs (LFCs) and Fabry-P\'erot devices produce ESPRESSO calibrations down to an accuracy of $24\,$m/s~\cite{Schmidt:2020ywz}. In principle the accuracy should be somewhat better than this value (and indeed 1 limit of 1-2 m/s has been reached for the High Accuracy Radial Velocity Planet Searcher (HARPS) instrument). Further work is anticipated. A limiting value of 24 m/s corresponds to an uncertainty of $\sim 10^{-6}$ on $\Delta\alpha/\alpha$.

\noindent{\it Magnesium isotopes:} The Mg{\sc\,ii} 2796/2803${\rm\,\AA}$ doublet is often strong in quasar absorption systems and is relatively insensitive to $\alpha$ variation. For these reasons, it was introduced as an "anchor" against which to measure relative shifts of more sensitive species such as Fe{\sc\,ii}~\cite{Webb:1998cq}. However, the low atomic mass of Mg means isotopic and hyperfine separations are non-negligible in the context of varying $\alpha$ measurements. Therefore, if we fit high redshift absorption lines using models that assume terrestrial isotopic abundances, and if relative isotopic abundances in quasar absorbers are not terrestrial, this may mimic a change in $\alpha$, as first pointed out in Ref.~\cite{Webb:1998cq}. Further discussions on this systematic are given in Refs.~\cite{Murphy:2000pz, Murphy:2002ve, Kozlov:2004fc, Ashenfelter:2004kk, Fenner:2005sr, Agafonova:2011sp, Berengut:2011hsa, Webb:2014lpa, Vangioni:2018zef}. This is an unsolved problem in the sense that the Mg relative isotopic abundances at high redshift are not yet well-determined, nor whether they may vary significantly from one quasar absorber to another. Despite that, this systematic is straightforward to quantify simply by treating the Mg isotopic abundances as unknown variables to be solved for in conjunction with modelling for $\alpha$ (the obvious penalty being a slightly weaker constraint on $\Delta\alpha/\alpha$). Of course, Mg is not the only element with this potential problem although it is perhaps the most important in the context of $\alpha$ measurements.

\begin{figure}
\centering
\includegraphics[width=0.9\linewidth]{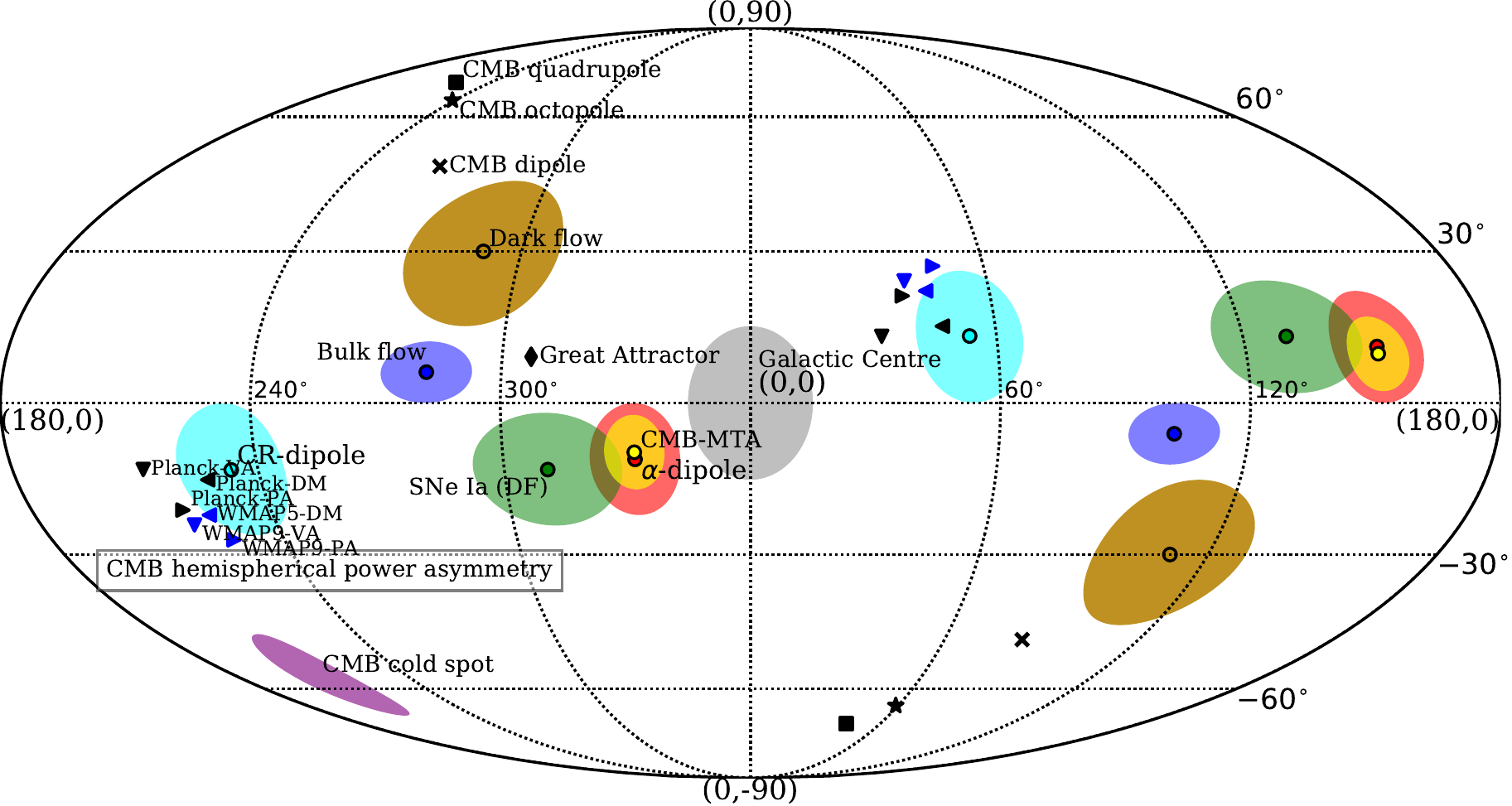}
\caption{Putative directions of anisotropy in the Universe in the galactic (l, b) coordinates with the galactic centre in the middle (grey ellipse). Directions from the literature are shown with different markers (if either no uncertainties were reported or the uncertainties are far too small to be seen) or ellipses (with the circle in the middle being the mean coordinates and the size of the ellipses corresponding to their 1$\sigma$ uncertainties) with text labels. The corresponding anti-directions ($l' = l + 180^{\circ}$, $b' = -b$) are shown with the same ellipse/marker without the text label. The following directions are shown: CMB kinematic dipole (cross); CMB octopole (star); CMB quadrupole (square); CMB hemispherical power asymmetry directions based on WMAP (blue triangles) and Planck (black triangles) data and measured with the dipole modulation (left-pointing), power asymmetry (right-pointing), and variance asymmetry (down-pointing) methods; the Great Attractor (black rhombus with no anti-direction shown); dark flow direction (brown ellipse); bulk flow direction (blue ellipse); SNe Ia dipole (green ellipse); fine-structure constant dipole (red ellipse); CMB maximum temperature asymmetry axis (red ellipse); high energy cosmic ray dipole (cyan ellipse); the Cold Spot (purple ellipse with no anti-direction shown). Figure taken, with permission, from Evgeny Zavarygin’s PhD thesis, 2020.
\label{fig:anisotropy_plot}}
\end{figure}

\noindent{\it Instrumental profile:} It is known that the intrinsic instrumental profile (IP) of the ESPRESSO instrument varies as a function of detector position and wavelength~\cite{2020PhDT........13M, Milakovic:2020ehq}. Since Voigt profile models must be convolved with the IP during the modelling process, it is imperative to have an accurate model of the IP. This problem has not yet been fully solved although numerical and analytic approximations have been made and help. A detailed study for HARPS~\cite{2021A&A...645A..23Z} has been carried out but not yet for ESPRESSO.

\paragraph{Current Status of the $\alpha$ Dipole.}
IR spectroscopy of very high redshift quasars allow us to extend the cosmological volume of the 2011 $\alpha$ measurements described above. At the time of writing, only one quasar spectrum obtained using the X-SHOOTER spectrograph on the VLT has been analysed, the $z=7.085$ quasar J1120+0641~\cite{Wilczynska:2020rxx}. The achievable spectral resolution is at present $\sim 1/4$ that of e.g.\ UVES/VLT and $\sim 1/10$ that of ESPRESSO/VLT. Whilst the X-SHOOTER $\Delta\alpha/\alpha$ error bars are substantially larger than the higher resolution HIRES/Keck or UVES/VLT data, at the high redshift end, these IR data are the best presently available and contribute 4 new measurements, increasing the sample size very little but significantly augmenting the redshift range to $5.5 < z_{\rm abs} < 7.1$. 

Further contributions come from measurements by Molaro and collaborators using UVES on the VLT. The Molaro team concentrated efforts on a small number of high signal to noise spectra, in order to reduce the $\Delta\alpha/\alpha$ statistical uncertainty. That sample of 21 measurements has been published in 10 papers and is summarised, with citations, in Table 1 of Ref.~\cite{Martins:2017qxd}. Additionally, another UVES/VLT sample of measurements are reported in Ref.~\cite{Wilczynska:2015una}, forming a combined total set of 323 measurements. As reported in Ref.~\cite{Wilczynska:2020rxx}, these additional measurements allow us to update the spatial dipole parameters: the updated dipole amplitude is $A = 0.72 \pm 0.16 \time 10^{-5}$ and the dipole sky location is right ascension $16.76 \pm 1.17$ hours and declination $-63.79^{\circ} \pm  10.30^{\circ}$. For that analysis, a dipole is preferred over a simpler monopole model at the 3.9$\sigma$ level. On the other hand, an independent analysis of essentially the same dataset as in Ref.~\cite{Martins:2017qxd}, using a different type of analysis (likelihood with marginalisation over certain parameters), arrives at a lower dipole significance level of 2.3$\sigma$. Given the similarity between the samples used by both groups, the origin of the difference between these significance levels is unclear. However, the error budget in a recent dipole estimate~\cite{Wilczynska:2020rxx} incorporates, in addition to the standard statistical uncertainty, an allowance for an "unknown" systematic $\sigma_{\rm rand}$ (as described in section 3.5.3 of Ref.~\cite{King:2012id}) and an additional uncertainty associated with possible long range distortion (even though none was found). The error estimate in Ref.~\cite{Wilczynska:2020rxx} is thus cautious, we are aware of no reason to think it unreliable, and suggest that measurement is the most reasonable. Whether a dipole will persist in new and forthcoming ESPRESSO/VLT spectra remains to be seen. Data is now beginning to accumulate in the ESO archive and new measurements using that higher quality data are underway.

The purpose of explaining at least some of the systematics associated with varying $\alpha$ measurements above was to highlight the importance of fully resolving those issues in order to facilitate rigorous tests of $\alpha$ variation, be they spatial or temporal. The tentative $\alpha$ dipole cannot be meaningfully tested until these issues are dealt with. Concerted efforts have been made, notably the development of AI-VPFIT, the new information criterion SpIC, detailed studies about model non-uniqueness, as well as extensive investigations into wavelength calibration, explicitly to be able to move ahead with stringent tests of $\alpha$ variation. A complete re-analysis of virtually all existing quasar absorption $\alpha$ measurements will soon be completed, together with new measurements from ESPRESSO. Finally, a number of possible anisotropic effects from a broad range of observations have been reported in the literature and are illustrated in the all-sky plot shown in Figure~\ref{fig:anisotropy_plot}.

\paragraph{Theoretical context.}
Theoretical models incorporating either temporal or spatial variations of fundamental constants require physics beyond General Relativity and hence beyond the standard FLRW cosmological model. Given the broad landscape of possible varying constant theories, the association between variations and the cosmological tensions reported extensively in this article is, as yet, undetermined. A generalised theory of varying $\alpha$~\citep{Barrow:2011kr} has been developed in which electromagnetism varies with the chamelionic scalar field potential $\phi$ via a variable coupling constant $\omega(\phi)$, the latter being constrained by quasar absorption observations. This theory extends the Bekenstein and Sandvik BSBM theory~\citep{Sandvik:2001rv,Magueijo:2002di} in which the observationally constrained coupling parameter does not depend on $\phi$. A further BSBM extension shows that if the kinetic energy in the scalar field drives the acceleration of the expansion, $\alpha$ need not asymptote to a constant value at the present epoch~\citep{Graham:2014hva}. If so, this motivates further precision increases in laboratory experiments to measure the time dependence of $\alpha$~\citep{Leefer:2013waa}. Various kinds of spatial variations in physics (i.e. not just fundamental constants) have been explored. For example~\cite{Audren:2014hza} examine a quantum gravity scenario in which a preferred time-direction occurs at each point in space-time, parameterised by a vector field coupled to dark matter. The same model could also explain spatial variations in fundamental constants. Interesting coincidences between different types of possible anisotropic phenomena have been pointed out in~\cite{Mariano:2012wx} (also see Figure~\ref{fig:anisotropy_plot} for a compilation of various claimed dipole effects). C. Martins and colleagues have explored a range of theoretical ideas showing how constraints on temporal or spatial variation of the fine structure constant provides important new requirements for dark energy models. Recent discussions from that group are given in~\cite{Leite:2017zmo} and in~\cite{Martins:2017yxk}.

There are also interesting ideas about extreme fine tuning of quantum corrections in theories with variation of $\alpha$ by O'Donoghue~\cite{Donoghue:2001cs} and Marsh~\cite{Marsh:2016ynw}. Self-consistent theories of gravity and electromagnetism, incorporating the fine structure constant as a self-gravitating scalar field, with self-consistent dynamics that couple to the geometry of spacetime, have been formulated in Refs.~\cite{Bekenstein:1982eu, Damour:1994zq, Sandvik:2001rv, Barrow:2008ju, Barrow:2011kr, Barrow:2013uza, Barrow:2014tua}, with extended to the Weinberg-Salam theory in Refs.~\cite{Kimberly:2003rz, Shaw:2004hk}. They generalise Maxwell's equations and general relativity in analogy to the way that Jordan-Brans-Dicke gravity theory~\cite{jor, Brans:1961sx} extends general relativity to include space or time variations of the Newtonian gravitational constant, $G$, by upgrading it to become a scalar field. This enables different constraints on a changing $\alpha$ at different redshifts to be coordinated; it supersedes the traditional approach~\cite{dyson2} to constraining varying $\alpha$ by simply allowing $\alpha$ to become a variable in the physical laws for constant $\alpha$. Further discussions relating spatial variations of $\alpha$ to inhomogeneous cosmological models can be found in~\cite{Dabrowski:2016lot,Balcerzak:2016azv}.

Direct measurements of $\alpha$ are also important for testing dynamical dark energy models, since they help to constrain the dynamics of the underlying scalar field~\cite{Martins:2017qxd} and thus dynamics can be constrained (through $\alpha$) even at epochs where dark energy is still not dominating the universe. Indeed, the possibility of doing these measurements deep into the matter era is particularly useful, since most other cosmological datasets (coming from type Ia supernovas, galaxy clustering, etc) are limited to lower redshifts. Recent reviews of the range of theoretical models so far explored are given in~\cite{Sola:2015xga} and in the comprehensive article by~\cite{Uzan:2010pm}. Such a broad theoretical landscape necessarily results in a correspondingly broad range of expectations, all with important implications, but few with tight quantitative predictions. In the absence of a first-principles fundamental theory, high precision experiments and observations are the only way forward if we wish to elucidate the true nature of the world.

\subsubsection{Galaxy Cluster Anisotropies and Anomalous Bulk Flows}
\label{sec:galaxy_cluster}

\paragraph{Scaling Relations in Galaxy Clusters.} It has been observed that scaling relations in galaxy clusters (see Ref.~\cite{Kaiser:1986ske}) at redshifts $z \lesssim 0.3$ detect anisotropies at high statistical significance ($> 4 \sigma$) in the direction of the CMB dipole~\cite{Migkas:2017vir, Migkas:2020fza, Migkas:2021zdo}. In particular, Refs.~\cite{Migkas:2017vir, Migkas:2020fza} studied the isotropy of the X-ray luminosity-temperature ($L_{X}-T$) relation and detected a $\sim 4.5 \sigma$ anisotropy in a direction consistent with the CMB dipole. The observation was subsequently extended to a suite of scaling relations, and by combining all available information, the authors reported an anisotropy and apparent 10\% spatial variation in the local $H_0$ in the direction $(l, b) \sim (280^{+35}_{-35}, -15^{+20}_{-20} )\,$deg and the rest of the sky~\cite{Migkas:2021zdo}.\footnote{Note that the analysis fixes $\Omega_{m}$ to a {\it Planck} value and finds that $H_0$ is \textit{lower} in the CMB dipole direction. This can be contrasted with Refs.~\cite{Krishnan:2021dyb, Krishnan:2021jmh, Luongo:2021nqh} where both $H_0$ and $\Omega_{m}$ are treated as free parameters and $H_0$ is \textit{higher} in the CMB dipole direction.} Given the robustness of the observation across scaling relations, unaccounted X-ray absorption can be discounted as an explanation, leaving i) bulk flows or ii) Hubble expansion anisotropies as potential explanations~\cite{Migkas:2020fza}. 

The first possibility is consistent with previous results in the literature, where anomalously large bulk flows exceeding the expectations of $\Lambda$CDM have been observed~\cite{1994ApJ...425..418L, Hudson:2004et, Kashlinsky:2008ut, Kashlinsky:2009dw, Atrio-Barandela:2014nda, Howlett:2022len}, but in contrast to Refs.~\cite{Migkas:2020fza, Migkas:2021zdo}, at smaller scales. Note that such large bulk flows have been questioned in the literature~\cite{Osborne:2010mf, Planck:2013rgv}, but such a motion, if confirmed, would require a major revision of large scale structure formation models. In tandem, anomalous bulk flows have been studied in SNIa, but the results are also contested~\cite{Colin:2010ds, Dai:2011xm, Turnbull:2011ty, Wiltshire:2012uh, Appleby:2014kea}. The second possibility is an anisotropy in the Hubble expansion and this would necessitate a $10\%$ difference in $H_0$ between the anisotropy direction and the rest of the sky. Here, one trades the normalization in the scaling relation with the Hubble constant $H_0$ on the assumption of a fixed matter density $\Omega_{m}$ consistent with {\it Planck} (see also related Secs.~\ref{sec:radio_dipole},~\ref{sec:qso_dipole},~\ref{sec:SN_dipole}, and~\ref{sec:H0_dipole}). Theoretically, attempts have been made to explain such low redshift effects through differences in the deceleration parameter in a "tilted Universe" model~\cite{Tsagas:2015mua, Asvesta:2022fts}.

\paragraph{Large Scale Bulk Flows.}The bulk flow on the scale of $\sim 100$Mpc is an important probe of the local Universe that can be compared to predictions from the $\Lambda$CDM model. In particular, the scale of $\sim 100$Mpc is large enough that the expectation for the bulk flow is sufficiently small, while being small enough that it is possible to estimate with some accuracy using peculiar velocity data, whose errors increase rapidly with distance. While some studies (for example~\cite{Turnbull:2011ty, Hoffman:2015waa, Qin:2018sxw}) have found results consistent with $\Lambda$CDM, others have found flows that are larger than expected~\cite{Watkins:2008hf,Feldman:2009es,Peery:2018wie}, with only $\sim 2\%$ chance of obtaining a bulk flow that is as large or larger. However, in comparing these studies it is important to keep in mind that they each probe the local velocity field in ways that emphasize different scales, and hence are not in fact comparable. Some progress has been made in developing an analysis method that estimates the bulk flow defined intuitively as the average velocity in a sphere with a fixed radius~\cite{Peery:2018wie}. The bulk flow estimated in this way is be comparable between studies and data sets. Bulk flow estimates made using this method are also independent of the value of $H_0$, thus decoupling the bulk flow from the uncertainty in its value.  With a large quantity of new peculiar velocity data on the horizon, we should soon be able to determine more precisely if there is tension between the bulk flow on these scales and the standard model.

\subsubsection{Radio Galaxy Cosmic Dipole}
\label{sec:radio_dipole}

There are two preferred frames in cosmology, based on the CMB and on the large-scale structure, and defined by the vanishing of the dipole in temperature or in number counts. This is a physical definition, independent of coordinates and space-time metric. The Cosmological Principle assumes that space-time is isotropic and homogeneous on large enough scales. It therefore requires that the CMB and matter frames must coincide -- otherwise, there will be a violation of large-scale isotropy.

This key test of the Cosmological Principle was  proposed by Ellis and Baldwin in 1984~\cite{1984MNRAS.206..377E}, who presented an expression for the dipole in radio galaxy number counts projected on the sky. Their proposal was later applied to the data from the NVSS and other radio galaxy surveys (see Refs.~\cite{Bengaly:2018ykb,Siewert:2020krp} and references cited therein). The striking point about these results, which cover a significant range of radio frequencies, is that they all find consistency of the radio dipole direction with that of the CMB -- but all of them also find that the velocity implied by the radio dipole is significantly higher than the velocity  derived from the CMB dipole. A recent analysis of the eBOSS quasar sample in the infrared produced qualitatively similar results~\cite{Secrest:2020has}, reporting a 4.9$\sigma$ disagreement in velocities.

The CMB defines the rest frame of the Universe because we have subtracted the CMB dipole on the assumption that it is purely kinematic, i.e.\ an effect due to relative motion. Note, given any dipole, one always has the freedom to choose a rest frame. Nevertheless, as a consequence of this choice, the sun is traveling at a precise velocity of $v = (369.8 \pm 0.1)\,$km/s in the direction (RA, DEC) = ($167.94 \pm 0.01, -6.944 \pm 0.005$)\,deg in equatorial coordinates (J2000) or $(l, b) = (264.02 \pm 0.01, 48.25 \pm 0.005)\,$deg in galactic coordinates~\cite{Planck:2018nkj}, with respect to the "CMB frame". This inference can and should be tested, especially since significant $\Lambda$CDM tensions exist. Following the proposal of Ellis and Baldwin~\cite{1984MNRAS.206..377E}, attempts have been made to check that we recover the same proper motion from counts of radio galaxies~\cite{Blake:2002gx, Singal:2011dy, Gibelyou:2012ri, Rubart:2013tx, Tiwari:2015tba, Colin:2017juj, Bengaly:2017slg, Siewert:2020krp}. Tellingly, across a number of radio galaxy catalogues, TGSS-ADR1~\cite{Intema:2016jhx}, WENSS~\cite{1997A&AS..124..259R}, SUMSS~\cite{Mauch:2003zh} and NVSS~\cite{Condon:1998iy}, the analysis returns a dipole that exceeds the CMB expectation in magnitude, but agrees in direction. For example, for the combined NVSS and SUMSS sources, Ref.~\cite{Colin:2017juj} infers our proper motion to be $v = (1729 \pm 187)\,$km/s in the direction (RA, DEC) = $(149 \pm 2, -17 \pm 12)$ deg. Interestingly, some frequency dependence with the radio dipole magnitude has been reported~\cite{Siewert:2020krp} (see also Ref.~\cite{Bengaly:2017slg}), which seems unlikely to be purely kinematic in origin, so it is imperative to tease apart any intrinsic clustering dipole from a kinematic dipole. One possible explanation for the discrepancy is a local (spherical) void with the observer sitting at the edge, e.g.\ Ref.~\cite{1990ApJ...364..341P}. This has been investigated~\cite{Rubart:2014lia}, but a single void model cannot give a sufficient contribution to the dipole to account for any mismatch. An interesting relevant "Ellipsoidal Universe" model was originally proposed to address the low quadrupole amplitude in the CMB~\cite{Campanelli:2006vb}, but it may also provide an explanation for $H_0$ and $S_8$ tensions~\cite{Cea:2022mtf}.

Are we seeing the emergence of another "tension" in the standard model of cosmology? Is there an anomalously large bulk flow in the Universe, possibly generated by a  super-Hubble primordial isocurvature mode (see e.g.\ Ref.~\cite{Tiwari:2021ikr})? There are intriguing possibilities to explain the apparent anomaly. However, the dipole in the matter is considerably more complex than the one in the CMB. It is sensitive to nonlinearities at low $z$ (e.g.\ Ref.~\cite{Bengaly:2018ykb}) and to astrophysical properties of the tracer being used~\cite{Maartens:2017qoa,Nadolny:2021hti}. It is also noteworthy that a recent determination of the dipole in the Pantheon SNIa sample, while finding agreement in direction with the CMB, also found that the velocity implied by the SNIa is {\em lower} than that from the CMB~\cite{Horstmann:2021jjg}.

The dipole formula presented by Ellis and Baldwin~\cite{1984MNRAS.206..377E} was based on a simplified assumption of no evolution in the source population properties. Since these are averaged over all redshifts, one might think that the assumption is reasonable. The redshift dependence of the kinematic dipole in galaxy (and $21\,$cm intensity mapping surveys) involves cosmic evolution, magnification bias and source population evolution, as shown in Ref.~\cite{Maartens:2017qoa}. This has recently been applied by Ref.~\cite{Dalang:2021ruy} to re-derive the radio galaxy dipole without the standard assumption of no-evolution. They find that it is certainly possible in principle for evolution to remove the tension between the matter and CMB velocities. Further work is needed, but this important result shows that it would be premature to conclude that there is a violation of large-scale isotropy.

\subsubsection{QSO Cosmic Dipole and Polarisation Alignments}
\label{sec:qso_dipole}

Recently, the Ellis and Baldwin test~\cite{1984MNRAS.206..377E} was repeated with 1.36 million QSOs from the CatWISE2020 catalogue~\cite{Eisenhardt_2020}. Using mid-infrared data from the Wide-field Infrared Survey Explorer (WISE)~\cite{Wright:2010qw} the authors of Ref.~\cite{Secrest:2020has} created reliable AGN/QSO catalogs and a custom QSO sample. 
It was shown that there is a statistically significant dipole in the counts of distant QSOs with direction $(l,b) =(238.2,28.8)\,$deg (galactic coordinates), which is $27.8\,$deg away from the direction of the CMB dipole, or within $2 \sigma$. However, the dipole amplitude was found to be $\mathcal{D} = 0.01554$, or $\sim2$ times larger than the value inferred from CMB, with a statistical significance at the $4.9\sigma$ level (or with a p-value of $5\times10^{-7}$) for a (single-sided) normal distribution. Taken at face value, this is a serious discrepancy, which is on par with $H_0$ tension, but is to all extent \textit{cosmological model independent}. It should be stressed that this new analysis leads to results consistent with earlier excesses from radio galaxies, but helps to address some systematic concerns. First, since WISE is a satellite, the observational systematics should differ from ground-based radio galaxy surveys. Secondly, the QSOs are expected to be deeper in redshift ($z \sim 1$), so the result is not expected to be significantly contaminated by local sources $(z < 0.1$), which can introduce an additional dipole~\cite{Tiwari:2015tba}. Separate studies of QSOs have reported much larger velocities, $v = (8100 \pm 1500)\,$km/s~\cite{Singal:2021kuu} (see also Ref.~\cite{Singal:2014wra}), yet once again in the expected direction.

The most naive interpretation of this result is that the CatWISE QSOs are tracing an anisotropic Universe on an axis aligned with the CMB dipole. Curiously, this same direction appears in the QSO literature in two separate contexts. First, there are documented Large Quasar Groups (LQG)~\cite{1991MNRAS.249..218C, Clowes:2012pn}, essentially the progenitors of today's superclusters, and also a documented preferred axis for QSO polarizations~\cite{Hutsemekers:2000fv, Hutsemekers:2005iz} in the same direction. Moreover, within the LQGs, the QSO polarizations are aligned~\cite{Pelgrims:2016zbr}. These coincidences warrant further study.

\subsubsection{Dipole in SNIa}
\label{sec:SN_dipole}

More recently, attempts have been made to study our motion with respect to Pantheon SNIa~\cite{Pan-STARRS1:2017jku}, leading to conflicting conclusions. As explained above, cosmology should be conducted in "CMB frame" for consistency, which means that observational redshifts need to be corrected for our motion with respect to the CMB. This is a small correction, but it is imperative to get this right at lower redshifts, especially with SNIa~\cite{Calcino:2016jpu, Davis:2019wet}. Indeed, one finds some criticism of Pantheon CMB redshifts in the literature~\cite{Rameez:2019nrd, Rameez:2019wdt, Steinhardt:2020kul}, and depending on the treatment of redshifts, diverging results have emerged.\footnote{Differences in the treatment of SNIa redshifts have even spilled over into a debate on the evidence for dark energy~\cite{Colin:2019opb, Rubin:2019ywt, Rahman:2021mti}.} In particular, Ref.~\cite{Singal:2021crs} infers our velocity with respect to Pantheon SNe to be $v = (1600 \pm 500)\,$km/s, or $\sim 4$ times larger than the CMB value, in the direction (RA, DEC) = $(173 \pm 12, 10 \pm 9)\,$deg. This value is consistent with radio and QSO galaxy dipoles. In contrast, Ref.~\cite{Horstmann:2021jjg}, which employs corrected Pantheon CMB frame redshifts~\cite{Steinhardt:2020kul}, arrives at the velocity $v = (249 \pm 51)\,$km/s towards (RA, DEC) = $(166 \pm 16, 10 \pm 19)\,$deg. This value is clearly lower than the CMB expectation, which makes it the first study inferring a smaller velocity. Moreover, the difference with Ref.~\cite{Singal:2021crs}, highlights the need to better understand SNIa redshifts, especially corrections for CMB frame.

\subsubsection{Emergent Dipole in $H_0$}
\label{sec:H0_dipole}

One can distinguish between bulk flows and Hubble expansion anisotropies through higher redshift data. The H0LiCOW collaboration are credited with one of the first results supporting a higher $H_0 > 70{\rm \,km\,s^{-1}\,Mpc^{-1}}$ determination~\cite{Wong:2019kwg}, but when one plots the H0LiCOW/TDCOSMO lenses on the sky, one finds that the largest $H_0$ values are oriented in the direction of the CMB dipole~\cite{Krishnan:2021dyb, Krishnan:2021jmh}. Given the small size of the sample, this is no more than a fluke; by running simulations, one assesses the probability of $H_0$ being higher in the CMB dipole direction at $p = 0.12$~\cite{Krishnan:2021jmh}. But is this trend there in other cosmological observables? One finds supporting trends in the Pantheon SNIa sample~\cite{Pan-STARRS1:2017jku} by performing hemisphere decomposition and inferences within flat $\Lambda$CDM across the sky~\cite{Krishnan:2021jmh}. This prompts the pertinent question: when the SNIa sample has been put in "CMB frame" from the outset, why is there still an emergent dipole in the direction of the CMB dipole? It is worth observing that since $H_0$ is a universal constant in any FLRW cosmology, this trend can be expected to be robust to changes in FLRW cosmologies. Similar configurations can arise by chance within the $\Lambda$CDM model with probability $p = 0.065$~\cite{Krishnan:2021jmh}. The same observation has been extended to \textit{standardisable} QSOs~\cite{Lusso:2020pdb} and GRBs~\cite{Demianski:2016zxi}, which places the observation in the $2-3 \sigma$ window~\cite{Luongo:2021nqh}. It should be stressed these samples are all at sufficiently high redshift that one would expect the sources to be in "CMB frame". While the strong lensing time delay and Pantheon SNIa result is expected to be solid at $1.7 \sigma$, attributing a higher significance to this tendency for cosmological $H_0$ to be larger in the CMB dipole direction depends on QSOs and GRBs, which are non-standard as distance indicators (see Ref.~\cite{Moresco:2022phi} for a  recent review).

\subsubsection{CMB Dipole: Intrinsic Versus Kinematic?}
\label{sec:CMB_dipole}

Given that the magnitude of the matter dipoles are currently discrepant with the magnitude of the CMB dipole, systematics aside, this implies that the Universe is not FLRW. Note that this discrepancy between the matter and radiation dipole, if true, is once again an example of an early Universe and late Universe discrepancy, reminiscent of $H_0$ and $S_8$ tension. In effect, a breakdown in FLRW could explain both $H_0$ and $S_8$ tension, as recently highlighted~\cite{Krishnan:2021dyb}. This focuses a spotlight on the cosmic dipole and its traditional purely kinematic interpretation, which is a working assumption. In principle, one should be able to determine if the CMB has an intrinsic dipole by studying Doppler-like and aberration-like couplings in the CMB. The first study in this direction constrains the intrinsic dipole to $v \sim (300 \pm 100)\,$km/s~\cite{Ferreira:2020aqa}. Moreover, one can study aberration effects at smaller scales (larger $\ell$)~\cite{Challinor:2002zh,Amendola:2010ty,Notari:2011sb,Roldan:2016ayx} and this has led to a determination~\cite{Planck:2013kqc} of
$v= 384\,$km/s $\pm 78\,$km/s ({\rm stat}) $\pm 115\,$km/s ({\rm sys}) in the direction $(l, b) = (264^\circ, 48^\circ)$ and more recently $v = (298.5 \pm 65.6)\,$km/s~\cite{Saha:2021bay} in the direction $(l, b) = (268.5 \pm 49.8, 61.8 \pm 12.3)\,$deg using the BipoSH formalism~\cite{Mukherjee:2013zbi}, which are consistent with the expected CMB value. These results are consistent with the kinematic interpretation. 

The small magnitude of the dipole-subtracted two-point temperature-temperature angular correlation function $C^{TT}(\theta)$, has been used to predict~\cite{Copi:2013cya} that the intrinsic CMB dipole is extremely small, see Sec.~\ref{sec:PredictionsfromCMBanomalies}, however given current CMB data, it is difficult to rule out an intrinsic component to the CMB dipole. So the question of whether any matter anisotropy is primordial or not is difficult to address.

\subsection{The Ly-$\alpha$ Forest BAO and CMB Anomalies}

\subsubsection{The Ly-$\alpha$ Forest BAO Anomaly}
\label{sec:lyman_alpha}

A $2.5-3\sigma$ discrepancy between the BAO peak in the Ly-$\alpha$ forest at $z \sim 2.34$ and the best fit {\it Planck} 2018 $\Lambda$CDM cosmology~\cite{Cuceu:2019for,Evslin:2016gre} has been observed.

Observations of baryon acoustic oscillations in the SDSS DR9 and DR11 have provided for the first time statistically independent measurements of $D_H/r_d$ and $D_A/r_d$ at intermediate redshifts of 0.57 (galaxy BAO) and higher redshifts of 2.34 (Ly-$\alpha$ BAO) using different tracers~\cite{BOSS:2014hwf,Aubourg:2014yra}. Here $D_{H}\equiv c/H(z)$ and $D_{A}\equiv\frac{c}{(1+z)}\int_0^z\frac{{\rm d}z'}{H(z')}$ are the Hubble and angular diameter distances, respectively. The combined constraints from the two correlation functions of the BAO observations in the Ly-$\alpha$ forest of BOSS DR11 quasars, implied that the estimated $D_A/r_d$ and $D_H/r_d$ are respectively $7\%$ lower and higher than the expectations of the best fit $\Lambda$CDM model to the {\it Planck} CMB data. This could be translated as a $2.5\sigma$ tension. In Ref.~\cite{Sahni:2014ooa}, it was shown that these measurements can be used to test the concordance $\Lambda$CDM model in a model independent manner using a modified version of the $Om$ diagnostic~\cite{Sahni:2008xx} called $Omh^2$. They confirmed a significant, 2 to 3$\sigma$, tension between the BAO Ly-$\alpha$ observations and {\it Planck} CMB data in the context of the concordance model. In Ref.~\cite{Sahni:2014ooa,Evslin:2016gre} it was suggested that this discrepancy arises not from the $\Lambda$CDM parameters but from the DE evolution itself at $0.57<z<2.34$, if there is no systematics in the observations. Further analysis done by~\cite{Zhao:2017cud} analyzing various tensions in the standard model of cosmology using Kullback-Leibler (KL) divergence reported this tension as one of the important inconsistencies. In the analysis of the cross-correlation of Ly-$\alpha$ absorption and quasars in eBOSS DR14~\cite{Blomqvist:2019rah,deSainteAgathe:2019voe}, the significance of the tension has been reduced to lower than $2\sigma$. Higher precision observations are needed to make a clear conclusion about this tension~\cite{DESI:2016fyo,Cuceu:2019for}. Currently, it is reduced to $\sim1.5\sigma$ in the final eBOSS (SDSS DR16) measurement, which combines all the data from eBOSS and BOSS~\cite{eBOSS:2020yzd,duMasdesBourboux:2020pck}. However, we note that this anomaly is still very important, especially when considered together with the $H_0$ and $S_8$ tensions. Similar to the situation with the Ly-$\alpha$ tension, reducing the $S_8$ tension within the concordance model and its minimal extensions tends to exacerbate the $H_0$ tension~\cite{DiValentino:2020vvd}; moreover, constraints on $S_8$ based on the Ly-$\alpha$ data are in agreement with the weak lensing surveys which probe similar late-time redshift scales as the Ly-$\alpha$ measurements~\cite{Palanque-Delabrouille:2019iyz} and the $S_8$ tension has also weakened with the latest observations~\cite{Hamana:2019etx,vanUitert:2017ieu}. These seem to imply that a simultaneous solution to the $H_0$ and Ly-$\alpha$ tensions might also address the $S_8$ tension, see Sec.~\ref{subsec:gDE}. The {\it Planck} Collaboration (2018)~\cite{Planck:2018vyg} does not include the Ly-$\alpha$ in their default BAO data compilation since for the concordance model and its simple/physically well motivated (e.g., spatial curvature, $w$CDM, quintessence, etc.) extensions, they do not provide significant constraints once the CMB and Galaxy BAO data are used, and they do not conform well with the rest of the data set within the framework of these models. This seems to suggest looking for non-trivial extensions to the $\Lambda$CDM model. The fact that Ly-$\alpha$ BAO prefers $D_{H}(2.33)$ larger than the Planck-$\Lambda$CDM, but $D_{A}(2.33)$ (is an integral over all $z<2.33$) less than the Planck-$\Lambda$CDM, implies correspondingly a preference of a smaller $H(z)$ at $z=2.33$ and a preference of a long enough larger $H(z)$ period before $z=2.33$, compared to the Planck-$\Lambda$CDM predictions. On the other hand, the BAO data from intermediate redshifts $z\sim0.5$, viz., galaxy BAO data, are consistent with the Planck-$\Lambda$CDM~\cite{eBOSS:2020yzd}. This picture can be explained with a DE density that decreases significantly with increasing redshift within the redshift range $0.5\lesssim z \lesssim2.3$, possibly, transitioning to the values below zero (implying a pole in the DE EoS parameter~\cite{Ozulker:2022slu}), which can give rise to a non-monotonic behavior of $H(z)$. This also finds support from model-independent (non-parametric) reconstructions of DE that consider Ly-$\alpha$ together with other data sets. In Ref.~\cite{Escamilla:2021uoj}, it is found that the current Ly-$\alpha$ data prefer a nearly null or negative DE density, or a transition from quintessence to a phantom DE for $z\gtrsim 1.5$; the DE EoS derived from its energy density reconstruction presents a discontinuity at redshift around $z\sim2$, which is necessary if the DE density transitioned to negative values. Finally, such DE models that fit better the Ly-$\alpha$ data can also relax the $H_0$ tension; they are expected to provide the same comoving angular diameter distance to last scattering, $D_M(z_*)=c\int_0^{z_{*}}\frac{{\rm d}z'}{H(z')}$ with $z_*\sim1090$, with the concordance model, therefore the decreased $H(z)$ at higher redshifts should be compensated by an increased $H(z)$ at lower redshifts (and hence an increased $H_0$) in order to keep the integral describing $D_M(z_*)$ unaltered (for such models, see Sec.~\ref{subsec:gDE} and references therein).

\subsubsection{Ly-$\alpha$--{\it Planck} 2018 Tension in $n_s$--$\Omega_m$}

In the last decade, observations of the Ly-$\alpha$ flux power spectrum have emerged as a very powerful tool to extract information on the late-time cosmological evolution. Among the many results, data gathered by MIKE and HIRES, which probe very small scales and high redshifts, have enabled to set stringent bounds on a variety of DM models predicting a suppression of the matter power spectrum at Ly-$\alpha$ scales, including warm DM~\cite{Viel:2013fqw} and other interacting models (see e.g.\ Ref.~\cite{Murgia:2017lwo} and references therein for a review). In addition, several evaluations of the many SDSS  data releases have led to very competitive constraints on cosmological parameters such as $n_s$, $\Omega_m$, and $\sigma_8$~\cite{Palanque-Delabrouille:2015pga, Palanque-Delabrouille:2019iyz}, some of which as currently the center of great attention due to the $S_8$ tension discussed in Sec.~\ref{sec:WG-S8measurements}.

Very recently, Ref.~\cite{Palanque-Delabrouille:2019iyz} interestingly noticed that, while the values of the cosmological parameters extracted from Ly-$\alpha$ are in overall very good agreement with weak leasing surveys, a ($2-3\sigma$) tension in the determination of the tilt of the matter power spectrum exists between Ly-$\alpha$ measurements and the inference from early-time probes such as {\it Planck} 2018. Specifically, they found that Ly-$\alpha$ prefers a sharper drop off of the spectrum with respect to what {\it Planck} would predict. This confirms previous findings by Ref.~\cite{Palanque-Delabrouille:2015pga}. Also, although in a much more qualitative way, a similar conclusion has been drown in Ref.~\cite{Hooper:2021rjc} for the case of MIKE/HIRES data.

These very intriguing results open the door to the possibility that some physics beyond of the standard $\Lambda$CDM model might be at play at Ly-$\alpha$ scales. This has been for instance investigated in the context of DM-(massive) neutrino interactions~\cite{Hooper:2021rjc}, finding a preference for a non-zero interaction strength at the 3$\sigma$ level. Although future work will be needed in order to confirm these conclusions (testing the solidity of the numerical setup and extending the cosmological probes employed), the fact that a model is preferred over $\Lambda$CDM with such a significance is already \textit{per se} very intriguing and indicative of the potential for discovery that the aforementioned tension might contain.

\subsection{Parity Violating Rotation of CMB Linear Polarization}

A parity-violating pseudoscalar field is a candidate of dark mater and dark energy. The field can couple to the electromagnetic tensor via a Chern-Simons interaction~\cite{Ni:1977zz,Turner:1987bw}, which rotates the plane of linear polarization of CMB photons by some angle $\beta$ as they travel from the last scattering surface to the present day~\cite{Carroll:1989vb,Carroll:1991zs,Harari:1992ea,Carroll:1998zi}, the so-called "cosmic birefringence" (see Ref.~\cite{Komatsu:2022nvu} for a review). Because polarization patterns of the CMB can be decomposed into parity-even $E$ modes and parity-odd $B$ modes, $\beta$ can be measured though the $EB$ cross correlation~\cite{Lue:1998mq}. In 2020, Minami and Komatsu~\cite{Minami:2020odp} reported $\beta = 0.35\pm 0.14\,\deg$ (68\%~C.L.) with the statistical significance of $2.4\sigma$, using the {\it Planck} high-frequency instrument (HFI) data at $\nu=100$, $143$, $217$, and $353\,$GHz released in 2018.

A previous measurement of cosmic birefringence with the {\it Planck} satellite had been limited by systematic uncertainties on miscalibration angles of polarization-sensitive detectors~\cite{Planck:2016soo}, which are degenerate with $\beta$.
Ref.~\cite{Minami:2020odp} lifted this degeneracy by determining the miscalibration angles with the polarized Galactic dust emission, assuming that the $EB$ cross correlation of the dust emission is null.
Recently, Diego-Palazuelos et al.~\cite{Diego-Palazuelos:2022dsq} relaxed this assumption and accounted for the impact of the foreground $EB$ by assuming that it is proportional to the $TB$ correlation of polarized dust emission.
They found $0.36 \pm 0.11\,\deg$ (68\%\,C.L.) from the latest {\it Planck} public data release 4 (PR4)~\cite{Planck:2020olo}, with
the statistical significance exceeding $3\sigma$.
Another important test is the frequency dependence of $\beta$, as astrophysical effects such as Faraday rotation give $\beta(\nu)\propto \nu^{-2}$.
Using all of the {\it Planck} data including low-frequency instrument (LFI) data at $\nu=30$, $44$, and $70$\,GHz, Eskilt~\cite{Eskilt:2022wav} found $\beta(\nu)\propto \nu^n$ with $n=-0.35^{+0.48}_{-0.47}$ (68\%~C.L.), which is consistent with no frequency dependence.

In future, analyses of data from the ongoing~\cite{Polarbear:2020lii, ACT:2020gnv,SPT-3G:2021eoc,BICEPKeck:2021gln, SPIDER:2021ncy,Dahal:2021uig} and future experiments~\cite{Westbrook:2018vod,SimonsObservatory:2018koc,SPO:2020,Abazajian:2019eic, LiteBIRD:2022cnt} using the same method as Ref.~\cite{Minami:2020odp} are expected to test the reported signal with higher statistical significance.
Since the method depends on the assumption of the nature of polarized dust emission, improvements on the knowledge of the $EB$ from the dust emission as well as on the calibration method of polarization sensitive detectors, e.g., calibration with Crab Nebula (Tau A)~\cite{Aumont:2018epb}, are needed.

\subsection{The Lithium Problem}

Big Bang nucleosynthesis (henceforth BBN) is one of the cornerstones of the modern cosmological paradigm. In its simplest form, and specifically assuming the standard particle cosmology model and that the relevant nuclear physics is known, it has a single free parameter (the baryon-to-photon ratio), and can therefore provide a consistency test of the overall paradigm, as well as stringent constraints on physics beyond the SM~\cite{Steigman:2007xt,Iocco:2008va,Pitrou:2018cgg}.

Nevertheless, this is not an unqualified success story due to the well-known lithium problem~\cite{Fields:2011zzb,Mathews:2019hbi}. Indeed, the theoretically expected abundance of lithium-7 (given our present knowledge nuclear and particle physics as well as generic  assumptions of astrophysics) seems to exceed the observed one, obtained from absorption spectroscopy in the photospheres of old, metal-poor Milky Way halo stars, by a factor of about $\sim 3.5$. When allowing for the statistical uncertainties on both sides, this represents a detection of ignorance (or missing physics) at more than five standard deviations. The canonical observational determination is that of Sbordone {et al.}~\cite{Sbordone:2010zi}, $^7Li/H=(1.6\pm 0.3)\times 10^{-10}$, while the most up-to-date theoretically expected value is that of Pitrou et al.~\cite{Pitrou:2020etk}, $^7Li/H=(5.464\pm 0.220)\times 10^{-10}$. The other canonical BBN nuclides ($D$, $^3He$, and $^4He$) have abundances that are in broad agreement with theoretical predictions (cf.\ the most recent PDG review of the observational status~\cite{Zyla:2020zbs}), although a possible weak (less than two standard deviations) has been identified for Deuterium, depending on the BBN code and nuclear cross sections that one uses therein~\cite{Pitrou:2020etk,Yeh:2020mgl}.

Attempts to solve the lithium problem can be broadly classified into four categories. First, a mundane solution would involve systematics in astrophysical observations, although none have been identified so far that would account for the large difference required for a solution. Second, one could analogously envisage systematics on the nuclear physics side, specifically in the measurements of the relevant cross-sections; however, the steady improvements in experimental techniques have all but closed this possible loophole~\cite{Iliadis:2020jtc,Hayakawa:2020bjr,Ishikawa:2020fbm,Mossa}.

A third possibility involves new physics beyond the standard paradigm, or a different early-time cosmological evolution. Many such attempts exist in the literature, so far without success. Examples of mechanisms considered include scalar-tensor theories, photon cooling, early- or late-time decaying particles (including decaying dark matter, or magnetic fields~\cite{Larena:2005tu,Kohri:2006cn,Kawasaki:2010yh,Yamazaki:2014fja,Kusakabe:2014moa,Poulin:2015woa,Goudelis:2015wpa,Sato:2016len,Salvati:2016jng,Hou:2017uap,Yamazaki:2017uvc,Luo:2018nth,Mori:2019cfo,Anchordoqui:2020djl}. They either fail by being incompatible with other constraints, require strong fine-tuning, or can only alleviate the problem (e.g., by increasing the error bars for the theoretically expected abundances).

As an example, consider the case of BBN in broad class of Grand Unified Theories with varying fundamental constants~\cite{Coc:2006sx,Berengut:2009js,Cheoun:2011yn}. Superficially this seems a good candidate explanation, because one expects the effects of varying constants to be stronger for heavier nuclei, and therefore one might envisage changing the primordial lithium-7 abundance without significantly changing those of lighter nuclei. A formalism enabling a self-consistent analysis of models where all such parameters (gauge and Yukawa couplings, fundamental particle masses, etc,) are allowed to vary was recently developed~\cite{Clara:2020efx}. The analysis of the most recent data~\cite{Deal:2021kjs} shows that such models can alleviate the lithium-7 problem, but not completely solve it. Specifically the value of the fine-structure constant at the BBN epoch is constrained to the level of parts per million of relative variation, as compared to the present-day laboratory value---a very stringent constraint, and not to far from the most stringent constraints on $\alpha$, which come from local laboratory tests with atomic clocks and from high-resolution astrophysical spectroscopy. Interestingly, a variation of $\alpha$ at this level of relative variation could explain the putative Deuterium discrepancy recently suggested in Ref.~\cite{Pitrou:2020etk}.

Finally, the simplest and currently most plausible solution to the lithium problem is the fourth, an astrophysical one. It is plausible that astrophysical measurements of lithium~\cite{Sbordone:2010zi,Melendez:2010kw} are not representative of the cosmological production mechanism~\cite{Spite:2012us}. Specifically, lithium is known the be both created and destroyed in stars~\cite{Pinsonneault:2001ub,Richard:2004pj,Korn:2006tv,2015MNRAS.452.3256F}. A recent detailed study~\cite{Deal:2021kjs} has quantified the amount of depletion needed to solve the lithium problem, and shown that transport processes of chemical elements in stars are able to account for it. Specifically, the combination of atomic diffusion, rotation and penetrative convection reproduces the lithium surface abundances of Population II stars, starting from the primordial Lithium abundance. More precise astrophysical measurements of abundances of lithium-7 and other nuclides, by a new generation of high-resolution ultra-stable spectrographs, such as ESPRESSO (already operational) and ANDES (forthcoming) will provide decisive tests of this scenario. All in all, it seems likely that the solution of the lithium-7 problem lies inside the stars and not in cosmology.

\subsection{Quasars Hubble Diagram Tension with Planck-$\Lambda$CDM}
\label{sec:qso_candles}

There are various proposals for standardising QSOs in the literature~\cite{Watson:2011um, Wang:2013ha, LaFranca:2014eba, Dai:2012wp, Solomon:2021jml}, but here we focus on a powerful method due to Risaliti and  Lusso~\cite{Risaliti:2015zla}. This analysis has indicated that there are large QSO data sets at odds with {\it Planck}-$\Lambda$CDM at $\sim 4 \sigma$~\cite{Risaliti:2018reu, Lusso:2019akb} and FLRW at $\sim 2 \sigma$~\cite{Luongo:2021nqh}. 

QSOs satisfy an empirical relation between luminosities in the UV and X-ray~\cite{1986ApJ...305...83A}, which serve as the basis for claims that they can be employed as standardizable candles~\cite{Risaliti:2015zla}. In recent years, this program has borne fruit and Risaliti and Lusso have succeeded in producing a compilation of $\sim 2000$ high redshift QSOs in the redshift range $0.01 \lesssim z \lesssim 7.5$ with suitably low internal scatter so that they may be useful for cosmology~\cite{Risaliti:2018reu, Lusso:2020pdb}. Remarkably, the original paper~\cite{Risaliti:2018reu}, and a subsequent follow-up~\cite{Lusso:2019akb}, which introduces GRB data, make the case for a strong deviation from the {\it Planck}-$\Lambda$CDM cosmological model beyond redshift $z \sim 2$.\footnote{In the original papers~\cite{Risaliti:2018reu, Lusso:2019akb}, the actual claim is that the data is inconsistent with $\Lambda$CDM for any value of $\Omega_{m}$. This strong result has been impacted by the cosmographic expansion~\cite{Yang:2019vgk, Banerjee:2020bjq}, however, the $\sim 4 \sigma$ discrepancy from {\it Planck}-$\Lambda$CDM is real~\cite{Yang:2019vgk}.}

Concretely, Risaliti and Lusso have combined SNIa with QSOs, but given that SNIa become sparse at higher redshifts, great care has been taken to ensure that the luminosity distances of binned QSOs and binned SNIa data agree in the redshift range where SNIa are relevant, namely, $0.01 \lesssim z \lesssim 1.4$. For this reason, the data below $z \sim 1.4$ is completely consistent with flat $\Lambda$CDM with $\Omega_{m} \sim 0.3$. Nevertheless, at higher redshift, one finds that the calibrated QSOs exhibit a strong deviation from the standard model. This is essentially driven by a preference for lower values of the luminosity distance $D_{L}(z)$ at higher redshifts, but it routinely gets interpreted in terms of matter density $\Omega_{m}$, curvature $\Omega_{k} < 0$, and dynamical dark energy~\cite{Yang:2019vgk, Velten:2019vwo, Khadka:2019njj, Khadka:2020tlm, Khadka:2021xcc, Bargiacchi:2021hdp, Demianski:2020bva, Bargiacchi:2021hdp,Bargiacchi:2021fow}.  

Interestingly, it has been shown that the QSO data set is consistent with the standard model, i.e.\ no tension, provided one does not cross-correlate with Type Ia supernovae~\cite{Melia:2019nev}. In other words, the $\sim 4 \sigma$ deviation from {\it Planck}-$\Lambda$CDM may rest upon the calibration. Indeed, in Ref.~\cite{Li:2021onq}, GP reconstruction is employed to perform the calibration and the discrepancy reduces to $\sim 2 \sigma$. However, calibration may not be the problem. It has been suggested that because the Risaliti-Lusso QSOs return different values of the constants entering the underlying UV-X-ray relation seems to vary across cosmological parameters~\cite{Khadka:2019njj, Khadka:2020tlm, Khadka:2021xcc}.
But, this may be a bit quick, since it is clear that cosmological parameters jump in the presence of (large) negative curvature, $\Omega_{k} < 0$, which is somewhat consistent with observations in Sec.~\ref{sec:WG-Curvature}, but this probably has a simple analytic explanation: one can lower $D_{L}(z)$ by invoking negative curvature~\cite{Luongo:2021nqh}. So, in some sense, it may be misleading to interpret the Risaliti-Lusso QSO data in any of the simple cosmological models, e.g.\ $\Lambda$CDM, $w$CDM, $w_0w_a$CDM. Now, the question is, since the Risaliti-Lusso QSOs are inconsistent with {\it Planck}-$\Lambda$CDM, are they a complete outlier? Surprisingly, the answer may be no. One should observe that Ly-$\alpha$ BAO~\cite{BOSS:2014hwf, duMasdesBourboux:2020pck} returns distances consistently lower than {\it Planck}, see Sec.~\ref{sec:lyman_alpha}. Moreover, there appears to be some evolution in $(H_0, \Omega_{m})$ within the HST SNIa~\cite{Dainotti:2021pqg} ($1 \lesssim z$), which prefers lower values of $H_0$ and higher values of $\Omega_{m}$, a behaviour that is once again consistent with the QSOs. Finally, Ref.~\cite{Solomon:2021jml} has recently also constructed a Hubble diagram from SDSS QSOs~\cite{MacLeod:2011fn}, which appears to show a similar drop off in $D_{L}(z)$ relative to {\it Planck} at $z \sim 2$. These comments may be speculative, but there appears to be a need to target a greater number of high redshift SNIa.  

Recently, Ref.~\cite{Hodgson:2020tuw} introduced a new method to measure the luminosity distances to QSO jets by equating the size of the flare as measured by Very Long Baseline Interferometry to the variability timescale via a light travel-time argument, effectively making AGN jets a standard ruler constrained by the speed of light. 
They applied the method to 3C~84, a local AGN at $z\simeq 0.078$, and obtained a luminosity distance of $72^{+5}_{-6}\,$Mpc, corresponding to $H_0 = 73^{+5}_{-6}{\rm \,km\,s^{-1}\,{Mpc}^{-1}}$.
The error-bars could be reduced when more sources are observed. An advantage of the method is that it is single rung and can be applied from the local Universe to $z\simeq 6$. However, the systematics still need to be understood further.

\subsection{Oscillating Force Signals in Short Range Gravity Experiments}

The scale of dark energy, required so that it starts dominating the Universe at recent cosmological times is $\lambda_{\rm DE} \equiv \sqrt[4]{\hbar c/\rho_{\rm DE}}\approx 0.085\,$mm, assuming $\Omega_m=0.3$ and $H_0=70{\rm\,km\,s^{-1}\,Mpc^{-1}}$. Therefore, if the origin of the accelerating expansion is geometrical ie due to modifications of GR, it is natural to expect the presence of signatures of modified gravity on scales of about $0.1\,$mm. 

A wide range of experiments has focused on this range of scales~\citep{Murata:2014nra,Kapner:2006si,Hoyle:2004cw,Hoyle:2000cv,Qvarfort:2021zrl} and constraints have been imposed on particular parametrizations of extensions of Newton's gravitational potential. Such parametrizations are motivated by viable extensions of GR and include Yukawa interactions leading to an effective gravitational potential
\begin{equation}
V_{\rm eff}= -G_N \frac{M}{r}(1+\alpha e^{- m r})\,.
\label{yukawaanz}
\end{equation}
For $m^2<0$ the Yukawa correction becomes generically oscillatory. Even though this type of behavior is associated with tachyonic instabilities in most theories, this is not the case for non-local gravity theories where such an oscillating correction to the Newtonian potential is consistent with stability. In particular, non-local (infinite derivative) gravity theories~\cite{Edholm:2016hbt,Kehagias:2014sda,Frolov:2015usa} predict such spatial oscillations of the Newtonian potential without the presence of ghosts (instabilities), while keeping a well-defined Newtonian limit. 

The analysis by Refs.~\cite{Perivolaropoulos:2016ucs,Antoniou:2017mhs,Perivolaropoulos:2019vkb} of short range gravity experiments has indicated the presence of an oscillating force signal with sub-millimeter wavelength. In particular Ref.~\cite{Perivolaropoulos:2016ucs} has indicated the presence of a signal at $2\sigma$ level of spatially oscillating new force residuals in the torsion balance data of the Washington experiment~\cite{Kapner:2006si}. As an extension of the previous analysis the study by Ref.~\cite{Antoniou:2017mhs} using  Monte Carlo simulation and analyzing the data of the Stanford Optically Levitated Microsphere Experiment (SOLME) which involves force measurements an optically levitated microsphere as a function of its distance $z$ from a gold  coated silicon cantilever~\cite{Rider:2016xaq} reports a oscillating signal at about $2\sigma$ level.

\subsection{$\Lambda$CDM and the Dark Matter Phenomenon at Galactic Scales}
 
There is no doubt that the nature of dark matter plays a fundamental role in Cosmology. There is observational evidence that this elusive unseen component mostly resides in and around galaxies of very different luminosities and morphologies that therefore become unique and fundamental cosmological laboratories.
In most galaxies there is overwhelming evidence for a dark spherical component, extended out to a radius fifteen times bigger than that of the stellar component~\cite{Rubin:1980zd, Persic:1995ru}, see also Ref.~\cite{Salucci:2018hqu}. From the theoretical point of view, it is well known that the $\Lambda$CDM scenario has established itself as the predominant one, in virtue of its simplicity, usefulness in accounting for open issues of the Standard Model and a very good agreement with many large-scale cosmological observations, see e.g.\ Ref.~\cite{Freese:2017idy}. This galaxy formation scenario, starring cold and collisionless Beyond Standard Model particles with a particular bottom up perturbation spectrum (e.g.\ Ref.~\cite{Profumo:2017hqp}), has been carefully investigated by large-N cosmological numerical simulations that have revealed an ubiquitous presence in the Universe of dark halos with a very large range in masses~\cite{Navarro:1996gj}. The crucial feature of these structures, formed in a bottom-up merging process, is their {\it universal} density profile, {\it cuspy} at the center: $\rho_{\rm NFW} \propto r^{-1}$, a marking feature of the $\Lambda$CDM collisionless dark particle scenario.
In detail, the above simulations show that, over 20 orders of magnitude in halo mass, the density profiles $\rho_{\rm NFW}(r)$ of the DM halos take the form~\cite{Navarro:1996gj}
\begin{equation}
    \label{NFW}
    \rho_{\rm NFW}(r)= \frac{\rho_s}{(r/r_s)(1+r/r_s)^2} =\frac{M_{\rm vir}}{4\pi R_{\rm vir}}\frac{c^2 g(c)}{\tilde{x}(1+c\tilde{x})^2} \,,
\end{equation}
with $g(c)=[\ln(1+c)-c/(1+c)]^{-1}$. In Eq.~\eqref{NFW} the density parameter $\rho_s$ and the scale radius $r_s$ vary from halo to halo in a strongly correlated way~\cite{Klypin:2010qw}, $\tilde{x} \equiv r/R_{\rm vir}$ and $c \equiv r_s/R_{\rm vir}$, with $M_{\rm vir}$ and $R_{\rm vir}$ the virial mass and the virial radius, respectively, with the latter defined as the radius inside which the DM halo mass is 200 times the critical density of the Universe times the volume inside this radius; finally, $c$ is weak function of the halo mass~\citep{Klypin:2010qw}: $c=9 \ [M_{\rm vir}/(10^{12}\, M_\odot)]^{-0.13}$. 

\begin{figure}
    \begin{center} 
    \includegraphics[width=0.49
    \textwidth]{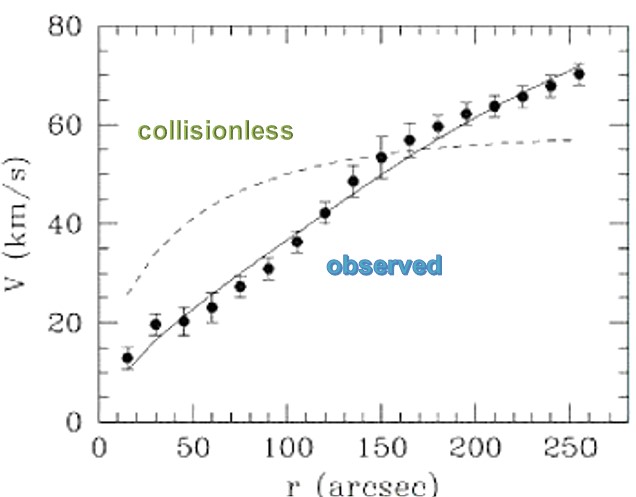}
    \caption{Observed Data (points with errorbars)  vs the  collisionless  NFW halo + stellar disk model (dashed line). Also shown the  cored Burkert halo + stellar disk model (solid line)    (see Ref.~\cite{Gentile:2006hv}). }
    \label{ddo47}
    \end{center}
\end{figure}

\begin{figure}
    \begin{center} 
    \includegraphics[width=0.8\textwidth]{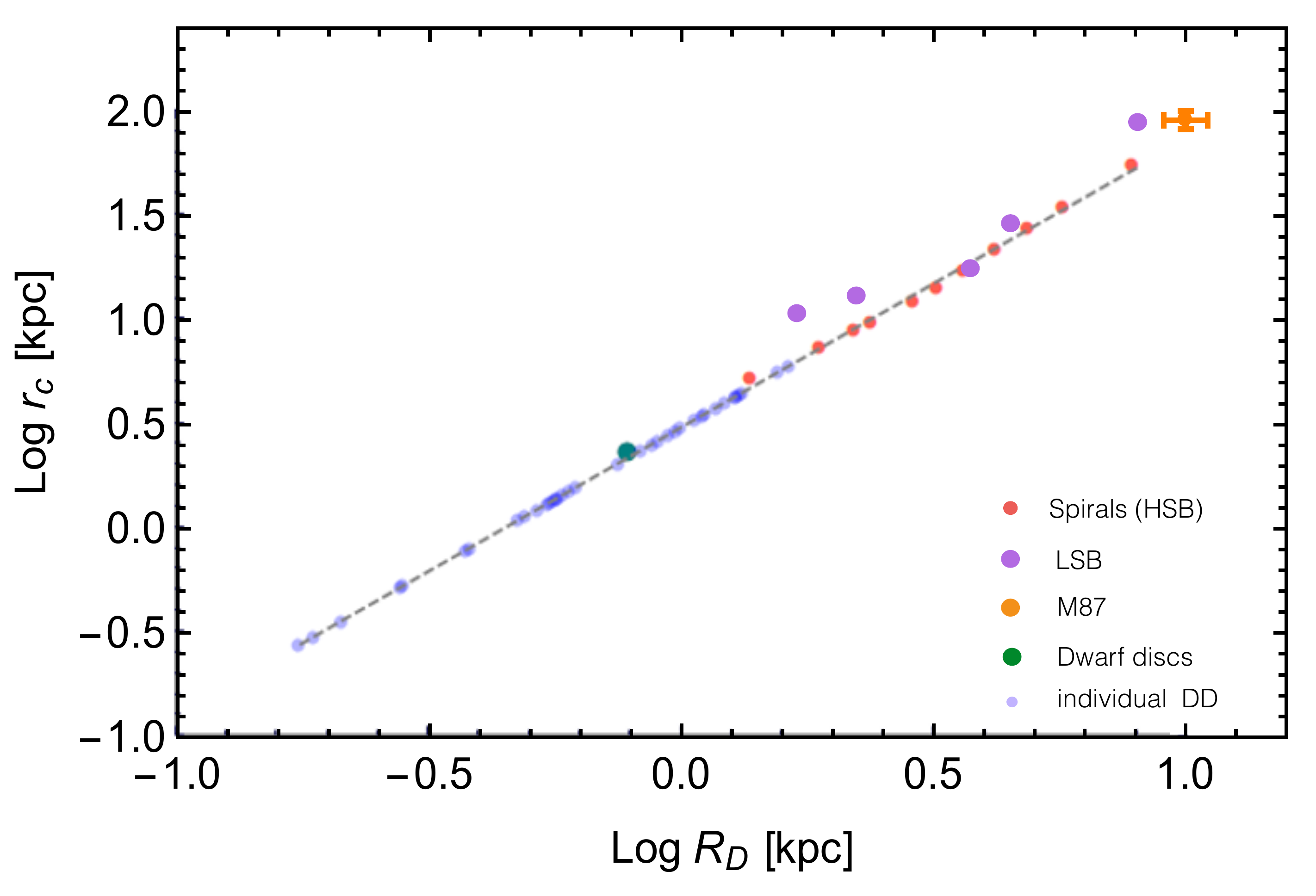}
    \caption{$\log_{10}\, r_c$ vs $\log_{10} \, R_D$ for galaxies of different stellar mass ($10^7M_\odot \leq M_\star \leq 10^{12}  M_\odot $) and morphology (High and Low Surface Brightness disks, the Giant Elliptical M87 and Dwarf Disks). Data come from the URC for HSBs, LSBs and DDs and from the individual kinematics for DDs and M87 (see Refs.~\cite{ Karukes:2017kne, DiPaolo:2019eib},\cite{Salucci:2020eqo}) }
    \label{rorsd}
\end{center}
\end{figure}

It is well known that the internal kinematics of galaxies endows us with a striking observational evidence that the dark matter halo density profiles do not comply with Eq.~\eqref{NFW}, see e.g.\ Refs.~\cite{deBlok:2002vgq, Bullock:2017xww, Salucci:2000tr, Gentile:2004tb, Oh:2010ea, Donato:2009ab}, and show a central core, i.e.\ a central region at about constant density of size $r_c$. Observational data, in fact, are very successfully represented by a cored Burkert profile~\citep{Burkert:1995yz, Salucci:2000ps} according to which
\begin{equation}
    \label{DM_density}
    \rho _{B} (r) = \frac{\rho _0 r_c ^3 }{(r+r_c)(r^2+r_c^2 )}\,,
\end{equation}
where $\rho_0$ is the central density and $r_c$ the core radius. Such profile only at large galactocentric distances ($r> 1/3 \ R_{\rm vir}$) converges to the (collisionless) NFW profile. Remarkably, this finding, which rules out the cuspy NFW profile in disk systems, is obtained from an accurate mass modeling of both the {\it individual }, see Fig.~\ref{ddo47}, and the {\it coadded } disk kinematics~\cite{Dehghani:2020cvl}. Furthermore, we stress that the presence of a cored halo distribution occurs in galaxies of: 1) different halo masses, (in the range $10^9 \ M_\odot -10^{13} \ M_\odot$) and 2) different Hubble Types (see Ref.~\cite{Salucci:2018hqu}). It is crucial to point out that,
as result of the inferred dark matter halo density profiles, we detect core radii ranging from $100\,$pc, in the dwarf galaxies, to $100\,$kpc, in the giant ones~\cite{Salucci:2018hqu}, as shown in Fig.~\ref{rorsd}. Thus, the sizes of the inner halo regions of constant DM density of different galaxies possess an impressive large range of values, ruling out the case in which all objects have a similar value for the size due to a dark particle-related physical process, as in the case of the (simplest) self-interacting DM scenario. 

Within the classic $\Lambda$CDM scenario, in which the density cores are obviously absent, the only reasonable core-forming process relies on the feedback mechanism involving supernovae explosions (e.g.\ Ref.~\cite{Pontzen:2014lma,DiCintio:2013qxa}) whose energy, injected in the interstellar gas, although by means of a very indirect and fine tuned process, is able to warm up the inner DM halo and create a constant DM density region. However, the recent evidence of the existence in dwarfs, giant and LSBs galaxies, see Refs.~\cite{Karukes:2017kne,DiPaolo:2019eib}, all systems for which this mechanism is inefficient due to the insufficient number of SN explosions occurring the unit of time, rules out the $\Lambda$CDM + baryonic feedback scenario as a viable explanation for the presence of the DM halo cores in the $\Lambda$CDM scenario. 

Therefore, in all disk systems, and in many spheroidal systems~\cite{Salucci:2011ee}, the total mass distribution consists in 1) a spherical dark matter halo with two structural parameters: a core radius and a 
central halo density, both varying among galaxies and 2) a stellar component (disk or spheroid) with two structural parameters: the stellar mass and its photometric radius $R_D$, also both varying among galaxies.

A further and perhaps more serious issue for the collisionless nature of the dark matter content of the Universe arises from the realization that all the above galaxy structural parameters are found related among themselves, in detail, those of the DM component: $\rho_0$, $r_c$ strongly correlate with those of the luminous component, $M_D$ and $R_D$ (see Ref.~\cite{Salucci:2020eqo} and references therein). In order to realize the importance of such evidence let us first remind that: $\rho_B(r)$ and $\mu(R) \propto \exp(-r/R_D)$ are the spherical DM density and the luminous surface density. Then, for galaxies and their DM halos, $r_c=-(3/2)\rho(r)/({\rm d} \rho(r)/{\rm d}r)|_{r_c}$ is a quantity deeply inserted in the {\it Dark} world, while $R_D= \frac {{\rm d}\ln \ \mu(R)}{{\rm d}\ln\ R}$ is a main quantity of the {\it Luminous} world. Surprisingly, these two quantities, that, incidentally, are obtained from observations in totally different ways, strongly correlate over three orders of magnitudes, see Fig.~\ref{rorsd}, giving rise to a tight relationship. The latter is the strongest evidence for a dark matter-luminous matter connection leading us to invoke a direct interaction between the dark particles and the Standard Model particles of the stellar disk/spheroid beyond the collisionless status of the pure gravitational interaction. However, one can claim a full entanglement between the dark and the luminous components of galaxies, evidenced by the existence of Universal Rotation Curve~\cite{Salucci:2007tm, Karukes:2017kne, DiPaolo:2019eib}.

Let us stress that such evidence in the dark and luminous mass distribution in galaxies, allied with a complete non-detection of the WIMP particle from accurate astrophysical observations and from properly devised experiments in underground laboratory or at LHC collider (e.g.\ Ref.~\cite{Arcadi:2017kky}), has recently opened the way for old and new scenarios in which the DM particles are other than cold and collisionless WIMPS, such as keV fermions, ultra light axions, self-interacting particles, boson condensates and interacting particles (e.g.\ Refs.~\cite{Spergel:1999mh, Hui:2016ltb, Drewes:2016upu, Visinelli:2017imh}).

Thus, the dark matter phenomenon at galactic scales seems to provide us with a portal towards the true nature of dark particle that clearly appears to be much more complex than cold, collisionless and particle mass independent as that of the $\Lambda$CDM Universe. Regarding this latter, the lack of coldness in the dark particles in galaxies and their emergent capability to sustain interactions with themselves or/and Standard Model particles, are likely to be fatal for its ambition to represent the actual cosmos~\cite{Salucci:2020eqo}.


\section{Stepping Up to the New Challenges} 
\label{sec:WG-perspectives}
\noindent \textbf{Coordinators: } Wendy Freedman, Adam Riess, and Arman Shafieloo.
\\

\noindent \textbf{Contributors: } Elcio Abdalla, Luis Anchordoqui, Nikki Arendse, Micol Benetti, Emanuele Berti, Kimberly K. Boddy, Alexander Bonilla, Erminia Calabrese, Bozena Czerny, Maria Dainotti, Eleonora Di Valentino, Celia Escamilla-Rivera, Ian Harrison, Dhiraj Kumar Hazra, J. Colin Hill, Daniel Holz, Ryan E. Keeley, Ruth Lazkoz, Benjamin L'Huillier,  Matteo Lucca, Roy Maartens, Dinko Milakovi{\'c}, Suvodip Mukherjee, Savvas Nesseris, Rafael Nunes, Antonella Palmese, Leandros Perivolaropoulos, Levon Pogosian, Dan Scolnic, Neelima Sehgal, Joseph Silk, Tommaso Treu, Jian-Min Wang, Amanda Weltman, Rados\l{}aw Wojtak, Gong-Bo Zhao. 
\bigskip

In order to address all the open questions, and to change the $\Lambda$CDM from an effective model to a physical model, testing the different predictions, the goals for the next decade will be to:
\begin{itemize}
\item improve our understanding of systematic uncertainties;
\item maximize the amount of information that can be extracted from the data by considering new analysis frameworks and exploring alternative connections between the different phenomena;
\item improve our understanding of the physics on non-linear scales;
\item de-standardize some of the $\Lambda$CDM assumptions, or carefully label them in the survey analysis pipelines, to pave the road to the beyond-$\Lambda$CDM models tests carried out by different groups.
\end{itemize}
This agenda is largely achievable in the next decade, thanks to a coordinated effort from the side of theory, data analysis, and observation. In 4 separate LoIs~\cite{DiValentino:2020vhf, DiValentino:2020zio, DiValentino:2020vvd, DiValentino:2020srs}, we provide a thorough discussion of these challenging questions, showing also the impossibility we have at the moment of solving all the tensions at the same time.

The next decade will provide a compelling and complementary view of the cosmos through a combination of enhanced statistics, refined analyses afforded by upgraded experiments and next-generation space missions and facilities on Earth:
\begin{itemize}
\item Local distance ladder observations will achieve a precision in the $H_0$ measurement of 1\%~\cite{Riess:2020xrj}.
\item Gravitational time delays will reach a $\sim 1.5\%$ precision on $H_0$ without relying on assumption on the radial mass density profiles~\cite{Birrer:2020jyr} with resolved stellar kinematics measurement from JWST or the next generation large ground based extremely large telescopes (ELTs).
\item CMB-S4 will constrain departures from the thermal history of the Universe predicted by the SM~\cite{Abazajian:2019eic,Abazajian:2019tiv}. The departures are usually conveniently quantified by the contribution of light relics to the effective number of relativistic species in the early Universe, $N_{\rm eff}$~\cite{Steigman:1977kc}. CMB-S4 will constrain $\Delta N_{\rm eff} \leq 0.06$ at the 95\% confidence level allowing detection of, or constraints on, a wide range of light relic particles even if they are too weakly interacting to be detected by lab-based experiments~\cite{Abazajian:2019eic}. CMB spectral distortions will be another possible avenue to test a variety of different cosmological models in the next decade~\cite{Chluba:2019kpb,Kogut:2019vqh,Lucca:2019rxf}
\item The {\it Euclid} space-based survey mission~\cite{EUCLID:2011zbd} and Dark Energy Spectroscopic Survey (DESI)~\cite{DESI:2016fyo} will use cosmological probes (gravitational lensing, baryon acoustic oscillations (BAO) and galaxy clustering) to investigate the nature of DE, DM, and gravity~\cite{Capozziello:2011et}.
\item The {\it Rubin} Observatory Legacy Survey of Space and Time (LSST~\cite{LSST:2008ijt}) is planned to undertake a 10-year survey beginning in 2022. {\it Rubin}/LSST will chart 20 billion galaxies, providing multiple simultaneous probes of DE, DM, and $\Lambda$CDM~\cite{LSSTScience:2009jmu,Zhan:2017uwu,Sahni:2006pa}.
\item The {\it Roman} Space Telescope (formerly known as WFIRST~\cite{akeson2019wide}) will be hundreds of times more efficient than the Hubble Space Telescope, investigating DE, cosmic acceleration, exoplanets, cosmic voids.
\item The combination of {\it Rubin}/LSST, {\it Euclid}, and {\it Roman}/WFIRST will improve another factor of ten the cosmological parameter bounds, allowing us to distinguish between models candidates to alleviate the tensions. 
\item The Square Kilometre Array Observatory (SKAO) will be a multi-purpose radio-interferometer, with up to 10 times more sensitivity, and 100 times faster survey capabilities than current radio-interferometers, providing leading edge science involving multiple science disciplines. SKAO will be able to probe DM properties (interactions, velocities and nature) through the detection of the redshifted 21 cm line in neutral hydrogen (HI), during the so-called Dark Ages, before the period of reionization. SKAO will also be able to test the DE properties and the difference between some MG and DE scenarios by detecting the $21$~cm HI emission line from around a billion galaxies over 3/4 of the sky, out to a redshift of $z \sim 2$.
\item GW coalescence events will provide a precise measurement of $H_0$~\cite{Schutz:1986gp,LIGOScientific:2017adf,LIGOScientific:2021aug}. With the LIGO-Virgo-KAGRA  network of GW detectors, it is expected to constrain $H_0$ to a precision of $\sim 2\%$ with about $50$ binary neutron star events~\cite{Chen:2017rfc, Feeney:2018mkj}. Among the sources without electromagnetic counterpart (called dark standard sirens), either by using the statistical host identification technique~\cite{Gray:2019ksv, Palmese:2019ehe}, or using the spatial cross-correlation~\cite{Oguri:2016dgk,Mukherjee:2020hyn,Bera:2020jhx} of the GW sources with spectroscopic galaxies detectable from DESI and SPHEREx~\cite{Diaz:2021pem}, one can achieve a similar precision with a few thousands of dark sirens from the future detector network. The dark siren measurement of the expansion history may also be possible from the mass spectrum of binary black holes by using a few hundreds of sources, if the mass spectrum can be standardized~\cite{Farr:2019twy, Mastrogiovanni:2021wsd, Mukherjee:2021rtw,Ezquiaga:2022zkx}.       
\item CERN's LHC experiments ATLAS and CMS will provide complementary information by searching for the elusive DM particle and hyperweak gauge interactions of light relics~\cite{Buchmueller:2017qhf,Penning:2017tmb,CidVidal:2018eel,Anchordoqui:2020znj}. In addition, the ForwArd Search ExpeRiment (FASER) will search for light hyperweakly-interacting particles produced in the LHC's high-energy collisions in the far-forward region~\cite{Ariga:2018pin,Ariga:2019ufm,Feng:2019bci}.
\end{itemize}

In what follows, we review the cosmological observations and challenges we will face in next coming years.

\subsection{Future Investigations in Cosmology}

\subsubsection{Supernova Cosmology}

SNe Ia remain the premier tool to link the very local Universe ($z<0.01$) to the Hubble flow and beyond.  Over the next decade, there will be a wealth of data from recently finished and upcoming surveys that will be released.  At low-$z$ ($z<0.1)$, the statistics will increase from current samples by a factor of $10\times$ from surveys like ZTF and YSE; at mid-z $0.1<z<1.0$, statistics will increase by a factor of $>300\times$ from {\it Rubin}/LSST; and at high-$z$ ($z>1.0$), statistics will increase by a factor of $100\times$ from the {\it Roman} Space Telescope.  While the predominant cosmological measurements with SNe Ia have been with the Hubble diagram to measure $w_{DE}$ or $\Omega_M$ or from its role in the distance ladder to measure $H_0$, the large increase in statistics will enable its usage for a variety of different cosmological measurements. These are described in Ref.~\cite{Scolnic:2019apa} and include measurements of $\sigma_8$ from peculiar velocities at low-$z$~\citep{Stahl:2021mat} and from the impact of lensing at high-$z$~\citep{Zhai:2020axw} and further measurements of $H_0$ from lensed SNe Ia~\cite{LSSTDarkEnergyScience:2019dgr}.

There are limitations to how well analyses will be able to leverage this boon in statistics.  First, for the direct distance ladder measurement of $H_0$, current measurements are limited by the number of SNe Ia that have exploded very nearby ($z<0.01$), which only occurs roughly once per year.  As current measurements utilize now $\sim40$ SNe Ia to calibrate the second rung of the distance ladder~\cite{Riess:2021jrx}, it will be difficult to significantly improve this measurement unless measurements of the Cepheids or TRGB can be made accurately at higher distances.  Second, as the statistics grow by orders of magnitude, pushing down the systematic floor will become an even higher priority.  For measurements of $H_0$, this means that the sample of SNe used to calibrate Cepheids/TRGB is consistent with the sample of SNe used in the Hubble flow.  For measurements of $w_{DE}$, this means improving calibration, constraints on evolution, etc.

\subsubsection{Local Distance Ladder}

 The local distance ladder remains the most precise and direct method to produce a high precision determination of the local value of the Hubble constant. There is much reason for optimism for continued progress due to the availability of new facilities like JWST, the {\it Roman} and {\it Rubin} Observatories, ELTs, and LIGO-Virgo-KAGRA as well as future data releases from Gaia. These facilities are likely to improve upon current measurements while improving the connections between the nearest links. While the distance ladder is approaching the ambitious goal of a 1\% determination of the Hubble constant, we may anticipate this goal being approached by a broader range of methods than those available now. Standard Sirens have the potential to provide a fully independent check on the local distance ladder though calibration uncertainties may produce an error floor short of the 1\% goal.

\subsubsection{Spectroscopic Galaxy Clustering}

The BAO and RSD measurements available in the next decade will provide key information of the expansion and growth history of the Universe, respectively. Compared to the precision offered by existing galaxy surveys such as SDSS BOSS~\cite{BOSS:2012dmf} and eBOSS~\cite{Dawson:2015wdb}, surveys in the next decade, including DESI~\cite{DESI:2016fyo}, PFS~\cite{PFSTeam:2012fqu}, Euclid~\cite{EUCLID:2011zbd}, {\it Roman}~\cite{Spergel:2015sza}, {\it Rubin}/LSST~\cite{LSSTScience:2009jmu} and so forth, are able to tighten the constraint on both BAO and RSD parameters significantly in a wide range of redshifts, thanks to the order-of-magnitude increase in the optical spectra to be collected.

Challenges exist to reach that precision though, as efforts are needed to mitigate both the theoretical and observational systematics when analyzing the clustering of galaxies. In this regard, the goal for the BAO measurement is relatively more straightforward to achieve, as the BAO signal is more robust against systematics~\cite{BOSS:2016apd,Ding:2017gad}. The measurement of RSD, however, requires more care to control the impact from both the observational and theoretical systematics. The observational systematics, for example, can root from issues of target selection from the imaging data, the effect of fibre collision, etc (see~\cite{deMattia:2020fkb} for a recent study on the eBOSS ELG sample). Theoretical systematics may include biases for inferring RSD parameters using a fixed template, as commonly used for RSD measurements from existing surveys. Recent development of the EFT-based approach can effectively remove the dependence on the template~\cite{Baumann:2010tm,Carrasco:2012cv,Perko:2016puo,Blas:2016sfa,Ivanov:2018gjr}, although more nuisance parameters are needed.

It is commonly known that the BAO data, on its own, can only measure the product $r_{\rm d}h$, where $r_{\rm d}$ is the sound horizon at the baryon decoupling. However, one can measure both $r_{\rm d}$ and $h$ separately in a way that is independent of the early Universe physics (e.g., physics prior and during recombination) if the BAO data is combined with a prior information on $\Omega_m h^2$~\cite{Pogosian:2020ded}. One possibility is to combine BAO with the CMB lensing and galaxy lensing data, which provide a handle on $\Omega_m h^2$ with practically no dependence on recombination physics. Alternatively, one could simply do a consistency test by taking the value of $\Omega_m h^2$ determined from CMB anisotropies and using it as a prior in the BAO analysis to see if the deduced values of $r_{\rm d}$ and $h$ agree with those obtained from CMB. Any disagreement would signal a missing ingredient in the model used to determine $r_{\rm d}$ when analyzing the CMB data. As forecasted in Ref.~\cite{Pogosian:2020ded}, the DESI BAO data, combined with the {\it Planck} prior on $\Omega_m h^2$, will be able to determine $H_0$ to 0.5\% accuracy, or $\sigma_{H_0} \sim 0.3{\rm\,km\,s^{-1}\,Mpc^{-1}}$.

Over the past five years, full-shape analyses of galaxy clustering datasets have proven their worth as a tool to constrain both the $\Lambda$CDM cosmological model and a wealth of extensions~\citep[e.g.,][]{Ivanov:2019hqk,DAmico:2020kxu,Chudaykin:2020ghx,Ivanov:2020ril,Xu:2021rwg,DAmico:2022gki,Cabass:2022wjy,Vagnozzi:2020rcz}. In the next decade, the power of such tools will only increase as the survey volume grows, and will eventually lead to large scale structure dominating the constraining power on a wide variety of parameters. 

Whilst significant improvements will undoubtedly be possible by applying the current analysis methods and pipelines (which have already been tested on simulations larger than the observable Universe~\citep{,Ivanov:2021kcd,Nishimichi:2020tvu,Chen:2020fxs}) to upcoming data~\citep[e.g.,][]{Chudaykin:2019ock}, improved theoretical modelling will sharpen the constraints further still.
The error on the Hubble parameter measurement 
from this method will reach 
$\sigma_{H_0} \sim 0.05{\rm\,km\,s^{-1}\,Mpc^{-1}}$ with the currently
available power spectrum and bispectrum models~\cite{Chudaykin:2020aoj,Chudaykin:2020hbf,Ivanov:2021kcd}.
However, even better results can be obtained from extending the EFT calculations up to higher-loop order, such as implementation of the two-loop
and three-loop power spectra~\citep{Carrasco:2013mua,Baldauf:2015aha,Konstandin:2019bay}, one-loop bispectrum~\citep{Eggemeier:2018qae}, or two-loop bispectrum~\citep{Baldauf:2021zlt}. This will extend the validity of the models, allowing smaller-scale information to be captured in a robust fashion, marginalizing over all relevant galaxy formation and hydrodynamic effects via symmetry arguments. Similar improvements can be obtained from simulation-based inferences~\citep{Kobayashi:2021oud,Kobayashi:2020zsw}, although care will be needed to avoid bias arising from hydrodynamic effects not captured in the modelling. Generalization to additional redshift components, such as the bispectrum multipoles, can also add a tranche of new data, and thereby constraining power~\citep{Gualdi:2020ymf,Philcox:2021eeh}.

Another promising avenue is via the inclusion of higher-order statistics, such as the galaxy trispectrum or four-point correlation function and beyond. Initial forecasts suggest that the statistics can yield impressive constraining power on $\sigma_8$~\citep{Gualdi:2021yvq}, arising due to degeneracy breaking, and recently, the first detections of such quantities have been made~\citep{Philcox:2021hbm,Gualdi:2022kwz}. Of course, their use will require the development of rigorous theoretical models~\citep{Steele:2021lnz} that must be tested on simulations (including systematic effects) before their application to data. $N$-point functions are not the only way to proceed, however: field-level inference provides a promising manner in which to analyze the galaxy survey without compression to summary statistics~\citep{Schmidt:2020tao,Cabass:2019lqx,Schmidt:2020viy,Seljak:2017rmr}, and could lead to strong constraints on parameters such as $\sigma_8$ from a perturbative- or simulation-based framework, once the relevant galactic effects are marginalized over.

Finally, it is important to discuss concerns on the applicability of the standard BAO template fitting method in the future. The fixed template method is currently proven to be biased and to result in loss of information for the galaxy RSD/AP measurements. The galaxy redshift clustering data, however, can be 
efficiently analyzed by means of the EFT-based full-shape method~\cite{Ivanov:2019pdj,DAmico:2019fhj}. As far as the BAO measurement is concerned, 
the dependence on the fiducial template is not so strong, and probably it will not cause a large systematic 
error even in the era of future galaxy surveys~\cite{Bernal:2020vbb}.
However, strictly speaking, it is not clear by default what fiducial cosmology should be used in the BAO template. One option is to fix it by first running the full-shape analysis without the post-reconstructed BAO data. This analysis will already constrain the cosmological parameters very tightly~\cite{Chudaykin:2019ock}. Then the BAO template can be calculated for the best-fit full-shape cosmology. All in all, all these arguments suggest that the BAO method does not require a major modification in order to match the precision 
of the future datasets.

\subsubsection{Cosmic Microwave Background}

Over three decades, with observational campaigns from space, balloons, and ground-based telescopes, the CMB has been the driving force in establishing the standard cosmological model. The most stringent limits on the $\Lambda$CDM parameters are obtained from {\it Planck} data or a combination of {\it Planck}/WMAP and ACT/SPT ground-based experiments. These experiments have accumulated observations in temperature and polarization over many frequencies and extended ranges of angular scales. In particular, with the latest {\it Planck} data we have hit the limit of cosmic variance over large and intermediate scales in temperature ({\it Planck} CMB temperature power spectrum measurements are cosmic-variance-limited for $\ell \lesssim 1600$).  However, much more information can be extracted from improved small-scale temperature measurements and high-accuracy polarization data at all scales, which are the key target of the next decade's CMB experiments. 

From the ground there are several ongoing, planned, and future experiments that will continue to improve CMB temperature and polarization observations. For the topics covered here, the most relevant experiments are those pushing observations over extended and high-resolution scales. These include future releases from the ongoing ACT and SPT experiments, as well as new experiments including the Simons Observatory~\cite{SimonsObservatory:2018koc}, the South Pole Observatory, CMB-S4~\cite{CMB-S4:2016ple,Abazajian:2019eic}, and the proposed CMB-HD~\cite{Sehgal:2019nmk,Sehgal:2019ewc,Sehgal:2020yja}. The new power spectra from these observatories will constrain $H_0$ with sub-percent precision in the context of $\Lambda$CDM (e.g.\ Ref.~\cite{SimonsObservatory:2018koc}). Even within more extended models, such as those proposed to resolve the tensions discussed earlier, the statistical power of these upcoming measurements will be sufficient to yield precise $H_0$ constraints and potentially discriminate amongst proposed resolutions, an extremely exciting prospect.  In addition, the extracted lensing signal will be revolutionary for constraining the growth of structure with percent level constraints on $\sigma_8$ when correlating CMB lensing with other tracers of the matter distribution.

Anomalies in the present CMB data and tensions between CMB temperature and other cosmological datasets in determining cosmological parameters can be tested with much better precision with upcoming CMB space- and ground-based surveys. {\it Planck} temperature data reveals an anomaly in the lensing amplitude.  While this lensing anomaly can be resolved with a closed Universe~\cite{Planck:2018vyg, DiValentino:2019qzk, Handley:2019tkm}, it aggravates the already existing tensions between the observations in determination of $H_0$ and $S_8$. Atacama Cosmology Telescope~\cite{ACT:2020gnv} temperature and polarization data, however, do not find this lensing anomaly and the standard model stays consistent with the data. Several new physics candidates and different possibilities of systematic uncertainties have been proposed to solve these discordances. These anomalies and disagreement between CMB datasets and the necessity of new physics will be tested with upcoming observations. Three major improvements in upcoming observations will be able to help in this regard. Firstly temperature anisotropy measurements from ground-based experiments such as Simons Observatory, CMB-S4, and CMB-HD will be able to explore the small differences between {\it Planck} and ACT at much better precision. Cosmic-variance-limited polarization measurements from LiteBIRD out to $\ell\sim800$ and small-scale polarization data from the ground-based surveys will help in identifying the potential source of anomalies and tensions. Finally, precise measurements of the lensing potential will also independently estimate the lensing amplitude and thereby can provide a statistically significant angle in the study of the lensing anomaly.

The most advanced proposal for a next-generation CMB satellite is the JAXA-led LiteBIRD satellite~\cite{LiteBIRD:2022cnt}.
With a predicted launch towards the end of the decade, LiteBIRD aims to measure the CMB temperature and polarization anisotropies with unprecedented precision and with the ultimate goal to deeply characterize the signal from inflation~\cite{Snowmass2021:Inflation}, ruling in or out well-motivated inflation models. The new maps will help shed light on existing parameter tensions in two ways: 1) LiteBIRD will provide a new, independent of {\it Planck} $H_0$ satellite measurement, which will be highly robust to systematic effects due to the use of data from a single, well calibrated CV limited polarization data up to 0.2 degree scales; 2) the new E-mode measurements will provide a new, cosmic-variance-limited measurement of $\tau$ reducing degeneracies with other parameters and $S_8$ in particular. \\

As already clear from the previous sections, the experimental effort invested in the cosmological landscape has been extremely prolific in the past decades and will continue to grow in the coming years. However, there are some intrinsic limitations in many observables that do not depend on the sensitivity or the duration of the given experiment, and can only be overcome with complementary probes. For instance, in the case of the CMB anisotropies, the presence of the cosmic variance or the well-known degeneracy between $A_s$ and $\tau_{\rm reio}$, as well as the limited observable scales, limit the amount of information that can be extracted from these measurements of the last-scattering surface.

In this regard, CMB spectral distortions (SDs)~\cite{Zeldovich:1969ff, Sunyaev:1970er, 1991A&A...246...49B, Hu:1992dc, Hu:1993gc, Hu:1995em, Chluba:2011hw} have been shown to have a remarkably deep synergy with the observation of the CMB power spectra~\cite{PRISM:2013fvg, Chluba:2013pya, Chluba:2019nxa, Chluba:2019kpb, Lucca:2019rxf, Fu:2020wkq, Schoneberg:2020nyg}. In fact, they are mainly produced in the pre-recombination era by any process that affects the energy spectrum of the CMB photons, and are therefore predicted to exist even within the standard $\Lambda$CDM model~\cite{Chluba:2011hw}. In this way, they enable to probe a large section of the history of the Universe which overlaps and complements the one already tested by means of the CMB power spectra. The applications in this direction range from $\Lambda$CDM~\cite{Chluba:2016bvg} to models invoking BSM physics~\cite{Chluba:2013wsa}, from inflation~\cite{Schoneberg:2020nyg} to recombination~\cite{Hart:2020voa}.

In particular, one of the main production mechanisms within $\Lambda$CDM is the dissipation of acoustic waves~\cite{Chluba:2012gq}. This effect takes place when the length scale of the density anisotropies is comparable to the mean free path of the CMB photons, so that the latter stream from the overdense to the underdense regions effectively creating a flow of energy injection which ultimately affects the global energy spectrum. Because of Thompson and Coulomb interactions, the photons end up dragging electrons and protons, which results in an isotropization of the plasma and a damping of the anisotropies at small scales, effect known as Silk damping. Since the amount of density anisotropies at a given scale directly depends on the value of primordial power spectrum (PPS) at that scale, SDs are very sensitive to the cosmological parameters ruling the PPS. This dependence and the fact that SDs are produced prior to recombination are of significant importance as they allow to probe the shape of the PPS (and the inflationary potential more in general) over more than four decades in Fourier space otherwise unexplored by other (complementary) measurements. In combination with, for instance, the CMB observation by {\it Planck} this would in principle allow to test the inflationary epoch to an unprecedented degree of precision (see e.g.\ Ref.~\cite{Schoneberg:2020nyg}). It is also worth nothing that, as shown in Ref.~\cite{Hart:2020voa}, also the eventual observation of the cosmic recombination radiation~\cite{Chluba:2015gta} would further increase the sensitivity of SDs to inflation, on top of the large amount of information it would provide on the recombination epoch. 

Additionally, as recently pointed out in Ref.~\cite{Lucca:2020fgp}, the fact that SDs can put strong constraints on the scalar spectral index $n_s$ can also play a significant role in the Hubble tension, see Sec.~\ref{sec:WG-H0measurements}. Indeed, many extensions of the $\Lambda$CDM model recently proposed to address this tension, such as the EDE and the SI$\nu$ models discussed in Secs. \ref{sec:ede} and \ref{sec:SInu}, respectively, introduce significant shifts in the standard cosmological parameters in the attempt to compensate for the new physics while still accurately fitting the CMB power spectra. This inevitably introduces strong degeneracies between the affected standard parameters and the model-specific quantities. Since one of the most common parameters that vary in these scenarios is $n_s$, SDs constitute an ideal independent probe to break the aforementioned degeneracies and potentially constrain the efficacy of these models as solutions to the Hubble tension. Moreover, in the context of the Hubble tension, SDs have also been advanced as a possible avenue to obtain an (almost) model-independent measurement of the $H_0$ value~\cite{Abitbol:2019ewx}. This can be achieved by precisely observing the evolution of the CMB monopole temperature over an extended timescale, thereby obtaining a measure for the cooling of the Universe and thus its expansion rate.

Interestingly, together with the dissipation of acoustic waves, SDs are also sensitive to the dissipation of tensor modes, making them a window to another large class of inflationary scenarios~\cite{Chluba:2014qia}. In this context, as recently discussed in Ref.~\cite{Kite:2020uix}, through this mechanism SDs are also able to probe primordial GWs over six orders of magnitude in frequency space. This can tightly bound a variety of scenarios involving phase transitions or cosmic strings~\cite{Kite:2020uix}, among many more.

Along with the spectral distortions on the CMB monopole, spatially fluctuating polarised CMB spectral distortion can also arise due to conversion of photons to axions (or axion like particles (ALPs)) in the presence of external magnetic field as proposed in Refs.~\cite{Mukherjee:2018oeb, Mukherjee:2019dsu}. The measurement of the polarised axion distortion will provide a new way to detect the coupling of photons with ALPs in the mass range $10^{-11}$--$10^{-14}\,$eV. This is possible from the upcoming ground based CMB experiment such as Simons Observatory, CMB-S4, CMB-HD and from space-based CMB telescope LiteBIRD.

The complementary between CMB anisotropies and SDs mentioned at the beginning of this section can also be applied to models requiring, for example, primordial magnetic fields (PMFs)~\cite{Jedamzik:1999bm, Kunze:2013uja, Jedamzik:2018itu}, DM with non-standard characteristics, such as decays~\cite{Hu:1993gc, Lucca:2019rxf} or interactions~\cite{Ali-Haimoud:2015pwa, Slatyer:2018aqg}, Axion-Like Particles~\cite{Mukherjee:2018oeb, Mukherjee:2019dsu} and dark photons, or primordial black holes~\cite{Tashiro:2008sf, Lucca:2019rxf, Acharya:2020jbv, Eroshenko:2021oeq,Papanikolaou:2021uhe}. Many of this scenarios have also been shown to solve either the Hubble tension, such as for PMFs~\cite{Jedamzik:2020krr}, the EDGES anomaly, such as for DM-b interactions~\cite{Kovetz:2018zan}, or the $\sigma_8$ tension, such as for DM-photon interactions~\cite{Stadler:2018jin, Becker:2020hzj}. Therefore, also in this context SDs could play a competitive and complementary role in the search for a solution to these modern problems.

However, profiting from the constraining power of SDs will only be possible with the advent of advanced dedicated missions such as SuperPIXIE~\cite{Kogut:2019vqh} and Voyage 2050 initiative~\cite{Chluba:2019nxa}, which, most likely, will have to be preceded by pathfinders from the ground or space. Therefore, the goal of the next decade in the context on CMB SDs will not only have to be in the theoretical and numerical development of the field, but will also have to be particularly devoted to the experimental realization of a concrete observational strategy to measure both temperature anisotropy and spectral distortions in a same setup~\cite{Mukherjee:2018fxd,Mukherjee:2019pcq}.

\subsubsection{Weak Lensing} 

A number of ongoing and planned weak lensing surveys promise to provide significantly tighter constraints on the amplitude of matter fluctuations $\sigma_8$ and the equation of state of dark energy (see Tables~\ref{tab:acronyms} and~\ref{timeline}).

These include ongoing surveys such as Kilo-Degree Survey (KiDS)~\cite{KIDSws}, Dark Energy Survey (DES)~\cite{DESws}, Hyper Suprime-Cam (HSC)~\cite{HSCws}, 
and planned to start surveys such as 
the {\it Rubin} Legacy Survey of Space and Time~\cite{LSSTws}, Euclid~\cite{Euclidws},
Nancy Grace {\it Roman} Space Telescope~\cite{NGRSTws}, and the The Ultraviolet Near- Infrared Optical Northern Survey (UNIONS) which is a collaboration between two the Hawai'ian observatories CFHT (Canada-France-Hawaii Telescope on Mauna Kea) and Pan-STARRS (Panoramic Survey Telescope and Rapid Response System on Maui)~\cite{UNIONS,CFISws,Pan-STARRSws}. 

These lensing surveys promise not only to map the photo-z sky and dark matter but also constrain the growth of large structures to unprecedented accuracy and precision.

\subsubsection{Time Lag Cosmography and Time Delay Cosmography}

In this Section we describe two methods based on "standard clock", i.e. the measurement of time intervals that can be converted into  a cosmic distance and thus a cosmological tool. First, we describe the use of the time lag between variability of the accretion disk surrounding black holes, which we refer to as "time lag cosmography". Then we describe the use of the time delay between multiple images of strongly gravitationally lensed sources, which we refer to as "time delay cosmography"~\citep{Treu:2016ljm}. 

\paragraph{$H_0$ Measurements from Quasars with {\it Rubin}/LSST Monitoring.}
The oldest method to use active galactic nuclei for cosmology is the method based on continuum time lag measurements~\cite{Collier:1998ev} in the accretion disk. The method is conceptually simple, it is based on the concept of accretion disk model of Ref.~\cite{Shakura:1972te}, which well describes the accretion disk at radii of order of 100 - 1000 gravitational radii, where the optical continuum forms. The method is not very sensitive to the innermost part of the accretion flow, close to the black hole, see e.g.\ Ref.~\citep{Kammoun:2020pno}, where the local approximation of the disk emission by a black body might not apply~\citep{Czerny:2003hj,Kubota:2018cuj}. However, the method was not successful so far since the so called disk size problem appeared: although the time measured delay $\tau$ scaled with the wavelength $\lambda$ as expected ($\tau \propto \lambda^{4/3}$,~ Ref.~\cite{Collier:1998ev}), the normalization of this relation did not follow the theoretical expectations, and the measured time delays were frequently longer implying larger disk size, see e.g.\ Refs.~\cite{Collier:1998ev, Cackett:2007kw, Shappee:2013mna, 2016ApJ...821...56F, Jiang:2016jua, DES:2017vao, 2020ApJS..246...16Y, 2021arXiv211107385F}, although part of the effect may be due to pre-selecting at the basis of S/N requirements and large intrinsic dispersion due to orientation, spin or accretion efficiency~\citep{Homayuni2019}. Disk sizes larger than expected were also found from microlensing measurements, see e.g.\ Ref.~\citep{Rauch1991}. The problems might have come from the considerable departure from the local black body emission, see e.g.\ Ref.~\citep{Hall:2017hnc}, but recent dense reverberation mapping proved that the problems with disk size come from the contribution to the continuum from the Broad Line Region, most notably Balmer continuum, but also Fe II continuum~\citep{Cackett:2017htp, Edelson:2018gbe, 2022MNRAS.509.2637N}. Identification of the problem opens a way to use the multicolor monitoring data from {\it Rubin}/LSST from the Dip Drilling Fields to cosmology~\citep{2020ApJS..246...16Y}, if the corrections due to Balmer continuum and Fe II are implemented. It will nicely complement the quasar method based on emission line time delays which allow to constrain other cosmological parameters, see e.g.\ Refs.~\citep{2021ApJ...912...10Z, Khadka:2021ukv} but is not yet applicable to measuring $H_0$ since it still requires external scaling.

\paragraph{ Constraining Cosmology with Multiply Imaged Supernovae and Quasars.} 
Time delay cosmography relies on the gravitational time delay between multiple images of a lensed variable source to constrain cosmology. 
The time delays between the images are primarily sensitive to the Hubble constant, but also weakly to several other cosmological parameters such as the matter density, curvature of the Universe and the EoS parameter for dark energy and its evolution over time. Other aspects that influence the time delays are the lens potential, microlensing due to substructure in the lens galaxy, and line-of sight-structures. If accounted for these effects properly, gravitationally lensed quasars and supernovae (glSNe) can be powerful probes to yield constraints on cosmological parameters, independently from the distance ladder or CMB. 

Multiply imaged quasars have been used for cosmology since their discovery in the late 70s. Since the year 2000 advances in sample size, data quality and analysis methods have shown that they can be used to determine the Hubble constant with high precision and accuracy. An account of the history of quasar based time delay cosmography and future prospects can be found in the review by Treu and Marshall~\cite{Treu:2016ljm}. Briefly, hundreds of multiply imaged quasars have already been discovered and future surveys by the Euclid, {\it Roman}, {\it Rubin}, and SKAO Telescopes are going to increase the samples by orders of magnitudes~\citep[e.g.][]{Oguri:2010ns}. The precision and accuracy of the inferred cosmology will not be limited by sample size, but rather by the quality of the follow-up data and understanding of systematic errors, which are in many case common to the study of glSNe, and therefore are discussed below.

When Refsdal predicted in 1964 how the phenomena of strong gravitational lensing could be used to calculate the cosmic expansion rate~\cite{Refsdal1964}, he initially suggested glSNe for this purpose. Only recently has this avenue become a reality.
The first discovery of a glSN was made by the Panoramic Survey Telescope and Rapid Response System 1 (Pan-STARRS1), when it detected an unusually red and bright SN. After spectroscopic confirmation of a foreground lensing galaxy, the transient was classified as a gravitationally lensed SNIa, although the imaging data had insufficient resolution to separate the images~\cite{XXX:2014xxi}. The Hubble Space Telescope (HST) provided data of the first resolved and multiply-lensed SN, a core-collapse SN appropriately named "Refsdal"~\cite{Kelly:2014mwa}. Four images were detected around an elliptical galaxy, and the prediction that another one would appear in a second host-galaxy image was verified one year later~\cite{Kelly:2015xvu}. SN Refsdal alone will provide a measurement of $H_0$ with a statistical error of 7$\%$~\cite{Grillo:2018ume}.

The intermediate Palomar Transient Factory (iPTF) identified glSN iPTF16geu, which was followed up by ESO VLT, Keck Observatory, and HST and spectroscopically identified as a type Ia SN~\cite{Goobar:2016uuf, Cano:2017rhv}.
The third resolved strongly lensed supernova AT2016jka ("SN Requiem") was discovered in Hubble Space Telescope data, and is predicted to host a fourth image in two decades time, which should allow for a sub-per cent precision measurement of the time delay~\cite{Rodney:2021keu}.

Current transient surveys capable of detecting glSNe, such as the Zwicky Transient Facility (ZTF) and Pan-STARRS, are only estimated to discover a few objects a year (6-9~\cite{Goldstein:2016fez, Goldstein:2018bue, Wojtak:2019hsc} and two~\cite{Wojtak:2019hsc} for ZTF and Pan-STARRS, respectively). The true potential of discovering and observing glSNe will be unveiled by future telescopes, which are predicted to discover orders of magnitude more. The {\it Rubin} Observatory Legacy Survey of Space and Time (LSST)~\cite{LSSTScience:2009jmu} is estimated to detect around 89 type Ia and 250 core collapse glSNe a year~\cite{Goldstein:2016fez, Goldstein:2018bue, Wojtak:2019hsc}, although the actual discovery rates may be lower when employing more restrictive criteria for selecting plausible candidates. 
The discovery rates of the {\it Roman} Space Telescope~\cite{Green:2012mj}, estimated with the image multiplicity method, are 9-56 Ias and 17-126 core collapse SNe a year, depending on the observing strategy~\cite{Pierel:2020tav}. The majority of the lensed core collapse SNe will be type IIn; however, this prediction is based on a relatively poor set of observed spectral templates of type IIn SNe and thus will likely be a subject to further improvement in the future.
Gravitationally lensed SNe appear as peculiarly bright (relative to type Ia brightness at redshift of the apparent host galaxy) and multiply imaged sources. These are the two main observational properties which can be used to select plausible glSN candidates. Novel techniques can be employed to help identify glSNe with partially blended images, such as neural networks trained on time series of images~\citep{Ramanah:2021bpb, Morgan:2021oxb}.
Another strategy for finding glSNe is to monitor known lensed galaxies~\citep{Craig:2021vld}. Precise measurements of gravitational time delays will strongly rely on follow-up observations compensating insufficient cadence of large transient surveys~\cite{LSSTDarkEnergyScience:2019dgr}.

Gravitationally lensed quasars and supernovae are highly complementary. In the remainder of this section we compare the pros and cons of each variable source and conclude by highlighting the common challenges.
Firstly, lensed quasars are around 60 times more numerous than lensed supernovae~\cite{Oguri:2010ns}.
In terms of the variability, supernovae (especially type Ia) have well-described light curves that can be fit with established templates, while quasars are characterized by a stochastic variability. 
Another important difference between glSNe and lensed quasars is that supernovae fade away over time. This poses a challenge, since follow-up observations are highly time-sensitive, but it also offers an advantage, as it allows for better kinematic follow-up observations of the lens and host galaxy once the supernova has faded away~\citep{Ding:2021bxs}. Such velocity dispersion measurements can help constrain the mass sheet degeneracy, which is a transformation of the potential that leaves all imaging observables the same, while changing the time delays~\cite{1985ApJ...289L...1F, 1988ApJ...327..693G, Saha:2000kn}. 
Type Ia supernovae offer another way of breaking the mass sheet degeneracy: their standard candle nature can provide an estimate of the absolute magnification~\cite{Foxley-Marrable:2018dzu, Birrer:2021use}.
However, it is worth noting that these constraints depend on the precision of type Ia SN standardisation, as current samples still yield intrinsic scatter in the Hubble diagram of around 0.12 mag (or 6 \% in distance)~\cite{Pan-STARRS1:2017jku}. The utility of the standard candle nature of lensed type Ia SNe can also be limited by the effect of microlensing, which gives rise to extra stochastic magnifications in every lens image~\cite{Yahalomi:2017ihe}.
Another difference is that glSNe typically have a smaller angular size and smaller image separations than lensed quasars, due to the lower mean redshifts of glSNe. Therefore, lensed SNe are prone to have very short time delays, of the order of several days or weeks. This means that time-delay measurements only require weeks of follow-up observations instead of year-long monitoring campaigns as is usual for lensed quasars. However, obtaining precise measurements of such short time delays can be highly challenging.
The small angular size of glSNe also leads to a higher sensitivity to microlensing from stars and other substructures in the lens galaxy. Microlensing can perturb light curves independently in each supernova image, which can lead to additional systematic errors in time delays of order 4$\%$~\cite{Goldstein:2017bny}. Fortunately, effects of microlensing can be minimised if multi-colour follow-up observations are obtained of the initial phase of the explosion. In this phase, the SN can be approximated as a point source and thus look identical in every filter, which leads to the same microlensing distortion across all wavelengths. This "achromatic" phase is expected to last until 3 rest-frame weeks after the explosion~\cite{Goldstein:2017bny, LSSTDarkEnergyScience:2019dgr}, in which 70$\%$ of {\it Rubin}/LSST glSN Ia images will be discovered. The useful implications are that the microlensing effects are expected to cancel out in early time colour curves, reducing the systematic uncertainties to 1$\%$ for typical {\it Rubin}/LSST systems~\cite{Goldstein:2017bny}.
Since microlensing is capable of altering the brightness of the multiple images, thereby changing whether they fall above or below our detection limits, it could impact the number of glSNe and lensed quasar we will be able to observe.

For both types of variable sources, the precision and accuracy of the inferred cosmology will likely not be limited by sample size but by the quality of follow-up data and systematic uncertainties associated with the analysis. Most of these are common between lensed quasars and supernovae. In addition to the time delays themselves, that will likely required dedicating monitoring capabilities to supplement the cadence of the survey themselves, e.g.\ through a dedicated 3-4m class telescope, the cosmological inference requires: i) angular resolution images at resolution of 10mas or better; ii) spatially resolved kinematics of the deflector galaxy or cluster galaxies to break the mass sheet degeneracy and the mass anisotropy degeneracy~\cite{Birrer:2020tax,Xu:2011ru}; iii) information about the structure along the line of sight. Whereas the information about the mass along the line of sight will likely be provided by the same surveys used for the discovery of the targets, in combination with high precision next generation cosmological simulations, the high angular resolution imaging and spatially resolved kinematics can only be obtained with the James Webb Space Telescope or the next generation of extremely large telescopes.

With proper follow-up, time delay cosmography will be able to deliver measurements of the Hubble constant with a precision and accuracy of $\sim$1\%, independent of all other methods, contributing to the resolution of the Hubble tension~\citep[e.g.,][]{Birrer:2020jyr}.

\subsubsection{Gravitational Wave Cosmology}
\label{GWsubsub}

It took almost 100 years for the prediction of gravitational waves to finally be directly confirmed~\cite{LIGOScientific:2016aoc}. Since the first detection of gravitational waves in 2015 by LIGO-Virgo Scientific Collaboration~\cite{LIGOScientific:2016aoc}, 90 events from compact object binary mergers have now been observed by the network of LIGO-Virgo detectors~\cite{LIGOScientific:2021djp}, and the field of GW astronomy is exploding.

The third generation ground-based interferometers (such as \ Cosmic Explorer~\cite{Reitze:2019iox, Evans:2021gyd}, Einstein Telescope~\cite{Maggiore:2019uih}) as well as space based gravitational wave detectors (e.g.\ LISA~\cite{LISA:2017pwj}, TianQin~\cite{Luo:2015ght}, Taiji~\cite{Hu:2017mde}, and DECIGO~\cite{Kawamura:2020pcg}), scheduled for the next decade, will open an even broader spectrum of gravitational-wave sources, providing novel cosmological probes with great precision. Also, possibly operating in conjunction, these experiments can improve our ability to investigate fundamental physics with GWs. In what follows, some perspectives on GWs for cosmology are presented.

In the next decade, gravitational wave standard sirens (GWSS)~\cite{Schutz:1986gp,Holz:2005df,Dalal:2006qt,Chen:2017rfc,DiValentino:2018jbh,Vitale:2018wlg,Palmese:2019ehe}, the GW analog of astronomical standard candles, are expected to play an important role in the context of cosmology. The amplitude of GWs is inversely related to the luminosity distance from the source, hence they can be used in conjunction with redshift information to the source to probe the distance-redshift relation.
Observations of the merger of the binary neutron-star system GW170817, along with the redshift from its host galaxy, identified thanks to the observation of an electromagnetic counterpart, yield $H_0=70_{-\,8}^{+12}{\rm km\,s^{-1}\,Mpc^{-1}}$ at 68\% CL~\cite{LIGOScientific:2017adf}. Though this constraint is broad, it does not require any cosmic "distance ladder" and is model-independent: the absolute luminosity distance is directly calibrated by the theory of general relativity~\cite{Schutz:1986gp}. In other words, GWSS of this kind are an ideal independent probe to weigh in on the Hubble tension problem. 
Around 50 additional observations of GWSS with electromagnetic counterparts are needed to discriminate between {\it Planck} and R19 measurements of $H_0$ with a precision of $1$--$2\%$~\cite{Nissanke:2013fka,Chen:2017rfc, Mortlock:2018azx}. Measurements of this kind can also play an important role in an extended parameter space~\cite{DiValentino:2017clw, DiValentino:2018jbh} in which CMB data are unable to strongly constrain $H_0$, in a way that breaks degeneracies amongst parameters. Standard sirens with EM counterparts are therefore expected to be extremely powerful cosmological probes, while the upgrades to the LIGO-Virgo second generation design sensitivities are critical to provide the above desired number of BNS merger detections.\footnote{Cosmic Explorer and Einstein Telescope are expected to observe far more binary neutron star mergers.} On the other hand, kilonovae associated with BNS mergers are expected to be close to isotropic, and are thus the most promising of all BNS EM counterparts. However, the kilonova associated with GW170817 was intrinsically faint, and nonetheless was thought to be on the bright end of the kilonova luminosity function.
In order to provide new associations, and hence a standard siren measurement that is informative for the Hubble tension, deep optical-to-near-infrared observations will be required to identify the fainter kilonovae, with more precise GW localizations playing a significant role in the success of these follow-up campaigns. Multi-object spectroscopic observations could play a central role in these campaigns through rapid classification of all the viable variable objects in the GW localization regions, which is currently a bottleneck for the identification of counterparts from imaging observations. Experiments such as DESI-II~\cite{Dawson:2020} could design a specific program aimed at standard siren cosmology through the identification of counterparts and observation of transients host galaxies’ redshifts, along with other key science projects.

Complementary dark GWSS (GW sources without EM counterparts) such as those used in Ref.~\cite{LIGOScientific:2018gmd, DES:2020nay,LIGOScientific:2021aug,Palmese:2019ehe} are expected to provide percent-level uncertainty on $H_0$ after combining a few hundreds to thousands of events from the statistical host identification technique~\cite{Gray:2019ksv} or by identifying the host galaxy for the nearby sources~\cite{Borhanian:2020vyr} using upgraded detector network in the future. In this case, a galaxy catalog overlapping with the GW events is necessary to provide the redshift information of the potential host galaxies. The implementation of this method depends on the completeness of the galaxy catalog up to high redshift, understanding the GW sources population, and better sky localization error of the GW sources. In order to avoid any biased measurement of the Hubble constant $H_0$~\cite{Trott:2021fnx}, a promising approach can be to use the statistical host identification technique as demonstrated by~\cite{LIGOScientific:2018gmd,Gray:2019ksv, Palmese:2019ehe, DES:2020nay,LIGOScientific:2021aug}. 
Another promising method to infer cosmological parameters from dark GWSS consists in cross-correlating GW sources with photometric or spectroscopic galaxy surveys to infer the clustering redshift of the GW sources~\cite{Oguri:2016dgk, Mukherjee:2018ebj, Scelfo:2018sny, Mukherjee:2019wcg, Mukherjee:2020hyn, Bera:2020jhx,Scelfo:2020jyw,Diaz:2021pem,Mukherjee:2022afz}. The cross-correlation technique is a promising avenue to achieve a $2\%$ measurement of the Hubble constant from the LIGO-Virgo-KAGRA network operational in its design sensitivity in synergy with galaxy surveys such as SPHEREx and DESI~\cite{Diaz:2021pem} after combining a few hundreds of well-localised GW sources. 
Moreover, this technique will also make it possible to measure the DE EoS and the connection between GW sources and dark matter from dark standard sirens. Though the implementation of the cross-correlation does not require the host galaxy of the GW source to be present in the catalog, its success will depend on the availability of GW sources with better sky localization error. This may be feasible by using multiple detectors, overlapping galaxy surveys with accurate redshift estimation, and marginalization over the redshift-dependent GW bias parameters~\cite{Mukherjee:2020hyn, Diaz:2021pem,Mukherjee:2022afz}. It follows that for the success of dark GWSS appropriate full sky galaxy catalogs that cover the redshift range of GW sources are required. Alternatively, it is possible to use features in the mass spectrum of binary black holes to infer redshift, providing a powerful and independent standard siren probe of cosmology~\cite{Chernoff:1993th, Farr:2019twy,Ezquiaga:2020tns,Mukherjee:2021rtw,LIGOScientific:2021aug, Ezquiaga:2022zkx}. These "spectral sirens’’ may provide percent-level constraints on the expansion history at $z\sim 1$ and beyond. However, for this technique to work, the mass distribution of the GW sources need to be standardised. Any redshift evolution~\cite{Mukherjee:2021rtw} of the GW mass distribution needs to be mitigated in order to avoid any systematic error in the inference of the cosmological parameters. Another dark siren approach is to use knowledge of the EoS of neutron stars to infer redshift. For third-generation detectors, this would allow us to use these systems as standard sirens to constrain cosmology even in the absence of an electromagnetic counterpart~\cite{Messenger:2011gi,Chatterjee:2021xrm}. Space-based, low-frequency GW detectors such as LISA may observe well-localized GW signals from SMBH binary mergers with EM counterparts. Depending on uncertainties in SMBH formation models and EM emission mechanisms, these SMBH standard sirens could provide useful constraints on the Hubble constant, $\Omega_M$ and $\Omega_\Lambda$ up to $z\sim 8$~\cite{Tamanini:2016zlh}.

Since standard sirens can be detected at cosmological redshifts, they can be sensitive to other cosmological parameters besides $H_0$ such as $\Omega_m$ and $w(z)$. As would be the case for any other probe, should one assume a specific model that were not true (e.g.\ $\Lambda$CDM), then the inferred parameters from that assumed model would be biased relative to the true ones. In the case of cosmological standard sirens that extend to $z\sim 1.0$ the bias on the inferred cosmological parameters exceeds $3\sigma$ when the true model was a Chevallier - Polarski - Linder parameterization (CPL)~\cite{Chevallier:2000qy,Linder:2002et} model that is consistent with current SN constraints~\cite{Shafieloo:2018qnc}. The biases that arise from assuming the wrong model can be mitigated, however, with the use of model-independent statistics such as Gaussian process regression~\cite{Keeley:2019hmw}. Because of their high redshifts, some GW sources are expected to be strongly lensed~\cite{Wang:1996as, Oguri:2018muv, Li:2018prc}; a catalog of multiply-imaged sources, and associated lensing rates, time-delays, and magnification distributions, may provide additional cosmological constraints~\cite{Oguri:2018muv, Xu:2021bfn}.

In the next decade, GWSS are expected to provide strong constraints on dark energy, modified gravity, dark matter and neutrinos, and shed light on several other important aspects in cosmology~\cite{Cai:2016sby,Pardo:2018ipy,Du:2018tia, Yang:2019bpr, Wei:2018cov,Zhang:2018byx,Wang:2018lun,Zhang:2019ple,Qi:2019spg, Yang:2019vni,Fu:2019oll,Yang:2020wby, Bonilla:2019mbm, Chen:2020zoq}. 

The physical degrees of freedom of GWs are imprints of the nature of gravity, and by modifying general relativity (GR), we have two major possibilities that can modify and leave fingerprint in the GW signal. 

\begin{itemize}

\item Propagation Effects: A common feature in almost all gravity theories beyond GR, considering the Universe on large scales, is that the new degrees of freedom in each theory modify the gravitational force/interaction on cosmological scales, mainly to explain the late-time acceleration of the Universe (dark-energy-dominated era). This roughly corresponds to an effective time-varying gravitational constant that will affect the propagation of GWs along the cosmic expansion, which in turn can induce amplitude and phase corrections on the GW signal over cosmological volumes~\cite{Cai:2016sby,Pardo:2018ipy,Du:2018tia,Yang:2019bpr,Wei:2018cov,Zhang:2018byx,Wang:2018lun,Zhang:2019ple,Qi:2019spg,Yang:2019vni,Fu:2019oll,Yang:2020wby,Bonilla:2019mbm,Cutler:2009qv,Nishizawa:2009jh,Nishizawa:2010xx,Yagi:2013du,Nunes:2018zot,Cai:2017aea,Saltas:2014dha,Nishizawa:2017nef,Belgacem:2017ihm,Belgacem:2018lbp,Arai:2017hxj,Belgacem:2019pkk,Lagos:2019kds,Nunes:2019bjq, Nunes:2018evm,Belgacem:2019lwx,DAgostino:2019hvh,Gao:2019liu,Belgacem:2019tbw,Calcagni:2019ngc,Dalang:2019rke,Wolf:2019hun,Nunes:2020rmr,Baker:2020apq,Mitra:2020vzq,Mastrogiovanni:2020mvm,Mukherjee:2020mha,Heisenberg:2022lob, Heisenberg:2022gqk, Lee:2022cyh}. The effective time variation of the gravitational constant could be measured with a few percent precision using a multi-messenger~\cite{Engel:2022yig} approach exploiting the unique relation between the GW luminosity distance, BAO angular scale and the sound horizon at decoupling~\cite{Mukherjee:2020mha}. We may also be able to test the propagation effect from gravitational lensing of GW sources~\cite{Holz:1997ic,Mukherjee:2019wcg, Mukherjee:2019wfw, Ezquiaga:2020dao, Ezquiaga:2021ler,Finke:2021znb}, which will provide new tests of general relativity~\cite{Bettoni:2016mij,Brax:2017hxh,Belgacem:2017ihm,Bonilla:2019mbm}.

\item Generation Mechanism: By changing the gravity interaction in the inspiral-merger-ringdown phases, we also change the generation mechanism of the gravitational radiation emitted by the binary systems. Such methodology can be quantified through the parameterized post-Einsteinian framework~\cite{Yunes:2009ke,Cornish:2011ys,Carson:2019fxr,Carson:2019kkh}. Future ground-based and LISA observations can lead to improvements of 2-4 orders of magnitude with respect to present constraints, while multiband observations can yield improvements of 1-6 orders of magnitude~\cite{Perkins:2020tra}.
\end{itemize}

An interesting and realistic approach might be to combine both effects on the waveform signal modeling. Some specific theoretical cases of these effects were recently investigated using the LIGO-Virgo-KAGRA GW Transient Catalog~\cite{LIGOScientific:2018mvr,LIGOScientific:2020ibl}. On the other hand, searching for polarization modes beyond the GR prediction provides a strong test of GR. This aspect can also be combined with the above mentioned effects. At a local level, the GW information from isolated or binary systems in the strong space-time curvature regime also provides several tests of GR~\cite{Berti:2015itd}. The improvement in amplitude of the spectral noise density in future detectors will be fundamental to investigate the limits of these theoretical aspects.

Another interesting possibility is the detection of {\it stochastic gravitational waves}.  Direct evidence of GWs has thus far been in the form of coherent measurements of resolved waveforms in processing the detected signals, which can be traced back to their origin as single points. Consequently, only a tiny fraction of these GW can be seen from a presumably much larger population spread throughout space at different scales. The extensive set of unresolved signals corresponding to multiple point sources or extended sources increases incoherently, leading to gravitational-wave backgrounds (GWBs). A variety of different backgrounds is expected, given the range of diverse GW sources in the Universe.
Most of these backgrounds are treated as stochastic, as they may be described by a non-deterministic strain signal and are hence referred to as stochastic gravitational-wave backgrounds (SGWBs). Indeed, the existence of a SGWB is a robust prediction of several well-motivated cosmological and astrophysical scenarios operating at both the early and late Universe~\cite{Maggiore:1999vm,Caprini:2018mtu,Giovannini:2019oii}. Below we present some of the most prominent observational projects to detect the SGWB across a broad range of frequencies of the GW spectrum.

\begin{itemize}
    \item {\it Very low frequencies}. Cosmic inflation is currently the leading paradigm to explain the initial conditions of our Universe. Single-field slow roll inflation, as well as alternative treatments~\cite{Benetti:2021uea,Myrzakulov:2015qaa,Odintsov:2021kup,Cannone:2014uqa,Graef:2015ova,Ricciardone:2016lym,Graef:2017cfy,Baldi:2005gk,Dimastrogiovanni:2016fuu,Obata:2016oym,Cook:2011hg,Mukohyama:2014gba,Cai:2016ldn,Cai:2020ovp,Stewart:2007fu,Brandenberger:2014faa,Hipolito-Ricaldi:2016kqq,Kinney:2018nny,Odintsov:2020sqy,Odintsov:2020zkl,Trivedi:2020wxf,Oikonomou:2020oex}, predict the existence of a SGWB. Such GWs induce a B-mode pattern in the polarization of the CMB. From the lack of a detection of primordial B-modes, CMB observations place upper limits on the amplitude of the primordial GW power spectrum~\cite{BICEP2:2018kqh,Planck:2018vyg}.
    For an angular resolution scale of $k \lesssim 0.1~\mathrm{Mpc}^{-1}$, these limits can be translated into limits on the amplitude of the SGWB for frequencies $f\lesssim 10^{-16}$~Hz.
    Current ground-based CMB experiments---ACT~\cite{ACT:2020gnv}, BICEP/Keck~\cite{BICEP2:2018kqh}, CLASS~\cite{Harrington:2016jrz}, POLARBEAR/Simons Array~\cite{POLARBEAR:2015ixw}, SPT~\cite{SPT:2019nip} ---measure the CMB polarization to high precision and will continue to better constrain primordial B-modes.
    Upcoming experiments, like Simons Observatory~\cite{SimonsObservatory:2018koc} and CMB-S4~\cite{Abazajian:2019eic,CMB-S4:2020lpa}, will achieve even greater sensitivity.
    In addition, satellite missions such as LiteBIRD~\cite{LiteBIRD:2022cnt} and balloon missions such as SPIDER~\cite{SPIDER:2017xxz} will complement ground-based efforts.\footnote{See Ref.~\cite{Caldwell:2022qsj} for a study of the potential of using an early Universe GW signal to probe fundamental physics.}

    \item {\it Low frequencies}. The high-precision timing of Galactic millisecond pulsars enables a search for nHz GWs, which are expected to be generated primarily by SMBH binaries. A GW transiting through the line of sight between Earth and the pulsar distorts the intervening spacetime, altering the expected time of arrival of pulses. A SGWB affects the timing of all pulsars in the Galaxy, causing spatially correlated time-delay signals between pulsar pairs, known as Hellings-Downs correlations~\cite{Hellings:1983fr}. These correlations are primarily quadrupolar, with subdominant contributions from higher multipoles, and are distinct from two main sources of spatially-correlated noise, errors in the terrestrial time standard (monopolar) and errors in the planetary ephemeris (dipolar)~\cite{1978SvA....22...36S,Detweiler:1979wn,1990ApJ...361..300F, Brazier:2019mmu, NANOGrav:2020gpb}.  World-wide efforts to search for a SGWB are underway: the North American Nanohertz Observatory for Gravitational Waves (NANOGrav)~\cite{NANOGrav:2020bcs}, the Parkes Pulsar Timing Array (PPTA)~\cite{Hobbs:2013aka,Kerr:2020qdo}, the European Pulsar Timing Array (EPTA)~\cite{Desvignes:2016yex}, and the Indian Pulsar Timing Array Project (InPTA)~\cite{Joshi:2018ogr}, all operate both individually and as a collective under the International Pulsar Timing Array (IPTA)~\cite{Perera:2019sca}. Recent evidence of a common-spectrum process has provided intriguing hints that a detection may be soon to come, but observing the Hellings-Downs spatial correlation will be key to interpreting any signal as a possible SGWB detection~\cite{NANOGrav:2020bcs,Goncharov:2021oub,Chen:2021rqp}.
    
    Distortions of spacetime affect not only the timing of pulsars, but also induce a shift in the apparent position of stars. These shifts may be observed in astrometric surveys~\cite{Braginsky:1989pv,Kaiser:1996wk} for GWs in the frequency range $10^{-18}~\mathrm{Hz} \lesssim f \lesssim 10^{-8}~\mathrm{Hz}$. Similar to PTAs, the shifts also exhibit particular angular correlations from a SGWB~\cite{Book:2010pf}. Indeed, there are expected cross correlations between PTAs and astrometric shifts that could permit complementary studies~\cite{Mihaylov:2018uqm,Qin:2018yhy,Mihaylov:2019lft,Qin:2020hfy}. Astrometric data from GAIA and extragalactic radio sources have set constraints on the energy density of the SGWB~\cite{Darling:2018hmc}. Future missions such as Theia could provide improved sensitivity~\cite{Theia:2017xtk}.
    
    \item {\it Intermediate frequencies}. The GWs are emitted mainly by individual binary systems, like, for instance, from binary black holes (BBH), binary neutron stars (BNS), and binary black hole-neutron stars (BBH-NS). The superposition of the signal from these sources over cosmological volumes is expected to form an  SGWB of astrophysical origin. This signal strongly depends on the sources that produce them, and we expect that signal to exist in the most diverse frequencies (in this case, for example $10^{-3} \geq f \geq 10^2$)~\cite{Zhu:2012xw,Mingarelli:2019mvk,Jenkins:2018uac}. Nevertheless, this specific type of signal has not been detected until the present moment, and only some upper limits on the SGWB signal have been obtained, for instance, the LIGO-VIRGO-KAGRA collaboration reported an astrophysical background with amplitude $< 4.8 \times 10^{-8}$~\cite{LIGOScientific:2019vic}. With the improvements in instrumental sensitivity in the coming years and the prospects of future detectors like the Einstein Telescope (ET)~\cite{Maggiore:2019uih}, Cosmic Explorer (CE)~\cite{Reitze:2019iox}, and LISA~\cite{LISA:2017pwj,Caprini:2019pxz}, it is expected to achieve enough sensitivity to detect SGWBs of astrophysical origin.
    
    \item {\it Ultra-high frequencies}. Recently, IPTA gave hints of GWs from cosmic strings~\cite{Antoniadis:2022pcn}, which has further encouraged the community to build a strong case for experiments that can tackle ultra-high-frequency gravitational waves (UHGW). This is because, on the one hand, cosmic strings are a great scenario that can be analyzed in the multiband frequency range and hence with experiments already covering the low, medium, and high-frequency range (e.g., PTA experiments, LISA and LIGO-Virgo-KAGRA, respectively). Further experiments that can cover the MHz to GHz region are missing. Their construction could then complement the study of cosmic string signals in the broadband frequency range. On the other hand, SGWB due to cosmic origin, for example, from inflation models, from the evaporation of primordial black holes or phase transitions generated when Grand Unified Theories break into the SM or other subgroup can have signals in the MHz to GHz range. Hence, the current GW experiments will not be able to detect them. Different initiatives are being carried out to promote the creation of a network of researchers to develop the science of GW's in the frequency range above $10\,$kHz.\footnote{Go to this home page, for example: \url{http://www.ctc.cam.ac.uk/activities/UHF-GW.php}}
\end{itemize}

Finally, several authors have pointed to the implementation of GW  standard sirens and GW stochastic backgrounds to explore the late and early Universe, in a broad spectrum of possibilities ranging from the Hubble tension~\cite{Li:2021htg} through the search for modified gravity~\cite{Nunes:2020rmr,Allahyari:2021enz}, variation of the speed of propagation of GWs~\cite{Bonilla:2019mbm}, inflation models~\cite{Li:2021htg} and even explore the exotic fundamental physical possibility as the variation of fundamental constants~\cite{Nunes:2016plz,Colaco:2022aut}.

\subsubsection{21 cm Cosmology}

Up to now, the key probes of large-scale structure in the Universe have been based on galaxy surveys,  and the upcoming generation of these surveys promises to  deliver a revolutionary advance in precision cosmology. 
Next-generation galaxy surveys
will be complemented by a new type of survey, which will map  the 21 cm emission line of neutral hydrogen (HI) in the post-reionization era ($z\lesssim 6$). The 21 cm emission of galaxies that contain HI provides a tracer of the dark matter distribution, with extremely precise redshift measurement as a direct by-product of detection. However, the faintness of the emission means that next-generation HI galaxy surveys with SKAO-MID will not be competitive with contemporaneous spectroscopic surveys in the  optical/ infra-red bands, although they will provide valuable complementary information and help to deal with systematics at low redshift~\cite{Bacon:2018dui}. 

An alternative approach foregoes the detection of individual galaxies and measures instead
the integrated emission from all galaxies in each 3-dimensional pixel. 21 cm intensity mapping in principle allows for spectroscopic surveys with   high effective number densities, that can measure large-scale fluctuations in huge cosmological volumes, over reasonable observing times~\cite{Bull:2014rha,Villaescusa-Navarro:2018vsg,Bacon:2018dui}.  
So far the signal has only been detected in cross-correlation with existing galaxy samples~\cite{Chang:2010jp,Masui:2012zc,Wolz:2021ofa,CHIME:2022kvg}, and upper limits on the power spectrum have been set by data from GBT~\cite{Switzer:2013ewa} and uGMRT~\cite{Chakraborty:2020zmx}.
Next-generation  21 cm intensity mapping surveys are capable of producing very high precision measurements of redshift-space distortions and BAO (e.g.~\cite{MeerKLASS:2017vgf, Bacon:2018dui,Castorina:2019zho,Cunnington:2020mnn,Kennedy:2021srz,Abdalla:2021nyj,Sailer:2021yzm,Costa:2021jsk}), provided that effective foreground-cleaning techniques are developed for the data pipeline. 

These foregrounds are orders of magnitude larger than the signal, but {their different properties can be exploited to separate them with various cleaning techniques}. In addition to foregrounds, other sources of contamination, such as polarization leakage, 1/f noise, and radio frequency interference (RFI), need to be precisely understood for the pipeline construction.
Currently major efforts are underway, in particular those based on recent measurements {with CHIME~\cite{CHIME:2022kvg} and MeerKAT}~\cite{Wang:2020lkn,Li:2020bcr,Matshawule:2020fjz,Cunnington:2020njn}, the 64-dish precursor of the SKAO-MID telescope array that will have 197 dishes upon completion. Other current work includes an SKAO blind foreground subtraction challenge~\cite{Spinelli:2021emp},  and methods preparing for science operations of OWFA~\cite{Chatterjee:2019ryp}, HIRAX~\cite{Crichton:2021hlc} and BINGO~\cite{Fornazier:2021ini}.

The interferometer arrays CHIME and HIRAX, and the large single-dish telescope BINGO are designed for cosmological constraints, especially via  BAO measurements. By contrast, SKAO-MID (and its precursor MeerKAT) were not designed for cosmology, but in fact forecasts show that SKAO-MID could deliver competitive constraints from a  21 cm intensity mapping survey ($0.3<z<3$ and 20,000\,deg$^2$)  in single-dish mode~\cite{Bacon:2018dui}.
In particular, the combination of the huge volume of an SKAO 21cm intensity mapping survey with next-generation galaxy and CMB surveys can potentially provide the tightest constraints on local
 primordial non-Gaussianity, using the multi-tracer technique~\cite{Alonso:2015sfa,Fonseca:2015laa,Bacon:2018dui,Ballardini:2019wxj,Viljoen:2021ypp} (the multi-tracer also provides tight constraints on neutrino mass using only linear scales~\cite{,Ballardini:2021frp}).
Interferometer arrays such as HIRAX (and the futuristic PUMA) are better suited to constrain other forms of primordial non-Gaussianity, using the power spectrum and bispectrum~\cite{Karagiannis:2019jjx}.  The 21 cm intensity bispectrum is an important tool for  cosmology with next-generation surveys, in interferometer or single-dish mode~\cite{Sarkar:2019ojl,Durrer:2020orn,Jolicoeur:2020eup,Cunnington:2021czb}.
 
Cross-correlation of upcoming 21 cm intensity maps with photometric galaxy samples can deliver measurements of weak lensing magnification~\cite{Jalilvand:2019bhk,Witzemann:2019ncy}.
In addition, this combination can be used to calibrate photometric redshifts through the clustering signal~\citep{Alonso:2017dgh,Cunnington:2018zxg,Guandalin:2021sxw}, potentially reducing one of the most significant systematic uncertainties in weak lensing experiments. Cross-correlation with a spectroscopic galaxy survey offers the exciting prospect of reconstructing the large-scale modes lost to foreground cleaning from their coupling to measured small-scale modes~\cite{Modi:2021okf}.

\subsection{Potential Probes of the Universe}

\subsubsection{Quasars as Geometric Rulers }

It has been a long pursuit to use quasars and active galactic nuclei (AGNs) for cosmology since the discovery of the first 
quasar 3C 273 in 1963. Most of the proposed approaches were found to be invalid or ineffective in the end. 
In recent years the situation changed thanks to the GRAVITY instrument onboard the Very Large Telescope Interferometer (VLTI), which provides an unprecedented spatial resolution down to 10\,$\mu$as in near-infrared bands with the spectroastrometry (SA) technique. Such a high spatial resolution immediately led to successful measurements of the characterized angular sizes ($\Delta\theta_{\rm BLR}$) of the broad-line regions (BLRs) in two quasars 3C~273 ($z\approx 0.158$, $\Delta\theta_{\rm BLR}\approx 46\pm 10\,\mu$as~\citep{2018Natur.563..657G}) and IRAS~09149-6206 ($z\approx 0.0573$, $\Delta\theta_{\rm BLR}\approx 65^{+30}_{-39}\,\mu$as~\citep{GRAVITY:2020cli}), as well as inferences on angular distributions of the ionized gas in the BLRs. Complementary to SA that measures angular sizes, the well-established reverberation mapping (RM) technique measures linear sizes ($\Delta R_{\rm BLR}$) of BLRs of quasars and AGNs through detecting time delays ($\Delta \tau_{\rm BLR}$) of variations between broad emission lines (BELs) and ionizing continuum, which transfer to BLR sizes simply as $\Delta R_{\rm BLR}=c\Delta \tau_{\rm BLR}$~\citep{1982ApJ...255..419B, Peterson:1993zz}, where $c$ is the speed of light. The delays can be routinely measured by cross-correlation analysis of the observed light curves of BELs and continuum. 

With SA and RM (SARM) observations, the angular-size distances of quasars and AGNs are then, 
in principle, directly determined by the geometric relation of 

\begin{equation}
D_{\rm A}=\frac{\Delta R_{\rm BLR}}{\Delta \theta_{\rm BLR}}
 =677.5\,\Delta R_{20}\Delta \theta_{50}^{-1}\,{\rm Mpc},
\end{equation}
 
where $\Delta R_{20}=\Delta R_{\rm BLR}/20\,{\rm ltd}$ and $\Delta\theta_{50}=\Delta\theta_{\rm BLR}/50\,{\mu}{\rm as}$ (here ltd is the unit of light-day and $\mu$as is $10^{-6}$ arcsecond). In practice, the spatial distributions of BLRs need to be taken into account to mimic real situations. The competitive advantages of the SARM approach for cosmic distances lies at: 1) it is fully free of extinction and reddening corrections, and 2) it also does not invoke standardized processes and calibrations of distance ladders. 

The first SARM application had been made successfully on the quasar 3C 273~\citep{Wang:2019gaq}, which simultaneously yielded the angular-size distance $D_{\rm A}=551.50_{-78.71}^{+97.31}\,{\rm Mpc}$
and the mass of the central supermassive black hole (SMBH) 
$M_{\bullet}=5.78_{-0.88}^{+1.11}\times 10^{8}\,M_{\odot}$, as well as the Hubble constant 
$H_{0}=71.5_{-10.6}^{+11.9}\,{\rm km\,s^{-1}\,Mpc^{-1}}$ with a statistic error of $\sim 15\%$. 
The second SARM application has been recently made by GRAVITY Collaboration~\citep{GRAVITY:2021ory} on NGC 3783. Those analysis only employed the light curves of the continuum and BEL in the RM data and therefore is usually denoted as 1D SARM.
As a comparison, 2D SARM analysis additionally employs profile variations of the BEL in the RM data and can make better use of the RM data. In particular, the 2D SARM analysis on 3C~273 carried out by Ref.~\citep{Li:2022agl} made use of the fact that H$\beta$ (from RM data) and Pa$\alpha$ (from SA data) regions are gravitationally bound by the same potential of the SMBH and also share the inclination. As such, 2D SARM approach can naturally remove the systematical arising from the possible differences between the H$\beta$ and Pa$\alpha$ line emission regions. 

In general, the systematic errors in SARM analysis are composed of three major sources
\citep[see details in][]{Wang:2019gaq}. First, RM measures variable parts of the BLR over 
a monitoring period while SA measures snapshots of the whole BLR in a specific epoch. This raises a potential discrepancy between SA and RM measurements, leading to a systematic error. Fortunately, this issue can be tested by comparing the mean spectrum with the root mean square (RMS) spectrum of BELs.
To be specific, if the two spectra have similar profiles and full widths half maximum ($V_{\rm FWHM}$), the variable 
parts are consistent with the whole BLR and therefore the resulting systematics can be neglected. A similarity of the mean and RMS spectra was indeed found in 
3C 273~\citep{2019ApJ...876...49Z}. Second, complicated structure and dynamics of BLRs constitutes another important source of systematics, considering the simple BLR modeling in present SARM analysis. Most of AGNs with RM observations showed disk-like structures from wavelength-resolved 
delays~\citep{2017ApJ...849..146G}. Multiple campaigns for one individual AGNs at different epochs
can directly testify the underlying dynamics of the BLR through the $\tau_{\rm BLR}-V_{\rm FWHM}$ relation. 
It has been demonstrated that several AGNs follow $\tau_{\rm BLR}\propto V_{\rm FWHM}^{-1/2}$, implying a simple Keplerian rotation~\citep{2014SSRv..183..253P}. Third, currently RM campaigns are 
conducted in optical band whereas GRAVITY operates in near-infrared band so that their corresponding BELs are different. This issue can be resolved by directly mapping near-infrared BELs in RM or by implementing 2D SARM analysis as in Ref.~\citep{Li-2D-SARM2022}. In a nutshell, all of the above systematics can be understood observationally.

In a few years, the next generation of GRAVITY, called GRAVITY+, will come available. It incorporates several key improvements with new technology and can detect much fainter quasars
up to redshifts $z=2-3$ (see detailed information about GRAVITY+). Moreover, the forthcoming Extremely Large Telescope has the inherent capability of doing spectroastrometry and therefore can also 
be employed for geometric measurements of cosmic distances. Recent Monte-Carlo simulations showed that the SARM approach is able to achieve a precision of $1-2\%$ for the Hubble constant with about 100 AGNs~\citep{Songsheng:2021erq}. SARM campaigns to measure cosmic distances at the noon of the Universe for the history of cosmic expansion therefore have a high merit.

\subsubsection{Redshift Drift}

In the standard cosmological model, a fixed observer will observe the redshift of objects of the Hubble flow to slowly change (or drift) systematically with time as the Universe expands ($\dot{z} \equiv \frac{\mathrm{d}z}{\mathrm{d}t_{\rm obs}} \neq 0$)~\cite{Sandage:1962,McVittie:1962}.
The drift at redshift $z$ is related to the values of the Hubble parameter at that redshift, $H(z)$, and today, $H_0$, as~\cite{McVittie:1962}:
\begin{equation}
    \dot{z}(z) = (1+z)H_0 - H(z).
\end{equation}
This can equivalently be expressed as acceleration, $\dot{v} = c\dot{z}(1+z)^{-1}$, where $c$ is the speed of light and $\dot{v}\equiv\frac{{\rm d}v}{{\rm d} t_{\rm obs}}$.

The measurement principle is simple: one compares the redshifts of an object in the Hubble flow derived from multi-epoch observations, where the epochs are sufficiently separated in time so that the measurement uncertainty is smaller than the expected cosmological signal. A significant advantage of this experiment is that the signal grows linearly with time and the measurements are straightforward to interpret, requiring no assumptions about astrophysical processes (unlike standard candle measurements for example).
Reconstructing $\dot{z}(z)$ is therefore a direct and model-independent way to measure the expansion history of the Universe~\cite{Heinesen:2020pms,Heinesen:2021qnl,Korzynski:2017nas} and can be used to constrain cosmological and dark energy models and extensions to GR~\cite{TeppaPannia:2013lbl,Martins:2016bbi,Heinesen:2021nrc}. It is also independent of other cosmological probes, such as supernovae, the CMB, and large scale structure measurements so its contribution to cosmology is unique.

Within $\Lambda$CDM, $\dot{z}$ depends on the densities of radiation, matter, and dark energy components in the Universe. Assuming $\Omega_m=0.3$ and $\Omega_\Lambda=0.7$, the value of $\dot{z}$ reaches $+0.2\,{\rm cm\,s}^{-1}$ per year at $z\sim1$ before becoming negative at $z>2$. If, however, $\Omega_\Lambda=0$, then $\dot{z}$ is always negative and quickly grows more negative with increasing redshift, c.f.\ Fig.~2 of Ref.~\cite{Liske:2008ph}. Redshift drift measurements can therefore provide an independent check on dark energy effects or other types of effects from alternative cosmologies. For example, the prediction of the scale invariant model of Ref.~\cite{Maeder:2017ksf} is similar to that of $\Lambda$CDM if $\Omega_m\approx0.3$ but otherwise diverges significantly (Maeder, private communication). The most interesting prediction of Ref.~\cite{Maeder:2017ksf} (in the context of redshift drift) is that, if $\Omega_m<0.2$, $\dot{z}$ is positive at least until $z=5$ -- in stark contrast with the predictions of $\Lambda$CDM. In fact, this model does not require the presence of dark matter so $\Omega_m$ is expected to be $\ll 0.2$, in which case $\dot{z}$ reaches at least $+1\,{\rm cms}^{-1}$ per year at $z=3$ and grows more positive with redshift. 

New, advanced astronomical instrumentation is being built to perform the redshift drift measurement in the near future. The SKAO aims to measure the drift by monitoring up to 1 billion neutral hydrogen 21cm emission lines for galaxies at redshifts $z\leq2$ in the radio-frequency domain~\cite{Klockner:2015rqa}, as first done for C3 273~\cite{Davis:1978}. Similarly, the high-resolution optical/NIR spectrograph ANDES on the future Extremely Large Telescope aims to obtain $\dot{z}$ measurements by monitoring up to one million absorption systems towards approximately a dozen bright quasar targets~\cite{Maiolino:2013bsa}. These optical measurements will probe redshifts $z>2$ so are complementary to the SKAO ones. Exploratory studies have demonstrated that, for the required measurement precision to be reached, monitoring should last for more than 10 years~\cite{Klockner:2015rqa,Liske:2008ph}. 

Observations of the 21cm line of neutral hydrogen for ten objects at $\langle z \rangle = 0.5$ over a period of 13.5 years found $\dot{z}=(-0.23\pm 0.8)\times10^{-8}$ or $\Delta v/\Delta t_{\rm obs}=(-5.5\pm 0.2)\,{\rm cm\,s^{-1}}$ per year, where the quoted numbers are averages over the ten objects~\cite{Darling:2012jf}. The uncertainty of this measurement is $\sim100$ times larger than the expected signal, making it obvious that there is a long way until the required precision is reached in practice. Regarding the optical measurements, an important issue that must be solved relates to the wavelength calibration of the instrument. Instrumental calibration must be stable to a level comparable to the expected cosmological signal, that is several ${\rm cm\,s}^{-1}$ per decade over the lifetime of the experiment. This level of precision is approximately equal to measuring the centres of absorption features with a relative precision of $\frac{\Delta\lambda}{\lambda}\sim 10^{-11}$ or only $3\times10^{-10}\,$m on the detector image plane -- less than the separation between individual Si atoms of the detector. Novel calibration methods using Laser Frequency Comb technology have been demonstrated to provide this stability, but only over periods of several hours~\cite{Milakovic:2020ehq,2020NatAs...4..603P}. 

Additional systematic effects arise from variations of the point spread function of the instrument across the detector, variations in pixel sizes and sensitivities, and data reduction process. Significant further effort is therefore required to ensure the technical feasibility of the experiment. Another topic of ongoing research concerns identifying optimal methods for measuring the redshift drift from quasar spectra. It is currently unclear whether existing methods (e.g.\ cross-correlation) are sufficient or more advanced methods, based on absorption system modelling using artificial intelligence~\cite{Lee:2020lof}, are required. Data collected using existing ultra-stable, high-resolution spectrographs such as HARPS and ESPRESSO is expected to provide useful input.

\subsubsection{Lunar Astronomy}

Lunar astronomy did not get a mention in the main text of the 2020 decadal review. Yet many of the goals of lunar astronomy overlap with traditional astronomy and high energy astrophysics. If our current decadal planning program does not get the ball rolling now for input into lunar telescopes, we expect that in the next decade the process will be more or less set in stone. There are many fields already vying for a slice of the lunar science pie. We need to penetrate new frontiers and open up new funding horizons that can best be achieved with unified planning. For large telescope projects, the planning should start now.

This is happening of course to some extent in the various lunar programs (NASA, China, ESA). But it is unfortunate that mainstream US astronomy planning (as exemplified in the recent Decadal Review) seems to be ignoring this potentially novel direction. Lunar telescopes are very much complementary to current and planned terrestrial and space projects. 

The funding might come in part from lunar exploration resources. This joint approach would open up the prospect of building otherwise unaffordably large telescopes and open up new science frontiers that are otherwise unattainable. A low frequency radio interferometer on the lunar far side is certainly the first major project one might develop incrementally, in part because we have mastered most of the relevant technology and are already building terrestrial counterparts such as SKAO-LOW. This would explore one of the last remaining great frontiers, the Dark Ages, the epoch before the first galaxies, stars and massive black holes.

Farside~\cite{Burns:2021pkx,Burns:2021ndk} is a NASA probe-class proposal with 128 dipole antennae deployed over 10 km envisaged for the 2030s. The base station correlator would relay the signal to an orbiting lunar satellite, to be transmitted back to Earth. This could be a pilot project for a much larger lunar low frequency radio interferometer with millions of antennae.

Clearly advance planning of a road map is essential for such projects. But there is equally more to do in the realms of infrared and GW astronomy~\cite{LGWA:2020mma}.\footnote{An attractive idea for lunar astronomy consists in detecting gravitational waves from the Moon~\cite{LGWA:2020mma,Jani:2020gnz} propose lunar GW detectors that would fill the gap in GW frequency between the LISA and the ground based detectors bands. Such experiments would be able to detect massive and intermediate mass BBH and stellar BBH and BNS inspirals, and they would therefore contribute to the GW cosmology probes explored in Section~\ref{GWsubsub}.} Planning for alternative lunar projects is required now while the lunar astronomy program is being developed. And the beauty of developing such studies is that the science case is very compelling and therefore very competitive. However there will be strong competition within the space agencies for lunar science projects. Support from the astronomical community could play a key role in advancing such telescopes.

The fact that this perspective is long term is not necessarily a disadvantage for astronomy. The perspective is no more long term than the key project in the decadal review~\cite{Voyage2050}, a UV/optical/IR telescope for launch in 2045+. Or to give an example from a different field, high energy physics, the proposed 100 TeV proton accelerator for 2050+.

We emphasize that lunar telescopes can highlight frontier science that is unique and that no other proposed telescope can do. This includes unprecedented scope for exploration of the dark ages, probing the birth of the Universe and imaging of the nearest exoplanets. These represent the most adventurous and exciting cosmology and astronomy prospects of all. Lunar telescopes have potential for opening new science frontiers, and represent the new frontier of human space exploration.

There are many issues to be discussed~\cite{RSLunar1,RSLunar2,RSLunar3,RSLunar4,RSLunar5,RSLunar6,RSLunar7,RSLunar8,RSLunar9,RSLunar10,RSLunar11,RSLunar12,RSLunar13,RSLunar14,RSLunar15,RSLunar16}, including how to discriminate the elusive science signals from pervasive low radio frequency and terahertz foregrounds, and handling the abrasive role of lunar dust. One can add imaging via mega-telescopes in dark and cold polar craters and making use of the seismic stability of the Moon for deployment of GW telescopes, using technology that dates back to, and goes far beyond, the Apollo-era era seismometers. We urge that a detailed planning effort be immediately initiated to examine the feasibility and competitiveness of such potentially break-through projects.

\subsubsection{Cosmology with Gamma-Ray Bursts and Fast-Radio Bursts}

Gamma-Ray Bursts (GRBs) are spectacular events in high-energy $\gamma$-rays. Their peculiarity is that in a few seconds they can emit the same amount of energy that our Sun emits in its entire lifetime. Because they are the most powerful events in the Universe after the Big Bang they can be observed at very large distances up to redshift 9.4~\citep{2011ApJ...736....7C} and in principle up to $z=20$~\citep{Lamb:2002vq}, since they are practically free from dust extinction in their $\gamma$-rays.

Because they span over a large redshift range from $z=0.0085$ with the closest GRB (GRB 980425) ever observed at a local distance and the largest redshift observed ($z=9.4$) are very appealing candidates to map the evolution history of the Universe and to investigate cosmological parameters beyond the epoch of re-ionization. GRBs are in principle excellent candidates for being standard candles, however their luminosities and energies span over $8-9$ order of magnitude from $10^{46}\,$erg to $10^{55}\,$erg, which can be contrasted with Supernova Ia (SNe Ia). 
However, similarly to SNe Ia, the Hubble diagram of GRBs is built using observed correlations between properties of the GRB light curve and the GRB luminosity, thus enabling the indirect measurement of a luminosity distance~\cite{Demianski:2016zxi, Demianski:2016dsa, Dirirsa:2019fcs}. 

In this realm, there have been attempts to use GRBs as standard candles by several groups by using the above mentioned correlations among properties of the prompt~\citep{Amati:2002ny, Yonetoku:2003gi, Amati:2008hq, Amati:2018tso, Dirirsa:2019fcs} and the afterglow emission~\citep{Dainotti:2008vw, Dainotti:2010ki, Dainotti:2011ue, Dainotti:2011yz, Dainotti:2013cta, Dainotti:2014tna, DelVecchio:2016vjn, Dainotti:2016iqn, Dainotti:2016yxl, Dainotti:2017fem, Dainotti:2017iqb,  Dai:2020rfo, Srinivasaragavan:2020isz, Dainotti:2020azn, Dainotti:2020jkj, Dainotti:2021lkl, Dainotti:2021pqg}.

More specifically, if we consider the prompt emission properties, one crucial variable is the observed photon energy of the peak spectral flux $E_{p,i}$. This quantity is observed to be correlated with the isotropic equivalent radiated energy $E_{\rm iso}$ through the so-called Amati relation~\citep{Amati:2002ny}. $E_{\rm iso}$ is written as follows:
\begin{equation}
    \ln \left( \frac{ E_{\rm iso} }{\rm 1\,erg} \right) = b+a\ln \left( \frac{ E_{p,i} }{\rm 1\,keV } \right)\,.
\end{equation}
Here, $E_{\rm iso}$ is related to the cosmological the luminosity distance ($d_L(z)$) via
\begin{equation}
    \label{Eiso}
    E_{\rm iso} = 4 \pi d_L^2(z) S_{\rm bolo} (1+z)^{-1}\,,
\end{equation}
where $S_{\rm bolo}$ is the bolometric fluence, which is analogous to the observed brightness but in gamma rays.
Thus measuring both $E_{p,i}$ and $S_{\rm bolo}$ from GRBs allows one to construct a Hubble diagram, with a given set of calibration parameters $a$ and $b$. This process is completely analogous to constructing a Hubble diagram from SNe.

Regarding the use of GRB afterglow the crucial quantity, object of study, is the rest frame end time of the plateau emission, $T^{*}_a$ which anti-correlates with the luminosity at the end of the plateau emission, $L_a$~\citep{Dainotti:2008vw}, which in turn correlates with the peak prompt luminosity, $L_{\rm peak}$~\citep{Dainotti:2011yz, Dainotti:2015gva}. Also here, the luminosity depends on the distance luminosity  similarly as in Eq.~\eqref{Eiso}.

However, the problem of using GRB relations as standardizable candles is that their errors are large compared to the relations used for SNIa, thus make GRBs less appealing. Indeed, GRBs as all sources observed at cosmological distances are hampered by the so-called Malmquist bias effects which allow us to detect only the brightest GRBs at high-z, and thus, we have a missing population of sources at low-z which are not detectable. In addition, we have to consider that the correlations which can be good candidates to be standardizable are also subjected to redshift evolution and selection biases. To overcome these problems we need to use reliable statistical methods such as the non-parametric methods~\cite{1992ApJ...399..345E} that can allow to define the slope of a given function which mimics the evolution. We can assume any function of the evolution such as a simple power law ($g(z)=(1+z)^k$) or a more complex one. Once the variables are corrected for the evolution and for selection biases we can then safely use these variables to determine if the correlations used for constraining cosmological parameters are due to the intrinsic physics or are due to the mere result of selection biases. If such evolutionary effects are not taken into account these can lead to a degeneracy in the cosmological parameter space which are induced by the biases rather than the underlying cosmology or to the physics of the correlations used. 
Another relevant point is that the correlation parameters even if they change for different cosmological setting, they sure must not change in a way which would change the physics of the underlying physics. For example if the so-called Dainotti relation in two dimensions can be interpreted within the magnetar model, than the slope of the correlation is expected to be $1$ within 1 $\sigma$.
For details on the topic of GRB correlations, on the selection biases they undergo and their application as cosmological tools see a series of review papers~\citep{Dainotti:2015gva, Dainotti:2016qxe, Dainotti:2017iqb, Dainotti:2021juc, Dainotti_2021_sub}

Recent works~\cite{Rezaei:2020lfy, Muccino:2020gqt, Demianski:2020bva} have used the Hubble diagram of GRBs to constrain the expansion history of the Universe and to measure the properties of dark energy.
These works rely on methodology developed earlier in Refs.~\citep{Cardone:2009mr,Cardone:2010rr, Postnikov:2014aua,Dainotti:2013cta}.

Besides the use of GRBs as cosmological tools, GRBs are similarly useful for tests of cosmic isotropy~\cite{Ripa:2017scm, Ripa:2018hak} (see also~\cite{Luongo:2021nqh}). Since these bright tracers are observed out to high redshift ($0.0085<z<9.4$), their distribution on the sky should be isotropic, should the Universe be. Using the Kolmogorov-Smirnov statistic $D$, Kuiper $V$, Anderson-Darling $AD$, and $\chi^2$ statistic, three of the standard GRB catalogs \textit{Fermi}/GBM \textit{CGRO}/BATSE \textit{Swift}BAT all show consistency with an isotropic distribution, both in terms of position on the sky and the properties of the GRBs.

Fast Radio Bursts (FRBs) are very bright ($\sim\,$Jy), brief ($\sim\,\mu$s to $\sim\,$ms scale) transients, observed in the radio part of the spectrum. The first FRB was discovered in 2007~\cite{Lorimer:2007qn} with many more discovered largely in archival data over the next decade. More recently, purpose built radio telescopes and arrays are on target to find thousands if not tens of thousands of bursts every year. Our understanding of the properties of FRBs is still evolving; it appears that many but not all repeat, with unknown possible periodicity, and it is likely that FRBs will eventually be categorised in classes based on some subset of their properties. Their other properties such as polarisation, rotation measures etc, are all still unknown as the observations do not all point to a single pattern and it is not always possible to disentangle host galaxy effects from propagation effects and intrinsic properties of the bursts. It does appear certain that they are found in host galaxies, and that they lie at cosmological distances~\cite{Tendulkar:2017vuq}, thus making FRBs potential candidates for cosmological measures as their propagation will probe the intergalactic medium, offering yet another tool to constrain cosmological parameters, for example~\cite{Walters:2017afr, Walters:2019cie,Weltman:2019cqv} and references therein. 

The progenitor mechanism driving FRBs is not yet known though there are strong hints that at least some are driven by magnetars through one of many physical possible mechanisms. A full database of theories~\cite{Platts:2018hiy}, once outnumbered the observations, but as further observations occur, the possibilities for theoretical explanations shrinks and so it is likely that soon we will have only a handful of possible contenders remaining, perhaps matching future classes of FRBs. 

There are a number of ways Fast Radio Bursts can be used as cosmological probes. Here we focus on a single application. After emission from the source, the photons from the burst travel to the observer through the intergalactic medium, and are slowed down as a function of their wavelength. The dispersion measure thus contains information about the distribution of electrons from the source to the observer,
\begin{equation}
    \label{eqn1}
    DM \simeq \int {\rm d}l\,n_e\,,
\end{equation}
where $n_e$ is the number of electrons along the line of sight and the integral d$l$ runs from the source to the observer. The quantity $n_e$ contains cosmological information and indeed this simple equation combined with precision cosmology constraints from the early Universe allows us to make a prediction for the fraction of baryons in the intergalactic medium~\cite{Walters:2019cie}, which can then be experimentally verified. Indeed, the so-called missing baryon problem is no longer an outstanding problem as the results of Ref.~\cite{Macquart:2020lln} used a handful of well located FRBs with the observed dispersion measures to show that the missing baryons are in the intergalactic medium and are playing the expected role of dispersing the FRB signal. This longstanding open problem is thus resolved not due to any great technological or theoretical breakthrough, but simply through the use of a few transient observations and an understanding of the contributions to the dispersion measure from our own galaxy. This brief example shows the power and potential use of FRBs as cosmological probes. There is immense and untapped potential for great discoveries with FRBs.

\subsubsection{The Forward Physics Facility at the High-Luminosity LHC}

Signals from the Universe could play significant role in the development of particle physics~\cite{Arkani-Hamed:2015bza}. Likewise, collider experiments offer a unique environment in which to search for the direct production of DM and DE particles, since they are sensitive to a variety of signatures and therefore to a wider array of possible DM and DE interactions with matter. Like other signals of new physics, DM and DE (if accessible at small scales) could manifest itself in high-energy particle collisions either through direct production or via modifications of electroweak observables induced by virtual DM/DE particles.

Future data to be collected at the CERN's LHC can  make accurate microscopic determinations of the cross sections that are important for DM detection. On the one hand, these determinations could give evidence that the observed particle does in fact make up the DM. On the other hand, they could be useful for astrophysics in estimating the local density of DM and in mapping its distribution in the Galaxy.

The newly proposed LHC Forward Physics Facility (FPF), planned to operate near the ATLAS interaction point, offers a complementary new terrain to test some of the many proposed models to address the $H_0$ and $S_8$ tensions~\cite{Anchordoqui:2021ghd,Feng:2022inv}. The FPF experiments to be operating during the LHC high-luminosity era will attain very large statistics to search for light and very weakly-interacting particles. The existence of such a light dark sector is predicted by many BSM models attempting to solve some of the biggest puzzles in fundamental physics, such as the nature of DM, the origin of neutrino masses, and the imbalance between matter and antimatter in the present-day Universe. New physics searches at the FPF would thus advance our understanding of the Universe on both the smallest and the largest distance scales.

\subsection{Data Analysis Challenges}

In recent decades, cosmology has established itself in the era of precision thanks to an interplay between theory, observation and statistics, where the latter has functioned as a glue for the first two, including highly refined numerical simulations. A large number of statistical methods have been developed to find a suitable way that can lead to answers to the most fundamental problems in this field. In this way, Bayesian methods have been widely used and they have been very successful, for example, in determining the main cosmological parameters that define the standard model of cosmology. However, the large amount of data that are extracted from current observational projects and its continuous increase in quantity and complexity, make the expected scientific return will be limited by the efficiency and sophistication of our statistical inference tools, which, if used improperly, can lead to biased or completely wrong conclusions~\cite{Trotta:2017wnx}.

The present section will review some of the most recent and prominent developments in statistical techniques, focusing on the determination of cosmological parameters and specifically on the most relevant tensions, such as $H_0$ and $S_8$, not pretending to give a concrete solution to this "hot topic", but trying to point to an alternative or independent analysis, given certain sets of astrophysical and cosmological data set that are at hand. The topics we have chosen cover only a small part of the current amount of statistical methods that we have at our disposal and are part of our opinion about what we believe is most relevant for the community of astrophysicists and cosmologists, at least for the next decade.

\subsubsection{The Inverse Problem and Inference in Cosmology}

The inverse problem in science is the process of estimating the causal factors that produced them from a set of experimental observations. In cosmology, given a set of observational data (SNe, CMB, BAO, LSS, among others), it is possible to calculate the value of the different parameters that characterize a specific model (i.e.\ the six parameters of $\Lambda$CDM),\footnote{This base model can be treated in two groups: the \textit{late-time} group formed by $\Omega_c$, $\Omega_b$, $\tau$ and $\Theta_{MC}$, which trace the dynamic evolution of the background and the \textit{primordial} given by $A_s$ and $n_s$, that describe the initial conditions of the perturbations produced by quantum fluctuation during inflation.} and in this way extrapolate, for example to the past, to find the initial conditions that gave rise to the Universe that we can perceive in the present and that in some measure reproduces what we can observe today.\footnote{In general, this methodology is used in numerical N-body simulations, i.e., given the value of a set of parameters, it is possible to run a simulation from the past (high redshifts) and estimate how similar it is to what we can observe at present, at least statistically, for example, through the power spectrum of matter.} So the situation is "given a data set \textbf{x}, what is the range of physical parameters $\boldsymbol{\theta}$ most probable?", and the answer to this question has traditionally been Bayesian statistics, determining the posterior probability distribution function (pdf) $p(\boldsymbol{\theta}\mid x)$, that encapsulates the final state of information about the parameters $\boldsymbol{\theta}$ starting from the prior $p(\boldsymbol{\theta})$ and including the full knowledge of the data \textbf{x} through likelihood and so on, the posterior is obtained via Bayes’ theorem, methodology generally known as Bayesian inference. However, the combination of CMB and LSS through this methodology demands the calculation of the theoretical power spectrum, obtained from Einstein-Boltzmann Solvers such as CLASS~\cite{Blas:2011rf},\footnote{\url{http://class-code.net.}}  and CAMB,\footnote{\url{https://camb.readthedocs.io/en/latest/}}, and the use of Markov Chain Monte Carlo (MCMC) and/or Nested sampling algorithms to sample the posterior distribution and, in this way, the observables are obtained computationally. Afterward, a comparison between the predictions at different parameter space points with the available data is made and based on the likelihood, a best-fit of the parameters is determined. In general, this process is computationally expensive for theoretical models that include many parameters or contain "slow parameters",\footnote{Most of them are late-time and delay the calculation of the power spectrum.} since the Einstein-Boltzmann Solver is executed at each step in the parameter space.

\subsubsection{Current Developments}

A promising approach for investigation of the cosmological parameters is to consider independent analysis of the physical model. In principle, this can be done via cosmographic approach~\cite{Kristian:1965sz,1985PhR...124..315E,Heinesen:2020bej,Heinesen:2021qnl}, which consists of performing a series expansion of a cosmological observable around $z = 0$, and then using the data to constrain the kinematic parameters. Such a procedure works well for lower values of $z$, but can be problematic at higher values of redshift; though see~\cite{Cattoen:2007sk,Hu:2022udt} for strategies of analysing high redshift data with cosmographic approaches.  Another possibility is to use the technique called \textit{machine learning} (ML), which in addition, can help us with the automated analysis of the great flood of data that can be produced in some observational project. In general, machine learning can be understood as a set of methods that automatically detect patterns in the data and thus use the discovered patterns to predict specific characteristics of interest for our analysis. ML can be supervised or unsupervised: in the latter case we can find problems of clustering, dimensionality reduction, model selection and processing, among others. Regarding the first class (supervised), we can find classification or regression problems, of which only the last ones are of interest to us for the present work. Data sets such as CMB, BAO, SNe, CC, H0LiCOW, among many others, are perfect candidates to be used in supervised regression problems and thus of great importance to the community of astrophysicists and cosmologists. ML has been widely used in cosmology, both for regression and classification problems, in order to estimate the main cosmological parameters of the standard model and beyond~\cite{Arjona:2019fwb, Munoz:2020gok, Arjona:2020kco}, detect gravitational lens systems and detect low-mass binaries on X-rays~\cite{refId0,Pattnaik:2020xvx}. In this subsection, we will deal with regression problems with ML (mainly focused on Gaussian Processes - GP, and Genetic Algorithms - GA) and deep learning (DL), being a subset of ML in which multilayered neural networks learn from the vast amount of data and what we will deal with in the BNNs paragraph. We will focus on the estimation of cosmological parameters, with special emphasis on $H_0$ and $S_8$. The most prevalent techniques are then: 

\begin{itemize}

\item Gaussian Processes (GPs): An interesting and robust alternative can be to consider a Gaussian process (GP) to reconstruct cosmological parameters in a physically model-independent way. The GP approach is a generic method of supervised learning (tasks to be learned and/or data training in GP terminology), which is implemented in regression problems and probabilistic classification. A GP is essentially a generalisation of the simple Gaussian distribution to the probability distributions of a function into the range of independent variables. In principle, this can be any stochastic process, however, it is much simpler in a Gaussian scenario and it is also more common, specifically for regression processes. The GP also provides a model independent smoothing method that can further reconstruct derivatives from data. In this sense, the GP is a non-parametric strategy because it does not depend on a set of free parameters of the particular model to be constrained, although it depends on the choice of the covariance function. The GP method has been used to reconstruct the dynamics of the DE, modified gravity, cosmic curvature, estimates of Hubble constant, and other perspectives in cosmology by several  authors~\cite{Seikel:2012uu,Shafieloo:2012ht,Seikel:2013fda,Zhang:2016tto,Busti:2014dua,Sahni:2014ooa,LHuillier:2016mtc,LHuillier:2017ani,Belgacem:2019zzu,Pinho:2018unz,Cai:2017yww,Haridasu:2018gqm,Zhang:2018gjb,Wang:2017jdm,Shafieloo:2018gin,Bengaly:2019oxx,Bengaly:2019ibu,LHuillier:2019imn,Sharma:2020unh,Nunes:2020hzy,Briffa:2020qli,Bahamonde:2021gfp,Li:2019nux,Simon:2004tf}. Estimates of $H_0$ can be found in Refs.~\cite{Liao:2019qoc,Gomez-Valent:2018hwc,OColgain:2021pyh,Krishnan:2020vaf} and a recent estimate yields a value of $H_0 = (73.78 \pm 0.84){\rm\,km\,s^{-1}\,Mpc^{-1}}$ constrained at 1.1\% precision at $1\sigma$ CL~\cite{Bonilla:2020wbn}, using data from SNe, BAO, CC and H0LiCOW, developing a joint analysis, which was also used in Ref.~\cite{Bonilla:2021dql} to estimate the interaction strength in a model of interaction between dark matter and dark energy (IDE), an analysis that also includes simulated data from GWs like standard sirens. Regarding $\sigma_8$ and/or $S_8$ measurements, authors report $S_8=0.707\pm 0.085$, using selected Red Shift Space Distortion (RSD) data set in the form of $f\sigma_8(z)$~\cite{Benisty:2020kdt}. Recently, Ref.~\cite{Avila:2022xad} found the following estimates: $\sigma_8(z=0)=0.766 \pm 0.116$, $S_8 (z=0)=0.732\pm 0.115$ and $\gamma (z=0) = 0.465\pm 0.140$ (growth index), using select independent data measurements of the growth rate $f(z)$ and of $[f\sigma_8](z)$ according to criteria of non-correlated data. 

\item An alternative to Gaussian process is the iterative smoothing method introduced by~\cite{Shafieloo:2005nd,Shafieloo:2007cs,Shafieloo:2018gin}. The idea is to start with an initial guess for the function to reconstruct, and convolve the residuals with a smoothing kernel. The resulting smooth function is then subtracted from the data at the next iteration to form new residuals, and reiterate for a given number of iteration. One can then vary the initial guess, and the reconstructions all converge to the same solution, which is thus independent from the initial guess and purely driven by the data. This method has been used to reconstruct the expansion history~\cite{Shafieloo:2005nd,Shafieloo:2007cs,LHuillier:2016mtc,Shafieloo:2018gin,Koo:2020wro,Koo:2021suo}, test the curvature, to perform model-selection and parameter estimation~\cite{LHuillier:2016mtc,Shafieloo:2018gin,Koo:2021suo}. Varying the initial guess makes the method less sensitive to the choice of the mean function in GP regression. However, the interpretation of the uncertainties is less straightforward than in GP regression. 

\item Genetic Algorithms (GA): The GA is an ML approach that can be used for unsupervised regression of data. Specifically, it allows for non-parametric reconstructions of data, using analytic functions of one or more variables. The GA is very loosely based on the idea of natural evolution, which is implemented via the genetic operations of crossover and mutation. In a nutshell, an initial set of randomly chosen functions evolves over time under the evolutionary pressure of the two aforementioned operators, namely mutation, i.e.\ a random change of the function, and crossover, i.e.\ the partial merging and swapping of parts of two or more functions of the population. This procedure is then iterated hundreds of times and with different random seeds, so as to explore the functional space and to ensure convergence. As the GA is a stochastic ML approach aiming to emulate evolution, the probability that a population of functions will obtain offspring, i.e.\ the next generation of functions, is taken to be proportional to its fitness to the data. The latter is usually measured by a $\chi^2$ statistic, which determines how well every function agrees with the data. Then, the probability to have offspring and the fitness of each individual is proportional to the likelihood causing an "evolutionary" pressure that favors the best-fitting functions in every population, hence directing the fit towards the minimum in a few generations. The GA have been extensively used in cosmology for several reconstructions of a plethora of different data, see for example Refs.~\cite{Bogdanos:2009ib,Nesseris:2010ep, Nesseris:2012tt,Nesseris:2013bia,Sapone:2014nna,Arjona:2020doi,Arjona:2020kco,Arjona:2019fwb,Arjona:2021hmg,Arjona:2020skf,Arjona:2020axn,Arjona:2021zac,Arjona:2021mzf} and also to perform forecasts for the {\it Euclid} survey, see Refs.~\cite{EUCLID:2020syl, Euclid:2021cfn, Euclid:2021frk}. Similar applications of the GA have also been performed in related areas such as particle physics~\cite{Abel:2018ekz,Allanach:2004my,Akrami:2009hp}, astronomy and astrophysics~\cite{2001A&A...379..115W,Rajpaul:2012wu,Ho:2019zap}. 

\item{Bayesian Neural Networks (BNNs):
Bayesian evidence method remains the preferred method compared with information criteria and GP in the literature. A full Bayesian inference for model selection –in the case we have a landscape in where we can discriminate a pivot model from a hypothesis– is computationally expensive and often suffers from multi-modal posteriors and parameter degeneracies. This latter issue leads to a large time consumption to obtain the final best fit for such parameters. As the study of LSS of the universe indicates, all our knowledge relies on state-of-the-art cosmological simulations to address several questions by constraining the cosmological parameters at hand using Bayesian techniques. Moreover, due to the computational complexity of these simulations, some studies look to remain computationally infeasible for the foreseeable future. It is at this point where computational techniques as machine learning (ML) can have some important uses, even for trying to understand our universe. The idea behind the ML is based in consider a Bayesian Neural Network (BNN) with a complex combination of neurons organised in nested layers. And every layer of a neural network thus transforms one input vector –or tensor depending the dimension– to another through a differentiable function, e.g. the BNN can learn the distribution of the distance moduli in the dark energy models, then feed the astrophysical surveys to the network to reconstruct the dark energy model and then discriminate between models.
With BNNs we can design numerical architecture cases in order to train data instead of simulate them. These kind of cases have been used in cosmology to train significant amount of information on several cosmological parameters without assuming a weight function as in comparison with GP regressions. Some interesting architectures have been treated in~\cite{Escamilla-Rivera:2019hqt,Escamilla-Rivera:2021vyw} in order to discuss dark energy models through parametrised equation of state. Also, it has been discussed the possibility to combine BNNs with MCMC methods to study anisotropies in the CMB data~\cite{Hortua:2019ryu}. As an extension, in~\cite{Hortua:2020qjr} BNNs architectures were used to estimate parameters from the 21 cm signal and in~\cite{Hortua:2020ljv} the method was extended to reconstruct the history of reionization.}

\end{itemize}

We leave the reader free to expand their knowledge of the previous topics, specifically in applications such as deep learning (DL)~\cite{Escamilla-Rivera:2020fxq},  
or iterative smoothing~\cite{Shafieloo:2005nd,Shafieloo:2007cs,Shafieloo:2009hi,LHuillier:2016mtc,Shafieloo:2018gin,LHuillier:2017ani,Koo:2021suo}, which are totally related to the present discussion. However, Bayesian inference will not stop being used in the coming decades since it is the most suitable for treating data in astrophysics and cosmology, whether it is used in combination with maximum likelihood or minimum chi-square (which is more traditional) or with the most recent techniques of a machine learning like the ones presented here. We can only finish by giving suggestions and recommendations for its correct use in future projects, pointing to some persistent difficulties.

\subsubsection{Checking the Conventional $\Lambda$CDM Assumptions for Extended Cosmology Analysis}

In response to the Hubble tension between late and early Universe measurements, a growing number of extended cosmological models beyond $\Lambda$CDM model have been proposed and examined (see Section~\ref{sec:WG-BothSolutions}). To obtain strong constraints on the model parameters and to explore their possible remedy to the tensions, many of these tests are done by combining published measurements from different surveys. However, the data analysis of current high-precision cosmological and astrophysical probes is complicated and uses many assumptions based on $\Lambda$CDM in the analysis pipeline to compress the large amount of data into limited degrees of freedom. Some of these assumptions are as obvious as the $\Lambda$CDM background geometry, while others are hidden in the details. The misuse of measured data based on $\Lambda$CDM assumptions on the beyond-$\Lambda$CDM models where these assumptions break down could result in misleading conclusions~\cite{Barreira:2016ovx,Hill:2020osr,Taruya:2013quf,Macpherson:2021gbh,Koksbang:2021qqc}. To combat this, theorists should be cautious when analyzing their models, and observers should be clear about the specific assumptions in published key paper results, if possible. 

Strictly speaking, there is no clear-cut boundary between the statement of "doing your theoretical model predictions right" and the point of "alleviating the $\Lambda$CDM assumptions in the surveys". Here, the focus on the latter aims to elucidate the relatively-easy-to-check but profound technical details that are risky to be overlooked by the working groups using the publicly published and deeply post-processed cosmological/astrophysical survey results on their extended models.

\paragraph{Baryonic Acoustic Oscillation.}
Recently the impact of fiducial cosmology assumptions on BAO measurements are discussed in several papers~\cite{Heinesen:2018hnh,Heinesen:2019phg,Carter:2019ulk, Sherwin:2018wbu}. A standard methodology for the recent BAO analysis is to reconstruct the galaxy power spectrum in a fiducial cosmology, then extract the sharpened BAO features from the reconstructed galaxy over-densities~\cite{Gil-Marin:2015nqa}. In this way, the BAO feature that has been smeared by the bulk flow of the astrophysical objects is recovered. The paper by Carter et al.~\cite{Carter:2019ulk} discussed the effect of fiducial cosmology assumption in the reconstruction process, and summarized three approximations that extended models should obey to avoid bias when using the fiducial BAO measurements. Among them, the first thing that we should pay attention to is the re-scaling of the redshift-distance relationship, at least on the scales relevant to the BAO survey. If the redshift-distance relationship rescaling is more complicated than a linear factor multiplication in the redshift range of the galaxy samples, the configuration space distortion introduced by placing the galaxy at wrong distance is likely be unable to be captured by the Alcock–Paczynski (AP) parameters. The other approximations are mainly related to whether the power spectrum template used in the BAO information compression is inclusive enough to mitigate the cosmology dependence. They need further detailed analytic or simulation validation per model. The reassuring news is that in most of the cases studied so far, BAO measurement bias is below current precision for models. Both Ref.~\cite{Sherwin:2018wbu,Carter:2019ulk} warned that this might be no longer true in higher precision next generation measurements due to fiducial cosmology assumption, even just in $\Lambda$CDM with varying cosmological parameters.

The inverse distance ladder approach of the BAO measurements is also worth mentioning. In this method, the sound horizon at drag scale $r_d$ is taken as a standard ruler, thus the BAO measured angle corresponding to $r_d$ at given redshift calibrates the distance to the observation. In this way we can get the constraint on $H_0$. Some literature cites this method as a "late Universe" or "cosmology independent" measurement of $H_0$, which is not exactly true. There is no way to constrain $H_0$ without any early Universe physics assumption on $r_d$, when the only information we know is $H_0\times r_d$ (maybe plus $\Omega_b h^2$, which still does not specify the sound speed completely). This is an indicator that $\Lambda$CDM or fixed early Universe is an essential presumption in inverse distance ladder $H_0$ constraints. Actually, in several strictly classical distance ladder BAO analysis~\cite{Aylor:2018drw,Wojtak:2019tkz} without any cosmological assumptions (not inverse, anchored to the late Universe supernovae or strong lensing), $r_d$ is found to be systematically smaller than the {\it Planck} CMB result. Given these, it is not surprising that the inverse distance ladder BAO constraint on $H_0$ is on the early Universe side of the current $H_0$ tension. 

\paragraph{Redshift Space Distortion.}
It has been discussed extensively for some relatively more established beyond-$\Lambda$CDM model, like $f(R)$ modified gravity, that the compressed RSD survey results typically reported in the form of $f\sigma_8(z)$ values might not be applicable to extended model constraints~\cite{Taruya:2013quf,Barreira:2016ovx,Bose:2017myh}. By definition, the growth factor and the fluctuation of matter field are scale insensitive in standard $\Lambda$CDM, 
\begin{align}
    f(z) & \equiv \frac{{\rm d}\ln D(z)}{{\rm d}\ln a}\,,\\
    \sigma^2_8(z) & \equiv \int_0^{\infty} \Delta^2(k,z) \left(\frac{3j_1(kR)}{kR}\right)^2 {\rm d}\ln k\,,
\end{align}
where $R = 8 h^{-1}\,$Mpc. The estimate of $f\sigma_8(z)$ is usually based on a fixed fiducial cosmology template of the redshift-space power spectrum~\cite{Taruya:2010mx, Sanchez:2012sg}, because we expect this quantity to be only dependent on the amplitude of the measured RSD. When the fiducial template cannot mitigate the scale-dependent features in the clustering, $f\sigma_8$ measurement could be biased. The robustness of RSD estimates of $f\sigma_8$ against certain modified gravity simulations has been tested by Refs.~\cite{Taruya:2013quf, Barreira:2016ovx, Bose:2017myh}. Specifically, these studies found that the bias tends to be non-negligible for the models with scale dependent growth rate deviating from standard $\Lambda$CDM. This might be due to the fact that the multiplicative template parameters cannot absorb such scale dependent deviations from the fiducial cosmology. This finding is likely to generalize to other beyond-$\Lambda$CDM models that introduce a  scale-dependent growth rate in the scale range sensible to the RSD measurements ( $0.01 h {\rm\,Mpc^{-1}} \lesssim k \lesssim 0.2 h{\rm\,Mpc^{-1}}$), and the RSD compression into $f\sigma_8$ constraints should be handled carefully.

\paragraph{Weak Lensing.}
In weak lensing, the main limitation of a fiducial cosmology assumption usually comes from the modeling of nonlinear matter power spectrum. Most halo model calculations used in fiducial pipelines for weak lensing analyses assume the same non-linear regime clustering physics as GR and $\Lambda$CDM. When testing a beyond-$\Lambda$CDM (and GR) model, the nonlinear regime should either be removed from the analysis or the non-linear modeling should be modified and validated -- e.g.\ by using N-body simulations or perturbation theory calculations~\cite{Cataneo:2018cic,Giblin:2019iit}. 

Another place where beyond-$\Lambda$CDM modeling is needed is in the default analysis pipeline of the weak lensing surveys, for example the one used in DES~\cite{DES:2017tss}, where the lensing kernel takes the form:
\begin{equation}
q_{\kappa} (\chi) = \frac{3H_0^2 \Omega_m}{2 c^2} \frac{\chi}{a(\chi)}\int^{\chi_h}_{\chi} {\rm d} \chi' \frac{n_{\kappa}(z(\chi'){\rm d}z/{\rm d}\chi'}{\bar{n}_{\kappa}}\frac{\chi'-\chi}{\chi'}.
\end{equation}

This is based on the late Universe where: 1. the matter is predominant contributor to the gravitational potential, and 2. gravity is described by standard General Relativity. The lensing kernel above thus relates the matter overdensity to the potential using the Poisson equation:
\begin{equation}
    \Phi = \frac{4 \pi G_N \rho_m a^2 \delta_m}{k^2}.
\end{equation}
However, when the two conditions above break down in extended cosmologies, for example in certain modified gravity theories, or when extra relativistic species in the late Universe are added, one should use the Weyl potential power spectrum and the corresponding lensing kernel to calculate the projected 2D correlations. This approach does not presume the Poisson equation, namely the matter power spectrum and the Weyl potential power spectrum are not necessarily linearly related by the factor $\left(\frac{4 \pi G_N \rho_m a^2}{k^2}\right)^2$. There are multiple ways to realize this adjustment to the analysis pipeline, the simplest being correcting the matter power spectrum used in $\Lambda$CDM pipeline by the ratio:
\begin{equation}
    R_{\rm Weyl}(k,z) = \left(\frac{k^2(1+z)^2}{4 \pi G_N \rho_m }\right)^2 \frac{P_{\rm Weyl}(k,z)}{P_m(k,z)},
\end{equation}
where $P_{\rm Weyl}(k,z)$ is the 3D k-space spectrum for Weyl potential perturbation for general gravitational potentials.

\paragraph{CMB.} The CMB is known for its "cleanness" in terms of cosmological theory predictions, as it happens in the fairly early era of the Universe. Most of the problems can be avoided as long as the Einstein and Boltzmann equations are carefully modified in the Boltzmann solving code. When using a CMB lensing likelihood and/or high-$\ell$ lensed $TT, EE, TE$ likelihoods, one should note that the non-linear matter power spectrum inducing the lensing effect is expected to be different from the $\Lambda$CDM modeling. The non-linear matter power spectrum or the Weyl potential power spectrum used for CMB lensing calculations should be treated as is described above in the 'Weak lensing' paragraph. Alternatively, we can limit our analysis to linear scales, for example the conservative multipole range ($L \leq 400$) of the {\it Planck} 2018 CMB lensing likelihood~\cite{Planck:2018lbu}. \\

\subsubsection{N-body Simulations Beyond $\Lambda$CDM }
 
N-body simulations have been a pillar of cosmology, allowing to predict structure formation in the non-linear regime. 
While the current trend is to simulate ever larger volumes at ever larger resolutions, to accompany the next generation cosmological surveys, in the last few years new codes have also appeared aiming to go beyond $\Lambda$CDM.
\begin{itemize}
	\item Dark energy: 
	Several groups have run suites of simulations of non-$\Lambda$ dark energy models, such as $w$CDM with the Quijote simulations~\cite{Villaescusa-Navarro:2019bje}, and CPL with the Mira-Titan simulations~\cite{Heitmann:2015xma}.
	\item Modified gravity: 
Modified gravity N-body codes can be divided into two main categories: Newtonian codes with modified Poisson equation~\cite{Puchwein:2013lza,Li:2011vk,Llinares:2013jza}, and relativistic codes~\cite{Adamek:2016zes,Hassani:2019lmy,Hassani:2019wed}.
See Ref.~\cite{Winther:2015wla} for a comparison of Modified Gravity N-body codes, and~\cite{Llinares:2018maz} for a recent review. Such simulations have been used for studying voids, clustering and  alignment~\cite{LHuillier:2017pdi}.
	\item Initial conditions: While simulations of primordial non-gaussianities have been carried out since the 1990s~\cite{Messina:1990ze}, a more recent development is the simulation of primordial features, i.e., deviation from power law, in the power spectrum~\cite{LHuillier:2017lgm,Beutler:2019ojk, Ballardini:2019tuc}. 
\end{itemize}

\subsubsection{Optimal Practices and Recommendations}

A consistent statistical methodology requires following a principle-based inference approach, using an incorrect or oversimplified sampling distribution, ignoring significant observational effects that affect data (for example, selection effects) or unrepresentative training data or simulations in machine learning. In all these circumstances and many others, best practices help achieve robust results, which are more likely to survive scrutiny and be corroborated by more extensive and more accurate future data. Astrophysics data often suffer from selection effects or bias. In statistical terms, the observed sample is non-representative of the total population of observed data whose global characteristics one might wish to infer (e.g., cosmological parameters). The latter translates into poor generalization onto the test from a non-representative training set in supervised learning as in GPs and BNNs. In machine learning, "Covariate shift" is a type of model drift that occurs when independent variables' distribution changes between the training environment and the live environment. Ignoring covariate shift leads to incorrect inferences on physical properties and affects the robustness of the results (e.g., for population studies for gravitational lensing classification, photon sources/x-ray, or GWs). Possible solutions include correcting selection effects using simulations, data increment, and unsupervised learning. Extensive use of data compression and summary statistics is common and persistent in cosmology and particle physics. Cosmology, in particular, typically analyses low-order correlation functions because often these carry most of the information, and model predictions are inestimable on small scales, where both non-linear gravity and baryonic effects become essential. Computational advances and a simulation-based approach construct sophisticated nonperturbative summary statistics both by hand and via neural compression methodology. Similar methods, along with more accurate physical modeling, will be essential in the coming years to extract complete scientific information from observational data sets whose volumes and complexity will increase dramatically.

Simultaneous modelling of cosmological data sets across multiple probes will be of crucial importance in both extracting the most precision possible from the data, and in ensuring accuracy is not lost to un-modelled systematics. Where different types of data are combined or compared, it is crucial that simulations used in their calibration and in calculating their likelihoods share the same underlying physical models. For example, combined modelling of radio and optical observations could assist both shear measurement~\citep[][]{Harrison:2016stv} and galaxy redshift measurements~\citep[][]{Alonso:2017dgh,Cunnington:2018zxg} and lead to improved and more robust measurements of the $S_8$ tension. New astrophysical discoveries such as GWs and Fast Radio Bursts (FRBs) will also provide new information for cosmology, with gains that will be maximised by modelling their sources alongside the large scale structures measured by other experiments. Meeting this challenge will require co-ordination and openness between different experiments and communities, the construction of modelling and inference pipelines that work simultaneously for multiple data types~\citep[see][for an early attempt at this]{skypy_collaboration_2020_4071945}, and infrastructure for sharing and accessing extremely large data sets.

\subsection{The Road Ahead: Beyond $\Lambda$CDM}

In view of the accumulation the $\Lambda$CDM tensions discussed in the previous sections it is becoming apparent that the need for a new standard cosmological model has been increasing during the last few years. The present time has similarities with the late nineties when it was becoming apparent that the standard model of that time (flat sCDM based on the Einstein-de Sitter model) had been accumulating a range of inconsistencies with data including indications for excess clustering on large scales, hints for $\Omega_{m}<1$, 
age of the Universe tension, velocity flows etc. It is interesting to recall the conditions that lead to the birth of the current standard cosmological model ($\Lambda$CDM) in order to identify similarities and use them as a possible guide for the construction the new standard model that may be approaching. These conditions can be summarized as follows
\begin{enumerate}
\item
The emergence of the tensions of sCDM. These tensions include the observations for excess of clustering on large scales $>10$ Mpc~\cite{Davis:1992a}, observations for low velocities on small scales ($1$ Mpc)~\cite{Davis:1985rj}, the ages of globular clusters which appeared to be larger than the predicted age of the Universe in the context of sCDM~\cite{Janes:1983jx} and indications for missing matter density $(\Omega_m<0.3)$~\cite{Bartlett:1992by,Davis:1992a}. 
\item
The use of simple new parameters to address these tensions. This lead to simple extensions of sCDM including the mixed dark matter (mixture of hot and cold dark matter), $\Lambda$CDM, titled primordial power spectrum, open Universe, loitering Universe etc. 
\item
Theoretical motivation and predictive power played an important role in the selection of specific parametrizations including $\Lambda$CDM. Mixed dark matter and $\Lambda$CDM were the two models based on simple and generic theoretically motivated mechanisms and as such they had an advantage over the other approaches which were purely phenomenological and required tuned potentials and/or initial conditions (e.g.\ the open or the loitering Universe models). $\Lambda$CDM had an additional advantage that it could make a smoking gun unique and testable prediction: accelerating expansion at low redshifts. This was the crucial feature of the $\Lambda$CDM model that when combined with the SNIa observations in and after 1998 established it as the new standard cosmological model.\footnote{The inertia of the sCDM was however not easy to overcome and it is well known that the first study of the Hubble diagram using SNIa did not favor $\Lambda$CDM~\cite{SupernovaCosmologyProject:1996grv} but sCDM albeit with large uncertainties.}
\end{enumerate}

In the context of an analogy with the above developments for the emergence of $\Lambda$CDM as a new standard cosmological model, it could be stated that we are currently at stage 2 in the process of the emergence of the new standard model beyond $\Lambda$CDM. Well defined tensions of $\Lambda$CDM with data have developed well beyond the level of the tensions that gave rise to it. These tensions include the $H_0$ tension ($5 \sigma$~\cite{Riess:2021jrx}), statistical anisotropy of the CMB temperature on large angles (multiple $>3\sigma$ anomalies), cosmic dipoles-anisotropies ($4-5\sigma$~\cite{Secrest:2020has,King:2012id}), growth $S_8$ ($2-3 \sigma$~\cite{Asgari:2019fkq,DES:2020ahh,Macaulay:2013swa,Skara:2019usd,Kazantzidis:2019dvk,Bull:2015stt,Kazantzidis:2018rnb,Nesseris:2017vor,Joudaki:2017zdt}) and others. 
Simple mechanisms and parametrizations have been proposed to address specific of the existing tensions.

In view of existing current status of events, the following question emerges: {\it What could be the features of the next standard cosmological model?}. Based also on the experience of the previous change of standard cosmological model that occurred in the late nineties, the following features of the next standard cosmological model could be anticipated:
\begin{itemize}
\item 
Consistency with main $\Lambda$CDM tensions ($H_0$, $S_8$, cosmic dipoles)
\item
Existence of simple parametrization (one or two more parameters) in the context of a specific physically, motivated mechanism.
\item
Simple and generic theoretical framework that justifies the parametrization.
\end{itemize}

Specific generic phenomenological parameters that may be considered in the above context include
\begin{itemize}
\item
The dark energy equation of state parameter $w_{DE}$.
\item
The effective parameters $\mu$ and $\Sigma$ connected with the strength of gravity with respect to matter and light; see Eq.~(\ref{mudef}). 
\item
The dipolar anisotropy parameter defined in Eq.~\eqref{eq:dipole} and the anisotropy parameter $S_{1/2}$ defined in Eq.~\eqref{eqn:Shalf}.
\item
The dimensionless fine structure constant connected with the strength of the electromagnetic force.
\end{itemize}

These (and other similar) effective parameters however have the following problems which make it difficult for them to replace $\Lambda$CDM soon:
\begin{itemize}
\item
They are purely phenomenological, usually motivated by fine tuned effective scalar field potentials, and a number of these scalar models can be shown to exacerbate tensions~\cite{Banerjee:2020xcn, Heisenberg:2022lob, Heisenberg:2022gqk, Lee:2022cyh}.
\item
In most cases they lack generic detectable smoking gun signatures in upcoming cosmological data as was the case for the accelerating expansion prediction of $\Lambda$CDM.
\item
These parameters appear to be disconnected and it seems unlikely that a single one of them will be able to simultaneously resolve more than one of the $\Lambda$CDM tensions. 
\end{itemize} 

The high accuracy and precision of the current cosmological data makes it probable that some type of an abrupt transition in space and/or in time either before recombination ($z>1100$)~\cite{Niedermann:2019olb,Niedermann:2020dwg,Niedermann:2020qbw} or after the end of the Hubble flow ($z<0.01$)~\cite{Marra:2021fvf,Alestas:2021nmi,Perivolaropoulos:2021bds}. Such an event would have the following advantages:
\begin{enumerate}
\item
It would leave intact the expansion history of the Universe from the time of recombination until the end of the Hubble flow, which is extremely well constrained by current data.
\item
If the transition is spatial (involving e.g.\ a first order phase transition) it has the potential to simultaneously address a wide range of tensions including the detection of cosmic dipoles, Hubble tension and growth tension.
\item
It has the potential for specific unique and generic signatures in cosmological and astrophysical data
\item
It may be based on a simple and generic physical mechanism: {\it The decay of the false vacuum} which has been well studied during the past few decades.
\end{enumerate}

In view of the above discussion, the strategic steps for the road ahead towards a possible discovery of a new standard model for cosmology could be the following:
\begin{itemize}
\item
The tuning of current missions towards verification or rejection of the existing non-standard signals and tensions of $\Lambda$CDM.
\item
The use of model-independent, data driven reconstructions of cosmological observables such as $(H(z),\mu(z),\Sigma(z),\dots)$ to identify simple forms for their possible evolution. 
\item
Use these forms to identify favored parametrizations of $H(z,w(z),r)$, $\mu(z,r)$, $\Sigma(z,r)$, $\alpha(z,r)$ and dipole anisotropy parameters assuming that at least some of the non-standard signals are physical.
\item
Identify theoretical models (field Lagrangians) that are consistent with these parametrizations.  Interestingly, for example only a small subset of modified gravity models is consistent with the weak gravity + $\Lambda$CDM background~\cite{Wittner:2020yfc,Pizzuti:2019wte,Gannouji:2020ylf,Gannouji:2018ncm} suggested in the context of the $S_8$ tension.
\end{itemize}
An exciting new era is probably approaching for cosmology which may soon lead to new discoveries in fundamental physics.


\begin{table*}[ht]
\begin{center}
\scalebox{0.9}{
\begin{tabular}{|l|l|l|}
\hline
Science & Facilities \\
\hline
21 cm & BINGO, CHIME, GBT, HIRAX, MeerKAT, OWFA, PUMA , SKAO, uGMRT\\
BAO and RSD & 4MOST, BINGO, CHIME, COMAP, DESI, Euclid, HIRAX, PFS, {\it Roman}, {\it Rubin}, SKAO, SPHEREx \\
CC & ATLAS, Euclid, SPHEREx\\
CMB & ACT, BICEP/Keck, CMB-HD, CMB-S4, LiteBIRD, SO, SPT \\
Distance ladder & ELTs, Gaia , GBT, JWST, LIGO, {\it Roman}, {\it Rubin}, VLA, VLBA \\
FRB & CHIME \\
GW & Cosmic Explorer, DECIGO , ET, LGWA, LIGO/Virgo/KAGRA/LIGO-India, LISA, Taiji, TianQin\\
Quasars & GRAVITY+ \\
Redshift drift & ANDES, ELTs, SKAO\\
SDs & SuperPIXIE \\
SNe & {\it Rubin}, {\it Roman}, YSE, ZTF  \\
Time Delay cosmography & Euclid, Pan-STARRS, {\it Roman}, {\it Rubin}, SKAO, ZTF\\
Time Lag cosmography & {\it Rubin} \\
Varying fundamental constant & ANDES, ELTs, ESPRESSO\\
WL & 4MOST, CFHT,DES, Euclid, HSC, KiDS,  Pan-STARRS , {\it Roman}, {\it Rubin}, SKAO, UNIONS\\
\hline
\end{tabular}}
\end{center}
\caption{List of Experiments per science topic. The timeline is instead reported in Table~\ref{tab:acronyms}.}
\label{timeline}
\end{table*}


\section{Conclusions} 

The present tensions and discrepancies among different measurements, in particular the $H_0$ tension as the most significant one, offer crucial insights in our understanding of the Universe. For example, the standard distance ladder result has many steps in common with the accelerating Universe discovery (which gave cosmology the evidence for DE). So, whatever the definite finding may be, whether about stars and their evolution, or DE, this is going to have far reaching consequences.

Solving the $H_0$ tension is very much an ongoing enterprise. The resolution of this conundrum will likely require a coordinated effort from the side of theory and interpretation (providing crucial tests of the exotic cosmologies), and data analysis and observation (expected to improve methods and disentangle systematics). This agenda will flourish in the next decade with future CMB and distance ladder experiments, that are expected to reach an uncertainty of, respectively, $\sim 0.15\%$ and $\sim1\%$ in the $H_0$ estimates, see Table~\ref{timeline} and Section~\ref{sec:WG-perspectives}. In other words, the next decade will test the $\Lambda$CDM model through next-generation experiments that will usher in a new era of cosmology.

In the near future, moreover, we expect precise measurements of the expansion and growth history over a large range of experiments (see Table~\ref{timeline} and Section~\ref{sec:WG-perspectives}), i.e.\ using maps of the Universe obtained by the {\it Euclid} satellite, measuring the peculiar motions of galaxies using Type Ia supernovae from {\it Rubin}/LSST~\cite{Howlett:2017asw,Scolnic:2019apa}, considering RSD with DESI and 4MOST, or using voids~\cite{Hamaus:2020cbu}. An important role will be played by the SKAO telescopes performing BAO surveys and measuring weak gravitational lensing using 21~cm intensity mapping~\cite{Pourtsidou:2014pra,Santos:2015gra,Bull:2015nra}. Additional upcoming 21~cm neutral hydrogen experiments measuring the expansion history will be CHIME and HIRAX. Finally, line-intensity mapping of emission from star-forming galaxies can be used to measure BAO, reaching percent-level constraints~\cite{Karkare:2018sar,Bernal:2019gfq} with the SPHEREx satellite or the ground-based COMAP instrument. All of these efforts will either reveal a systematic cause or harden the tension to strong statistical significance, thus informing the theories mentioned above and guiding any extension/overhaul of the standard model.

Meanwhile, the CMB temperature map has a number of independent pieces of evidence, each with $p$-values less than $0.3\%$ (i.e. less likely than a $3\sigma$ fluctuation), that the $\Lambda$CDM prediction of statistical isotropy is violated (see Sec.~\ref{sec:Largeangleanomalies}). These cannot be addressed by mere changes in the background scale factor dynamics. They demand early universe or large scale departures from homogeneity and/or isotropy. A theoretical priority is exploring potential physical models, such as cosmic topology. Existing data, ongoing and future observational programs can increase our confidence that these are not mere statistical flukes by exploring conditional predictions of $\Lambda$CDM, especially those for which we would expect physical models to disagree.

In the four LoIs~\cite{DiValentino:2020vhf, DiValentino:2020zio, DiValentino:2020vvd, DiValentino:2020srs} before, and now in this White Paper, we presented a snapshot, at the beginning of the SNOWMASS process, of the concordance $\Lambda$CDM model and its connections with the experiments, and monitored the new advances in the field to come out with a clear roadmap for the coming decades. This is a cutting-edge field in the area of cosmology, with unrestrained growth over the last decade. On the experimental side, we have learned that it is really important to have multiple precise and robust measurements of the same observable, with experiments conducted blindly in regard to the expected outcome. This provides a unique opportunity to study similar physics from various points of view. While on the theory side, it is really important to have robust and testable predictions for the proposed physical models that can be probed with the data. With the synergy between these two sides, significant progress can be made to answer fundamental physics questions.


\color{black}

\section*{Acknowledgments}

\begin{enumerate}

\item[$\bullet$] Amin Aboubrahim is supported by the BMBF under contract 05P21PMCAA and by the DFG through the Research Training Network 2149 "Strong and Weak Interactions - from Hadrons to Dark Matter".

\item[$\bullet$] Adriano Agnello is supported by a Villum Experiment Grant, project number 36225.

\item[$\bullet$] \"{O}zg\"{u}r Akarsu acknowledges the support by the Turkish Academy of Sciences in the scheme of the Outstanding Young Scientist Award (T\"{U}BA-GEB\.{I}P).

\item[$\bullet$]Yashar Akrami is supported by Richard S. Morrison Fellowship and LabEx ENS-ICFP: ANR-10-LABX-0010/ANR-10-IDEX-0001-02 PSL*.

\item[$\bullet$]George Alestas is supported by the project "Dioni: Computing Infrastructure for Big-Data Processing and Analysis" (MIS No. 5047222) co-funded by European Union (ERDF) and Greece through Operational Program "Competitiveness, Entrepreneurship and Innovation", NSRF 2014-2020.

\item[$\bullet$]Luca Amendola  acknowledges support from DFG project  456622116 and  from the CAPES-DAAD bilateral project  "Data Analysis and Model Testing in the Era of Precision Cosmology".

\item[$\bullet$] Luis A. Anchordoqui and Jorge F. Soriano are supported by the 
  U.S. National Science Foundation (NSF Grant PHY-2112527).

\item[$\bullet$] Mario Ballardini acknowledges financial support from the contract ASI/ INAF for the {\it Euclid} mission n.2018-23-HH.0. 

\item[$\bullet$] Micol Benetti acknowledges the Istituto Nazionale di Fisica Nucleare (INFN), sezione di Napoli, iniziativa specifica QGSKY.

\item[$\bullet$]David Benisty acknowledges the support the supports of the Blavatnik and the Rothschild fellowships.

\item[$\bullet$] John Blakeslee is supported by NOIRLab, which is managed by the Association of Universities for Research in Astronomy (AURA) under a cooperative agreement with the US National Science Foundation.

\item[$\bullet$] Thomas Buchert has received funding from the European Research Council (ERC) under the European Union's Horizon 2020 research and innovation programme (grant agreement ERC advanced grant 740021--ARTHUS, PI: Thomas Buchert).

\item[$\bullet$] Erminia Calabrese acknowledges support from the STFC Ernest Rutherford Fellowship ST/M004856/2, STFC Consolidated Grant ST/S00033X/1 and from the European Research Council (ERC) under the European Union’s Horizon 2020 research and innovation programme (Grant agreement No. 849169).

\item[$\bullet$] Salvatore Capozziello  acknowledges the Istituto Nazionale di Fisica Nucleare (INFN), sezione di Napoli, iniziative specifiche QGSKY and MOONLIGHT2.

\item[$\bullet$]Javier de Cruz P\'erez is supported by a FPI fellowship associated to the project FPA2016-76005-C2-1-P.

\item[$\bullet$] Peter Denton acknowledges support from the US Department of Energy under Grant Contract DE-SC0012704.

\item[$\bullet$]Eleonora Di Valentino is supported by a Royal Society Dorothy Hodgkin Research Fellowship.

\item[$\bullet$]Keith R. Dienes was supported in part by the U.S.\ Department of Energy under Grant DE-FG02-13ER41976 / DE-SC0009913, and also by the U.S.\ National Science Foundation through its employee IR/D program. 

\item[$\bullet$]Celia Escamilla-Rivera is supported by DGAPA-PAPIIT UNAM Project TA100122 and acknowledges the Royal Astronomical Society as FRAS 10147 and the Cosmostatistics National Group (\href{https://www.nucleares.unam.mx/CosmoNag/index.html}{CosmoNag}) project.

\item[$\bullet$]Noemi Frusciante is  supported by Funda\c{c}\~{a}o para a Ci\^{e}ncia e a Tecnologia (FCT) through the research grants UIDB/04434/2020, UIDP/04434/2020, PTDC/FIS-OUT/29048/2017, CERN/FIS-PAR/0037/2019 and the personal FCT grant "CosmoTests -- Cosmological tests of gravity theories beyond General Relativity" with ref.\ number CEECIND/00017/2018 and the FCT project "BEYLA--BEYond LAmbda" with ref.\ number PTDC/FIS-AST/0054/2021.

\item[$\bullet$] Adri\`a G\'omez-Valent is funded by the Instituto Nazionale di Fisica Nucleare (INFN) through the project "Dark Energy and Modified Gravity Models in the light of Low-Redshift Observations" (n. 22425/2020). 

\item[$\bullet$] Asta Heinesen has received funding from the European Research Council (ERC) under the European Union's Horizon 2020 research and innovation programme (grant agreement ERC advanced grant 740021--ARTHUS, PI: Thomas Buchert).

\item[$\bullet$] J.~Colin Hill acknowledges support from NSF grant AST-2108536.  The Flatiron Institute is supported by the Simons Foundation.

\item[$\bullet$]Mustapha Ishak acknowledges that this material is based upon work supported in part by the Department of Energy, Office of Science, under Award Number DE-SC0022184.

\item[$\bullet$] Michael Klasen is supported by the BMBF under contract 05P21PMCAA and by the DFG through the Research Training Network 2149 "Strong and Weak Interactions - from Hadrons to Dark Matter".

\item[$\bullet$]  Suresh Kumar gratefully acknowledges support from the Science and Engineering Research Board (SERB), Govt. of India (File No.~CRG/2021/004658).

\item[$\bullet$] Ruth Lazkoz is supported by the Spanish Ministry of Science and Innovation through research projects FIS2017-85076-P (comprising FEDER funds), and also by the Basque Government and Generalitat Valenciana through research projects GIC17/116-IT956-16 and PROMETEO/2020/079 respectively.

\item[$\bullet$]Benjamin L'Huillier would like to acknowledge the support of the National Research Foundation of Korea (NRF-2019R1I1A1A01063740) and the support of the Korea Institute for Advanced Study (KIAS) grant funded by the government of Korea.

\item[$\bullet$]Jackson Levi Said would like to acknowledge support from Cosmology@MALTA which is supported by the University of Malta.

\item[$\bullet$] Roy Maartens is supported by the South African Radio Astronomy Observatory and the National Research Foundation (Grant No. 75415).

\item[$\bullet$]Valerio Marra thanks CNPq (Brazil) and FAPES (Brazil) for partial financial support. 
The work of Yuto Minami was supported in part by the Japan Society for the Promotion of Science (JSPS) KAKENHI, Grants No.~JP20K14497. 

\item[$\bullet$]The work of Carlos Martins was financed by FEDER---Fundo Europeu de Desenvolvimento Regional funds through the COMPETE 2020---Operational Programme for Competitiveness and Internationalisation (POCI), and by Portuguese funds through FCT - Funda\c c\~ao para a Ci\^encia e a Tecnologia in the framework of the project POCI-01-0145-FEDER-028987 and PTDC/FIS-AST/28987/2017.

\item[$\bullet$] Olga Mena is supported by the Spanish grants PID2020-113644GB-I00, PROMETEO/2019/083 and by the European ITN project HIDDeN (H2020-MSCA-ITN-2019//860881-HIDDeN).

\item[$\bullet$]Cristian Moreno-Pulido is funded by  PID2019-105614GB-C21 and FPA2016-76005-C2-1-P (MINECO, Spain), 2017-SGR-929 (Generalitat de Catalunya) and CEX2019-000918-M (ICCUB) and partially supported  by  the fellowship 2019 FI$_{-}$B 00351.

\item[$\bullet$] Michele Moresco acknowledges support from MIUR, PRIN 2017 (grant 20179ZF5KS) and grants ASI n.I/023/12/0 and ASI n.2018-23-HH.0.

\item[$\bullet$]Suvodip Mukherjee is supported by the Simons Foundation. Research at Perimeter Institute is supported in part by the Government of Canada through the Department of Innovation, Science and Economic Development and by the Province of Ontario through the Ministry of Colleges and Universities.

\item[$\bullet$] Pran Nath is supported in part by the NSF Grant PHY-1913328.

\item[$\bullet$]Savvas Nesseris acknowledges support from the Research Project No. PGC2018-094773-B-C32 and the Centro de Excelencia Severo Ochoa Program No.\ CEX2020-001007-S.

\item[$\bullet$] Rafael Nunes acknowledges support from the Funda\c{c}{\~a}o de Amparo \`a Pesquisa do Estado de S{\~a}o Paulo (FAPESP, S{\~a}o Paulo Research Foundation) under the project No.\ 2018/18036-5.

\item[$\bullet$]Eoin \'O Colg\'ain was supported by the National Research Foundation of Korea grant funded by the Korea government (MSIT) (NRF-2020R1A2C1102899). 

\item[$\bullet$]Supriya Pan acknowledges the financial supports from the Science and Engineering Research Board, Govt. of India, under Mathematical Research Impact-Centric Support Scheme (File No. MTR/2018/000940) and the Department of Science and Technology (DST), Govt. of India, under the Scheme "Fund for Improvement of S\&T Infrastructure (FIST)" [File No. SR/FST/MS-I/2019/41]. 

\item[$\bullet$]Santiago E. Perez Bergliaffa acknowledges partial support from Coordena\c c\~ao de Aperfei\c coamento de Pessoal de N\'ivel Superior (CAPES)- C\'odigo de Financiamento 001, and Universidade do Estado do Rio de Janeiro (Brazil).

\item[$\bullet$] Leandros Perivolaropoulos acknowledges support by the Hellenic Foundation for Research and Innovation (H.F.R.I.), under the "First call for H.F.R.I. Research Projects to support Faculty members and Researchers and the procurement of high-cost research equipment Grant" (Project Number: 789).

\item[$\bullet$]Fabrizio Renzi is supported by the NWO and the Dutch Ministry of Education, Culture and Science (OCW), and from the D-ITP consortium, a program of the NWO that is funded by the OCW.

\item[$\bullet$]Nils Sch\"oneberg acknowledges the support of the following Maria de Maetzu fellowship grant: Esta publicaci\'on es parte de la ayuda CEX2019-000918-M, financiado por MCIN/AEI/10.13039/501100011033.

\item[$\bullet$]Anjan A Sen acknowledges the funding from SERB, Govt of India under the research grants no: CRG/2020/004347 and MTR/20l9/000599.

\item[$\bullet$]Arman Shafieloo would like to acknowledge the support by National Research Foundation of Korea NRF-2021M3F7A1082053, and the support of the Korea Institute for Advanced Study (KIAS) grant funded by the government of Korea. 

\item[$\bullet$] M.M. Sheikh-Jabbari acknowledges the support by SarAmadan grant No. ISEF/M/400122.

\item[$\bullet$]Joan Sol\`a Peracaula is funded by  PID2019-105614GB-C21 and FPA2016-76005-C2-1-P (MINECO, Spain), 2017-SGR-929 (Generalitat de Catalunya), CEX2019-000918-M (ICCUB) and also partially supported by the COST Association Action CA18108  "Quantum Gravity Phenomenology in the Multimessenger Approach  (QG-MM)".

\item[$\bullet$]Denitsa Staicova is supported by Bulgarian NSF grant KP-06-N 38/11.

\item[$\bullet$]Glenn Starkman is partly supported by a Department of Energy grant DESC0009946 to the particle astrophysics theory group at CWRU.

\item[$\bullet$]Brooks Thomas is supported in part by the National Science Foundation under Grant PHY-2014104.

\item[$\bullet$]Luca Visinelli has received support from the European Union's Horizon 2020 research and innovation programme under the Marie Sk{\l}odowska-Curie grant agreement "TALeNT" No.~754496 (H2020-MSCA-COFUND-2016 FELLINI).

\item[$\bullet$]Shao-Jiang Wang is supported by the National Key Research and Development Program of China Grant No. 2021YFC2203004, No. 2021YFA0718304, the National Natural Science Foundation of China Grant No. 12105344, and the China Manned Space Project with NO.CMS-CSST-2021-B01.

\item[$\bullet$]Weiqiang Yang has been supported by the National Natural Science Foundation of China under Grants No. 12175096 and No. 11705079, and Liaoning Revitalization Talents Program under Grant no. XLYC1907098. 

\item[$\bullet$]Gong-Bo Zhao is supported by the National Key Basic Research and Development Program of China (No. 2018YFA0404503), NSFC Grants 11925303, 11720101004, and a grant of CAS Interdisciplinary Innovation Team.

\end{enumerate}


\bibliographystyle{utphys} 
\bibliography{main}



\end{document}